\documentclass[%
 reprint,
unsortedaddress,
 amsmath,amssymb,
 aps,
]{revtex4-1}

\usepackage{graphicx}
\usepackage{dcolumn}
\usepackage{bm}

\usepackage{soul}
\usepackage{url}
\usepackage{hyperref}

\usepackage{multirow} 
\usepackage{siunitx} 
\usepackage{float} 

\begin{document}

\preprint{NIMA }

\title{Ultrafast Radiographic Imaging and Tracking: \\ An overview of instruments, methods, data, and applications}


\author{Zhehui Wang}
\email{Correspondence: zwang@lanl.gov (ZW)}
\affiliation{Los Alamos National Laboratory, Los Alamos, NM 87545, USA}%

\author{Andrew F. T. Leong}
\affiliation{Los Alamos National Laboratory, Los Alamos, NM 87545, USA}%

\author{Angelo Dragone}
\affiliation{SLAC National Accelerator Laboratory,  Menlo Park,  CA  94025, USA}%

\author{Arianna E. Gleason}
\affiliation{SLAC National Accelerator Laboratory,  Menlo Park,  CA  94025, USA}%

\author{Rafael Ballabriga}
\affiliation{CERN, 1211 Geneva 23, Switzerland}

\author{Christopher Campbell}
\affiliation{Los Alamos National Laboratory, Los Alamos, NM 87545, USA}%

\author{Michael Campbell} 
\affiliation{CERN, 1211 Geneva 23, Switzerland}

\author{Samuel J. Clark}
\affiliation{X-ray Science Division, Advanced Photon Source, Argonne National Laboratory, Lemont, IL 60439, USA}%

\author{Cinzia Da Vi\`a}
\affiliation{The University of Manchester, Manchester, M13 9PL, UK}%

\author{Dana M. Dattelbaum}
\affiliation{Los Alamos National Laboratory, Los Alamos, NM 87545, USA}%

\author{Marcel Demarteau}
\affiliation{Oak Ridge National Laboratory, Oak Ridge, TN 37831, USA}%

\author{Lorenzo Fabris}%
\affiliation{Oak Ridge National Laboratory, Oak Ridge, TN 37831, USA}%

\author{Kamel Fezzaa}%
\affiliation{X-ray Science Division, Advanced Photon Source, Argonne National Laboratory, Lemont, IL 60439, USA}%

\author{Eric R. Fossum}
\affiliation{Thayer school of engineering at Dartmouth, Dartmouth college,
Hanover, NH 03755, USA}

\author{Sol M. Gruner}%
\affiliation{Department of Physics, Cornell University, Ithaca, NY 14853, USA}%

\author{Todd Hufnagel}
\affiliation{Department of Materials Science and Engineering, Johns Hopkins University, Baltimore, MD 21218, USA}%

\author{Xiaolu Ju}
\altaffiliation[Also at ]{\it Shanghai Institute of Applied Physics, Chinese Academy of Sciences, Shanghai 201800, China; {\rm and}\\
University of Chinese Academy of Sciences, Beijing 100049, China}

\affiliation{Shanghai Synchrotron Radiation Facility, Shanghai Advanced Research Institute, \\
Chinese Academy of Sciences, Shanghai 201204, China}

\author{Ke Li}
\affiliation{Shanghai Synchrotron Radiation Facility, Shanghai Advanced Research Institute, \\
Chinese Academy of Sciences, Shanghai 201204, China}

\author{Xavier Llopart}
\affiliation{CERN, 1211 Geneva 23, Switzerland}

\author{Bratislav Luki\'c}
\affiliation{ESRF -- The European Synchrotron, 71 avenue des Martyrs - \\
CS 40220, 38043 Grenoble Cedex 9, France}

\author{Alexander Rack}
\affiliation{ESRF -- The European Synchrotron, 71 avenue des Martyrs - \\
CS 40220, 38043 Grenoble Cedex 9, France}

\author{Joseph Strehlow}
\affiliation{Los Alamos National Laboratory, Los Alamos, NM 87545, USA}%

\author{Audrey C. Therrien}
\affiliation{Interdisciplinary Institute for Technological Innovation, Université de Sherbrooke, 3000 boulevard de l'Université, Sherbrooke, J1K 0A5, Québec, Canada}

\author{Julia Thom-Levy}%
\affiliation{Department of Physics, Cornell University, Ithaca, NY 14853, USA}%

\author{Feixiang Wang}
\affiliation{Shanghai Synchrotron Radiation Facility, Shanghai Advanced Research Institute, \\
Chinese Academy of Sciences, Shanghai 201204, China}

\author{Tiqiao Xiao}
\altaffiliation[Also at ]{\it Shanghai Institute of Applied Physics, Chinese Academy of Sciences, Shanghai 201800, China; {\rm and}\\
University of Chinese Academy of Sciences, Beijing 100049, China}

\affiliation{Shanghai Synchrotron Radiation Facility, Shanghai Advanced Research Institute, \\
Chinese Academy of Sciences, Shanghai 201204, China}

\author{Mingwei Xu }
\altaffiliation[Also at ]{\it Shanghai Institute of Applied Physics, Chinese Academy of Sciences, Shanghai 201800, China; {\rm and}\\
University of Chinese Academy of Sciences, Beijing 100049, China}

\affiliation{Shanghai Synchrotron Radiation Facility, Shanghai Advanced Research Institute, \\
Chinese Academy of Sciences, Shanghai 201204, China}

\author{Xin Yue}
\affiliation{Thayer school of engineering at Dartmouth, Dartmouth college,
Hanover, NH 03755, USA}

\date{\today}

\begin{abstract}

Ultrafast radiographic imaging and tracking (U-RadIT) use state-of-the-art ionizing particle and light sources to experimentally study {\it sub-nanosecond} transients or dynamic processes in physics, chemistry, biology, geology, materials science and other fields. These processes are fundamental to modern technologies and applications, such as nuclear fusion energy, advanced manufacturing, communication, and green transportation, which often involve one mole or more atoms and elementary particles, and thus are challenging to compute by using the first principles of quantum physics or other forward models. 
One of the central problems in U-RadIT is to optimize information yield through, {\it e.g.} high-luminosity X-ray and particle sources, efficient imaging and tracking detectors, novel methods to collect data, and large-bandwidth online and offline data processing, regulated by the  underlying physics, statistics, and computing power. We review and highlight recent progress in: a.) Detectors such as high-speed complementary metal-oxide semiconductor (CMOS) cameras, hybrid pixelated array detectors integrated with Timepix4 and other application-specific integrated circuits (ASICs), and digital photon detectors; b.) U-RadIT modalities such as dynamic phase contrast imaging, dynamic diffractive imaging, and four-dimensional (4D) particle tracking; c.) U-RadIT data and algorithms such as neural networks and machine learning, and d.) Applications in ultrafast dynamic material science using XFELs, synchrotrons and laser-driven sources.  Hardware-centric approaches to U-RadIT optimization are constrained by detector material properties, low signal-to-noise ratio, high cost and long development cycles of critical hardware components such as ASICs.  Interpretation of experimental data, including comparisons with forward models, is frequently hindered by sparse measurements, model  and measurement uncertainties, and noise.  Alternatively, U-RadIT make increasing use of data science and machine learning algorithms, including experimental implementations of compressed sensing. Machine learning and artificial intelligence approaches, refined by physics and materials information, may also contribute significantly to data interpretation, uncertainty quantification and U-RadIT optimization. 

\end{abstract}

\pacs{Valid PACS appear here}
\maketitle

\begin{widetext}
\tableofcontents
\end{widetext}


\section{Introduction \label{sec:1}} 

Ultrafast Radiographic Imaging and Tracking (U-RadIT) use sub-nanosecond (sub-ns) pulses of ionizing radiation such as X-rays or energetic particles with mass (protons, electrons, neutrons, {\it etc.}) for high-speed imaging and tomography (IT). U-RadIT, as the ultrafast version of RadIT~\cite{Wan:2022},  complement ultrafast IT by using visible light as in traditional ultrafast photography~\cite{EK:1979,LW:2018, YCQ:2020}, magnetic fields as in magnetic resonance imaging~\cite{tsa:2010}, and ultrasound~\cite{VBF:2020}. As the peak intensities of the short-pulse lasers at visible and longer wavelengths continue to increase towards the critical intensity or the Schwinger limit of 4$\times$ 10$^{29}$ W/cm$^2$, with a recent record exceeding $10^{23}$ W/cm$^2$~\cite{YKC:2021}, visible and longer wavelength high power lasers can also become U-RadIT tools through multi-photon ionization and secondary X-ray, neutron and energetic charged particle production. Ultrafast imaging collects two-dimensional (2D) information at high speed, and ultrafast tomography gathers three-dimensional (3D) data very promptly. 

Due to the penetrating power of X-rays and energetic particles with mass, and some unique interaction physics such as large inner-shell electron cross section, U-RadIT can interrogate ultrafast phenomena or ultrafast evolution of a physical quantity of almost any materials, at both macroscopic (meter size and larger) and microscopic (down to individual atoms) length scales, and reveal transient dynamic details down to the individual molecular, atomic and even sub-atomic events such as electron transition from one quantum state to another at the same time. Such individual transient events are usually described by the laws of quantum mechanics, and happen very fast ($< 1$ ps). One class of such universal ultrafast processes is femto-chemistry, or molecular scale atom and electron motion, pioneered by A. H. Zewail and collaborators~\cite{Zew:2000}, but there are many others, which may require U-RadIT and other ultrafast imaging and measurement methods. A recent review on electron microscopy with applications to biology and nanoscale systems may be found in~\cite{HP:2020}. 2018 roadmap of ultrafast X-ray atomic and molecular physics was given in~\cite{YUG:2018}.

The duration of an event ($\tau$) may be estimated classically by $\tau \sim l/v$, where $l$ is the characteristic length and $v$ is the characteristic speed of the transient event. Since $v \leq c$ according to the theory of relativity, with $c$ being the speed of light, ultrafast processes naturally occur on small length scales at highest speeds up to the speed of light. IT of microscopic processes involving photons and electrons over atom scales requires attosecond (1 as = 10$^{-18}$ seconds) time resolution~\cite{KI:2009,ORN:2018}. Attosecond photography and attosecond RadIT are currently limited by the availability of a bright strobe illumination source of visible light, X-rays and other ionizing radiation. In a laboratory setting, since the speeds of electrons, atoms, molecules, nanoparticles  and larger objects are usually below 100 km/s (compression of mm-size targets by high power lasers in inertial confinement fusion can reach hundreds of km/s), studying transients over a molecule ($\sim$ 1 nm) and longer lengths is sufficient by using `femtosecond photography'~\cite{Zew:2000} or femtosecond RadIT. X-ray free electron lasers (XFEL) now can deliver an intense pulse of X-rays to make femtosecond RadIT practical. When making movies of mass compression dynamics as in  inertial confinement fusion (ICF), with $l \sim$ 1 mm, $v \sim$ 300 km/s, `picosecond photography' or picosecond RadIT is sufficient. Electron and proton accelerators, XFELs, synchrotron light sources, together with laser-produced plasmas offer many options for picosecond RadIT. 

High speed processes on the mesoscale ($<$ 10 micrometers) are not always required to be ultrafast. The emerging field of `macroscopic quantum systems and phenomena'  can potentially qualify and these quantum phenomena may also use U-RadIT methods. An object, large or small, traveling slowly but over an ultrashort distance, may also be treated as `ultrafast'. For example, the Laser Interferometer Gravitational-Wave Observatory (LIGO) can probe astrophysics and gravitational waves with attometer precision. The time for an object such as a LIGO mirror to traverse one-attometer distance, $l \sim$ 10$^{-18}$ m,  can be very short and to make a movie of such an extremely boring (from classical-physics point of view) process may also need ultrafast photography or U-RadIT. From quantum-physics point of view, 10$^{-18}$ m may allow us to see quantum vacuum fluctuations such as the Casimir effect. In other words, imaging quantum phenomena such as quantum fluctuations is a potentially new frontier for ultrafast photography and U-RadIT.

U-RadIT come in several different flavors. First is ultrafast detection of individual X-rays, protons, other ionizing radiation such as fast electrons, and the ionizing-radiation-induced secondary particles including neutrons (when nuclear reactions are induced by the primary beam of photons or particles with mass), electron-hole pairs in semiconductors or visible light in scintillators~\cite{Wan:2023}. Second is high-speed imaging or high-speed tomography of ultrafast processes such as photosynthesis on the molecular level, many other phenomena in femto-chemistry~\cite{Zew:2000}, or dynamic objects such as ultrafastly compressed millimeter and smaller targets in ICF. Third is a time-resolved high-resolution 2D or 3D measurement of physical quantities such as density, velocity, temperature, pressure, or their correlations such as the equation of state (EOS). 

Here we give an overview of recent advances in U-RadIT with an emphasis on sub-ns time-resolved RadIT. 
In Sec.~\ref{sec:OP1}, we discuss the physics and computational foundations of U-RadIT, U-RadIT hardware metrics, and frame U-RadIT as an information-yield optimization problem. In Sec.~\ref{sec:inst}, we summarize the U-RadIT instrument advances in terms of  `10H' frontiers, and highlight development in Timepix ASICs, hybrid pixelated array detectors (PADs), and ultrafast CMOS cameras, 3D digital-to-photon converters, and possible options beyond CMOS technology. Sec.~\ref{sec:mm} on U-RadIT modalities includes phase contrast, diffraction and 4D tracking methods, as well as approaches to improve image contrast through, {\it e.g.} motion contrast imaging. In Sec.~\ref{sec:da1}, we recognize that, while there are plentiful of data available, experimental data acquisition are usually sparse. Compressed sensing can be used for offline data processing as well as real-time data acquisition. Neural networks are used for growing number of data workflows, including phase retrieval and uncertainty quantification. In Sec.~\ref{sec:ap} on applications, we highlight high-repetition-rate high-power experiments, dynamic experiments using LCLS XFEL, ESRF and APS synchrotrons. 
\section{ Open problems in U-RadIT optimization \label{sec:OP1}}
Modern technologies and applications usually require a certain minimal amount of materials or mass, one mole or more of atoms being typical, to work. An important application of U-RadIT is to understand and predict {\it time-dependent} or {\it dynamic} properties of such macroscopic (millimeters and larger) bulk materials, through measurements of their mass, electric charge, and energy flows at the electronic, atomic and molecular levels (nanometers and smaller). Elementary processes of electron, atom and molecular motion dictate the temporal resolution at sub-ns. The material thickness or areal density (thickness integrated mass density) dictates the use of ionizing radiation to penetrate inside the materials and reveal their internal structures. In addition to natural materials aging, materials under extreme pressure, temperature, ionizing radiation, high energy density requires U-RadIT measurements in-situ and in real time. The experimental data and information collected from U-RadIT can then be used by, for example, coupling to molecular dynamics (MD) simulations, to guide new materials discovery and design including synthesis on a large scale beyond research laboratories. 

\begin{figure}[thbp] 
   \centering
   \includegraphics[width=0.45\textwidth]{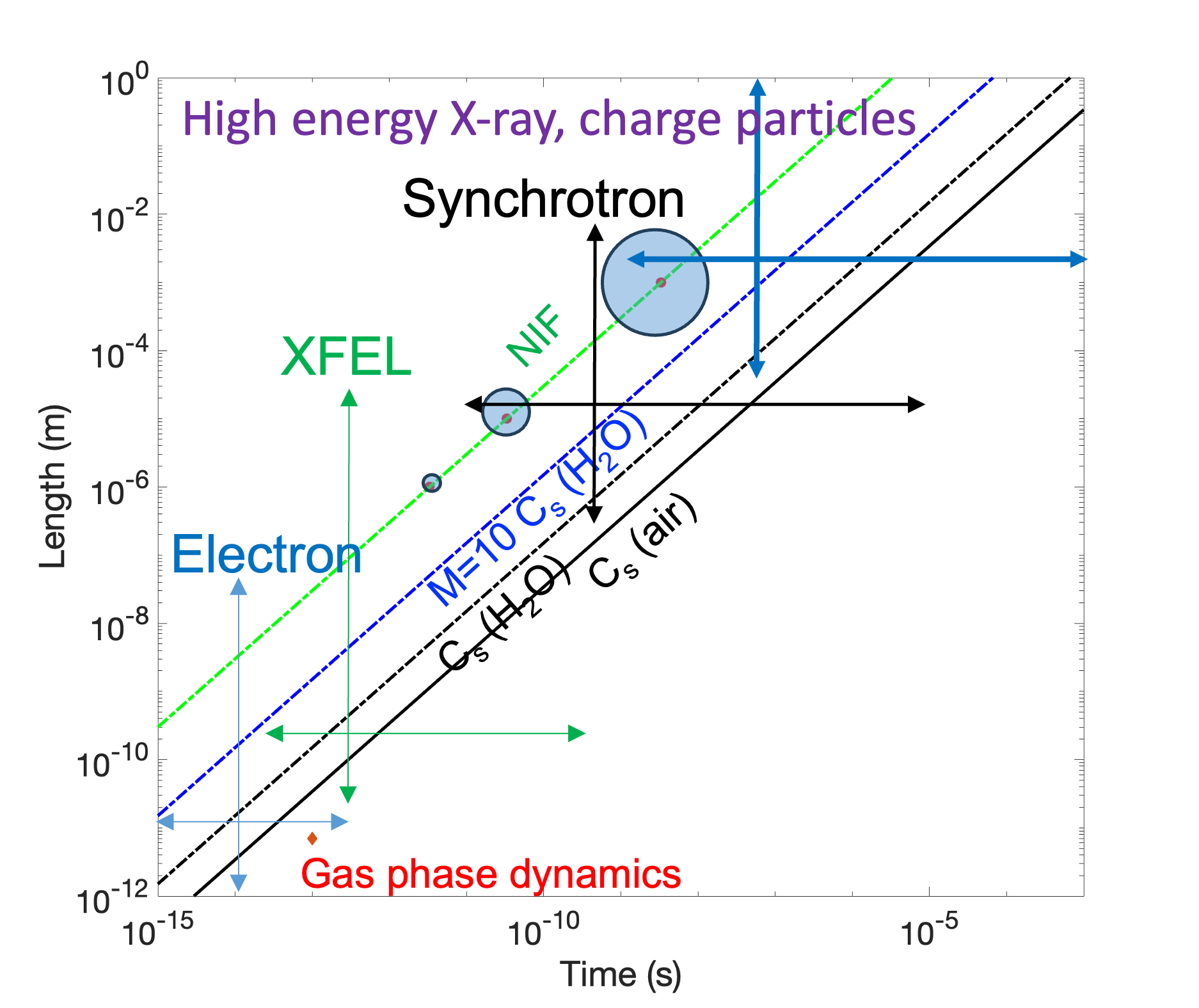} 
   \caption{A comparison of various U-RadIT modalities, electrons, X-rays at different energies,  and charged particles such as protons, and their applicable temporal and spatial scales. The four sloped lines correspond to, respectively, the sound speed in air ($C_s$(air)), the sound speed in water ($C_s$(H$_2$O)), 10 times the $C_s$(H$_2$O), and the hypervelocity at 300 km/s as in the National Ignition Facility (NIF) experiments. }
   \label{fig2:TS1}
\end{figure}

U-RadIT modalities now include electrons, X-rays of different energies (XFELs, synchrotrons, and high-energy X-rays), energetic charged particles such as protons, as summarized in Fig.~\ref{fig2:TS1}. Neutrons, limited by the available flux, may complement other U-RadIT modalities. Ultrafast electron methods have been very successful in femtochemistry due to in part the strong interactions of electrons and the compact source size~\cite{Zew:2000}. X-rays of different energies are now routinely used for ultrafast imaging due to the growing number of synchrotron and XFEL facilities. Heavy charged particles such as protons, together with high-energy X-rays above 100 keV, are usually used to examine larger objects than by electrons and photons from XFELs and synchrotrons, with reduced spatial and temporal resolution.

\subsection{Physics principles and computation}
The fundamental physics principles of U-RadIT, which describe the probabilistic interactions between atoms and ionizing radiation such as X-rays, $\gamma$-rays, electrons, protons, other charged particles, and neutrons, are now complete in the laws of quantum physics, {\it i.e.} many-body Schr\"odinger's equation in the non-relativistic regime or equivalent formulations~\cite{SBB:2002}. However, knowing such principles and how to use them for calculations, which are the basis of forward models, are not enough to make quantitative and accurate predictions in chemistry, materials science, fusion energy, nor to interpret U-RadIT measurements. The difficulty was recognized as early as 1929 by Paul Dirac~\cite{Dirac:1929}. In modern terms, the challenges are known as the `curse of large dimensionality' or the `curse of large numbers' that easily overwhelms the memories and processing capacities of the state-of-the-art computers.

More practical approaches to forward modeling can be broken down to a hierarchy of temporal and spatial scales, parallel to the hierarchy of measurements shown in Fig.~\ref{fig2:TS1}. Quantum chemistry calculations involve individual electrons and nuclei~\cite{Sim:2023}. The most sophisticated models based on a many-body Schr\"odinger's equation can simulate a few hundred atoms for a duration of sub-ns~\cite{OKCW:2016}. Use of quantum computers for quantum chemistry is emerging~\cite{Rubin:2020} and the problem complexity (10s of atoms) is still lagging behind classical computers. 

The next level in simulation hierarchy (sub-ns to ms, nm to \textmu m) is molecular dynamics (MD)~\cite{HOG:2002,Rap:2004} or molecular mechanics, which bridges the quantum regime with the classical regime (dynamics is described by Newton's laws of motion), and includes many applications to materials science, biology, fusion energy~\cite{GAB:2021}. In highly ionized plasmas, the fourth state of matter, the equivalents to MD simulations are simulations using kinetic equations. Significant improvements in simulation speed, accuracy through availability of computing power, and novel algorithms accelerated by neural networks~\cite{BP:2007}, together with experimental data from synchrotrons, XFELs, electron microscopy and proton facilities, have led to growingly adoption of MD beyond the materials science community such as molecular biology~\cite{DM:2011}, drug discovery~\cite{HD:2018}, and nuclear fusion energy~\cite{LRD:2011, SWT:2016}. 

The last level in simulation belongs to the continuum or `macroscopic regime' (sub-ns and above in time, sub-\textmu m and above in length). Transport coefficients and other emergent material properties such as thermal conductivity~\cite{HCG:2016}, electric conductivity, opacity, and equation of state~\cite{HMG:2011} are used to predict material evolutions on the longer temporal and spatial scales using continuum or fluid approximation. Coupling to the MD and kinetic simulations, through the transport coefficients and different moments of non-isotropic material properties, is an important feature in simulations at this level. A recent trend is that, as the continuum simulation models continue to improve in temporal and spatial resolution, their overlaps with MD and kinetic models also grow with time. It is likely the boundary between the macroscopic and mesoscale models such as MD may disappear, depending on the available computing resource. Another feature is that these simulations allow direct comparisons with measurements including U-RadIT. In inertial fusion experiments, for example, xRAGE~\cite{HKL:2022} and HYDRA~\cite{MKG:2001} are often used to generate synthetic X-ray and neutron data including images for comparison with measurements.

For U-RadIT experimental data interpretation, the simulations described above need to couple with radiation transport and radiation interactions with matter. The detector models are also needed to account for detector responses such as noise, the point spread function and other effects in the experimental data. There are now a growing number of multi-physics codes available for detector modeling, {\it e.g.} Allpix squared multi-physics simulation framework~\cite{SS:2022}, which has been used to generate synthetic data for a silicon detector~\cite{YL:2023}. One simplification to detector modeling is that the material composition and structures, such as charge collection, defects, and storage capacitors, may be assumed to be constant and do not change with time. However, as the ionizing radiation sources such as X-rays and charged particles continue to become brighter and operate at higher repetition rate, time-dependent detector responses may need to be accounted for data interpretation.


\subsection{U-RadIT hardware metrics \label{sec:met} }

\begin{figure*}[th!] 
   \centering
   \includegraphics[width=0.65\textwidth]{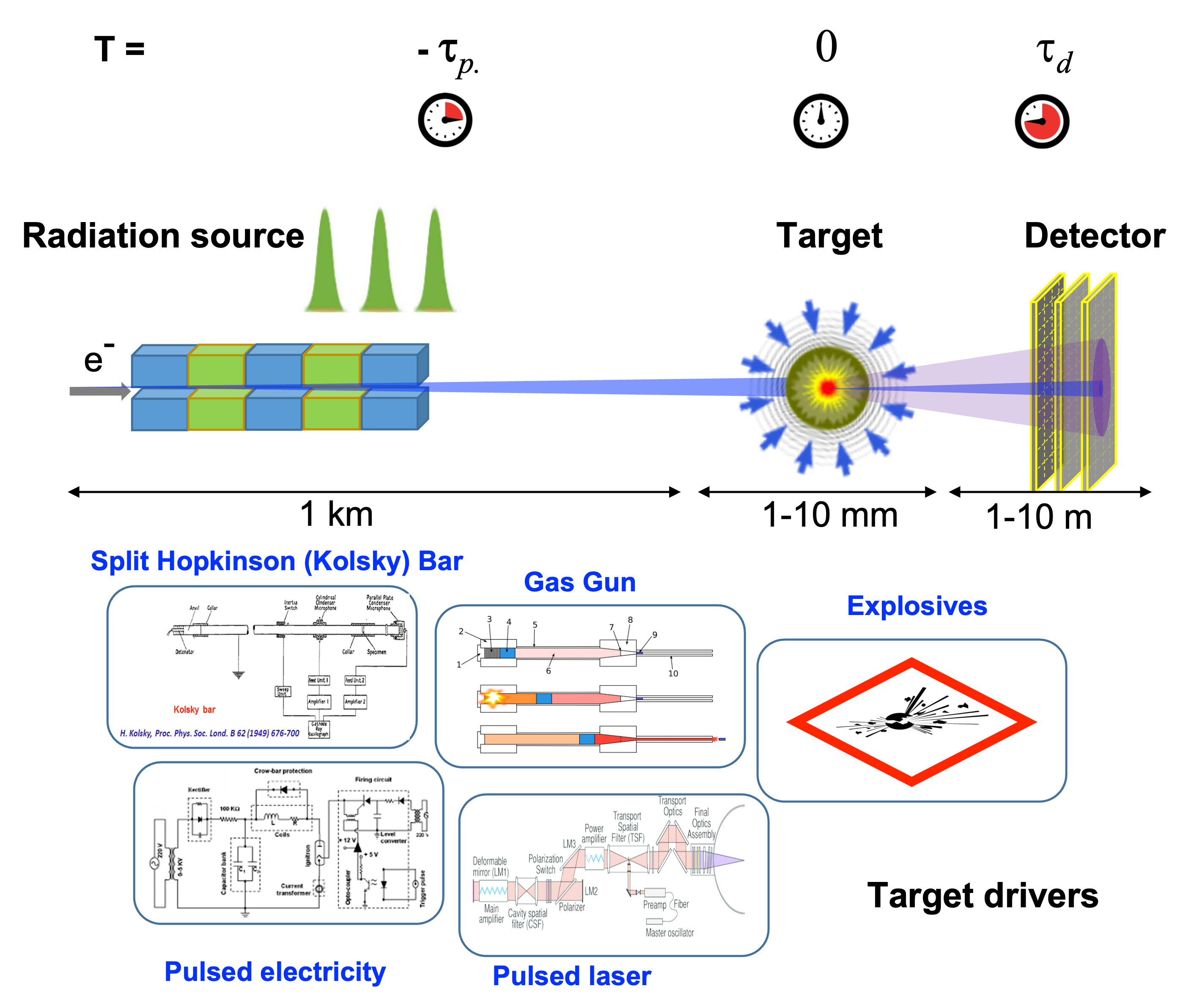} 
   \caption{A simplified schematic U-RadIT setup consisting of a radiation source, a target that interacts with a driver (energy source), and the detector. Mechanical, kinetic, chemical, electric, and electromagnetic (including lasers) energy are possible sources of driver energy.}
   \label{fig1:setup}
\end{figure*}

Both traditional forward models as described above and recent machine learning (ML) models driven by data will continue to rely on U-RadIT and other experimental methods for model validation and verification, in particular, as the complexity of the experiments grows in terms of temporal and spatial durations as well as in temporal and spatial resolutions, or spatial and temporal dynamic ranges.

A simplified lensless U-RadIT setup is shown in Fig.~\ref{fig1:setup}. The largest footprint of the overall system typically comes from the radiation source that generates prompt X-ray or energetic particles for illumination.  For example, the Advanced Photon Source (APS) and its upgrade (APS-U) have an electron storage ring circumference 1.1 km long. Some other synchrotron facilities listed in Table.~\ref{tab:table3} range from 0.518 km (TPS storage ring in Taiwan) to 2.3 km (PETRA-III/PETRA-IV in Germany). The LANSCE proton linear accelerator at Los Alamos is about 800 m long, which can be further augmented by a proton storage ring, delivers 800 MeV proton bunches or micropulses at a rate up to 120 pulses per second. The minimum micropulse separation is about 4.96 ns. A typical proton micropulse has a few times 10$^8$ protons.

\begin{table*}[!htb]
\caption{\label{tab:table3}%
A summary of different radiation sources for U-RadIT. In comparison,  a CPA Ti:sapphire laser (800 nm) can deliver a short pulse in the range of 5-20 fs, 1-10 mJ pulse energy at a repetition rate in the range of 1-2$\times$10$^5$ Hz~\cite{MGC:2020}.}
\begin{ruledtabular}
\begin{tabular}{lccccc}
\textrm{Source} &
\textrm{Particle (GeV)} &
\textrm{Particles}&
\textrm{Energy}&
\textrm{Pulse} &
\textrm{Emittance }
\\
 & \textrm{[Photon (keV)]} &
\textrm{[Photons]}&
\textrm{(mJ)}&
\textrm{width} & x/(y)
\\
 & Energy & per pulse & per pulse  & \textrm{(ps)}& (pm$\cdot$rad) \\ 
\colrule
APS~\cite{DBB:2022} & 7 GeV &9.6$\times$10$^{10}$ &108 J & 34 & 3110/(40.5) \\
& \textrm{[3-100 keV]} & [$<$ 10$^{20}$]~\footnote{brillance, in ph/(s$\cdot$mm$^2$$\cdot$mrad$^2\cdot$0.1\%BW) \label{fnote1}}&&41& \\
APS-U~\cite{DBB:2022} & 6 GeV & 9.6$\times$10$^{10}$ &92.2 J & 104 & 42/(4.2) \\
& \textrm{[1-120 keV]} & [$<$ 10$^{23}$]~\footref{fnote1}&&100& \\
CHESS~\footnote{positron}& 6 GeV & 3.5-36$\times$10$^{10}$ && 46.7& 29210/292  \\
& [10-100 keV] & [$<$10$^{20}$]~\footref{fnote1} &&& \\
DIAMOND & 3 GeV & 3.7$\times$10$^9$ && 17.1 & 3140/8 \\
ESRF~\cite{RBB:2023} & 6 GeV & 4.4$\times$10$^{10}$ && $<$100 & 3985/4\\
ESRF-EBS~\cite{LCC:2016,RBB:2023} & 6 GeV & 4.4$\times$10$^{10}$ && 60 & 133/1\\
& \textrm{[10-50 keV]} & [6.6 - 1.37 $\times$10$^{21}$]~\footref{fnote1}&&20-55& \\
Eu-XFEL~\cite{BBG:2012} & 8.5-17.5 GeV & 0.1-6.9$\times$10$^9$& & 1.2-76.6 fs& 0.77779/0.72763~\footnote{in mm $\cdot$ mrad}\\
& [12.4 keV] & [10$^{12}$]~\footnote{equivalent brilliance = 10$^{33}$ ph/(s$\cdot$mm$^2$$\cdot$mrad$^2\cdot$0.1\%BW) } & 4 & $\leq$0.1 & \\
LANSCE~\footnote{proton \label{fnote2}}~\cite{WBD:2022} & 0.8 GeV & 3.1$\times$10$^{8}$~\footnote{10 mA equivalent current. Typically 16 bunches, spread over 80 ns are used in proton radiography} & & $<$100 & \\
LCLS & [4.5-11 keV] & [0.15-14$\times$10$^{12}$] &0.6-2.0& 10-50 fs & \\
LCLS & [0.4-1.2 keV] & [3.1-47$\times$10$^{12}$] &1.5-2.5& 10-250 fs & \\
MaRIE~\cite{CAB:2019} & [42 keV] & [5$\times$10$^{10}$] && $<$1 & \\
MAX-IV& 3 GeV & 2.2$\times$10$^{10}$ &&29 & 330/2-8 \\
NIF~\footnote{neutron}& 14.1 MeV & 6$\times$10$^{17}$~\footnote{The NIF neutron yield record as of Aug. 2023 is higher.} &1.35 MJ & $<$100 & 100-120 \textmu m $\cdot$ (4 $\pi$)~\footnote{NIF compressed target around 100 to 120 \textmu m diameter at the peak compression. Neutron emission into 4 $\pi$ solid angle.} \\
NIF ARC & [1 MeV] & [$>10^{10}$]~\footnote{ph/cm$^2$} && 1-50 & (4 $\pi$) \\
NSLS-II& 3 GeV & 6$\times$10$^{10}$ && 15 - 30 & 550/ \\
& [0.1-23 keV] & $<$10$^{21}$~\footref{fnote1}  & & & \\
PETRA-III  & 6 GeV & 1.2$\times$ 10$^{11}$~\footnote{40 bunch mode} && 44 & 1200/12 \\
& [0.15-200 keV] & [$>$ 10$^{21}$]~\footref{fnote1} &&  & \\
PETRA-IV~\cite{SAB:2018} & 6 GeV &  && 65 (75) & 10-30/($<$10) \\
SACLA & [10 keV] & [3 $\times$ 10$^{11}$]& 0.5 & $<$10 fs & \\
SHINE~\cite{HDL:2021} & 8 GeV & &1.8 & $<$50 fs & \\
& [0.4-25 keV] & & & & \\
SNS~\footref{fnote2}& 1 GeV & 1.5 $\times$ 10$^{14}$& 24 kJ & 695 ns & \\
Spring-8 & 8 GeV & 5$\times$10$^{10}$ && $\sim$60 & 3400/6.8\\
& [0.3-300 keV] & $[ <$10$^{20}$] ~\footref{fnote1} & & & \\
SSRF & 3-5 GeV & 4$\times$10$^{10}$ && 11  & 3900/\\
& [0.04-200 keV]& 10$^9$  & & 80 & \\
TPS & 3 GeV & && 9.5 & 1500/15 \\
& [0.1-30 keV] & [$<$10$^{21}$]~\footref{fnote1}  & & & \\
\end{tabular}
\end{ruledtabular}
\end{table*}

Besides the radiation sources (more than one radiation source are used in, for example, multi-modal U-RadIT), the other critical hardware component of a U-RadIT system is detectors and especially pixelated image sensors.
Photography films as an analog image sensors~\cite{Ray:1997} have now been mostly replaced by digital sensors since the invention of charge coupled device (CCD) at the Bell Labs in 1969. Imaging plate and speciality plastics such as CR-39 are still used because of their simplicity and robustness against transient electromagnetic pulse (EMP) associated with radiation source operation. The introduction of active pixel sensors in the 1990s~\cite{Dart:1} ushered in the era of CMOS image sensors, which now dominate over CCD image sensors in commercial applications such as in cell phones, drones, automobiles and other smart or autonomous systems. Pixel detectors were developed for High Energy Physics~\cite{DVB:1997} and in the Medipix Collaborations~\cite{Camp:2011}. Hybridized direct x-ray imagers, in which a pixelated x-ray absorbing layer is electrically bonded pixel-by-pixel to a pixelated CMOS ASIC, were first used for synchrotron science applications at the turn of the Millenium. Graafsma~\cite{Gra:2018} provides a good history of the early hybrid detectors used for synchrotron science. Pioneering work on burst-rate hybrid imagers was done by the Cornell Detector Group at CHESS using a microsecond rate burst-mode hybrid imager consisting of a Si sensor bonded to an ASIC~\cite{RRE:1999}. Burst-mode imagers operate by storing a limited number of successive images in the ASIC pixel in analog form for later digitization and readout. This avoids the temporal bottleneck involved in analog-to-digital conversion and readout of the images into computer memory. Cornell Keck-PAD was such a burst-mode imager that stored 8 frames at a frame rate of 10 MHz~\cite{G1}. This was followed by CS-PAD, an X-ray hybrid CMOS image sensor for LCLS~\cite{PKH:2007, HBC:2012}. Many other hybrid CMOS cameras such as AGIPD~\cite{BBG:2012}, Jungfrau~\cite{LMB:2020}, MM-PAD~\cite{G2} have since been introduced and some of them such as Pilatus, Eiger are commercialized~\cite{HG:2015}. There are now a growing number of image sensors and pixelated detectors to choose from, see Sec.~\ref{sec:inst} for further discussions. We summarize the metrics, features, and common terminology of imaging detectors first.

{\it Direct and Indirect sensors}~\hspace{0.2 cm} Most commercial image sensors are for visible light, with relatively fewer for other photon wavelengths, such as infrared, UV, X-rays and $\gamma$-rays. Hybrid CMOS devices such as the CS-PAD and AGIPD are known as direct sensors since they convert X-rays directly into electron-hole pairs that are collected and stored as signals. On the other hand, indirect sensors use a 2-step process where scintillators convert ionizing radiation into visible light, which is then detected by visible light cameras. These setups offer more flexibility and radiation hardness, but may offer lower energy or spatial resolution.

{\it Frame rate and record length}~\hspace{0.2 cm} The frame rate measures how frequently multiple images can be taken in sequential imaging or ``movie mode". Record length measures how many frames of images can be taken and recorded, limited by temporally storage memory (commonly used in burst mode imaging) or data transmission bandwidth (in continuous mode imaging). In U-RadIT, the frame rate of an image sensor is dictated by the repetition rate of radiation source, , as well as by the detector. Synchrotrons and XFELs can now or will soon (as in APS-U) deliver bright sub-ns pulses at a rate above 10 MHz, which exceeds the highest frame rates of existing cameras, {\it e.g.} Shimadzu HPV-X2. Burst-mode imaging may also be achieved by time-multiplexing several visible light CCD or CMOS cameras to the light emitted by a scintillator screen via use of beam-splitters, perhaps with intermediate image intensification. The multiplexing frame rate may be limited by the scintillator decay time and brightness, as well as the response time of the image intensifiers. When images are sparse, an alternative approach of data-driven hit streaming can be used, see Sec.~\ref{sec:tmpx}.

{\it Pixel resolution and number of pixels}~\hspace{0.2 cm} Pixel resolution or pitch is the size of the smallest sensing unit (usually a square or rectangle shape, although hexagonal shapes have also been used) from which the signals are collected and digitized. Pixel resolution determines the spatial resolution ($\delta$) of the image sensor. Direct image sensors typically have 10s to 100s of micrometers pitch, and indirect image sensors have less than 10 micrometers pitch. The spatial resolution is not always determined by pixel size but rather by the charge deposition process within the active sensor material, {\it e.g.} fluorescence in high-Z direct detectors or light propagation and spreading in scintillators. If multiple, slightly displaced images are used, and the point spread function is smaller than the pixel pitch, one can actually recover images with higher resolution than the pixel pitch. Individual direct image sensors may have less than 1 million pixels (mega-pixels). Multiple such units, through butting or tiling, may be arranged to form larger areas, often with minimal inter-module gaps. Indirect image sensors commonly have more than 10 mega-pixels. 100 mega-pixel sensors are also now available commercially. The small pixel pitches that are involved allow 10s of millions or more pixels to be fabricated on a single silicon die. 

{\it Quantum efficiency and sensitivity}~\hspace{0.2 cm} Quantum efficiency (QE) measures the fraction of X-rays and other ionizing radiation that impinge on a detector are detected. Sensitivity measures whether individual quanta can be distinguished from background and noise. In direct detection, quantum efficiency is often energy and particle dependent. 500-\textmu m thick silicon sensors are sufficient to stop 10 keV or less X-rays completely and achieve close to 100\% quantum efficiency. The sensitivity to X-rays in silicon is also very high. The energy to create one electron-hole pair in silicon (bandgap 1.12 eV) is only about 3.63 eV at room temperature~\cite{Lutz:2007,Spi:2005}, corresponding to about 275 $\pm$ 6 charge pairs per keV.  The Fano factor of 0.11 was applied to estimate the charge fluctuation of 6~\cite{Spi:2005,LS:2007,Mazz:2008}. Charge sharing among neighboring pixels and electronic noise can lower the sensitivity. QE and sensitivity of indirect detection depend on scintillator light yield and photodetectors, with more details given in, {\it e.g.}~\cite{Knoll2000}.

{\it Gain and noise}~\hspace{0.2 cm}  For improved sensitivity for individual quantum detection, charge amplifiers are used within pixels to multiply the raw charge collected. Direct image sensor such as AGIPD 1.0 has a noise around 300 e$^-$~\cite{BBG:2012}. A balance between the amount of gain and the dynamic range is sometimes needed. The state-of-the-art CMOS image sensors now have a noise per pixel below 1 $e^-$ at room temperature, such sensors are often used for indirect imaging in RadIT applications.

{\it Data bit depth and dynamic range}~\hspace{0.2 cm} 8 to 16 data bits are common in digital image sensors. Logarithm of the dynamic range is proportional to the bit depth. Lowering the noise floor potentially makes the full data bits available for dynamic range or maximizing the number of quanta detectable per pixel.

{\it Power consumption per mm$^2$}~\hspace{0.2 cm} Signal generation (turning an ionizing quantum into electron-hole pairs), analog-to-digital conversion, capacitor storage and  charge removal all consume power. Dark and leak current also contribute to power consumption and sensor and ASIC heating. ePix100 (50 \textmu m pitch, 352 $\times 384$ pixels per sensor, readout speed 120-240 Hz, 5-10 MHz pixel clock), a more recent direct detector for LCLS, consumes about 12 \textmu W/pixel.  In high-speed imaging, the power consumption also depends on the frame rate and radiation source repetition rate and intensity. Assuming that the resolution or contrast give a certain number of $\mathcal{N}_0$ of quanta for an image, faster frame rate $1/T_0$ corresponds to higher power of illumination for ultrafast imaging,
\begin{equation}
P_0 = \frac{\mathcal{N}_0 E_0}{\eta T_0},
\end{equation}
assuming a monochromatic light or mono-energetic particle. The factor $\eta$, with $0 < \eta < 1$,  accounts for parasitic power consumptions. Various models have been derived for $\mathcal{N}_0$ as a function of resolution ($\delta / M$, with $M$ being the magnification of the object after projection onto the detector, and $\delta$ the pixel resolution of the detector),
\begin{equation}
\mathcal{N}_0 = \frac{c_0 M^\alpha}{\delta^\alpha},
\end{equation}
with $\alpha \sim 4$.

{\it Fabrication technology and cost} CMOS technology is widely used for image sensor fabrication. The feature size of the image CMOS fabrication continue to decline with time, allowing smaller pixel sizes and more functions for the same sensor area. CS-PAD used 0.25-\textmu m TSMC CMOS process, ePix100, a more recent direct detector for LCLS also uses 0.25-\textmu m TSMC CMOS process, AGIPD used IBM 130 nm process. One of the open question is radiation hardness of the CMOS image sensors as the radiation sources get brighter, faster, or emits higher energy photons. The High Energy Physics community has led the way in understanding and mitigating the effects of radiation in CMOS circuits~\cite{ACD:1999} and has designed ASICs capable of withstanding 100’s of MRads~\cite{PDP:2015}.

{\it In-situ data storage} In ultrafast imaging, in-situ data storage is used to improve the frame rate by removing the time burden of data transfer. For example, HPV-X2 has 128 storage cells per pixel. AGIPD has 352 storage cells because each pixel is 200 \textmu m. The Keck-PAD can store 8 frames of data at up to 10 MHz frame-rate. The in-situ or temporary storage capacity limits the maximum number of frames that can be taken before a pause is required to transfer the stored images off the detector. 

\subsection{Open problems and opportunities}

Since the primary purpose of U-RadIT is to collect time-dependent data, such as 2D images, and to extract information from the data about the dynamic experiments such as protein unfolding, implosion of mass densities, shockwave propagation, defect and void generation and migration in materials, {\it etc.}, one of the central questions is how to optimize the information yield from a U-RadIT measurement. In the case of implosion in an ICF experiment, for example, the highest information content corresponds to the highest spatial and temporal resolution of mass density over the full implosion length (initial radius $\sim$ 1 mm). Since the number of voxels is proportional to the resolution ($\delta$) to the third power, $\delta^3$, at $\delta = 10$ \textmu m (the state of the art both in terms of the experiments and computation), the number of voxels is 8$\times 10^6$. Improving the experimental imaging resolution to $\delta = $1 \textmu m, which corresponds to 10$^3$ times more voxels, is an open problem believed to be pivotal in further advancing controlled experiments of ignited fusion plasmas. Protein unfolding, shockwave propagation, defect and void generation and migration in materials, {\it etc.} have similar degrees of difficulty since high spatial and temporal resolution requirements are recurring themes for U-RadIT measurements.

\begin{figure*}[thbp] 
   \centering
   \includegraphics[width=0.85\textwidth]{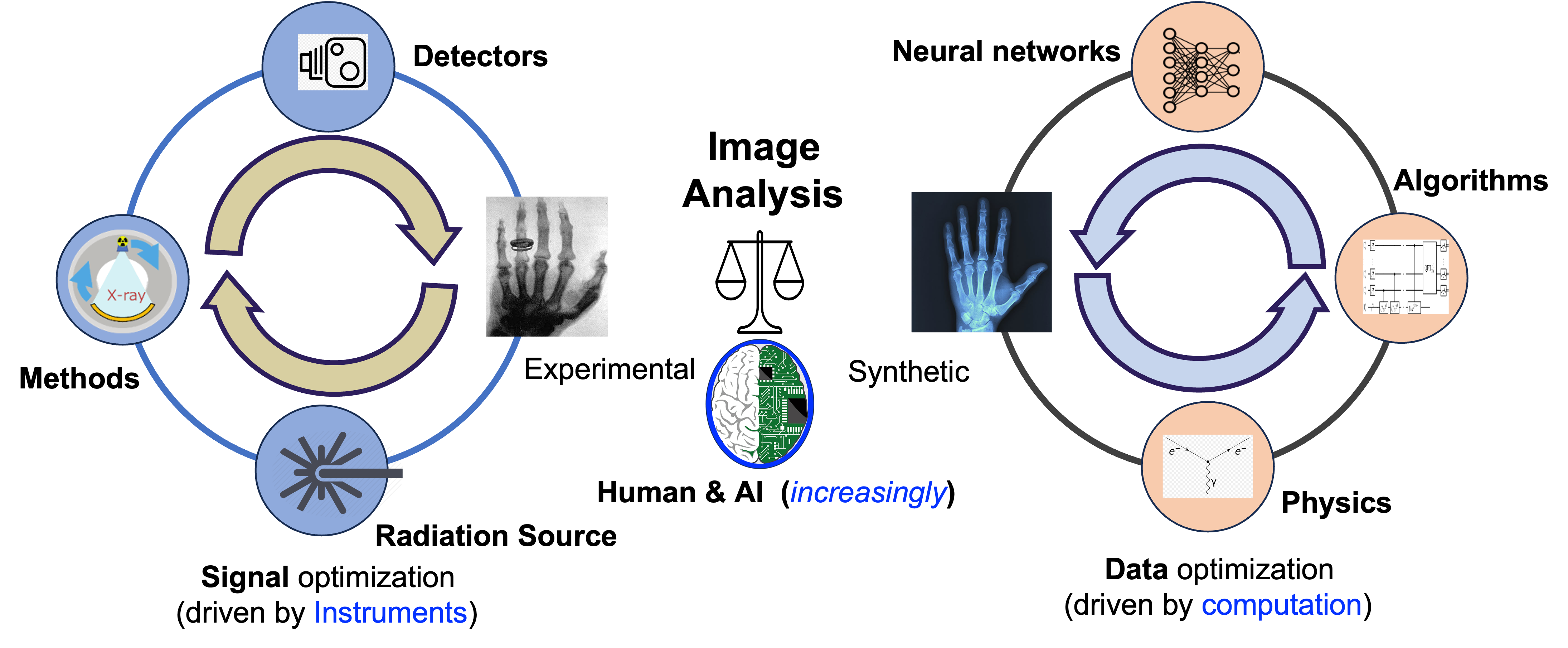} 
   \caption{A holistic framework to optimize U-RadIT information yield includes two correlated loops: The hardware loop on the left is driven by signal optimization, which may include  the radiation source (low emittance, coherence, adjustable spectrum), the detectors, and detection methods. The data or `digital twin' loop on the right is driven by computation towards interpretable synthetic data that can be directly compared with experimental data including images.}    \label{fig2:Opt}
\end{figure*}
 
A framework for the information-yield optimization is given in Fig.~\ref{fig2:Opt}, which consists of a signal optimization loop for signal generation and recording through radiation source(s), data collection methods or imaging modalities with examples given in Sec.~\ref{sec:mm}, and detectors (Sec.~\ref{sec:inst}), and a  data optimization loop based on the physics inspired forward modeling or data methods such as deep learning, and implemented through computation (traditional grid-based or particle-tracking algorithms and new neural network algorithms). Image analysis algorithms, more details in Sec.~\ref{sec:da1},  that compare experimental data with synthetic data may also be optimized, potentially adding a third loop, which bridges the signal optimization loop in hardware space and the data loop in the `virtual reality'  or `digital twin' space and is not explicitly shown in Fig.~\ref{fig2:Opt}. 

The optimization process may be described in terms of searching for a high-dimensional vector ({\bf x}), such as the 3D positions of many atoms involved in the protein unfolding, the mass density voxel map in an ICF implosion experiment, or a 3D velocity field in a shock experiment, which satisfies the following 
\begin{eqnarray}
({\bf x}, \tilde {\bf x}) & \equiv & {\rm argmin }_{f({\bf x}) =0; \tilde f(\tilde {\bf x}) =0} \mathcal{L} ({\bf x}, \tilde {\bf x}), \label{eq:op1}\\
\mathcal{L} ({\bf x}, \tilde {\bf x}) & \equiv &  \| I_{exp} ({\bf x}) - \tilde I_{syn} (\tilde {\bf x})\|_p +\mathcal{R} ({\bf x}, \tilde {\bf x}). \label{eq:op2}
\end{eqnarray} 
`$\equiv$' symbolizes definition or `is identical to'. Here we define the cost function, also known as loss function, $\mathcal{L}$  to be the difference between one (or more) experimental images $ I_{exp} ({\bf x})$ (each is a two-dimensional intensity map) and the corresponding synthetic image(s) $  \tilde I_{syn} (\tilde {\bf x})$. $\tilde {\bf x}$ is the corresponding theoretical vector of {\bf x}. The regularization functions $f ({\bf x}) = 0$, and $\tilde f ( \tilde {\bf x})$ =0  are further discussed next. Additional regularization to $\mathcal{L}$ through $\mathcal{R}$ is often used as well. If $p = 2$, the difference is calculated by using the so-called $l^2$ norm or the Euclidean  distance~\cite{BV:2009}. In some cases, $p = 0$ may be desired, which is, however, computationally hard and related to an open Millennium Prize problem (P {\it vs.} NP) ~\cite{Tao:2008}, see some details in Sec.~\ref{sec:cmr}. In practice, $p=1$ is often used.

The regularization functions $f ({\bf x}) = 0$, and $\tilde f ( \tilde {\bf x})$ =0 are necessary for a number of reasons. First, {\bf x} may only be known statistically due to a number of reasons.  Probabilistic interactions between an object and ionizing radiation field imply that, for the same setup, the same object, the same ionizing radiation source, and the same detector, $I_{exp} ({\bf x})$ is not unique. For repetitive experiments of the same setup, objects, radiation source and detector may not be reproducible at the highest resolution (at atomic resolution, for example). Only sparse measurement through U-RadIT is possible in practice, which implies that the number of unknowns ({\bf x}) is greater than the number of equations, as further explained in Sec.~\ref{sec:cmr}, limited by the source intensity, and detectors. The background and noise can fluctuate from experiment to experiment, which further compound the recovery of {\bf x}.

In the hardware and instrument loop, here are some additional factors that may contribute to $f ({\bf x}) = 0$ regularization. The X-ray and particle source intensity, emittance (angular distribution), and spectrum control of the sources. The detector optimization through the metrics described in Sec.~\ref{sec:met}, which may be constrained by chip clock speed, power consumption, memory, sensor thickness, radiation-induced electron or photon transport. Additional imaging and tracking detectors as described in Sec.~\ref{sec:inst}, and novel methods to collect data, Sec.~\ref{sec:mm}.  

In the data loop and $\tilde f (\tilde {\bf x}) = 0$ regularization, conversation laws of physics (mass, energy, and momentum, for instance), traditional forward modeling algorithms, and more recently data-driven algorithms are the building blocks of the optimization schemes, regulated by the underlying physics, statistics, bandwidth of the online and offline data processing, and computing power.  Some of the recent trends are towards physics-informed machine learning and data-driven models, and also towards using data-driven models such as neural networks as surrogates for traditional first-principle or derivative forward models for accelerated computing. 

In short, U-RadIT information-yield optimization is the central problem in U-RadIT applications, and the problem is known to be difficult due to high-dimensionality. Integrated approaches to radiation source, instrumentation, physics, statistics, data and algorithms offer many opportunities that may overcome the limitations in hardware for data acquisition or computing power for data processing.

\section{Instruments \label{sec:inst}}

We may summarize the overall trend in detector and instrument requirements as `10H' frontiers (`H' stands for {\it higher}), which combines the requirements derived from the radiation sources and applications. The following trends in radiation sources drive the U-RadIT detector instruments and methods (Further discussions in Sec.~\ref{sec:mm}) development: {\it higher} photon or particle flux, {\it higher} photon energy, {\it higher} photon source coherence, {\it higher} source repetition rate, simultaneous use of more than one radiation source, such as charged particles together with photons~\cite{CAB:2019}, charged particles with neutrons, or neutrons with photons, for multi-modal U-RadIT -- {\it higher} detection versatility.  The following application needs place additional requirements on instruments and detectors: {\it higher} detection efficiency or sensitivity, {\it higher} spatial or position resolution, {\it higher} detection dynamic range, {\it higher} radiation resistance (or radiation hardness), and {\it higher} data or information yield. Higher information yield can be obtained through, {\it e.g.} on-board machine learning (ML).

Not all `10H' features or requirements can be met simultaneously in a single experiment or a detector, and tradeoffs are often adopted in practice. For example, a tradeoff between efficiency and temporal resolution (frame-rate) may be necessary for ultrafast X-ray measurements. Another example of tradeoff is the spatial resolution and data yield due to the real estate constraint on a wafer. 
Cost reduction, including CMOS prototyping cost, is yet another important driver for tradeoffs in hardware optimization. For example, even though the production costs per wafer are more-or-less constant with time, the masking costs have exploded recently.

\subsection{Sensors for ionizing radiation }

A basic construction of modern radiation detectors consists of a sensor frontend and electronics backend, similar to back-illuminated optical detectors.
The sensor converts ionizing radiation into electron-hole pairs or visible light as recordable signals by electronics. Many elements in the periodic table have now found sensor applications, either as semiconductor diode sensors, when electron-hole pairs are created, or as scintillator sensors, when visible light is first induced by the ionizing radiation and then detected by photodetectors, such as semiconductor diode sensors.  Majority of the sensors are in solid state. Gas and liquid sensors are usually used in large volumes and when solid sensors and electronics become too expensive. Electronics are progressively miniaturized, and can perform specialized radiation detection functions, such as charge amplification (gain), analog-to-digital conversion, noise rejection, and therefore are called application specific integrated circuits (ASICs). ASICs, through large scale complementary metal-oxide semiconductor (CMOS) integration process,  also allow large area sensor diodes to function as large pixelated arrays (as in cameras) with individual pixels perform exactly the same functions.  Timepix ASICs are highlighted in Sec.~\ref{sec:tmpx}.

Ionizing radiation may penetrate many hundreds of \textmu ms into a detector sensor. Especially in the case of hybrid modules this may require use of sensors fabricated from unusually thick ($>$ 1 mm) semiconductor wafers. Higher-Z (than silicon) sensors, such as GaAs, CdTe, CZT, or thicker silicons have been used as synchrotrons and XFEL detectors in the so-called hybrid configuration, which uses wafer-scale bump bonding to integrate the sensor diodes with ASICs. Examples include Keck-PAD, CS-PAD, AGIPD, ePix, Timepix, Medipix, and MM-PAD in Sec.~\ref{sec:hybrid}. When scintillators are used, the stopping power and thickness can be adjusted without changing the optical detectors, and therefore scintillators offer more flexibility. However. the spatial resolution by the scintillator approach is usually worse than hybrid detectors due to isotropic emission of light from a thick scintillator. The energy resolution by scintillators is also worse due to the loss of light due to, for example, refractive index mismatch at the scintillator boundary.

One trend in semiconductor sensor and ASIC innovation is in gain control and noise reduction on the pixel scale ($\sim$ 10 \textmu m), so that single-photon sensitivity, similar to photomultiplier detectors (PMTs), which is too bulky to build mega-pixel arrays, can be obtained for millions or more pixels. Silicon photo-multipliers (SiPMs), single-photon avalanche diodes (SPADs)~\cite{Char:2014}, monolithic active pixel sensors (MAPS)~\cite{TBC:2001}, and 3D photon-to-digital converters (in Sec.~\ref{sec:PDC}) are some examples. A trend in scintillator sensor innovation is material structural engineering through, {\it e.g.} metasurfaces and bulk metastructures, so that light can emit anisotropically and be collected more efficiently. 

Sensors traditionally are regarded as `analog' devices due to their low detection sensitivity and the need for a large gain. As many detectors now reach single-visible-photon sensitivity, and with very compact (10 \textmu m or smaller footprint per pixel) solid-state designs, the state-of-the-art radiation detectors are quantum devices that can readily distinguish individual particles and X-ray photons. It may now be anticipated that photon counting with high energy resolution, or `spectroscopic photon counting' for ionizing radiation, and quantum detection with imbedded machine learning (ML) algorithms are forthcoming. 

\subsection{Timepix ASICs \label{sec:tmpx}}

The Timepix4 ASIC~\cite{LAB:2022} has recently been added to the Timepix family of hybrid pixel detector readout chips~\cite{BCL:2018}. Here we summarize the latest results on Timepix4 in the context of their predecessors. 

Table~\ref{tab2:timepix} summarizes the Timepix family of hybrid pixel detector readout chips. The Timepix chip~\cite{LBC:2007}, which was designed on 250-nm CMOS, became available in 2005 and was the first large area hybrid pixel detector readout chip which could be programmed at the pixel level to count photons, measure Time of Arrival (ToA) with respect to an external shutter or measure Time-over-Threshold (ToT) providing an indication of the charge deposited per pixel. It was this third mode of operation which led to the extensive use of Timepix in a multitude of applications~\cite{BCL:2018}. In particular, the development of a miniaturized USB readout system~\cite{VJP:2006} with the Pixelman software~\cite{THJ:2011} allowed for turnkey usage of the device for monitoring of background radiation as well as numerous scientific applications. Following the formation of the Advacam company in Prague, the Minipix system became widely available at a relatively low price and was used extensively in space and in schools. 

\begin{table*}[!htb]
\caption{\label{tab2:timepix}%
Summary of the main characteristics of the Timepix ASIC family.}
\begin{ruledtabular}
\begin{tabular}{lcccc}
\textrm{} &
\textrm{Timepix} &
\textrm{Timepix2}&
\textrm{Timepix3}&
\textrm{Timepix4 }
\\
\colrule
Year & 2005 & 2018 & 2014 & 2020  \\
Tech. node (nm) & 250 & 130 & 130 & 65  \\
Pixel size (\textmu m) & 55 & 55  & 55 & 55  \\
\# pixels (x $\times$ y)& 256 $\times$ 256 & 256 $\times$ 256 & 256 $\times$ 256 & 448 $\times$ 512\\
time bin resolution (ns)& 10 & 10 & 1.5 & 0.2 \\
readout & Frame-based &Frame-based &Data-driven  & Data-driven \\
architecture &(sequential R\&W) &(sequential or & or Frame-based & or Frame-based\\
&  &  continuous R/W) & (sequential R\&W) & (sequential or \\
&& && continuous R/W)\\
\# sides & 3 & 3 & 3 & 4 \\
for tiling &&&& \\
\end{tabular}
\end{ruledtabular}
\end{table*}

Timepix2 (developed in 2018 in a 130nm CMOS process) is a replacement for the Timepix device which went out of production as the 250nm process was ended by the foundry. Timepix2~\cite{Camp:6} addresses a number of known limitations of Timepix. When very large charges are deposited within one pixel of Timepix the ToT measurement becomes non monotonic with the input charge and collapses. This became known as the ‘volcano effect’ because heavily ionizing particles which stopped in the sensors would yield ToT profiles over many pixels which, instead of being mountain shaped, had a characteristic crater shape. In Timepix2 the monotonicity of the ToT with energy is maintained until very high input charges and saturates around 300 k$e^-$. Another limitation in the Timepix chip was related to the shutter control logic. When the shutter is opened any ToT output which is already high (coming from a preceding hit) would be recorded as a truncated hit. Equally the closure of the shutter would result in the truncation of ToT values for hits which occurred near the end of the shutter. In Timepix2 hits arriving before the shutter are ignored and hits which arrive just before the shutter closes are registered with the correct ToT.  Moreover, because Timepix2 uses a much denser technology than Timepix, it has a total of 48 bits per pixel (instead of 14) and these can be configured to permit data taking while readout is underway and permit recording both time and energy simultaneously.

Timepix3 was produced in 2014 and introduced data driven readout to pixel electronics for the first time~\cite{Camp:7}. Each time a pixel is hit a 48-bit packet of information is produced containing the address of the hit pixel, the ToA information with a precision of 1.6ns and up to 10 bits of ToT. A number of circuit innovations were required in order to make the high precision time tagging possible at a reasonable power consumption and avoiding the coherent noise which would be produced by using a conventional clock tree on the large area ASIC. A super pixel architecture was used, 
grouping pixels of 2 x 4 pixels in the column direction. Each super pixel contains one VCO whose oscillation frequency (at 640MHz) is locked to a PLL at the periphery of the chip with a VCO identical to the one distributed across the pixel matrix. The VCO in the super pixel starts to oscillate when the discriminator fires and stops when with the next the rising edge of the 40MHz master clock. A fast 4-bit counter counts the number of VCO clock ticks. Each pixel in the super pixel records the fast counter value at the rising edge of its own discriminator and the fast counter value when the clock edge rises. This allows for a precision of 1.6ns in measurement without the need for a 640MHz VCO to be running continuously across the entire chip. The 40MHz master clock is buffered from super pixel to super pixel in each double column smoothing out the power supply bounce at the column level. Moreover, the peripheral electronics can be programmed to produce up to 16 master clocks delayed by $\sim$ 1.6 ns with respect to each other and these can be applied to different groups of columns also with the intention of reducing power supply bounce.

\begin{figure}[thbp] 
   \centering
   \includegraphics[width=0.45\textwidth]{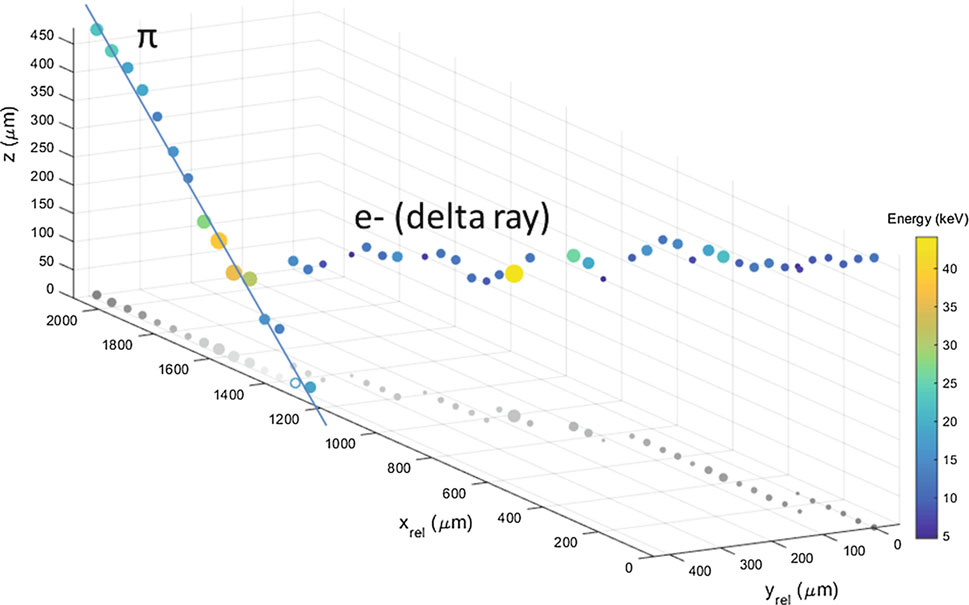} 
   \caption{A 120GeV/c muon track which produces a delta electron reconstructed from the ToT and ToA information provided by a Timepix3 chip~\cite{Camp:8}. The colors and diameters of the points represent the charge detected in that voxel.}
   \label{fig3:Camp1}
\end{figure}

The provision of a 1.6ns timestamp at the pixel level opens many new applications in high energy physics and beyond. Figure~\ref{fig3:Camp1} shows the reconstruction of a 120 GeV/c pion traversing a 500 \textmu m thick silicon detector at the CERN SPS~\cite{Camp:8}. The depth of interaction of the pion is measured to a precision of $\sim$ 28 \textmu m using the precise ToA information. The pion traverses the sensor while ejecting a delta electron. The amplitude of the detected charge is indicated by the color and diameter of the circles.  A more unusual example of the on-pixel time stamping and data-driven readout is the optical readout of a liquid Argon Time Projection Chamber (TPC) using a Timepix3-based camera~\cite{Camp:9}. The timestamp precision required in this application is much less constrained than the case of the silicon TPC but the big advantage of this approach is that, by using an optical readout, a large area (0.5 m $\times$ 0.5 m) can be focused on a single ASIC strongly reducing the number of readout channels. The readout system runs at room temperature and the hit data comes from the ASIC in one continuous zero-suppressed stream. Returning to the semiconductor readout and looking beyond HEP an exciting development is taking place in the Czech Republic led by the spin out company, Advacam. A 2-mm thick CdTe sensor is connected to the Timepix3 and, because Timepix3 can record multiple hits within the same sensor layer, a single layer Compton camera becomes feasible~\cite{Camp:10}.  This could lead in time to the development of a new apparatus for thyroid imaging with substantially improved spatial resolution and sensitivity compared with conventional techniques.

In parallel with the development of Timepix3, Through Silicon Via (TSV) processing was explored as a means by which to reduce the dead area when covering large areas with pixel detector tiles~\cite{Camp:11}. In this work Medipix3 and Timepix3 readout wafers were processed using a TSV-last process from CEA-LETI, Grenoble, France. IO pads on the pixel periphery were accessed from the rear side of the chip and a copper ReDistribution Layer (RDL) was used to bring the IO pads to a matrix of large pads suited to subsequent Ball Grid Array (BGA) assembly. This proved the feasibility of reading out pixel chips via TSVs from the rear side but, to be fully sensitive over a large tiled area, the peripheral electronics would have to be ‘hidden’ beneath the bump bonding pads to the sensor. This was the background which led to the development of the Timepix4 ASIC~\cite{LAB:2022}.

\begin{table*}[]
\centering
\caption{Comparison of the main characteristics of the Timepix3 and Timepix4 ASICs.}
\begin{tabular}{|lll|cc|}
\hline
\multicolumn{3}{|l|}{}                                                                                                                                                     & \multicolumn{1}{c|}{\textbf{Timepix3 (2013)}}                                            & \textbf{Timepix4 (2019)}                                            \\ \hline
\multicolumn{3}{|l|}{\textbf{Technology}}                                                                                                                                  & \multicolumn{1}{c|}{130nm - 8 metal}                                                     & 65nm - 10 metal                                                     \\ \hline
\multicolumn{3}{|l|}{\textbf{Pixel Size}}                                                                                                                                  & \multicolumn{1}{c|}{55 x 55$\ \si{\micro\meter}$}                                        & 55 x 55$\ \si{\micro\meter}$                                        \\ \hline
\multicolumn{3}{|l|}{\textbf{Pixel Arrangement}}                                                                                                                           & \multicolumn{1}{c|}{\begin{tabular}[c]{@{}c@{}}3-side buttable\\ 256 x 256\end{tabular}} & \begin{tabular}[c]{@{}c@{}}4-side buttable\\ 512 x 448\end{tabular} \\ \hline
\multicolumn{3}{|l|}{\textbf{Sensitive Area}}                                                                                                                              & \multicolumn{1}{c|}{$1.28\ \si{\centi\meter^2}$}                                         & $6.94\ \si{\centi\meter^2}$                                         \\ \hline
\multicolumn{1}{|c|}{\multirow{7}{*}{\textbf{ \rotatebox{90}{Readout Modes}}}} & \multicolumn{1}{l|}{\multirow{4}{*}{\begin{tabular}[c]{@{}l@{}}Data driven\\ (Tracking) \end{tabular}}} & \textbf{Mode}           & \multicolumn{2}{c|}{TOT and TOA}                                                                                                                               \\ \cline{3-5} 
\multicolumn{1}{|c|}{}                           & \multicolumn{1}{l|}{}                                                                                  & Event Packet   & \multicolumn{1}{c|}{48-bit}                                                              & 64-bit                                                              \\ \cline{3-5} 
\multicolumn{1}{|c|}{}                           & \multicolumn{1}{l|}{}                                                                                  & Max rate      & \multicolumn{1}{c|}{$0.43 \times 10^6$ hits/mm$^2$/s}                                    & $3.58 \times 10^6$ hits/mm$^2$/s                                    \\ \cline{3-5} 
\multicolumn{1}{|c|}{}                           & \multicolumn{1}{l|}{}                                                                                  & Max Pix rate   & \multicolumn{1}{c|}{1.3 kHz/pixel}                                                       & 10.8 kHz/pixel                                                      \\ \cline{2-5} 
\multicolumn{1}{|c|}{}                           & \multicolumn{1}{l|}{\multirow{3}{*}{\begin{tabular}[c]{@{}l@{}} Frame based \\ (Imaging) \end{tabular}}} & \textbf{Mode}           & \multicolumn{1}{c|}{PC (10-bit) and iTOT (14-bit)}                                       & CRW: PC (8 or 16-bit)                                               \\ \cline{3-5} 
\multicolumn{1}{|c|}{}                           & \multicolumn{1}{l|}{}                                                                                  & Frame          & \multicolumn{1}{c|}{Zero-suppressed (with pixel addr)}                                   & Full Frame (without pixel addr)                                     \\ \cline{3-5} 
\multicolumn{1}{|c|}{}                           & \multicolumn{1}{l|}{}                                                                                  & Max count rate & \multicolumn{1}{c|}{$\sim 0.82 \times 10^9$ hits/mm$^2$/s}                               & $\sim 5 \times 10^9$ hits/mm$^2$/s                                  \\ \hline
\multicolumn{3}{|l|}{\textbf{TOT energy resolution}}                                                                                                                       & \multicolumn{1}{c|}{\textless{}2 keV}                                                    & \textless{}1 keV                                                    \\ \hline
\multicolumn{3}{|l|}{\textbf{TOA binning resolution}}                                                                                                                      & \multicolumn{1}{c|}{1.56 ns}                                                             & 195 ps                                                              \\ \hline
\multicolumn{3}{|l|}{\textbf{TOA dynamic range}}                                                                                                                           & \multicolumn{1}{c|}{409.6 $\si{\micro\second}$ (14-bits @ 40 MHz)}                        & 1.6384 ms (16 bits @ 40 MHz)                                        \\ \hline
\multicolumn{3}{|l|}{\textbf{Readout bandwidth}}                                                                                                                           & \multicolumn{1}{c|}{$\leq$5.12 Gbps (8x SLVS @ 640 Mbps)}                                & $\leq$163.84 Gbps (16x @ 10.24 Gbps)                                \\ \hline
\multicolumn{3}{|l|}{\textbf{Target global minimum threshold}}                                                                                                             & \multicolumn{1}{c|}{\textless{} 500 e$^-$}                                                & \textless{} 500 e$^-$                                                \\ \hline
\end{tabular}
\label{Tb4:Camp2}
\end{table*}

Table~\ref{Tb4:Camp2} compares the detailed characteristics of the Timepix3 and Timepix4 readout ASICs. Apart from the possibility of tiling chips on 4 sides, a number of major improvements have been incorporated. In particular, hits which are well above threshold can now be tagged to a bin of 200ps and the maximum flux in data-driven mode has been increased by a factor of $\sim$ 8. The ASIC itself fills the entire reticle and has 448 $\times$ 512 pixels, divided into 2 matrices of 448 $\times$ 256 pixels. The readout bandwidth has been increased by a factor of $\sim$ 30 to cope with the increased hit rate and larger number of pixels. Once again, a particular design challenge came with the need for precise time tagging at the pixel level. Pixels are organized in 2 $\times$ 4 super pixels and 64 super pixels form a double column in one matrix. Each pixel contains a VCO which has a similar behavior to those of Timepix3 oscillating at 640MHz when the discriminator fires. The finer time tagging is achieved by recording the state of the internal inverters which make up the VCO. The 40MHz master clock which is propagated up and down the pixel double columns is delayed by buffers placed every 4 super pixels. A digital delay-locked loop is used to very precisely fix the delays between the super pixel blocks, ensuring a precise timestamp reference with minimum power supply bounce. 

Another design challenge was associated with the need for a front-side RDL to connect a regular array of 448 $\times$ 512 bump bonding pads spaced at 55 \textmu m pitch to the two underlying pixel matrices each composed of 224 $\times$ 256 with a pitch of 55 \textmu m $\times$ 51.2 \textmu m. In the first version of the chip the input capacitance was equalized across the entire pixel matrix. Unfortunately, the shielding of the traces connecting the pixels at the upper and lower edges of the matrix proved inadequate leading to a slightly elevated minimum threshold (which is set at roughly 5 times the quadratic sum of noise and threshold variation). In the subsequent versions of the ASIC (v1 and v2) the shielding above the peripheral regions was improved leading to marginal increase in noise for those pixels ($\sim$10 $e^-$ rms). Figure 2 shows the histogram of the noise measurements for the 3 versions of the chip as well as the geographical distribution of the noise for v2.    

\begin{figure}[thbp] 
   \centering
   \includegraphics[width=0.45\textwidth]{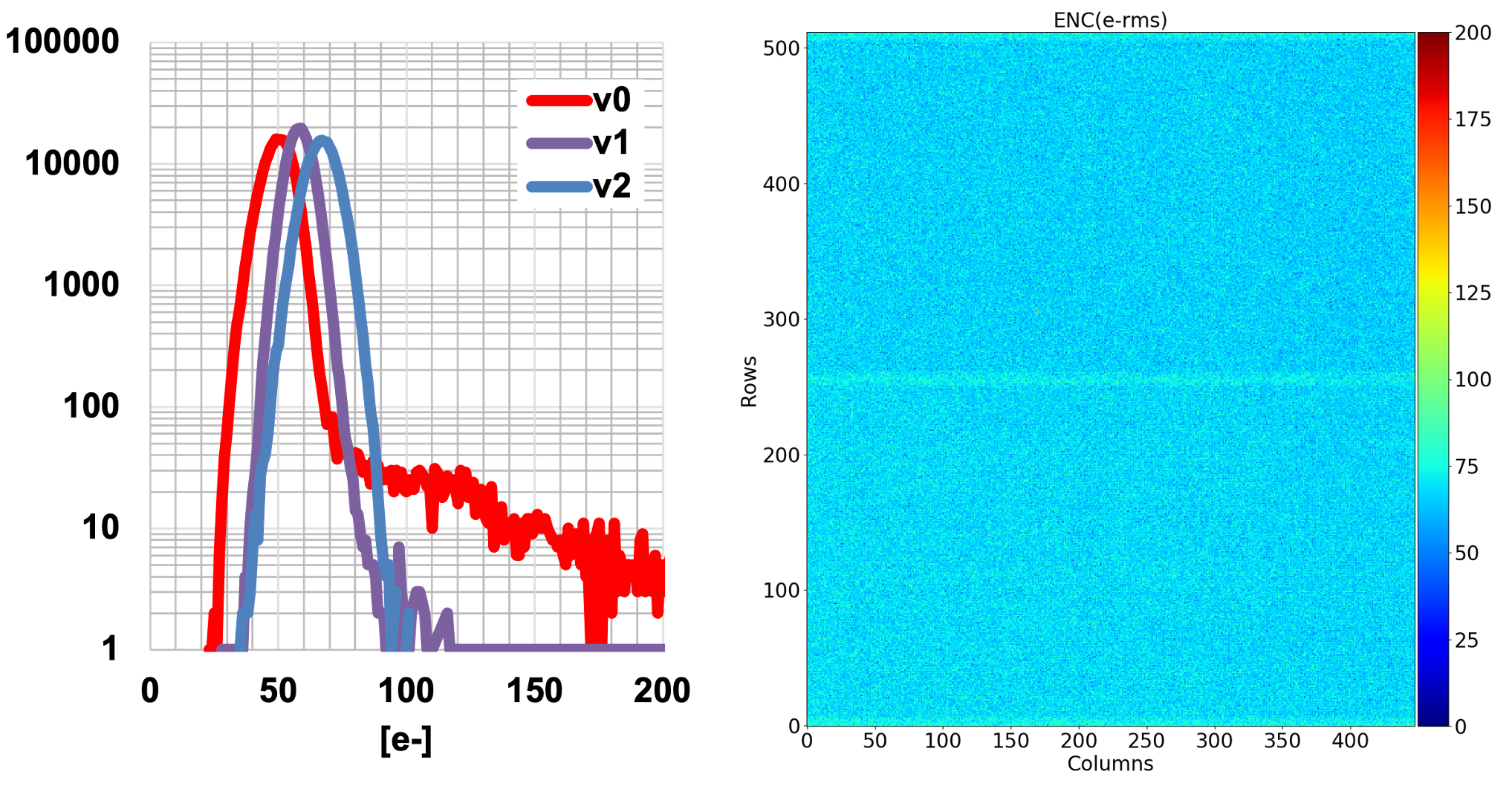} 
   \caption{Left: histogram showing the number of pixels for a given noise in e- rms. Right: the geographical distribution of noise across the ASIC. Pixels in regions above the peripheral and control logic have slightly elevated noise values ($\sim$10 $e^-$ rms more) because of the shielding added to prevent the injection of digital noise.   }
   \label{fig5:Camp3}
\end{figure}

In conclusion, the Timepix family of hybrid pixel detector readout chips has proven to be extremely versatile and adaptable to a number of widely varying applications. Timepix was the original device, using the same frame-based readout as Medipix2. Timepix2 is the direct successor to Timepix and maintains the straightforward frame-based readout approach but brings considerably enhanced functionality including the option of continuous R/W. Timepix3 introduced data driven readout as well as time stamping to a bin of 1.6 ns at the pixel level. Finally, Timepix4 provides on-pixel time stamping at 200ps while including an architecture that allows to tile ASICs seamlessly on 4 sides, a first in the field for large area hybrid pixel detectors.

\subsection{Hybrid pixelated array detectors (PADs) \label{sec:hybrid}}
X-ray experiments at synchrotron radiation (SR) sources demand many type of hard x-ray imaging detectors. For the purposes of categorization most imaging SR experiments may be crudely divided into two groups: Experiments where the path of the incident beam deviates only very slightly when passing through the sample. Examples include most radiographies; let’s call the detectors used in these cases “radiographic” imagers. Given the small footprint typical of SR beams, radiographic detectors usually only need small detective areas and, hence, very small pixels typically of \textmu m-sized dimensions. Examples include thin scintillator crystals optically coupled to visible light imaging cameras. In the second category of experiment x-rays scatter from a sample through a relatively large angle. In these cases, the scattered x-rays diverge from the footprint of the incident beam SR on the sample; hence, the experiment can accommodate “diffraction imagers” with larger pixels by increasing the sample to detector distance. 

Some experiments allow use of detectors that digitally count individual x-rays (“photon counting imagers”) while still maintaining a high detective quantum efficiency (DQE). In other experiments multiple x-rays arrive at a given pixel at too fast a rate for photon counting (e.g., at x-ray free electron lasers), thereby demanding the use of detectors that integrate the x-ray energy per pixel for an exposure before digitizing the signal (``integrating imagers”). 

The x-ray energy strongly influences detector fabrication if a high DQE is required. As a practical matter, detectors based on silicon x-ray sensors may be nearly ideal for x-rays below $\sim$20 keV but too transparent for higher energies, thereby requiring high atomic weight sensors (``hi-Z sensors”). Some experiments demand acquisition of a limited number of x-ray images at MHz frame rates (“burst-rate imagers”) while other experiments need “continuous framing imagers”. Yet other experiments deliver images with many orders of magnitude of x-ray flux/pixel across the image, thereby demanding “single photon sensitive, wide dynamic range imagers” that do not saturate the high intensity pixels where x-rays may be arriving at rate greater than hundreds of MHz, yet provide single x-ray sensitivity in the low flux areas of the same image.
The above distinctions by no means exhaust the variability of experimental detector requirements, but it does arguably cover a majority of current SR imaging experiments.
The Cornell Detector Group, now led by Sol Gruner and Julia Thom-Levy, has for many decades been a leader in developing x-ray imagers for SR research applications. Recent focus has been on experimental needs that are still ill-served by most commercially available detectors. These include burst-rate diffraction imagers framing at $>$10 MHz rates based on variations of the Keck-PAD family~\cite{G1}. Versions now in development for use at the Dynamic Compression Sector at the Advanced Photon Source at ANL will image at SR bunch separations of $<$ 77 ns. These will be equipped with hi-Z sensors (e.g., CdTe) for x-ray energies $>$ 36 keV. 

The Cornell Detector Group is also developing a family of single photon sensitive, wide dynamic range imagers based on the Mixed-Mode Pixel Array Detector (MM-PAD) approach~\cite{G2,G3}. MM-PADs are integrating imagers that achieve simultaneous high sensitivity at low flux and a very wide dynamic range by using a dynamic charge removal concept: The size of the integrating amplifier feedback capacitor is set sufficiently small as to achieve excellent single x-ray sensitivity. As integrated charge pixel approaches saturation on the integrating amplifier, a fixed bolus of charge is removed from the feedback capacitor, thereby partially resetting the amplifier. The number of charge removals during an exposure are digitally tallied. At the end of the exposure the integrated charge is computed as the sum of (number of charge removals) $\times $(charge per removal bolus) + (remaining charge in the amplifier). The amount of charge removed per bolus is typically set as equivalent to that from several hundred x-rays, and the maximum removal rate is 100 MHz. This allows $>$10$^8$ x-rays/pixel/second incident flux without amplifier saturation. The most current detectors frame continuously at 10 kHz~\cite{G2}. Adaptations of the MM-PAD concept have proven to be very useful in advancing detection for scanning transmission electron microscopy~\cite{G4, G5, G6}. 

It is important to recognize that all hi-Z sensors currently in use are not nearly as ideal as silicon sensors. Available hi-Z sensors tend to suffer from defects, lag, non-linear response, polarization or practical matters  of availability in appropriate areas and thicknesses. Accordingly, another area of focus of the Cornell Detector Group, in collaboration with colleagues at BNL, ANL, SLAC and MIT- Lincoln Labs, is on researching hi-Z x-ray sensors for use in hybrid PADs. These include CdTe, Germanium, CZT, and perovskite sensor materials. 

\subsection {Ultrafast CMOS cameras~\label{sec:Xin}}
Since the invention of active CMOS image sensors in 1993~\cite{Dart:1}, high-speed, ultra-high-speed (UHS) or ultrafast CMOS image sensors have revolutionized the field of high-speed imaging by taking advantage of advanced CMOS technology and solid-state imaging technology~\cite{EDF:2009,Etoh:2011}. In this section, we provide some historical background and discuss the state-of-the-art of UHS CMOS image sensors, including the key features and advantages of these sensors, as well as their implementation and limitations.

\begin{figure}[thbp] 
   \centering
   \includegraphics[width=0.46\textwidth]{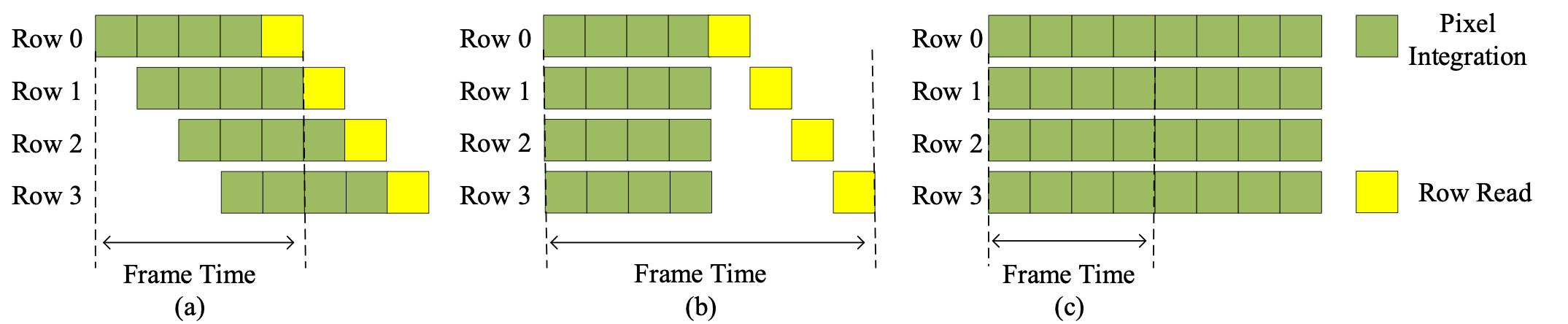} 
   \caption{Conceptual operation timing diagram of the rolling shutter image sensor(a), global shutter image sensor(b), and burst-mode image sensor.}
   \label{fig:Dart1}
\end{figure}

The frame rate of typical rolling shutter CMOS image sensors has remained around a few tens to a few hundred frames per second (fps). A typical operation timing of a rolling shutter image sensor is shown in Figure~\ref{fig:Dart1}(a), where each row of pixels starts to integrate incident photo-generated electrons at different times, creating a rolling shutter artifact when shooting high-speed moving objects~\cite{Dart:2}. To solve this issue, global shutter image sensors were proposed by~\cite{Dart:3}, where the whole frame of pixels starts to integrate electrons at the same time for the same amount of duration, as shown in Figure~\ref{fig:Dart1}(b). 

However, as shown in Figure~\ref{fig:Dart1}(a) and (b), regardless of the rolling or global shutter image sensor, the on-chip readout circuits continuously read pixel information and transmit the digitized data through high-speed data links to the receiver side. Due to the limited data rates of transmitters, typically ranging from a few gigabits per second (Gbps) to a few tens of gigabits per second in modern CMOS technologies, and the power budget of image sensors, achieving a frame rate of over tens of millions per second in continuous-mode CMOS image sensors is challenging. Two of the highest reported frame rate continuous mode CMOS image sensors are~\cite{Dart:4} and~\cite{Dart:5}, which operate at 80 kfps and 7.6 kfps, respectively.

To overcome the speed bottleneck of readout circuits and data transmitters, Ref.~\cite{Dart:6} introduced the burst mode image sensor. Figure~\ref{fig:Dart1}(c) shows the conceptual operation trimming of the burst-mode image sensor, where the entire frame of pixels samples and stores incident photon information into on-chip memories in voltage or charge domains simultaneously and continuously. Once all on-chip memories are filled, the readout circuit starts to read out the stored images. Since the sample and hold phase does not involve ADC conversion and data transmission, the frame operation duration of the burst-mode image sensor is most likely determined by the charge transfer speed, which can be as short as nanoseconds~\cite{Dart:7}. Therefore, it is feasible to achieve a frame rate of over tens of millions per second in burst-mode operation.

\begin{figure*}[thbp] 
   \centering
   \includegraphics[width=0.84\textwidth]{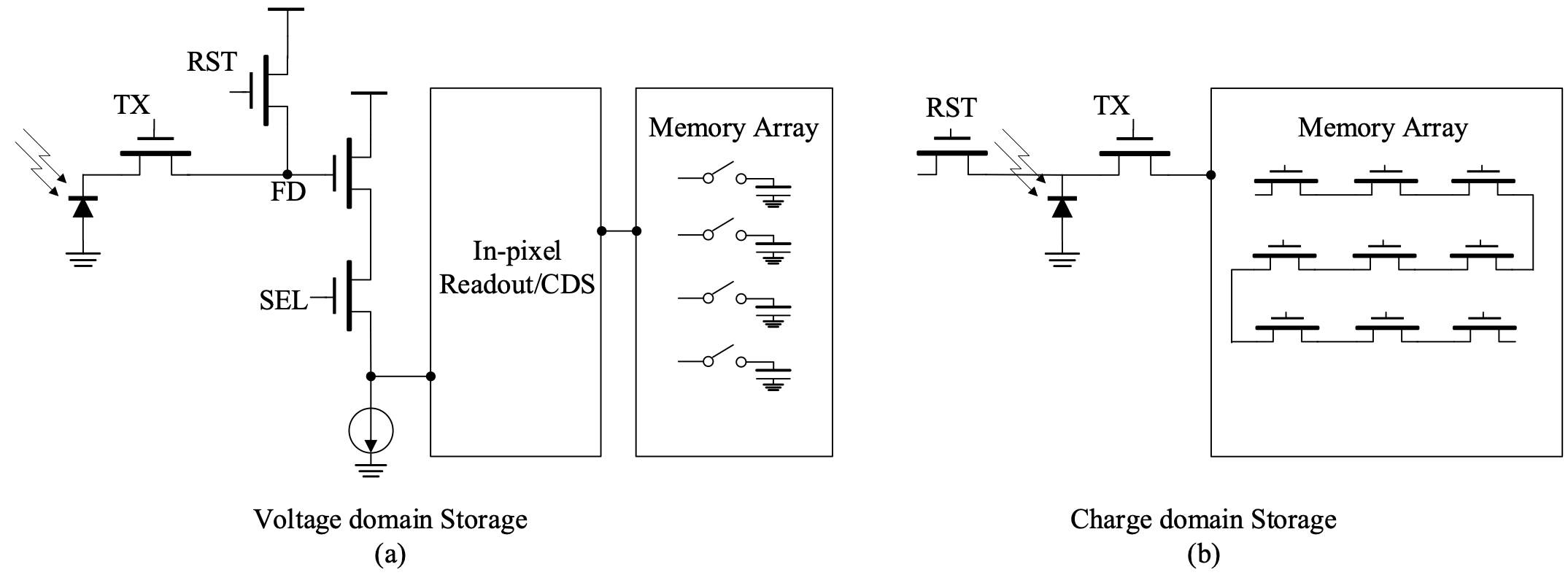} 
   \caption{Conceptual structure of voltage domain storage burst-mode image sensor(a) and charge domain storage burst-mode image sensor(b).}
   \label{fig:Dart2}
\end{figure*}

In recent years, several UHS CMOS image sensors have been developed, which have demonstrated frame rates of up to several hundred million frames per second (Mfps) in voltage domain storage, as illustrated in Figure~\ref{fig:Dart2}(a). Some of these sensors can even achieve giga frames per second (Gfps) in charge domain storage, as shown in Figure~\ref{fig:Dart2}(b). 

In a study by~\cite{Dart:7}, a 96 $\times$128 pixel image sensor operating at 10 Mfps based on a 180nm process was reported. Benefitting from high-density vertical capacitor technology, this image sensor is capable of storing 960 frames of the image in the in-pixel capacitor array. Another study by~\cite{Dart:10} reported a 32$\times$42 pixel image sensor operating at 20 Mfps based on a 130 nm process. By introducing capacitor-based passive amplifiers, this image sensor can achieve 8.4 $e^-$ input-referred noise. Similarly, Ref.~\cite{Dart:8} reported a 64$\times$64 pixel image sensor operating at over 20 Mfps based on a standard 180nm process. With the aid of a novel charge-sweep transfer gate, this image sensor reported the lowest input-referred noise, at 5.8$e^-$, by simulation. By optimizing the in-pixel sample and hold circuit and charge transfer time, Ref.~\cite{Dart:9} reported a 50 $\times$108 pixel image sensor operating over 100 Mfps based on a customized 180nm process.
Unlike voltage-domain storage, the charge-domain storage method eliminates the settling requirement of sample and hold capacitors, making it potentially capable of achieving a much higher frame rate. For instance, Ref.~\cite{Dart:11} reported a 64$\times$64 pixel CCD-based image sensor that ran at 100 Mfps with a 16-frame record length, while Ref.~\cite{Dart:12} introduced multi-collection gates and reported the possibility of achieving 1 Gfps. Moreover, based on the multi-collection gates, Ref.~\cite{Dart:13} simulated the potential of achieving a 50 fps frame interval, which translates to a frame rate of 20 Gfps. Typically, the design of a burst mode image sensor consists of three critical parts: a) charge transfer, b) in-pixel readout and c) frame memory unit. This section will focus on these three parts.
	
{\it High-speed Charge transfer} It is well-known that electrons can achieve a higher velocity in a strong electrical field. Therefore, a strong electrical field needs to be established in the pixel. Eq.~(\ref{eq:dart1}) provides a simplified relationship between the maximum electrostatic potential in a photodiode ($\psi$), the elementary charge ($q$), the doping concentration of the photodiode ($N_D$), the doping concentration of the substrate ($N_A$), and the photodiode half width ($X_n$),

\begin{equation}
\psi_{\max} \sim \frac{q N_D X_n^2}{2 \epsilon_0 \epsilon_r} (1 +\frac{N_D}{N_A}),
\label{eq:dart1}
\end{equation}
with $\epsilon_0 = 8.85 \times 10^{-12}$ F/m  being the vacuum  permittivity, and $\epsilon_r$ the relative permittivity.

\begin{figure}[thbp] 
   \centering
   \includegraphics[width=0.46\textwidth]{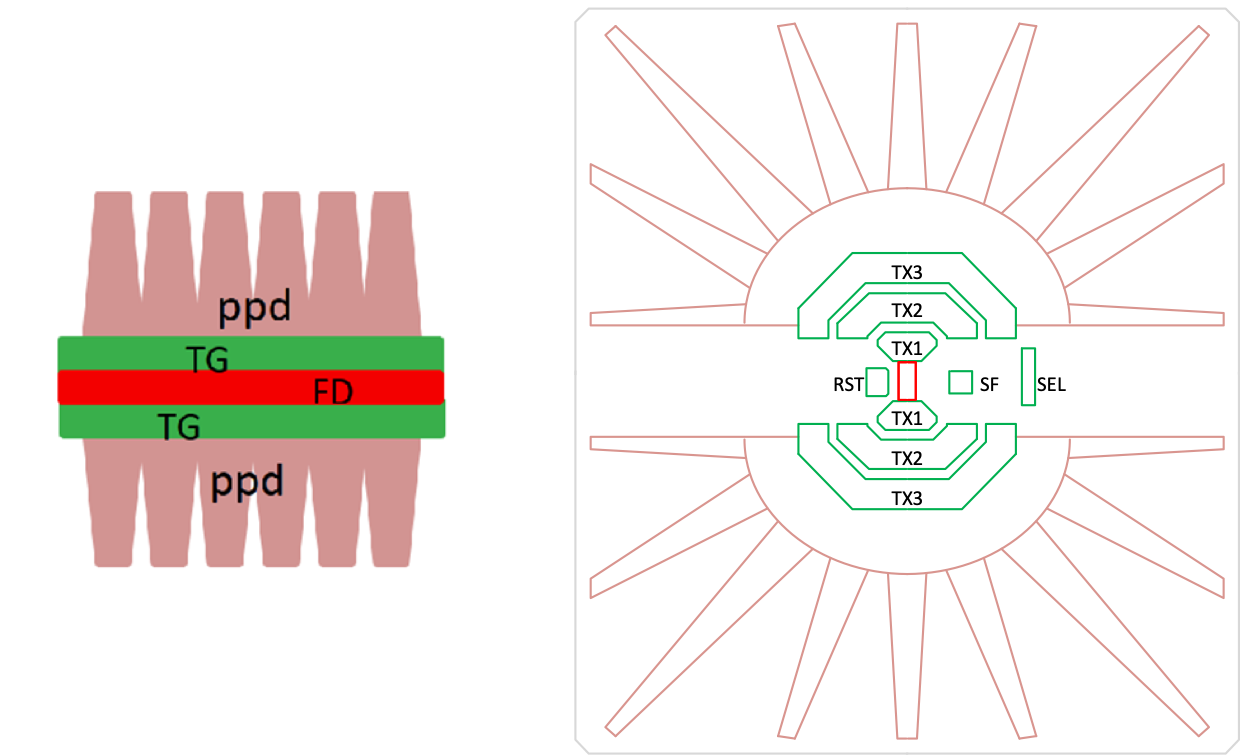} 
   \caption{The layout of the high-speed pixel from Ref.~\cite{Dart:14} (left) and high-speed charge-sweep pixel from Ref.~\cite{Dart:8} (right).}
   \label{fig:Dart4}
\end{figure}

\begin{figure}[thbp] 
   \centering
   \includegraphics[width=0.46\textwidth]{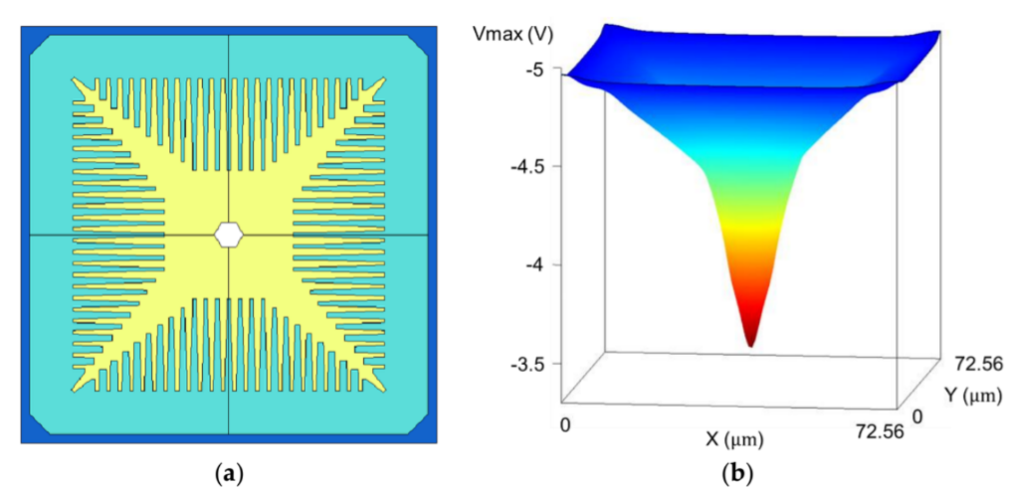} 
   \caption{Conceptual layout of the high-speed pixel from Ref.~\cite{Dart:15} (a) and its electrostatic potential diagram (b).}
   \label{fig:Dart5}
\end{figure}

\begin{figure}[thbp] 
   \centering
   \includegraphics[width=0.2\textwidth]{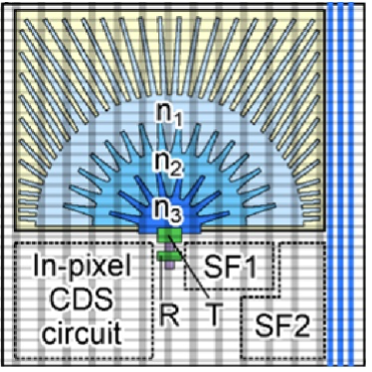} 
   \caption{Conceptual layout of the high-speed pixel from Ref.~\cite{Dart:9}.}
   \label{fig:Dart6}
\end{figure}

To create an electrical field that enables smooth and fast charge transfer, the maximum electrostatic potential in the photodiode needs to increase as it approaches the transfer gate. Figure~\ref{fig:Dart4} shows that Ref.~\cite{Dart:14} and~\cite{Dart:8} were able to create a lateral electrical field from the tip of fingers to the center of the pixel by varying the photodiode depletion width ($X_n$). Additionally, Ref.~\cite{Dart:15} used comb-shaped marks and three different implantation energies for photodiode dopants to create an inversed-pyramid shape electrostatic potential, as shown in Figure~\ref{fig:Dart5}, which can quickly transfer electrons from the backside of the substrate to the TX gate. To create a strong electrical field pointing to the TX gate, Ref.~\cite{Dart:9} combined multi-step doping techniques with varying depletion widths, as illustrated in Figure~\ref{fig:Dart6}, where the n-dopant concentration increases as $n_1<n_2<n_3$.

\begin{figure}[thbp] 
   \centering
   \includegraphics[width=0.46\textwidth]{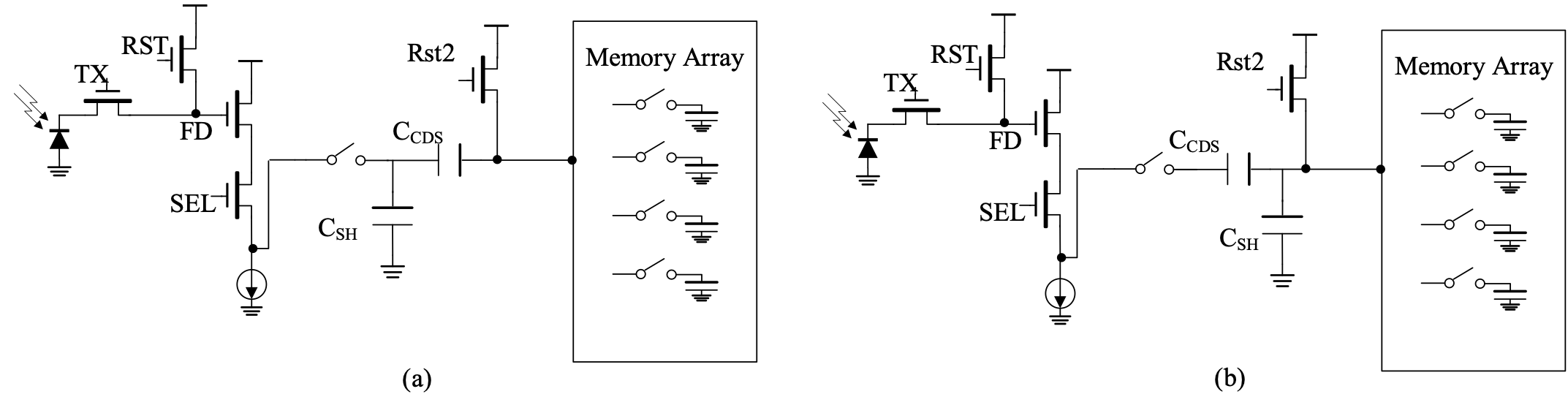} 
   \caption{Two major implementations of the in-pixel readout circuit.}
   \label{fig:Dart7}
\end{figure}

{\it In-pixel Readout} To implement voltage domain storage in ultra-high-speed image sensors, an in-pixel readout circuit that acts as a correlated-double-sampling (CDS) circuit is widely used. This circuit buffers the voltage difference between the pixel reset voltage and the pixel signal voltage to a capacitor. Figure~\ref{fig:Dart7} illustrates two major ways to implement this circuit. In Figure~\ref{fig:Dart7}(a), $C_{SH}$ acts as a decoupling capacitor connected to the output of the first pixel SF output, which does not introduce additional voltage attenuation to the pixel SF output swing. However, from the Rst2 switch's point of view, it will only see the $C_{CDS}$ capacitor. Therefore, the thermal noise introduced by the Rst2 switch is $\sqrt{kT/C_{CDS}}$. 

On the other hand, in Figure~\ref{fig:Dart7}(b), $C_{SH}$ is connected to the input of the second pixel SF input. $C_{CDS}$ and $C_{SH}$ form a capacitor voltage divider and attenuate the 1st stage pixel SF output swing, which may increase the total input referred noise. However, from the Rst2 switch's point of view, it will see the shunt connection of CCDS and CSH, and the thermal noise introduced is reduced to $\sqrt{Kt/(C_{CDS}+C_{SH})}$. Both of these circuits are reported in designs based on the image sensor noise budget and frame rate requirement.

\begin{figure}[thbp] 
   \centering
   \includegraphics[width=0.35\textwidth]{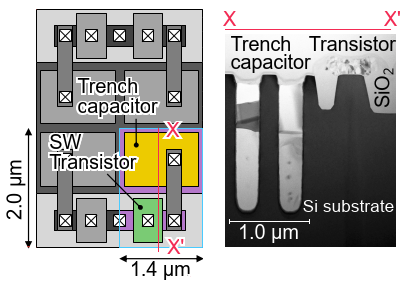} 
   \caption{High capacitance density vertical capacitor and its cross-section reported in Ref.~\cite{Dart:9}.}
   \label{fig:Dart8}
\end{figure}

{\it Frame Memory Unit} Capacitors are the most popular option for implementing frame memory in voltage-domain-storage burst-mode image sensors. In modern CMOS processes, capacitors are typically implemented as metal-insulator-metal (MIM) capacitors, metal-oxide-metal (MOM) capacitors, and poly-gate capacitors. Typically, in the same process, CPoly exhibits higher capacitance per unit area than CMOM and CMIM. A report by Ref.~\cite{Dart:10} states that a 10 fF per unit poly-gate capacitor is used in the design. To reduce the CMOS switch PN junction leakage and increase the unit cell capacitance, Ref.~\cite{Dart:8} introduced a combination of a 1.8V low-voltage poly gate capacitor and a hand-layout MOM capacitor. Furthermore, Ref.~\cite{Dart:9} has developed a novel vertical high-density poly capacitor that achieved 50 fF per unit cell in a 1.4 \textmu m $\times$ 2 \textmu m area, resulting in a fourfold improvement in capacitance density compared to typical poly gate capacitors, as illustrated in Figure~\ref{fig:Dart8}.

\begin{figure}[thbp] 
   \centering
   \includegraphics[width=0.26\textwidth]{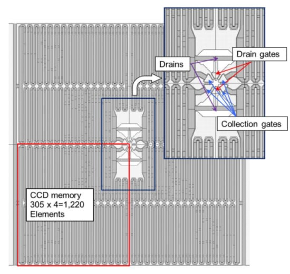} 
   \caption{layout of the CCD memories in pixel reported in Ref.~\cite{Dart:15}.}
   \label{fig:Dart9}
\end{figure}

\begin{figure}[thbp] 
   \centering
   \includegraphics[width=0.26\textwidth]{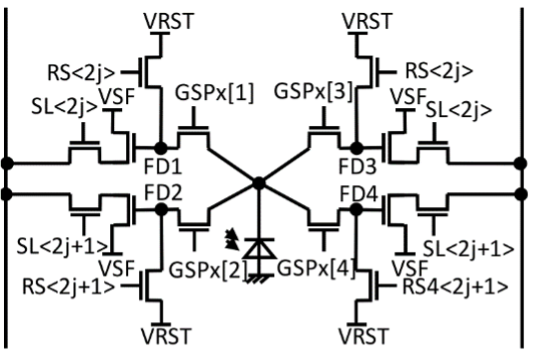} 
   \caption{Schematic of pixel based on FD storage reported in Ref.~\cite{Dart:18}.}
   \label{fig:Dart10}
\end{figure}

In charge domain storage, the charge-coupled device (CCD) is the dominant technology for frame memory design. As shown in Figure~\ref{fig:Dart9}, Ref.~\cite{Dart:15} reported the implementation of 1220 in-pixel frame memory units based on CCD. However, CCD cells typically suffer from high operation voltage and large power dissipation. As mentioned by Ref.~\cite{Dart:18}, another way to implement charge domain storage is by using the floating diffusion node as a frame memory, as FD1~FD4 shown in Figure~\ref{fig:Dart10}. However, the frame recording length is typically limited to a few frames due to physical implementation limitations. 

\subsection{3D photon-to-digital converters \label{sec:PDC}}
Many scintillation detectors show very fast response to various types of radiation (X-rays, gamma-rays, particles, etc.) via scintillation light with wavelengths that span from visible light to ultraviolet. Such responses are as fast as sub-ns; thus, these scintillators offer the potential for very fast timing. The scintillation light emitted upon incidence of a radiation event is typically read out with photodetectors that transduce the emitted light into electrical signals. The first popular photodetectors were the ubiquitous photomultipliers (PMTs). These devices offered great reliability and excellent timing. More recently, silicon photomultipliers (SiPMs) made their way into instrumentation designs with the intent of replacing PMTs~\cite{AG:2019,BIV:2022}. Their attractiveness stems from having good photodetection efficiency (PDE), being immune to magnetic fields, requiring lower bias voltage, and being physically lighter and less bulky because they are made from silicon. Over the course of the last ten years, however, SiPMs have shown limitations intrinsic to their design that make their use less straightforward than originally assumed. In particular, their dark current is relatively high because it corresponds to first approximation to the leakage of a reverse-biased semiconductor junction; they exhibit spurious responses that are correlated to the presence of a preceding event (this is known as after pulsing); their fill-factor (the amount of active surface covered by the detector versus the actual size of the device) is lower than PMTs, although their PDE can make up for it. Another characteristic that makes SiPMs less than ideal is their intrinsically high capacitance. A high capacitance detector offers, in general, challenges in designing precise instrumentation. These challenges originate from the fact that high capacitance has the effect of distorting signals, increasing electronic noise (especially for large areas), and requiring more power in the readout electronics. Another noticeable effect introduced by capacitance can be found in the SiPM output signal. While in PMTs, the output signal is somewhat close to reproducing the light flash, in SiPMs their capacitance has the effect of producing a tail with a decay of the order of tens to hundred ns in their response. The tail is due to the recharge mechanism of the device which is limited by the internal architecture. Also, when read out by a current amplifier, the capacitance has the undesired effect of limiting the rise time.

Silicon photomultipliers achieve their goal of detecting scintillation photons by using arrays of single-photon avalanche diodes (SPADs). In the device, each SPAD is biased slightly above its breakdown voltage; thus, when a photon interacts within it, the SPAD enters the avalanche region and produces an amount of charge which is limited from being destructive by a series resistor (known as quenching resistor). All SPADs have a common terminal, or node, so that if multiple SPADs avalanche, their individual charge is summed into the output node. Since each SPAD responds to scintillation light with the same amount of charge, the sum charge at the output node is proportional to the number of detected photons which is, in turn, proportional to the energy deposited in the scintillator.  This process certainly works but is also the origin of some of the SiPM’s limitations. In particular, the capacitance of each SPAD is set by its geometric characteristics at depletion, and these capacitors are connected in parallel. Since in a standard device, thousands of SPADs are connected to form the SiPM, the SPAD array capacitance can reach values between 30 and 90 pF/mm2. Common SiPM capacitances range from a few hundred pF to a few nF in the sizes of interest. Since this is an intrinsic property of the devices, there are no practical mitigations. 

\begin{figure}[thbp] 
   \centering
   \includegraphics[width=0.45\textwidth]{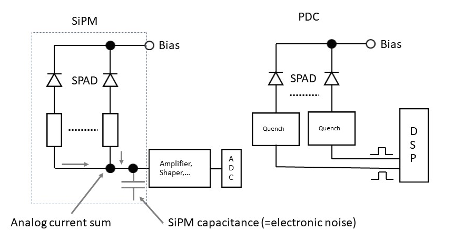} 
   \caption{ Conceptual difference between a SiPM and a PDC. PDCs are intrinsically digital devices and avoid the need for expensive analog processing and digital conversion. }
   \label{fig:L1}
\end{figure}

In recent years, a different paradigm for reading out the charge produced by the SPAD array has been envisioned by a few research groups around the world. About a decade after the birth of the SiPM by Saveliev and Golovin~\cite{L1}, Haemisch et. al.~\cite{L2} came up with a different concept for implementing SPAD quenching that would eventually change the way SiPMs work. In the new work an important observation was made: the true nature of the information generated by the SPAD array is binary. In fact, even in conventional SiPMs, a SPAD either avalanches or it does not. It is the knowledge of how many elements avalanched that contains information on the incident radiation, hence the knowledge of the total charge produced is superfluous. Haemisch and his colleagues envisioned a technique for resetting the SPADs that relies on active switches (CMOS transistors). The reset transistors restore charge in the SPAD with high speed and produce at the same time a fast signal. From such signal it can be inferred that a given SPAD fired; additionally, fast timing (of the order of tens of ps) can be extracted from the same signal. The reset signals are digital, thus, a sum of the number of avalanching SPADs can be easily derived by means of simple digital counting circuitry. These devices have become known as “digital” SiPMs (dSiPMs) or, more appropriately, as photon-to-digital converters (PDC). A simplified explanation of the difference between SiPMs and PDCs is shown in Figure~\ref{fig:L1}.

The advantages offered by PDCs are: 
a)	The SPAD becomes effectively a digital device, thus, there is no need for analog processing at all.
b)	The digital nature of the readout is such that the device capacitance is simply irrelevant.
c)	There is no need to “match” device amplitude response across devices because there is no such thing as amplitude measurements.
d)	Excellent timing characteristics are achievable starting from the single photoelectron and up.
e)	Dark-count behavior is improved by the possibility of turning off particularly noisy SPADs. 
f)	Afterpulsing phenomena can be mitigated because the presence of a pulse is known with high precision and can be used to veto afterpulse events.

Early efforts in developing these devices resulted in their availability in some commercial medical imaging instruments. However, early developments suffered from the constraint of having to integrate the digital CMOS quenching and readout circuits into the same substrate as the SPAD array, resulting in low fill-factors. SPAD performance is, in principle, also affected due to the need to integrate the SPAD devices using a process meant for CMOS devices rather than detectors.  

\begin{figure}[thbp] 
   \centering
   \includegraphics[width=0.45\textwidth]{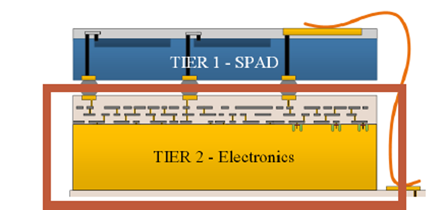} 
   \caption{conceptual cross-section of a PDC. The top tier is made by using a dedicated optoelectronic process. The bottom tier has all the digital electronics needed to read out each SPAD. }
   \label{fig:L2}
\end{figure}

Recent efforts by Pratte {\it et al.} at Sherbrooke University~\cite{L3}, and later by Torilla et. al. at INFN, Italy~\cite{L4}, address the above limitations. The work by Torilla {\it et al.} aims at improving PDC technology by using modern digital techniques in the readout. The work by Pratte {\it et al.} has seen more development and has already produced functioning PDC devices. At present, they are evolving the state-of-the-art by moving the integration of the devices in the vertical direction. In this implementation, the SPAD array is implemented in a dedicated process designed to optimize performance of the photon detector itself. The CMOS readout is developed independently on a different substrate, using CMOS-specific processes and technologies. The wafers are then processed according to a specific recipe and the layers are vertically integrated using specially developed 3D integration methods similar to those currently used in state-of-the-art interconnection techniques by the semiconductor industry at-large. Such development is bound to greatly improve the PDC fill factor and to allow tiling of many PDC devices into large to very large areas (up to square meters) while maintaining excellent timing (below 100 ps FWHM for single photoelectrons), low power consumption, and high PDE. Some of the demonstrated performance is discussed further in~\cite{L3}. Currently, each PDC consists of a 5 $\times$ 5 mm SPAD array and readout electronics such as the device shown in cross-section in Figure~\ref{fig:L2}. 

\begin{figure}[thbp] 
   \centering
   \includegraphics[width=0.45\textwidth]{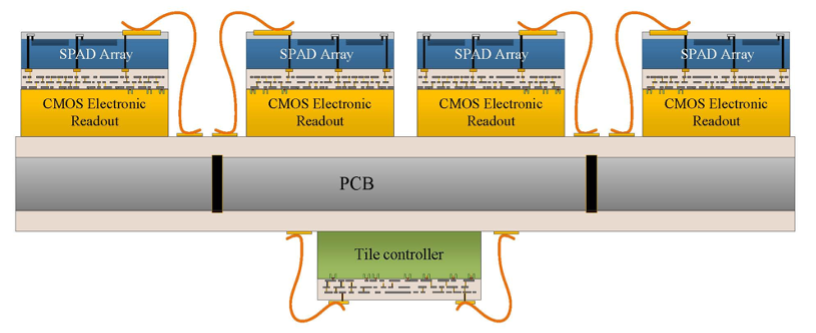} 
   \caption{ PDC tile concept. A tile controller manages currently up to 64 PDC chips. Note that wire bonds are in the process of being phased out in favor of through-silicon vias to improve fill-factor.}
   \label{fig:L3}
\end{figure}

It should be noted that an additional circuit to manage each PDC and to communicate with the external world is required. Such a device, known as a “tile controller” is also a fully digital circuit. A full conceptual tile is shown in Figure~\ref{fig:L3}.

Having fully demonstrated the technique’s applicability in fast neutron radiography, the effort is now focused on the production of PDCs using 3D integration techniques. These devices will be the basic building block for light readout instrumentation. A broader effort is planned for the near future where the 3D device design can be tailored to specific applications spaces where certain characteristics are more desired than others. For example, some application will require the fastest achievable timing but will not need exceedingly precise amplitude measurements or the lowest achievable power. Other applications will require different wavelengths than currently explored, etc. Families of devices can be constructed given the intrinsic modularity and simplicity of PDCs, making them a very desirable scintillation light detector for a wide variety of scientific and commercial applications.

\subsection{Beyond silicon sensors and CMOS integration}
It has been predicted that the theoretical limit in temporal resolution for 550 nm light is 11.1 ps in silicon~\cite{Dart:13}. Silicon (Si) sensors have some known limitations in U-RadIT applications. For example, relatively small atomic number (low-Z) of Si makes Si detectors less ideal for high-energy ($>40$ keV) ultrafast X-ray applications, since very thick Si or equivalently thousands of sensor layers may be needed for high detection efficiency~\cite{Wan:2015}.  The electron and hole drift speed in the photodiode, a critical factor in charge transport and collection, are intrinsic material properties of Si. Radiation hardness of silicon may limit the silicon detector lifetime in a high-repetition-rate and high luminosity settings of XFELs and synchrotrons.

Based on the recent studies from the ATTRACT-ERDIT initiative, fast-timing-based U-RadIT require spatial resolution of 10 \textmu m and temporal resolution of 10 ps. Another requirement is to cover large detection areas~\cite{Cin:1}. Fast and low-cost scintillators are widely used. Metamaterials, graphene and 2D materials are also being pursued by various groups. 
Metamaterials define a family of compounds which are a combination of repeating patterns of plastic, metals and crystals at scales that are smaller than the wavelengths of the phenomena they influence. Meta-surfaces are mainly divided into resonance and waveguide but can be further classified into several classes depending on their operational wavelengths and application design. Resonance meta-surfaces are classified into plasmonic and Mie resonance-based on the all-dielectric types. Their precise shape, geometry, size, orientation, and arrangement give them their smart properties capable of manipulating electromagnetic waves by blocking, absorbing, enhancing, or bending waves, to achieve benefits that go beyond what is possible with conventional materials. The application of metamaterials for radiation sensing is at the emerging stage. An application which is already commercially available is the use of metalenses in combination with silicon photomultipliers’ arrays proposed by Hamamatsu, to improve their performance~\cite{Cin:11}. Their presence allows focusing the light generated by scintillators to the central section of the SPADS allowing an increase of fill factor from 50\% to ~82\% by using circular transmission nanopillars made of Hafnium Oxide (HfO2) on a 300 \textmu m thick glass substrate to minimize absorption in the Near UV. The interest in using such configuration is because circular transmission nanopillar varies the locally effective index by changing the diameter of the meta-surface and does not depend on incident polarization.

After the work on graphene which awarded the Nobel prize to Geim and Novoselov in 2010, a lot of work have been dedicated on so called 2D materials, or monolayers of elements with exceptional properties. One of the work lines have been the combination by layering of such materials to create new 3D materials with exceptional properties useful for macroscopic applications. One of the most interesting advancements for detectors and electronics was the creation of semiconducting multi-layers with large enough bandgaps and sufficiently large areas to be used at room temperature. Hexagonal Boron Nitride (hBN) is one of the most promising candidates combining “2D materials beyond graphene” with a bandgap of 5eV~\cite{Cin:12}. There are several strategies to assemble 2D materials into integrated functional nanostructures. An example is given for optoelectronics applications, where two graphene layers are separated by several layers of boron-nitride, which serve as a tunneling barrier.  A built-in electric field, created by the proximity of one of the graphene layers to a monolayer of MoS$_2$ (molybdenum di-sulfite), separates the electron–hole pair, which is created by an incoming particle~\cite{Cin:13, Cin:14}. Another interesting application of graphene was proposed where the property of graphene to exhibit a sharp change in resistance as a function of applied field, near the charge neutrality point ("Dirac point") is used. The authors proved the principle using different substrates (Si, SiC, GaAs) with a layer of graphene deposited on its top surface and connected electrically to the substrate’s electrical circuit. They proved that the variation of electric field produced by the release of energy from an impinging particle in the diode would produce the sharp change in resistance in graphene and therefore generate a signal~\cite{Cin:15}. The response of a layer of graphene deposited on 20nm of SiO2 as a photodetector was also investigated with a 1.55nm laser beam. The intrinsic response time of the generated photocurrent was $\sim$ 2.1ps using a second order interference generated using a second laser~\cite{Cin:16}.

Non-scintillator materials such as bismuth silicon oxide (BSO) and cadmium telluride (CdTe) have recent been reported for fast timing applications~\cite{TCJ:2023}. The complex refractive indices of these materials can changed on the order of femtoseconds when exposed to ionizing radiation, as revealed by the LCLS measurements. In addition to materials research and discovery, other enabling elements include additive manufacturing, better known as 3-D printing,  micro- and nanofabrication and novel integration and packaging methods. 

{\it 3D printing} is an emerging strategy in the fabrication of particle detections systems. There are three fundamental printing methodologies used depending on the material being used. Fused Deposition Modelling (FDM), a good option for open space devices, involves the extrusion by heating of solid thermoplastic filaments, usually polylactic acid (PLA) or acrylonitrile butadiene styrene (ABS). Composite materials are also available~\cite{Cin:2}. Vat polymerization Stereolithography (SLA) and Digital Light Processing (DLP) enable fabrication of both open or closed devices and employ UV-curable epoxy or acrylic-based resins. The liquid resins are layer-by-layer photopolymerized on a movable solid platform by a scanning laser or a fixed UV-light source for SLA ad DLP, respectively~\cite{Cin:3}. Photopolymer inkjet printing (PIP, often referred to as PolyJet or Multi\&ProJet work well for both open and closed geometries and use SLA/DLP photocurable liquid resins, but with the option of one-step multi-material printing~\cite{Cin:4}. The CERN based ``3D printed Detectors” (3DET) collaboration was formed in 2019 with the goal of investigating and developing additive manufacturing as a new production technique for future organic scintillator-based particle detectors. Natural applications for such production in big scintillating detectors with complex geometries, for example future neutrino detectors, calorimeters, or neutron detectors. Preliminary results of produced scintillators are showing attenuation lengths of about 20cm, and the possible simultaneous fabrication of reflective coatings used to enhance the light output~\cite{Cin:5}. Plastic scintillators were also used for gamma-ray detection at 477 keV. This energy is particularly interesting since it is close to the 511 keV energy of the gamma ray emitted during positron emission tomography. The results presented in~\cite{Cin:6} show a light output of 67\% and a transmittance of 74\% when relative to a commercial BC408 scintillator. Other measured parameters include an average decay time constant of 15.6 ns, an intrinsic energy resolution of 13.2 keV at 477 keV and intrinsic detection efficiency of 6.81\% at 477 keV. Another material used successfully as an additive detector is perovskite. It was reported that the growth of perovskite onto a graphene substrate created an ultrasensitive x-ray detector. The results reported in~\cite{Cin:7} show that detection was possible at dose rates below 1 \textmu Gy/s with photocurrent responses as a function of time at 100mV bias voltage. The key for this extraordinarily high sensitivity was due to the combination of 600 \textmu ms walls of aerosol jet printed perovskite methylammonium lead iodide (MAPbI$_3$) on graphene. 3D printed was also attempted for silicon sensors. In~\cite{Cin:8} the authors report successful results in printing micro and nano silicon structures without the need of a cleanroom. In this case the layer-by-layer fabrication was based on alternating steps of chemical vapor deposition of silicon and local implantation of gallium ions by focused ion beam (FIB) writing. In a final step, the defined 3D structures were formed by etching the silicon in potassium hydroxide (KOH), in which the local ion implantation provides the etching selectivity. The method was finally demonstrated by fabricating 3D structures made of two and three silicon layers, including suspended beams that were 40 nm thick, 500 nm wide, and 4 \textmu m long, and patterned lines 33 nm wide. 3D printed complex inorganic polycrystalline scintillators based on Yttrium Aluminium Garnet doped with Cerium (YAG:Ce) was also successful following data presented in~\cite{Cin:9}. A green glowing body was printed using a stereo-photolithography approach from co-precipitated powders which were then sintered at 1600 °C in air to afford translucent ceramics. The paper reports that the scintillation light yield using 5.5 MeV $\alpha$-particle excitation was more than 60\% higher than that of the reference YAG:Ce single crystal. This was possible due to the higher scintillation light yield due to high activator (Ce) concentration possible during the preparation of the YAG:Ce powder prior to 3D printing and which is impossible in monocrystalline YAG-Ce, the concentration being only  0.1-0.5\%. High resolution 3D printing is also opening interesting venues in electronics fabrication. The authors of~\cite{Cin:10} obtained electronic materials in conductors, semiconductors, and insulators, in 3D printing methods by using ink-jet printing, direct writing and photocuring and in 3D printing of device components with interconnects, batteries, antennas and sensors being covered. A particularly interesting and timely methodology is the one used in liquid-metal-based flexible and stretchable electronics with potential applications in, multilayer circuits, soft sensors, and pressing-on switch. 

{\it Microfabrication} Microfabricated 3D sensors, where electrodes penetrate the silicon bulk due to cylindrical holes, started a new field in fast and radiation hard silicon at the beginning of the years 2000 with applications in high energy physics ~\cite{Cin:17} and were used in the construction of the Insertable B-Layer in ATLAS~\cite{Cin:18}. Many recent developments include unprecedented radiation tolerance in the 10$^{17}$ neutron equivalent per square cm ~\cite{Cin:19} and 11ps time response using trench electrodes~\cite{Cin:20}. Microfabrication technology is also used in many other applications beside radiation detection. Micro channels for thermal management or for microfluidic testing in biology are among few examples. Thermal management in very dense electronics interconnects also used microfabricated vias in the attempt to reducing voltage loss and excess heat by moving the power delivery network on the wafer’s back side in electronics packaging~\cite{Cin:21}. Nanofabrication is also playing a key role in reducing the dimensions in fin-fet transistors and nano-electro-mechanical switches~\cite{Cin:22}. 

{\it Integration.} The ultimate speed, however, was reached in data processing using wafer interconnectivity for the generation of superfast chips where cores are not separated after processing and are interconnected at wafer level.  The so called “silicon interconnect fabric” allow bare chips also known as “chiplets” or “dielets” interconnections faster at larger dimensions by using Silicon wafers as support rather than PCB-SoC System on Chip~\cite{Cin:23}. The Cerebra AI supercomputer is an example of such innovative methodology.  The dimension of the wafer is of 46,255 square millimetres with 1.2 trillion transistors, 400,000 processor cores, 18 gigabytes of SRAM, interconnects capable of moving 10$^{17}$ bits per second making this chip more than 10$^4$ times faster than a GPU. AI neural networks that previously took months to train can now train in minutes. As a comparison, the Joule Supercomputer costs tens of millions of dollars to build, with 84,000 CPU cores spread over dozens of racks, and it consumes 450 kilowatts of power. The Cerebra computer is 200 times faster, costs several million dollars and uses 20 kilowatts of power~\cite{Cin:24}.

\section{Methods and U-RadIT modalities \label{sec:mm}}
Here we emphasize `lens-less' modalities in U-RadIT since a.) For X-ray energies above 20 keV (wavelength 0.062 nm), focusing lens is technically difficult to fabricate;  b.) Use of lenses always introduces image blur and aberration that are hard to correct in data analysis; and c.) Data methods such as computational imaging potentially allow `refocusing' after the image data have been collected, or `digital refocusing'~\cite{PDT:2015}. We first discuss different forms of dynamic X-ray phase contrast imaging (X-PCI) in Sec.~\ref{sec:XPCI}, followed by X-ray dynamic diffractive imaging, Sec.~\ref{sec:diff}, and move contrast imaging, Sec.~\ref{sec:mci}.  Other possible novel methods may include the use of collimators, coded apertures, hard X-ray beam splitting through Laue diffraction~\cite{VPP:2018}, kinoforms, ultrafast (300 ps) photonic micro-systems to manipulate hard X-rays~\cite{CJW:2019}, incoherent diffraction imaging at high photon energies~\cite{TAC:2020}, and three-dimensional X-ray micro-velocimetry~\cite{LFU:2010}. For charged particle and neutron modalities, we highlight the recent advances in solid-state (primarily silicon) ultrafast detectors that enabled 4D charged particle tracking, Sec.~\ref{sec:4D}, as an alternative to focusing (magnetic, for example) lenses and flux integration detectors~\cite{MKK:2013}.%

\subsection{Dynamic X-ray Phase Constrast Imaging \label{sec:XPCI}}
\begin{figure*}[thbp] 
   \centering
   \includegraphics[width=0.85\textwidth]{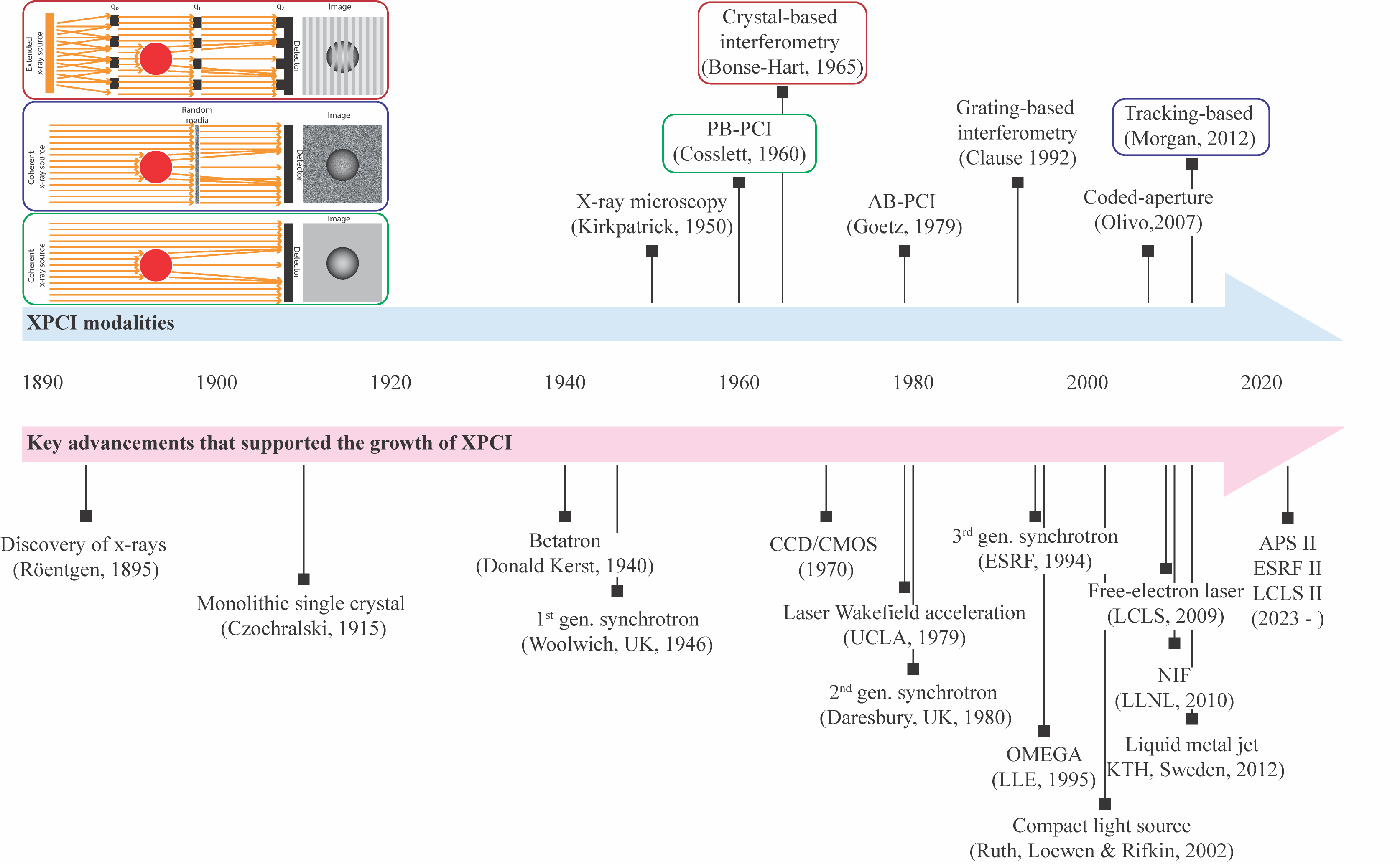} 
   \caption{Evolution of XPCI modalities enabled by the advances in X-ray sources. Three popular modalities of XPCI: a.) Grating-based interferometric method; b.) Speckle-based XPCI; and c.) propagation-based XPCI are highlighted in the upper left corner.}
   \label{fig:A1}
\end{figure*}

X-ray phase contrast imaging (XPCI) replies on X-ray attenuation and phase modulation by an object to form image contrast. A growing number of XPCI modalities has been reported as shown in Fig.~\ref{fig:A1}. XPCI is highly sensitive to biological soft tissues and organic chemicals, which generate low contrast by X-ray absorption methods~\cite{AM:2005}. X-ray phase information can not be measured directly. However, X-ray phases can manifest as intensity variations due to interferences at the imaging detector location. The synchrotron and XFEL sources allow ultrafast or dynamic XPCI with very high spatial and temporal resolution that are hard to obtain with compact laboratory sources~\cite{WNG:2015}, with a possible exception of compact high-power laser-driven sources~\cite{ESB:2015,barbato2019quantitative}, which will continue to get brighter with higher laser power.

X-ray wavefront modulation by an object is described by the photon-energy-dependent complex refractive index at position {\bf r}, $n (\nu, {\bf r}) = 1 -\delta (\nu, {\bf r}) - i\beta (\nu, {\bf r}) $, with $E = h \nu$ being the X-ray photon energy, and real numbers $\delta (\nu, {\bf r}) $, $\beta (\nu, {\bf r}) \ll 1$. XPCI is therefore sensitive to both the real (phase shift) and imaginary (absorption) part of the complex refractive index.   One of the requirements of XPCI is partial spatial coherence, which is readily met by synchrotrons and XFELs. Otherwise, the use of gratings, coded-apertures, or speckle patterns can also modulate the wavefronts and produce phase contrast. Here we highlight several XPCI configurations and {\it quantitative image analysis methods} actively utilized for studying dynamic events: (a.) grating-based interferometric XPCI; (b.) speckle-based XPCI, and (c.) propagation-based XPCI. 
Other state-of-the-art XPCI modalities are coded-aperture imaging~\cite{A58}, ghost imaging~\cite{A59} and time-resolved XPCI microscopy~\cite{A60}. 

Grating-based interferometric methods exploit the Talbot effect and employ one or more gratings to visualize differential phase gradients. In the three-grating geometry~\cite{Clauser:1992,PWB:2006}, Fig.~\ref{fig:A1}, 
the source grating converts an incoherent X-ray source into an array of coherent X-ray beams before the second grating (beam-splitter) transforms the x-ray beam into a Talbot pattern. The Talbot pattern recurs periodically along the X-ray beam path due to Fresnel propagation. A detector records the Talbot pattern downstream. Inserting an object in the X-ray path distorts the Talbot pattern that can then be measured to retrieve the object phase. However, detectors often cannot resolve the distortion and thus the third grating (analyzer) is positioned in front of the detector to create a Moiré pattern. The detector can resolve the Moiré pattern but requires phase stepping, where multiple images are acquired, to recover the object phase.  This makes dynamic imaging challenging. 

Valdivia {\it et al.}~\cite{A2} developed a method combining deflectometry with interferometry so that only a single image is needed to recover the object phase, attenuation and dark field image. Grating-based method is also being utilized at synchrotrons and XFELs for wavefront sensing~\cite{A9, A10, A11}. Since these X-ray sources are sufficiently coherent and brilliant, only a single grid is needed to create the Talbot pattern.  Use of interferometry for imaging samples has not yet become widespread mainly because of phase wrapping. Objects possessing jumps in mass density (e.g., shock waves) and shape (e.g., microcracks) produce phase gradients larger than 2$\pi$ per detector pixel. Work is ongoing to address this limitation through a number of avenues including: (a) developing higher resolution detectors along with fabricating small pitch grids~\cite{A12}, and (b) incorporating iterative and/or machine learning into the phase reconstruction process to unwrap the phase~\cite{TKF:2017,A13}.

Speckle-based XPCI use a single random mask, such as a paper, to modulate the X-ray wavefront from a synchrotron or a laboratory X-ray source~\cite{A15, A19}. A speckled pattern is created at the detector plane, and the pattern is distorted when an object is inserted. Speckle-based XPCI operates within the near-field regime where sample-induced phase gradients diffeomorphically displaces the speckle pattern. The existing iterative and other methods such as neural networks for image analysis and phase retrieval can be time consuming~\cite{A19}. Furthermore, current reconstruction methods assume that the second or higher order derivatives of the object phase are zero, which restrict speckle-based XPCI  to studying slowly varying objects. Wang et al.~\cite{A22} addressed this restriction by recording an additional image of the sample without the random mask, but this may still not be usable for single-shot dynamic studies. Iterative and neural network methods are actively being pursued to ameliorate this limitation~\cite{A19,Qiao:2022} 

Propagation-based XPCI (PB-XPCI) has been by the far most popular XPCI modality for studying dynamic events due to its simplicity. It requires only a spatially coherent source behind the sample and uses free-space propagation to convert the sample phase into intensity modulations. Aside from its simple setup, PB-XPCI images directly represents the object with the addition of edge enhancements that allows direct interpretation possible without needing the images to be reconstructed. On the other hand, interferometry and speckle-based XPCI images requires phase retrieval to separate the speckle or grating pattern from the object before the images can be interpreted. As a result, PB-XPCI has been deployed in preclinical~\cite{Morgan:2020}, materials science~\cite{Chen:2014, Parab:2016, Leong:2018, Wainwright:2019}, fuel injection~\cite{Sforzo:2019, Zhao:2021}, fusion energy~\cite{Montgomery:2016, Hodge:2021, Hodge:2022, montgomery2023}, and shock physics~\cite{BIC:2017,Yanuka:2019,D3}.  
\subsection{Dynamic Diffractive imaging \label{sec:diff}}
{\it X-ray topography} based on Bragg diffraction (reflection geometry) and Laue diffraction (transmission geometry) are widely used in synchrotrons to study crystal defects such as dislocations, stacking faults, inclusions, and surface damage~\cite{BlackNIST,DanilewskyCRT:20}. \mbox{X-ray} topography using laboratory \mbox{x-ray} sources for \emph{in situ} studies of materials  dates to the 1960s~\cite{ChikawaAPL:68}. Synchrotron sources and indirect detection schemes using optical cameras viewing a scintillator crystal 
enabled time-resolved or dynamic X-ray topography~\cite{TuomiNIM:83,RackJXST:10,DanilewskyJCG:11}. In time-resolved imaging both the exposure time and the frame rate are important. Exposure time (together with the effective pixel size of the detector) places an upper limit on the speed of features that can be observed without unacceptable motion blur. Frame rate sets a limit on the duration of events for which the velocity can be accurately measured. If the duration of an event is shorter than the time between two frames, then the velocity calculated from two successive images will be the mean velocity over the frame time and is thus a lower bound on the instantaneous velocity of the event. 

An example of these considerations is the observation by Rack and coworkers of crack propagation in silicon due to thermal strains ~\cite{RackIUCRJ:16}. Their images were formed using \mbox{x-rays} from single electron bunches from the synchrotron with a duration of around \SI{100}{\pico\second} and an effective pixel size of \SI{62}{\micro\meter}, so motion blur would only be expected to be a problem at speeds around $\SI{62}{\micro\meter}/\SI{100}{\pico\second}\simeq\SI{6e5}{\meter\per\second}$, comfortably below the expected speeds of around \SI{3e3}{\meter\per\second}\cite{ShermanIJF:06}. On the other hand, with a frame time of \SI{28}{\micro s} the fastest speed they could measure was around $\SI{62}{\micro\meter}/\SI{28}{\micro s}\simeq\SI{2}{\meter\per\second}$, although they could show that the motion was intermittent. Indeed, in subsequent work crack propagation at speeds of up to $\SI{2.5e3}{\meter\per\second}$ was observed with single-bunch diffractive imaging at frame rates of $\sim\SI{1}{\mega\hertz}$~\cite{PetitJAC:22}. Similar considerations will apply to imaging of dynamic phenomena such as propagation of cracks, shock waves, and phase transformation fronts at speeds of $\qtyrange{e3}{e4}{m.s^{-1}}$ (or higher). It is worth noting, however, that because the features observed in \mbox{x-ray} topography can be rather diffuse, motion blur may not be as important here as it is in radiography or phase contrast imaging.

\mbox{X-ray} topography offers some intriguing possibilities for imaging of dynamic phenomena in crystalline solids. First, because topography is sensitive to lattice strains around defects, it could be invaluable in situations where the defect itself is invisible to, or cannot be resolved by, radiography or phase-contrast imaging. Second, it may be possible to obtain information from several diffracted beams simultaneously, providing new insights into phenomena such as the dynamics of crack propagation. By providing multiple points of view of the phenomenon of interest, diffractive imaging with several beams simultaneously could, perhaps, provide a kind of back-door approach to determining the 3D structure of rapidly-evolving features. A first step along these lines is in the experiments mentioned above on fracture of silicon, in which both topographic images and in-line phase-contrast images were recorded simultaneously~\cite{RackIUCRJ:16}.

{\it Dark-field X-ray microscopy} One limitation of \mbox{x-ray} topography 
is that the spatial resolution of the measurement is determined by the effective pixel size of the detector system (accounting for magnification by objective lenses and geometrical projection effects). For single-bunch radiography and phase-contrast imaging this is about \qtyrange{2}{3}{\micro\meter}~\cite{LuoRSI:12}, but for diffractive imaging performed to date it has typically been somewhat larger (\qtyrange{150}{300}{\micro\meter}~\cite{PetitJAC:22}), presumably due to the lower intensity of diffracted beams compared to the direct beam. 

\begin{figure}
\centering
\includegraphics{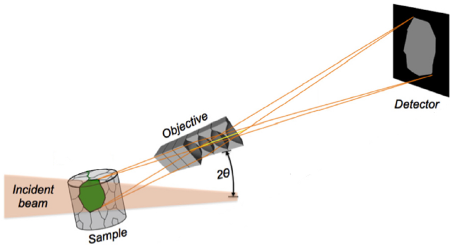}
\caption{In dark-field x-ray microscopy an objective is placed in the diffracted beam to provide a magnified view of a crystal grain. (Adapted from Ref.~\cite{PoulsenCOSSMS:20}.)
\label{fig:dfxm}}
\end{figure}

One way to overcome this is to place an objective lens in the diffracted beam, leading to the technique known as dark-field \mbox{x-ray} microscopy (DFXM) (Fig.~\ref{fig:dfxm})~\cite{SimonsNC:15,PoulsenCOSSMS:20}. Besides magnification, DXFM allows isolation of an image from a single grain embedded in a polycrystalline solid and imaging of specific components of strain. For example, Laanait and coworkers achieved spatial resolution of $\sim\SI{70}{nm}$ and temporal resolution of a few tens of milliseconds in their study of ferroelectric response of thin films~\cite{LaanaitASCI:17}, while Bucsek and coworkers used DFXM to study thermally-induced martensite formation in a single buried grain in a Ni-Ti polycrystalline specimen with similar spatial and temporal resolution~\cite{BucsekAM:19}.

As an example the use of DFXM to study dynamic processes, Dresselhaus-Marais and coworkers imaged the evolution of dislocation structures in a buried aluminum grain as the temperature was increased towards the melting point. They achieved $\sim\SI{300}{nm}$ spatial resolution, integration time of \SI{100}{ms}, and frame time of \SI{250}{ms}~\cite{DresselhausMaraisSA:21}. One limitation of DFXM is that the relatively low efficiency of the objective lens, coupled with the use of narrow-bandpass monochromatic \mbox{x-rays}, limits the opportunity for time-resolved studies using synchrotron sources. The recent development of multilayer Laue lenses as objectives (replacing compound refractive lenses) has improved this situation~\cite{KustalIOP:19}, and continued advances in \mbox{x-ray} sources and detectors will make possible experiments with better temporal resolution. For example, recent simulations~\cite{HolstadJAC:22} suggest that it should be possible to image strain wave propagation in diamond using an XFEL as the source, and proof-of-concept experiments have demonstrated DFXM image formation with single XFEL pulses~\cite{DresselhausMaraisARXIV:22}.

{\it CDI} Fraunhoffer or far-field diffraction of coherent X-rays (and electrons and neutrons) leads to coherent diffraction imaging (CDI), corresponding to a small Fresnel number, 
\begin{equation}
\mathcal{F} \equiv \frac{a^2}{z\lambda} \ll 1,
\end{equation}
where $\lambda$ is the wavelength of the particle/photon, $a$ is the characteristic interaction length that can change the direction of particle/photon propagation, and $z$ the distance between the location of interaction and the detector. For electrons, the interaction length is very small due to Coulomb force, and therefore electron diffraction pattern in an electron microscope can be measured at a relatively short distance. The diffraction distance $z > 100 $ m may be necessary for high-energy photons $E >  20 $ keV, assuming $a = 100$ \textmu m. 

One or more oversampled diffraction patterns from an object of interest are recorded in CDI and the real-space structure determined by Fourier transformation. This requires of both the amplitude and phase of the diffracted \mbox{x-rays}. The amplitude can be calculated directly from the measured intensity, while the phase is recovered separately by means of an iterative algorithm~\cite{ChapmanNMat:09,NugentAdvPhys:10,MiaoScience:15}. This gives CDI the ability to reveal the 3D structure of individual objets with sub-nanometer resolution~\cite{MiaoScience:15}.

There are several variations of CDI, but for dynamic studies the most relevant is plane-wave CDI which requires only a single \mbox{x-ray} diffraction pattern to recover a 2D projected image. 
For dynamic studies, Chapman and coworkers provided proof-of-concept demonstrations of image formation using single XFEL pulses, including synchronization with pulses from a pump laser for time-resolved studies~\cite{ChapmanNP:06,BatyNPhot:08}. The extreme power density (\SI{4e13}{\watt\per\centi\meter^2}) in these experiments destroyed the samples --- but not before diffraction patterns could be recorded, allowing reconstruction of 2D images. At lower power densities (with synchrotron radiation) time-resolved CDI can be used to study reversible phenomena in a pump-probe scheme, for example in imaging of surface acoustic waves~\cite{NicolasJAC:14}. Further dynamic studies along these lines will benefit greatly from the development of fourth-generation synchrotron sources with improved coherent flux, as well as new XFEL sources with higher repetition rates~\cite{KovalchukCR:22}.

In Bragg coherent diffraction imaging (BCDI) multiple diffraction patterns in the wide-angle regime are collected as the crystal is rotated, to develop a 3D map of diffracted intensity around (usually) a single reciprocal lattice point. Again the phase is recovered computationally, permitting reconstruction of the 3D real-space structure including (for example) spatially-resolved determination of the lattice strain as well as the presence of defects such as dislocations, a capability that is useful for \emph{operando} studies of battery cathode materials~\cite{SingerNatE:18,LiuNature:22}, ferroelectrics~\cite{SimonsNMat:18}, and in catalysis~\cite{KimNatComm:18}. With pump-probe techniques BCDI can achieve picosecond temporal resolution, for example in imaging acoustic phonons in gold nanocrystals~\cite{ClarkScience:13}. An important recent development is the demonstration of optical trapping to hold particles in suspension during BCDI, instead of immobilized on a substrate or embedded in a solid~\cite{GaoPNAS:19}. However, the need to collect multiple diffraction patterns (to sample the 3D shape of the reciprocal lattice points) will continue to restrict BCDI to either pump-probe experiments of reversible phenomena, or to quasi-static or slowly-evolving situations.
\subsection{ Move contrast X-ray imaging \label{sec:mci} }
Even with the modern light sources such as synchrotrons and XFELs, further improvements in small feature detection sensitivity and contrast are still needed in imaging of complex systems and their dynamics~\cite{CC1,CC3}. Methods such as phase contrast~\cite{CC4}, contrast labeling~\cite{CC5}, K-edge subtraction~\cite{CC6}, and time subtraction~\cite{CC7} are often used for contrast enhancement and extraction of weak signals in complex systems, but they are still prone to motional blur and high-frequency noise. 

Move contrast X-ray imaging (MCXI)~\cite{WZL:2020}, which takes advantage of the time evolution of modulation of each moving component to incident light field in a complex system, can differentiate the components and image them separately. Accordingly, the mutual interference between components is eliminated and the sensitivity to weak signals is improved significantly. Experimental results of angiography with low agent dose~\cite{WZL:2020}, agent-free imaging of water refilling along microvessels in plant branch~\cite{XLX:2023}, sensitive tracking to ion migration in an electrolytic cell~\cite{JLY:2022}, and ultrafast imaging of cavitation evolution, demonstrate the practicability of move contrast X-ray imaging of weak signals in complex systems while traditional methods fail.

\begin{figure}[thbp] 
   \centering
   \includegraphics[width=0.45\textwidth]{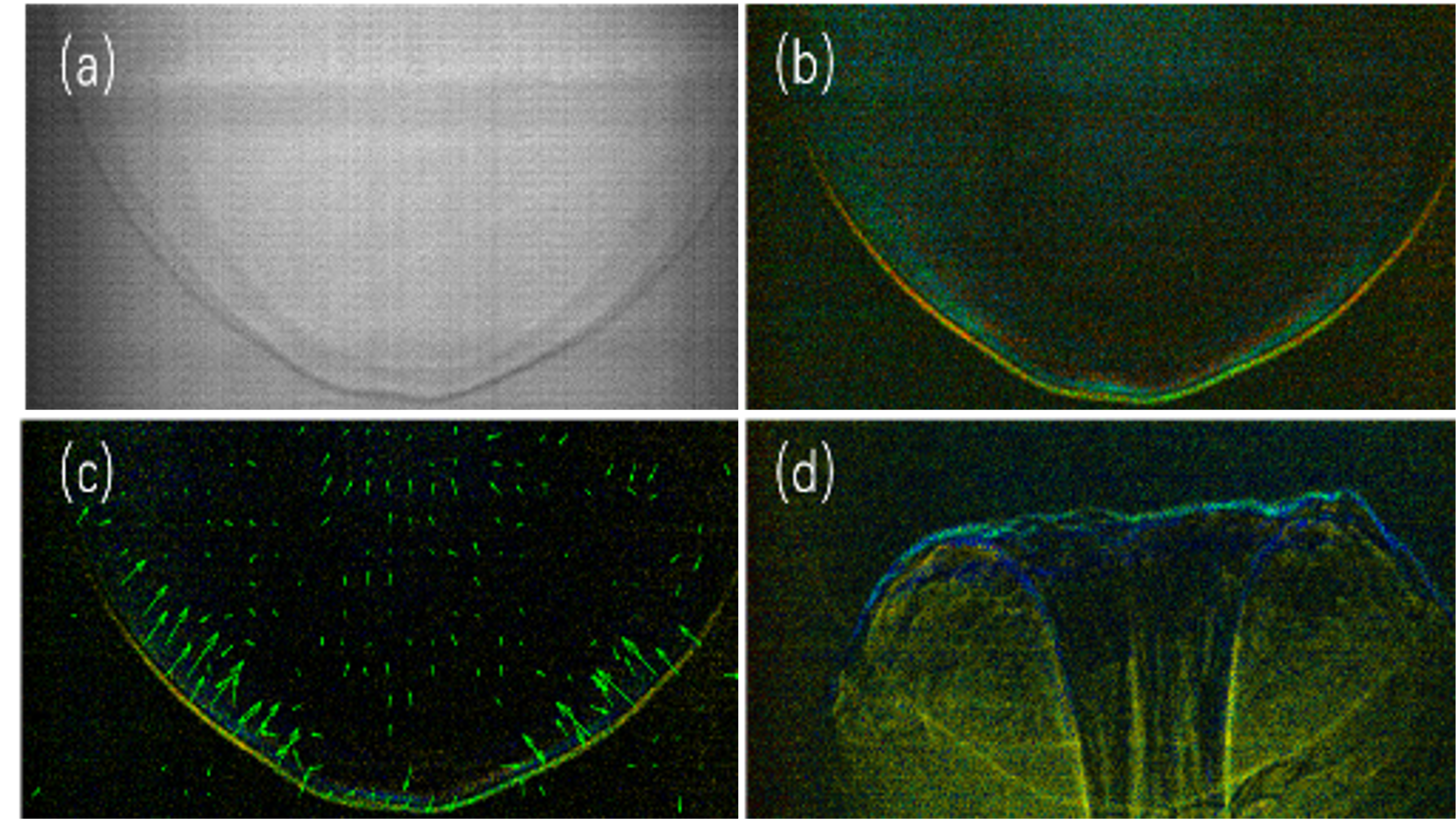} 
   \caption{MCXI of evolution of a laser-induced cavitation. Where (a) frame No.115 of the original image of cavitation evolution; (b) cavitation contraction and center re-excitation stage, starting frame No.97; (c) movement trend diagram of cavitation contraction stage; (d) cavitation rupture and jet flow stage, starting frame No.133.}
   \label{fig:CC1}
\end{figure}

{\it Laser induced cavitation} The pulsation of vacuoles is a complex physicochemical process, which is the key point in the study of fluid mechanics and cavitation dynamics~\cite{BBP:2023}. At present, laser-induced cavitation is an important means to study the cavitation mechanism~\cite{CC14}. Using synchrotron radiation X-ray as the illumination source will greatly improve the poor contrast of other imaging methods for cavitation evolution imaging, especially for X-ray imaging facility with high photon flux density where the undulator is used as an insert, and with detectors with high-performance scintillators and high optical magnifying lenses. As a result, the spatiotemporal resolution of X-ray imaging can be greatly improved. The experiment was carried out at the ultrafast X-ray imaging beamline (16U2) of SSRF, in which the pulsed laser was focused in water. White X-ray beam was used to ensure the SNR of the high-speed X-ray camera. The X-ray detector system has an effective pixel size of 3 \textmu m and records a total of 180 frames at a frame rate of 0.5 Mfps. Since this process changes very fast, the period of 30 frames of 58 ms already contains rich dynamic changes. Therefore, to avoid the complex confusion of motion features during data processing, a specific frame was selected as the starting frame, and 30 frames was continuously selected for MCXI processing. The key to the study of cavitation mechanism lies in the study of cavitation pulsation, which in liquid mainly goes through growth-rupture-jet stages~\cite{CC15}.
In the MCXI results, the contraction of the vacuole and the final excitation pulse of the vacuole core were observed to be a simultaneous process. That is, the outer wall of the vacuole began to rebound and contract, and at the same time, the core area of the vacuole stimulated new pulsations, as shown in Fig.~\ref{fig:CC1} (b), where the original data starting frame was the 97th frame. The phase parameters of the move contrast were pseudocolorized from red to green to blue, that is, different colors were used to represent the temporal characteristics of the material motion. It can be seen from the figure that the contraction of the vacuole also starts from both sides. When the contraction wave front of the outer wall meets and collides with the excitation wave front of the center, the energy steady state in the vacuole disappears and the jet is generated, as shown in Fig.~\ref{fig:CC1} (d), where the original data starting frame was the 133rd frame.

However, in the original image at the corresponding time shown in Fig.~\ref{fig:CC1} (a), it is difficult to distinguish the re-excitation process at the core of the vacuole. Moreover, the faster imaging frame rate leads to a lower image SNR. This feature may cause the subsequent extraction and analysis of cavitation dynamic evolution parameters to be hindered. The move contrast phase parameter represents the temporal information of the material motion, so the derivative of the phase parameter can characterize the direction of the material motion. The direction information can be added into the move contrast image in the form of an arrow to more directly observe the changing trend of the system. Fig.~\ref{fig:CC1} (c) shows the direction of motion when the cavity begins to contract. This figure further shows that the energy distribution on the outer wall of the cavity is not uniform. The cavity first contracts from both sides, and then the jet is ejected from the weak point of energy, pointing to the area of the contraction lag of the outer wall, as shown in Fig.~\ref{fig:CC1} (d). In the future, more exciting advantages and effects are expected in more MCXI practices of ultra-fast processes.

The angiography of animal model, water refilling along microvessels of willow branch, electrolytic reaction and laser induced cavitation are all typical representatives of complex systems. The successful imaging of weak signals in the above systems by MCXI method shows that this method has significant advantages in improving imaging sensitivity and SNR compared with traditional direct imaging methods. However, this method is still being improved, optimized and tried in more complex systems in different fields and with different characteristics. At present, MCXI methods mainly focus on time-domain signals of imaging systems. However, material motion not only has time-domain properties, but also has rich spatial characteristics, so space-time fusion MCXI method should be the main direction of future development. 
\subsection{4D particle tracking \label{sec:4D}}

Ultra-fast silicon detectors with tens of millions of pixels are now available to track charged particles in four dimensions (4D), or 3D in position + 1D in time~\cite{SSC:2017}. Timing information is essential for applications in {\it e.g.} high-luminosity large hadron collider (HL-LHC) since the density of the particle tracks is very high and significant track overlaps in 3D space is expected. The system can measure trajectories of charged particles with 10-30 ps in timing resolution and about 10 \textmu m in position accuracy~\cite{CAB:2017}. Different silicon designs, such as low-gain avalanche avalanche diodes (LGADs) and CMOS monlithic active pixel sensors (MAPS), are available for further detector optimization, which includes gain control, power consumption, pile-up reduction, radiation hardness, high fill factor up to 100\%, and investigation of the timing limits.

Even though such 4D particle tracking technology has not yet been used directly in U-RadIT, it is possible to deploy the technology for U-RadIT in several ways by tracking charged particles (ions) and neutrons. Proton radiography based on, for example,  the LANSCE 800 MeV linear accelerator, can deliver proton pulses with a timing spread less than 100 ps. Use of the tracking methods to reconstruct individual proton tracks, similar to situation in HL-LHC, after the protons passing through an object can achieve sufficient spatial resolution for proton radiography while reducing proton dose from the current approach, when the scintillator light intensity from protons are measured as signals. When laser-produced protons are used, even though protons are no longer mono-energetic, but the small spot and precise timing when the protons are emitted can help to reconstruct proton tracks for imaging applications. It may also be possible to use the 4D tracking technology for fast neutron tracking~\cite{WM:2013,CJW:2022}, by using a spallation source of fast neutrons from a high-current accelerator like LANSCE, laser-produced neutrons, or nuclear fusion neutrons recently produced in the National Ignition Facility.




\section{Data and Algorithms \label{sec:da1}}
Higher brilliance and  repetition rate in XFELs, synchrotrons, particle sources, together with commensurate higher recording rate by detectors, all contribute to the trend in generating huge volumes of data, large varieties of data, and higher rate of data generation~\cite{xfel:2020}, collectively known as `big data'. 
Harnessing big data from the state-of-the-art imaging modalities might accelerate the design and realization of
advanced functional materials~\cite{KSA:2015}.
Here we discuss data trends in U-RadIT, motivated by high-resolution ($\sim$ nm or better, such as individual transistors and other nanostructures~\cite{BSF:2019}) measurements of macroscopic structures (a fraction of 1 mm or larger, such as an ICF target), enabled by the high-repetition-rate X-ray and particle sources and high-data-yield detectors, Sec.~\ref{sec:dt}. Followed by a discussion on data compression and compressed sensing, Sec.~\ref{sec:cmr}, and example applications in sparse image capture, Sec.~\ref{sec:sic}. We then discuss the use of neural networks for phase retrieval, Sec.~\ref{sec:NN}, and end the section with a discussion on uncertainty quantification of data processing, Sec.~\ref{sec:uq}.

\subsection{Data sets: large and small \label{sec:dt}}
According to Fig.~\ref{fig2:Opt}, we may have experimental data sets, which are collected by detectors, and synthetic data sets, which are generated from theory and computation to emulate the experimental data. There are additional data sets, sometimes called `{\it meta data}', from experiments such as calibration data, and for theory and computation, such as material property data, collisional cross section data~\cite{NIST:23}, equation of state, {\it etc.} For example, to set up a NIF experiment requires about 18,000 parameters (corresponding to about 66,000 controlled devices), about 2 million software operations per shot, and 13.5 million lines of codes through about 2500 computers~\cite{NIF:22}.

There is no upper limit in the amount of synthetic data that can be generated. In practice, the amount of synthetic data is generated on demand based on the need in experimental data analysis and interpretation; {\it e.g.} to train neural networks for experimental data processing.
The amount of experimental data that is generated from an object depends on the field of view of the detector(s), spatial resolution of the detector(s), the duration of the experiment, experimental repetition rate and temporal resolution of the detector(s). For example, in a recent high-resolution mapping of a microchip~\cite{LWK:2022}, scanning an area of 82 \textmu m $\times$ 80 \textmu m took about 9861 scan positions for a duration of 1 hour and 14 minutes. The detector used was a commercial Eiger 4M detector (pixel size 75 \textmu m, array format 2070 $\times$ 2167 pixels, image bit depth 16 or 32, maximum frame rate 0.75 kHz, or a maximum data rate of  6.7 GB/s [16 bit] or 13.4 GB/s [32 bit]). Some other detectors in Table.~\ref{tab4:data} also collect data at similar rates for the continuous repetitive mode. 

XFEL facilities such as LCLS/LCLS-II are undergoing major upgrades that will enable them to operate at 1 MHz frame rate with raw data acquisition rate exceeding 1 TB/s by using megapixel and larger format cameras. LCLS-II presents several challenges for novel detectors, including expanded higher energy, increased radiation fluence, and the need for full-frame readout at the same rate as the X-ray source~\cite{HCF:2023}. In another example, APS-U will have $\sim$ 77 ns pulse spacing in the 48-bunch mode~\cite{DBB:2022}. In addition to higher data rate, higher repetition rate high-energy photon sources like APS-U may also need new brighter and faster scintillators for efficient and expedient data acquisition~\cite{HZZ:2018,HZZ:2019}. These upgrades will result in increased production of data, which require real-time analysis and automation of experiments, see Sec.~\ref{sec:cmr}.  Other U-RadIT data are collected using the burst mode, see examples in Table.~\ref{tab4:data}. The instantaneous data rate in burst can exceed 1 TB/s or 1 Tpixel/s for a 8-bit pixel.  The two data trends from imaging detectors perspective are: in continuous repetitive mode of imaging, the use of increasingly large format cameras, in conjunction with multi-beam illumination~\cite{LWK:2022}, will continue to boost data rate from 10s of GB/s today towards 1 TB/s. In burst mode imaging, increases in the number of image frames and in the solid angle of data collection will continue to push the instantaneous data rate from around 1 TB/s today to  higher rates, even though the overall volume of the burst-mode data may be small than the continuous repetitive mode of data acquisition.


\begin{table*}[!htb]
\caption{\label{tab4:data}%
A comparison of different camera data rate. The state of the art is  $>$10 Gpixel/s in continuous mode imaging. The burst mode imaging is $>$ 1 Tpixel/s~\cite{Turch:2017}.}
\begin{ruledtabular}
\begin{tabular}{lccccc}
\textrm{Detector} &
\textrm{Facility} &
\textrm{Det. Mode}&
\textrm{Array format}&
\textrm{frame-rate } &
\textrm{data}
\\
 (camera) & (particle/ &
\textrm{({\bf D}irect/}&
\textrm{(voxel size, \textmu m$^3$ /}&
\textrm{(fps/} & bits
\\
 &  photon)&
\textrm{ {\bf inD}irect)}&
\textrm{ pixel size, \textmu m$^2$ )}&
\textrm{Hz)} & 
\\

\colrule
AGIPD~\cite{BBG:2012} & Eu-XFEL & {\bf D} & 512 $\times$ 128~\footnote{AGIPD is deployed as mega-pixel/voxel cameras through tiling.} & 16 k/6.5 M~\footnote{burst mode for 352 stored frames.} & 14 \\
& (12.4 keV)&&(200$^2 \times$ 500 )&& \\
CS-PAD~\cite{PKH:2007} & LCLS & {\bf D} & 194 $\times$ 370~\footnote{CS-PAD is deployed as tiled 2, 8, and 32 modules with up to 2.3 M voxels.} & 120 & 14 \\
&(8.3 keV)&&(110$^2 \times$ 500 )&& \\
ePix100~\cite{CAB:2016}  & LCLS & {\bf D} & 384 $\times$ 352~\footnote{ ePix100 is deployed as tiled 4 modules with about 0.5 M voxels. } &  120 &  14\\
&(8.3 keV)&&(50$^2 \times$ 500)& ($\leq$ 240) &\\
ePix10k  & LCLS & {\bf D} & 384 $\times$ 352~\footnote{ ePix10K replaces CS-PAD, and  is deployed as a single, or tiled 16 modules with about 2.2 M voxels. } &  120 &  21?\\
&(8.3 keV)&&(100$^2 \times$ 500)& ($\leq$ 10$^3$) &\\
EIGER2  & APS \& others & {\bf D} & 1028 $\times$ 512~\footnote{ Eiger2 is deployed as a single, or tiled modules with more than 10 M voxels. } &  2.25 k &  16\\
(Dectris) & &&(75$^2 \times$ 450)& (4.5 k) & (8)\\
HEXITEC~\cite{VSW:2018} & DIAMOND&{\bf D} &80$^2$& 6.3 - 8.9 k & 14 \\
&2-200 keV &&(250$^2 \times$ 1000~\footnote{also 2 mm CdZnTe})&&\\
Icarus~\cite{CEF:2017} & NIF, Z & {\bf D} & 1024 $\times$ 512 & $\geq$ 250 M~\footnote{ in burst mode for 4 frames.} & 10\\
(Advanced & 0.7 - 10 keV &&(25$^2 \times$ 25)& & \\
hCMOS Sys.) &  && & & \\
Keck-PAD~\cite{G1} & CHESS & {\bf D} & 128$^2$& 10 k/10 M~\footnote{In burst-mode for 8 frames. Frames are later read out at ~1kHz. Modules are tiled into a 256 x 384 pixel and larger array.} & 12\\
&  $>$20 keV && (150$^2 \times$ 500)~\footnote{Both 500 \textmu m thick Si (for $<$ 20 keV) and 750 \textmu m CdTe (for $>$ 20 keV) versions are available.} & & \\
MM-PAD~\cite{G2} & CHESS & {\bf D} & 128$^2$& 1.1 k & 32\\
&  $>$20 keV~\footnote{Both 500 \textmu m thick Si (for $<$ 20 keV) and 750 \textmu m CdTe (for $>$ 20 keV) versions are available.} && (150$^2 \times$ 500) & & \\
SOPHIAS & SACLA & {\bf D} & 891 $\times$ 2157 & 60 & 12\\
& &&(30$^2 \times$ 500)& &  \\ \hline
HC-4502 & & {\bf inD}  & & 100 M~\footnote{burst mode, 8 frames.} & \\
(Astro)&&&&&\\
HPV-X2~\cite{Dart:5} & APS \& others & {\bf inD} & 400 $\times$ 250 & 7.8 k/5 M~\footnote{in burst mode for 128 stored frames; or 10 M Hz frame rate and 256 stored frames possible by reducing the number of pixels by half.} & 10 \\
(Shimadzu) &  10-40 keV & & (32$^2$)& &  \\
Kraken~\cite{LBC:2021} & NNSS & {\bf inD} & 800 $\times$ 800 & 20 M~\footnote{ in burst mode for 8 frames. Read noise 157 $e^-$, Full Well 4.0$\times$10$^5$ $e^-$. Buttable to larger array 2$\times$2.} & 12\\
 &  && (30$^2$) & & \\
MX170-HS & LCLS  & {\bf inD} & 3840$^2$ & 2.5~\footnote{higher frame rate can be obtained through pixel binning, at 10$\times$10 binning, the frame rate increases to 120 Hz} & 16 \\
(Rayonix) &  8-12 keV & & (44$^2$)& &  \\
PI MAX 4 & APS  & {\bf inD} & 1024$^2$ & 26~\footnote{higher frame rate can be obtained through pixel binning, at 4$\times$4 binning, the frame rate increases to 95 Hz} & 16 \\
(Teledyne) &  10-40 keV & & (12.8$^2$)& &  \\
pRAD-2~\cite{KDB:2014} &  LANSCE &  {\bf inD} & 1100$^2$& 4M~\footnote{burst mode for 10 frames.} & 12 \\
& 800 MeV && (40$^2$) & & \\
SA-Z & & {\bf inD} & 1024$^2$& 20k/120k~\footnote{reduced RoI with 512 $\times$ 256 pixels} & 12 \\
(Photron) &&&(20$^2$)&& \\
TMX 7510 & & {\bf inD} & 1024$^2$& 76k/456k~\footnote{reduced RoI with 640 $\times$ 256 pixels} & 12 \\
(Phantom) &&&(18.5$^2$)&& \\
\end{tabular}
\end{ruledtabular}
\end{table*}

\subsection{Data compression and reduction \label{sec:cmr} }
Raw data compression and reduction are becoming increasingly important to ultrafast imaging including U-RadIT as the data acquisition rate now exceeds 10s of GB/s in continuous repetitive mode or 1 TB/s burst mode, respectively. Limited by the temporary memories of imaging cameras or data transmission bandwidth, the traditional data pipeline, acquire -- transmit -- store (or store -- transmit, in burst mode) -- analyze (off-line), may no longer be feasible to handle the rapid increases in data volume, variety and rate (also called `velocity'). To address these challenges, automated data compression, automated data reduction through {\it e.g.} machine learning (ML) techniques coupled to on-detector data processing early in the data flow chain~\cite{HKO:2023}, are being developed to reduce the load on the back-end infrastructure. The modified pipelines could be acquire -- {\it reduce (compress)} -- transmit -- store (or store -- transmit, in burst mode) -- analyze, or {\it reduce (compress)} -- acquire -- transmit -- store (or store -- transmit, in burst mode) -- analyze~\cite{KTF:2023}.

In addition to classical and popular data compression algorithms such as JPEG, JPEG-2000, Fourier and wavelet transforms, the use of compressed sensing (CS)~\cite{CRT:2006, Don:2006,CT:2006}, also known as compressive sensing~\cite{Bar:2007}, compressive sampling~\cite{CW:2008}, is growing rapidly for computation-enhanced imaging, or simply computational imaging, as well as for real-time data acquisition and compression. This is in part due to the recognition that many natural signals including images are intrinsically compressible or sparse. In other words, the number of non-redundant parameters needed to describe the signals or objects in an imaging scene are relatively small compared to the degree of the freedom that the signals or images reside in. For example, for a one-million pixel image, the number of non-zero pixel value could be only 10$^4$, or only 1\% of the total number of pixels. Ideally, only the 10$^4$ pixels should be acquired (digitized), stored and transmitted. This down-sampling is indeed possible for X-ray imaging, {\it e.g.} as described in a compressed sensing X-ray camera design with a multilayer architecture~\cite{WIL:2018}. A second motivation for CS is that in many U-RadIT experiments, as shown in Fig.~\ref{fig1:setup}, the number of the line-of-sight and field of view are very sparse, and often limited to a single line of sight, and a small solid angle due to the high cost of accelerator-driven sources and the experiments themselves. A third motivation for CS application in U-RadIT is that the growing number of algorithms and more accessible computing power that allow computational RadIT.

A central problem in CS is to invert a classical underdetermined matrix equation in linear algebra,
\begin{equation}
A x + b = y,
\label{eq:sparse}
\end{equation}
where $x$, a $n \times 1$ matrix (vector), is the unknown.  Here $A$ is a $m\times n$ measurement matrix, $b$, a $m \times 1$ matrix, is the noise (also unknown on most occasions, or at least hard to describe quantitatively due to statistical and random fluctuations), and  $y$ is a $m \times 1$ matrix, representing the results of the sparse measurement. $n = 10^6$ for a mega-pixel image, and $m \ll n$ (the number of equations is far less than the number of unknowns). Here we use an example of a sparse 2D image to illustrate the CS framework and algorithms. Object recovery in U-RadIT usually involves a 3D object or a 4D object (time-dependent 3D object), and the corresponding measurement and noise are 2D images. There is no difficulty in expressing a 3D or 4D object in terms of a $n \times 1$ matrix and the corresponding 2D measurements and noise in terms of $m \times 1$ matrices, with $n$ being a much larger number than 10$^6$ (2D case), depending on the spatial and temporal resolution.

When the measurement noise is ignored, $b =0$, the least square solution ($x^\sharp$) as given by minimizing the $l^2$-norm, $\|x \|_2$~\cite{KTF:2023}, subject to $Ax = y$ (or $\| Ax-y \|_2 \leq \epsilon$ for $b \neq$ 0), is
\begin{equation}
x^\sharp \equiv {\rm argmin}_{x:Ax=y} \| x\|_2 =  (A A^*)^{-1} A^* y,
\end{equation}
with $A^*$ being the hermitian transpose. $x^\sharp $ based on minimizing the least square is not always effective in practice, see {\it e.g.}~\cite{JPC:2012}. If the most sparse solution ($x^*$) is needed, {\it e.g.} as motivated by physics or other considerations, the solution to $x$ is given by minimizing the $l^0$-norm, $\|x \|_0$, subject to $Ax = y$,
\begin{equation}
x^* \equiv {\rm argmin}_{x:Ax=y} \| x \|_0.
\label{eq:sparse1}
\end{equation}
However, solving Eq.~(\ref{eq:sparse1}) is equivalent to solving the {\it sub-set sum problem}, an NP-complete problem~\cite{Tao:2008}. Therefore, Eq.~(\ref{eq:sparse1}) is computationally hard for $n$ as small as 10$^3$. A blind and exhaustive search of $m$ = 10 sparse terms among $n$ = 10$^3$, assuming no prior knowledge for example, would give rise to $\displaystyle{C_n^m = \frac{n!}{m! (n-m)!}} \sim$ 2.6$\times$ 10$^{23}$ possibilities. A compromise is to minimize $l^1$-norm, $\|x \|_1$, subject to $Ax = y$ (or $\| Ax-y \|_2 \leq \epsilon$ for $b \neq$ 0),
\begin{equation}
x^S \equiv {\rm argmin}_{x:Ax=y} \| x \|_1,
\end{equation}
when many practical sparse inversion algorithms~\cite{TW:2010, RDD:2018},  such as  basis pursuit, orthogonal matching pursuit, gradient descent, iterative hard thresholding, Bayesian-based algorithms, iterative methods such as PIE and ePIE, and difference MAP algorithm exist~\cite{TDB:2009}. It was also shown that another requirement for CS is that the measurement matrix $A$ should satisfy  restricted isometry properties (RIP) condition~\cite{CT:2005}, and varifiable by coherence checks~\cite{KTF:2023}. In other words, the measurement matrix $A$ should not be reducible to  a lower rank matrix. Wavelet transform, Fourier transform, discrete cosine transform, and peudo-random masks are possible options to construct $A$.
More recently, deep neural-network algorithms such as CNN, U-Net were also introduced to CS~\cite{MJU:2017,MP:2022}. 



CS has been used successfully in different RadIT modalities. In electron modalities such as electron tomography~\cite{LSM:2013}, electron microscopy~\cite{BDD:2012}, and transmission electron microscopy, addition of CS has delivered impressive results such as removal of reconstruction artifacts, reduced number of projections needed for 3D reconstruction, and real-time reconstruction of sparsely sampled images~\cite{LDK:2018}. In X-ray modalities, a first-order method based on Nesterov's algorithm was used in a cone-beam CT application~\cite{CWZ:2010}. 
In another example, it was shown that if the object to be reconstructed are piece-wise constant in density, total variation minimization can lead to accurate reconstruction~\cite{YW:2009}.

Physics models can be implemented to enhance CS, {\it e.g.} through designs of measurement matrix $A$ for high-speed imaging~\cite{KTF:2023}.  A few additional examples are briefly mentioned here. Prior image constrained compressed sensing (PICCS) was developed for in-vivo dynamic CT~\cite{CTL:2008}, and resulted in a potential radiation dose reduction by a factor of 32. Several statistical physics methods have found CS applications. The replica method  has been used in CS algorithms~\cite{RFG:2012, KZ:2022}, including basis pursuit, least absolute shrinkage and selection operator (LASSO), linear estimation with thresholding, and zero norm-regularized estimation. Physics-informed CS (PiCS) is currently under utilized in U-RadIT, which motivates further growth through, {\it e.g.} incorporation of physics-informed generative models~\cite{BJP:2017}. 

Multi-domain sampling is frequently encountered in U-RadIT. Similar to PiCS, applications of CS to multi-domain is in a relatively early stage that motivates further development. Examples of multi-domain include time-frequency domain, intensity-phase domain, and space and spatial frequency domain. Time-frequency domain sampling may be described by Wigner function~\cite{Boa:1988}. Some advantages of using Wigner function for electron microscopy were reported~\cite{RB:1992}, including that influences of the instrumentation function can be entirely separated from the information from the object, super-resolution limited by the wavelength of the electron or photon can be achieved, and object thickness or the source coherence requirements can also be relaxed.  Phase-intensity domain sampling, which is a mixture of X-ray intensity variations due to both X-ray absorption and phase-induced intensity variations, is described in the Sec.~\ref{sec:XPCI} and~\ref{sec:diff} above. 2D ptychograms combine spatial and spatial-frequency sampling in X-ray ptychography~\cite{Pfe:2018, RM:2019}. Different sampling domains can be reached experimentally by placing the detectors at different distance relative to the target and the X-ray source, as in Fig.~\ref{fig1:setup}. The pixel size (also called pitch) determines the spatial resolution, the overall detector size determines the highest frequency that can be measured. Arrangement of the pixels can be optimized by compressed sensing concepts, such as pseudo-random masks, to maximize the useful information collected. Phase retrieval through X-ray ptychography was demonstrated in 1996~\cite{Cha:1996}, which was partly limited by the detectors and computing power at the time. The combination of improved detector arrays and commercial graphic processing units (GPUs) have since allowed real-time 3D tomographic CT reconstruction for a volume with 512$^3$ voxels~\cite{XM:2007}.

\subsection{Sparse image capture \label{sec:sic}}
\begin{figure}[thbp] 
   \centering
   \includegraphics[width=0.45\textwidth]{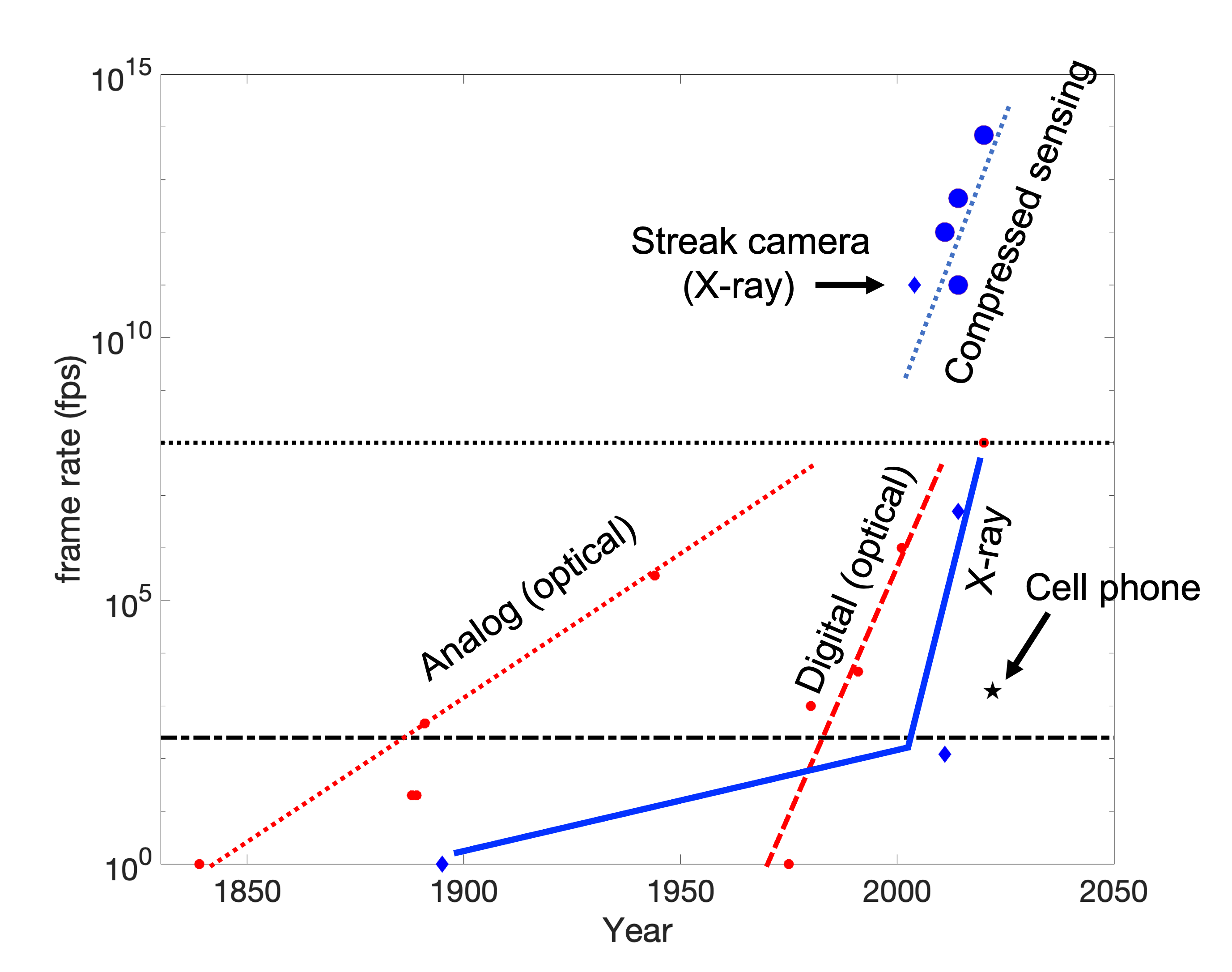} 
   \caption{ Evolution of high-speed imaging frame rate, including high-speed X-ray imaging. The large blue dots in the upper right corner corresponds to high frame rates reached by using different compressed sensing methods.}
   \label{fig:His2}
\end{figure}

 Here we emphasize ultrafast image-capture applications of compressed sensing (CS) that bypass the requirements of the Nyquist-Shannon sampling theorem. As recognized in~\cite{KTF:2023}, image compression before recording may relax the bandwidth requirement of the digital-to-analog converters in a camera, reduce the power consumption in image acquisition. A growing number of optical compressed sensing methods has led to ultra high-speed imaging at a frame rate exceeding 10$^{9}$ fps, or sub-ns temporal resolution, as shown in Fig.~\ref{fig:His2}, and imaging cameras with as little as a single pixels.  A sparsely sampled optical Fourier ptychography was able to reduce the acquisition time by about 50\%~\cite{DBS:2014}.

For X-rays and ionizing radiation, some additional benefits of the compressed sensing include faster imaging frame rate at reduced dose or source intensity, or improved image quality (signal-to-noise) for the same radiation dose. A traditional medical diagnostic CT may need more than one hundred single projections for good reconstruction. In ultrafast imaging as shown in Fig.~\ref{fig1:setup}, multiple view setup may not be practical because of the high cost of the accelerator-driven source, which usually only deliver a single projection from each experiment.

\begin{figure}[thbp] 
   \centering
   \includegraphics[width=0.46\textwidth]{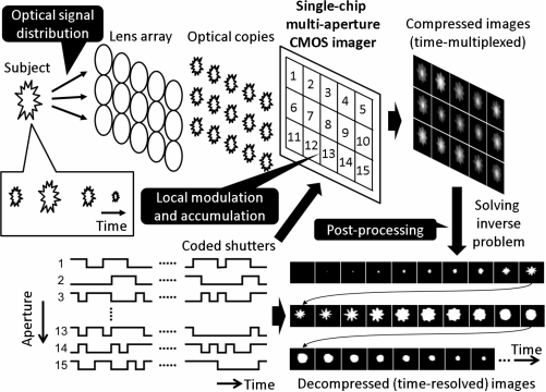} 
   \caption{The overall flow of compressive imager from~\cite{Dart:16}.}
   \label{fig:Dart3}
\end{figure}

Aside from direct imaging sensors that exhibit dependency and redundancy between adjacent pixels as discussed in Sec.~\ref{sec:Xin}, computational image sensors present another option. A computational CMOS sensor architecture was described in~\cite{RGC:2010}, which implemented pseudorandom vectors called noiselets as measurement basis before digitization. Figure~\ref{fig:Dart3} from~\cite{Dart:16} illustrates the concept of compressive-imaging-based ultra-high-speed image sensors. The frame rate is determined solely by the charge transfer speed from a pixel to a storage node, which could be only a few nanoseconds. Ref.~\cite{Dart:17} reported a 5$\times$3 aperture compressive imager running at 200 Mfps, with each pixel serving as a frame memory. Similarly, Ref.~\cite{Dart:18} used compressive imaging and reported a frame rate of 303 Mfps.

\subsection{Neural networks for phase retrieval \label{sec:NN}}
 Sec.~\ref{sec:XPCI} summarized different dynamic XPCI modalities that convert sample-induced phase modulations into intensity variations to improve the contrast of samples. However, these intensity modulations are not direct representations of the sample densities, and it is therefore often desired to convert the image intensity back into the sample-induced phase perturbations to perform quantitative analysis of the sample properties such as mass density. This process is known as phase retrieval~\cite{BLT:2011, JEH:2015}. 
 
 XPCI imaging can be described as an X-ray wavefield  $\psi (x,y,z)=\sqrt{I(x,y,z)} \exp\left[ i\phi(x,y,z)\right]$, with amplitude $\sqrt{(I(x,y,z)}$ and phase $\phi(x,y,z)$, that span Cartesian coordinates $x$, $y$ and propagate along the optical axis $z$. It is modulated by an object located at z=0 and its intensity $I(x,y,L)$ recorded at the detector plane $z=L$ is given by
 \begin{eqnarray}
I(x,y,L)& = &\lvert \psi (x,y,L) \rvert^2  \nonumber \\
&=& \lvert \mathfrak{H}(x,y,L) \psi (x,y,0) \rvert ^2,  
\label{eq:ph5}
\end{eqnarray}
where $\mathfrak{H}(x,y,L)$ denotes the forward operator that propagates the wavefield from $z$ = 0 to L. $\mathfrak{H}(x,y,L)$ can contain different transfer functions such as monochromators, gratings and free-space to describe different XPCI imaging setups. Phase retrieval can then be seen as an inverse problem solving for the sample-induce phase perturbation $\phi(x,y,0)$ (‘object phase’) from measured image intensity $I(x,y,L)$. The challenge is the nonlinear operator acting on $\phi (x,y,0)$ makes the solution non-unique and depends discontinuously on $I(x,y,L)$. Moreover, this problem is compounded by the image intensity corrupted by the detector point spread function, X-ray beam fluctuations, quantum mottle and finite X-ray coherence.

Neural networks have progressed rapidly both in demand and applicability in recent times due to exponential increases in computing power and experimental data. Image data are acquired at increasingly rapid rates as a result of highly brilliant light sources, high rate detectors and data transfer speeds, see Sec.~\ref{sec:dt}. Neural networks offer significant advantages over traditional computational modeling by analyzing images in real-time, modeling complex physical processes of image formation, and incorporating priors from other imaging modalities. Over the last decade, neural networks have solved a variety of difficult computational imaging problems such as super-resolution imaging, denoising and phase recovery~\cite{Zuo:2022}. Here, we focus on phase retrieval from single or a small number of recorded images, which is highly sought after for dynamic experiments. 

We start with the current state of physics-based phase retrieval models, and then discuss notable recent advancements in phase recovery using neural networks for (i) analyzers-based (gratings, monochromator, mask), and (ii) propagation-based XPCI methods.

Traditional single-shot phase retrieval techniques for analyzer-based XPCI methods linearize the local fringe/rocking curve~\cite{Kitchen:2011,Wang:2019} or use Fourier methods~\cite{Valdivia:2020}. Broadly speaking, these methods assume the image intensity with ($I_O$) and without ($I_R$) the object are related by: $\displaystyle{I_O=I_R ((x,y)-f(\nabla \phi (x,y,0)),L)}$, which can be derived from Eq.~(\ref{eq:ph5}). Consequently, by tracking and integrating the distortion of the image intensity between $I_O$ and $I_R$, $\phi(x,y,0)$ can be computed. However, they suffer from numerical integration-induced low-frequency noise amplification, phase wrapping, unresolved large phase jumps and fringe artifacts. 

Qiao et al.~\cite{Qiao:2022} recently addresses these issues by developing a model-based X-ray phase signal extraction multi-network comprised of a: (1) feature extractor for performing a feature-based image registration to track the displacement of the speckle pattern, (2) estimator to estimate the phase gradient along the vertical and horizontal direction, and (3) refiner to remove additional noise and improve the overall accuracy of the phase gradient. The network significantly improved the quality of the reconstructed phase particularly at large phase gradients while also achieving two orders of magnitude faster processing speed compared to digital image correlation on a flour bug. A major drawback, however, is it requires ground truth data (free of noise and artifacts) for training. While experimental data in general are abundant, training data to reconstruct particular classes of objects are not always available and synthetic training data may not always accurately model experimental data. Oh et al.~\cite{Oh:2022} avoids this issue by adopting the Noise2Noise deep learning framework that uses noisy images to train its network. This network was applied to noisy grating-based interferometry images and was able to output denoised phase gradient images, which can then be integrated to recover the object phase. However, it is assumed that the expected values of the true images are the same and the noise is zero mean.

For propagated-based XPCI (also known as in-line holography or lensless imaging), reversing the diffractive effects of free-space Fresnel propagation have traditionally employed the near-field approximation to linearize Eq.~(\ref{eq:ph5}), thus providing unique analytical solutions to the object phase. Most recently, this has supported the nuclear fusion program by reconstructing the areal mass density map from PB-XPCI images of shock-induced void collapse~\cite{Hodge:2022}. Voids created, either by design or inadvertently, in target inertial confinement fusion target ablators, can prevent ignition. Reconstruction of the areal density can improve the accuracy of hydrodynamic models used to understand and devise strategies to mitigate or even leverage the voids. Phase retrieval algorithms have also been developed to quantify dynamic behavior of porous and granular structures. For example, pore and grain size distributions have been measured  from lung- and material-induced XPCI speckles, respectively~\cite{Fouras:2009,Leong:2018,Leong:2019}. Very recently, the near-field approximation has been applied to the Fokker-Planck equation to recover, in addition to the object phase, the dark field, providing information of structures below the detector resolution~\cite{Paganin:2019}. 

Despite the wide range of applications, the near-field approximation is restricted to weakly scattering materials such as soft tissue and short sample-to-detector propagation distances. As the application of XPCI expands to imaging higher Z materials and images become increasingly radiation-starved in order to reach greater temporal resolution, it becomes necessary to image outside the near-field and into the holographic regime. Moreover, with the advent of powerful laboratory-based X-ray source (e.g., liquid-metal jets~\cite{A19}, compact light source~\cite{Morgan:2020} and wakefield accelerators~\cite{Wood:2018}), each carrying their own sources of noise and image artifacts, it becomes increasingly difficult to incorporate these effects into and retrieve the phase from Eq.~(\ref{eq:ph5}). While iterative-based optimization methods provides the flexibility to integrate stochastic models~\cite{A22b,A23,A24,A26,A27}, they are generally time-consuming and use indiscriminate priors, making them inadequate to handle today’s increasing rate of images collected and achieve optimal task-based reconstructions. As a result, neural networks have paved the way for the latest phase retrieval techniques. Broadly speaking, artificial neural networks attempt to replace the nonlinear operator in Eq.~(\ref{eq:ph5}) to achieve greater accuracy and computation speed. Rivenson et al. (2018)~\cite{A28} developed an end-to-end deep convolutional neural network that inputs the hologram and outputs the object phase. However, it requires pairs of hologram and object phase images for supervised training and does not incorporate imaging physics. Ground truths are not always available and having the network to perform both forward and backward propagation makes training difficult. Zhang et al.~\cite{A29} combined the Gerchberg-Saxton iterative algorithm with a complex-valued U-Net, while a deep image prior (DIP) neural network that is incorporated into an unconstrained objective function containing the fidelity term was developed by Li et al.~\cite{A30}. Galande et al.~\cite{NN31} expanded on this work by including a denoise term. These DIP-based methods do not require training; instead, the weights from the neural network are randomly initialized and then optimized by iteratively minimizing the objective function. Effectively, the phase is retrieved indirectly from the fitted neural network parameters. 


\subsection{Uncertainty Quantification \label{sec:uq}}
Uncertainty quantification (UQ)~\cite{smith2013uncertainty} may be applied to the existing and future new data with different emphasis. For the existing data, UQ is to answer the question, {\it how good are the data}~\cite{Jen:2017}. UQ uses quantitative, systematic, and increasingly automated approaches to characterize, estimate, and bound the uncertainties or errors in the existing data~\cite{higdon2008cmc}. Symbolically, for any $x$ in experiment (or $\tilde x $ in simulations) as given in Eqs.~(\ref{eq:op1}) and (\ref{eq:op2}), there is an uncertainty function $\sigma (x)$, root mean square error being most common,  that describes the possible error range. If $\sigma (x) \le \epsilon_0$, an acceptable error bound (for example, 5\% accuracy in the reconstructed density $x$ at a certain position and time in a dynamic compression experiment), then $x$ is accepted. In practice, this is rarely the case (error of $x$ in some regime could be more than 10\%, for example), which motivate UQ to guide further optimization workflow and iterations to search for a better $x$, ideally the optimal solution through a probability distribution, $p(x)$, and a smaller $\sigma (x) $. For new data that do not yet exist, the emphasis of UQ shifts to identifying or predicting the optimal value or values from the probability distribution $p(x)$, which can then be used to design new experiments to collect the data and confirm the prediction. When prior information ($\theta$) exists about $x$, searching for a constrained probability $p(x; \theta)$ through Bayesian optimization is often used.

Several possible U-RadIT scenarios motivate different approaches to UQ workflow: ($a$) UQ of the existing experimental data collection, $ \mathcal{I}_0 = \{I_{exp}^j (x), j = 1, 2, \cdots, N \}$, which usually consists of one or more datasets generated under nearly identical {\it macroscopic} or closely related conditions, and no additional experimental data are assumed. Here we use $I_{exp}^j (x)$ to represent a specific ($j$th) experimental image, as in Eqs.~(\ref{eq:op1}) and (\ref{eq:op2}). ($b$) A finite number of additional new experiments can be performed, which will add to the existing experimental data collection, and  $\mathcal{I}_0$ becomes $\mathcal{I}_0^+ $.  ($c$) A large number of new experiments are possible, so that $\mathcal{I}_0$ turns into $\mathcal{I}_0^\infty$, and extra experimental data can be generated on demand until a sufficiently large number $N^\infty$ is found. $\sigma (x)$ decreases in proportion to $(N^\infty)^{-1/2} $, according to the law of large numbers in statistical theory.

Scenario ($a$) is usually the starting point of UQ. Assuming that an algorithm to find $x_j$ from measurement $I_{exp}^j (x)$ is known, for example, through inverse algorithms~\cite{CS:2008, Bar:2012,BLS:2020}, then $\sigma (x)$ is found from the standard deviation from the experimental mean, $\bar{x}$ = $\displaystyle{ \sum_{j=1}^N} x_j$. However, when $\sigma (x) > \epsilon_0$, the desired error bound for UQ for $x$, there is more work left. Scenario ($c$) is simpler due to the freedom to conduct more measurements until the desired error bound is reached, and UQ reduces to collection of a sufficiently large number ($N^\infty$) of new experimental data.  If a sufficiently accurate theoretical or computational model exists~\citep{kenn:ohag:2001}, then we may enhance the experimental data collection through synthetic data generation from the highly accurate models, so that $ \mathcal{I}_0$ turns into $ \tilde{\mathcal{I}_0}$ = $\mathcal{I}_0 + \tilde{\mathcal{I}}_{syn}^\infty$, with $ \tilde{\mathcal{I}}_{syn}^\infty = \{\tilde I_{syn}^j (\tilde {\bf x}), j = 1, 2, \cdots, \tilde{N}^\infty \}$ and scenario ($a$) is turned into scenario ($c$)  through synthetic data augmentation.

In scenario ($a$) with $\sigma (x) > \epsilon_0$, and that the corresponding computational model is either insufficiently accurate or the inputs to the computer models are not completely known, here we may symbolize the problem as ($a^-$) for convenience of discussion, then further reduction in $\sigma (x)$ may not be fruitful using the standard statistical approach alone. The uncertainty function $\sigma (x)$ may depend on a large number of variables in U-RadIT: in experimental data generation, in synthetic data generation, and in image analysis, as shown in Fig.~\ref{fig2:Opt}. In the signal optimization loop, the radiation sources have intrinsic Poisson fluctuations and can change with time as in, {\it e.g.} surface electron emissivity or laser beam intensity. Methods to set up the experiment are subject to uncertainties in the material composition and structures of the target in Fig.~\ref{fig1:setup}, and environmental fluctuations, for example. Radiation-target interactions are governed by the laws of quantum physics, which are probabilistic; {\it e.g.} individual X-rays or neutrons can be absorbed or scattered, individual protons or electrons are subject to different scattering mechanisms at once, and such absorption and scattering events can not be predicted ahead of time. The detectors to collect data have intrinsic noise from different sources. In the data optimization loop shown, the state-of-the-art multi-physics computer codes to model a dynamic experiments such as xRAGE and HYDRA have known model uncertainties. The algorithms and neural networks are subject to inference errors~\cite{APH:2021}. Synthetic data from physics models or data models can therefore not avoid model uncertainties~\cite{BAJ:2022}. In the image analysis stage,  data sets may not be big enough for models with a large number of tunable parameters.

A recent trend for ($a^-$) class of problems is to combine data analysis and UQ through probabilistic modeling and probabilistic algorithms~\cite{KTR:2017,XCL:2019}.  Probabilistic modeling has long been used in both classical and quantum physics~\cite{BH:2011}. Traditional forward and physics models are combined with statistical analysis for data interpretation and predictions.  Data-driven neural network models, when combined with statistical inferences, give rise to probabilistic data models and algorithms such as Bayesian neural networks~\cite{OSG:2021, APH:2021}. Probabilistic approaches allow scientific intuitions to be incorporated in the models through dimension reduction, principle component analysis, image and other knowledge priors, and likelihood for training of neural networks. One of the potential concerns is overfitting the data or false discovery rate, which may require careful selection of $\epsilon_0$ and ultimately, experimental validation.

We may also frame the problem ($a^-$) as `{\it small N, large x}'~\cite{Spi:2014}. Here we translate the terminology in~\cite{Spi:2014} into our context, $N$, the size of the data, and $x$, an unknown vector in a high dimensional space. This formulation of the UQ problem is now strikingly similar to the sparse inversion problem described by Eq.~(\ref{eq:sparse}) above, which may not be surprising. If we rewrite $x$ in Eq.~(\ref{eq:sparse}) as $x + \delta x$, with $\delta x$ standing for the uncertainty, 

\begin{equation}
A x + B\delta x = y
\label{eq:UQ}
\end{equation}
by introducing a new matrix $B$ so that $B\delta x = A \delta x + b$. Since the noise term $b$ is independent of $x$, matrix $B$ can have the same number of rows and columns as $A$; {\it i.e.} the uncertainty $\delta x$ is just as under-determined as $x$, yet $A$ and $B$ can be totally independent of each other due to random noise $b$. 

There are several implications from  Eq.~(\ref{eq:UQ}), which reformulates UQ as a sparse or under-determined problem. First, searching for an optimal $x$ and UQ (minimizing $\delta x$) now complement each other.  Second, optimization and UQ can now share mathematical methods and algorithms that were previously developed independently of each other. For example,  compressed sensing for $x$ may give rise to compressed UQ methods, and vice versa. Third, both underdetermined UQ and $x$-optimization favor generative models. $x$-optimization and UQ may both include statistical analysis of a large number of additional hypothesis~\cite{Spi:2014}, and down-select certain new hypotheses and the corresponding generative models to augment the existing (usually incomplete) data, and supplement the missing information. 

Scenario (b) comes up naturally as the next step to address the insufficient-experimental-data problems encountered in ($a^-$). By expanding existing experimental set from $\mathcal{I}_0$ to $\mathcal{I}_0^+ $, scenario (b) provides room to validate probabilistic models, algorithms and prediction associated with the optimization and UQ in ($a^-$). On the other hand, the number of additional experiments is usually limited. For example, experiments can be costly, or the possible experiments reside in a very large dimensions, `curse of dimensionality' mandates careful experimental design and selection, constrained by the allowable experimental time, or the repetition rate of an experiment, or other experimental resources, see Sec.~\ref{sec:ap} for examples. 
Combining optimization and UQ according to Eq.~(\ref{eq:UQ}) may therefore minimize the number of new experiments in  $\mathcal{I}_0^+ $. 

\section{Applications \label{sec:ap}}
In a typical application illustrated in Fig.~\ref{fig1:setup}, the target size is in the range of 1-10 mm, determined partly by the attenuation length of X-rays at synchrotrons and XFELs. If energetic protons and neutrons are used as the radiation sources, the need to obtain high spatial resolution rather than the adequate penetration length, which the ranges of energetic protons and neutrons readily exceed, motivates a compact target $< 10$ mm. From material-properties point of view, 1-10 mm objects are usually large enough to represent larger bulk materials and structures. Below, we highlight ultrafast imaging of shocks in liquids at ESRF in Sec.~\ref{sec:esrf},  followed by a study of structural dynamics of 3D-printed polymers in the Dynamic Compression Sector (DCS) at APS, Sec.~\ref{sec:DCS}. Dynamic material properties measured by MEC end-station at LCLS are discussed in Sec.~\ref{sec:MEC}. The section ends with a discussion on 10-Hz and higher repetition rate U-RadIT experiments to accelerate the inertial confinement fusion energy research, in Sec.\ref{sec:10Hz}, by using laser-produced X-rays and other ionizing particles. Co-locating a synchrotron or XFEL with a NIF-class implosion facility may seem to be too expensive for now.

\subsection{Ultrafast imaging at ESRF beamline ID19 \label{sec:esrf}}

The European Synchrotron ESRF (Grenoble, France) runs an ultra-high speed full-field X-ray imaging
program at the microtomography and radiography beamline ID19.
The beamline operates an experimental hutch 150~m downstream of its insertion device sources:
the corresponding partial coherent illumination at the position of the sample is highly beneficial for hard X-ray
imaging as it allows for enhanced sensitivity, so-called (propagation-based) phase
contrast. The latter has been widely exploited in the past for high-fidelity applications
of microtomography. In recent years, the combination of polychromatic illumination for
X-ray imaging with fast (indirect) acquisition schemes using CMOS-based cameras has lead
to drastically reduced exposure times \cite{rack2010}. Here, hard X-ray phase contrast
is not only beneficial to enhance the contrast, but due to its edge-enhancing nature it can
also be exploited to beat noise limitations related to the finite amount of available photon flux
density. Especially for weakly attenuating objects the effective exposure time can be
reduced such that in the so-called timing modes of the ESRF (16-bunch filling mode with
176 ns bunch separation and 5.6 MHz repetition rate and 4-bunch mode with 704 ns separation
and 1.4 MHz repetition rate) individual flashes from isolated bunches in the storage ring
are used: so-called single-bunch imaging operates at ESRF with effective exposure times
down to 60~ps \cite{rack2014}. Radiation-hard indirect detectors equipped with short decay
scintillators and frame-transfer CMOS cameras allow for acquiring series of images
following the MHz-based time structure of the storage ring flashes in a continuous
manner \cite{olbinado2017}. The versatile X-ray optical layout of beamline ID19 allows for
working with beam sizes up to several square centimeters.
It is worth mentioning that the high-speed imaging program of beamline ID19 operates
on a broad range of acquisition rates, $i.\,e.$ fast acquisition with several FPS up to hundreds
of FPS is frequently used for example to study solidification in metals (in 2D and 3D), several 10$^3$~FPS up
to several 10$^4$~FPS is highly beneficial for laser-welding of metals and studies of
battery abuse testing, the range of 10$^5$~FPS is frequently requested to study
additive manufacturing of metals while 10$^6$~FPS and more is used to study materials
under high strain rates and impact. For frequently requested experiments, $in~situ$
environments are made available at the beamline, including rigs for laser processing
of metals, chambers for battery abuse testing and for experiments with energetic materials as well as a mesoscale gas launcher, ns-pulsed
shock laser, and a Split-Hopkinson pressure bar.

In order to fully exploit the potential of the above mentioned, highly sophisticated
experimental setups at beamline ID19, new access modes are required: installation for
example of a mesoscale gas launcher requires a substantial amount of (beam)time and 
hence, is rarely efficient for isolated experiments. In the frame of its upgrade
program, ESRF has made several new access modes available for the user community
\cite{mccarthy2022}. One of them is the Beamtime Allocation Group (BAG)
proposals. At beamline ID19 the so-called ``Shock" BAG brings together experts in shock physics and
dynamic behaviour of materials in order to study matter under a plethora of extreme scenarios. 
The community-driven scientific topics tackle the growing demand
for developing novel engineering materials with the ability to sustain the high strain rate and shock
as well as fundamental physical questions of material phase change and instabilities of shocked matter. The ``Shock" BAG allows for beamtime access in a routine manner, $i.\,e.$ to prepare experiments ahead as well as to follow detailed studies rather than single-shot experimental campaigns.

In this section, two highlight examples from recent work of the shock community at beamline
ID19 exploiting single-bunch imaging with MHz acquisition rates will be shown. Both consider
shock wave propagation in opaque and light materials: cavity collapse induced by high-speed impact 
as well as hydrodynamic instabilities driven by pulsed power wire explosion \cite{escauriza2020, strucka2023}.  

\begin{figure}
    \centering
    \includegraphics[width=0.85\columnwidth]{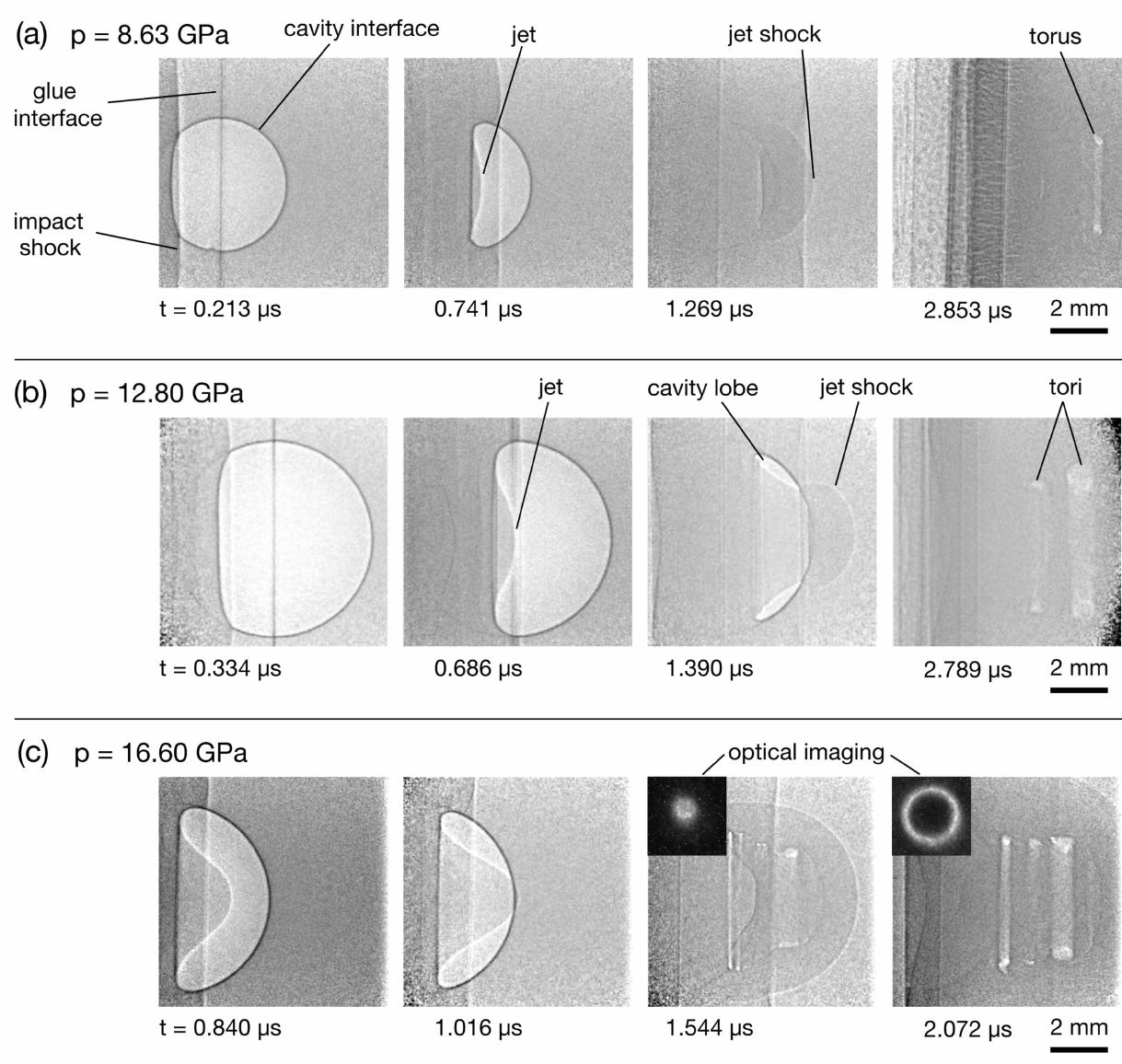}
    \caption{Series of radiography images showing shock-cavity interactions: (a) 4~mm cavity with 8.63~GPa shock, (b) 6~mm cavity with 12.80~GPa shock, (c) 6~mm cavity with 16.60~GPa shock imaged with 3.8~million FPS. The insets in (c) show the rear-surface optical images of the toroidal plasma emission. Reproduced from \cite{escauriza2020}. CC BY 4.0.
    }
    \label{fig:cavitycollapse}
\end{figure}

The first example is a study of impulsively driven cavity collapse which is directly linked to inertial fusion research: nuclear fusion has the potential
to complement renewable energies such as solar or wind which face challenges concerning the requirement for base 
load \cite{ball2021}. Here, one approach is impact-generated inertial confinement by using hydrodynamic pressure
amplification where the implosion velocity into the imparted fuel exceeds substantially the original impact and 
therefore lowers the ignition threshold. For the development of the amplifier, ground-truth data is required to trim 
numerical models and here ultra-high speed radiography delivers highly valuable input. As a model system for the collapse
process in a solid, cavities in a polymethyl methacrylate medium were impacted with different velocities in order to 
apply a range of dynamic stress states. For impact both, a single-stage and two-stage gas launcher were used at
beamline ID19. The resulting shock pressures are ranging from 0.49 to 16.90~GPa. The experiment was carried out at
ESRF using the above mentioned 16-bunch timing mode. An indirect detector consisting of a LYSO:Ce single-crystal
scintillator lens-coupled (1$\times$ magnification) $via$ pellicle mirrors to two MHz cameras type HPV-X2 (Shimadzu, Japan) 
acquired images at a repetition rate of 3.8~million FPS in a continous manner \cite{escauriza2018, montgomery2023}. The results are shown as series
of images depicting fluid-dominated dynamics of cavity collapse in Figure~\ref{fig:cavitycollapse}.

\begin{figure}
    \centering
    \includegraphics[width=0.85\columnwidth]{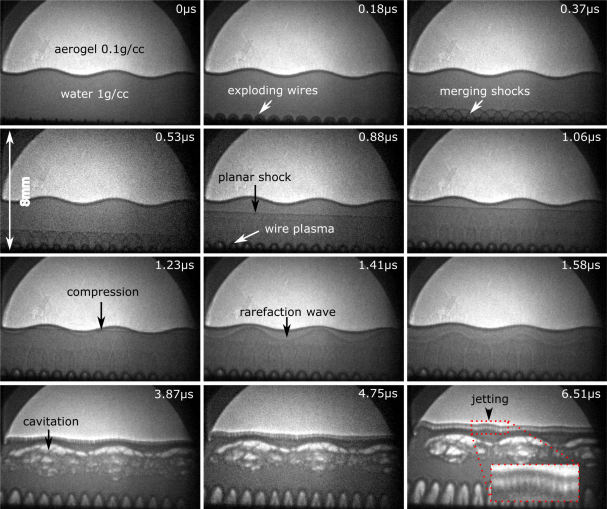}
    \caption{Series of phase-contrast radiographs showing the pulsed-power induced planar shock wave from an array of wires, and its interactions at the interface of different density media. Important features are marked within the images.      
    Reprinted from \cite{strucka2023}, with the permission of AIP Publishing.
    }
    \label{fig:pulsepower}
\end{figure}

The second example considers hydrodynamic instabilities which are ubiquitous in nature 
and adresses fundamental questions related to geophysical and astrophysical flows, 
high energy density physics, and confinement fusion, as well as engineering applications such as rock fracking and fluid-structure interaction. 
Instabilities occurring at the abruptly accelerated interface between fluids of different densities (\emph{e.g.} due to the passage of a shock wave) such as the Richtmyer-Meshkov (RM) and Kelvin-Helmholtz (KH) instability are a common focus of the investigation, where full-field experimental measurements are paramount for developing advanced multi-scale-multi-physics models which can reliably capture these phenomena. 
In accord, an experimental platform based on a pulsed power-driven resistive wire array (\emph{i.e.} discharge current 30~kA with deposited energy of 350~J in 1~\textmu s), aimed at producing convergent shock waves within both solids and liquids, was tailored to the ID19 beamline MHz-radiography capabilities \cite{strucka2023}. 
The high versatility of the platform allows for introducing shock-induced density discontinuities in arbitrary geometries which are then measured through the acquired X-ray imaging data. 
An example experiment measuring the planar RM instability between high-density (\emph{i.e.}, water) and low-density (\emph{i.e.}, aerogel) media, in the linear regime, serves as a benchmark model while highlighting the capabilities offered. 
The experiment was conducted using the above-mentioned 16-bunch timing mode (Figure~\ref{fig:pulsepower}). 
Cylindrical shocks from wire array explosion form a planar interface (v~= 2.2~km/s) in water, which hydrodynamically compresses the aerogel with resulting interfacial instability, but also induces cavitation in the denser medium through the rarefaction wave. 
The results are shown as discrete series of radiographs, capturing all the underlining phenomena involved during 4~\textmu s, due to the sufficient temporal resolution over a wide field of view.      

%
%
%
%
%
%
\subsection{High-speed capabilities at 32-ID of the APS}
Since its inception 20 years ago, XSD beamline 32-ID of the Advanced Photon Source (APS) at Argonne National Laboratory has been at the forefront of the dynamic x-ray imaging field, by pioneering the first in vivo functional imaging of small animals~\cite{Fez1}, and then with the first use of white beam to probe ultrafast, sub-microsecond fluid dynamics~\cite{Fez2} and, more recently, the use of single-pulse techniques to probe shock dynamics in real and reciprocal space~\cite{LuoRSI:12,Fez4,Fez5}. In order to achieve the current exquisite operational parameters (80-ps exposure time, a frame rate of up to 6.5 MHz, a spatial resolution of $\sim$ 1 \textmu m, and a field of view of $\sim$ 2 mm$^2$), the program takes full advantage of what the APS X-ray source has to offer in terms of flux, energy, and time structure. The full white X-ray beam from a standard APS Undulator A with a 33-mm period (U33) was initially used, but such a powerful and polychromatic beam imposes many limitations on the data quality and the experimental possibilities. The more recent acquisition of a second undulator with a much shorter period of 18 mm (U18) addressed a number of these important limitations. First, the reduced heat load allowed for imaging over longer intervals without damaging or reducing the efficiency of the scintillator crystals. Second, the suppressed higher harmonics permitted operation with a quasi-single-line beam, reducing the background and allowing for quantitative measurements, especially for X-ray diffraction and wide/small angle X-ray scattering (W/SAXS). Finally, the available intensity increased dramatically at the first harmonic energy of 24 keV, so that thick and high Z materials could be studied.

Ultrafast full-field imaging is the workhorse technique at 32-ID beamline, but many other complementary techniques and platforms are constantly being developed as well. Below are some research applications that highlight the multimodality and multiscale aspects of this program.  

Magnesium alloys show a remarkable potential as structural components for their low density, high specific stiffness, and high specific strength. However, wide applications of magnesium alloys are hindered by their poor formability at room temperature, and overcoming such a deficiency requires better understanding of deformation mechanisms, and microstructural effects on mechanical properties of these alloys~\cite{Fez6}. In this first example, speckled full-field X-ray images were combined with digital image correlation analysis (thus the development of XDIC) to map strain fields in bulk samples with very high spatial (\textmu m) and temporal (\textmu s) resolutions. This study of the anisotropic deformation of an extruded magnesium alloy AZ31 under uniaxial compression along two different directions, multiscale measurements including stress-strain curves (macroscale), X-ray digital image correlation (mesoscale), and diffraction (microscale) were obtained simultaneously. Preliminary results showed that the rapid increase in strain hardening rate is attributed to marked {$101\bar{1}2$} extension twinning and subsequent homogenization of deformation, while dislocation motion leads to inhomogeneous deformation and a decrease in strain hardening rate.

\begin{figure}
    \centering
    \includegraphics[width=1.0\columnwidth]{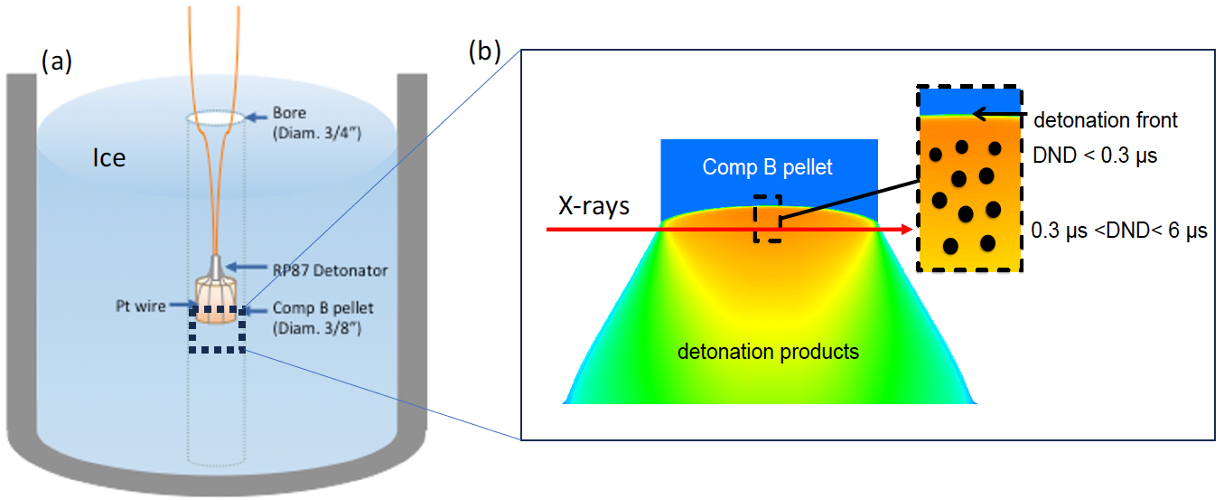}
    \caption{Simplified schematic of the detonation experiment. (a) The composite explosives mixture (comp. B) pallet is placed in an ice block that is cut open to recover the detonation products. (b) inset of the different regions of the detonation front that the X-ray beam is probing. Figure from Ref.~\cite{Fez7} reused with permission.
    }
    \label{fig:Fezzaa1}
\end{figure}

Characterization of the initial morphology of detonation nanodiamond (DND) has been the focus of many research studies that aim to develop a fundamental understanding of carbon condensation under extreme conditions. Identifying the pathways of DND formation has the potential for significant impact on many of the controlled synthesis of nanoscale carbon with a tailored functionality; currently, a wide range of possible (and conflicting) mechanisms of nucleation and growth have been proposed, and further research is needed. Building a comprehensive understanding of DND formation is challenging because it requires {\it in situ} characterization on the sub-microsecond timescale during a high-explosive detonation. In this second example, Time-Resolved Small-Angle X-ray Scattering (TR-SAXS) was used to reveal the early-stage DND morphology from $<$ 0.1 to 6 \textmu s after the detonation front passes through the X-ray beam path~\cite{Fez7}. Figure~\ref{fig:Fezzaa1} shows schematics of the detonation setup. In these experiments, scattered X-rays from a single line X-ray beam were collected using a four-camera (PiMAX4) detector system, with each camera capable of two consecutive frames. The full first harmonic of X-ray energies from the short period undulator was used and had a full width at half-maximum of $\sim$10\% and mean energy of $\sim$ 24 keV. The exposure time was 80 ps, and the time between sequential frames was a multiple of the inter-bunch gap (153.4 ns). 

The SAXS from both late-time ($>$1 \textmu s) {\it in situ} and recovered DND exhibits consistent features in the $I(q)$ curve. Such a close similarity allows a high-fidelity SAXS model derived from the ex-situ SAXS and TEM measurements to be applied to the {\it in-situ} data, which yields new insight into the early-stage ($<$1 \textmu s) morphology of DND. Results indicate that during detonation, carbon is condensed into nanoscale diamond much faster than that previously reported in other studies. Furthermore, the surface texture of the DND is shown to arise during condensation rather than via subsequent graphitization. 

\begin{figure}
    \centering
    \includegraphics[width=0.8\columnwidth]{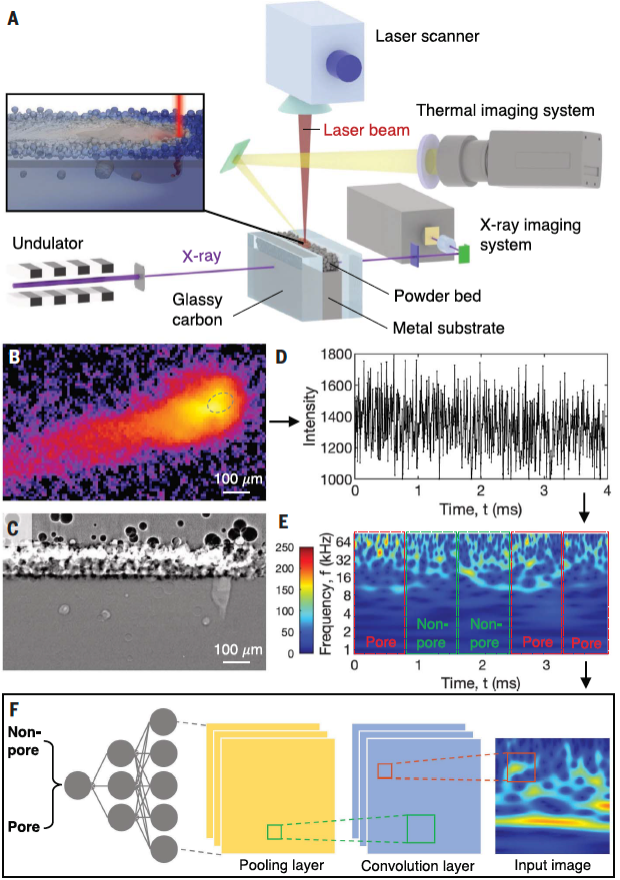}
    \caption{Real-time keyhole porosity detection in LPBF. (A) Schematic of the simultaneous synchrotron x-ray and thermal imaging experiment on scanning laser melting of Ti-6Al-4V. (B) A representative angle top-view thermal image. (C) A representative side-view x-ray image. (D) Typical time-series signal of the average emission intensity from the keyhole region [(B), dashed oval] extracted from the thermal image sequence. (E) Wavelet analysis performed over the time-series signals in (D). The scalogram is sectioned into a few windows, which are then labeled as either ``Non-pore” or ``Pore” on the basis of the operando X-ray imaging result. (F) Machine-learning approach with sectioned scalograms as input data. A CNN was used, which is composed of a series of alternating convolution and pooling layers and a final layer. Each convolution layer extracts features from its previous layer, using filters learned from the trained model, to form a feature map. The feature map is then down-sampled by a pooling layer to reduce the number of parameters to learn. The final layer of the CNN classifies the input scalogram as either ``Non-pore” or ``Pore.”  Figure from Ref.~\cite{Fez8} reused with permission.
    }
    \label{fig:Fezzaa2}
\end{figure}

In the third example, high resolution MHz X-ray imaging is combined with high-speed thermal imaging and machine learning to predict with near certainty defects formation during laser powder bed fusion (LPBF) additive manufacturing (AM)~\cite{Fez8}.  In a typical LPBF process, a high-power laser beam is used to locally melt and consolidate metal powder to form three-dimensional (3D) objects layer by layer. The extreme thermal conditions involved in the printing process trigger transient phenomena and complex structural dynamics. Their interplay often leads to structural defects, such as porosity, which is a major factor that hinders the widespread adoption of AM technologies. One common porosity is caused by the momentary collapse of the vapor depression zone, known as keyhole porosity. With simultaneous high-speed synchrotron X-ray imaging and thermal imaging, coupled with multi-physics simulations, two types of keyhole oscillation in LPBF of Ti-6Al-4V were discovered: one intrinsic and the other perturbative. Amplifying this understanding with machine learning, a breakthrough approach was developed to detect the stochastic keyhole porosity generation events with sub-millisecond temporal resolution and near-perfect prediction rate. The highly accurate data labeling enabled by operando X-ray imaging allowed the demonstration of a facile and practical way to adopt the new approach in commercial systems. Figure~\ref{fig:Fezzaa2} shows the setup schematic and the data analysis workflow.

32-ID beamline will undergo a substantial enhancement as part of the APS Upgrade project. It will benefit from the new source characteristics and new instruments, including an improved spatial resolution ($\sim$ 100 nm), dual-beam and multi-modal capabilities, without sacrifice in time resolution.
\subsection{Structural dynamics of 3D-printed polymers \label{sec:DCS}}
A supersonic source of energy such as explosion, implosion and hypervelocity impact induces shock waves in solid, liquid and other states of matter due to compressibility of the medium~\cite{LL:1987}. Shock waves produce discontinuities in materials properties, {\it e.g.} density, pressure, temperature and material phase, which can be understood by the conservation of mass, momentum and energy~\cite{DG:1979, For:2012}. The continuities or the shock wave thicknesses are comparable to collisional mean free paths of the atoms and molecules. In room-temperature air, for example, the thicknesses would be around 200 nm~\cite{FM:1992}, and the mean free path at the standard temperature pressure is 60-70 nm in air. Sound speeds in polymers range from less than 1 km/s to several km/s, depending on the structure and density, which make them good material platforms to shock loading experiments that can be directly compared with the results from the finite-element computer codes such as Abaqus for design of new structured polymers.

Additive manufacturing (AM), also known as 3D printing, has created a new paradigm shift in structure-based control of material properties for a wide variety of materials including polymers~\cite{DC:2019}.  AM offers polymer structure control potentially down to nanometer scales. On the nanometer scale, the control options include network chemistry, crosslinking and crystallinity, filler particles and polymer-filler interactions through AM feedstock materials and fillers. On the micrometer scale, individual polymer ligaments can be varied in both scale and geometry around connection or node points.  On the millimeter scale, layer symmetries can be tailored to affect deformation or compaction mechanisms, and to alter wave propagation through the structures.  AM polymer Menger structure, as a new type of porous metamaterials, may possess properties quite different from the base stochastic polymers free of pores~\cite{BIC:2017, D1}. Many applications of AM polymer structures have been recognized, including vibration and acoustic damping, thermal management, and shockwave localization~\cite{WKY:2018}.  

\begin{figure}
    \centering
    \includegraphics[width=1.0\columnwidth]{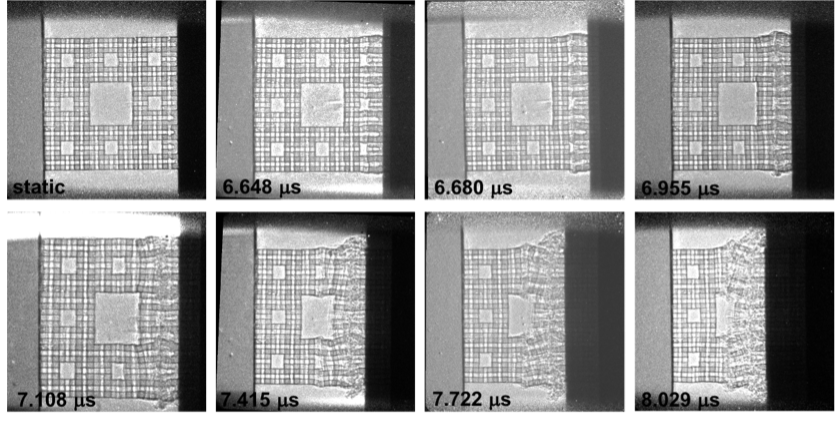}
    \caption{Pre-shock (static) and 7 dynamic frames from x-ray phase contrast imaging of a 3rd-order Menger structure during shockwave loading following projectile impact at 331( or 318?) m/s (Shot 19-2-017).  The frames are timed to the 24-bunch mode of the Advanced Photon Source.  In the phase contrast images, the shock travels from the baseplate (right side, no X-ray transmission) into the structure, resulting in substantial lateral displacement and dissipation of the shock. Reprinted from~\cite{D1}, with the permission of AIP Publishing.
    }
    \label{fig:dana1}
\end{figure}

Dynamic X-ray phase contrast imaging was recently used to study impact and shockwave responses of 3D-printed polymer Menger structures (with the third order length $L_3$ = 126 \textmu m) at the Dynamic Compression Sector (DCS) using the Advanced Photon Source (APS), Fig.~\ref{fig:dana1}. Shockwaves were generated using the IMPact system for the ULtrafast Synchrotron Experiments (IMPULSE). X-ray energy was about 25 keV with a pulse-to-pulse time of 153.4 ns. The imaging system consisted of a LuAG:Ce scintillator frontend and four independently triggered ICCD-4 (Princeton Instruments) cameras to collect the eight images shown. 

As the shock traverses the first few layers, rarefaction waves from free surfaces of the cubic voids interact, resulting in significant lateral deformation and buckling~\cite{D3}. The kinetic energy imparted to the structure from the impact event is partitioned into elastic and viscoplastic energy, with viscoplastic energy increasing with time/distance (in this experiment leading to a temperature rise $\Delta$T = +170 K).

\begin{figure}
    \centering
    \includegraphics[width=1.0\columnwidth]{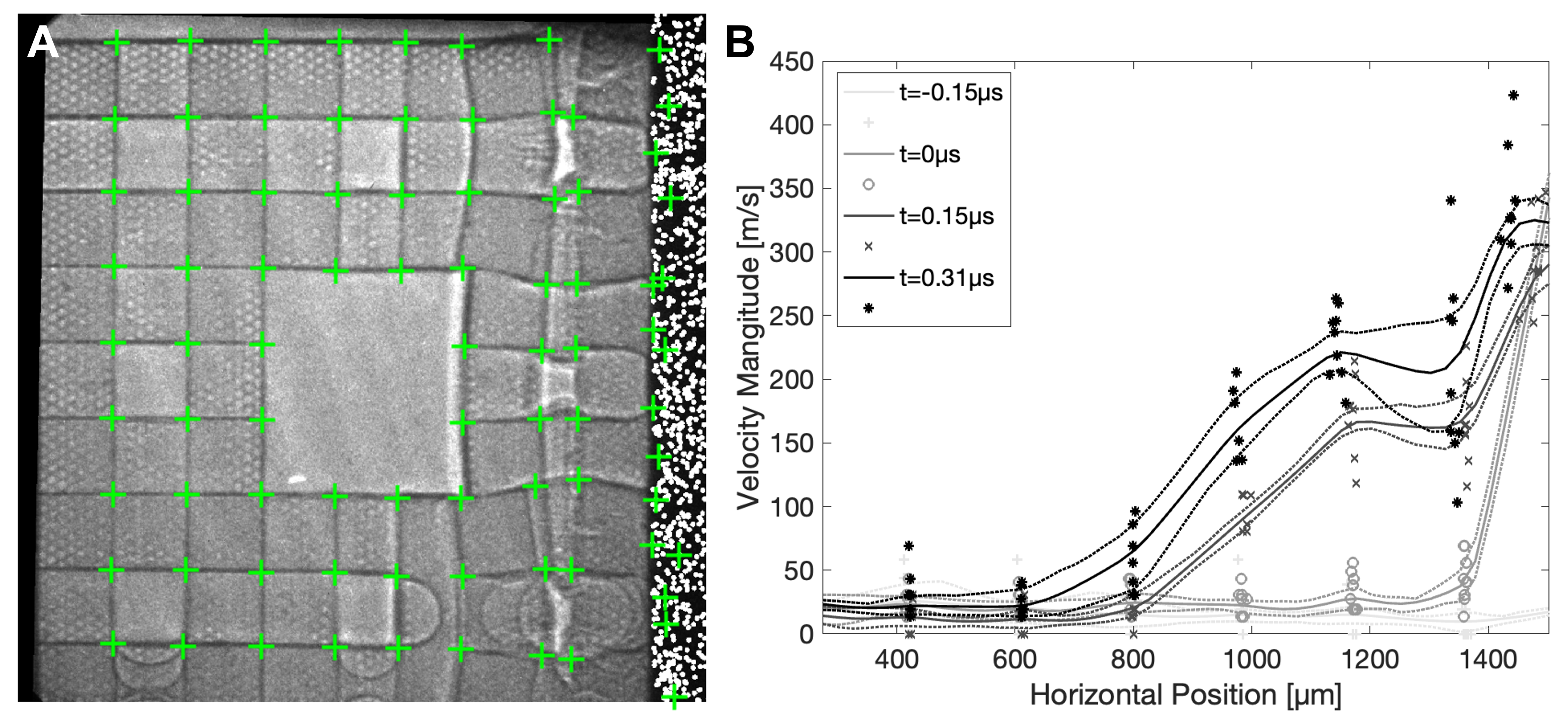}
    \caption{Single frame analysis from shot 20-2-081 using the ground truth approach.  Here, the 4th frame is shown, in which shock localization led to "cell" rotation and material extrusion lateral to the shock direction.  b) Corner velocity as a function of horizontal position stacked by frame shows a broad velocity disturbance with a maximum of ~250 m/s moving into the structure.  The velocity increase is not discontinuous (shock-like), but is spread over nearly 0.5 mm. Figure adopted from~\cite{D2}.
    }
    \label{fig:dana2}
\end{figure}

We have examined several methods of velocity field analysis for the X-ray phase contrast images, or “X-ray velocimetry”. Two of the critical steps of X-ray velocimetry, similar to optical velocimetry, are feature recognition or object identification from individual movie frames to another, and feature matching or object tracking from one frame to another. Even though X-ray velocimetry is not yet widely practiced in the literature, we have used existing algorithms for optical imaging and velocimetry for the X-ray data. Several challenges in X-ray velocimetry are recognized: a) Low signal-to-noise ratio in the raw data limited by the X-ray source intensity and camera hardware; b) Shockwave or impact on the porous structure can modify the features significantly and frequently destroy the features completely from one movie frame to another; i.e., many features are not recognizable after the impact or the shockwave front; so manual tuning of the existing algorithms is necessary for X-ray velocimetry to improve the reliability of the velocity estimation; c) It is difficult to obtain the 2D velocity uniformly across the images because of the limited number of features; d) There is limit amount of information on ‘ground-truths’ to validate the algorithms; i.e., except for some estimates of the shock velocity based on the bulk material properties; and e) The velocity field is intrinsically 3D, while the X-ray image only captures the projected information. 
%
%
%
%
%


\subsection{Dynamic material properties by MEC Endstation at LCLS \label{sec:MEC}}

X-ray Coherent diffractive imaging (XCDI), together with its variants such as BCDI~\cite{Xiong:2014}, as discussed in Sec.~\ref{sec:diff}, has emerged in the last decade, also known as the first decade of XFELs~\cite{xfel:2020},
as a sub-ps time-resolved workhorse to characterize a wide variety of
materials under extreme conditions, {\it e.g.} 10s of GPa pressure and above (100 GPa = 1 Mbar) ~\cite{MBW:2013}, thousands of degree K temperature, harsh X-ray and neutron radiation environment. The Matter in Extreme Conditions (MEC) endstation at
LCLS~\cite{Nagler:2015, GFG:2016} can deliver quasi-monochromatic ($\Delta$E/E =  0.1–2\%), fully transverse coherent,
energy-tunable (0.25 to 25 keV) X-ray pulses of 10s-of-fs duration with an average of $\sim $10$^{12}$ photons per pulse. The focused X-ray beam spot can be adjusted using a series of beryllium-focusing lenses to the range of 10 to 200-\textmu m (1 mm unfocused at 8 keV) in diameter on the target package. The experiments can be repeated at up to 5 Hz by a 1-J 25-TW laser to make ultrafast movies of non-thermal melting, phase transition kinetics,  chemistry important to life, crystal growth~\cite{Gleason:2015}, compression freezing kinetics of water to ice~\cite{Gleason:2017}, and metalization under high pressure ramp compression and shocks. Here, several application examples are highlighted.

\begin{figure}
    \centering
    \includegraphics[width=0.85\columnwidth]{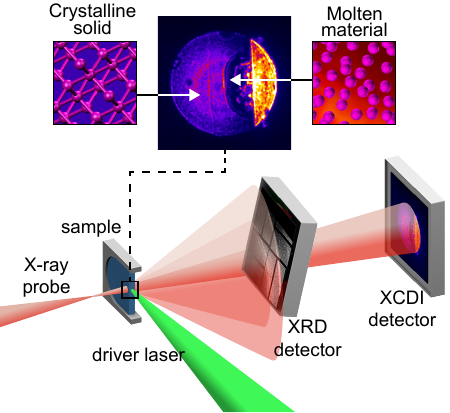}
    \caption{Schematic of laser-driven shock compression of crystalline phase using simultaneous {\it in situ} imaging and X-ray diffraction. Dynamic X-ray imaging diagnostics at the MEC, LCLS End-station can visualize the shock compression process of crystalline materials phase transforming into a molten state. X-ray imaging and diffraction fidelity can resolve lattice level transformations, provide phase fraction information and data quality suitable for Reitveld refinements and {\it in situ} imaging resolution to 400 nm within 60 femtoseconds. Schematic courtesy of G. Stewart, SLAC Graphic Artist.  }
    \label{fig:Ari3}
\end{figure}

XRD and XCDI at LCLS are used to make the ultrafast movies at 10-ps intervals for the transition from elastic to plastic regime in polycrystalline Cu~\cite{MBW:2013}. These results validated predictions of the yield stress of atomistic simulations. In another example, ultrafast lattice dynamics of silicon have been measured by using the simultaneous {\it in situ} imaging and X-ray diffraction~\cite{BGG:2019}. The setup is shown in Fig.~\ref{fig:Ari3}. A 100-J 527-nm laser ablates plastic from the front of the target and drives  a shockwave through the sample. The optical drive pulse has a temporal
profile that can give rise to 7.5 $\times$ 10$^{11}$ W/cm$^2$ within 300 ps and increase
linearly to 2.5 $\times$ 10$^{12}$ W/cm$^2$ by the end of the 15-ns pulse duration. Diffraction of the probe X-ray is captured
on Cornell–SLAC pixel array detectors (CS PAD) or ePix10k detectors. One example using this technique to examine silicon shows diffraction
of multiple broad peaks and the first silicon melt feature
located between 28 and 32.5 nm$^{-1}$. The primary beam continues to the phase contrast
imaging detector located 4 m down stream the target location, with a field of view of 200-\textmu m diameter and a spatial resolution about 5 \textmu m down to 100s of nm. The measurements answered some long-standing questions about the silicon dynamics of high-pressure elastic,  inelastic phases and melt. In the third example, {\it in situ}
ultrafast X-ray diffraction was used to study the plasticity of hexagonal-close-packed (hcp)-Fe~\cite{MHB:2021}. Laser-induced shock compressed Fe up to 187(10) GPa and
4070(285) K at 10$^8$ s$^{-1}$ in strain rate. It is revealed that $\{ 10 \bar{1}2\}$ deformation twinning controls the polycrystalline Fe microstructures and
occurs within 1 ns. The fourth example is high-resolution imaging of an inertial confinement fusion (ICF) target and {\it in situ} void collapse within the target. Recent efforts at LCLS
are enabling a new nanoscale imaging instrument based on a ptychographic method for the
collection of static, high resolution 2D and 3D images, down to 10s nm resolution, {\it e.g.} could be
used in support of ICF and inertial fusion energy (IFE) capsule or sample inspection/studies. See Sec.~\ref{sec:10Hz} for further discussions. Preliminary efforts
at LCLS have inspected Cu-foam from LLNL with unprecedented 30 nm resolution over a few
\textmu m$^3$ volume providing 2D and partial 3D reconstructions. Follow on work using novel X-ray optics may remove the need to rotate the sample~\cite{WPL:2019}, and this 
opens up the possibility to perform 2D and 3D imaging {\it in situ},
during dynamic compression. At MEC endstation, near-field propagation-based phase contrast
imaging (PCI) and Talbot-CDI are proving very successful in measuring 2D void collapse in ICF ablator
materials to 200 GPa. These images of a single collapsing void with sub-\textmu m spatial
resolution and picosecond temporal resolution will allow us to study high pressure (several
Mbar) shock interaction for insights on hot spots and instability mitigation in a single void.
These data are being compared with 2D xRAGE simulations. In addition, the Talbot-CDI method, which can reconstruct
the actual wavefront and be applied to reconstruct the image of a shocked sample, {\it i.e.} the
wavefront after the sample interaction, results in a novel single-shot imaging technique. Prior
campaigns on single void collapse have leveraged the LCLS pulse train and UXI cameras to
measure a shock traversing a single void in a single sample~\cite{Hodge:2021}.  %

It has also been recognized that further improvements in XCDI spatial resolution, together with advances in dynamic 3D X-ray diffraction, including high-energy diffraction and microscopy~\cite{BSR:2020},
would allow predictive modeling capabilities
for material strength and plasticity in extreme
environments. This includes collecting time-resolved structure factor measurements extending to a broader class of materials such as liquids, glasses~\cite{MHG:2020}, other amorphous and non-periodic materials and structures, with applications to geology, planetary science, life science, and to the
design of novel materials.
MEC at LCLS is now pushing spatial resolution below 10 nm
and can provide accurate phase projections to yield detailed quantitative 2D and 3D
reconstructions with a resolution that is not limited by imaging optics. 


\subsection{Towards 10$^+$ Hz Inertial Fusion Energy (IFE) experiments \label{sec:10Hz}}
The recent breakthroughs in laser-driven inertial confinement fusion at the National Ignition Facility (NIF) came after more than half a century of coherent efforts~\cite{NWT:1972,KZC:2022}. Now IFE experiments at 10 Hz and above (10$^+$) repetition-rate are being pursued towards economically viable electricity generation from controlled release of fusion energy. Some of the open problems related to 10$^+$-Hz experiments include laser driver, target fabrication at low cost, target defect controls, dynamic properties of the target, neutron yield optimization, and fusion energy capture~\cite{Ma:2023}. U-RadIT can play important roles in studying the dynamic properties of IFE targets and shed light (X-rays) on the effects of defects such as void, as discussed in Sec.~\ref{sec:MEC}.

\begin{figure}[!hbt]
    \centering
    \includegraphics[width=0.95\columnwidth]{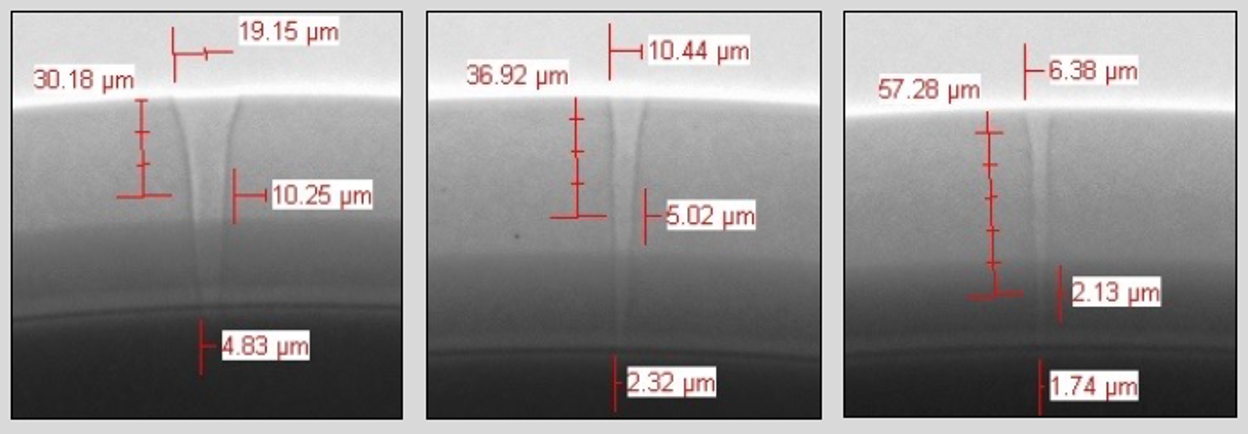}
    \caption{ An example of the reduction of fill-tube hole diameter with time in NIF experiments~\cite{Ma:2023}. The smallest fill-tube design was deployed in the ignited NIF experiments. Image credit: General Atomics (GA) and Lawrence Livermore National Laboratory (LLNL).}
    \label{fig:Tammy1}
\end{figure}

It has been recognized that mesoscale pores, voids and other defects, as well as built-in structures such as a fill-tube can have an significant impact on the mechanical response of the target under intense laser or X-ray compression~\cite{PCH:2022}. A fill-tube can also seed a perturbation that injects the ablator material into the target center, radiating away some of the hot-spot energy~\cite{WCA:2020}. A smaller fill-tube diameter has been shown to be beneficial to the recent NIF ignition. An example of reducing fill-tube size is included in Fig.~\ref{fig:Tammy1}. The effects of a fill-tube are complicated, besides the dimensions of the fill-tube, other factors such as ablator materials, target structures, laser driver  energy and timing, hohlraum designs, and so on, also come into play. Coupling high-resolution U-RadIT measurement with high-fidelity modeling is a must for data interpretation and target optimization. Until the targets can be fabricated exactly the same at sub-\textmu m or even nanometer precision, single-shot measurements at facilities such as LCLS may be necessary to aid the target optimization. Due to the size of an ICF target ($\sim 1$ mm) and the use of high-Z materials (W, Cu, Au), high energy X-rays ($>$ 20 keV) may also be required. For example, MEC experiments at LCLS showed best contrast at about 18 keV, with a resolution of 400 nm and 10s of fs in space and time, respectively~\cite{Hodge:2021}. Higher energy radiographic imaging using synchrotrons can complement the XFEL studies, at a lower temporal resolution but potentially higher spatial resolution~\cite{escauriza2020}. 

It is not yet clear whether all the target implosion and material studies towards IFE target optimization can be done by using XFEL and synchrotron facilities exclusively. In other words, in-situ NIF or high-repetition IFE measurement may still be necessary, not by using a synchrotron or an XFEL source of X-rays, rather by one or more high-power lasers co-located with the NIF or an IFE. One such an example is the NIF Advanced Radiographic Capability (NIF-ARC)~\cite{DNY:2015, SMK:2021,WCA:2021}. In these experiments, high laser power above 10$^{17}$ W/cm$^2$ and 1.5 kJ NIF beamlet may be focused onto different materials to produce an intense flash of X-rays ($>$ 50 keV) and energetic protons ($>$ 10 MeV) for ultrafast radiographic imaging with a temporal resolution down to 1 ps (30 ps laser FWHM). In 2020, a multi-pulse imaging technique was executed with 4 NIF ARC beamlines, two of them each on a separate Au wire target to produce bremsstrahlung X-rays from 50-200 keV \cite{tommasini2020time}. Higher energy X-rays pave way towards U-RadIT applications for denser materials. Even though the demonstrated spatial resolution ($\sim$ 10 \textmu m) is less than in synchrotrons and XFELs, further hardware optimizations such as beam spot size reduction (currently 150 \textmu m), target materials, laser profiles, and imaging detectors are possible.
 
\begin{figure}[!htb]
    \centering
    \includegraphics[width=0.85\columnwidth]{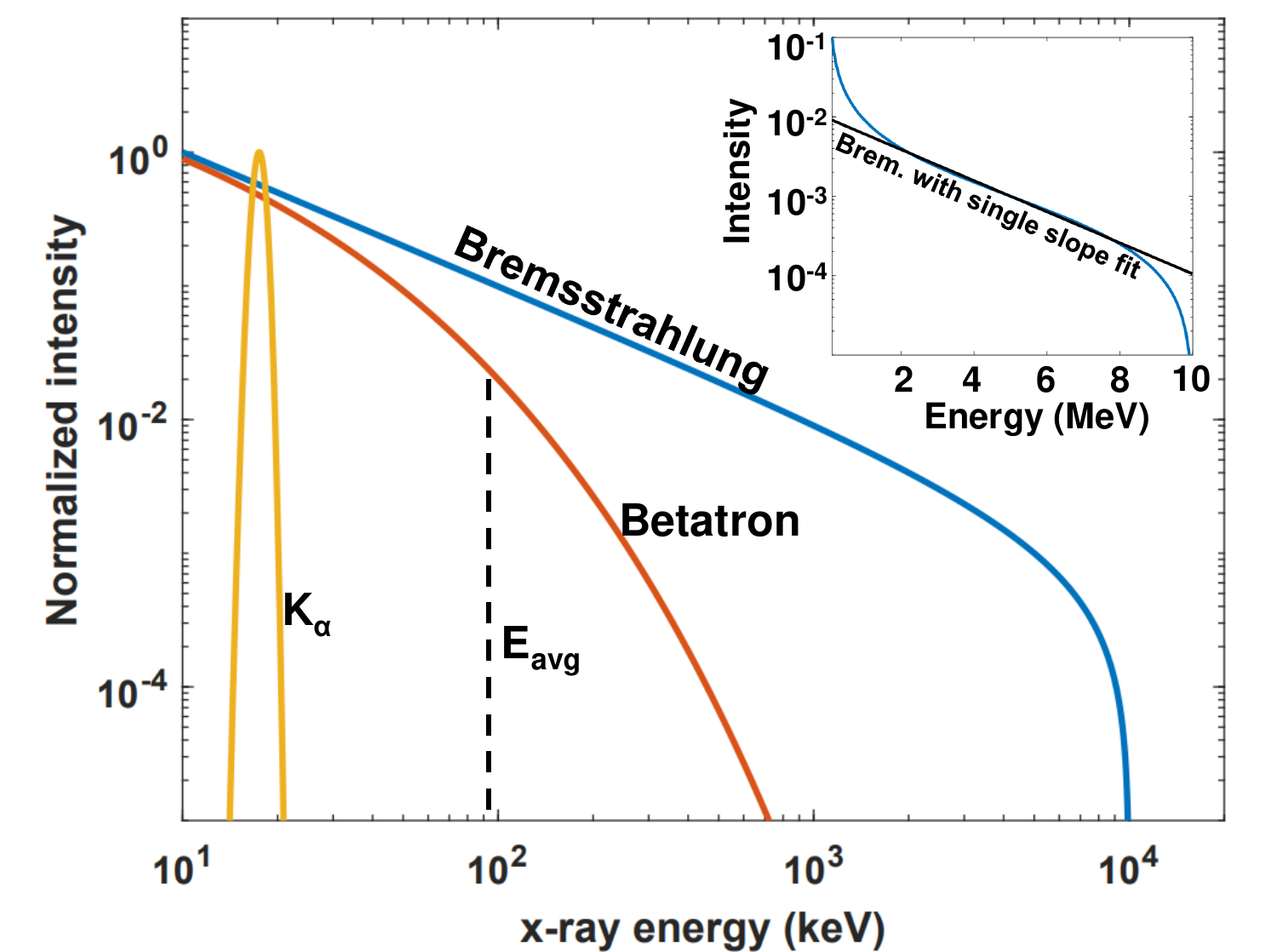}
    \caption{The spectral shapes of the laser-driven sources and their representative energies. Idealized $K_\alpha$ emission is a monoenergetic peak, shown here at 17.5 keV to represent a molybdenum $K_\alpha$ line. The betatron spectrum has a mean energy $E_{avg}\approx$ 100 keV, identified by the dashed line. The bremsstrahlung spectrum is characterized by a single slope, or "temperature," in the MeV range (inset). The fit for the intensity ($I_0$) as a function of energy ($E$) is of the form $I_0 \propto e^{-E/T}$, with temperature $T$ = 2.2 MeV. }
    \label{fig:StrSS}
\end{figure}

NIF-ARC-like capability also exists elsewhere~\cite{schwarz2008activation, maywar2008omega}, which opens door to a broad range of U-RadIT applications with even higher X-ray energies exceeding 40 keV. Ultrafast bursts of bremsstrahlung x-rays for Compton radiography of fusion implosions was first investigated as a proof-of-principle experiment on the Titan laser in 2008~\cite{tommasini2008development}.  In 2009, Brambrick {\it et al.}~\cite{brambrink2009direct} reported $>$ 40 keV X-ray emission from 18 \textmu m wires used to probe iron shocked by a ns laser driver. Areal densities were obtained within 10\% error. Chu {\it et al.} extended this type of study to near-MeV photons in 2018~\cite{chu2018high}. He {\it et al.} reported using the ultrafast probe to distinguish between samples in the solid state and melt-on-release state \cite{he2019high}. 

\begin{figure}[!htb]
    \centering
    \includegraphics[width=0.85\columnwidth]{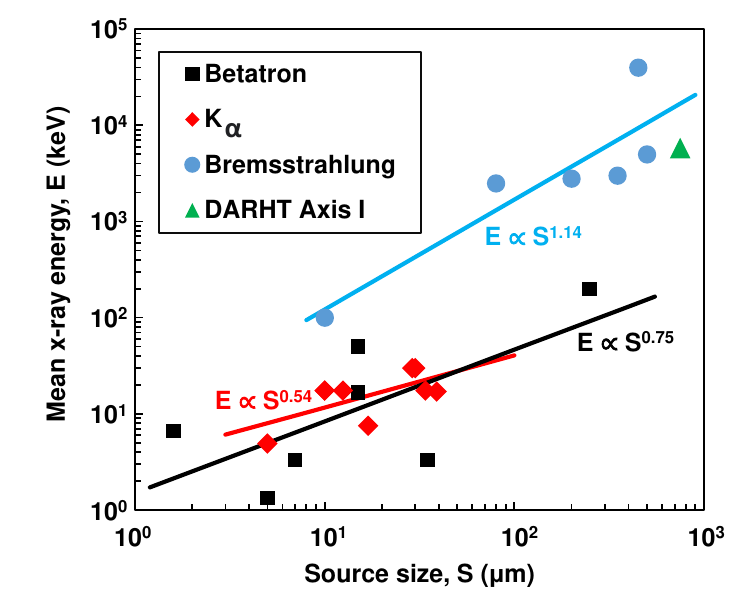}
    \caption{X-ray energy (defined in Figure \ref{fig:StrSS}) vs source size as delivered via a wide range of high intensity, short pulse laser-target configurations. For each source type, a least-squares regression (solid lines) establishes the relationship between energy $E$ and source size $S$. An additional data point for a conventional MeV X-ray source, the 60 ns DARHT Axis I accelerator (triangle)~\cite{Darht:2008,Nath:2010}, is shown for comparison.}
    \label{fig:Str1}
\end{figure}

Over the past two decades, energetic X-rays from high-intensity lasers (peak power $>$1 PW, pulse duration 10's of fs to ps)  have gone from early characterization of their properties and origin \cite{hatchett2000electron}, to practical radiography sources with enough dose ($>$ 10 Rad or $>$10$^{10}$ MeV photons/cm$^2$ at 1m) of MeV photons to deeply penetrate high areal density materials. Such sources provide new avenues to U-RadIT. Three distinct X-ray generation mechanisms have emerged with practical application in ultrafast imaging: bremsstrahlung, K$_\alpha$, and betatron, Fig.~\ref{fig:StrSS}, each with their own unique source characteristics, Fig.~\ref{fig:Str1}. Xin {\it et al.}~\cite{xin2019x} characterized the bremsstrahlung spectrum post-experiment, finding it to fit a simple exponential of $T$ = 336 keV between 0.1-0.9 MeV. In Fig.~\ref{fig:Str1}, two critical parameters, X-ray energy ($E$) and source size ($S$), are found to moderately correlate via a $E \sim S^{c_1}$ power law with $c_1$ = 0.54, 0.75, and 1.14 for $K_\alpha$, betatron, and bremsstrahlung, respectively. These power law relationships point to a tradeoff between the penetrating power of the laser-driven source, measured by the X-ray energy, and the practical resolution limit, often constrained by source size. 

The data for Fig.~\ref{fig:Str1} compiled include bremsstrahlung \cite{palaniyappan2018mev, tommasini2011development, glinec2005high, kerr2023development, courtois2011high, courtois2009effect, courtois2013characterisation, tommasini2017short}, K$_{\alpha}$ \cite{park2006high, workman2010phase, toth2007evaluation, antonelli2019x, gambari2020exploring}, and betatron \cite{wood2018ultrafast, cipiccia2011gamma, albert2017observation, albert2018betatron, ferri2016electron, rosmej2021bright} sources. An additional data point from a conventional source, the Dual-Axis Radiographic Hydrodynamic Test Facility (DARHT), also lies in agreement of this trend \cite{espy2021spectral}. Though the pulse duration is 60 ns rather than the ps-scale laser-driven sources \cite{Nath:2010}, the X-ray dose is $\sim$50$\times$ greater than the best-performing laser-driven sources \cite{kerr2023development}. Note there are techniques that may induce some deviation, such as using mass limited targets (i.e. wires) to constrain the source size, relative to foils with transverse dimensions much larger than the laser focal spot.

Another important feature of laser-driven sources is the sub-ns duration for a wide range of X-ray energies. With the 10-ps Titan laser incident on a 10 $\mu$m thick Au wire, a 12 ps (FWHM) burst of X-rays was produced, measured in the 3-10 keV range \cite{tommasini2011development}. When incident on 50 $\mu$m Au foils, the 22 ps ARC laser produces a pulse of 8 - 25 keV x-rays for $<$ 40 ps \cite{chen2017high}. Recent measurements with the Gamma Reaction History diagnostic on the ARC laser \cite{meaney2021multi, kerr2023development} constrained the pulse duration for a laser-driven X-rays $>$ 3 MeV. 

There are some known challenges for further advancing laser-driven U-RadIT. The fractions of laser energy absorption by various populations of electrons need better quantitative understanding for optimization. Some studies suggested a few 10's of one percent at intensities of $\sim10^{20}$ W/cm$^2$~\cite{ping2008absorption, park2021absolute}. Many high-power lasers contain an intensity pedestal that ramps up to the peak intensity, creating a pre-plasma before the arrival of the main $\sim$ps laser pulse. Controlling for pre-plasma remains a significant challenge, and it is important because instabilities such as hosing~\cite{ceurvorst2016mitigating} can be activated. As hosing modifies the laser trajectory in the pre-plasma, the resulting X-ray source stability, {\it i.e.} highly reproducible spectra and yield, is brought into question. 

\section{Summary} 
Ultrafast radiographic imaging and tracking (U-RadIT) use sub-nanosecond pulses of X-rays, $\gamma$-rays (high-energy X-rays $>$ 100 keV), and ionizing particles with mass such as electrons, protons and neutrons to collect information about material structures, densities, mass flow, other quasi-static and dynamic properties. As the light and particle sources become brighter (higher brilliance), narrower in pulse width towards $\sim$ 1 fs (in XFELs), higher repetition rate above 10 MHz (in 4th generation synchrotorons), and as new sources such as laser-driven ultrashort multi-species sources, which emit a broad spectrum of X-rays and particles simultaneously, become available, U-RadIT are important IT tools to study dynamic processes in physics, chemistry, biology, geology, materials science and others, including quantum fluctuations in emerging macroscopic quantum systems and phenomena.

The state-of-the-art computational forward models, as approximations to accelerate the calculations of the first-principle quantum physics models, can still not reliably predict dynamic properties of materials at high resolution when one mole or more atoms are involved. U-RadIT measurements are thus essential to validate model approximations, model predictions, and aid further model refinements, by providing high-quality experimental data for traditional physics-driven forward models, emerging data-driven models such as deep neural networks, or the hybrid models that merge physics with data. One of the central problems in U-RadIT is to optimize information yield from experiments through, {\it e.g.} high-luminosity X-ray and particle sources, efficient imaging and tracking detectors, novel IT modalities to collect data, and high-bandwidth online and offline data processing, regularized by the  underlying physics, geometry, statistics, and computing power. 

Steady progress in high-speed sensors and detector electronics in `10H' frontiers has led to a large number of high-data-yield detector technologies for U-RadIT optimization. The highlighted examples are ultrafast CMOS cameras, hybrid pixelated array detectors with flexible frontends, 3D photon-to-digital converters, Timepix4 ASICs that can be used for ultrafast particle counting, LGADs for 4D particle tracking.  As many detectors now reach single-visible-photon sensitivity, and with very compact (10 \textmu m or smaller pitch) solid-state designs, the state-of-the-art radiation detectors are quantum devices that can readily distinguish individual energetic particles and X-ray photons. It may be anticipated that photon counting with high energy resolution, or `spectroscopic photon counting' for ionizing radiation, and quantum detection with imbedded machine learning (ML) algorithms are forthcoming. Such advances may also benefit from alternate fabrication methods to CMOS integration such as 3D printing.

Hardware-centric approaches to U-RadIT optimization, which are sometimes constrained by detector sensor material properties, low signal-to-noise ratio, high cost and long development cycles of critical hardware components such as application-specific integrated circuits (ASICs), are now complemented by data methods, such as synthetic data generation from forward models or trained neural network models. New compressed sensing framework can relax the requirements of Nyquist-Shannon sampling theory and leads to sparse U-RadIT modalities to efficiently collect data from experiments. 

Data science and machine learning algorithms are also growingly applied to  post-processing of U-RadIT experimental data, including new phase retrieval algorithms in dynamic phase contrast imaging, moving contrast imaging and dynamic diffractive imaging. Uncertainty quantification (UQ) is important to data interpretation, predictions of new results, and guiding new experimental designs. Machine learning and artificial intelligence approaches, when enhanced by UQ, physics, and material information, may contribute significantly to data interpretation and overall U-RadIT optimization. 

Some of the exciting applications of U-RadIT are pushing the limits in temporal and spatial resolution on one end, and trying to reach to the largest spatial and temporal dynamic range on the other, {\it i.e.} by extending the measurement to the largest spatial scale and longest time duration possible, in order to shed light (or X-rays, or particles) on microscopic (down to atomic scales) processes while simultaneously revealing macroscopic functionality or emergent properties {\it in situ}. A few examples are included: cavity dynamics related to implosion and shock propagation, materials dynamics in novel 3D-printed structures under an impulse of energy, additive manufacturing optimization, and high-repetition-rate inertial confinement fusion energy experiments. 

\begin{acknowledgments}
We would like to thank Ms. Samantha Thurman (SLAC) and Mr. Jack Heyer (SLAC) for helping with the organization of  Ultrafast Imaging and Tracking Instrumentation, Methods and Application (ULITIMA 2023) Conference, March 13-16, 2023, Menlo Park, CA, USA. The special ULITIMA 2023 issue of Nuclear Instruments and Methods in Physics Research - section A (NIM-A) was made possible by many people from Elsevier, and especially Ms. M. Priyadharsini, Ms. Xinyi Xu, and Dr. William Barletta. ZW also wishes to thank Drs. Tammy Ma (Lawrence Livermore National Laboratory), Yuri K. Batygin (LANL), and Prof. Mark Foster (Johns Hopkins University) for stimulating discussions,  and Drs. Bob Reinovsky (LANL), Ann Satsangi (LANL), Rich Sheffield (LANL), Dmitry Yarotski (LANL) for encouragement and support to carry out the work. LANL work was performed under the auspices of the U.S. Department of Energy (DOE) by Triad National Security, LLC, operator of the Los Alamos National Laboratory under Contract No. 89233218CNA000001, including LANL Laboratory Directed Research and Development (LDRD) Program. This research used resources of the Advanced Photon Source, a U.S. Department of Energy (DOE) Office of Science user facility operated for the DOE Office of Science by Argonne National Laboratory under Contract No. DE-AC02-06CH11357.
\end{acknowledgments}





\bibliography{UFRadITv5noline}

\providecommand{\noopsort}[1]{}\providecommand{\singleletter}[1]{#1}%
\begin{thebibliography}{376}%
\makeatletter
\providecommand \@ifxundefined [1]{%
 \@ifx{#1\undefined}
}%
\providecommand \@ifnum [1]{%
 \ifnum #1\expandafter \@firstoftwo
 \else \expandafter \@secondoftwo
 \fi
}%
\providecommand \@ifx [1]{%
 \ifx #1\expandafter \@firstoftwo
 \else \expandafter \@secondoftwo
 \fi
}%
\providecommand \natexlab [1]{#1}%
\providecommand \enquote  [1]{``#1''}%
\providecommand \bibnamefont  [1]{#1}%
\providecommand \bibfnamefont [1]{#1}%
\providecommand \citenamefont [1]{#1}%
\providecommand \href@noop [0]{\@secondoftwo}%
\providecommand \href [0]{\begingroup \@sanitize@url \@href}%
\providecommand \@href[1]{\@@startlink{#1}\@@href}%
\providecommand \@@href[1]{\endgroup#1\@@endlink}%
\providecommand \@sanitize@url [0]{\catcode `\\12\catcode `\$12\catcode
  `\&12\catcode `\#12\catcode `\^12\catcode `\_12\catcode `\%12\relax}%
\providecommand \@@startlink[1]{}%
\providecommand \@@endlink[0]{}%
\providecommand \url  [0]{\begingroup\@sanitize@url \@url }%
\providecommand \@url [1]{\endgroup\@href {#1}{\urlprefix }}%
\providecommand \urlprefix  [0]{URL }%
\providecommand \Eprint [0]{\href }%
\providecommand \doibase [0]{http://dx.doi.org/}%
\providecommand \selectlanguage [0]{\@gobble}%
\providecommand \bibinfo  [0]{\@secondoftwo}%
\providecommand \bibfield  [0]{\@secondoftwo}%
\providecommand \translation [1]{[#1]}%
\providecommand \BibitemOpen [0]{}%
\providecommand \bibitemStop [0]{}%
\providecommand \bibitemNoStop [0]{.\EOS\space}%
\providecommand \EOS [0]{\spacefactor3000\relax}%
\providecommand \BibitemShut  [1]{\csname bibitem#1\endcsname}%
\let\auto@bib@innerbib\@empty
\bibitem [{\citenamefont {Wang}(2022)}]{Wan:2022}%
  \BibitemOpen
  \bibfield  {author} {\bibinfo {author} {\bibfnamefont {Z.}~\bibnamefont
  {Wang}},\ }\href@noop {} {\bibfield  {journal} {\bibinfo  {journal} {Appl.
  Opt.}\ }\textbf {\bibinfo {volume} {61}},\ \bibinfo {pages} {RDS1} (\bibinfo
  {year} {2022})},\ \bibinfo {note}
  {https://doi.org/10.1364/AO.455628}\BibitemShut {NoStop}%
\bibitem [{\citenamefont {Edgerton}\ and\ \citenamefont
  {Killian}(1979)}]{EK:1979}%
  \BibitemOpen
  \bibfield  {author} {\bibinfo {author} {\bibfnamefont {H.~E.}\ \bibnamefont
  {Edgerton}}\ and\ \bibinfo {author} {\bibfnamefont {J.~R.}\ \bibnamefont
  {Killian}},\ }\href@noop {} {\emph {\bibinfo {title} {Moments of Vision - the
  Stroboscopic revolution in photography}}}\ (\bibinfo  {publisher} {MIT
  Press},\ \bibinfo {year} {1979})\BibitemShut {NoStop}%
\bibitem [{\citenamefont {Liang}\ and\ \citenamefont {Wang}(2018)}]{LW:2018}%
  \BibitemOpen
  \bibfield  {author} {\bibinfo {author} {\bibfnamefont {J.}~\bibnamefont
  {Liang}}\ and\ \bibinfo {author} {\bibfnamefont {L.~V.}\ \bibnamefont
  {Wang}},\ }\href@noop {} {\bibfield  {journal} {\bibinfo  {journal} {Optica}\
  }\textbf {\bibinfo {volume} {5 (9)}},\ \bibinfo {pages} {1113 } (\bibinfo
  {year} {2018})}\BibitemShut {NoStop}%
\bibitem [{\citenamefont {Yang}\ \emph {et~al.}(2020)\citenamefont {Yang},
  \citenamefont {Cao}, \citenamefont {Qi}, \citenamefont {He}, \citenamefont
  {Ding}, \citenamefont {Yao}, \citenamefont {Jia}, \citenamefont {Sun},\ and\
  \citenamefont {Zhang}}]{YCQ:2020}%
  \BibitemOpen
  \bibfield  {author} {\bibinfo {author} {\bibfnamefont {C.}~\bibnamefont
  {Yang}}, \bibinfo {author} {\bibfnamefont {F.}~\bibnamefont {Cao}}, \bibinfo
  {author} {\bibfnamefont {D.}~\bibnamefont {Qi}}, \bibinfo {author}
  {\bibfnamefont {Y.}~\bibnamefont {He}}, \bibinfo {author} {\bibfnamefont
  {P.}~\bibnamefont {Ding}}, \bibinfo {author} {\bibfnamefont {J.}~\bibnamefont
  {Yao}}, \bibinfo {author} {\bibfnamefont {T.}~\bibnamefont {Jia}}, \bibinfo
  {author} {\bibfnamefont {Z.}~\bibnamefont {Sun}}, \ and\ \bibinfo {author}
  {\bibfnamefont {S.}~\bibnamefont {Zhang}},\ }\href@noop {} {\bibfield
  {journal} {\bibinfo  {journal} {Phys.\ Rev.\ Lett.}\ }\textbf {\bibinfo
  {volume} {124}},\ \bibinfo {pages} {023902} (\bibinfo {year}
  {2020})}\BibitemShut {NoStop}%
\bibitem [{\citenamefont {Tsao}(2010)}]{tsa:2010}%
  \BibitemOpen
  \bibfield  {author} {\bibinfo {author} {\bibfnamefont {J.}~\bibnamefont
  {Tsao}},\ }\href@noop {} {\bibfield  {journal} {\bibinfo  {journal} {J.\
  Mag.\ Res.\ Imag.}\ }\textbf {\bibinfo {volume} {32}},\ \bibinfo {pages} {252
  } (\bibinfo {year} {2010})}\BibitemShut {NoStop}%
\bibitem [{\citenamefont {Villemain}\ \emph {et~al.}(2020)\citenamefont
  {Villemain}, \citenamefont {Baranger}, \citenamefont {Friedberg},
  \citenamefont {Papadacci}, \citenamefont {Dizeux}, \citenamefont {Messas},
  \citenamefont {Tanter}, \citenamefont {Pernot},\ and\ \citenamefont
  {Mertens}}]{VBF:2020}%
  \BibitemOpen
  \bibfield  {author} {\bibinfo {author} {\bibfnamefont {O.}~\bibnamefont
  {Villemain}}, \bibinfo {author} {\bibfnamefont {J.}~\bibnamefont {Baranger}},
  \bibinfo {author} {\bibfnamefont {M.~K.}\ \bibnamefont {Friedberg}}, \bibinfo
  {author} {\bibfnamefont {C.}~\bibnamefont {Papadacci}}, \bibinfo {author}
  {\bibfnamefont {A.}~\bibnamefont {Dizeux}}, \bibinfo {author} {\bibfnamefont
  {E.}~\bibnamefont {Messas}}, \bibinfo {author} {\bibfnamefont
  {M.}~\bibnamefont {Tanter}}, \bibinfo {author} {\bibfnamefont
  {M.}~\bibnamefont {Pernot}}, \ and\ \bibinfo {author} {\bibfnamefont
  {L.}~\bibnamefont {Mertens}},\ }\href@noop {} {\bibfield  {journal} {\bibinfo
   {journal} {JACC Cardiovasc Imag.}\ }\textbf {\bibinfo {volume} {13 (8)}},\
  \bibinfo {pages} {1771 } (\bibinfo {year} {2020})}\BibitemShut {NoStop}%
\bibitem [{\citenamefont {Yoon}\ \emph {et~al.}(2021)\citenamefont {Yoon},
  \citenamefont {Kim}, \citenamefont {Choi}, \citenamefont {Sung},
  \citenamefont {Lee}, \citenamefont {Lee},\ and\ \citenamefont
  {Nam}}]{YKC:2021}%
  \BibitemOpen
  \bibfield  {author} {\bibinfo {author} {\bibfnamefont {J.~W.}\ \bibnamefont
  {Yoon}}, \bibinfo {author} {\bibfnamefont {Y.~G.}\ \bibnamefont {Kim}},
  \bibinfo {author} {\bibfnamefont {I.~W.}\ \bibnamefont {Choi}}, \bibinfo
  {author} {\bibfnamefont {J.~H.}\ \bibnamefont {Sung}}, \bibinfo {author}
  {\bibfnamefont {H.~W.}\ \bibnamefont {Lee}}, \bibinfo {author} {\bibfnamefont
  {S.~K.}\ \bibnamefont {Lee}}, \ and\ \bibinfo {author} {\bibfnamefont
  {C.~H.}\ \bibnamefont {Nam}},\ }\href@noop {} {\bibfield  {journal} {\bibinfo
   {journal} {Optica}\ }\textbf {\bibinfo {volume} {8(5)}},\ \bibinfo {pages}
  {630} (\bibinfo {year} {2021})}\BibitemShut {NoStop}%
\bibitem [{\citenamefont {Zewail}(2000)}]{Zew:2000}%
  \BibitemOpen
  \bibfield  {author} {\bibinfo {author} {\bibfnamefont {A.~H.}\ \bibnamefont
  {Zewail}},\ }\href@noop {} {\bibfield  {journal} {\bibinfo  {journal} {J.\
  Phys.\ Chem.\ A.}\ }\textbf {\bibinfo {volume} {104 (24)}},\ \bibinfo {pages}
  {5660} (\bibinfo {year} {2000})}\BibitemShut {NoStop}%
\bibitem [{\citenamefont {Han}\ and\ \citenamefont {Porter}(2020)}]{HP:2020}%
  \BibitemOpen
  \bibfield  {author} {\bibinfo {author} {\bibfnamefont {Z.}~\bibnamefont
  {Han}}\ and\ \bibinfo {author} {\bibfnamefont {A.~E.}\ \bibnamefont
  {Porter}},\ }\href@noop {} {\bibfield  {journal} {\bibinfo  {journal} {Front.
  Nanotechnol.}\ }\textbf {\bibinfo {volume} {2}},\ \bibinfo {pages} {606253}
  (\bibinfo {year} {2020})},\ \bibinfo {note}
  {doi:10.3389/fnano.2020.606253}\BibitemShut {NoStop}%
\bibitem [{\citenamefont {Young}\ \emph {et~al.}(2018)\citenamefont {Young},
  \citenamefont {Ueda}, \citenamefont {Gühr}, \citenamefont {Bucksbaum},
  \citenamefont {Simon}, \citenamefont {Mukamel}, \citenamefont {Rohringer},
  \citenamefont {Prince}, \citenamefont {Masciovecchio}, \citenamefont {Meyer},
  \citenamefont {Rudenko}, \citenamefont {Rolles}, \citenamefont {Bostedt},
  \citenamefont {Fuchs}, \citenamefont {Reis}, \citenamefont {Santra},
  \citenamefont {Kapteyn}, \citenamefont {Murnane}, \citenamefont {Ibrahim},
  \citenamefont {Légaré}, \citenamefont {Vrakking}, \citenamefont {Isinger},
  \citenamefont {Kroon}, \citenamefont {Gisselbrecht}, \citenamefont
  {L'Huillier}, \citenamefont {Wörner},\ and\ \citenamefont
  {Leone}}]{YUG:2018}%
  \BibitemOpen
  \bibfield  {author} {\bibinfo {author} {\bibfnamefont {L.}~\bibnamefont
  {Young}}, \bibinfo {author} {\bibfnamefont {K.}~\bibnamefont {Ueda}},
  \bibinfo {author} {\bibfnamefont {M.}~\bibnamefont {Gühr}}, \bibinfo
  {author} {\bibfnamefont {P.~H.}\ \bibnamefont {Bucksbaum}}, \bibinfo {author}
  {\bibfnamefont {M.}~\bibnamefont {Simon}}, \bibinfo {author} {\bibfnamefont
  {S.}~\bibnamefont {Mukamel}}, \bibinfo {author} {\bibfnamefont
  {N.}~\bibnamefont {Rohringer}}, \bibinfo {author} {\bibfnamefont {K.~C.}\
  \bibnamefont {Prince}}, \bibinfo {author} {\bibfnamefont {C.}~\bibnamefont
  {Masciovecchio}}, \bibinfo {author} {\bibfnamefont {M.}~\bibnamefont
  {Meyer}}, \bibinfo {author} {\bibfnamefont {A.}~\bibnamefont {Rudenko}},
  \bibinfo {author} {\bibfnamefont {D.}~\bibnamefont {Rolles}}, \bibinfo
  {author} {\bibfnamefont {C.}~\bibnamefont {Bostedt}}, \bibinfo {author}
  {\bibfnamefont {M.}~\bibnamefont {Fuchs}}, \bibinfo {author} {\bibfnamefont
  {D.~A.}\ \bibnamefont {Reis}}, \bibinfo {author} {\bibfnamefont
  {R.}~\bibnamefont {Santra}}, \bibinfo {author} {\bibfnamefont
  {H.}~\bibnamefont {Kapteyn}}, \bibinfo {author} {\bibfnamefont
  {M.}~\bibnamefont {Murnane}}, \bibinfo {author} {\bibfnamefont
  {H.}~\bibnamefont {Ibrahim}}, \bibinfo {author} {\bibfnamefont
  {F.}~\bibnamefont {Légaré}}, \bibinfo {author} {\bibfnamefont
  {M.}~\bibnamefont {Vrakking}}, \bibinfo {author} {\bibfnamefont
  {M.}~\bibnamefont {Isinger}}, \bibinfo {author} {\bibfnamefont
  {D.}~\bibnamefont {Kroon}}, \bibinfo {author} {\bibfnamefont
  {M.}~\bibnamefont {Gisselbrecht}}, \bibinfo {author} {\bibfnamefont
  {A.}~\bibnamefont {L'Huillier}}, \bibinfo {author} {\bibfnamefont {H.~J.}\
  \bibnamefont {Wörner}}, \ and\ \bibinfo {author} {\bibfnamefont {S.~R.}\
  \bibnamefont {Leone}},\ }\href@noop {} {\bibfield  {journal} {\bibinfo
  {journal} {J. Phys. B: At. Mol. Opt. Phys.}\ }\textbf {\bibinfo {volume}
  {51}},\ \bibinfo {pages} {032003} (\bibinfo {year} {2018})}\BibitemShut
  {NoStop}%
\bibitem [{\citenamefont {Krausz}\ and\ \citenamefont
  {Ivanov}(2009)}]{KI:2009}%
  \BibitemOpen
  \bibfield  {author} {\bibinfo {author} {\bibfnamefont {F.}~\bibnamefont
  {Krausz}}\ and\ \bibinfo {author} {\bibfnamefont {M.}~\bibnamefont
  {Ivanov}},\ }\href@noop {} {\bibfield  {journal} {\bibinfo  {journal} {Rev.
  Mod. Phys.}\ }\textbf {\bibinfo {volume} {81}},\ \bibinfo {pages} {163}
  (\bibinfo {year} {2009})}\BibitemShut {NoStop}%
\bibitem [{\citenamefont {Ossiander}\ \emph {et~al.}(2018)\citenamefont
  {Ossiander}, \citenamefont {Riemensberger}, \citenamefont {Neppl},
  \citenamefont {Mittermair}, \citenamefont {Schäffer}, \citenamefont
  {Duensing}, \citenamefont {Wagner}, \citenamefont {Heider}, \citenamefont
  {Wurzer}, \citenamefont {Gerl}, \citenamefont {Schnitzenbaumer},
  \citenamefont {Barth}, \citenamefont {Libisch}, \citenamefont {Lemell},
  \citenamefont {Burgdörfer}, \citenamefont {Feulner},\ and\ \citenamefont
  {Kienberger}}]{ORN:2018}%
  \BibitemOpen
  \bibfield  {author} {\bibinfo {author} {\bibfnamefont {M.}~\bibnamefont
  {Ossiander}}, \bibinfo {author} {\bibfnamefont {J.}~\bibnamefont
  {Riemensberger}}, \bibinfo {author} {\bibfnamefont {S.}~\bibnamefont
  {Neppl}}, \bibinfo {author} {\bibfnamefont {M.}~\bibnamefont {Mittermair}},
  \bibinfo {author} {\bibfnamefont {M.}~\bibnamefont {Schäffer}}, \bibinfo
  {author} {\bibfnamefont {A.}~\bibnamefont {Duensing}}, \bibinfo {author}
  {\bibfnamefont {M.~S.}\ \bibnamefont {Wagner}}, \bibinfo {author}
  {\bibfnamefont {R.}~\bibnamefont {Heider}}, \bibinfo {author} {\bibfnamefont
  {M.}~\bibnamefont {Wurzer}}, \bibinfo {author} {\bibfnamefont
  {M.}~\bibnamefont {Gerl}}, \bibinfo {author} {\bibfnamefont {M.}~\bibnamefont
  {Schnitzenbaumer}}, \bibinfo {author} {\bibfnamefont {J.~V.}\ \bibnamefont
  {Barth}}, \bibinfo {author} {\bibfnamefont {F.}~\bibnamefont {Libisch}},
  \bibinfo {author} {\bibfnamefont {C.}~\bibnamefont {Lemell}}, \bibinfo
  {author} {\bibfnamefont {J.}~\bibnamefont {Burgdörfer}}, \bibinfo {author}
  {\bibfnamefont {P.}~\bibnamefont {Feulner}}, \ and\ \bibinfo {author}
  {\bibfnamefont {R.}~\bibnamefont {Kienberger}},\ }\href@noop {} {\bibfield
  {journal} {\bibinfo  {journal} {Nature}\ }\textbf {\bibinfo {volume} {561}},\
  \bibinfo {pages} {374} (\bibinfo {year} {2018})}\BibitemShut {NoStop}%
\bibitem [{\citenamefont {Wang}\ \emph {et~al.}(2023)\citenamefont {Wang},
  \citenamefont {Dujardin}, \citenamefont {Freeman}, \citenamefont {Gehring},
  \citenamefont {Hunter}, \citenamefont {Lecoq}, \citenamefont {Liu},
  \citenamefont {Melcher}, \citenamefont {Morris}, \citenamefont {Nikl},
  \citenamefont {Pilania}, \citenamefont {Pokharel}, \citenamefont {Robertson},
  \citenamefont {Rutstrom}, \citenamefont {Sjue}, \citenamefont {Tremsin},
  \citenamefont {Watson}, \citenamefont {Wiggins}, \citenamefont {Winch},\ and\
  \citenamefont {Zhuravleva}}]{Wan:2023}%
  \BibitemOpen
  \bibfield  {author} {\bibinfo {author} {\bibfnamefont {Z.}~\bibnamefont
  {Wang}}, \bibinfo {author} {\bibfnamefont {C.}~\bibnamefont {Dujardin}},
  \bibinfo {author} {\bibfnamefont {M.~S.}\ \bibnamefont {Freeman}}, \bibinfo
  {author} {\bibfnamefont {A.~E.}\ \bibnamefont {Gehring}}, \bibinfo {author}
  {\bibfnamefont {J.~F.}\ \bibnamefont {Hunter}}, \bibinfo {author}
  {\bibfnamefont {P.}~\bibnamefont {Lecoq}}, \bibinfo {author} {\bibfnamefont
  {W.}~\bibnamefont {Liu}}, \bibinfo {author} {\bibfnamefont {C.~L.}\
  \bibnamefont {Melcher}}, \bibinfo {author} {\bibfnamefont {C.~L.}\
  \bibnamefont {Morris}}, \bibinfo {author} {\bibfnamefont {M.}~\bibnamefont
  {Nikl}}, \bibinfo {author} {\bibfnamefont {G.}~\bibnamefont {Pilania}},
  \bibinfo {author} {\bibfnamefont {R.}~\bibnamefont {Pokharel}}, \bibinfo
  {author} {\bibfnamefont {D.~G.}\ \bibnamefont {Robertson}}, \bibinfo {author}
  {\bibfnamefont {D.~J.}\ \bibnamefont {Rutstrom}}, \bibinfo {author}
  {\bibfnamefont {S.~K.}\ \bibnamefont {Sjue}}, \bibinfo {author}
  {\bibfnamefont {A.~S.}\ \bibnamefont {Tremsin}}, \bibinfo {author}
  {\bibfnamefont {S.}~\bibnamefont {Watson}}, \bibinfo {author} {\bibfnamefont
  {B.~W.}\ \bibnamefont {Wiggins}}, \bibinfo {author} {\bibfnamefont {N.~M.}\
  \bibnamefont {Winch}}, \ and\ \bibinfo {author} {\bibfnamefont
  {M.}~\bibnamefont {Zhuravleva}},\ }\href@noop {} {\bibfield  {journal}
  {\bibinfo  {journal} {IEEE Trans. Nucl. Sci.}\ }\textbf {\bibinfo {volume}
  {70}},\ \bibinfo {pages} {1244 } (\bibinfo {year} {2023})},\ \bibinfo {note}
  {https://doi.org/10.1109/TNS.2023.3290826}\BibitemShut {NoStop}%
\bibitem [{\citenamefont {Styer}\ \emph {et~al.}(2002)\citenamefont {Styer},
  \citenamefont {Balkin}, \citenamefont {Becker}, \citenamefont {Burns},
  \citenamefont {Dudley}, \citenamefont {Forth}, \citenamefont {Gaumer},
  \citenamefont {Kramer}, \citenamefont {Oertel}, \citenamefont {Park},
  \citenamefont {Rinkoski}, \citenamefont {Smith},\ and\ \citenamefont
  {Wotherspoon}}]{SBB:2002}%
  \BibitemOpen
  \bibfield  {author} {\bibinfo {author} {\bibfnamefont {D.~F.}\ \bibnamefont
  {Styer}}, \bibinfo {author} {\bibfnamefont {M.~S.}\ \bibnamefont {Balkin}},
  \bibinfo {author} {\bibfnamefont {K.~M.}\ \bibnamefont {Becker}}, \bibinfo
  {author} {\bibfnamefont {M.~R.}\ \bibnamefont {Burns}}, \bibinfo {author}
  {\bibfnamefont {C.~E.}\ \bibnamefont {Dudley}}, \bibinfo {author}
  {\bibfnamefont {S.~T.}\ \bibnamefont {Forth}}, \bibinfo {author}
  {\bibfnamefont {J.~S.}\ \bibnamefont {Gaumer}}, \bibinfo {author}
  {\bibfnamefont {M.~A.}\ \bibnamefont {Kramer}}, \bibinfo {author}
  {\bibfnamefont {D.~C.}\ \bibnamefont {Oertel}}, \bibinfo {author}
  {\bibfnamefont {L.~H.}\ \bibnamefont {Park}}, \bibinfo {author}
  {\bibfnamefont {M.~T.}\ \bibnamefont {Rinkoski}}, \bibinfo {author}
  {\bibfnamefont {C.~T.}\ \bibnamefont {Smith}}, \ and\ \bibinfo {author}
  {\bibfnamefont {T.~D.}\ \bibnamefont {Wotherspoon}},\ }\href@noop {}
  {\bibfield  {journal} {\bibinfo  {journal} {Am. J. Phys.}\ }\textbf {\bibinfo
  {volume} {70}},\ \bibinfo {pages} {288} (\bibinfo {year} {2002})}\BibitemShut
  {NoStop}%
\bibitem [{\citenamefont {Dirac}(1929)}]{Dirac:1929}%
  \BibitemOpen
  \bibfield  {author} {\bibinfo {author} {\bibfnamefont {P.~A.~M.}\
  \bibnamefont {Dirac}},\ }\href@noop {} {\bibfield  {journal} {\bibinfo
  {journal} {Proc. R. Soc. Lond. A}\ }\textbf {\bibinfo {volume} {123}},\
  \bibinfo {pages} {714} (\bibinfo {year} {1929})}\BibitemShut {NoStop}%
\bibitem [{\citenamefont {Simons}(2023)}]{Sim:2023}%
  \BibitemOpen
  \bibfield  {author} {\bibinfo {author} {\bibfnamefont {J.}~\bibnamefont
  {Simons}},\ }\href@noop {} {\bibfield  {journal} {\bibinfo  {journal} {J. Am.
  Chem. Soc.}\ }\textbf {\bibinfo {volume} {145}},\ \bibinfo {pages} {4343}
  (\bibinfo {year} {2023})}\BibitemShut {NoStop}%
\bibitem [{\citenamefont {Ozboyaci}\ \emph {et~al.}(2016)\citenamefont
  {Ozboyaci}, \citenamefont {Kokh}, \citenamefont {Corni},\ and\ \citenamefont
  {Wade}}]{OKCW:2016}%
  \BibitemOpen
  \bibfield  {author} {\bibinfo {author} {\bibfnamefont {M.}~\bibnamefont
  {Ozboyaci}}, \bibinfo {author} {\bibfnamefont {D.~B.}\ \bibnamefont {Kokh}},
  \bibinfo {author} {\bibfnamefont {S.}~\bibnamefont {Corni}}, \ and\ \bibinfo
  {author} {\bibfnamefont {R.~C.}\ \bibnamefont {Wade}},\ }\href@noop {}
  {\bibfield  {journal} {\bibinfo  {journal} {Quart. Rev. Biophys.}\ }\textbf
  {\bibinfo {volume} {49}},\ \bibinfo {pages} {e4} (\bibinfo {year}
  {2016})}\BibitemShut {NoStop}%
\bibitem [{\citenamefont {{Google AI quantum and
  collaborators}}(2020)}]{Rubin:2020}%
  \BibitemOpen
  \bibfield  {author} {\bibinfo {author} {\bibnamefont {{Google AI quantum and
  collaborators}}},\ }\href@noop {} {\bibfield  {journal} {\bibinfo  {journal}
  {Science}\ }\textbf {\bibinfo {volume} {369}},\ \bibinfo {pages} {1084}
  (\bibinfo {year} {2020})}\BibitemShut {NoStop}%
\bibitem [{\citenamefont {Hansson}\ \emph {et~al.}(2002)\citenamefont
  {Hansson}, \citenamefont {Oostenbrink},\ and\ \citenamefont {van
  Gunsteren}}]{HOG:2002}%
  \BibitemOpen
  \bibfield  {author} {\bibinfo {author} {\bibfnamefont {T.}~\bibnamefont
  {Hansson}}, \bibinfo {author} {\bibfnamefont {C.}~\bibnamefont
  {Oostenbrink}}, \ and\ \bibinfo {author} {\bibfnamefont {W.}~\bibnamefont
  {van Gunsteren}},\ }\href@noop {} {\bibfield  {journal} {\bibinfo  {journal}
  {Current Opinion Struct. Bio.}\ }\textbf {\bibinfo {volume} {12}},\ \bibinfo
  {pages} {(2), 190} (\bibinfo {year} {2002})}\BibitemShut {NoStop}%
\bibitem [{\citenamefont {Rapaport}(2004)}]{Rap:2004}%
  \BibitemOpen
  \bibfield  {author} {\bibinfo {author} {\bibfnamefont {D.~C.}\ \bibnamefont
  {Rapaport}},\ }\href@noop {} {\emph {\bibinfo {title} {The art of molecular
  dynamics simulation}}},\ \bibinfo {edition} {2nd}\ ed.\ (\bibinfo
  {publisher} {Cambridge University Press},\ \bibinfo {year}
  {2004})\BibitemShut {NoStop}%
\bibitem [{\citenamefont {Gilbert}\ and\ \citenamefont {{\it et
  al.}}(2021)}]{GAB:2021}%
  \BibitemOpen
  \bibfield  {author} {\bibinfo {author} {\bibfnamefont {M.~R.}\ \bibnamefont
  {Gilbert}}\ and\ \bibinfo {author} {\bibnamefont {{\it et al.}}},\
  }\href@noop {} {\bibfield  {journal} {\bibinfo  {journal} {J. Nucl. Mater.}\
  }\textbf {\bibinfo {volume} {554}},\ \bibinfo {pages} {153113} (\bibinfo
  {year} {2021})}\BibitemShut {NoStop}%
\bibitem [{\citenamefont {Behler}\ and\ \citenamefont
  {Parrinello}(2007)}]{BP:2007}%
  \BibitemOpen
  \bibfield  {author} {\bibinfo {author} {\bibfnamefont {J.}~\bibnamefont
  {Behler}}\ and\ \bibinfo {author} {\bibfnamefont {M.}~\bibnamefont
  {Parrinello}},\ }\href@noop {} {\bibfield  {journal} {\bibinfo  {journal}
  {Phys. Rev. Lett.}\ }\textbf {\bibinfo {volume} {98}},\ \bibinfo {pages}
  {(14), 146401} (\bibinfo {year} {2007})}\BibitemShut {NoStop}%
\bibitem [{\citenamefont {Durrant}\ and\ \citenamefont
  {McCammon}(2011)}]{DM:2011}%
  \BibitemOpen
  \bibfield  {author} {\bibinfo {author} {\bibfnamefont {J.}~\bibnamefont
  {Durrant}}\ and\ \bibinfo {author} {\bibfnamefont {J.~A.}\ \bibnamefont
  {McCammon}},\ }\href@noop {} {\bibfield  {journal} {\bibinfo  {journal} {BMC
  Biology}\ }\textbf {\bibinfo {volume} {9}},\ \bibinfo {pages} {art. No. 71}
  (\bibinfo {year} {2011})}\BibitemShut {NoStop}%
\bibitem [{\citenamefont {Hollingsworth}\ and\ \citenamefont
  {Dror}(2018)}]{HD:2018}%
  \BibitemOpen
  \bibfield  {author} {\bibinfo {author} {\bibfnamefont {S.~A.}\ \bibnamefont
  {Hollingsworth}}\ and\ \bibinfo {author} {\bibfnamefont {R.~O.}\ \bibnamefont
  {Dror}},\ }\href@noop {} {\bibfield  {journal} {\bibinfo  {journal}
  {Neuron.}\ }\textbf {\bibinfo {volume} {99(6)}},\ \bibinfo {pages} {1129}
  (\bibinfo {year} {2018})}\BibitemShut {NoStop}%
\bibitem [{\citenamefont {Lambert}\ \emph {et~al.}(2011)\citenamefont
  {Lambert}, \citenamefont {Recoules}, \citenamefont {Decoster}, \citenamefont
  {Cl\'erouin},\ and\ \citenamefont {Desjarlais}}]{LRD:2011}%
  \BibitemOpen
  \bibfield  {author} {\bibinfo {author} {\bibfnamefont {F.}~\bibnamefont
  {Lambert}}, \bibinfo {author} {\bibfnamefont {V.}~\bibnamefont {Recoules}},
  \bibinfo {author} {\bibfnamefont {A.}~\bibnamefont {Decoster}}, \bibinfo
  {author} {\bibfnamefont {J.}~\bibnamefont {Cl\'erouin}}, \ and\ \bibinfo
  {author} {\bibfnamefont {M.}~\bibnamefont {Desjarlais}},\ }\href@noop {}
  {\bibfield  {journal} {\bibinfo  {journal} {Phys, Plasma.}\ }\textbf
  {\bibinfo {volume} {18}},\ \bibinfo {pages} {056306} (\bibinfo {year}
  {2011})}\BibitemShut {NoStop}%
\bibitem [{\citenamefont {Nan}\ \emph {et~al.}(2016)\citenamefont {Nan},
  \citenamefont {Yuan}, \citenamefont {Wang}, \citenamefont {Peng},
  \citenamefont {Chen}, \citenamefont {Du}, \citenamefont {Zhang},\ and\
  \citenamefont {Lv}}]{SWT:2016}%
  \BibitemOpen
  \bibfield  {author} {\bibinfo {author} {\bibfnamefont {S.}~\bibnamefont
  {Nan}}, \bibinfo {author} {\bibfnamefont {W.}~\bibnamefont {Yuan}}, \bibinfo
  {author} {\bibfnamefont {T.}~\bibnamefont {Wang}}, \bibinfo {author}
  {\bibfnamefont {H.}~\bibnamefont {Peng}}, \bibinfo {author} {\bibfnamefont
  {L.}~\bibnamefont {Chen}}, \bibinfo {author} {\bibfnamefont {X.}~\bibnamefont
  {Du}}, \bibinfo {author} {\bibfnamefont {D.}~\bibnamefont {Zhang}}, \ and\
  \bibinfo {author} {\bibfnamefont {P.}~\bibnamefont {Lv}},\ }\href@noop {}
  {\bibfield  {journal} {\bibinfo  {journal} {High Power Laser and Particle
  Beams.}\ }\textbf {\bibinfo {volume} {28}},\ \bibinfo {pages} {092001}
  (\bibinfo {year} {2016})}\BibitemShut {NoStop}%
\bibitem [{\citenamefont {Hu}\ \emph {et~al.}(2016)\citenamefont {Hu},
  \citenamefont {Collins}, \citenamefont {Goncharov}, \citenamefont {Kress},
  \citenamefont {McCrory},\ and\ \citenamefont {Skupsky}}]{HCG:2016}%
  \BibitemOpen
  \bibfield  {author} {\bibinfo {author} {\bibfnamefont {S.~X.}\ \bibnamefont
  {Hu}}, \bibinfo {author} {\bibfnamefont {L.~A.}\ \bibnamefont {Collins}},
  \bibinfo {author} {\bibfnamefont {V.~N.}\ \bibnamefont {Goncharov}}, \bibinfo
  {author} {\bibfnamefont {J.~D.}\ \bibnamefont {Kress}}, \bibinfo {author}
  {\bibfnamefont {R.~L.}\ \bibnamefont {McCrory}}, \ and\ \bibinfo {author}
  {\bibfnamefont {S.}~\bibnamefont {Skupsky}},\ }\href@noop {} {\bibfield
  {journal} {\bibinfo  {journal} {Phys. Plasma}\ }\textbf {\bibinfo {volume}
  {23}},\ \bibinfo {pages} {042704} (\bibinfo {year} {2016})}\BibitemShut
  {NoStop}%
\bibitem [{\citenamefont {Hu}\ \emph {et~al.}(2011)\citenamefont {Hu},
  \citenamefont {Militzer}, \citenamefont {Goncharov},\ and\ \citenamefont
  {Skupsky}}]{HMG:2011}%
  \BibitemOpen
  \bibfield  {author} {\bibinfo {author} {\bibfnamefont {S.~X.}\ \bibnamefont
  {Hu}}, \bibinfo {author} {\bibfnamefont {B.}~\bibnamefont {Militzer}},
  \bibinfo {author} {\bibfnamefont {V.~N.}\ \bibnamefont {Goncharov}}, \ and\
  \bibinfo {author} {\bibfnamefont {S.}~\bibnamefont {Skupsky}},\ }\href@noop
  {} {\bibfield  {journal} {\bibinfo  {journal} {Phys. Rev. B}\ }\textbf
  {\bibinfo {volume} {84}},\ \bibinfo {pages} {224109} (\bibinfo {year}
  {2011})}\BibitemShut {NoStop}%
\bibitem [{\citenamefont {Haines}\ \emph {et~al.}(2022)\citenamefont {Haines},
  \citenamefont {Keller}, \citenamefont {Long}, \citenamefont {{M. D. McKay,
  Jr.}}, \citenamefont {Medin}, \citenamefont {Park}, \citenamefont
  {Rauenzahn}, \citenamefont {Scott}, \citenamefont {Anderson}, \citenamefont
  {Collins}, \citenamefont {Green}, \citenamefont {Marozas}, \citenamefont
  {McKenty}, \citenamefont {Peterson}, \citenamefont {Vold}, \citenamefont
  {Stefano}, \citenamefont {Lester}, \citenamefont {Sauppe}, \citenamefont
  {Stark},\ and\ \citenamefont {Velechovsky}}]{HKL:2022}%
  \BibitemOpen
  \bibfield  {author} {\bibinfo {author} {\bibfnamefont {B.~M.}\ \bibnamefont
  {Haines}}, \bibinfo {author} {\bibfnamefont {D.~E.}\ \bibnamefont {Keller}},
  \bibinfo {author} {\bibfnamefont {K.~P.}\ \bibnamefont {Long}}, \bibinfo
  {author} {\bibnamefont {{M. D. McKay, Jr.}}}, \bibinfo {author}
  {\bibfnamefont {Z.~J.}\ \bibnamefont {Medin}}, \bibinfo {author}
  {\bibfnamefont {H.}~\bibnamefont {Park}}, \bibinfo {author} {\bibfnamefont
  {R.~M.}\ \bibnamefont {Rauenzahn}}, \bibinfo {author} {\bibfnamefont {H.~A.}\
  \bibnamefont {Scott}}, \bibinfo {author} {\bibfnamefont {K.~S.}\ \bibnamefont
  {Anderson}}, \bibinfo {author} {\bibfnamefont {T.~J.~B.}\ \bibnamefont
  {Collins}}, \bibinfo {author} {\bibfnamefont {L.~M.}\ \bibnamefont {Green}},
  \bibinfo {author} {\bibfnamefont {J.~A.}\ \bibnamefont {Marozas}}, \bibinfo
  {author} {\bibfnamefont {P.~W.}\ \bibnamefont {McKenty}}, \bibinfo {author}
  {\bibfnamefont {J.~H.}\ \bibnamefont {Peterson}}, \bibinfo {author}
  {\bibfnamefont {E.~L.}\ \bibnamefont {Vold}}, \bibinfo {author}
  {\bibfnamefont {C.~D.}\ \bibnamefont {Stefano}}, \bibinfo {author}
  {\bibfnamefont {R.~S.}\ \bibnamefont {Lester}}, \bibinfo {author}
  {\bibfnamefont {J.~P.}\ \bibnamefont {Sauppe}}, \bibinfo {author}
  {\bibfnamefont {D.~J.}\ \bibnamefont {Stark}}, \ and\ \bibinfo {author}
  {\bibfnamefont {J.}~\bibnamefont {Velechovsky}},\ }\href@noop {} {\bibfield
  {journal} {\bibinfo  {journal} {Phys. Plasma.}\ }\textbf {\bibinfo {volume}
  {29}},\ \bibinfo {pages} {083901} (\bibinfo {year} {2022})}\BibitemShut
  {NoStop}%
\bibitem [{\citenamefont {Marinak}\ \emph {et~al.}(2001)\citenamefont
  {Marinak}, \citenamefont {Kerbel}, \citenamefont {Gentile}, \citenamefont
  {Jones}, \citenamefont {Munro}, \citenamefont {Pollaine}, \citenamefont
  {Dittrich},\ and\ \citenamefont {Haan}}]{MKG:2001}%
  \BibitemOpen
  \bibfield  {author} {\bibinfo {author} {\bibfnamefont {M.~M.}\ \bibnamefont
  {Marinak}}, \bibinfo {author} {\bibfnamefont {G.~D.}\ \bibnamefont {Kerbel}},
  \bibinfo {author} {\bibfnamefont {N.~A.}\ \bibnamefont {Gentile}}, \bibinfo
  {author} {\bibfnamefont {O.}~\bibnamefont {Jones}}, \bibinfo {author}
  {\bibfnamefont {D.}~\bibnamefont {Munro}}, \bibinfo {author} {\bibfnamefont
  {S.}~\bibnamefont {Pollaine}}, \bibinfo {author} {\bibfnamefont {T.~R.}\
  \bibnamefont {Dittrich}}, \ and\ \bibinfo {author} {\bibfnamefont {S.~W.}\
  \bibnamefont {Haan}},\ }\href@noop {} {\bibfield  {journal} {\bibinfo
  {journal} {Phys. Plasma.}\ }\textbf {\bibinfo {volume} {8}},\ \bibinfo
  {pages} {2275} (\bibinfo {year} {2001})}\BibitemShut {NoStop}%
\bibitem [{\citenamefont {Spannagel}\ \emph {et~al.}(2018)\citenamefont
  {Spannagel}, \citenamefont {Wolters}, \citenamefont {Hynds}, \citenamefont
  {Tehrani}, \citenamefont {Benoit}, \citenamefont {Dannheim}, \citenamefont
  {Gauvin}, \citenamefont {N{\"u}rnberg}, \citenamefont {Sch{\"u}tze},\ and\
  \citenamefont {Vincente}}]{SS:2022}%
  \BibitemOpen
  \bibfield  {author} {\bibinfo {author} {\bibfnamefont {S.}~\bibnamefont
  {Spannagel}}, \bibinfo {author} {\bibfnamefont {K.}~\bibnamefont {Wolters}},
  \bibinfo {author} {\bibfnamefont {D.}~\bibnamefont {Hynds}}, \bibinfo
  {author} {\bibfnamefont {N.~A.}\ \bibnamefont {Tehrani}}, \bibinfo {author}
  {\bibfnamefont {M.}~\bibnamefont {Benoit}}, \bibinfo {author} {\bibfnamefont
  {D.}~\bibnamefont {Dannheim}}, \bibinfo {author} {\bibfnamefont
  {N.}~\bibnamefont {Gauvin}}, \bibinfo {author} {\bibfnamefont
  {A.}~\bibnamefont {N{\"u}rnberg}}, \bibinfo {author} {\bibfnamefont
  {P.}~\bibnamefont {Sch{\"u}tze}}, \ and\ \bibinfo {author} {\bibfnamefont
  {M.}~\bibnamefont {Vincente}},\ }\href@noop {} {\bibfield  {journal}
  {\bibinfo  {journal} {Nucl. Instrum. Meth. A}\ }\textbf {\bibinfo {volume}
  {901}},\ \bibinfo {pages} {164} (\bibinfo {year} {2018})}\BibitemShut
  {NoStop}%
\bibitem [{\citenamefont {Yue}\ \emph {et~al.}(2022)\citenamefont {Yue},
  \citenamefont {Lin}, \citenamefont {Li}, \citenamefont {Wolfe}, \citenamefont
  {Clayton}, \citenamefont {Makela}, \citenamefont {Morris}, \citenamefont
  {Spannagel}, \citenamefont {Ramberg}, \citenamefont {Estrada}, \citenamefont
  {Zhu}, \citenamefont {Liu}, \citenamefont {Fossum},\ and\ \citenamefont
  {Wang}}]{YL:2023}%
  \BibitemOpen
  \bibfield  {author} {\bibinfo {author} {\bibfnamefont {X.}~\bibnamefont
  {Yue}}, \bibinfo {author} {\bibfnamefont {S.}~\bibnamefont {Lin}}, \bibinfo
  {author} {\bibfnamefont {W.}~\bibnamefont {Li}}, \bibinfo {author}
  {\bibfnamefont {B.~T.}\ \bibnamefont {Wolfe}}, \bibinfo {author}
  {\bibfnamefont {S.}~\bibnamefont {Clayton}}, \bibinfo {author} {\bibfnamefont
  {M.}~\bibnamefont {Makela}}, \bibinfo {author} {\bibfnamefont {C.~L.}\
  \bibnamefont {Morris}}, \bibinfo {author} {\bibfnamefont {S.}~\bibnamefont
  {Spannagel}}, \bibinfo {author} {\bibfnamefont {E.}~\bibnamefont {Ramberg}},
  \bibinfo {author} {\bibfnamefont {J.}~\bibnamefont {Estrada}}, \bibinfo
  {author} {\bibfnamefont {H.}~\bibnamefont {Zhu}}, \bibinfo {author}
  {\bibfnamefont {J.}~\bibnamefont {Liu}}, \bibinfo {author} {\bibfnamefont
  {E.~R.}\ \bibnamefont {Fossum}}, \ and\ \bibinfo {author} {\bibfnamefont
  {Z.}~\bibnamefont {Wang}},\ }\href@noop {} {\bibfield  {journal} {\bibinfo
  {journal} {PoS}\ }\textbf {\bibinfo {volume} {{(Pixel 2022)}}},\ \bibinfo
  {pages} {041} (\bibinfo {year} {2022})}\BibitemShut {NoStop}%
\bibitem [{\citenamefont {Maiuri}\ \emph {et~al.}(2020)\citenamefont {Maiuri},
  \citenamefont {Garavelli},\ and\ \citenamefont {Cerullo}}]{MGC:2020}%
  \BibitemOpen
  \bibfield  {author} {\bibinfo {author} {\bibfnamefont {M.}~\bibnamefont
  {Maiuri}}, \bibinfo {author} {\bibfnamefont {M.}~\bibnamefont {Garavelli}}, \
  and\ \bibinfo {author} {\bibfnamefont {G.}~\bibnamefont {Cerullo}},\
  }\href@noop {} {\bibfield  {journal} {\bibinfo  {journal} {J. Am. Chem.
  Soc.}\ }\textbf {\bibinfo {volume} {142}},\ \bibinfo {pages} {3} (\bibinfo
  {year} {2020})}\BibitemShut {NoStop}%
\bibitem [{\citenamefont {Dooling}\ \emph {et~al.}(2022)\citenamefont
  {Dooling}, \citenamefont {Borland}, \citenamefont {Berg}, \citenamefont
  {Calvey}, \citenamefont {Decker}, \citenamefont {Emery}, \citenamefont
  {Harkay}, \citenamefont {Lindberg}, \citenamefont {Navrotksi}, \citenamefont
  {Sajaev}, \citenamefont {Shoaf}, \citenamefont {Sun}, \citenamefont
  {Wootton}, \citenamefont {Xiao}, \citenamefont {Grannan},\ and\ \citenamefont
  {Lumpkin}}]{DBB:2022}%
  \BibitemOpen
  \bibfield  {author} {\bibinfo {author} {\bibfnamefont {J.}~\bibnamefont
  {Dooling}}, \bibinfo {author} {\bibfnamefont {M.}~\bibnamefont {Borland}},
  \bibinfo {author} {\bibfnamefont {W.}~\bibnamefont {Berg}}, \bibinfo {author}
  {\bibfnamefont {J.}~\bibnamefont {Calvey}}, \bibinfo {author} {\bibfnamefont
  {G.}~\bibnamefont {Decker}}, \bibinfo {author} {\bibfnamefont
  {L.}~\bibnamefont {Emery}}, \bibinfo {author} {\bibfnamefont
  {K.}~\bibnamefont {Harkay}}, \bibinfo {author} {\bibfnamefont
  {R.}~\bibnamefont {Lindberg}}, \bibinfo {author} {\bibfnamefont
  {G.}~\bibnamefont {Navrotksi}}, \bibinfo {author} {\bibfnamefont
  {V.}~\bibnamefont {Sajaev}}, \bibinfo {author} {\bibfnamefont
  {S.}~\bibnamefont {Shoaf}}, \bibinfo {author} {\bibfnamefont {Y.~P.}\
  \bibnamefont {Sun}}, \bibinfo {author} {\bibfnamefont {K.~P.}\ \bibnamefont
  {Wootton}}, \bibinfo {author} {\bibfnamefont {A.}~\bibnamefont {Xiao}},
  \bibinfo {author} {\bibfnamefont {A.}~\bibnamefont {Grannan}}, \ and\
  \bibinfo {author} {\bibfnamefont {A.~H.}\ \bibnamefont {Lumpkin}},\
  }\href@noop {} {\bibfield  {journal} {\bibinfo  {journal} {Phys. Rev. Accel.
  Beams}\ }\textbf {\bibinfo {volume} {25}},\ \bibinfo {pages} {043001}
  (\bibinfo {year} {2022})}\BibitemShut {NoStop}%
\bibitem [{\citenamefont {Raimondi}\ \emph {et~al.}(2023)\citenamefont
  {Raimondi}, \citenamefont {Benabderrahmane}, \citenamefont {Berkvens},
  \citenamefont {Biasci}, \citenamefont {Borowiec}, \citenamefont {Bouteille},
  \citenamefont {Brochard}, \citenamefont {Brookes}, \citenamefont
  {Carmignani}, \citenamefont {Carver}, \citenamefont {Chaize}, \citenamefont
  {Chavanne}, \citenamefont {Checchia}, \citenamefont {Chushkin}, \citenamefont
  {Cianciosi}, \citenamefont {Michiel}, \citenamefont {Dimper}, \citenamefont
  {D’Elia}, \citenamefont {Einfeld}, \citenamefont {Ewald}, \citenamefont
  {Farvacque}, \citenamefont {Goirand}, \citenamefont {Hardy}, \citenamefont
  {Jacob}, \citenamefont {Jolly}, \citenamefont {Krisch}, \citenamefont {Bec},
  \citenamefont {Leconte}, \citenamefont {Liuzzo}, \citenamefont {Maccarrone},
  \citenamefont {Marchial}, \citenamefont {Martin}, \citenamefont {Mezouar},
  \citenamefont {Nevo}, \citenamefont {Perron}, \citenamefont {Plouviez},
  \citenamefont {Reichert}, \citenamefont {Renaud}, \citenamefont {Revol},
  \citenamefont {Roche}, \citenamefont {Scheidt}, \citenamefont {Serriere},
  \citenamefont {Sette}, \citenamefont {Susini}, \citenamefont {Torino},
  \citenamefont {Versteegen}, \citenamefont {White},\ and\ \citenamefont
  {Zontone}}]{RBB:2023}%
  \BibitemOpen
  \bibfield  {author} {\bibinfo {author} {\bibfnamefont {P.}~\bibnamefont
  {Raimondi}}, \bibinfo {author} {\bibfnamefont {C.}~\bibnamefont
  {Benabderrahmane}}, \bibinfo {author} {\bibfnamefont {P.}~\bibnamefont
  {Berkvens}}, \bibinfo {author} {\bibfnamefont {J.~C.}\ \bibnamefont
  {Biasci}}, \bibinfo {author} {\bibfnamefont {P.}~\bibnamefont {Borowiec}},
  \bibinfo {author} {\bibfnamefont {J.-F.}\ \bibnamefont {Bouteille}}, \bibinfo
  {author} {\bibfnamefont {T.}~\bibnamefont {Brochard}}, \bibinfo {author}
  {\bibfnamefont {N.~B.}\ \bibnamefont {Brookes}}, \bibinfo {author}
  {\bibfnamefont {N.}~\bibnamefont {Carmignani}}, \bibinfo {author}
  {\bibfnamefont {L.~R.}\ \bibnamefont {Carver}}, \bibinfo {author}
  {\bibfnamefont {J.-M.}\ \bibnamefont {Chaize}}, \bibinfo {author}
  {\bibfnamefont {J.}~\bibnamefont {Chavanne}}, \bibinfo {author}
  {\bibfnamefont {S.}~\bibnamefont {Checchia}}, \bibinfo {author}
  {\bibfnamefont {Y.}~\bibnamefont {Chushkin}}, \bibinfo {author}
  {\bibfnamefont {F.}~\bibnamefont {Cianciosi}}, \bibinfo {author}
  {\bibfnamefont {M.~D.}\ \bibnamefont {Michiel}}, \bibinfo {author}
  {\bibfnamefont {R.}~\bibnamefont {Dimper}}, \bibinfo {author} {\bibfnamefont
  {A.}~\bibnamefont {D’Elia}}, \bibinfo {author} {\bibfnamefont
  {D.}~\bibnamefont {Einfeld}}, \bibinfo {author} {\bibfnamefont
  {F.}~\bibnamefont {Ewald}}, \bibinfo {author} {\bibfnamefont
  {L.}~\bibnamefont {Farvacque}}, \bibinfo {author} {\bibfnamefont
  {L.}~\bibnamefont {Goirand}}, \bibinfo {author} {\bibfnamefont
  {L.}~\bibnamefont {Hardy}}, \bibinfo {author} {\bibfnamefont
  {J.}~\bibnamefont {Jacob}}, \bibinfo {author} {\bibfnamefont
  {L.}~\bibnamefont {Jolly}}, \bibinfo {author} {\bibfnamefont
  {M.}~\bibnamefont {Krisch}}, \bibinfo {author} {\bibfnamefont {G.~L.}\
  \bibnamefont {Bec}}, \bibinfo {author} {\bibfnamefont {I.}~\bibnamefont
  {Leconte}}, \bibinfo {author} {\bibfnamefont {S.~M.}\ \bibnamefont {Liuzzo}},
  \bibinfo {author} {\bibfnamefont {C.}~\bibnamefont {Maccarrone}}, \bibinfo
  {author} {\bibfnamefont {T.}~\bibnamefont {Marchial}}, \bibinfo {author}
  {\bibfnamefont {D.}~\bibnamefont {Martin}}, \bibinfo {author} {\bibfnamefont
  {M.}~\bibnamefont {Mezouar}}, \bibinfo {author} {\bibfnamefont
  {C.}~\bibnamefont {Nevo}}, \bibinfo {author} {\bibfnamefont {T.}~\bibnamefont
  {Perron}}, \bibinfo {author} {\bibfnamefont {E.}~\bibnamefont {Plouviez}},
  \bibinfo {author} {\bibfnamefont {H.}~\bibnamefont {Reichert}}, \bibinfo
  {author} {\bibfnamefont {P.}~\bibnamefont {Renaud}}, \bibinfo {author}
  {\bibfnamefont {J.-L.}\ \bibnamefont {Revol}}, \bibinfo {author}
  {\bibfnamefont {B.}~\bibnamefont {Roche}}, \bibinfo {author} {\bibfnamefont
  {K.-B.}\ \bibnamefont {Scheidt}}, \bibinfo {author} {\bibfnamefont
  {V.}~\bibnamefont {Serriere}}, \bibinfo {author} {\bibfnamefont
  {F.}~\bibnamefont {Sette}}, \bibinfo {author} {\bibfnamefont
  {J.}~\bibnamefont {Susini}}, \bibinfo {author} {\bibfnamefont
  {L.}~\bibnamefont {Torino}}, \bibinfo {author} {\bibfnamefont
  {R.}~\bibnamefont {Versteegen}}, \bibinfo {author} {\bibfnamefont
  {S.}~\bibnamefont {White}}, \ and\ \bibinfo {author} {\bibfnamefont
  {F.}~\bibnamefont {Zontone}},\ }\href@noop {} {\bibfield  {journal} {\bibinfo
   {journal} {Communications Physics}\ }\textbf {\bibinfo {volume} {6}},\
  \bibinfo {pages} {1} (\bibinfo {year} {2023})}\BibitemShut {NoStop}%
\bibitem [{\citenamefont {Liuzzo}\ \emph {et~al.}(2016)\citenamefont {Liuzzo},
  \citenamefont {Carmignani}, \citenamefont {Chavanne}, \citenamefont {Bec},
  \citenamefont {Nash}, \citenamefont {Raimondi}, \citenamefont {Versteegen},\
  and\ \citenamefont {White}}]{LCC:2016}%
  \BibitemOpen
  \bibfield  {author} {\bibinfo {author} {\bibfnamefont {S.}~\bibnamefont
  {Liuzzo}}, \bibinfo {author} {\bibfnamefont {N.}~\bibnamefont {Carmignani}},
  \bibinfo {author} {\bibfnamefont {J.}~\bibnamefont {Chavanne}}, \bibinfo
  {author} {\bibfnamefont {L.~F. G.~L.}\ \bibnamefont {Bec}}, \bibinfo {author}
  {\bibfnamefont {B.}~\bibnamefont {Nash}}, \bibinfo {author} {\bibfnamefont
  {P.}~\bibnamefont {Raimondi}}, \bibinfo {author} {\bibfnamefont
  {R.}~\bibnamefont {Versteegen}}, \ and\ \bibinfo {author} {\bibfnamefont
  {S.~M.}\ \bibnamefont {White}},\ }in\ \href@noop {} {\emph {\bibinfo
  {booktitle} {Proc. 7th Int. Part. Accel. Conf. (IPAC'2016)}}}\ (\bibinfo
  {address} {Busan, Korea},\ \bibinfo {year} {May 8-13, 2016})\ pp.\ \bibinfo
  {pages} {2818--2821}\BibitemShut {NoStop}%
\bibitem [{\citenamefont {Allahgholi}\ \emph {et~al.}(2019)\citenamefont
  {Allahgholi}, \citenamefont {Becker}, \citenamefont {Delfs}, \citenamefont
  {Dinapoli}, \citenamefont {G{\"o}ttlicher}, \citenamefont {Graafsma},
  \citenamefont {Greiffenberg}, \citenamefont {Hirsemann}, \citenamefont
  {Jack}, \citenamefont {Klyuev}, \citenamefont {Kr{\"u}ger}, \citenamefont
  {Kuhn}, \citenamefont {Laurus}, \citenamefont {Marras}, \citenamefont
  {Mezza}, \citenamefont {Mozzanica}, \citenamefont {Poehlsen}, \citenamefont
  {Shalev}, \citenamefont {Sheviakov}, \citenamefont {Schmitt}, \citenamefont
  {Schwandt}, \citenamefont {Shi}, \citenamefont {Smoljanin}, \citenamefont
  {Trunk}, \citenamefont {Zhang},\ and\ \citenamefont {Zimmer}}]{BBG:2012}%
  \BibitemOpen
  \bibfield  {author} {\bibinfo {author} {\bibfnamefont {A.}~\bibnamefont
  {Allahgholi}}, \bibinfo {author} {\bibfnamefont {J.}~\bibnamefont {Becker}},
  \bibinfo {author} {\bibfnamefont {A.}~\bibnamefont {Delfs}}, \bibinfo
  {author} {\bibfnamefont {R.}~\bibnamefont {Dinapoli}}, \bibinfo {author}
  {\bibfnamefont {P.}~\bibnamefont {G{\"o}ttlicher}}, \bibinfo {author}
  {\bibfnamefont {H.}~\bibnamefont {Graafsma}}, \bibinfo {author}
  {\bibfnamefont {D.}~\bibnamefont {Greiffenberg}}, \bibinfo {author}
  {\bibfnamefont {H.}~\bibnamefont {Hirsemann}}, \bibinfo {author}
  {\bibfnamefont {S.}~\bibnamefont {Jack}}, \bibinfo {author} {\bibfnamefont
  {A.}~\bibnamefont {Klyuev}}, \bibinfo {author} {\bibfnamefont
  {H.}~\bibnamefont {Kr{\"u}ger}}, \bibinfo {author} {\bibfnamefont
  {M.}~\bibnamefont {Kuhn}}, \bibinfo {author} {\bibfnamefont {T.}~\bibnamefont
  {Laurus}}, \bibinfo {author} {\bibfnamefont {A.}~\bibnamefont {Marras}},
  \bibinfo {author} {\bibfnamefont {D.}~\bibnamefont {Mezza}}, \bibinfo
  {author} {\bibfnamefont {A.}~\bibnamefont {Mozzanica}}, \bibinfo {author}
  {\bibfnamefont {J.}~\bibnamefont {Poehlsen}}, \bibinfo {author}
  {\bibfnamefont {O.~S.}\ \bibnamefont {Shalev}}, \bibinfo {author}
  {\bibfnamefont {I.}~\bibnamefont {Sheviakov}}, \bibinfo {author}
  {\bibfnamefont {B.}~\bibnamefont {Schmitt}}, \bibinfo {author} {\bibfnamefont
  {J.}~\bibnamefont {Schwandt}}, \bibinfo {author} {\bibfnamefont
  {X.}~\bibnamefont {Shi}}, \bibinfo {author} {\bibfnamefont {S.}~\bibnamefont
  {Smoljanin}}, \bibinfo {author} {\bibfnamefont {U.}~\bibnamefont {Trunk}},
  \bibinfo {author} {\bibfnamefont {J.}~\bibnamefont {Zhang}}, \ and\ \bibinfo
  {author} {\bibfnamefont {M.}~\bibnamefont {Zimmer}},\ }\href@noop {}
  {\bibfield  {journal} {\bibinfo  {journal} {Nucl. Instrum. Meth. A}\ }\textbf
  {\bibinfo {volume} {942}},\ \bibinfo {pages} {162324} (\bibinfo {year}
  {2019})}\BibitemShut {NoStop}%
\bibitem [{\citenamefont {Welton}\ \emph {et~al.}(2022)\citenamefont {Welton},
  \citenamefont {Bollinger}, \citenamefont {Dehnel}, \citenamefont {Draganic},
  \citenamefont {Faircloth}, \citenamefont {Han}, \citenamefont {Lettry},
  \citenamefont {Stockli}, \citenamefont {Tarvainen},\ and\ \citenamefont
  {Ueno}}]{WBD:2022}%
  \BibitemOpen
  \bibfield  {author} {\bibinfo {author} {\bibfnamefont {R.}~\bibnamefont
  {Welton}}, \bibinfo {author} {\bibfnamefont {D.}~\bibnamefont {Bollinger}},
  \bibinfo {author} {\bibfnamefont {M.}~\bibnamefont {Dehnel}}, \bibinfo
  {author} {\bibfnamefont {I.}~\bibnamefont {Draganic}}, \bibinfo {author}
  {\bibfnamefont {D.}~\bibnamefont {Faircloth}}, \bibinfo {author}
  {\bibfnamefont {B.}~\bibnamefont {Han}}, \bibinfo {author} {\bibfnamefont
  {J.}~\bibnamefont {Lettry}}, \bibinfo {author} {\bibfnamefont
  {M.}~\bibnamefont {Stockli}}, \bibinfo {author} {\bibfnamefont
  {O.}~\bibnamefont {Tarvainen}}, \ and\ \bibinfo {author} {\bibfnamefont
  {A.}~\bibnamefont {Ueno}},\ }\href {\doibase 10.1088/1742-6596/2244/1/012045}
  {\bibfield  {journal} {\bibinfo  {journal} {J. Physics: Conf. Ser.}\ }\textbf
  {\bibinfo {volume} {2244}},\ \bibinfo {pages} {1} (\bibinfo {year} {2022})},\
  \bibinfo {note} {iCIS2021, FERMILAB-CONF-22-386-AD}\BibitemShut {NoStop}%
\bibitem [{\citenamefont {Carlsten}\ \emph {et~al.}(2019)\citenamefont
  {Carlsten}, \citenamefont {Anisimov}, \citenamefont {Barnes}, \citenamefont
  {Marksteiner}, \citenamefont {Robles},\ and\ \citenamefont
  {Yampolsky}}]{CAB:2019}%
  \BibitemOpen
  \bibfield  {author} {\bibinfo {author} {\bibfnamefont {B.~E.}\ \bibnamefont
  {Carlsten}}, \bibinfo {author} {\bibfnamefont {P.~M.}\ \bibnamefont
  {Anisimov}}, \bibinfo {author} {\bibfnamefont {C.~W.}\ \bibnamefont
  {Barnes}}, \bibinfo {author} {\bibfnamefont {Q.~R.}\ \bibnamefont
  {Marksteiner}}, \bibinfo {author} {\bibfnamefont {R.~R.}\ \bibnamefont
  {Robles}}, \ and\ \bibinfo {author} {\bibfnamefont {N.}~\bibnamefont
  {Yampolsky}},\ }\href@noop {} {\bibfield  {journal} {\bibinfo  {journal}
  {Instruments}\ }\textbf {\bibinfo {volume} {3}},\ \bibinfo {pages} {52}
  (\bibinfo {year} {2019})}\BibitemShut {NoStop}%
\bibitem [{\citenamefont {Schroer}\ \emph {et~al.}(2018)\citenamefont
  {Schroer}, \citenamefont {Agapov}, \citenamefont {Brefeld}, \citenamefont
  {Brinkmann}, \citenamefont {Chae}, \citenamefont {Chao}, \citenamefont
  {Eriksson}, \citenamefont {Keil}, \citenamefont {Gavald\`a}, \citenamefont
  {R{\"o}hlsberger}, \citenamefont {Seeck}, \citenamefont {Sprung},
  \citenamefont {Tischer}, \citenamefont {Wanzenberg},\ and\ \citenamefont
  {Weckert}}]{SAB:2018}%
  \BibitemOpen
  \bibfield  {author} {\bibinfo {author} {\bibfnamefont {C.~G.}\ \bibnamefont
  {Schroer}}, \bibinfo {author} {\bibfnamefont {I.}~\bibnamefont {Agapov}},
  \bibinfo {author} {\bibfnamefont {W.}~\bibnamefont {Brefeld}}, \bibinfo
  {author} {\bibfnamefont {R.}~\bibnamefont {Brinkmann}}, \bibinfo {author}
  {\bibfnamefont {Y.-C.}\ \bibnamefont {Chae}}, \bibinfo {author}
  {\bibfnamefont {H.-C.}\ \bibnamefont {Chao}}, \bibinfo {author}
  {\bibfnamefont {M.}~\bibnamefont {Eriksson}}, \bibinfo {author}
  {\bibfnamefont {J.}~\bibnamefont {Keil}}, \bibinfo {author} {\bibfnamefont
  {X.~N.}\ \bibnamefont {Gavald\`a}}, \bibinfo {author} {\bibfnamefont
  {R.}~\bibnamefont {R{\"o}hlsberger}}, \bibinfo {author} {\bibfnamefont
  {O.~H.}\ \bibnamefont {Seeck}}, \bibinfo {author} {\bibfnamefont
  {M.}~\bibnamefont {Sprung}}, \bibinfo {author} {\bibfnamefont
  {M.}~\bibnamefont {Tischer}}, \bibinfo {author} {\bibfnamefont
  {R.}~\bibnamefont {Wanzenberg}}, \ and\ \bibinfo {author} {\bibfnamefont
  {E.}~\bibnamefont {Weckert}},\ }\href@noop {} {\bibfield  {journal} {\bibinfo
   {journal} {J. Synchr. Rad.}\ }\textbf {\bibinfo {volume} {25}},\ \bibinfo
  {pages} {1277} (\bibinfo {year} {2018})}\BibitemShut {NoStop}%
\bibitem [{\citenamefont {Huang}\ \emph {et~al.}(2021)\citenamefont {Huang},
  \citenamefont {Deng}, \citenamefont {Liu}, \citenamefont {Wang},\ and\
  \citenamefont {Zhao}}]{HDL:2021}%
  \BibitemOpen
  \bibfield  {author} {\bibinfo {author} {\bibfnamefont {N.}~\bibnamefont
  {Huang}}, \bibinfo {author} {\bibfnamefont {H.}~\bibnamefont {Deng}},
  \bibinfo {author} {\bibfnamefont {B.}~\bibnamefont {Liu}}, \bibinfo {author}
  {\bibfnamefont {D.}~\bibnamefont {Wang}}, \ and\ \bibinfo {author}
  {\bibfnamefont {Z.}~\bibnamefont {Zhao}},\ }\href@noop {} {\bibfield
  {journal} {\bibinfo  {journal} {The Innovation}\ }\textbf {\bibinfo {volume}
  {2(2)}},\ \bibinfo {pages} {100097} (\bibinfo {year} {2021})}\BibitemShut
  {NoStop}%
\bibitem [{\citenamefont {Ray}(1997)}]{Ray:1997}%
  \BibitemOpen
  \bibinfo {editor} {\bibfnamefont {S.~F.}\ \bibnamefont {Ray}},\ ed.,\
  \href@noop {} {\emph {\bibinfo {title} {High-speed photography and
  photonics}}}\ (\bibinfo  {publisher} {SPIE Press},\ \bibinfo {address}
  {Bellingham, Washington, USA},\ \bibinfo {year} {1997})\BibitemShut {NoStop}%
\bibitem [{\citenamefont {Fossum}(1993)}]{Dart:1}%
  \BibitemOpen
  \bibfield  {author} {\bibinfo {author} {\bibfnamefont {E.~R.}\ \bibnamefont
  {Fossum}},\ }\href@noop {} {\bibfield  {journal} {\bibinfo  {journal} {SPIE
  Proc.}\ }\textbf {\bibinfo {volume} {1900}} (\bibinfo {year} {1993})},\
  \bibinfo {note} {{``Active pixel sensors: Are CCDS dinosaurs?"}, in {\it
  Charge-Coupled Devices and Solid State Optical Sensors III};
  https://doi.org/10.1117/12.148585.}\BibitemShut {Stop}%
\bibitem [{\citenamefont {Via}\ \emph {et~al.}(1997)\citenamefont {Via},
  \citenamefont {Bates}, \citenamefont {Bertolucci}, \citenamefont {Bottigli},
  \citenamefont {Campbell}, \citenamefont {Chesi}, \citenamefont {Conti},
  \citenamefont {Auria}, \citenamefont {Delpapa}, \citenamefont {Fantacci},
  \citenamefont {Grossi}, \citenamefont {Heijne}, \citenamefont {Mancini},
  \citenamefont {Middelkamp}, \citenamefont {Raine}, \citenamefont {Russo},
  \citenamefont {Shea}, \citenamefont {Scharfetter}, \citenamefont {Smith},
  \citenamefont {Snoeys},\ and\ \citenamefont {Stefanini}}]{DVB:1997}%
  \BibitemOpen
  \bibfield  {author} {\bibinfo {author} {\bibfnamefont {C.~D.}\ \bibnamefont
  {Via}}, \bibinfo {author} {\bibfnamefont {R.}~\bibnamefont {Bates}}, \bibinfo
  {author} {\bibfnamefont {E.}~\bibnamefont {Bertolucci}}, \bibinfo {author}
  {\bibfnamefont {U.}~\bibnamefont {Bottigli}}, \bibinfo {author}
  {\bibfnamefont {M.}~\bibnamefont {Campbell}}, \bibinfo {author}
  {\bibfnamefont {E.}~\bibnamefont {Chesi}}, \bibinfo {author} {\bibfnamefont
  {M.}~\bibnamefont {Conti}}, \bibinfo {author} {\bibfnamefont
  {S.}~\bibnamefont {Auria}}, \bibinfo {author} {\bibfnamefont
  {C.}~\bibnamefont {Delpapa}}, \bibinfo {author} {\bibfnamefont
  {M.}~\bibnamefont {Fantacci}}, \bibinfo {author} {\bibfnamefont
  {G.}~\bibnamefont {Grossi}}, \bibinfo {author} {\bibfnamefont
  {E.}~\bibnamefont {Heijne}}, \bibinfo {author} {\bibfnamefont
  {E.}~\bibnamefont {Mancini}}, \bibinfo {author} {\bibfnamefont
  {P.}~\bibnamefont {Middelkamp}}, \bibinfo {author} {\bibfnamefont
  {C.}~\bibnamefont {Raine}}, \bibinfo {author} {\bibfnamefont
  {P.}~\bibnamefont {Russo}}, \bibinfo {author} {\bibfnamefont
  {V.}~\bibnamefont {Shea}}, \bibinfo {author} {\bibfnamefont {L.}~\bibnamefont
  {Scharfetter}}, \bibinfo {author} {\bibfnamefont {K.}~\bibnamefont {Smith}},
  \bibinfo {author} {\bibfnamefont {W.}~\bibnamefont {Snoeys}}, \ and\ \bibinfo
  {author} {\bibfnamefont {A.}~\bibnamefont {Stefanini}},\ }\href@noop {}
  {\bibfield  {journal} {\bibinfo  {journal} {Nucl. Instrum. Meth. Phys. Res.
  Sec. A.}\ }\textbf {\bibinfo {volume} {395}},\ \bibinfo {pages} {148}
  (\bibinfo {year} {1997})}\BibitemShut {NoStop}%
\bibitem [{\citenamefont {Campbell}(2011)}]{Camp:2011}%
  \BibitemOpen
  \bibfield  {author} {\bibinfo {author} {\bibfnamefont {M.}~\bibnamefont
  {Campbell}},\ }\href {\doibase 10.1016/j.nima.2010.06.106} {\bibfield
  {journal} {\bibinfo  {journal} {Nucl. Instrum. Meth. Phys. Res. Sec. A.}\
  }\textbf {\bibinfo {volume} {633}},\ \bibinfo {pages} {S1} (\bibinfo {year}
  {2011})}\BibitemShut {NoStop}%
\bibitem [{\citenamefont {Graafsma}(2018)}]{Gra:2018}%
  \BibitemOpen
  \bibfield  {author} {\bibinfo {author} {\bibfnamefont {H.}~\bibnamefont
  {Graafsma}},\ }in\ \href@noop {} {\emph {\bibinfo {booktitle} {Semiconductor
  Radiation Systems}}},\ \bibinfo {editor} {edited by\ \bibinfo {editor}
  {\bibfnamefont {K.}~\bibnamefont {Iniewski}}}\ (\bibinfo  {publisher} {CRC
  Press},\ \bibinfo {address} {{Boca Raton, FL, USA}},\ \bibinfo {year}
  {2018})\ pp.\ \bibinfo {pages} {217--236},\ \bibinfo {note} {{\it `Hybrid
  Pixel Array Detectors for Photon Science'}}\BibitemShut {NoStop}%
\bibitem [{\citenamefont {Rossi}\ \emph {et~al.}(1999)\citenamefont {Rossi},
  \citenamefont {Renzi}, \citenamefont {Eikenberry}, \citenamefont
  {Bilderback}, \citenamefont {Fontes}, \citenamefont {Wixted}, \citenamefont
  {Barna},\ and\ \citenamefont {Gruner}}]{RRE:1999}%
  \BibitemOpen
  \bibfield  {author} {\bibinfo {author} {\bibfnamefont {G.}~\bibnamefont
  {Rossi}}, \bibinfo {author} {\bibfnamefont {M.~J.}\ \bibnamefont {Renzi}},
  \bibinfo {author} {\bibfnamefont {M.~W.}\ \bibnamefont {Eikenberry},
  \bibfnamefont {E.~F.~andTate}}, \bibinfo {author} {\bibfnamefont
  {D.}~\bibnamefont {Bilderback}}, \bibinfo {author} {\bibfnamefont
  {E.}~\bibnamefont {Fontes}}, \bibinfo {author} {\bibfnamefont
  {R.}~\bibnamefont {Wixted}}, \bibinfo {author} {\bibfnamefont
  {S.}~\bibnamefont {Barna}}, \ and\ \bibinfo {author} {\bibfnamefont {S.~M.}\
  \bibnamefont {Gruner}},\ }\href {\doibase 10.1107/S0909049599009802}
  {\bibfield  {journal} {\bibinfo  {journal} {Journal of Synchrotron
  Radiation}\ }\textbf {\bibinfo {volume} {6}},\ \bibinfo {pages} {1096}
  (\bibinfo {year} {1999})}\BibitemShut {NoStop}%
\bibitem [{\citenamefont {Philipp}\ \emph {et~al.}(2016)\citenamefont
  {Philipp}, \citenamefont {Tate}, \citenamefont {Purohit}, \citenamefont
  {Shanks}, \citenamefont {Weiss},\ and\ \citenamefont {Gruner}}]{G1}%
  \BibitemOpen
  \bibfield  {author} {\bibinfo {author} {\bibfnamefont {H.~T.}\ \bibnamefont
  {Philipp}}, \bibinfo {author} {\bibfnamefont {M.~W.}\ \bibnamefont {Tate}},
  \bibinfo {author} {\bibfnamefont {P.}~\bibnamefont {Purohit}}, \bibinfo
  {author} {\bibfnamefont {K.~S.}\ \bibnamefont {Shanks}}, \bibinfo {author}
  {\bibfnamefont {J.~T.}\ \bibnamefont {Weiss}}, \ and\ \bibinfo {author}
  {\bibfnamefont {S.~M.}\ \bibnamefont {Gruner}},\ }\href@noop {} {\bibfield
  {journal} {\bibinfo  {journal} {J. Synchrotron Rad.}\ }\textbf {\bibinfo
  {volume} {23}},\ \bibinfo {pages} {395} (\bibinfo {year} {2016})},\ \bibinfo
  {note} {doi:10.1107/S1600577515022754}\BibitemShut {NoStop}%
\bibitem [{\citenamefont {Philipp}\ \emph {et~al.}(2011)\citenamefont
  {Philipp}, \citenamefont {Hromalik}, \citenamefont {Tate}, \citenamefont
  {Koerner},\ and\ \citenamefont {Gruner}}]{PKH:2007}%
  \BibitemOpen
  \bibfield  {author} {\bibinfo {author} {\bibfnamefont {H.~T.}\ \bibnamefont
  {Philipp}}, \bibinfo {author} {\bibfnamefont {M.}~\bibnamefont {Hromalik}},
  \bibinfo {author} {\bibfnamefont {M.}~\bibnamefont {Tate}}, \bibinfo {author}
  {\bibfnamefont {L.}~\bibnamefont {Koerner}}, \ and\ \bibinfo {author}
  {\bibfnamefont {S.~M.}\ \bibnamefont {Gruner}},\ }\href@noop {} {\bibfield
  {journal} {\bibinfo  {journal} {Nucl. Instrum. Meth. A}\ }\textbf {\bibinfo
  {volume} {649}},\ \bibinfo {pages} {67} (\bibinfo {year} {2011})}\BibitemShut
  {NoStop}%
\bibitem [{\citenamefont {Hart}\ \emph {et~al.}(2012)\citenamefont {Hart},
  \citenamefont {Boutet}, \citenamefont {Carini}, \citenamefont {Dragone},
  \citenamefont {Duda}, \citenamefont {Freytag}, \citenamefont {Haller},
  \citenamefont {Herbst}, \citenamefont {Herrmann}, \citenamefont {Kenney},
  \citenamefont {Morse}, \citenamefont {Nordby}, \citenamefont {Pines},
  \citenamefont {van Bakel}, \citenamefont {Weaver},\ and\ \citenamefont
  {Williams}}]{HBC:2012}%
  \BibitemOpen
  \bibfield  {author} {\bibinfo {author} {\bibfnamefont {P.}~\bibnamefont
  {Hart}}, \bibinfo {author} {\bibfnamefont {S.}~\bibnamefont {Boutet}},
  \bibinfo {author} {\bibfnamefont {G.}~\bibnamefont {Carini}}, \bibinfo
  {author} {\bibfnamefont {A.}~\bibnamefont {Dragone}}, \bibinfo {author}
  {\bibfnamefont {B.}~\bibnamefont {Duda}}, \bibinfo {author} {\bibfnamefont
  {D.}~\bibnamefont {Freytag}}, \bibinfo {author} {\bibfnamefont
  {G.}~\bibnamefont {Haller}}, \bibinfo {author} {\bibfnamefont
  {R.}~\bibnamefont {Herbst}}, \bibinfo {author} {\bibfnamefont
  {S.}~\bibnamefont {Herrmann}}, \bibinfo {author} {\bibfnamefont
  {C.}~\bibnamefont {Kenney}}, \bibinfo {author} {\bibfnamefont
  {J.}~\bibnamefont {Morse}}, \bibinfo {author} {\bibfnamefont
  {M.}~\bibnamefont {Nordby}}, \bibinfo {author} {\bibfnamefont
  {J.}~\bibnamefont {Pines}}, \bibinfo {author} {\bibfnamefont
  {N.}~\bibnamefont {van Bakel}}, \bibinfo {author} {\bibfnamefont
  {M.}~\bibnamefont {Weaver}}, \ and\ \bibinfo {author} {\bibfnamefont
  {G.}~\bibnamefont {Williams}},\ }\href@noop {} {\enquote {\bibinfo {title}
  {The cornell-slac pixel array detector at lcls},}\ } (\bibinfo {year}
  {2012}),\ \bibinfo {note} {{SLAC-PUB-15284}}\BibitemShut {NoStop}%
\bibitem [{\citenamefont {Mozzanica}\ \emph {et~al.}(2018)\citenamefont
  {Mozzanica}, \citenamefont {Andr{\"a}}, \citenamefont {Barten}, \citenamefont
  {Bergamaschi}, \citenamefont {Chiriotti}, \citenamefont {Br{\"u}ckner},
  \citenamefont {Dinapoli}, \citenamefont {Fr{\"o}jdh}, \citenamefont
  {Greiffenberg}, \citenamefont {Leonarski}, \citenamefont {Lopez-Cuenca},
  \citenamefont {D.Mezza}, \citenamefont {Redford}, \citenamefont {Ruder},
  \citenamefont {Schmitt}, \citenamefont {Shi}, \citenamefont {Thattil},
  \citenamefont {Tinti}, \citenamefont {Vetter},\ and\ \citenamefont
  {Zhang}}]{LMB:2020}%
  \BibitemOpen
  \bibfield  {author} {\bibinfo {author} {\bibfnamefont {A.}~\bibnamefont
  {Mozzanica}}, \bibinfo {author} {\bibfnamefont {M.}~\bibnamefont
  {Andr{\"a}}}, \bibinfo {author} {\bibfnamefont {R.}~\bibnamefont {Barten}},
  \bibinfo {author} {\bibfnamefont {A.}~\bibnamefont {Bergamaschi}}, \bibinfo
  {author} {\bibfnamefont {S.}~\bibnamefont {Chiriotti}}, \bibinfo {author}
  {\bibfnamefont {M.}~\bibnamefont {Br{\"u}ckner}}, \bibinfo {author}
  {\bibfnamefont {R.}~\bibnamefont {Dinapoli}}, \bibinfo {author}
  {\bibfnamefont {E.}~\bibnamefont {Fr{\"o}jdh}}, \bibinfo {author}
  {\bibfnamefont {D.}~\bibnamefont {Greiffenberg}}, \bibinfo {author}
  {\bibfnamefont {F.}~\bibnamefont {Leonarski}}, \bibinfo {author}
  {\bibfnamefont {C.}~\bibnamefont {Lopez-Cuenca}}, \bibinfo {author}
  {\bibnamefont {D.Mezza}}, \bibinfo {author} {\bibfnamefont {S.}~\bibnamefont
  {Redford}}, \bibinfo {author} {\bibfnamefont {C.}~\bibnamefont {Ruder}},
  \bibinfo {author} {\bibfnamefont {B.}~\bibnamefont {Schmitt}}, \bibinfo
  {author} {\bibfnamefont {X.}~\bibnamefont {Shi}}, \bibinfo {author}
  {\bibfnamefont {D.}~\bibnamefont {Thattil}}, \bibinfo {author} {\bibfnamefont
  {G.}~\bibnamefont {Tinti}}, \bibinfo {author} {\bibfnamefont
  {S.}~\bibnamefont {Vetter}}, \ and\ \bibinfo {author} {\bibfnamefont
  {J.}~\bibnamefont {Zhang}},\ }\href {\doibase 10.1080/08940886.2018.1528429}
  {\bibfield  {journal} {\bibinfo  {journal} {Synchrotron Radiation News}\
  }\textbf {\bibinfo {volume} {31}},\ \bibinfo {pages} {16} (\bibinfo {year}
  {2018})}\BibitemShut {NoStop}%
\bibitem [{\citenamefont {Gadkari}\ \emph {et~al.}(2022)\citenamefont
  {Gadkari}, \citenamefont {Shanks}, \citenamefont {Hu}, \citenamefont
  {Philipp}, \citenamefont {Tate}, \citenamefont {Thom-Levy},\ and\
  \citenamefont {Gruner}}]{G2}%
  \BibitemOpen
  \bibfield  {author} {\bibinfo {author} {\bibfnamefont {D.}~\bibnamefont
  {Gadkari}}, \bibinfo {author} {\bibfnamefont {K.~S.}\ \bibnamefont {Shanks}},
  \bibinfo {author} {\bibfnamefont {H.}~\bibnamefont {Hu}}, \bibinfo {author}
  {\bibfnamefont {H.~T.}\ \bibnamefont {Philipp}}, \bibinfo {author}
  {\bibfnamefont {M.~W.}\ \bibnamefont {Tate}}, \bibinfo {author}
  {\bibfnamefont {J.}~\bibnamefont {Thom-Levy}}, \ and\ \bibinfo {author}
  {\bibfnamefont {S.~M.}\ \bibnamefont {Gruner}},\ }\href {\doibase DOI:
  10.1088/1748-0221/17/03/P03003} {\bibfield  {journal} {\bibinfo  {journal}
  {Journal of Instrumentation}\ }\textbf {\bibinfo {volume} {17}},\ \bibinfo
  {pages} {P03003} (\bibinfo {year} {2022})}\BibitemShut {NoStop}%
\bibitem [{\citenamefont {Hatsui}\ and\ \citenamefont
  {Graafsma}(2015)}]{HG:2015}%
  \BibitemOpen
  \bibfield  {author} {\bibinfo {author} {\bibfnamefont {T.}~\bibnamefont
  {Hatsui}}\ and\ \bibinfo {author} {\bibfnamefont {H.}~\bibnamefont
  {Graafsma}},\ }\href@noop {} {\bibfield  {journal} {\bibinfo  {journal}
  {IUCrJ}\ }\textbf {\bibinfo {volume} {2(3)}},\ \bibinfo {pages} {371}
  (\bibinfo {year} {2015})}\BibitemShut {NoStop}%
\bibitem [{\citenamefont {Lutz}(2007)}]{Lutz:2007}%
  \BibitemOpen
  \bibfield  {author} {\bibinfo {author} {\bibfnamefont {G.}~\bibnamefont
  {Lutz}},\ }\href {\doibase 10.1007/978-3-540-71679-2} {\emph {\bibinfo
  {title} {{Semiconductor Radiation Detectors}}}},\ \bibinfo {edition} {2nd}\
  ed.\ (\bibinfo  {publisher} {Springer},\ \bibinfo {year} {2007})\BibitemShut
  {NoStop}%
\bibitem [{\citenamefont {Spieler}(2005)}]{Spi:2005}%
  \BibitemOpen
  \bibfield  {author} {\bibinfo {author} {\bibfnamefont {H.}~\bibnamefont
  {Spieler}},\ }\href {\doibase 10.1093/acprof:oso/9780198527848.001.0001}
  {\emph {\bibinfo {title} {{Semiconductor Detector Systems}}}}\ (\bibinfo
  {publisher} {Oxford University Press},\ \bibinfo {year} {2005})\BibitemShut
  {NoStop}%
\bibitem [{\citenamefont {Lowe}\ and\ \citenamefont {Sareen}(2007)}]{LS:2007}%
  \BibitemOpen
  \bibfield  {author} {\bibinfo {author} {\bibfnamefont {B.}~\bibnamefont
  {Lowe}}\ and\ \bibinfo {author} {\bibfnamefont {R.}~\bibnamefont {Sareen}},\
  }\href {\doibase 10.1016/j.nima.2007.03.020} {\bibfield  {journal} {\bibinfo
  {journal} {Nucl. Instrum. Meth. Phys. Res. Sec. A.}\ }\textbf {\bibinfo
  {volume} {576}},\ \bibinfo {pages} {367} (\bibinfo {year}
  {2007})}\BibitemShut {NoStop}%
\bibitem [{\citenamefont {Mazziotta}(2008)}]{Mazz:2008}%
  \BibitemOpen
  \bibfield  {author} {\bibinfo {author} {\bibfnamefont {M.}~\bibnamefont
  {Mazziotta}},\ }\href {\doibase 10.1016/j.nima.2007.10.043} {\bibfield
  {journal} {\bibinfo  {journal} {Nucl. Instrum. Meth. Phys. Res. Sec. A.}\
  }\textbf {\bibinfo {volume} {584}},\ \bibinfo {pages} {436} (\bibinfo {year}
  {2008})}\BibitemShut {NoStop}%
\bibitem [{\citenamefont {Knoll}(2010)}]{Knoll2000}%
  \BibitemOpen
  \bibfield  {author} {\bibinfo {author} {\bibfnamefont {G.~F.}\ \bibnamefont
  {Knoll}},\ }\href@noop {} {\emph {\bibinfo {title} {Radiation Detection and
  Measurement}}},\ \bibinfo {edition} {4th}\ ed.\ (\bibinfo  {publisher} {John
  Wiley \& Sons},\ \bibinfo {year} {2010})\BibitemShut {NoStop}%
\bibitem [{\citenamefont {Anelli}\ \emph {et~al.}(1999)\citenamefont {Anelli},
  \citenamefont {Campbell}, \citenamefont {Delmastro}, \citenamefont {Faccio},
  \citenamefont {Florian}, \citenamefont {Giraldo}, \citenamefont {Heijne},
  \citenamefont {Jarron}, \citenamefont {Kloukinas}, \citenamefont {Marchioro},
  \citenamefont {Moreira},\ and\ \citenamefont {Snoeys}}]{ACD:1999}%
  \BibitemOpen
  \bibfield  {author} {\bibinfo {author} {\bibfnamefont {G.}~\bibnamefont
  {Anelli}}, \bibinfo {author} {\bibfnamefont {M.}~\bibnamefont {Campbell}},
  \bibinfo {author} {\bibfnamefont {M.}~\bibnamefont {Delmastro}}, \bibinfo
  {author} {\bibfnamefont {F.}~\bibnamefont {Faccio}}, \bibinfo {author}
  {\bibfnamefont {S.}~\bibnamefont {Florian}}, \bibinfo {author} {\bibfnamefont
  {A.}~\bibnamefont {Giraldo}}, \bibinfo {author} {\bibfnamefont
  {E.}~\bibnamefont {Heijne}}, \bibinfo {author} {\bibfnamefont
  {P.}~\bibnamefont {Jarron}}, \bibinfo {author} {\bibfnamefont
  {K.}~\bibnamefont {Kloukinas}}, \bibinfo {author} {\bibfnamefont
  {A.}~\bibnamefont {Marchioro}}, \bibinfo {author} {\bibfnamefont
  {P.}~\bibnamefont {Moreira}}, \ and\ \bibinfo {author} {\bibfnamefont
  {W.}~\bibnamefont {Snoeys}},\ }\href@noop {} {\bibfield  {journal} {\bibinfo
  {journal} {IEEE Trans. Nucl. Sci.}\ }\textbf {\bibinfo {volume} {46}},\
  \bibinfo {pages} {1690 } (\bibinfo {year} {1999})}\BibitemShut {NoStop}%
\bibitem [{\citenamefont {Poikela}\ \emph {et~al.}(2015)\citenamefont
  {Poikela}, \citenamefont {Gaspari}, \citenamefont {Plosila}, \citenamefont
  {Westerlund}, \citenamefont {Ballabriga}, \citenamefont {Buytaert},
  \citenamefont {Campbell}, \citenamefont {Llopart}, \citenamefont {Wyllie},
  \citenamefont {Gromov}, \citenamefont {van Beuzekom},\ and\ \citenamefont
  {Zivkovic}}]{PDP:2015}%
  \BibitemOpen
  \bibfield  {author} {\bibinfo {author} {\bibfnamefont {T.}~\bibnamefont
  {Poikela}}, \bibinfo {author} {\bibfnamefont {M.~D.}\ \bibnamefont
  {Gaspari}}, \bibinfo {author} {\bibfnamefont {J.}~\bibnamefont {Plosila}},
  \bibinfo {author} {\bibfnamefont {T.}~\bibnamefont {Westerlund}}, \bibinfo
  {author} {\bibfnamefont {R.}~\bibnamefont {Ballabriga}}, \bibinfo {author}
  {\bibfnamefont {J.}~\bibnamefont {Buytaert}}, \bibinfo {author}
  {\bibfnamefont {M.}~\bibnamefont {Campbell}}, \bibinfo {author}
  {\bibfnamefont {X.}~\bibnamefont {Llopart}}, \bibinfo {author} {\bibfnamefont
  {K.}~\bibnamefont {Wyllie}}, \bibinfo {author} {\bibfnamefont
  {V.}~\bibnamefont {Gromov}}, \bibinfo {author} {\bibfnamefont
  {M.}~\bibnamefont {van Beuzekom}}, \ and\ \bibinfo {author} {\bibfnamefont
  {V.}~\bibnamefont {Zivkovic}},\ }\href {\doibase
  10.1088/1748-0221/10/01/C01057} {\bibfield  {journal} {\bibinfo  {journal}
  {J. Instrum.}\ }\textbf {\bibinfo {volume} {10}},\ \bibinfo {pages} {C01057}
  (\bibinfo {year} {2015})}\BibitemShut {NoStop}%
\bibitem [{\citenamefont {Boyd}\ and\ \citenamefont
  {Vanderberghe}(2009)}]{BV:2009}%
  \BibitemOpen
  \bibfield  {author} {\bibinfo {author} {\bibfnamefont {S.}~\bibnamefont
  {Boyd}}\ and\ \bibinfo {author} {\bibfnamefont {L.}~\bibnamefont
  {Vanderberghe}},\ }\href@noop {} {\emph {\bibinfo {title} {Convex
  Optimization}}},\ \bibinfo {edition} {7th}\ ed.\ (\bibinfo  {publisher}
  {Cambridge University Press},\ \bibinfo {address} {Cambridge, UK},\ \bibinfo
  {year} {2009})\BibitemShut {NoStop}%
\bibitem [{\citenamefont {Tao}(2008)}]{Tao:2008}%
  \BibitemOpen
  \bibfield  {author} {\bibinfo {author} {\bibfnamefont {T.}~\bibnamefont
  {Tao}},\ }\href@noop {} {\enquote {\bibinfo {title} {{The Onsager lecture on
  compressed sensing}},}\ } (\bibinfo {year} {2008})\BibitemShut {NoStop}%
\bibitem [{\citenamefont {Charbon}(2014)}]{Char:2014}%
  \BibitemOpen
  \bibfield  {author} {\bibinfo {author} {\bibfnamefont {E.}~\bibnamefont
  {Charbon}},\ }\href@noop {} {\bibfield  {journal} {\bibinfo  {journal} {Phil.
  Trans. R. Soc. A.}\ }\textbf {\bibinfo {volume} {372}},\ \bibinfo {pages}
  {20130100} (\bibinfo {year} {2014})},\ \bibinfo {note}
  {https://doi.org/10.1098/rsta.2013.0100}\BibitemShut {NoStop}%
\bibitem [{\citenamefont {Turchetta}\ \emph {et~al.}(2001)\citenamefont
  {Turchetta}, \citenamefont {Berst}, \citenamefont {Casadei}, \citenamefont
  {Claus}, \citenamefont {Colledani}, \citenamefont {Dulinski}, \citenamefont
  {Hu}, \citenamefont {Husson}, \citenamefont {Normand}, \citenamefont
  {Riester}, \citenamefont {Deptuch}, \citenamefont {Goerlach}, \citenamefont
  {Higueret},\ and\ \citenamefont {Winter}}]{TBC:2001}%
  \BibitemOpen
  \bibfield  {author} {\bibinfo {author} {\bibfnamefont {R.}~\bibnamefont
  {Turchetta}}, \bibinfo {author} {\bibfnamefont {J.}~\bibnamefont {Berst}},
  \bibinfo {author} {\bibfnamefont {B.}~\bibnamefont {Casadei}}, \bibinfo
  {author} {\bibfnamefont {G.}~\bibnamefont {Claus}}, \bibinfo {author}
  {\bibfnamefont {C.}~\bibnamefont {Colledani}}, \bibinfo {author}
  {\bibfnamefont {W.}~\bibnamefont {Dulinski}}, \bibinfo {author}
  {\bibfnamefont {Y.}~\bibnamefont {Hu}}, \bibinfo {author} {\bibfnamefont
  {D.}~\bibnamefont {Husson}}, \bibinfo {author} {\bibfnamefont {J.~L.}\
  \bibnamefont {Normand}}, \bibinfo {author} {\bibfnamefont {J.}~\bibnamefont
  {Riester}}, \bibinfo {author} {\bibfnamefont {G.}~\bibnamefont {Deptuch}},
  \bibinfo {author} {\bibfnamefont {U.}~\bibnamefont {Goerlach}}, \bibinfo
  {author} {\bibfnamefont {S.}~\bibnamefont {Higueret}}, \ and\ \bibinfo
  {author} {\bibfnamefont {M.}~\bibnamefont {Winter}},\ }\href@noop {}
  {\bibfield  {journal} {\bibinfo  {journal} {Nucl. Instrum. Meth. A}\ }\textbf
  {\bibinfo {volume} {458}},\ \bibinfo {pages} {677} (\bibinfo {year} {Feb.
  2001})},\ \bibinfo {note}
  {https://doi.org/10.1016/S0168-9002(00)00893-7}\BibitemShut {NoStop}%
\bibitem [{\citenamefont {Llopart}\ \emph {et~al.}(2022)\citenamefont
  {Llopart}, \citenamefont {Alozy}, \citenamefont {Ballabriga}, \citenamefont
  {Campbell}, \citenamefont {Casanova}, \citenamefont {Gromov}, \citenamefont
  {Heijne}, \citenamefont {Poikela}, \citenamefont {Santin}, \citenamefont
  {Sriskaran},\ and\ \citenamefont {Tlustos}}]{LAB:2022}%
  \BibitemOpen
  \bibfield  {author} {\bibinfo {author} {\bibfnamefont {X.}~\bibnamefont
  {Llopart}}, \bibinfo {author} {\bibfnamefont {J.}~\bibnamefont {Alozy}},
  \bibinfo {author} {\bibfnamefont {R.}~\bibnamefont {Ballabriga}}, \bibinfo
  {author} {\bibfnamefont {M.}~\bibnamefont {Campbell}}, \bibinfo {author}
  {\bibfnamefont {R.}~\bibnamefont {Casanova}}, \bibinfo {author}
  {\bibfnamefont {V.}~\bibnamefont {Gromov}}, \bibinfo {author} {\bibfnamefont
  {E.}~\bibnamefont {Heijne}}, \bibinfo {author} {\bibfnamefont
  {T.}~\bibnamefont {Poikela}}, \bibinfo {author} {\bibfnamefont
  {E.}~\bibnamefont {Santin}}, \bibinfo {author} {\bibfnamefont
  {V.}~\bibnamefont {Sriskaran}}, \ and\ \bibinfo {author} {\bibfnamefont
  {L.}~\bibnamefont {Tlustos}},\ }\href {\doibase DOI
  10.1088/1748-0221/17/01/C01044} {\bibfield  {journal} {\bibinfo  {journal}
  {J. Instrum.}\ }\textbf {\bibinfo {volume} {17}},\ \bibinfo {pages} {C01044}
  (\bibinfo {year} {2022})}\BibitemShut {NoStop}%
\bibitem [{\citenamefont {Ballabriga}\ \emph {et~al.}(2018)\citenamefont
  {Ballabriga}, \citenamefont {Llopart},\ and\ \citenamefont
  {Campbell}}]{BCL:2018}%
  \BibitemOpen
  \bibfield  {author} {\bibinfo {author} {\bibfnamefont {R.}~\bibnamefont
  {Ballabriga}}, \bibinfo {author} {\bibfnamefont {X.}~\bibnamefont {Llopart}},
  \ and\ \bibinfo {author} {\bibfnamefont {M.}~\bibnamefont {Campbell}},\
  }\href@noop {} {\bibfield  {journal} {\bibinfo  {journal} {Nucl. Instrum.
  Meth. A}\ }\textbf {\bibinfo {volume} {878}},\ \bibinfo {pages} {10}
  (\bibinfo {year} {2018})}\BibitemShut {NoStop}%
\bibitem [{\citenamefont {Llopart}\ \emph {et~al.}(2007)\citenamefont
  {Llopart}, \citenamefont {Ballabriga}, \citenamefont {Campbell},
  \citenamefont {Tlustos},\ and\ \citenamefont {Wong}}]{LBC:2007}%
  \BibitemOpen
  \bibfield  {author} {\bibinfo {author} {\bibfnamefont {X.}~\bibnamefont
  {Llopart}}, \bibinfo {author} {\bibfnamefont {R.}~\bibnamefont {Ballabriga}},
  \bibinfo {author} {\bibfnamefont {M.}~\bibnamefont {Campbell}}, \bibinfo
  {author} {\bibfnamefont {L.}~\bibnamefont {Tlustos}}, \ and\ \bibinfo
  {author} {\bibfnamefont {W.}~\bibnamefont {Wong}},\ }\href@noop {} {\bibfield
   {journal} {\bibinfo  {journal} {Nucl. Instrum. Meth. A}\ }\textbf {\bibinfo
  {volume} {581}},\ \bibinfo {pages} {485} (\bibinfo {year}
  {2007})}\BibitemShut {NoStop}%
\bibitem [{\citenamefont {Vykydal}\ \emph {et~al.}(2006)\citenamefont
  {Vykydal}, \citenamefont {Jakubek},\ and\ \citenamefont
  {Pospisil}}]{VJP:2006}%
  \BibitemOpen
  \bibfield  {author} {\bibinfo {author} {\bibfnamefont {Z.}~\bibnamefont
  {Vykydal}}, \bibinfo {author} {\bibfnamefont {J.}~\bibnamefont {Jakubek}}, \
  and\ \bibinfo {author} {\bibfnamefont {S.}~\bibnamefont {Pospisil}},\
  }\href@noop {} {\bibfield  {journal} {\bibinfo  {journal} {Nucl. Instrum.
  Meth. A}\ }\textbf {\bibinfo {volume} {878}},\ \bibinfo {pages} {10}
  (\bibinfo {year} {2006})}\BibitemShut {NoStop}%
\bibitem [{\citenamefont {Tureček}\ \emph {et~al.}(2011)\citenamefont
  {Tureček}, \citenamefont {Holy}, \citenamefont {Jakubek}, \citenamefont
  {Pospisil},\ and\ \citenamefont {Vykydal}}]{THJ:2011}%
  \BibitemOpen
  \bibfield  {author} {\bibinfo {author} {\bibfnamefont {D.}~\bibnamefont
  {Tureček}}, \bibinfo {author} {\bibfnamefont {T.}~\bibnamefont {Holy}},
  \bibinfo {author} {\bibfnamefont {J.}~\bibnamefont {Jakubek}}, \bibinfo
  {author} {\bibfnamefont {S.}~\bibnamefont {Pospisil}}, \ and\ \bibinfo
  {author} {\bibfnamefont {Z.}~\bibnamefont {Vykydal}},\ }\href@noop {}
  {\bibfield  {journal} {\bibinfo  {journal} {J. Instrum.}\ }\textbf {\bibinfo
  {volume} {6}},\ \bibinfo {pages} {C01046} (\bibinfo {year}
  {2011})}\BibitemShut {NoStop}%
\bibitem [{\citenamefont {Wong}\ \emph {et~al.}(2020)\citenamefont {Wong},
  \citenamefont {Alozy}, \citenamefont {Ballabriga}, \citenamefont {Campbell},
  \citenamefont {Kremastiotis}, \citenamefont {Llopart}, \citenamefont
  {Poikela}, \citenamefont {Sriskaran}, \citenamefont {Tlustos},\ and\
  \citenamefont {Turecek}}]{Camp:6}%
  \BibitemOpen
  \bibfield  {author} {\bibinfo {author} {\bibfnamefont {W.~S.}\ \bibnamefont
  {Wong}}, \bibinfo {author} {\bibfnamefont {J.}~\bibnamefont {Alozy}},
  \bibinfo {author} {\bibfnamefont {R.}~\bibnamefont {Ballabriga}}, \bibinfo
  {author} {\bibfnamefont {M.}~\bibnamefont {Campbell}}, \bibinfo {author}
  {\bibfnamefont {I.}~\bibnamefont {Kremastiotis}}, \bibinfo {author}
  {\bibfnamefont {X.}~\bibnamefont {Llopart}}, \bibinfo {author} {\bibfnamefont
  {T.}~\bibnamefont {Poikela}}, \bibinfo {author} {\bibfnamefont
  {V.}~\bibnamefont {Sriskaran}}, \bibinfo {author} {\bibfnamefont
  {L.}~\bibnamefont {Tlustos}}, \ and\ \bibinfo {author} {\bibfnamefont
  {D.}~\bibnamefont {Turecek}},\ }\href@noop {} {\bibfield  {journal} {\bibinfo
   {journal} {Radiat Meas.}\ }\textbf {\bibinfo {volume} {131}},\ \bibinfo
  {pages} {106230} (\bibinfo {year} {2020})}\BibitemShut {NoStop}%
\bibitem [{\citenamefont {Poikela}\ \emph {et~al.}(2014)\citenamefont
  {Poikela}, \citenamefont {Plosila}, \citenamefont {Westerlund}, \citenamefont
  {Campbell}, \citenamefont {Gaspari}, \citenamefont {Llopart}, \citenamefont
  {Gromov}, \citenamefont {Kluit}, \citenamefont {Beuzekom}, \citenamefont
  {Zappon}, \citenamefont {Zivkovic}, \citenamefont {Brezina}, \citenamefont
  {Desch}, \citenamefont {Fu},\ and\ \citenamefont {Kruth}}]{Camp:7}%
  \BibitemOpen
  \bibfield  {author} {\bibinfo {author} {\bibfnamefont {T.}~\bibnamefont
  {Poikela}}, \bibinfo {author} {\bibfnamefont {J.}~\bibnamefont {Plosila}},
  \bibinfo {author} {\bibfnamefont {T.}~\bibnamefont {Westerlund}}, \bibinfo
  {author} {\bibfnamefont {M.}~\bibnamefont {Campbell}}, \bibinfo {author}
  {\bibfnamefont {M.}~\bibnamefont {Gaspari}}, \bibinfo {author} {\bibfnamefont
  {X.}~\bibnamefont {Llopart}}, \bibinfo {author} {\bibfnamefont
  {V.}~\bibnamefont {Gromov}}, \bibinfo {author} {\bibfnamefont
  {R.}~\bibnamefont {Kluit}}, \bibinfo {author} {\bibfnamefont
  {M.}~\bibnamefont {Beuzekom}}, \bibinfo {author} {\bibfnamefont
  {F.}~\bibnamefont {Zappon}}, \bibinfo {author} {\bibfnamefont
  {V.}~\bibnamefont {Zivkovic}}, \bibinfo {author} {\bibfnamefont
  {C.}~\bibnamefont {Brezina}}, \bibinfo {author} {\bibfnamefont
  {K.}~\bibnamefont {Desch}}, \bibinfo {author} {\bibfnamefont
  {Y.}~\bibnamefont {Fu}}, \ and\ \bibinfo {author} {\bibfnamefont
  {A.}~\bibnamefont {Kruth}},\ }\href@noop {} {\bibfield  {journal} {\bibinfo
  {journal} {J. Instrum.}\ }\textbf {\bibinfo {volume} {9}},\ \bibinfo {pages}
  {C05013} (\bibinfo {year} {2014})}\BibitemShut {NoStop}%
\bibitem [{\citenamefont {Bergmann}\ \emph {et~al.}(2017)\citenamefont
  {Bergmann}, \citenamefont {Pichotka}, \citenamefont {Pospisil}, \citenamefont
  {Vycpalek}, \citenamefont {Burian}, \citenamefont {Broulim},\ and\
  \citenamefont {Jakubek}}]{Camp:8}%
  \BibitemOpen
  \bibfield  {author} {\bibinfo {author} {\bibfnamefont {T.}~\bibnamefont
  {Bergmann}}, \bibinfo {author} {\bibfnamefont {M.}~\bibnamefont {Pichotka}},
  \bibinfo {author} {\bibfnamefont {S.}~\bibnamefont {Pospisil}}, \bibinfo
  {author} {\bibfnamefont {J.}~\bibnamefont {Vycpalek}}, \bibinfo {author}
  {\bibfnamefont {P.}~\bibnamefont {Burian}}, \bibinfo {author} {\bibfnamefont
  {P.}~\bibnamefont {Broulim}}, \ and\ \bibinfo {author} {\bibfnamefont
  {J.}~\bibnamefont {Jakubek}},\ }\href@noop {} {\bibfield  {journal} {\bibinfo
   {journal} {Eur. Phys. J. C}\ }\textbf {\bibinfo {volume} {77}},\ \bibinfo
  {pages} {1} (\bibinfo {year} {2017})}\BibitemShut {NoStop}%
\bibitem [{\citenamefont {Lowe}\ \emph {et~al.}(2020)\citenamefont {Lowe},
  \citenamefont {Majumdar}, \citenamefont {Mavrokoridis}, \citenamefont
  {Philippou}, \citenamefont {Roberts}, \citenamefont {Touramanis},\ and\
  \citenamefont {Vann}}]{Camp:9}%
  \BibitemOpen
  \bibfield  {author} {\bibinfo {author} {\bibfnamefont {A.}~\bibnamefont
  {Lowe}}, \bibinfo {author} {\bibfnamefont {K.}~\bibnamefont {Majumdar}},
  \bibinfo {author} {\bibfnamefont {K.}~\bibnamefont {Mavrokoridis}}, \bibinfo
  {author} {\bibfnamefont {B.}~\bibnamefont {Philippou}}, \bibinfo {author}
  {\bibfnamefont {A.}~\bibnamefont {Roberts}}, \bibinfo {author} {\bibfnamefont
  {C.}~\bibnamefont {Touramanis}}, \ and\ \bibinfo {author} {\bibfnamefont
  {J.}~\bibnamefont {Vann}},\ }\href@noop {} {\bibfield  {journal} {\bibinfo
  {journal} {Instruments}\ }\textbf {\bibinfo {volume} {4}},\ \bibinfo {pages}
  {1} (\bibinfo {year} {2020})}\BibitemShut {NoStop}%
\bibitem [{\citenamefont {Turecek}\ \emph {et~al.}(2020)\citenamefont
  {Turecek}, \citenamefont {Jakubek}, \citenamefont {Trojanova},\ and\
  \citenamefont {Sefc}}]{Camp:10}%
  \BibitemOpen
  \bibfield  {author} {\bibinfo {author} {\bibfnamefont {D.}~\bibnamefont
  {Turecek}}, \bibinfo {author} {\bibfnamefont {J.}~\bibnamefont {Jakubek}},
  \bibinfo {author} {\bibfnamefont {E.}~\bibnamefont {Trojanova}}, \ and\
  \bibinfo {author} {\bibfnamefont {L.}~\bibnamefont {Sefc}},\ }\href@noop {}
  {\bibfield  {journal} {\bibinfo  {journal} {J. Instrum.}\ }\textbf {\bibinfo
  {volume} {15}},\ \bibinfo {pages} {C01014} (\bibinfo {year}
  {2020})}\BibitemShut {NoStop}%
\bibitem [{\citenamefont {Campbell}\ \emph {et~al.}(2016)\citenamefont
  {Campbell}, \citenamefont {Alozy}, \citenamefont {Ballabriga}, \citenamefont
  {Frojdh}, \citenamefont {Heijne}, \citenamefont {Llopart}, \citenamefont
  {Poikela}, \citenamefont {Tlustos}, \citenamefont {Valerio},\ and\
  \citenamefont {Wong}}]{Camp:11}%
  \BibitemOpen
  \bibfield  {author} {\bibinfo {author} {\bibfnamefont {M.}~\bibnamefont
  {Campbell}}, \bibinfo {author} {\bibfnamefont {J.}~\bibnamefont {Alozy}},
  \bibinfo {author} {\bibfnamefont {R.}~\bibnamefont {Ballabriga}}, \bibinfo
  {author} {\bibfnamefont {E.}~\bibnamefont {Frojdh}}, \bibinfo {author}
  {\bibfnamefont {E.}~\bibnamefont {Heijne}}, \bibinfo {author} {\bibfnamefont
  {X.}~\bibnamefont {Llopart}}, \bibinfo {author} {\bibfnamefont
  {T.}~\bibnamefont {Poikela}}, \bibinfo {author} {\bibfnamefont
  {L.}~\bibnamefont {Tlustos}}, \bibinfo {author} {\bibfnamefont
  {P.}~\bibnamefont {Valerio}}, \ and\ \bibinfo {author} {\bibfnamefont
  {W.}~\bibnamefont {Wong}},\ }\href@noop {} {\bibfield  {journal} {\bibinfo
  {journal} {J. Instrum.}\ }\textbf {\bibinfo {volume} {11}},\ \bibinfo {pages}
  {C01007} (\bibinfo {year} {2016})}\BibitemShut {NoStop}%
\bibitem [{\citenamefont {Tate}\ \emph {et~al.}(2013)\citenamefont {Tate},
  \citenamefont {Chamberlain}, \citenamefont {Green}, \citenamefont {Philipp},
  \citenamefont {Purohit}, \citenamefont {Strohman},\ and\ \citenamefont
  {Gruner}}]{G3}%
  \BibitemOpen
  \bibfield  {author} {\bibinfo {author} {\bibfnamefont {M.}~\bibnamefont
  {Tate}}, \bibinfo {author} {\bibfnamefont {D.}~\bibnamefont {Chamberlain}},
  \bibinfo {author} {\bibfnamefont {K.}~\bibnamefont {Green}}, \bibinfo
  {author} {\bibfnamefont {H.}~\bibnamefont {Philipp}}, \bibinfo {author}
  {\bibfnamefont {P.}~\bibnamefont {Purohit}}, \bibinfo {author} {\bibfnamefont
  {C.}~\bibnamefont {Strohman}}, \ and\ \bibinfo {author} {\bibfnamefont
  {S.~M.}\ \bibnamefont {Gruner}},\ }\href@noop {} {\bibfield  {journal}
  {\bibinfo  {journal} {J. Physics: Conf. Ser.}\ }\textbf {\bibinfo {volume}
  {425}},\ \bibinfo {pages} {062009} (\bibinfo {year} {2013})},\ \bibinfo
  {note} {{DOI: 10.1088/1742-6596/425/6/062004}}\BibitemShut {NoStop}%
\bibitem [{\citenamefont {Tate}\ \emph {et~al.}(2016)\citenamefont {Tate},
  \citenamefont {Purohit}, \citenamefont {Chamberlain}, \citenamefont {Nguyen},
  \citenamefont {Hovden}, \citenamefont {Chang}, \citenamefont {Deb},
  \citenamefont {Turgut}, \citenamefont {Heron}, \citenamefont {Schlom},
  \citenamefont {Ralph}, \citenamefont {Fuchs}, \citenamefont {Shanks},
  \citenamefont {Philipp}, \citenamefont {Muller},\ and\ \citenamefont
  {Gruner}}]{G4}%
  \BibitemOpen
  \bibfield  {author} {\bibinfo {author} {\bibfnamefont {M.~W.}\ \bibnamefont
  {Tate}}, \bibinfo {author} {\bibfnamefont {P.}~\bibnamefont {Purohit}},
  \bibinfo {author} {\bibfnamefont {D.}~\bibnamefont {Chamberlain}}, \bibinfo
  {author} {\bibfnamefont {K.~X.}\ \bibnamefont {Nguyen}}, \bibinfo {author}
  {\bibfnamefont {R.}~\bibnamefont {Hovden}}, \bibinfo {author} {\bibfnamefont
  {C.~S.}\ \bibnamefont {Chang}}, \bibinfo {author} {\bibfnamefont
  {P.}~\bibnamefont {Deb}}, \bibinfo {author} {\bibfnamefont {E.}~\bibnamefont
  {Turgut}}, \bibinfo {author} {\bibfnamefont {J.~T.}\ \bibnamefont {Heron}},
  \bibinfo {author} {\bibfnamefont {D.~G.}\ \bibnamefont {Schlom}}, \bibinfo
  {author} {\bibfnamefont {D.~C.}\ \bibnamefont {Ralph}}, \bibinfo {author}
  {\bibfnamefont {G.~D.}\ \bibnamefont {Fuchs}}, \bibinfo {author}
  {\bibfnamefont {K.~S.}\ \bibnamefont {Shanks}}, \bibinfo {author}
  {\bibfnamefont {H.~T.}\ \bibnamefont {Philipp}}, \bibinfo {author}
  {\bibfnamefont {D.~A.}\ \bibnamefont {Muller}}, \ and\ \bibinfo {author}
  {\bibfnamefont {S.~M.}\ \bibnamefont {Gruner}},\ }\href@noop {} {\bibfield
  {journal} {\bibinfo  {journal} {Microscopy and Microanalysis}\ }\textbf
  {\bibinfo {volume} {22}},\ \bibinfo {pages} {237} (\bibinfo {year} {2016})},\
  \bibinfo {note} {{DOI: 10.1017/S1431927615015664}}\BibitemShut {NoStop}%
\bibitem [{\citenamefont {Jiang}\ \emph {et~al.}(2018)\citenamefont {Jiang},
  \citenamefont {Chen}, \citenamefont {Han}, \citenamefont {Deb}, \citenamefont
  {Gao}, \citenamefont {Xie}, \citenamefont {Purohit}, \citenamefont {Tate},
  \citenamefont {Park}, \citenamefont {Gruner}, \citenamefont {Elser},\ and\
  \citenamefont {Muller}}]{G5}%
  \BibitemOpen
  \bibfield  {author} {\bibinfo {author} {\bibfnamefont {Y.}~\bibnamefont
  {Jiang}}, \bibinfo {author} {\bibfnamefont {Z.}~\bibnamefont {Chen}},
  \bibinfo {author} {\bibfnamefont {Y.}~\bibnamefont {Han}}, \bibinfo {author}
  {\bibfnamefont {P.}~\bibnamefont {Deb}}, \bibinfo {author} {\bibfnamefont
  {H.}~\bibnamefont {Gao}}, \bibinfo {author} {\bibfnamefont {S.}~\bibnamefont
  {Xie}}, \bibinfo {author} {\bibfnamefont {P.}~\bibnamefont {Purohit}},
  \bibinfo {author} {\bibfnamefont {M.~W.}\ \bibnamefont {Tate}}, \bibinfo
  {author} {\bibfnamefont {J.}~\bibnamefont {Park}}, \bibinfo {author}
  {\bibfnamefont {S.~M.}\ \bibnamefont {Gruner}}, \bibinfo {author}
  {\bibfnamefont {V.}~\bibnamefont {Elser}}, \ and\ \bibinfo {author}
  {\bibfnamefont {D.~A.}\ \bibnamefont {Muller}},\ }\href@noop {} {\bibfield
  {journal} {\bibinfo  {journal} {Nature}\ }\textbf {\bibinfo {volume} {559}},\
  \bibinfo {pages} {343} (\bibinfo {year} {2018})},\ \bibinfo {note} {{DOI:
  10.1038/s41586-018-0298-5}}\BibitemShut {NoStop}%
\bibitem [{\citenamefont {Philipp}\ \emph {et~al.}(2022)\citenamefont
  {Philipp}, \citenamefont {Tate}, \citenamefont {Shanks}, \citenamefont
  {Mele}, \citenamefont {Peemen}, \citenamefont {Dona}, \citenamefont
  {Hartong}, \citenamefont {van Veen}, \citenamefont {Shao}, \citenamefont
  {Chen}, \citenamefont {Thom-Levy}, \citenamefont {Muller},\ and\
  \citenamefont {Gruner}}]{G6}%
  \BibitemOpen
  \bibfield  {author} {\bibinfo {author} {\bibfnamefont {H.~T.}\ \bibnamefont
  {Philipp}}, \bibinfo {author} {\bibfnamefont {M.~W.}\ \bibnamefont {Tate}},
  \bibinfo {author} {\bibfnamefont {K.~S.}\ \bibnamefont {Shanks}}, \bibinfo
  {author} {\bibfnamefont {L.}~\bibnamefont {Mele}}, \bibinfo {author}
  {\bibfnamefont {M.}~\bibnamefont {Peemen}}, \bibinfo {author} {\bibfnamefont
  {P.}~\bibnamefont {Dona}}, \bibinfo {author} {\bibfnamefont {R.}~\bibnamefont
  {Hartong}}, \bibinfo {author} {\bibfnamefont {G.}~\bibnamefont {van Veen}},
  \bibinfo {author} {\bibfnamefont {Y.-T.}\ \bibnamefont {Shao}}, \bibinfo
  {author} {\bibfnamefont {Z.}~\bibnamefont {Chen}}, \bibinfo {author}
  {\bibfnamefont {J.}~\bibnamefont {Thom-Levy}}, \bibinfo {author}
  {\bibfnamefont {D.~A.}\ \bibnamefont {Muller}}, \ and\ \bibinfo {author}
  {\bibfnamefont {S.~M.}\ \bibnamefont {Gruner}},\ }\href@noop {} {\bibfield
  {journal} {\bibinfo  {journal} {Microsc. Microanal.}\ }\textbf {\bibinfo
  {volume} {28}},\ \bibinfo {pages} {425} (\bibinfo {year} {2022})},\ \bibinfo
  {note} {{DOI: 10.1017.S1431927622000174}}\BibitemShut {NoStop}%
\bibitem [{\citenamefont {El-Desouki}\ \emph {et~al.}(2009)\citenamefont
  {El-Desouki}, \citenamefont {Deen}, \citenamefont {Fang}, \citenamefont
  {Liu}, \citenamefont {Tse},\ and\ \citenamefont {Armstrong}}]{EDF:2009}%
  \BibitemOpen
  \bibfield  {author} {\bibinfo {author} {\bibfnamefont {M.}~\bibnamefont
  {El-Desouki}}, \bibinfo {author} {\bibfnamefont {M.~J.}\ \bibnamefont
  {Deen}}, \bibinfo {author} {\bibfnamefont {Q.}~\bibnamefont {Fang}}, \bibinfo
  {author} {\bibfnamefont {L.}~\bibnamefont {Liu}}, \bibinfo {author}
  {\bibfnamefont {F.}~\bibnamefont {Tse}}, \ and\ \bibinfo {author}
  {\bibfnamefont {D.}~\bibnamefont {Armstrong}},\ }\href@noop {} {\bibfield
  {journal} {\bibinfo  {journal} {Sensors}\ }\textbf {\bibinfo {volume} {9}},\
  \bibinfo {pages} {430} (\bibinfo {year} {2009})},\ \bibinfo {note}
  {https://doi.org/10.3390/s90100430.}\BibitemShut {Stop}%
\bibitem [{\citenamefont {Etoh}\ \emph {et~al.}(2011)\citenamefont {Etoh},
  \citenamefont {Dao}, \citenamefont {Nguyen}, \citenamefont {Fife},
  \citenamefont {Kureta}, \citenamefont {Segawa}, \citenamefont {Arai},\ and\
  \citenamefont {Shinohar}}]{Etoh:2011}%
  \BibitemOpen
  \bibfield  {author} {\bibinfo {author} {\bibfnamefont {T.~G.}\ \bibnamefont
  {Etoh}}, \bibinfo {author} {\bibfnamefont {V.~T.~S.}\ \bibnamefont {Dao}},
  \bibinfo {author} {\bibfnamefont {H.~D.}\ \bibnamefont {Nguyen}}, \bibinfo
  {author} {\bibfnamefont {K.}~\bibnamefont {Fife}}, \bibinfo {author}
  {\bibfnamefont {M.}~\bibnamefont {Kureta}}, \bibinfo {author} {\bibfnamefont
  {M.}~\bibnamefont {Segawa}}, \bibinfo {author} {\bibfnamefont
  {M.}~\bibnamefont {Arai}}, \ and\ \bibinfo {author} {\bibfnamefont
  {T.}~\bibnamefont {Shinohar}},\ }\href@noop {} {\enquote {\bibinfo {title}
  {{Progress of Ultra-high-speed Image Sensors with In-situ CCD Storage}},}\ }
  (\bibinfo {year} {2011}),\ \bibinfo {note} {{https://imagesensors.org}, Paper
  \# R57.pdf}\BibitemShut {NoStop}%
\bibitem [{\citenamefont {Liang}\ \emph {et~al.}(2008)\citenamefont {Liang},
  \citenamefont {Chang},\ and\ \citenamefont {Chen}}]{Dart:2}%
  \BibitemOpen
  \bibfield  {author} {\bibinfo {author} {\bibfnamefont {C.-K.}\ \bibnamefont
  {Liang}}, \bibinfo {author} {\bibfnamefont {L.-W.}\ \bibnamefont {Chang}}, \
  and\ \bibinfo {author} {\bibfnamefont {H.~H.}\ \bibnamefont {Chen}},\
  }\href@noop {} {\bibfield  {journal} {\bibinfo  {journal} {IEEE Trans. Imag.
  Proc.}\ }\textbf {\bibinfo {volume} {17}},\ \bibinfo {pages} {1323} (\bibinfo
  {year} {2008})},\ \bibinfo {note} {doi:
  https://doi.org/10.1109/tip.2008.925384.}\BibitemShut {Stop}%
\bibitem [{\citenamefont {Fossum}(1994)}]{Dart:3}%
  \BibitemOpen
  \bibfield  {author} {\bibinfo {author} {\bibfnamefont {E.~R.}\ \bibnamefont
  {Fossum}},\ }\href@noop {} {\enquote {\bibinfo {title} {Active pixel sensor
  array with electronic shuttering},}\ } (\bibinfo {year} {Jan. 1994}),\
  \bibinfo {note} {{U.S. Patent 6,486,503}}\BibitemShut {NoStop}%
\bibitem [{\citenamefont {Blerkom}\ \emph {et~al.}(2021)\citenamefont
  {Blerkom}, \citenamefont {Truong}, \citenamefont {Rysinski}, \citenamefont
  {Corlan}, \citenamefont {Venkatesan}, \citenamefont {Bagwell}, \citenamefont
  {Oniciuc},\ and\ \citenamefont {Bergey}}]{Dart:4}%
  \BibitemOpen
  \bibfield  {author} {\bibinfo {author} {\bibfnamefont {D.~V.}\ \bibnamefont
  {Blerkom}}, \bibinfo {author} {\bibfnamefont {L.}~\bibnamefont {Truong}},
  \bibinfo {author} {\bibfnamefont {J.}~\bibnamefont {Rysinski}}, \bibinfo
  {author} {\bibfnamefont {R.}~\bibnamefont {Corlan}}, \bibinfo {author}
  {\bibfnamefont {K.}~\bibnamefont {Venkatesan}}, \bibinfo {author}
  {\bibfnamefont {S.}~\bibnamefont {Bagwell}}, \bibinfo {author} {\bibfnamefont
  {L.}~\bibnamefont {Oniciuc}}, \ and\ \bibinfo {author} {\bibfnamefont
  {J.}~\bibnamefont {Bergey}},\ }\href@noop {} {\enquote {\bibinfo {title} {A
  1mpixel, 80k fps global shutter cmos image sensor for high speed imaging},}\
  } (\bibinfo {year} {2021}),\ \bibinfo {note} {{https://imagesensors.org},
  Paper \# R38.pdf}\BibitemShut {NoStop}%
\bibitem [{\citenamefont {Tochigi}\ \emph {et~al.}(2013)\citenamefont
  {Tochigi}, \citenamefont {Hanzawa}, \citenamefont {Kato}, \citenamefont
  {Kuroda}, \citenamefont {Mutoh}, \citenamefont {Hirose}, \citenamefont
  {Tominaga}, \citenamefont {Takubo}, \citenamefont {Kondo},\ and\
  \citenamefont {Sugawa}}]{Dart:5}%
  \BibitemOpen
  \bibfield  {author} {\bibinfo {author} {\bibfnamefont {Y.}~\bibnamefont
  {Tochigi}}, \bibinfo {author} {\bibfnamefont {K.}~\bibnamefont {Hanzawa}},
  \bibinfo {author} {\bibfnamefont {Y.}~\bibnamefont {Kato}}, \bibinfo {author}
  {\bibfnamefont {R.}~\bibnamefont {Kuroda}}, \bibinfo {author} {\bibfnamefont
  {H.}~\bibnamefont {Mutoh}}, \bibinfo {author} {\bibfnamefont
  {R.}~\bibnamefont {Hirose}}, \bibinfo {author} {\bibfnamefont
  {H.}~\bibnamefont {Tominaga}}, \bibinfo {author} {\bibfnamefont
  {K.}~\bibnamefont {Takubo}}, \bibinfo {author} {\bibfnamefont
  {Y.}~\bibnamefont {Kondo}}, \ and\ \bibinfo {author} {\bibfnamefont
  {S.}~\bibnamefont {Sugawa}},\ }\href@noop {} {\bibfield  {journal} {\bibinfo
  {journal} {IEEE J. Solid-State Circuits}\ }\textbf {\bibinfo {volume} {48}},\
  \bibinfo {pages} {329} (\bibinfo {year} {Jan. 2013})}\BibitemShut {NoStop}%
\bibitem [{\citenamefont {Kosonocky}\ \emph {et~al.}(1996)\citenamefont
  {Kosonocky}, \citenamefont {Yang}, \citenamefont {Ye}, \citenamefont {Kabra},
  \citenamefont {Xie}, \citenamefont {Lawrence}, \citenamefont {Mastrocolla},
  \citenamefont {Shallcross},\ and\ \citenamefont {Patel}}]{Dart:6}%
  \BibitemOpen
  \bibfield  {author} {\bibinfo {author} {\bibfnamefont {W.~F.}\ \bibnamefont
  {Kosonocky}}, \bibinfo {author} {\bibfnamefont {G.}~\bibnamefont {Yang}},
  \bibinfo {author} {\bibfnamefont {C.}~\bibnamefont {Ye}}, \bibinfo {author}
  {\bibfnamefont {R.~K.}\ \bibnamefont {Kabra}}, \bibinfo {author}
  {\bibfnamefont {L.}~\bibnamefont {Xie}}, \bibinfo {author} {\bibfnamefont
  {J.~L.}\ \bibnamefont {Lawrence}}, \bibinfo {author} {\bibfnamefont
  {V.}~\bibnamefont {Mastrocolla}}, \bibinfo {author} {\bibfnamefont {F.~V.}\
  \bibnamefont {Shallcross}}, \ and\ \bibinfo {author} {\bibfnamefont
  {V.}~\bibnamefont {Patel}},\ }\href@noop {} {\bibfield  {journal} {\bibinfo
  {journal} {IEEE Xplore}\ } (\bibinfo {year} {Feb. 1996})},\ \bibinfo {note}
  {https://doi.org/10.1109/ISSCC.1996.488562}\BibitemShut {NoStop}%
\bibitem [{\citenamefont {Suzuki}\ \emph {et~al.}(2017)\citenamefont {Suzuki},
  \citenamefont {Suzukia}, \citenamefont {Kurodaa}, \citenamefont {Kumagai},
  \citenamefont {Chibab}, \citenamefont {Miurab}, \citenamefont {Kuriyama},\
  and\ \citenamefont {Sugawa}}]{Dart:7}%
  \BibitemOpen
  \bibfield  {author} {\bibinfo {author} {\bibfnamefont {M.}~\bibnamefont
  {Suzuki}}, \bibinfo {author} {\bibfnamefont {M.}~\bibnamefont {Suzukia}},
  \bibinfo {author} {\bibfnamefont {R.}~\bibnamefont {Kurodaa}}, \bibinfo
  {author} {\bibfnamefont {Y.}~\bibnamefont {Kumagai}}, \bibinfo {author}
  {\bibfnamefont {A.}~\bibnamefont {Chibab}}, \bibinfo {author} {\bibfnamefont
  {N.}~\bibnamefont {Miurab}}, \bibinfo {author} {\bibfnamefont
  {N.}~\bibnamefont {Kuriyama}}, \ and\ \bibinfo {author} {\bibfnamefont
  {S.}~\bibnamefont {Sugawa}},\ }\href@noop {} {\enquote {\bibinfo {title} {{10
  Mfps 960 Frames Video Capturing Using a UHS Global Shutter CMOS Image Sensor
  with High Density Analog Memories}},}\ } (\bibinfo {year} {2017}),\ \bibinfo
  {note} {{https://imagesensors.org}, Paper \# R37.pdf}\BibitemShut {NoStop}%
\bibitem [{\citenamefont {Wu}\ \emph {et~al.}(2018)\citenamefont {Wu},
  \citenamefont {Bello}, \citenamefont {Coppejans}, \citenamefont {Craninckx},
  \citenamefont {Süss}, \citenamefont {Rosmeulen}, \citenamefont {Wambacq},\
  and\ \citenamefont {Borremans}}]{Dart:10}%
  \BibitemOpen
  \bibfield  {author} {\bibinfo {author} {\bibfnamefont {L.}~\bibnamefont
  {Wu}}, \bibinfo {author} {\bibfnamefont {D.~S.~S.}\ \bibnamefont {Bello}},
  \bibinfo {author} {\bibfnamefont {P.}~\bibnamefont {Coppejans}}, \bibinfo
  {author} {\bibfnamefont {J.}~\bibnamefont {Craninckx}}, \bibinfo {author}
  {\bibfnamefont {A.}~\bibnamefont {Süss}}, \bibinfo {author} {\bibfnamefont
  {M.}~\bibnamefont {Rosmeulen}}, \bibinfo {author} {\bibfnamefont
  {P.}~\bibnamefont {Wambacq}}, \ and\ \bibinfo {author} {\bibfnamefont
  {J.}~\bibnamefont {Borremans}},\ }\href@noop {} {\bibfield  {journal}
  {\bibinfo  {journal} {Sensors}\ }\textbf {\bibinfo {volume} {18}},\ \bibinfo
  {pages} {3683} (\bibinfo {year} {Oct. 2018})},\ \bibinfo {note}
  {https://doi.org/10.3390/s18113683}\BibitemShut {NoStop}%
\bibitem [{\citenamefont {Yue}\ and\ \citenamefont {Fossum}(2023)}]{Dart:8}%
  \BibitemOpen
  \bibfield  {author} {\bibinfo {author} {\bibfnamefont {X.}~\bibnamefont
  {Yue}}\ and\ \bibinfo {author} {\bibfnamefont {E.~R.}\ \bibnamefont
  {Fossum}},\ }\href@noop {} {\bibfield  {journal} {\bibinfo  {journal}
  {Electronic Imaging}\ }\textbf {\bibinfo {volume} {35}},\ \bibinfo {pages}
  {328} (\bibinfo {year} {Jan. 2023})},\ \bibinfo {note}
  {https://doi.org/10.2352/ei.2023.35.6.iss-328}\BibitemShut {NoStop}%
\bibitem [{\citenamefont {Suzuki}\ \emph {et~al.}(2020)\citenamefont {Suzuki},
  \citenamefont {Sugama}, \citenamefont {Kuroda},\ and\ \citenamefont
  {Sugawa}}]{Dart:9}%
  \BibitemOpen
  \bibfield  {author} {\bibinfo {author} {\bibfnamefont {M.}~\bibnamefont
  {Suzuki}}, \bibinfo {author} {\bibfnamefont {Y.}~\bibnamefont {Sugama}},
  \bibinfo {author} {\bibfnamefont {R.}~\bibnamefont {Kuroda}}, \ and\ \bibinfo
  {author} {\bibfnamefont {S.}~\bibnamefont {Sugawa}},\ }\href@noop {}
  {\bibfield  {journal} {\bibinfo  {journal} {Sensors}\ }\textbf {\bibinfo
  {volume} {20}},\ \bibinfo {pages} {1086} (\bibinfo {year} {Jan, 2020})},\
  \bibinfo {note} {doi: https://doi.org/10.3390/s20041086.}\BibitemShut {Stop}%
\bibitem [{\citenamefont {La\`zovsky}\ and\ \citenamefont
  {et~al}(2005)}]{Dart:11}%
  \BibitemOpen
  \bibfield  {author} {\bibinfo {author} {\bibfnamefont {L.}~\bibnamefont
  {La\`zovsky}}\ and\ \bibinfo {author} {\bibnamefont {et~al}},\ }\href@noop {}
  {\bibfield  {journal} {\bibinfo  {journal} {Proceedings of Airborne
  Intelligence, Surveillance, Reconnaissance (ISR) Systems and Applications
  II}\ ,\ \bibinfo {pages} {184}} (\bibinfo {year} {2005})}\BibitemShut
  {NoStop}%
\bibitem [{\citenamefont {Etoh}\ \emph {et~al.}(2013)\citenamefont {Etoh},
  \citenamefont {Son}, \citenamefont {Yamada},\ and\ \citenamefont
  {Charbon}}]{Dart:12}%
  \BibitemOpen
  \bibfield  {author} {\bibinfo {author} {\bibfnamefont {T.}~\bibnamefont
  {Etoh}}, \bibinfo {author} {\bibfnamefont {D.}~\bibnamefont {Son}}, \bibinfo
  {author} {\bibfnamefont {T.}~\bibnamefont {Yamada}}, \ and\ \bibinfo {author}
  {\bibfnamefont {E.}~\bibnamefont {Charbon}},\ }\href@noop {} {\bibfield
  {journal} {\bibinfo  {journal} {Sensors}\ }\textbf {\bibinfo {volume} {13}},\
  \bibinfo {pages} {4640–4658} (\bibinfo {year} {Apr. 2013})},\ \bibinfo
  {note} {doi: https://doi.org/10.3390/s130404640}\BibitemShut {NoStop}%
\bibitem [{\citenamefont {Etoh}\ \emph {et~al.}(2019)\citenamefont {Etoh},
  \citenamefont {Okinaka}, \citenamefont {Takano}, \citenamefont {Takehara},
  \citenamefont {Nakano}, \citenamefont {Shimonomura}, \citenamefont {Ando},
  \citenamefont {Ngo}, \citenamefont {Kamakura}, \citenamefont {Dao},
  \citenamefont {Nguyen}, \citenamefont {Charbon}, \citenamefont {Zhang},
  \citenamefont {Moor}, \citenamefont {Goetschalckx},\ and\ \citenamefont
  {Haspeslagh}}]{Dart:13}%
  \BibitemOpen
  \bibfield  {author} {\bibinfo {author} {\bibfnamefont {T.}~\bibnamefont
  {Etoh}}, \bibinfo {author} {\bibfnamefont {T.}~\bibnamefont {Okinaka}},
  \bibinfo {author} {\bibfnamefont {Y.}~\bibnamefont {Takano}}, \bibinfo
  {author} {\bibfnamefont {K.}~\bibnamefont {Takehara}}, \bibinfo {author}
  {\bibfnamefont {H.}~\bibnamefont {Nakano}}, \bibinfo {author} {\bibfnamefont
  {K.}~\bibnamefont {Shimonomura}}, \bibinfo {author} {\bibfnamefont
  {T.}~\bibnamefont {Ando}}, \bibinfo {author} {\bibfnamefont {N.}~\bibnamefont
  {Ngo}}, \bibinfo {author} {\bibfnamefont {Y.}~\bibnamefont {Kamakura}},
  \bibinfo {author} {\bibfnamefont {V.~T.~S.}\ \bibnamefont {Dao}}, \bibinfo
  {author} {\bibfnamefont {A.~Q.}\ \bibnamefont {Nguyen}}, \bibinfo {author}
  {\bibfnamefont {E.}~\bibnamefont {Charbon}}, \bibinfo {author} {\bibfnamefont
  {C.}~\bibnamefont {Zhang}}, \bibinfo {author} {\bibfnamefont {P.~D.}\
  \bibnamefont {Moor}}, \bibinfo {author} {\bibfnamefont {P.}~\bibnamefont
  {Goetschalckx}}, \ and\ \bibinfo {author} {\bibfnamefont {L.}~\bibnamefont
  {Haspeslagh}},\ }\href@noop {} {\bibfield  {journal} {\bibinfo  {journal}
  {Sensors}\ }\textbf {\bibinfo {volume} {19}},\ \bibinfo {pages} {2247}
  (\bibinfo {year} {Oct. 2019})},\ \bibinfo {note} {doi:
  https://doi.org/10.3390/s19102247}\BibitemShut {NoStop}%
\bibitem [{\citenamefont {Cao}\ \emph {et~al.}(2015)\citenamefont {Cao},
  \citenamefont {Gäbler}, \citenamefont {Lee}, \citenamefont {Ling},
  \citenamefont {Jarau}, \citenamefont {Tien}, \citenamefont {Chuan},\ and\
  \citenamefont {Bold}}]{Dart:14}%
  \BibitemOpen
  \bibfield  {author} {\bibinfo {author} {\bibfnamefont {X.}~\bibnamefont
  {Cao}}, \bibinfo {author} {\bibfnamefont {D.}~\bibnamefont {Gäbler}},
  \bibinfo {author} {\bibfnamefont {C.}~\bibnamefont {Lee}}, \bibinfo {author}
  {\bibfnamefont {T.~P.}\ \bibnamefont {Ling}}, \bibinfo {author}
  {\bibfnamefont {D.~A.}\ \bibnamefont {Jarau}}, \bibinfo {author}
  {\bibfnamefont {D.~K.~C.}\ \bibnamefont {Tien}}, \bibinfo {author}
  {\bibfnamefont {T.~B.}\ \bibnamefont {Chuan}}, \ and\ \bibinfo {author}
  {\bibfnamefont {B.}~\bibnamefont {Bold}},\ }\href@noop {} {\enquote {\bibinfo
  {title} {{Design and Optimisation of Large 4T Pixel}},}\ } (\bibinfo {year}
  {2015}),\ \bibinfo {note} {{https://imagesensors.org}, Poster in
  Session}\BibitemShut {NoStop}%
\bibitem [{\citenamefont {Dao}\ \emph {et~al.}(2018)\citenamefont {Dao},
  \citenamefont {Ngo}, \citenamefont {Nguyen}, \citenamefont {Morimoto},
  \citenamefont {Shimonomura}, \citenamefont {Goetschalckx}, \citenamefont
  {Haspeslagh}, \citenamefont {Moor}, \citenamefont {Takehara},\ and\
  \citenamefont {Etoh}}]{Dart:15}%
  \BibitemOpen
  \bibfield  {author} {\bibinfo {author} {\bibfnamefont {V.~T.~S.}\
  \bibnamefont {Dao}}, \bibinfo {author} {\bibfnamefont {N.}~\bibnamefont
  {Ngo}}, \bibinfo {author} {\bibfnamefont {A.~Q.}\ \bibnamefont {Nguyen}},
  \bibinfo {author} {\bibfnamefont {K.}~\bibnamefont {Morimoto}}, \bibinfo
  {author} {\bibfnamefont {K.}~\bibnamefont {Shimonomura}}, \bibinfo {author}
  {\bibfnamefont {P.}~\bibnamefont {Goetschalckx}}, \bibinfo {author}
  {\bibfnamefont {L.}~\bibnamefont {Haspeslagh}}, \bibinfo {author}
  {\bibfnamefont {P.~D.}\ \bibnamefont {Moor}}, \bibinfo {author}
  {\bibfnamefont {K.}~\bibnamefont {Takehara}}, \ and\ \bibinfo {author}
  {\bibfnamefont {T.~G.}\ \bibnamefont {Etoh}},\ }\href@noop {} {\bibfield
  {journal} {\bibinfo  {journal} {Sensors}\ }\textbf {\bibinfo {volume} {18}},\
  \bibinfo {pages} {E3112} (\bibinfo {year} {Sep. 2018})},\ \bibinfo {note}
  {doi: https://doi.org/10.3390/s18093112}\BibitemShut {NoStop}%
\bibitem [{\citenamefont {Kagawa}\ \emph {et~al.}(2022)\citenamefont {Kagawa},
  \citenamefont {Horio}, \citenamefont {Pham}, \citenamefont {Ibrahim},
  \citenamefont {ichiro Okihara}, \citenamefont {Furuhashi}, \citenamefont
  {Takasawa}, \citenamefont {Yasutomi}, \citenamefont {Kawahito},\ and\
  \citenamefont {Nagahara}}]{Dart:18}%
  \BibitemOpen
  \bibfield  {author} {\bibinfo {author} {\bibfnamefont {K.}~\bibnamefont
  {Kagawa}}, \bibinfo {author} {\bibfnamefont {M.}~\bibnamefont {Horio}},
  \bibinfo {author} {\bibfnamefont {A.~N.}\ \bibnamefont {Pham}}, \bibinfo
  {author} {\bibfnamefont {T.}~\bibnamefont {Ibrahim}}, \bibinfo {author}
  {\bibfnamefont {S.}~\bibnamefont {ichiro Okihara}}, \bibinfo {author}
  {\bibfnamefont {T.}~\bibnamefont {Furuhashi}}, \bibinfo {author}
  {\bibfnamefont {T.}~\bibnamefont {Takasawa}}, \bibinfo {author}
  {\bibfnamefont {K.}~\bibnamefont {Yasutomi}}, \bibinfo {author}
  {\bibfnamefont {S.}~\bibnamefont {Kawahito}}, \ and\ \bibinfo {author}
  {\bibfnamefont {H.}~\bibnamefont {Nagahara}},\ }\href@noop {} {\bibfield
  {journal} {\bibinfo  {journal} {Sensors}\ }\textbf {\bibinfo {volume} {22}},\
  \bibinfo {pages} {1953} (\bibinfo {year} {Jan. 2022})},\ \bibinfo {note}
  {doi: https://doi.org/10.3390/s22051953}\BibitemShut {NoStop}%
\bibitem [{\citenamefont {Acerbi}\ and\ \citenamefont
  {Gundacker}(2019)}]{AG:2019}%
  \BibitemOpen
  \bibfield  {author} {\bibinfo {author} {\bibfnamefont {F.}~\bibnamefont
  {Acerbi}}\ and\ \bibinfo {author} {\bibfnamefont {S.}~\bibnamefont
  {Gundacker}},\ }\href@noop {} {\bibfield  {journal} {\bibinfo  {journal}
  {Nucl. Instrum. Meth. A}\ }\textbf {\bibinfo {volume} {926}},\ \bibinfo
  {pages} {16} (\bibinfo {year} {2019})},\ \bibinfo {note}
  {https://doi.org/10.1016/j.nima.2018.11.118}\BibitemShut {NoStop}%
\bibitem [{\citenamefont {Bandi}\ \emph {et~al.}(2022)\citenamefont {Bandi},
  \citenamefont {Ilisie}, \citenamefont {Vornicu}, \citenamefont
  {Carmona-Galán}, \citenamefont {Benlloch},\ and\ \citenamefont {Ángel
  Rodríguez-Vázquez}}]{BIV:2022}%
  \BibitemOpen
  \bibfield  {author} {\bibinfo {author} {\bibfnamefont {F.}~\bibnamefont
  {Bandi}}, \bibinfo {author} {\bibfnamefont {V.}~\bibnamefont {Ilisie}},
  \bibinfo {author} {\bibfnamefont {I.}~\bibnamefont {Vornicu}}, \bibinfo
  {author} {\bibfnamefont {R.}~\bibnamefont {Carmona-Galán}}, \bibinfo
  {author} {\bibfnamefont {J.~M.}\ \bibnamefont {Benlloch}}, \ and\ \bibinfo
  {author} {\bibnamefont {Ángel Rodríguez-Vázquez}},\ }\href@noop {}
  {\bibfield  {journal} {\bibinfo  {journal} {Sensors}\ }\textbf {\bibinfo
  {volume} {22}},\ \bibinfo {pages} {122} (\bibinfo {year} {2022})},\ \bibinfo
  {note} {https://doi.org/10.3390/s22010122}\BibitemShut {NoStop}%
\bibitem [{\citenamefont {Saveliev}\ and\ \citenamefont {Golovin}(2000)}]{L1}%
  \BibitemOpen
  \bibfield  {author} {\bibinfo {author} {\bibfnamefont {V.}~\bibnamefont
  {Saveliev}}\ and\ \bibinfo {author} {\bibfnamefont {V.}~\bibnamefont
  {Golovin}},\ }\href@noop {} {\bibfield  {journal} {\bibinfo  {journal} {Nucl.
  Instrum. Meth. A}\ }\textbf {\bibinfo {volume} {442}},\ \bibinfo {pages}
  {223} (\bibinfo {year} {2000})},\ \bibinfo {note}
  {https://doi.org/10.1016/S0168-9002(99)01225-5}\BibitemShut {NoStop}%
\bibitem [{\citenamefont {Haemisch}\ \emph {et~al.}(2012)\citenamefont
  {Haemisch}, \citenamefont {Frach}, \citenamefont {Degenhardt},\ and\
  \citenamefont {Thon}}]{L2}%
  \BibitemOpen
  \bibfield  {author} {\bibinfo {author} {\bibfnamefont {Y.}~\bibnamefont
  {Haemisch}}, \bibinfo {author} {\bibfnamefont {T.}~\bibnamefont {Frach}},
  \bibinfo {author} {\bibfnamefont {C.}~\bibnamefont {Degenhardt}}, \ and\
  \bibinfo {author} {\bibfnamefont {A.}~\bibnamefont {Thon}},\ }\href@noop {}
  {\bibfield  {journal} {\bibinfo  {journal} {Physics Procedia}\ }\textbf
  {\bibinfo {volume} {37}},\ \bibinfo {pages} {1546} (\bibinfo {year}
  {2012})},\ \bibinfo {note} {doi: 10.1016/j.phpro.2012.03.749}\BibitemShut
  {NoStop}%
\bibitem [{\citenamefont {Pratte}\ \emph {et~al.}(2021)\citenamefont {Pratte},
  \citenamefont {Nolet}, \citenamefont {Parent}, \citenamefont {Vachon},
  \citenamefont {andTommy Rossignol}, \citenamefont {Deslandes}, \citenamefont
  {Dautet}, \citenamefont {Fontaine},\ and\ \citenamefont {Charlebois}}]{L3}%
  \BibitemOpen
  \bibfield  {author} {\bibinfo {author} {\bibfnamefont {J.-F.}\ \bibnamefont
  {Pratte}}, \bibinfo {author} {\bibfnamefont {F.}~\bibnamefont {Nolet}},
  \bibinfo {author} {\bibfnamefont {S.}~\bibnamefont {Parent}}, \bibinfo
  {author} {\bibfnamefont {F.}~\bibnamefont {Vachon}}, \bibinfo {author}
  {\bibfnamefont {N.~R.}\ \bibnamefont {andTommy Rossignol}}, \bibinfo {author}
  {\bibfnamefont {K.}~\bibnamefont {Deslandes}}, \bibinfo {author}
  {\bibfnamefont {H.}~\bibnamefont {Dautet}}, \bibinfo {author} {\bibfnamefont
  {R.}~\bibnamefont {Fontaine}}, \ and\ \bibinfo {author} {\bibfnamefont
  {S.~A.}\ \bibnamefont {Charlebois}},\ }\href@noop {} {\bibfield  {journal}
  {\bibinfo  {journal} {Sensors}\ }\textbf {\bibinfo {volume} {21}},\ \bibinfo
  {pages} {598} (\bibinfo {year} {2021})},\ \bibinfo {note}
  {https://doi.org/10.3390/s21020598}\BibitemShut {NoStop}%
\bibitem [{\citenamefont {Torilla}\ \emph {et~al.}(2022)\citenamefont
  {Torilla}, \citenamefont {Giroletti}, \citenamefont {Brogi}, \citenamefont
  {Collazuol}, \citenamefont {Betta}, \citenamefont {Marrocchesi},
  \citenamefont {Morsani}, \citenamefont {Pancheri}, \citenamefont {Ratti},
  \citenamefont {Minga},\ and\ \citenamefont {Vacchi}}]{L4}%
  \BibitemOpen
  \bibfield  {author} {\bibinfo {author} {\bibfnamefont {G.}~\bibnamefont
  {Torilla}}, \bibinfo {author} {\bibfnamefont {S.}~\bibnamefont {Giroletti}},
  \bibinfo {author} {\bibfnamefont {P.}~\bibnamefont {Brogi}}, \bibinfo
  {author} {\bibfnamefont {G.}~\bibnamefont {Collazuol}}, \bibinfo {author}
  {\bibfnamefont {G.-F.~D.}\ \bibnamefont {Betta}}, \bibinfo {author}
  {\bibfnamefont {P.}~\bibnamefont {Marrocchesi}}, \bibinfo {author}
  {\bibfnamefont {F.}~\bibnamefont {Morsani}}, \bibinfo {author} {\bibfnamefont
  {L.}~\bibnamefont {Pancheri}}, \bibinfo {author} {\bibfnamefont
  {L.}~\bibnamefont {Ratti}}, \bibinfo {author} {\bibfnamefont
  {J.}~\bibnamefont {Minga}}, \ and\ \bibinfo {author} {\bibfnamefont
  {C.}~\bibnamefont {Vacchi}},\ }\href@noop {} {\bibfield  {journal} {\bibinfo
  {journal} {IEEE Nuclear Science Symposium and Medical Imaging Conference}\ }
  (\bibinfo {year} {2022})},\ \bibinfo {note} {(Milan, Italy)}\BibitemShut
  {NoStop}%
\bibitem [{\citenamefont {Wang}(2015)}]{Wan:2015}%
  \BibitemOpen
  \bibfield  {author} {\bibinfo {author} {\bibfnamefont {Z.}~\bibnamefont
  {Wang}},\ }\href@noop {} {\bibfield  {journal} {\bibinfo  {journal} {J.
  Instrum.}\ }\textbf {\bibinfo {volume} {10(12)}},\ \bibinfo {pages} {C12013}
  (\bibinfo {year} {2015})}\BibitemShut {NoStop}%
\bibitem [{\citenamefont {{ATTRACT Consortium website}}(2018)}]{Cin:1}%
  \BibitemOpen
  \bibfield  {author} {\bibinfo {author} {\bibnamefont {{ATTRACT Consortium
  website}}},\ }\href@noop {} {} (\bibinfo {year} {2018}),\ \bibinfo {note}
  {{https://attract-eu.com/consortium/}}\BibitemShut {NoStop}%
\bibitem [{\citenamefont {Uenoyama}\ and\ \citenamefont {Ota}(2021)}]{Cin:11}%
  \BibitemOpen
  \bibfield  {author} {\bibinfo {author} {\bibfnamefont {S.}~\bibnamefont
  {Uenoyama}}\ and\ \bibinfo {author} {\bibfnamefont {R.}~\bibnamefont {Ota}},\
  }\href@noop {} {\bibfield  {journal} {\bibinfo  {journal} {ACS Photonics}\
  }\textbf {\bibinfo {volume} {8}},\ \bibinfo {pages} {1548} (\bibinfo {year}
  {2021})},\ \bibinfo {note} {{
  https://doi.org/10.1021/acsphotonics.1c00257}}\BibitemShut {NoStop}%
\bibitem [{\citenamefont {Shavanova}\ \emph {et~al.}(2016)\citenamefont
  {Shavanova}, \citenamefont {Bakakina}, \citenamefont {Burkova}, \citenamefont
  {Shtepliuk}, \citenamefont {Viter}, \citenamefont {Ubelis}, \citenamefont
  {Beni}, \citenamefont {Starodub}, \citenamefont {Yakimova},\ and\
  \citenamefont {Khranovskyy}}]{Cin:12}%
  \BibitemOpen
  \bibfield  {author} {\bibinfo {author} {\bibfnamefont {K.}~\bibnamefont
  {Shavanova}}, \bibinfo {author} {\bibfnamefont {Y.}~\bibnamefont {Bakakina}},
  \bibinfo {author} {\bibfnamefont {I.}~\bibnamefont {Burkova}}, \bibinfo
  {author} {\bibfnamefont {I.}~\bibnamefont {Shtepliuk}}, \bibinfo {author}
  {\bibfnamefont {R.}~\bibnamefont {Viter}}, \bibinfo {author} {\bibfnamefont
  {A.}~\bibnamefont {Ubelis}}, \bibinfo {author} {\bibfnamefont
  {V.}~\bibnamefont {Beni}}, \bibinfo {author} {\bibfnamefont {N.}~\bibnamefont
  {Starodub}}, \bibinfo {author} {\bibfnamefont {R.}~\bibnamefont {Yakimova}},
  \ and\ \bibinfo {author} {\bibfnamefont {V.}~\bibnamefont {Khranovskyy}},\
  }\href@noop {} {\bibfield  {journal} {\bibinfo  {journal} {Sensors}\ }\textbf
  {\bibinfo {volume} {16}},\ \bibinfo {pages} {223} (\bibinfo {year} {2016})},\
  \bibinfo {note} {{ https://doi.org/10.3390/s16020223}}\BibitemShut {NoStop}%
\bibitem [{\citenamefont {{ K. S. Novoselov and V. Fal'ko and L. Colombo and P.
  R. Gellert and M. G. Schwab and K. Kim}}(2012)}]{Cin:13}%
  \BibitemOpen
  \bibfield  {author} {\bibinfo {author} {\bibnamefont {{ K. S. Novoselov and
  V. Fal'ko and L. Colombo and P. R. Gellert and M. G. Schwab and K. Kim}}},\
  }\href@noop {} {\bibfield  {journal} {\bibinfo  {journal} {Nature}\ }\textbf
  {\bibinfo {volume} {490}},\ \bibinfo {pages} {192} (\bibinfo {year}
  {2012})},\ \bibinfo {note}
  {{https://doi.org/10.1038/nature11458}}\BibitemShut {NoStop}%
\bibitem [{\citenamefont {{Hui Cai, Yiling Yu, Yu-Chuan Lin, Alex A Puretzky,
  David B Geohegan, Kai Xiao}}(2021)}]{Cin:14}%
  \BibitemOpen
  \bibfield  {author} {\bibinfo {author} {\bibnamefont {{Hui Cai, Yiling Yu,
  Yu-Chuan Lin, Alex A Puretzky, David B Geohegan, Kai Xiao}}},\ }\href@noop {}
  {\bibfield  {journal} {\bibinfo  {journal} {Nano Res.}\ }\textbf {\bibinfo
  {volume} {14}},\ \bibinfo {pages} {1625} (\bibinfo {year} {2021})},\ \bibinfo
  {note} {{https://doi.org/10.1007/s12274-020-3047-7}}\BibitemShut {NoStop}%
\bibitem [{\citenamefont {Foxe}\ \emph {et~al.}(2012)\citenamefont {Foxe},
  \citenamefont {Lopez}, \citenamefont {Childres}, \citenamefont {Jalilian},
  \citenamefont {Patil}, \citenamefont {Roecker}, \citenamefont {Boguski},
  \citenamefont {Jovanovic},\ and\ \citenamefont {Chen}}]{Cin:15}%
  \BibitemOpen
  \bibfield  {author} {\bibinfo {author} {\bibfnamefont {M.}~\bibnamefont
  {Foxe}}, \bibinfo {author} {\bibfnamefont {G.}~\bibnamefont {Lopez}},
  \bibinfo {author} {\bibfnamefont {I.}~\bibnamefont {Childres}}, \bibinfo
  {author} {\bibfnamefont {R.}~\bibnamefont {Jalilian}}, \bibinfo {author}
  {\bibfnamefont {A.}~\bibnamefont {Patil}}, \bibinfo {author} {\bibfnamefont
  {C.}~\bibnamefont {Roecker}}, \bibinfo {author} {\bibfnamefont
  {J.}~\bibnamefont {Boguski}}, \bibinfo {author} {\bibfnamefont
  {I.}~\bibnamefont {Jovanovic}}, \ and\ \bibinfo {author} {\bibfnamefont
  {Y.~P.}\ \bibnamefont {Chen}},\ }\href@noop {} {\bibfield  {journal}
  {\bibinfo  {journal} {IEEE Trans. Nanotechnol.}\ }\textbf {\bibinfo {volume}
  {11}},\ \bibinfo {pages} {581} (\bibinfo {year} {2012})},\ \bibinfo {note}
  {{http://dx.doi.org/10.1109/TNANO.2012.2186312}}\BibitemShut {NoStop}%
\bibitem [{\citenamefont {Urich}\ \emph {et~al.}(2011)\citenamefont {Urich},
  \citenamefont {Unterrainer},\ and\ \citenamefont {Mueller}}]{Cin:16}%
  \BibitemOpen
  \bibfield  {author} {\bibinfo {author} {\bibfnamefont {A.}~\bibnamefont
  {Urich}}, \bibinfo {author} {\bibfnamefont {K.}~\bibnamefont {Unterrainer}},
  \ and\ \bibinfo {author} {\bibfnamefont {T.}~\bibnamefont {Mueller}},\
  }\href@noop {} {\bibfield  {journal} {\bibinfo  {journal} {Nano Lett.}\
  }\textbf {\bibinfo {volume} {11}},\ \bibinfo {pages} {2804} (\bibinfo {year}
  {2011})},\ \bibinfo {note} {{https://doi.org/10.1021/nl2011388}}\BibitemShut
  {NoStop}%
\bibitem [{\citenamefont {Tao}\ \emph {et~al.}(2023)\citenamefont {Tao},
  \citenamefont {Coffee}, \citenamefont {Jeong},\ and\ \citenamefont
  {Levin}}]{TCJ:2023}%
  \BibitemOpen
  \bibfield  {author} {\bibinfo {author} {\bibfnamefont {L.}~\bibnamefont
  {Tao}}, \bibinfo {author} {\bibfnamefont {R.~N.}\ \bibnamefont {Coffee}},
  \bibinfo {author} {\bibfnamefont {D.}~\bibnamefont {Jeong}}, \ and\ \bibinfo
  {author} {\bibfnamefont {C.~S.}\ \bibnamefont {Levin}},\ }\href@noop {}
  {\bibfield  {journal} {\bibinfo  {journal} {Phys. Med. Biol.}\ }\textbf
  {\bibinfo {volume} {66}},\ \bibinfo {pages} {045032} (\bibinfo {year}
  {2023})},\ \bibinfo {note} {doi:10.1088/1361-6560/abd951}\BibitemShut
  {NoStop}%
\bibitem [{\citenamefont {Salentijn}\ \emph {et~al.}(2017)\citenamefont
  {Salentijn}, \citenamefont {Oomen}, \citenamefont {Grajewski},\ and\
  \citenamefont {Verpoorte}}]{Cin:2}%
  \BibitemOpen
  \bibfield  {author} {\bibinfo {author} {\bibfnamefont {G.~I.}\ \bibnamefont
  {Salentijn}}, \bibinfo {author} {\bibfnamefont {P.~E.}\ \bibnamefont
  {Oomen}}, \bibinfo {author} {\bibfnamefont {M.}~\bibnamefont {Grajewski}}, \
  and\ \bibinfo {author} {\bibfnamefont {E.}~\bibnamefont {Verpoorte}},\
  }\href@noop {} {\bibfield  {journal} {\bibinfo  {journal} {Anal Chem.}\
  }\textbf {\bibinfo {volume} {89}},\ \bibinfo {pages} {7053} (\bibinfo {year}
  {2017})},\ \bibinfo {note} {{doi: 10.1021/acs.analchem.7b00828}}\BibitemShut
  {NoStop}%
\bibitem [{\citenamefont {Gross}\ \emph {et~al.}(2017)\citenamefont {Gross},
  \citenamefont {Lockwood},\ and\ \citenamefont {Spence}}]{Cin:3}%
  \BibitemOpen
  \bibfield  {author} {\bibinfo {author} {\bibfnamefont {B.}~\bibnamefont
  {Gross}}, \bibinfo {author} {\bibfnamefont {S.~Y.}\ \bibnamefont {Lockwood}},
  \ and\ \bibinfo {author} {\bibfnamefont {D.~M.}\ \bibnamefont {Spence}},\
  }\href@noop {} {\bibfield  {journal} {\bibinfo  {journal} {Anal Chem.}\
  }\textbf {\bibinfo {volume} {89}},\ \bibinfo {pages} {57} (\bibinfo {year}
  {2017})},\ \bibinfo {note} {{doi: 10.1021/acs.analchem.6b04344}}\BibitemShut
  {NoStop}%
\bibitem [{\citenamefont {Capel}\ \emph {et~al.}(2018)\citenamefont {Capel},
  \citenamefont {Rimington}, \citenamefont {Lewis},\ and\ \citenamefont
  {Christie}}]{Cin:4}%
  \BibitemOpen
  \bibfield  {author} {\bibinfo {author} {\bibfnamefont {A.~J.}\ \bibnamefont
  {Capel}}, \bibinfo {author} {\bibfnamefont {R.~P.}\ \bibnamefont
  {Rimington}}, \bibinfo {author} {\bibfnamefont {M.~P.}\ \bibnamefont
  {Lewis}}, \ and\ \bibinfo {author} {\bibfnamefont {S.~D.~R.}\ \bibnamefont
  {Christie}},\ }\href@noop {} {\bibfield  {journal} {\bibinfo  {journal} {Nat.
  Rev. Chem.}\ }\textbf {\bibinfo {volume} {2}},\ \bibinfo {pages} {422}
  (\bibinfo {year} {2018})},\ \bibinfo {note}
  {{https://doi.org/10.1038/s41570-018-0058-y}}\BibitemShut {NoStop}%
\bibitem [{\citenamefont {{CERN, ETH Zurich, HEIG-VD and ISMA}}(2019)}]{Cin:5}%
  \BibitemOpen
  \bibfield  {author} {\bibinfo {author} {\bibnamefont {{CERN, ETH Zurich,
  HEIG-VD and ISMA}}},\ }\href@noop {} {\enquote {\bibinfo {title} {The 3d
  printed detector (3det) project},}\ } (\bibinfo {year} {2019}),\ \bibinfo
  {note}
  {{https://ep-news.web.cern.ch/content/3d-printed-detector-3det-project}}\BibitemShut
  {NoStop}%
\bibitem [{\citenamefont {Kim}\ and\ \citenamefont {et~al.}(2020)}]{Cin:6}%
  \BibitemOpen
  \bibfield  {author} {\bibinfo {author} {\bibfnamefont {D.}~\bibnamefont
  {Kim}}\ and\ \bibinfo {author} {\bibnamefont {et~al.}},\ }\href@noop {}
  {\bibfield  {journal} {\bibinfo  {journal} {Nucl. Eng. Technol.}\ }\textbf
  {\bibinfo {volume} {52}},\ \bibinfo {pages} {2910} (\bibinfo {year}
  {2020})}\BibitemShut {NoStop}%
\bibitem [{\citenamefont {Glushkova}\ \emph {et~al.}(2021)\citenamefont
  {Glushkova}, \citenamefont {Andričević}, \citenamefont {Smajda},
  \citenamefont {Náfrádi}, \citenamefont {Kollár}, \citenamefont {Djokić},
  \citenamefont {Arakcheeva}, \citenamefont {Forró}, \citenamefont {Pugin},\
  and\ \citenamefont {Horváth}}]{Cin:7}%
  \BibitemOpen
  \bibfield  {author} {\bibinfo {author} {\bibfnamefont {A.}~\bibnamefont
  {Glushkova}}, \bibinfo {author} {\bibfnamefont {P.}~\bibnamefont
  {Andričević}}, \bibinfo {author} {\bibfnamefont {R.}~\bibnamefont
  {Smajda}}, \bibinfo {author} {\bibfnamefont {B.}~\bibnamefont {Náfrádi}},
  \bibinfo {author} {\bibfnamefont {M.}~\bibnamefont {Kollár}}, \bibinfo
  {author} {\bibfnamefont {V.}~\bibnamefont {Djokić}}, \bibinfo {author}
  {\bibfnamefont {A.}~\bibnamefont {Arakcheeva}}, \bibinfo {author}
  {\bibfnamefont {L.}~\bibnamefont {Forró}}, \bibinfo {author} {\bibfnamefont
  {R.}~\bibnamefont {Pugin}}, \ and\ \bibinfo {author} {\bibfnamefont
  {E.}~\bibnamefont {Horváth}},\ }\href@noop {} {\bibfield  {journal}
  {\bibinfo  {journal} {ACS Nano}\ }\textbf {\bibinfo {volume} {15}},\ \bibinfo
  {pages} {4077} (\bibinfo {year} {2021})},\ \bibinfo {note}
  {{https://doi.org/10.1021/acsnano.0c07993}}\BibitemShut {NoStop}%
\bibitem [{\citenamefont {Fischer}\ \emph {et~al.}(2012)\citenamefont
  {Fischer}, \citenamefont {Belova}, \citenamefont {Rikers}, \citenamefont
  {Malm}, \citenamefont {Radamson}, \citenamefont {Kolahdouz}, \citenamefont
  {Gylfason}, \citenamefont {Stemme},\ and\ \citenamefont {Niklaus}}]{Cin:8}%
  \BibitemOpen
  \bibfield  {author} {\bibinfo {author} {\bibfnamefont {A.~C.}\ \bibnamefont
  {Fischer}}, \bibinfo {author} {\bibfnamefont {L.~M.}\ \bibnamefont {Belova}},
  \bibinfo {author} {\bibfnamefont {Y.~G.~M.}\ \bibnamefont {Rikers}}, \bibinfo
  {author} {\bibfnamefont {B.~G.}\ \bibnamefont {Malm}}, \bibinfo {author}
  {\bibfnamefont {H.~H.}\ \bibnamefont {Radamson}}, \bibinfo {author}
  {\bibfnamefont {M.}~\bibnamefont {Kolahdouz}}, \bibinfo {author}
  {\bibfnamefont {K.~B.}\ \bibnamefont {Gylfason}}, \bibinfo {author}
  {\bibfnamefont {G.}~\bibnamefont {Stemme}}, \ and\ \bibinfo {author}
  {\bibfnamefont {F.}~\bibnamefont {Niklaus}},\ }\href@noop {} {\bibfield
  {journal} {\bibinfo  {journal} {{Adv. Funct. Mater.}}\ }\textbf {\bibinfo
  {volume} {22}},\ \bibinfo {pages} {4004} (\bibinfo {year} {2012})},\ \bibinfo
  {note} {{https://doi.org/10.1002/adfm.201200845}}\BibitemShut {NoStop}%
\bibitem [{\citenamefont {Dosovitskiy}\ \emph {et~al.}(2017)\citenamefont
  {Dosovitskiy}, \citenamefont {Karpyuk}, \citenamefont {Evdokimov},
  \citenamefont {Kuznetsova}, \citenamefont {Mechinsky}, \citenamefont
  {Borisevich}, \citenamefont {Fedorov}, \citenamefont {Putlayev},
  \citenamefont {Dosovitskiye},\ and\ \citenamefont {Korjik}}]{Cin:9}%
  \BibitemOpen
  \bibfield  {author} {\bibinfo {author} {\bibfnamefont {G.~A.}\ \bibnamefont
  {Dosovitskiy}}, \bibinfo {author} {\bibfnamefont {P.~V.}\ \bibnamefont
  {Karpyuk}}, \bibinfo {author} {\bibfnamefont {P.~V.}\ \bibnamefont
  {Evdokimov}}, \bibinfo {author} {\bibfnamefont {D.~E.}\ \bibnamefont
  {Kuznetsova}}, \bibinfo {author} {\bibfnamefont {V.~A.}\ \bibnamefont
  {Mechinsky}}, \bibinfo {author} {\bibfnamefont {A.~E.}\ \bibnamefont
  {Borisevich}}, \bibinfo {author} {\bibfnamefont {A.~A.}\ \bibnamefont
  {Fedorov}}, \bibinfo {author} {\bibfnamefont {V.~I.}\ \bibnamefont
  {Putlayev}}, \bibinfo {author} {\bibfnamefont {A.~E.}\ \bibnamefont
  {Dosovitskiye}}, \ and\ \bibinfo {author} {\bibfnamefont {M.~V.}\
  \bibnamefont {Korjik}},\ }\href@noop {} {\bibfield  {journal} {\bibinfo
  {journal} {CrystEngComm}\ }\textbf {\bibinfo {volume} {19}},\ \bibinfo
  {pages} {4260} (\bibinfo {year} {2017})},\ \bibinfo {note}
  {{https://doi.org/10.1039/C7CE00541E}}\BibitemShut {NoStop}%
\bibitem [{\citenamefont {Park}\ \emph {et~al.}(2022)\citenamefont {Park},
  \citenamefont {Yun}, \citenamefont {Chung}, \citenamefont {Park},
  \citenamefont {Lee},\ and\ \citenamefont {Park}}]{Cin:10}%
  \BibitemOpen
  \bibfield  {author} {\bibinfo {author} {\bibfnamefont {Y.-G.}\ \bibnamefont
  {Park}}, \bibinfo {author} {\bibfnamefont {I.}~\bibnamefont {Yun}}, \bibinfo
  {author} {\bibfnamefont {W.~G.}\ \bibnamefont {Chung}}, \bibinfo {author}
  {\bibfnamefont {W.}~\bibnamefont {Park}}, \bibinfo {author} {\bibfnamefont
  {D.~H.}\ \bibnamefont {Lee}}, \ and\ \bibinfo {author} {\bibfnamefont
  {J.-U.}\ \bibnamefont {Park}},\ }\href@noop {} {\bibfield  {journal}
  {\bibinfo  {journal} {Ad. Sci.}\ }\textbf {\bibinfo {volume} {9}},\ \bibinfo
  {pages} {2104623} (\bibinfo {year} {2022})},\ \bibinfo {note} {{
  https://doi.org/10.1002/advs.202104623}}\BibitemShut {NoStop}%
\bibitem [{\citenamefont {{Cinzia Da Vià, Gian-Franco Dalla Betta, Sherwood
  Parker}}(2019)}]{Cin:17}%
  \BibitemOpen
  \bibfield  {author} {\bibinfo {author} {\bibnamefont {{Cinzia Da Vià,
  Gian-Franco Dalla Betta, Sherwood Parker}}},\ }\href@noop {} {\emph {\bibinfo
  {title} {Radiation Sensors with 3D Electrodes}}}\ (\bibinfo  {publisher} {CRC
  Press},\ \bibinfo {year} {2019})\ \bibinfo {note}
  {https://doi.org/10.1201/9780429055324}\BibitemShut {NoStop}%
\bibitem [{\citenamefont {Da~Vià}\ \emph {et~al.}(2013)\citenamefont
  {Da~Vià}, \citenamefont {Boscardil}, \citenamefont {Dalla~Betta},
  \citenamefont {Darbo}, \citenamefont {Fleta}, \citenamefont {Gemme},
  \citenamefont {Giacomini}, \citenamefont {Grenier}, \citenamefont
  {Grinstein}, \citenamefont {Hansen}, \citenamefont {Hasi}, \citenamefont
  {Kenney}, \citenamefont {Kok}, \citenamefont {La~Rosa}, \citenamefont
  {Micelli}, \citenamefont {Parker}, \citenamefont {Pellegrini}, \citenamefont
  {Pohl}, \citenamefont {Povoli}, \citenamefont {Vianello}, \citenamefont
  {Zorzi},\ and\ \citenamefont {Watts}}]{Cin:18}%
  \BibitemOpen
  \bibfield  {author} {\bibinfo {author} {\bibfnamefont {C.}~\bibnamefont
  {Da~Vià}}, \bibinfo {author} {\bibfnamefont {M.}~\bibnamefont {Boscardil}},
  \bibinfo {author} {\bibfnamefont {G.}~\bibnamefont {Dalla~Betta}}, \bibinfo
  {author} {\bibfnamefont {G.}~\bibnamefont {Darbo}}, \bibinfo {author}
  {\bibfnamefont {C.}~\bibnamefont {Fleta}}, \bibinfo {author} {\bibfnamefont
  {C.}~\bibnamefont {Gemme}}, \bibinfo {author} {\bibfnamefont
  {G.}~\bibnamefont {Giacomini}}, \bibinfo {author} {\bibfnamefont
  {P.}~\bibnamefont {Grenier}}, \bibinfo {author} {\bibfnamefont
  {S.}~\bibnamefont {Grinstein}}, \bibinfo {author} {\bibfnamefont {T.-E.}\
  \bibnamefont {Hansen}}, \bibinfo {author} {\bibfnamefont {J.}~\bibnamefont
  {Hasi}}, \bibinfo {author} {\bibfnamefont {C.}~\bibnamefont {Kenney}},
  \bibinfo {author} {\bibfnamefont {A.}~\bibnamefont {Kok}}, \bibinfo {author}
  {\bibfnamefont {A.}~\bibnamefont {La~Rosa}}, \bibinfo {author} {\bibfnamefont
  {A.}~\bibnamefont {Micelli}}, \bibinfo {author} {\bibfnamefont
  {S.}~\bibnamefont {Parker}}, \bibinfo {author} {\bibfnamefont
  {G.}~\bibnamefont {Pellegrini}}, \bibinfo {author} {\bibfnamefont {D.-L.}\
  \bibnamefont {Pohl}}, \bibinfo {author} {\bibfnamefont {M.}~\bibnamefont
  {Povoli}}, \bibinfo {author} {\bibfnamefont {E.}~\bibnamefont {Vianello}},
  \bibinfo {author} {\bibfnamefont {N.}~\bibnamefont {Zorzi}}, \ and\ \bibinfo
  {author} {\bibfnamefont {S.~J.}\ \bibnamefont {Watts}},\ }\href@noop {}
  {\bibfield  {journal} {\bibinfo  {journal} {Nucl. Instrum. Meth. A}\ }\textbf
  {\bibinfo {volume} {699}},\ \bibinfo {pages} {18} (\bibinfo {year} {Jan.
  2013})},\ \bibinfo {note}
  {https://doi.org/10.1016/j.nima.2012.05.070}\BibitemShut {NoStop}%
\bibitem [{\citenamefont {Wang}\ \emph
  {et~al.}(2020{\natexlab{a}})\citenamefont {Wang}, \citenamefont {Wang},
  \citenamefont {He}, \citenamefont {Li}, \citenamefont {Luo},\ and\
  \citenamefont {Chen}}]{Cin:19}%
  \BibitemOpen
  \bibfield  {author} {\bibinfo {author} {\bibfnamefont {Z.}~\bibnamefont
  {Wang}}, \bibinfo {author} {\bibfnamefont {W.}~\bibnamefont {Wang}}, \bibinfo
  {author} {\bibfnamefont {X.}~\bibnamefont {He}}, \bibinfo {author}
  {\bibfnamefont {X.}~\bibnamefont {Li}}, \bibinfo {author} {\bibfnamefont
  {T.}~\bibnamefont {Luo}}, \ and\ \bibinfo {author} {\bibfnamefont
  {J.}~\bibnamefont {Chen}},\ }\href@noop {} {\bibfield  {journal} {\bibinfo
  {journal} {Nucl. Instrum. Meth. A}\ }\textbf {\bibinfo {volume} {979}},\
  \bibinfo {pages} {164468} (\bibinfo {year} {Nov. 2020}{\natexlab{a}})},\
  \bibinfo {note} {https://doi.org/10.1016/j.nima.2020.164468}\BibitemShut
  {NoStop}%
\bibitem [{\citenamefont {Lampis}\ \emph {et~al.}(2023)\citenamefont {Lampis},
  \citenamefont {Borgato}, \citenamefont {Brundu}, \citenamefont {Cardini},
  \citenamefont {Cossu}, \citenamefont {Betta}, \citenamefont {Garau},
  \citenamefont {Delfa}, \citenamefont {Lai}, \citenamefont {Loi},
  \citenamefont {Obertino}, \citenamefont {Simi},\ and\ \citenamefont
  {Vecchi}}]{Cin:20}%
  \BibitemOpen
  \bibfield  {author} {\bibinfo {author} {\bibfnamefont {A.}~\bibnamefont
  {Lampis}}, \bibinfo {author} {\bibfnamefont {F.}~\bibnamefont {Borgato}},
  \bibinfo {author} {\bibfnamefont {D.}~\bibnamefont {Brundu}}, \bibinfo
  {author} {\bibfnamefont {A.}~\bibnamefont {Cardini}}, \bibinfo {author}
  {\bibfnamefont {G.}~\bibnamefont {Cossu}}, \bibinfo {author} {\bibfnamefont
  {G.-F.~D.}\ \bibnamefont {Betta}}, \bibinfo {author} {\bibfnamefont
  {M.}~\bibnamefont {Garau}}, \bibinfo {author} {\bibfnamefont {L.~L.}\
  \bibnamefont {Delfa}}, \bibinfo {author} {\bibfnamefont {A.}~\bibnamefont
  {Lai}}, \bibinfo {author} {\bibfnamefont {A.}~\bibnamefont {Loi}}, \bibinfo
  {author} {\bibfnamefont {M.}~\bibnamefont {Obertino}}, \bibinfo {author}
  {\bibfnamefont {G.}~\bibnamefont {Simi}}, \ and\ \bibinfo {author}
  {\bibfnamefont {S.}~\bibnamefont {Vecchi}},\ }\href {\doibase
  10.1088/1748-0221/18/01/C01051} {\bibfield  {journal} {\bibinfo  {journal}
  {Journal of Instrumentation}\ }\textbf {\bibinfo {volume} {18}},\ \bibinfo
  {pages} {C01051} (\bibinfo {year} {2023})}\BibitemShut {NoStop}%
\bibitem [{\citenamefont {Da~Vià}\ \emph {et~al.}(2021)\citenamefont
  {Da~Vià}, \citenamefont {Petagna}, \citenamefont {Romagnoli}, \citenamefont
  {Hellenschmidt}, \citenamefont {Munoz-Sanchez},\ and\ \citenamefont
  {Dann}}]{Cin:21}%
  \BibitemOpen
  \bibfield  {author} {\bibinfo {author} {\bibfnamefont {C.}~\bibnamefont
  {Da~Vià}}, \bibinfo {author} {\bibfnamefont {P.}~\bibnamefont {Petagna}},
  \bibinfo {author} {\bibfnamefont {G.}~\bibnamefont {Romagnoli}}, \bibinfo
  {author} {\bibfnamefont {D.}~\bibnamefont {Hellenschmidt}}, \bibinfo {author}
  {\bibfnamefont {F.}~\bibnamefont {Munoz-Sanchez}}, \ and\ \bibinfo {author}
  {\bibfnamefont {N.}~\bibnamefont {Dann}},\ }\href@noop {} {\bibfield
  {journal} {\bibinfo  {journal} {Frontiers in Physics}\ }\textbf {\bibinfo
  {volume} {9}},\ \bibinfo {pages} {633970} (\bibinfo {year} {Apr. 2021})},\
  \bibinfo {note} {doi: 10.3389/fphy.2021.633970}\BibitemShut {NoStop}%
\bibitem [{\citenamefont {Radamson}\ \emph {et~al.}(2021)\citenamefont
  {Radamson}, \citenamefont {Zhu}, \citenamefont {Wu}, \citenamefont {He},
  \citenamefont {Lin}, \citenamefont {Liu}, \citenamefont {Xiang},
  \citenamefont {Kong}, \citenamefont {Xiong}, \citenamefont {Li},
  \citenamefont {Cui}, \citenamefont {Gao}, \citenamefont {Yang}, \citenamefont
  {Du}, \citenamefont {Xu}, \citenamefont {Li}, \citenamefont {Zhao},
  \citenamefont {Yu}, \citenamefont {Dong},\ and\ \citenamefont
  {Wang}}]{Cin:22}%
  \BibitemOpen
  \bibfield  {author} {\bibinfo {author} {\bibfnamefont {H.~H.}\ \bibnamefont
  {Radamson}}, \bibinfo {author} {\bibfnamefont {H.}~\bibnamefont {Zhu}},
  \bibinfo {author} {\bibfnamefont {Z.}~\bibnamefont {Wu}}, \bibinfo {author}
  {\bibfnamefont {X.}~\bibnamefont {He}}, \bibinfo {author} {\bibfnamefont
  {H.}~\bibnamefont {Lin}}, \bibinfo {author} {\bibfnamefont {J.}~\bibnamefont
  {Liu}}, \bibinfo {author} {\bibfnamefont {J.}~\bibnamefont {Xiang}}, \bibinfo
  {author} {\bibfnamefont {Z.}~\bibnamefont {Kong}}, \bibinfo {author}
  {\bibfnamefont {W.}~\bibnamefont {Xiong}}, \bibinfo {author} {\bibfnamefont
  {J.}~\bibnamefont {Li}}, \bibinfo {author} {\bibfnamefont {H.}~\bibnamefont
  {Cui}}, \bibinfo {author} {\bibfnamefont {J.}~\bibnamefont {Gao}}, \bibinfo
  {author} {\bibfnamefont {H.}~\bibnamefont {Yang}}, \bibinfo {author}
  {\bibfnamefont {Y.}~\bibnamefont {Du}}, \bibinfo {author} {\bibfnamefont
  {B.}~\bibnamefont {Xu}}, \bibinfo {author} {\bibfnamefont {B.}~\bibnamefont
  {Li}}, \bibinfo {author} {\bibfnamefont {X.}~\bibnamefont {Zhao}}, \bibinfo
  {author} {\bibfnamefont {J.}~\bibnamefont {Yu}}, \bibinfo {author}
  {\bibfnamefont {Y.}~\bibnamefont {Dong}}, \ and\ \bibinfo {author}
  {\bibfnamefont {G.}~\bibnamefont {Wang}},\ }\href@noop {} {\bibfield
  {journal} {\bibinfo  {journal} {Nanomater.}\ }\textbf {\bibinfo {volume}
  {10}},\ \bibinfo {pages} {1555} (\bibinfo {year} {Apr. 2021})},\ \bibinfo
  {note} {doi:10.3390/nano10081555}\BibitemShut {NoStop}%
\bibitem [{\citenamefont {Iyer}\ \emph {et~al.}(2019)\citenamefont {Iyer},
  \citenamefont {Jangam},\ and\ \citenamefont {Vaisband}}]{Cin:23}%
  \BibitemOpen
  \bibfield  {author} {\bibinfo {author} {\bibfnamefont {S.~S.}\ \bibnamefont
  {Iyer}}, \bibinfo {author} {\bibfnamefont {S.}~\bibnamefont {Jangam}}, \ and\
  \bibinfo {author} {\bibfnamefont {B.}~\bibnamefont {Vaisband}},\ }\href
  {\doibase 10.1147/JRD.2019.2940427} {\bibfield  {journal} {\bibinfo
  {journal} {IBM Journal of Research and Development}\ }\textbf {\bibinfo
  {volume} {63}},\ \bibinfo {pages} {5:1} (\bibinfo {year} {2019})}\BibitemShut
  {NoStop}%
\bibitem [{\citenamefont {Moore}(2020)}]{Cin:24}%
  \BibitemOpen
  \bibfield  {author} {\bibinfo {author} {\bibfnamefont {S.~K.}\ \bibnamefont
  {Moore}},\ }\href@noop {} {\enquote {\bibinfo {title} {Cerebras’s giant
  chip will smash deep learning’s speed barrier},}\ } (\bibinfo {year}
  {2020}),\ \bibinfo {note}
  {{https://spectrum.ieee.org/u/samuel-k-moore}}\BibitemShut {NoStop}%
\bibitem [{\citenamefont {Phillips}\ \emph {et~al.}(2015)\citenamefont
  {Phillips}, \citenamefont {D'Ambrosio}, \citenamefont {Tian}, \citenamefont
  {Rulison}, \citenamefont {Patel}, \citenamefont {Sadras}, \citenamefont
  {Gande}, \citenamefont {Switz}, \citenamefont {Fletcher},\ and\ \citenamefont
  {Waller}}]{PDT:2015}%
  \BibitemOpen
  \bibfield  {author} {\bibinfo {author} {\bibfnamefont {Z.~F.}\ \bibnamefont
  {Phillips}}, \bibinfo {author} {\bibfnamefont {M.~V.}\ \bibnamefont
  {D'Ambrosio}}, \bibinfo {author} {\bibfnamefont {L.}~\bibnamefont {Tian}},
  \bibinfo {author} {\bibfnamefont {J.~J.}\ \bibnamefont {Rulison}}, \bibinfo
  {author} {\bibfnamefont {H.~S.}\ \bibnamefont {Patel}}, \bibinfo {author}
  {\bibfnamefont {N.}~\bibnamefont {Sadras}}, \bibinfo {author} {\bibfnamefont
  {A.~V.}\ \bibnamefont {Gande}}, \bibinfo {author} {\bibfnamefont {N.~A.}\
  \bibnamefont {Switz}}, \bibinfo {author} {\bibfnamefont {D.~A.}\ \bibnamefont
  {Fletcher}}, \ and\ \bibinfo {author} {\bibfnamefont {L.}~\bibnamefont
  {Waller}},\ }\href {\doibase 10.1371/journal.pone.0124938} {\bibfield
  {journal} {\bibinfo  {journal} {PLoS ONE}\ }\textbf {\bibinfo {volume}
  {10}},\ \bibinfo {pages} {e0124938} (\bibinfo {year} {2015})}\BibitemShut
  {NoStop}%
\bibitem [{\citenamefont {Villanueva-Perez}\ \emph {et~al.}(2018)\citenamefont
  {Villanueva-Perez}, \citenamefont {Pedrini}, \citenamefont {Mokso},
  \citenamefont {Vagovic}, \citenamefont {Guzenko}, \citenamefont {Leake},
  \citenamefont {Willmott}, \citenamefont {Oberta}, \citenamefont {David},
  \citenamefont {Chapman},\ and\ \citenamefont {Stampanoni}}]{VPP:2018}%
  \BibitemOpen
  \bibfield  {author} {\bibinfo {author} {\bibfnamefont {P.}~\bibnamefont
  {Villanueva-Perez}}, \bibinfo {author} {\bibfnamefont {B.}~\bibnamefont
  {Pedrini}}, \bibinfo {author} {\bibfnamefont {R.}~\bibnamefont {Mokso}},
  \bibinfo {author} {\bibfnamefont {P.}~\bibnamefont {Vagovic}}, \bibinfo
  {author} {\bibfnamefont {V.~A.}\ \bibnamefont {Guzenko}}, \bibinfo {author}
  {\bibfnamefont {S.~J.}\ \bibnamefont {Leake}}, \bibinfo {author}
  {\bibfnamefont {P.~R.}\ \bibnamefont {Willmott}}, \bibinfo {author}
  {\bibfnamefont {P.}~\bibnamefont {Oberta}}, \bibinfo {author} {\bibfnamefont
  {C.}~\bibnamefont {David}}, \bibinfo {author} {\bibfnamefont {H.~N.}\
  \bibnamefont {Chapman}}, \ and\ \bibinfo {author} {\bibfnamefont
  {M.}~\bibnamefont {Stampanoni}},\ }\href@noop {} {\bibfield  {journal}
  {\bibinfo  {journal} {Optica}\ }\textbf {\bibinfo {volume} {5}},\ \bibinfo
  {pages} {1521} (\bibinfo {year} {2018})}\BibitemShut {NoStop}%
\bibitem [{\citenamefont {Chen}\ \emph {et~al.}(2019)\citenamefont {Chen},
  \citenamefont {Jung}, \citenamefont {Walko}, \citenamefont {Li},
  \citenamefont {Gao}, \citenamefont {Shenoy}, \citenamefont {L\'opez},\ and\
  \citenamefont {Wang}}]{CJW:2019}%
  \BibitemOpen
  \bibfield  {author} {\bibinfo {author} {\bibfnamefont {P.}~\bibnamefont
  {Chen}}, \bibinfo {author} {\bibfnamefont {I.~W.}\ \bibnamefont {Jung}},
  \bibinfo {author} {\bibfnamefont {D.~A.}\ \bibnamefont {Walko}}, \bibinfo
  {author} {\bibfnamefont {Z.}~\bibnamefont {Li}}, \bibinfo {author}
  {\bibfnamefont {Y.}~\bibnamefont {Gao}}, \bibinfo {author} {\bibfnamefont
  {G.~K.}\ \bibnamefont {Shenoy}}, \bibinfo {author} {\bibfnamefont
  {D.}~\bibnamefont {L\'opez}}, \ and\ \bibinfo {author} {\bibfnamefont
  {J.}~\bibnamefont {Wang}},\ }\href@noop {} {\bibfield  {journal} {\bibinfo
  {journal} {Nat. Comm.}\ }\textbf {\bibinfo {volume} {10}},\ \bibinfo {pages}
  {{art. No. 1158}} (\bibinfo {year} {2019})}\BibitemShut {NoStop}%
\bibitem [{\citenamefont {Trost}\ \emph {et~al.}(2020)\citenamefont {Trost},
  \citenamefont {Ayyer},\ and\ \citenamefont {Chapman}}]{TAC:2020}%
  \BibitemOpen
  \bibfield  {author} {\bibinfo {author} {\bibfnamefont {F.}~\bibnamefont
  {Trost}}, \bibinfo {author} {\bibfnamefont {K.}~\bibnamefont {Ayyer}}, \ and\
  \bibinfo {author} {\bibfnamefont {H.~N.}\ \bibnamefont {Chapman}},\ }\href
  {\doibase 10.1088/1367-2630/aba85c} {\bibfield  {journal} {\bibinfo
  {journal} {New J. Phys.}\ }\textbf {\bibinfo {volume} {22}},\ \bibinfo
  {pages} {083070} (\bibinfo {year} {2020})}\BibitemShut {NoStop}%
\bibitem [{\citenamefont {Lee}\ \emph {et~al.}(2010)\citenamefont {Lee},
  \citenamefont {Fezzaa},\ and\ \citenamefont {Uemura}}]{LFU:2010}%
  \BibitemOpen
  \bibfield  {author} {\bibinfo {author} {\bibfnamefont {W.-K.}\ \bibnamefont
  {Lee}}, \bibinfo {author} {\bibfnamefont {K.}~\bibnamefont {Fezzaa}}, \ and\
  \bibinfo {author} {\bibfnamefont {T.}~\bibnamefont {Uemura}},\ }\href
  {\doibase 10.1107/S0909049510040434} {\bibfield  {journal} {\bibinfo
  {journal} {Journal of Synchrotron Radiation}\ }\textbf {\bibinfo {volume}
  {18}},\ \bibinfo {pages} {302} (\bibinfo {year} {2010})}\BibitemShut
  {NoStop}%
\bibitem [{\citenamefont {Morris}\ \emph {et~al.}(2013)\citenamefont {Morris},
  \citenamefont {King}, \citenamefont {Kwiatkowski}, \citenamefont {Mariam},
  \citenamefont {Merrill},\ and\ \citenamefont {Saunders}}]{MKK:2013}%
  \BibitemOpen
  \bibfield  {author} {\bibinfo {author} {\bibfnamefont {C.~L.}\ \bibnamefont
  {Morris}}, \bibinfo {author} {\bibfnamefont {N.~S.~P.}\ \bibnamefont {King}},
  \bibinfo {author} {\bibfnamefont {K.}~\bibnamefont {Kwiatkowski}}, \bibinfo
  {author} {\bibfnamefont {F.~G.}\ \bibnamefont {Mariam}}, \bibinfo {author}
  {\bibfnamefont {F.~E.}\ \bibnamefont {Merrill}}, \ and\ \bibinfo {author}
  {\bibfnamefont {A.}~\bibnamefont {Saunders}},\ }\href@noop {} {\bibfield
  {journal} {\bibinfo  {journal} {Rep. Prog. Phys.}\ }\textbf {\bibinfo
  {volume} {76}},\ \bibinfo {pages} {046301} (\bibinfo {year}
  {2013})}\BibitemShut {NoStop}%
\bibitem [{\citenamefont {Momose}(2005)}]{AM:2005}%
  \BibitemOpen
  \bibfield  {author} {\bibinfo {author} {\bibfnamefont {A.}~\bibnamefont
  {Momose}},\ }\href@noop {} {\bibfield  {journal} {\bibinfo  {journal} {Japn.
  J. Appl. Phys.}\ }\textbf {\bibinfo {volume} {44}},\ \bibinfo {pages} {6355}
  (\bibinfo {year} {2005})}\BibitemShut {NoStop}%
\bibitem [{\citenamefont {Wilkins}\ \emph {et~al.}(2015)\citenamefont
  {Wilkins}, \citenamefont {Nesterets}, \citenamefont {Gureyev}, \citenamefont
  {Mayo}, \citenamefont {Pogany},\ and\ \citenamefont {Stevenson}}]{WNG:2015}%
  \BibitemOpen
  \bibfield  {author} {\bibinfo {author} {\bibfnamefont {S.~W.}\ \bibnamefont
  {Wilkins}}, \bibinfo {author} {\bibfnamefont {Y.~I.}\ \bibnamefont
  {Nesterets}}, \bibinfo {author} {\bibfnamefont {T.~E.}\ \bibnamefont
  {Gureyev}}, \bibinfo {author} {\bibfnamefont {S.~C.}\ \bibnamefont {Mayo}},
  \bibinfo {author} {\bibfnamefont {A.}~\bibnamefont {Pogany}}, \ and\ \bibinfo
  {author} {\bibfnamefont {A.~W.}\ \bibnamefont {Stevenson}},\ }\href@noop {}
  {\bibfield  {journal} {\bibinfo  {journal} {Phil. Trans. R. Soc. A}\ }\textbf
  {\bibinfo {volume} {372}},\ \bibinfo {pages} {20130021} (\bibinfo {year}
  {2015})}\BibitemShut {NoStop}%
\bibitem [{\citenamefont {Eggl}\ \emph {et~al.}(2015)\citenamefont {Eggl},
  \citenamefont {Schleede}, \citenamefont {Bech}, \citenamefont {Achterhold},
  \citenamefont {Loewen}, \citenamefont {Ruth},\ and\ \citenamefont
  {Pfeiffer}}]{ESB:2015}%
  \BibitemOpen
  \bibfield  {author} {\bibinfo {author} {\bibfnamefont {E.}~\bibnamefont
  {Eggl}}, \bibinfo {author} {\bibfnamefont {S.}~\bibnamefont {Schleede}},
  \bibinfo {author} {\bibfnamefont {M.}~\bibnamefont {Bech}}, \bibinfo {author}
  {\bibfnamefont {K.}~\bibnamefont {Achterhold}}, \bibinfo {author}
  {\bibfnamefont {R.}~\bibnamefont {Loewen}}, \bibinfo {author} {\bibfnamefont
  {R.~D.}\ \bibnamefont {Ruth}}, \ and\ \bibinfo {author} {\bibfnamefont
  {F.}~\bibnamefont {Pfeiffer}},\ }\href@noop {} {\bibfield  {journal}
  {\bibinfo  {journal} {PNAS}\ }\textbf {\bibinfo {volume} {112}},\ \bibinfo
  {pages} {5567} (\bibinfo {year} {2015})}\BibitemShut {NoStop}%
\bibitem [{\citenamefont {Barbato}\ \emph {et~al.}(2019)\citenamefont
  {Barbato}, \citenamefont {Atzeni}, \citenamefont {Batani}, \citenamefont
  {Bleiner}, \citenamefont {Boutoux}, \citenamefont {Brabetz}, \citenamefont
  {Bradford}, \citenamefont {Mancelli}, \citenamefont {Neumayer}, \citenamefont
  {Schiavi} \emph {et~al.}}]{barbato2019quantitative}%
  \BibitemOpen
  \bibfield  {author} {\bibinfo {author} {\bibfnamefont {F.}~\bibnamefont
  {Barbato}}, \bibinfo {author} {\bibfnamefont {S.}~\bibnamefont {Atzeni}},
  \bibinfo {author} {\bibfnamefont {D.}~\bibnamefont {Batani}}, \bibinfo
  {author} {\bibfnamefont {D.}~\bibnamefont {Bleiner}}, \bibinfo {author}
  {\bibfnamefont {G.}~\bibnamefont {Boutoux}}, \bibinfo {author} {\bibfnamefont
  {C.}~\bibnamefont {Brabetz}}, \bibinfo {author} {\bibfnamefont
  {P.}~\bibnamefont {Bradford}}, \bibinfo {author} {\bibfnamefont
  {D.}~\bibnamefont {Mancelli}}, \bibinfo {author} {\bibfnamefont
  {P.}~\bibnamefont {Neumayer}}, \bibinfo {author} {\bibfnamefont
  {A.}~\bibnamefont {Schiavi}},  \emph {et~al.},\ }\href@noop {} {\bibfield
  {journal} {\bibinfo  {journal} {Scientific reports}\ }\textbf {\bibinfo
  {volume} {9}},\ \bibinfo {pages} {18805} (\bibinfo {year}
  {2019})}\BibitemShut {NoStop}%
\bibitem [{\citenamefont {Olivo}(2021)}]{A58}%
  \BibitemOpen
  \bibfield  {author} {\bibinfo {author} {\bibfnamefont {A.}~\bibnamefont
  {Olivo}},\ }\href {\doibase 10.1088/1361-648X/ac0e6e} {\bibfield  {journal}
  {\bibinfo  {journal} {J. Phys.: Condensed Matter}\ }\textbf {\bibinfo
  {volume} {33}},\ \bibinfo {pages} {363002} (\bibinfo {year}
  {2021})}\BibitemShut {NoStop}%
\bibitem [{\citenamefont {Olbinado}\ \emph {et~al.}(2021)\citenamefont
  {Olbinado}, \citenamefont {Paganin}, \citenamefont {Cheng},\ and\
  \citenamefont {Rack}}]{A59}%
  \BibitemOpen
  \bibfield  {author} {\bibinfo {author} {\bibfnamefont {M.~P.}\ \bibnamefont
  {Olbinado}}, \bibinfo {author} {\bibfnamefont {D.~M.}\ \bibnamefont
  {Paganin}}, \bibinfo {author} {\bibfnamefont {Y.}~\bibnamefont {Cheng}}, \
  and\ \bibinfo {author} {\bibfnamefont {A.}~\bibnamefont {Rack}},\ }\href
  {\doibase 10.1364/optica.437481} {\bibfield  {journal} {\bibinfo  {journal}
  {Optica}\ }\textbf {\bibinfo {volume} {8}},\ \bibinfo {pages} {1538}
  (\bibinfo {year} {2021})}\BibitemShut {NoStop}%
\bibitem [{\citenamefont {Wen}\ \emph {et~al.}(2019)\citenamefont {Wen},
  \citenamefont {Cherukara},\ and\ \citenamefont {Holt}}]{A60}%
  \BibitemOpen
  \bibfield  {author} {\bibinfo {author} {\bibfnamefont {H.}~\bibnamefont
  {Wen}}, \bibinfo {author} {\bibfnamefont {M.~J.}\ \bibnamefont {Cherukara}},
  \ and\ \bibinfo {author} {\bibfnamefont {M.~V.}\ \bibnamefont {Holt}},\
  }\href {\doibase 10.1146/annurev-matsci-070616-124014} {\bibfield  {journal}
  {\bibinfo  {journal} {Annual Review of Materials Research}\ }\textbf
  {\bibinfo {volume} {49}},\ \bibinfo {pages} {389} (\bibinfo {year} {2019})},\
  \Eprint {http://arxiv.org/abs/1811.03785} {arXiv:1811.03785} \BibitemShut
  {NoStop}%
\bibitem [{\citenamefont {Clauser}\ and\ \citenamefont
  {Reinsch}(1992)}]{Clauser:1992}%
  \BibitemOpen
  \bibfield  {author} {\bibinfo {author} {\bibfnamefont {J.~F.}\ \bibnamefont
  {Clauser}}\ and\ \bibinfo {author} {\bibfnamefont {M.~W.}\ \bibnamefont
  {Reinsch}},\ }\href {\doibase 10.1007/BF00325384} {\bibfield  {journal}
  {\bibinfo  {journal} {Applied Physics B: Photophysics and Laser Chemistry}\
  }\textbf {\bibinfo {volume} {54}},\ \bibinfo {pages} {380} (\bibinfo {year}
  {1992})}\BibitemShut {NoStop}%
\bibitem [{\citenamefont {Pfeiffer}\ \emph {et~al.}(2006)\citenamefont
  {Pfeiffer}, \citenamefont {Weitkamp}, \citenamefont {Bunk},\ and\
  \citenamefont {David}}]{PWB:2006}%
  \BibitemOpen
  \bibfield  {author} {\bibinfo {author} {\bibfnamefont {F.}~\bibnamefont
  {Pfeiffer}}, \bibinfo {author} {\bibfnamefont {T.}~\bibnamefont {Weitkamp}},
  \bibinfo {author} {\bibfnamefont {O.}~\bibnamefont {Bunk}}, \ and\ \bibinfo
  {author} {\bibfnamefont {C.}~\bibnamefont {David}},\ }\href@noop {}
  {\bibfield  {journal} {\bibinfo  {journal} {Nat. Phys.}\ }\textbf {\bibinfo
  {volume} {2}},\ \bibinfo {pages} {258} (\bibinfo {year} {2006})}\BibitemShut
  {NoStop}%
\bibitem [{\citenamefont {Valdivia}\ \emph {et~al.}(2022)\citenamefont
  {Valdivia}, \citenamefont {Perez-Callejo}, \citenamefont {Bouffetier},
  \citenamefont {Collins}, \citenamefont {Stoeckl}, \citenamefont {Filkins},
  \citenamefont {Mileham}, \citenamefont {Romanofsky}, \citenamefont
  {Begishev}, \citenamefont {Theobald}, \citenamefont {Klein}, \citenamefont
  {Schneider}, \citenamefont {Beg}, \citenamefont {Casner},\ and\ \citenamefont
  {Stutman}}]{A2}%
  \BibitemOpen
  \bibfield  {author} {\bibinfo {author} {\bibfnamefont {M.~P.}\ \bibnamefont
  {Valdivia}}, \bibinfo {author} {\bibfnamefont {G.}~\bibnamefont
  {Perez-Callejo}}, \bibinfo {author} {\bibfnamefont {V.}~\bibnamefont
  {Bouffetier}}, \bibinfo {author} {\bibfnamefont {G.~W.}\ \bibnamefont
  {Collins}}, \bibinfo {author} {\bibfnamefont {C.}~\bibnamefont {Stoeckl}},
  \bibinfo {author} {\bibfnamefont {T.}~\bibnamefont {Filkins}}, \bibinfo
  {author} {\bibfnamefont {C.}~\bibnamefont {Mileham}}, \bibinfo {author}
  {\bibfnamefont {M.}~\bibnamefont {Romanofsky}}, \bibinfo {author}
  {\bibfnamefont {I.~A.}\ \bibnamefont {Begishev}}, \bibinfo {author}
  {\bibfnamefont {W.}~\bibnamefont {Theobald}}, \bibinfo {author}
  {\bibfnamefont {S.~R.}\ \bibnamefont {Klein}}, \bibinfo {author}
  {\bibfnamefont {M.~K.}\ \bibnamefont {Schneider}}, \bibinfo {author}
  {\bibfnamefont {F.~N.}\ \bibnamefont {Beg}}, \bibinfo {author} {\bibfnamefont
  {A.}~\bibnamefont {Casner}}, \ and\ \bibinfo {author} {\bibfnamefont
  {D.}~\bibnamefont {Stutman}},\ }\href {\doibase 10.1063/5.0101865} {\bibfield
   {journal} {\bibinfo  {journal} {Review of Scientific Instruments}\ }\textbf
  {\bibinfo {volume} {93}},\ \bibinfo {pages} {115102} (\bibinfo {year}
  {2022})}\BibitemShut {NoStop}%
\bibitem [{\citenamefont {Yamada}\ \emph {et~al.}(2020)\citenamefont {Yamada},
  \citenamefont {Inoue}, \citenamefont {Nakamura}, \citenamefont {Kameshima},
  \citenamefont {Yamauchi}, \citenamefont {Matsuyama},\ and\ \citenamefont
  {Yabashi}}]{A9}%
  \BibitemOpen
  \bibfield  {author} {\bibinfo {author} {\bibfnamefont {J.}~\bibnamefont
  {Yamada}}, \bibinfo {author} {\bibfnamefont {T.}~\bibnamefont {Inoue}},
  \bibinfo {author} {\bibfnamefont {N.}~\bibnamefont {Nakamura}}, \bibinfo
  {author} {\bibfnamefont {T.}~\bibnamefont {Kameshima}}, \bibinfo {author}
  {\bibfnamefont {K.}~\bibnamefont {Yamauchi}}, \bibinfo {author}
  {\bibfnamefont {S.}~\bibnamefont {Matsuyama}}, \ and\ \bibinfo {author}
  {\bibfnamefont {M.}~\bibnamefont {Yabashi}},\ }\href {\doibase
  10.3390/s20247356} {\bibfield  {journal} {\bibinfo  {journal} {Sensors}\
  }\textbf {\bibinfo {volume} {20}},\ \bibinfo {pages} {1} (\bibinfo {year}
  {2020})}\BibitemShut {NoStop}%
\bibitem [{\citenamefont {Liu}\ \emph {et~al.}(2018)\citenamefont {Liu},
  \citenamefont {Seaberg}, \citenamefont {Zhu}, \citenamefont {Krzywinski},
  \citenamefont {Seiboth}, \citenamefont {Hardin}, \citenamefont {Cocco},
  \citenamefont {Aquila}, \citenamefont {Nagler}, \citenamefont {Lee},
  \citenamefont {Boutet}, \citenamefont {Feng}, \citenamefont {Ding},
  \citenamefont {Marcus},\ and\ \citenamefont {Sakdinawat}}]{A10}%
  \BibitemOpen
  \bibfield  {author} {\bibinfo {author} {\bibfnamefont {Y.}~\bibnamefont
  {Liu}}, \bibinfo {author} {\bibfnamefont {M.}~\bibnamefont {Seaberg}},
  \bibinfo {author} {\bibfnamefont {D.}~\bibnamefont {Zhu}}, \bibinfo {author}
  {\bibfnamefont {J.}~\bibnamefont {Krzywinski}}, \bibinfo {author}
  {\bibfnamefont {F.}~\bibnamefont {Seiboth}}, \bibinfo {author} {\bibfnamefont
  {C.}~\bibnamefont {Hardin}}, \bibinfo {author} {\bibfnamefont
  {D.}~\bibnamefont {Cocco}}, \bibinfo {author} {\bibfnamefont
  {A.}~\bibnamefont {Aquila}}, \bibinfo {author} {\bibfnamefont
  {B.}~\bibnamefont {Nagler}}, \bibinfo {author} {\bibfnamefont {H.~J.}\
  \bibnamefont {Lee}}, \bibinfo {author} {\bibfnamefont {S.}~\bibnamefont
  {Boutet}}, \bibinfo {author} {\bibfnamefont {Y.}~\bibnamefont {Feng}},
  \bibinfo {author} {\bibfnamefont {Y.}~\bibnamefont {Ding}}, \bibinfo {author}
  {\bibfnamefont {G.}~\bibnamefont {Marcus}}, \ and\ \bibinfo {author}
  {\bibfnamefont {A.}~\bibnamefont {Sakdinawat}},\ }\href {\doibase
  10.1364/OPTICA.5.000967} {\bibfield  {journal} {\bibinfo  {journal} {Optica}\
  }\textbf {\bibinfo {volume} {5}},\ \bibinfo {pages} {967} (\bibinfo {year}
  {2018})}\BibitemShut {NoStop}%
\bibitem [{\citenamefont {Grizolli}\ \emph {et~al.}(2017)\citenamefont
  {Grizolli}, \citenamefont {Shi}, \citenamefont {Kolodziej}, \citenamefont
  {Shvyd'ko},\ and\ \citenamefont {Assoufid}}]{A11}%
  \BibitemOpen
  \bibfield  {author} {\bibinfo {author} {\bibfnamefont {W.~C.}\ \bibnamefont
  {Grizolli}}, \bibinfo {author} {\bibfnamefont {X.}~\bibnamefont {Shi}},
  \bibinfo {author} {\bibfnamefont {T.}~\bibnamefont {Kolodziej}}, \bibinfo
  {author} {\bibfnamefont {Y.}~\bibnamefont {Shvyd'ko}}, \ and\ \bibinfo
  {author} {\bibfnamefont {L.}~\bibnamefont {Assoufid}},\ }\href {\doibase
  10.1117/12.2274023} {\bibfield  {journal} {\bibinfo  {journal} {SPIE Proc.}\
  }\textbf {\bibinfo {volume} {10385}},\ \bibinfo {pages} {1038502} (\bibinfo
  {year} {2017})}\BibitemShut {NoStop}%
\bibitem [{\citenamefont {Jefimovs}\ \emph {et~al.}(2021)\citenamefont
  {Jefimovs}, \citenamefont {Vila-Comamala}, \citenamefont {Arboleda},
  \citenamefont {Wang}, \citenamefont {Romano}, \citenamefont {Shi},
  \citenamefont {Kagias},\ and\ \citenamefont {Stampanoni}}]{A12}%
  \BibitemOpen
  \bibfield  {author} {\bibinfo {author} {\bibfnamefont {K.}~\bibnamefont
  {Jefimovs}}, \bibinfo {author} {\bibfnamefont {J.}~\bibnamefont
  {Vila-Comamala}}, \bibinfo {author} {\bibfnamefont {C.}~\bibnamefont
  {Arboleda}}, \bibinfo {author} {\bibfnamefont {Z.}~\bibnamefont {Wang}},
  \bibinfo {author} {\bibfnamefont {L.}~\bibnamefont {Romano}}, \bibinfo
  {author} {\bibfnamefont {Z.}~\bibnamefont {Shi}}, \bibinfo {author}
  {\bibfnamefont {M.}~\bibnamefont {Kagias}}, \ and\ \bibinfo {author}
  {\bibfnamefont {M.}~\bibnamefont {Stampanoni}},\ }\href@noop {} {\bibfield
  {journal} {\bibinfo  {journal} {Micromachines}\ }\textbf {\bibinfo {volume}
  {12}},\ \bibinfo {pages} {517} (\bibinfo {year} {2021})}\BibitemShut
  {NoStop}%
\bibitem [{\citenamefont {{von Teuffenbach}}\ \emph {et~al.}(2017)\citenamefont
  {{von Teuffenbach}}, \citenamefont {Koehler}, \citenamefont {Fehringer},
  \citenamefont {Viermetz}, \citenamefont {Brendel}, \citenamefont {Herzen},
  \citenamefont {Prokda}, \citenamefont {Rummeny}, \citenamefont {Pfeiffer},\
  and\ \citenamefont {No{\"e}l}}]{TKF:2017}%
  \BibitemOpen
  \bibfield  {author} {\bibinfo {author} {\bibfnamefont {M.}~\bibnamefont {{von
  Teuffenbach}}}, \bibinfo {author} {\bibfnamefont {T.}~\bibnamefont
  {Koehler}}, \bibinfo {author} {\bibfnamefont {A.}~\bibnamefont {Fehringer}},
  \bibinfo {author} {\bibfnamefont {M.}~\bibnamefont {Viermetz}}, \bibinfo
  {author} {\bibfnamefont {B.}~\bibnamefont {Brendel}}, \bibinfo {author}
  {\bibfnamefont {J.}~\bibnamefont {Herzen}}, \bibinfo {author} {\bibfnamefont
  {R.}~\bibnamefont {Prokda}}, \bibinfo {author} {\bibfnamefont {E.~J.}\
  \bibnamefont {Rummeny}}, \bibinfo {author} {\bibfnamefont {F.}~\bibnamefont
  {Pfeiffer}}, \ and\ \bibinfo {author} {\bibfnamefont {P.~B.}\ \bibnamefont
  {No{\"e}l}},\ }\href@noop {} {\bibfield  {journal} {\bibinfo  {journal} {Sci.
  Reports}\ }\textbf {\bibinfo {volume} {7}},\ \bibinfo {pages} {7476}
  (\bibinfo {year} {2017})}\BibitemShut {NoStop}%
\bibitem [{\citenamefont {Xu}\ \emph {et~al.}(2022)\citenamefont {Xu},
  \citenamefont {Tao}, \citenamefont {Bian}, \citenamefont {Bai}, \citenamefont
  {Tian}, \citenamefont {Hao}, \citenamefont {Kuang},\ and\ \citenamefont
  {Liu}}]{A13}%
  \BibitemOpen
  \bibfield  {author} {\bibinfo {author} {\bibfnamefont {Y.}~\bibnamefont
  {Xu}}, \bibinfo {author} {\bibfnamefont {S.}~\bibnamefont {Tao}}, \bibinfo
  {author} {\bibfnamefont {Y.}~\bibnamefont {Bian}}, \bibinfo {author}
  {\bibfnamefont {L.}~\bibnamefont {Bai}}, \bibinfo {author} {\bibfnamefont
  {Z.}~\bibnamefont {Tian}}, \bibinfo {author} {\bibfnamefont {X.}~\bibnamefont
  {Hao}}, \bibinfo {author} {\bibfnamefont {C.}~\bibnamefont {Kuang}}, \ and\
  \bibinfo {author} {\bibfnamefont {X.}~\bibnamefont {Liu}},\ }\href@noop {}
  {\bibfield  {journal} {\bibinfo  {journal} {Opt. Lasers Eng.}\ }\textbf
  {\bibinfo {volume} {152}},\ \bibinfo {pages} {106960} (\bibinfo {year}
  {2022})}\BibitemShut {NoStop}%
\bibitem [{\citenamefont {Aloisio}\ \emph {et~al.}(2015)\citenamefont
  {Aloisio}, \citenamefont {Paganin}, \citenamefont {Wright},\ and\
  \citenamefont {Morgan}}]{A15}%
  \BibitemOpen
  \bibfield  {author} {\bibinfo {author} {\bibfnamefont {I.~A.}\ \bibnamefont
  {Aloisio}}, \bibinfo {author} {\bibfnamefont {D.~M.}\ \bibnamefont
  {Paganin}}, \bibinfo {author} {\bibfnamefont {C.~A.}\ \bibnamefont {Wright}},
  \ and\ \bibinfo {author} {\bibfnamefont {K.~S.}\ \bibnamefont {Morgan}},\
  }\href {\doibase 10.1107/S1600577515011406} {\bibfield  {journal} {\bibinfo
  {journal} {Journal of Synchrotron Radiation}\ }\textbf {\bibinfo {volume}
  {22}},\ \bibinfo {pages} {1279} (\bibinfo {year} {2015})}\BibitemShut
  {NoStop}%
\bibitem [{\citenamefont {Zanette}\ \emph {et~al.}(2014)\citenamefont
  {Zanette}, \citenamefont {Zhou}, \citenamefont {Burvall}, \citenamefont
  {Lundstr{\"o}m}, \citenamefont {Larsson}, \citenamefont {Zdora},
  \citenamefont {Thibault}, \citenamefont {Pfeiffer},\ and\ \citenamefont
  {Hertz}}]{A19}%
  \BibitemOpen
  \bibfield  {author} {\bibinfo {author} {\bibfnamefont {I.}~\bibnamefont
  {Zanette}}, \bibinfo {author} {\bibfnamefont {T.}~\bibnamefont {Zhou}},
  \bibinfo {author} {\bibfnamefont {A.}~\bibnamefont {Burvall}}, \bibinfo
  {author} {\bibfnamefont {U.}~\bibnamefont {Lundstr{\"o}m}}, \bibinfo {author}
  {\bibfnamefont {D.~H.}\ \bibnamefont {Larsson}}, \bibinfo {author}
  {\bibfnamefont {M.}~\bibnamefont {Zdora}}, \bibinfo {author} {\bibfnamefont
  {P.}~\bibnamefont {Thibault}}, \bibinfo {author} {\bibfnamefont
  {F.}~\bibnamefont {Pfeiffer}}, \ and\ \bibinfo {author} {\bibfnamefont
  {H.~M.}\ \bibnamefont {Hertz}},\ }\href@noop {} {\bibfield  {journal}
  {\bibinfo  {journal} {Phys. Rev. Lett.}\ }\textbf {\bibinfo {volume} {112}},\
  \bibinfo {pages} {253903} (\bibinfo {year} {2014})}\BibitemShut {NoStop}%
\bibitem [{\citenamefont {Wang}\ \emph {et~al.}(2017)\citenamefont {Wang},
  \citenamefont {Wang}, \citenamefont {Wei}, \citenamefont {Du}, \citenamefont
  {Xue}, \citenamefont {Hu}, \citenamefont {Li}, \citenamefont {Deng},
  \citenamefont {Xie},\ and\ \citenamefont {Xiao}}]{A22}%
  \BibitemOpen
  \bibfield  {author} {\bibinfo {author} {\bibfnamefont {F.}~\bibnamefont
  {Wang}}, \bibinfo {author} {\bibfnamefont {Y.}~\bibnamefont {Wang}}, \bibinfo
  {author} {\bibfnamefont {G.}~\bibnamefont {Wei}}, \bibinfo {author}
  {\bibfnamefont {G.}~\bibnamefont {Du}}, \bibinfo {author} {\bibfnamefont
  {Y.}~\bibnamefont {Xue}}, \bibinfo {author} {\bibfnamefont {T.}~\bibnamefont
  {Hu}}, \bibinfo {author} {\bibfnamefont {K.}~\bibnamefont {Li}}, \bibinfo
  {author} {\bibfnamefont {B.}~\bibnamefont {Deng}}, \bibinfo {author}
  {\bibfnamefont {H.}~\bibnamefont {Xie}}, \ and\ \bibinfo {author}
  {\bibfnamefont {T.}~\bibnamefont {Xiao}},\ }\href@noop {} {\bibfield
  {journal} {\bibinfo  {journal} {Appl. Phys. Lett.}\ }\textbf {\bibinfo
  {volume} {111}},\ \bibinfo {pages} {174101} (\bibinfo {year}
  {2017})}\BibitemShut {NoStop}%
\bibitem [{\citenamefont {Qiao}\ \emph {et~al.}(2022)\citenamefont {Qiao},
  \citenamefont {Shi}, \citenamefont {Yao}, \citenamefont {Wojcik},
  \citenamefont {Rebuffi}, \citenamefont {Cherukara},\ and\ \citenamefont
  {Assoufid}}]{Qiao:2022}%
  \BibitemOpen
  \bibfield  {author} {\bibinfo {author} {\bibfnamefont {Z.}~\bibnamefont
  {Qiao}}, \bibinfo {author} {\bibfnamefont {X.}~\bibnamefont {Shi}}, \bibinfo
  {author} {\bibfnamefont {Y.}~\bibnamefont {Yao}}, \bibinfo {author}
  {\bibfnamefont {M.~J.}\ \bibnamefont {Wojcik}}, \bibinfo {author}
  {\bibfnamefont {L.}~\bibnamefont {Rebuffi}}, \bibinfo {author} {\bibfnamefont
  {M.~J.}\ \bibnamefont {Cherukara}}, \ and\ \bibinfo {author} {\bibfnamefont
  {L.}~\bibnamefont {Assoufid}},\ }\href {\doibase 10.1364/optica.453748}
  {\bibfield  {journal} {\bibinfo  {journal} {Optica}\ }\textbf {\bibinfo
  {volume} {9}},\ \bibinfo {pages} {391} (\bibinfo {year} {2022})},\ \Eprint
  {http://arxiv.org/abs/2201.07232} {arXiv:2201.07232} \BibitemShut {NoStop}%
\bibitem [{\citenamefont {Morgan}\ \emph {et~al.}(2020)\citenamefont {Morgan},
  \citenamefont {Parsons}, \citenamefont {Cmielewski}, \citenamefont
  {McCarron}, \citenamefont {Gradl}, \citenamefont {Farrow}, \citenamefont
  {Siu}, \citenamefont {Takeuchi}, \citenamefont {Suzuki}, \citenamefont
  {Uesugi}, \citenamefont {Uesugi}, \citenamefont {Yagi}, \citenamefont {Hall},
  \citenamefont {Klein}, \citenamefont {Maksimenko}, \citenamefont {Stevenson},
  \citenamefont {Hausermann}, \citenamefont {Dierolf}, \citenamefont
  {Pfeifferb},\ and\ \citenamefont {Donnelleyd}}]{Morgan:2020}%
  \BibitemOpen
  \bibfield  {author} {\bibinfo {author} {\bibfnamefont {K.~S.}\ \bibnamefont
  {Morgan}}, \bibinfo {author} {\bibfnamefont {D.}~\bibnamefont {Parsons}},
  \bibinfo {author} {\bibfnamefont {P.}~\bibnamefont {Cmielewski}}, \bibinfo
  {author} {\bibfnamefont {A.}~\bibnamefont {McCarron}}, \bibinfo {author}
  {\bibfnamefont {R.}~\bibnamefont {Gradl}}, \bibinfo {author} {\bibfnamefont
  {N.}~\bibnamefont {Farrow}}, \bibinfo {author} {\bibfnamefont
  {K.}~\bibnamefont {Siu}}, \bibinfo {author} {\bibfnamefont {A.}~\bibnamefont
  {Takeuchi}}, \bibinfo {author} {\bibfnamefont {Y.}~\bibnamefont {Suzuki}},
  \bibinfo {author} {\bibfnamefont {K.}~\bibnamefont {Uesugi}}, \bibinfo
  {author} {\bibfnamefont {M.}~\bibnamefont {Uesugi}}, \bibinfo {author}
  {\bibfnamefont {N.}~\bibnamefont {Yagi}}, \bibinfo {author} {\bibfnamefont
  {C.}~\bibnamefont {Hall}}, \bibinfo {author} {\bibfnamefont {M.}~\bibnamefont
  {Klein}}, \bibinfo {author} {\bibfnamefont {A.}~\bibnamefont {Maksimenko}},
  \bibinfo {author} {\bibfnamefont {A.}~\bibnamefont {Stevenson}}, \bibinfo
  {author} {\bibfnamefont {D.}~\bibnamefont {Hausermann}}, \bibinfo {author}
  {\bibfnamefont {M.}~\bibnamefont {Dierolf}}, \bibinfo {author} {\bibfnamefont
  {F.}~\bibnamefont {Pfeifferb}}, \ and\ \bibinfo {author} {\bibfnamefont
  {M.}~\bibnamefont {Donnelleyd}},\ }\href {\doibase 10.1107/S1600577519014863}
  {\bibfield  {journal} {\bibinfo  {journal} {Journal of Synchrotron
  Radiation}\ }\textbf {\bibinfo {volume} {27}},\ \bibinfo {pages} {164}
  (\bibinfo {year} {2020})}\BibitemShut {NoStop}%
\bibitem [{\citenamefont {Chen}\ \emph {et~al.}(2014)\citenamefont {Chen},
  \citenamefont {Hudspeth}, \citenamefont {Claus}, \citenamefont {Parab},
  \citenamefont {Black}, \citenamefont {Fezzaa},\ and\ \citenamefont
  {Luo}}]{Chen:2014}%
  \BibitemOpen
  \bibfield  {author} {\bibinfo {author} {\bibfnamefont {W.~W.}\ \bibnamefont
  {Chen}}, \bibinfo {author} {\bibfnamefont {M.~C.}\ \bibnamefont {Hudspeth}},
  \bibinfo {author} {\bibfnamefont {B.}~\bibnamefont {Claus}}, \bibinfo
  {author} {\bibfnamefont {N.~D.}\ \bibnamefont {Parab}}, \bibinfo {author}
  {\bibfnamefont {J.~T.}\ \bibnamefont {Black}}, \bibinfo {author}
  {\bibfnamefont {K.}~\bibnamefont {Fezzaa}}, \ and\ \bibinfo {author}
  {\bibfnamefont {S.~N.}\ \bibnamefont {Luo}},\ }\href {\doibase
  10.1098/rsta.2013.0191} {\bibfield  {journal} {\bibinfo  {journal}
  {Philosophical transactions. Series A, Mathematical, physical, and
  engineering sciences}\ }\textbf {\bibinfo {volume} {372}},\ \bibinfo {pages}
  {20130191} (\bibinfo {year} {2014})}\BibitemShut {NoStop}%
\bibitem [{\citenamefont {Parab}\ \emph {et~al.}(2016)\citenamefont {Parab},
  \citenamefont {Roberts}, \citenamefont {Harr}, \citenamefont {Mares},
  \citenamefont {Casey}, \citenamefont {Gunduz}, \citenamefont {Hudspeth},
  \citenamefont {Claus}, \citenamefont {Sun}, \citenamefont {Fezzaa},
  \citenamefont {Son},\ and\ \citenamefont {Chen}}]{Parab:2016}%
  \BibitemOpen
  \bibfield  {author} {\bibinfo {author} {\bibfnamefont {N.~D.}\ \bibnamefont
  {Parab}}, \bibinfo {author} {\bibfnamefont {Z.~A.}\ \bibnamefont {Roberts}},
  \bibinfo {author} {\bibfnamefont {M.~H.}\ \bibnamefont {Harr}}, \bibinfo
  {author} {\bibfnamefont {J.~O.}\ \bibnamefont {Mares}}, \bibinfo {author}
  {\bibfnamefont {A.~D.}\ \bibnamefont {Casey}}, \bibinfo {author}
  {\bibfnamefont {I.~E.}\ \bibnamefont {Gunduz}}, \bibinfo {author}
  {\bibfnamefont {M.}~\bibnamefont {Hudspeth}}, \bibinfo {author}
  {\bibfnamefont {B.}~\bibnamefont {Claus}}, \bibinfo {author} {\bibfnamefont
  {T.}~\bibnamefont {Sun}}, \bibinfo {author} {\bibfnamefont {K.}~\bibnamefont
  {Fezzaa}}, \bibinfo {author} {\bibfnamefont {S.~F.}\ \bibnamefont {Son}}, \
  and\ \bibinfo {author} {\bibfnamefont {W.~W.}\ \bibnamefont {Chen}},\ }\href
  {\doibase 10.1063/1.4963137} {\bibfield  {journal} {\bibinfo  {journal}
  {Applied Physics Letters}\ }\textbf {\bibinfo {volume} {109}},\ \bibinfo
  {pages} {131903} (\bibinfo {year} {2016})}\BibitemShut {NoStop}%
\bibitem [{\citenamefont {Leong}\ \emph {et~al.}(2018)\citenamefont {Leong},
  \citenamefont {Robinson}, \citenamefont {Fezzaa}, \citenamefont {Sun},
  \citenamefont {Sinclair}, \citenamefont {Casem}, \citenamefont {Lambert},
  \citenamefont {Hustedt}, \citenamefont {Daphalapurkar}, \citenamefont
  {Ramesh},\ and\ \citenamefont {Hufnagel}}]{Leong:2018}%
  \BibitemOpen
  \bibfield  {author} {\bibinfo {author} {\bibfnamefont {A.~F.~T.}\
  \bibnamefont {Leong}}, \bibinfo {author} {\bibfnamefont {A.~K.}\ \bibnamefont
  {Robinson}}, \bibinfo {author} {\bibfnamefont {K.}~\bibnamefont {Fezzaa}},
  \bibinfo {author} {\bibfnamefont {T.}~\bibnamefont {Sun}}, \bibinfo {author}
  {\bibfnamefont {N.}~\bibnamefont {Sinclair}}, \bibinfo {author}
  {\bibfnamefont {D.~T.}\ \bibnamefont {Casem}}, \bibinfo {author}
  {\bibfnamefont {P.~K.}\ \bibnamefont {Lambert}}, \bibinfo {author}
  {\bibfnamefont {C.~J.}\ \bibnamefont {Hustedt}}, \bibinfo {author}
  {\bibfnamefont {N.~P.}\ \bibnamefont {Daphalapurkar}}, \bibinfo {author}
  {\bibfnamefont {K.~T.}\ \bibnamefont {Ramesh}}, \ and\ \bibinfo {author}
  {\bibfnamefont {T.~C.}\ \bibnamefont {Hufnagel}},\ }\href {\doibase
  10.1007/s11340-018-0414-3} {\bibfield  {journal} {\bibinfo  {journal}
  {Experimental Mechanics}\ }\textbf {\bibinfo {volume} {58}},\ \bibinfo
  {pages} {1423} (\bibinfo {year} {2018})}\BibitemShut {NoStop}%
\bibitem [{\citenamefont {Wainwright}\ \emph {et~al.}(2019)\citenamefont
  {Wainwright}, \citenamefont {Lakshman}, \citenamefont {Leong}, \citenamefont
  {Kinsey}, \citenamefont {Gibbins}, \citenamefont {Arlington}, \citenamefont
  {Sun}, \citenamefont {Fezzaa}, \citenamefont {Hufnagel},\ and\ \citenamefont
  {Weihs}}]{Wainwright:2019}%
  \BibitemOpen
  \bibfield  {author} {\bibinfo {author} {\bibfnamefont {E.~R.}\ \bibnamefont
  {Wainwright}}, \bibinfo {author} {\bibfnamefont {S.~V.}\ \bibnamefont
  {Lakshman}}, \bibinfo {author} {\bibfnamefont {A.~F.}\ \bibnamefont {Leong}},
  \bibinfo {author} {\bibfnamefont {A.~H.}\ \bibnamefont {Kinsey}}, \bibinfo
  {author} {\bibfnamefont {J.~D.}\ \bibnamefont {Gibbins}}, \bibinfo {author}
  {\bibfnamefont {S.~Q.}\ \bibnamefont {Arlington}}, \bibinfo {author}
  {\bibfnamefont {T.}~\bibnamefont {Sun}}, \bibinfo {author} {\bibfnamefont
  {K.}~\bibnamefont {Fezzaa}}, \bibinfo {author} {\bibfnamefont {T.~C.}\
  \bibnamefont {Hufnagel}}, \ and\ \bibinfo {author} {\bibfnamefont {T.~P.}\
  \bibnamefont {Weihs}},\ }\href {\doibase 10.1016/j.combustflame.2018.10.019}
  {\bibfield  {journal} {\bibinfo  {journal} {Combustion and Flame}\ }\textbf
  {\bibinfo {volume} {199}},\ \bibinfo {pages} {194} (\bibinfo {year}
  {2019})}\BibitemShut {NoStop}%
\bibitem [{\citenamefont {Sforzo}\ \emph {et~al.}(2019)\citenamefont {Sforzo},
  \citenamefont {Kastengren}, \citenamefont {Matusik}, \citenamefont {{Gomez
  Del Campo}},\ and\ \citenamefont {Powell}}]{Sforzo:2019}%
  \BibitemOpen
  \bibfield  {author} {\bibinfo {author} {\bibfnamefont {B.~A.}\ \bibnamefont
  {Sforzo}}, \bibinfo {author} {\bibfnamefont {A.~L.}\ \bibnamefont
  {Kastengren}}, \bibinfo {author} {\bibfnamefont {K.~E.}\ \bibnamefont
  {Matusik}}, \bibinfo {author} {\bibfnamefont {F.}~\bibnamefont {{Gomez Del
  Campo}}}, \ and\ \bibinfo {author} {\bibfnamefont {C.~F.}\ \bibnamefont
  {Powell}},\ }\href {\doibase 10.1115/1.4045217} {\bibfield  {journal}
  {\bibinfo  {journal} {Journal of Engineering for Gas Turbines and Power}\
  }\textbf {\bibinfo {volume} {141}} (\bibinfo {year} {2019}),\
  10.1115/1.4045217}\BibitemShut {NoStop}%
\bibitem [{\citenamefont {Zhao}\ \emph {et~al.}(2021)\citenamefont {Zhao},
  \citenamefont {Li}, \citenamefont {Deng}, \citenamefont {Li},\ and\
  \citenamefont {Wu}}]{Zhao:2021}%
  \BibitemOpen
  \bibfield  {author} {\bibinfo {author} {\bibfnamefont {W.}~\bibnamefont
  {Zhao}}, \bibinfo {author} {\bibfnamefont {Z.}~\bibnamefont {Li}}, \bibinfo
  {author} {\bibfnamefont {J.}~\bibnamefont {Deng}}, \bibinfo {author}
  {\bibfnamefont {L.}~\bibnamefont {Li}}, \ and\ \bibinfo {author}
  {\bibfnamefont {Z.}~\bibnamefont {Wu}},\ }\href {\doibase
  10.1615/ATOMIZSPR.2021036903} {\bibfield  {journal} {\bibinfo  {journal}
  {Atomization and Sprays}\ }\textbf {\bibinfo {volume} {31}},\ \bibinfo
  {pages} {67} (\bibinfo {year} {2021})}\BibitemShut {NoStop}%
\bibitem [{\citenamefont {Montgomery}(2016)}]{Montgomery:2016}%
  \BibitemOpen
  \bibfield  {author} {\bibinfo {author} {\bibfnamefont {D.~S.}\ \bibnamefont
  {Montgomery}},\ }\href {\doibase 10.1063/1.4946016} {\bibfield  {journal}
  {\bibinfo  {journal} {Physics of Plasmas}\ }\textbf {\bibinfo {volume}
  {23}},\ \bibinfo {pages} {055601} (\bibinfo {year} {2016})}\BibitemShut
  {NoStop}%
\bibitem [{\citenamefont {Hodge}\ \emph {et~al.}(2021)\citenamefont {Hodge},
  \citenamefont {Pandolfi}, \citenamefont {Liu}, \citenamefont {Li},
  \citenamefont {Sakdinawat}, \citenamefont {Seaberg}, \citenamefont {Hart},
  \citenamefont {Galtier}, \citenamefont {Khaghani}, \citenamefont {Nagler},
  \citenamefont {Lee}, \citenamefont {Bolme}, \citenamefont {Ramos},
  \citenamefont {Kozlowski}, \citenamefont {Montgomery}, \citenamefont
  {Dayton}, \citenamefont {Dresselhaus-Marias}, \citenamefont {Curry},
  \citenamefont {Carver}, \citenamefont {Ali}, \citenamefont {Sandberg},\ and\
  \citenamefont {Gleason}}]{Hodge:2021}%
  \BibitemOpen
  \bibfield  {author} {\bibinfo {author} {\bibfnamefont {D.}~\bibnamefont
  {Hodge}}, \bibinfo {author} {\bibfnamefont {S.}~\bibnamefont {Pandolfi}},
  \bibinfo {author} {\bibfnamefont {Y.}~\bibnamefont {Liu}}, \bibinfo {author}
  {\bibfnamefont {K.}~\bibnamefont {Li}}, \bibinfo {author} {\bibfnamefont
  {A.}~\bibnamefont {Sakdinawat}}, \bibinfo {author} {\bibfnamefont
  {M.}~\bibnamefont {Seaberg}}, \bibinfo {author} {\bibfnamefont
  {P.}~\bibnamefont {Hart}}, \bibinfo {author} {\bibfnamefont {E.}~\bibnamefont
  {Galtier}}, \bibinfo {author} {\bibfnamefont {D.}~\bibnamefont {Khaghani}},
  \bibinfo {author} {\bibfnamefont {B.}~\bibnamefont {Nagler}}, \bibinfo
  {author} {\bibfnamefont {H.~J.}\ \bibnamefont {Lee}}, \bibinfo {author}
  {\bibfnamefont {C.}~\bibnamefont {Bolme}}, \bibinfo {author} {\bibfnamefont
  {K.}~\bibnamefont {Ramos}}, \bibinfo {author} {\bibfnamefont
  {P.}~\bibnamefont {Kozlowski}}, \bibinfo {author} {\bibfnamefont
  {D.}~\bibnamefont {Montgomery}}, \bibinfo {author} {\bibfnamefont
  {M.}~\bibnamefont {Dayton}}, \bibinfo {author} {\bibfnamefont
  {L.}~\bibnamefont {Dresselhaus-Marias}}, \bibinfo {author} {\bibfnamefont
  {C.}~\bibnamefont {Curry}}, \bibinfo {author} {\bibfnamefont
  {T.}~\bibnamefont {Carver}}, \bibinfo {author} {\bibfnamefont
  {S.}~\bibnamefont {Ali}}, \bibinfo {author} {\bibfnamefont {R.}~\bibnamefont
  {Sandberg}}, \ and\ \bibinfo {author} {\bibfnamefont {A.}~\bibnamefont
  {Gleason}},\ }\href@noop {} {\bibfield  {journal} {\bibinfo  {journal} {SPIE
  Proc.}\ }\textbf {\bibinfo {volume} {11839}},\ \bibinfo {pages} {1183908}
  (\bibinfo {year} {2021})}\BibitemShut {NoStop}%
\bibitem [{\citenamefont {Hodge}\ \emph {et~al.}(2022)\citenamefont {Hodge},
  \citenamefont {Leong}, \citenamefont {Pandolfi}, \citenamefont {Kurzer-Ogul},
  \citenamefont {Montgomery}, \citenamefont {Aluie}, \citenamefont {Bolme},
  \citenamefont {Carver}, \citenamefont {Cunningham}, \citenamefont {Curry},
  \citenamefont {Dayton}, \citenamefont {Decker}, \citenamefont {Galtier},
  \citenamefont {Hart}, \citenamefont {Khaghani}, \citenamefont {{Ja Lee}},
  \citenamefont {Li}, \citenamefont {Liu}, \citenamefont {Ramos}, \citenamefont
  {Shang}, \citenamefont {Vetter}, \citenamefont {Nagler}, \citenamefont
  {Sandberg},\ and\ \citenamefont {Gleason}}]{Hodge:2022}%
  \BibitemOpen
  \bibfield  {author} {\bibinfo {author} {\bibfnamefont {D.~S.}\ \bibnamefont
  {Hodge}}, \bibinfo {author} {\bibfnamefont {A.~F.~T.}\ \bibnamefont {Leong}},
  \bibinfo {author} {\bibfnamefont {S.}~\bibnamefont {Pandolfi}}, \bibinfo
  {author} {\bibfnamefont {K.}~\bibnamefont {Kurzer-Ogul}}, \bibinfo {author}
  {\bibfnamefont {D.~S.}\ \bibnamefont {Montgomery}}, \bibinfo {author}
  {\bibfnamefont {H.}~\bibnamefont {Aluie}}, \bibinfo {author} {\bibfnamefont
  {C.}~\bibnamefont {Bolme}}, \bibinfo {author} {\bibfnamefont
  {T.}~\bibnamefont {Carver}}, \bibinfo {author} {\bibfnamefont
  {E.}~\bibnamefont {Cunningham}}, \bibinfo {author} {\bibfnamefont {C.~B.}\
  \bibnamefont {Curry}}, \bibinfo {author} {\bibfnamefont {M.}~\bibnamefont
  {Dayton}}, \bibinfo {author} {\bibfnamefont {F.-J.}\ \bibnamefont {Decker}},
  \bibinfo {author} {\bibfnamefont {E.}~\bibnamefont {Galtier}}, \bibinfo
  {author} {\bibfnamefont {P.}~\bibnamefont {Hart}}, \bibinfo {author}
  {\bibfnamefont {D.}~\bibnamefont {Khaghani}}, \bibinfo {author}
  {\bibfnamefont {H.}~\bibnamefont {{Ja Lee}}}, \bibinfo {author}
  {\bibfnamefont {K.}~\bibnamefont {Li}}, \bibinfo {author} {\bibfnamefont
  {Y.}~\bibnamefont {Liu}}, \bibinfo {author} {\bibfnamefont {K.}~\bibnamefont
  {Ramos}}, \bibinfo {author} {\bibfnamefont {J.}~\bibnamefont {Shang}},
  \bibinfo {author} {\bibfnamefont {S.}~\bibnamefont {Vetter}}, \bibinfo
  {author} {\bibfnamefont {B.}~\bibnamefont {Nagler}}, \bibinfo {author}
  {\bibfnamefont {R.~L.}\ \bibnamefont {Sandberg}}, \ and\ \bibinfo {author}
  {\bibfnamefont {A.~E.}\ \bibnamefont {Gleason}},\ }\href {\doibase
  10.1364/oe.472275} {\bibfield  {journal} {\bibinfo  {journal} {Optics
  Express}\ }\textbf {\bibinfo {volume} {30}},\ \bibinfo {pages} {38405}
  (\bibinfo {year} {2022})}\BibitemShut {NoStop}%
\bibitem [{\citenamefont {Montgomery}(2023)}]{montgomery2023}%
  \BibitemOpen
  \bibfield  {author} {\bibinfo {author} {\bibfnamefont {D.~S.}\ \bibnamefont
  {Montgomery}},\ }\href {\doibase 10.1063/5.0127497} {\bibfield  {journal}
  {\bibinfo  {journal} {Rev. Sci. Instrum.}\ }\textbf {\bibinfo {volume}
  {94}},\ \bibinfo {pages} {021103} (\bibinfo {year} {2023})}\BibitemShut
  {NoStop}%
\bibitem [{\citenamefont {Branch}\ \emph {et~al.}(2017)\citenamefont {Branch},
  \citenamefont {Ionita}, \citenamefont {Clements}, \citenamefont {Montgomery},
  \citenamefont {Jensen}, \citenamefont {Patterson}, \citenamefont
  {Schmalzer},\ and\ \citenamefont {Dattelbaum}}]{BIC:2017}%
  \BibitemOpen
  \bibfield  {author} {\bibinfo {author} {\bibfnamefont {B.}~\bibnamefont
  {Branch}}, \bibinfo {author} {\bibfnamefont {A.}~\bibnamefont {Ionita}},
  \bibinfo {author} {\bibfnamefont {B.~E.}\ \bibnamefont {Clements}}, \bibinfo
  {author} {\bibfnamefont {D.~S.}\ \bibnamefont {Montgomery}}, \bibinfo
  {author} {\bibfnamefont {B.~J.}\ \bibnamefont {Jensen}}, \bibinfo {author}
  {\bibfnamefont {B.}~\bibnamefont {Patterson}}, \bibinfo {author}
  {\bibfnamefont {A.}~\bibnamefont {Schmalzer}}, \ and\ \bibinfo {author}
  {\bibfnamefont {A.~M. D.~M.}\ \bibnamefont {Dattelbaum}},\ }\href@noop {}
  {\bibfield  {journal} {\bibinfo  {journal} {J. Appl. Phys.}\ }\textbf
  {\bibinfo {volume} {121}},\ \bibinfo {pages} {135102} (\bibinfo {year}
  {2017})}\BibitemShut {NoStop}%
\bibitem [{\citenamefont {Yanuka}\ \emph {et~al.}(2019)\citenamefont {Yanuka},
  \citenamefont {Theocharous}, \citenamefont {Efimov}, \citenamefont {Bland},
  \citenamefont {Rososhek}, \citenamefont {Krasik}, \citenamefont {Olbinado},\
  and\ \citenamefont {Rack}}]{Yanuka:2019}%
  \BibitemOpen
  \bibfield  {author} {\bibinfo {author} {\bibfnamefont {D.}~\bibnamefont
  {Yanuka}}, \bibinfo {author} {\bibfnamefont {S.}~\bibnamefont {Theocharous}},
  \bibinfo {author} {\bibfnamefont {S.}~\bibnamefont {Efimov}}, \bibinfo
  {author} {\bibfnamefont {S.~N.}\ \bibnamefont {Bland}}, \bibinfo {author}
  {\bibfnamefont {A.}~\bibnamefont {Rososhek}}, \bibinfo {author}
  {\bibfnamefont {Y.~E.}\ \bibnamefont {Krasik}}, \bibinfo {author}
  {\bibfnamefont {M.~P.}\ \bibnamefont {Olbinado}}, \ and\ \bibinfo {author}
  {\bibfnamefont {A.}~\bibnamefont {Rack}},\ }\href {\doibase
  10.1063/1.5089011} {\bibfield  {journal} {\bibinfo  {journal} {Journal of
  Applied Physics}\ }\textbf {\bibinfo {volume} {125}},\ \bibinfo {pages}
  {093301} (\bibinfo {year} {2019})}\BibitemShut {NoStop}%
\bibitem [{\citenamefont {Dattelbaum}\ \emph
  {et~al.}(2020{\natexlab{a}})\citenamefont {Dattelbaum}, \citenamefont
  {Branch}, \citenamefont {Ionita}, \citenamefont {Patterson}, \citenamefont
  {Kuettner},\ and\ \citenamefont {Herman}}]{D3}%
  \BibitemOpen
  \bibfield  {author} {\bibinfo {author} {\bibfnamefont {D.~M.}\ \bibnamefont
  {Dattelbaum}}, \bibinfo {author} {\bibfnamefont {B.~A.}\ \bibnamefont
  {Branch}}, \bibinfo {author} {\bibfnamefont {A.}~\bibnamefont {Ionita}},
  \bibinfo {author} {\bibfnamefont {B.~M.}\ \bibnamefont {Patterson}}, \bibinfo
  {author} {\bibfnamefont {L.}~\bibnamefont {Kuettner}}, \ and\ \bibinfo
  {author} {\bibfnamefont {M.}~\bibnamefont {Herman}},\ }\href@noop {}
  {\bibfield  {journal} {\bibinfo  {journal} {AIP Conf. Proc.}\ }\textbf
  {\bibinfo {volume} {2272}},\ \bibinfo {pages} {040002} (\bibinfo {year}
  {2020}{\natexlab{a}})}\BibitemShut {NoStop}%
\bibitem [{\citenamefont {Black}\ and\ \citenamefont {Long}(2004)}]{BlackNIST}%
  \BibitemOpen
  \bibfield  {author} {\bibinfo {author} {\bibfnamefont {D.~R.}\ \bibnamefont
  {Black}}\ and\ \bibinfo {author} {\bibfnamefont {G.~G.}\ \bibnamefont
  {Long}},\ }\href
  {https://tsapps.nist.gov/publication/get\_pdf.cfm?pub\_id=901559} {\emph
  {\bibinfo {title} {{X-Ray Topography}}}},\ \bibinfo {type} {Tech. Rep.}\
  (\bibinfo  {institution} {National Institute of Standards and Technology,
  U.S. Dept. Comm.},\ \bibinfo {year} {2004})\BibitemShut {NoStop}%
\bibitem [{\citenamefont {Danilewsky}(2020)}]{DanilewskyCRT:20}%
  \BibitemOpen
  \bibfield  {author} {\bibinfo {author} {\bibfnamefont {A.~N.}\ \bibnamefont
  {Danilewsky}},\ }\href {\doibase 10.1002/crat.202000012} {\bibfield
  {journal} {\bibinfo  {journal} {Crystal Research and Technology}\ }\textbf
  {\bibinfo {volume} {55}},\ \bibinfo {pages} {2000012} (\bibinfo {year}
  {2020})}\BibitemShut {NoStop}%
\bibitem [{\citenamefont {Chikawa}\ and\ \citenamefont
  {Fujimoto}(1968)}]{ChikawaAPL:68}%
  \BibitemOpen
  \bibfield  {author} {\bibinfo {author} {\bibfnamefont {J.-I.}\ \bibnamefont
  {Chikawa}}\ and\ \bibinfo {author} {\bibfnamefont {I.}~\bibnamefont
  {Fujimoto}},\ }\href {\doibase 10.1063/1.1652483} {\bibfield  {journal}
  {\bibinfo  {journal} {Applied Physics Letters}\ }\textbf {\bibinfo {volume}
  {13}},\ \bibinfo {pages} {387} (\bibinfo {year} {1968})}\BibitemShut
  {NoStop}%
\bibitem [{\citenamefont {Tuomi}\ \emph {et~al.}(1983)\citenamefont {Tuomi},
  \citenamefont {Kelhä},\ and\ \citenamefont {Blomberg}}]{TuomiNIM:83}%
  \BibitemOpen
  \bibfield  {author} {\bibinfo {author} {\bibfnamefont {T.}~\bibnamefont
  {Tuomi}}, \bibinfo {author} {\bibfnamefont {V.}~\bibnamefont {Kelhä}}, \
  and\ \bibinfo {author} {\bibfnamefont {M.}~\bibnamefont {Blomberg}},\ }\href
  {\doibase 10.1016/0167-5087(83)91206-1} {\bibfield  {journal} {\bibinfo
  {journal} {Nuclear Instruments and Methods in Physics Research}\ }\textbf
  {\bibinfo {volume} {208}},\ \bibinfo {pages} {697} (\bibinfo {year}
  {1983})}\BibitemShut {NoStop}%
\bibitem [{\citenamefont {Rack}\ \emph
  {et~al.}(2010{\natexlab{a}})\citenamefont {Rack}, \citenamefont
  {Garcia-Moreno}, \citenamefont {Schmitt}, \citenamefont {Betz}, \citenamefont
  {Cecilia}, \citenamefont {Ershov}, \citenamefont {Rack}, \citenamefont
  {Banhart},\ and\ \citenamefont {Zabler}}]{RackJXST:10}%
  \BibitemOpen
  \bibfield  {author} {\bibinfo {author} {\bibfnamefont {A.}~\bibnamefont
  {Rack}}, \bibinfo {author} {\bibfnamefont {F.}~\bibnamefont {Garcia-Moreno}},
  \bibinfo {author} {\bibfnamefont {C.}~\bibnamefont {Schmitt}}, \bibinfo
  {author} {\bibfnamefont {O.}~\bibnamefont {Betz}}, \bibinfo {author}
  {\bibfnamefont {A.}~\bibnamefont {Cecilia}}, \bibinfo {author} {\bibfnamefont
  {A.}~\bibnamefont {Ershov}}, \bibinfo {author} {\bibfnamefont
  {T.}~\bibnamefont {Rack}}, \bibinfo {author} {\bibfnamefont {J.}~\bibnamefont
  {Banhart}}, \ and\ \bibinfo {author} {\bibfnamefont {S.}~\bibnamefont
  {Zabler}},\ }\href {\doibase 10.3233/xst-2010-0273} {\bibfield  {journal}
  {\bibinfo  {journal} {Journal of X-Ray Science and Technology}\ }\textbf
  {\bibinfo {volume} {18}},\ \bibinfo {pages} {429} (\bibinfo {year}
  {2010}{\natexlab{a}})}\BibitemShut {NoStop}%
\bibitem [{\citenamefont {Danilewsky}\ \emph {et~al.}(2011)\citenamefont
  {Danilewsky}, \citenamefont {Wittge}, \citenamefont {Croell}, \citenamefont
  {Allen}, \citenamefont {McNally}, \citenamefont {Vagovič}, \citenamefont
  {Rolo}, \citenamefont {Li}, \citenamefont {Baumbach}, \citenamefont
  {Gorostegui-Colinas}, \citenamefont {Garagorri}, \citenamefont {Elizalde},
  \citenamefont {Fossati}, \citenamefont {Bowen},\ and\ \citenamefont
  {Tanner}}]{DanilewskyJCG:11}%
  \BibitemOpen
  \bibfield  {author} {\bibinfo {author} {\bibfnamefont {A.~N.}\ \bibnamefont
  {Danilewsky}}, \bibinfo {author} {\bibfnamefont {J.}~\bibnamefont {Wittge}},
  \bibinfo {author} {\bibfnamefont {A.}~\bibnamefont {Croell}}, \bibinfo
  {author} {\bibfnamefont {D.}~\bibnamefont {Allen}}, \bibinfo {author}
  {\bibfnamefont {P.}~\bibnamefont {McNally}}, \bibinfo {author} {\bibfnamefont
  {P.}~\bibnamefont {Vagovič}}, \bibinfo {author} {\bibfnamefont {T.~d.~S.}\
  \bibnamefont {Rolo}}, \bibinfo {author} {\bibfnamefont {Z.}~\bibnamefont
  {Li}}, \bibinfo {author} {\bibfnamefont {T.}~\bibnamefont {Baumbach}},
  \bibinfo {author} {\bibfnamefont {E.}~\bibnamefont {Gorostegui-Colinas}},
  \bibinfo {author} {\bibfnamefont {J.}~\bibnamefont {Garagorri}}, \bibinfo
  {author} {\bibfnamefont {M.~R.}\ \bibnamefont {Elizalde}}, \bibinfo {author}
  {\bibfnamefont {M.~C.}\ \bibnamefont {Fossati}}, \bibinfo {author}
  {\bibfnamefont {D.~K.}\ \bibnamefont {Bowen}}, \ and\ \bibinfo {author}
  {\bibfnamefont {B.~K.}\ \bibnamefont {Tanner}},\ }\href {\doibase
  10.1016/j.jcrysgro.2010.10.199} {\bibfield  {journal} {\bibinfo  {journal}
  {Journal of Crystal Growth}\ }\textbf {\bibinfo {volume} {318}},\ \bibinfo
  {pages} {1157} (\bibinfo {year} {2011})}\BibitemShut {NoStop}%
\bibitem [{\citenamefont {Rack}\ \emph {et~al.}(2016)\citenamefont {Rack},
  \citenamefont {Scheel},\ and\ \citenamefont {Danilewsky}}]{RackIUCRJ:16}%
  \BibitemOpen
  \bibfield  {author} {\bibinfo {author} {\bibfnamefont {A.}~\bibnamefont
  {Rack}}, \bibinfo {author} {\bibfnamefont {M.}~\bibnamefont {Scheel}}, \ and\
  \bibinfo {author} {\bibfnamefont {A.}~\bibnamefont {Danilewsky}},\ }\href
  {\doibase 10.1107/s205225251502271x} {\bibfield  {journal} {\bibinfo
  {journal} {IUCrJ}\ }\textbf {\bibinfo {volume} {3}},\ \bibinfo {pages} {108}
  (\bibinfo {year} {2016})}\BibitemShut {NoStop}%
\bibitem [{\citenamefont {Sherman}(2006)}]{ShermanIJF:06}%
  \BibitemOpen
  \bibfield  {author} {\bibinfo {author} {\bibfnamefont {D.}~\bibnamefont
  {Sherman}},\ }\href {\doibase 10.1007/s10704-006-0048-9} {\bibfield
  {journal} {\bibinfo  {journal} {International Journal of Fracture}\ }\textbf
  {\bibinfo {volume} {140}},\ \bibinfo {pages} {125} (\bibinfo {year}
  {2006})}\BibitemShut {NoStop}%
\bibitem [{\citenamefont {Petit}\ \emph {et~al.}(2022)\citenamefont {Petit},
  \citenamefont {Pokam}, \citenamefont {Mazen}, \citenamefont {Tardif},
  \citenamefont {Landru}, \citenamefont {Kononchuk}, \citenamefont {Mohamed},
  \citenamefont {Olbinado}, \citenamefont {Rack},\ and\ \citenamefont
  {Rieutord}}]{PetitJAC:22}%
  \BibitemOpen
  \bibfield  {author} {\bibinfo {author} {\bibfnamefont {A.}~\bibnamefont
  {Petit}}, \bibinfo {author} {\bibfnamefont {S.}~\bibnamefont {Pokam}},
  \bibinfo {author} {\bibfnamefont {F.}~\bibnamefont {Mazen}}, \bibinfo
  {author} {\bibfnamefont {S.}~\bibnamefont {Tardif}}, \bibinfo {author}
  {\bibfnamefont {D.}~\bibnamefont {Landru}}, \bibinfo {author} {\bibfnamefont
  {O.}~\bibnamefont {Kononchuk}}, \bibinfo {author} {\bibfnamefont {N.~B.}\
  \bibnamefont {Mohamed}}, \bibinfo {author} {\bibfnamefont {M.~P.}\
  \bibnamefont {Olbinado}}, \bibinfo {author} {\bibfnamefont {A.}~\bibnamefont
  {Rack}}, \ and\ \bibinfo {author} {\bibfnamefont {F.}~\bibnamefont
  {Rieutord}},\ }\href {\doibase 10.1107/s1600576722006537} {\bibfield
  {journal} {\bibinfo  {journal} {Journal of Applied Crystallography}\ }\textbf
  {\bibinfo {volume} {55}},\ \bibinfo {pages} {911} (\bibinfo {year}
  {2022})}\BibitemShut {NoStop}%
\bibitem [{\citenamefont {Luo}\ \emph {et~al.}(2012)\citenamefont {Luo},
  \citenamefont {Jensen}, \citenamefont {Hooks}, \citenamefont {Fezzaa},
  \citenamefont {Ramos}, \citenamefont {Yeager}, \citenamefont {Kwiatkowski},
  \citenamefont {1},\ and\ \citenamefont {Shimada}}]{LuoRSI:12}%
  \BibitemOpen
  \bibfield  {author} {\bibinfo {author} {\bibfnamefont {S.~N.}\ \bibnamefont
  {Luo}}, \bibinfo {author} {\bibfnamefont {B.~J.}\ \bibnamefont {Jensen}},
  \bibinfo {author} {\bibfnamefont {D.~E.}\ \bibnamefont {Hooks}}, \bibinfo
  {author} {\bibfnamefont {K.}~\bibnamefont {Fezzaa}}, \bibinfo {author}
  {\bibfnamefont {K.~J.}\ \bibnamefont {Ramos}}, \bibinfo {author}
  {\bibfnamefont {J.~D.}\ \bibnamefont {Yeager}}, \bibinfo {author}
  {\bibfnamefont {K.}~\bibnamefont {Kwiatkowski}}, \bibinfo {author}
  {\bibnamefont {1}}, \ and\ \bibinfo {author} {\bibfnamefont {T.}~\bibnamefont
  {Shimada}},\ }\href {\doibase 10.1063/1.4733704} {\bibfield  {journal}
  {\bibinfo  {journal} {Review of Scientific Instruments}\ }\textbf {\bibinfo
  {volume} {83}},\ \bibinfo {pages} {073903} (\bibinfo {year} {2012})},\
  \Eprint {http://arxiv.org/abs/1204.6071} {1204.6071} \BibitemShut {NoStop}%
\bibitem [{\citenamefont {Simons}\ \emph {et~al.}(2015)\citenamefont {Simons},
  \citenamefont {King}, \citenamefont {Ludwig}, \citenamefont {Detlefs},
  \citenamefont {Pantleon}, \citenamefont {Schmidt}, \citenamefont {Stöhr},
  \citenamefont {Snigireva}, \citenamefont {Snigirev},\ and\ \citenamefont
  {Poulsen}}]{SimonsNC:15}%
  \BibitemOpen
  \bibfield  {author} {\bibinfo {author} {\bibfnamefont {H.}~\bibnamefont
  {Simons}}, \bibinfo {author} {\bibfnamefont {A.}~\bibnamefont {King}},
  \bibinfo {author} {\bibfnamefont {W.}~\bibnamefont {Ludwig}}, \bibinfo
  {author} {\bibfnamefont {C.}~\bibnamefont {Detlefs}}, \bibinfo {author}
  {\bibfnamefont {W.}~\bibnamefont {Pantleon}}, \bibinfo {author}
  {\bibfnamefont {S.}~\bibnamefont {Schmidt}}, \bibinfo {author} {\bibfnamefont
  {F.}~\bibnamefont {Stöhr}}, \bibinfo {author} {\bibfnamefont
  {I.}~\bibnamefont {Snigireva}}, \bibinfo {author} {\bibfnamefont
  {A.}~\bibnamefont {Snigirev}}, \ and\ \bibinfo {author} {\bibfnamefont
  {H.~F.}\ \bibnamefont {Poulsen}},\ }\href {\doibase 10.1038/ncomms7098}
  {\bibfield  {journal} {\bibinfo  {journal} {Nature Communications}\ }\textbf
  {\bibinfo {volume} {6}},\ \bibinfo {pages} {6098} (\bibinfo {year}
  {2015})}\BibitemShut {NoStop}%
\bibitem [{\citenamefont {Poulsen}(2020)}]{PoulsenCOSSMS:20}%
  \BibitemOpen
  \bibfield  {author} {\bibinfo {author} {\bibfnamefont {H.}~\bibnamefont
  {Poulsen}},\ }\href {\doibase 10.1016/j.cossms.2020.100820} {\bibfield
  {journal} {\bibinfo  {journal} {Current Opinion in Solid State and Materials
  Science}\ }\textbf {\bibinfo {volume} {24}},\ \bibinfo {pages} {100820}
  (\bibinfo {year} {2020})}\BibitemShut {NoStop}%
\bibitem [{\citenamefont {Laanait}\ \emph {et~al.}(2017)\citenamefont
  {Laanait}, \citenamefont {Saenrang}, \citenamefont {Zhou}, \citenamefont
  {Eom},\ and\ \citenamefont {Zhang}}]{LaanaitASCI:17}%
  \BibitemOpen
  \bibfield  {author} {\bibinfo {author} {\bibfnamefont {N.}~\bibnamefont
  {Laanait}}, \bibinfo {author} {\bibfnamefont {W.}~\bibnamefont {Saenrang}},
  \bibinfo {author} {\bibfnamefont {H.}~\bibnamefont {Zhou}}, \bibinfo {author}
  {\bibfnamefont {C.-B.}\ \bibnamefont {Eom}}, \ and\ \bibinfo {author}
  {\bibfnamefont {Z.}~\bibnamefont {Zhang}},\ }\href {\doibase
  10.1186/s40679-017-0044-3} {\bibfield  {journal} {\bibinfo  {journal}
  {Advanced Structural and Chemical Imaging}\ }\textbf {\bibinfo {volume}
  {3}},\ \bibinfo {pages} {11} (\bibinfo {year} {2017})}\BibitemShut {NoStop}%
\bibitem [{\citenamefont {Bucsek}\ \emph {et~al.}(2019)\citenamefont {Bucsek},
  \citenamefont {Seiner}, \citenamefont {Simons}, \citenamefont {Yildirim},
  \citenamefont {Cook}, \citenamefont {Chumlyakov}, \citenamefont {Detlefs},\
  and\ \citenamefont {Stebner}}]{BucsekAM:19}%
  \BibitemOpen
  \bibfield  {author} {\bibinfo {author} {\bibfnamefont {A.}~\bibnamefont
  {Bucsek}}, \bibinfo {author} {\bibfnamefont {H.}~\bibnamefont {Seiner}},
  \bibinfo {author} {\bibfnamefont {H.}~\bibnamefont {Simons}}, \bibinfo
  {author} {\bibfnamefont {C.}~\bibnamefont {Yildirim}}, \bibinfo {author}
  {\bibfnamefont {P.}~\bibnamefont {Cook}}, \bibinfo {author} {\bibfnamefont
  {Y.}~\bibnamefont {Chumlyakov}}, \bibinfo {author} {\bibfnamefont
  {C.}~\bibnamefont {Detlefs}}, \ and\ \bibinfo {author} {\bibfnamefont
  {A.~P.}\ \bibnamefont {Stebner}},\ }\href {\doibase
  10.1016/j.actamat.2019.08.036} {\bibfield  {journal} {\bibinfo  {journal}
  {Acta Materialia}\ }\textbf {\bibinfo {volume} {179}},\ \bibinfo {pages}
  {273} (\bibinfo {year} {2019})}\BibitemShut {NoStop}%
\bibitem [{\citenamefont {Dresselhaus-Marais}\ \emph
  {et~al.}(2021)\citenamefont {Dresselhaus-Marais}, \citenamefont {Winther},
  \citenamefont {Howard}, \citenamefont {Gonzalez}, \citenamefont {Breckling},
  \citenamefont {Yildirim}, \citenamefont {Cook}, \citenamefont {Kutsal},
  \citenamefont {Simons}, \citenamefont {Detlefs}, \citenamefont {Eggert},\
  and\ \citenamefont {Poulsen}}]{DresselhausMaraisSA:21}%
  \BibitemOpen
  \bibfield  {author} {\bibinfo {author} {\bibfnamefont {L.~E.}\ \bibnamefont
  {Dresselhaus-Marais}}, \bibinfo {author} {\bibfnamefont {G.}~\bibnamefont
  {Winther}}, \bibinfo {author} {\bibfnamefont {M.}~\bibnamefont {Howard}},
  \bibinfo {author} {\bibfnamefont {A.}~\bibnamefont {Gonzalez}}, \bibinfo
  {author} {\bibfnamefont {S.~R.}\ \bibnamefont {Breckling}}, \bibinfo {author}
  {\bibfnamefont {C.}~\bibnamefont {Yildirim}}, \bibinfo {author}
  {\bibfnamefont {P.~K.}\ \bibnamefont {Cook}}, \bibinfo {author}
  {\bibfnamefont {M.}~\bibnamefont {Kutsal}}, \bibinfo {author} {\bibfnamefont
  {H.}~\bibnamefont {Simons}}, \bibinfo {author} {\bibfnamefont
  {C.}~\bibnamefont {Detlefs}}, \bibinfo {author} {\bibfnamefont {J.~H.}\
  \bibnamefont {Eggert}}, \ and\ \bibinfo {author} {\bibfnamefont {H.~F.}\
  \bibnamefont {Poulsen}},\ }\href {\doibase 10.1126/sciadv.abe8311} {\bibfield
   {journal} {\bibinfo  {journal} {Science Advances}\ }\textbf {\bibinfo
  {volume} {7}},\ \bibinfo {pages} {eabe8311} (\bibinfo {year}
  {2021})}\BibitemShut {NoStop}%
\bibitem [{\citenamefont {Kutsal}\ \emph {et~al.}(2019)\citenamefont {Kutsal},
  \citenamefont {Bernard}, \citenamefont {Berruyer}, \citenamefont {Cook},
  \citenamefont {Hino}, \citenamefont {Jakobsen}, \citenamefont {Ludwig},
  \citenamefont {Ormstrup}, \citenamefont {Roth}, \citenamefont {Simons},
  \citenamefont {Smets}, \citenamefont {Sierra}, \citenamefont {Wade},
  \citenamefont {Wattecamps}, \citenamefont {Yildirim}, \citenamefont
  {Poulsen},\ and\ \citenamefont {Detlefs}}]{KustalIOP:19}%
  \BibitemOpen
  \bibfield  {author} {\bibinfo {author} {\bibfnamefont {M.}~\bibnamefont
  {Kutsal}}, \bibinfo {author} {\bibfnamefont {P.}~\bibnamefont {Bernard}},
  \bibinfo {author} {\bibfnamefont {G.}~\bibnamefont {Berruyer}}, \bibinfo
  {author} {\bibfnamefont {P.~K.}\ \bibnamefont {Cook}}, \bibinfo {author}
  {\bibfnamefont {R.}~\bibnamefont {Hino}}, \bibinfo {author} {\bibfnamefont
  {A.~C.}\ \bibnamefont {Jakobsen}}, \bibinfo {author} {\bibfnamefont
  {W.}~\bibnamefont {Ludwig}}, \bibinfo {author} {\bibfnamefont
  {J.}~\bibnamefont {Ormstrup}}, \bibinfo {author} {\bibfnamefont
  {T.}~\bibnamefont {Roth}}, \bibinfo {author} {\bibfnamefont {H.}~\bibnamefont
  {Simons}}, \bibinfo {author} {\bibfnamefont {K.}~\bibnamefont {Smets}},
  \bibinfo {author} {\bibfnamefont {J.~X.}\ \bibnamefont {Sierra}}, \bibinfo
  {author} {\bibfnamefont {J.}~\bibnamefont {Wade}}, \bibinfo {author}
  {\bibfnamefont {P.}~\bibnamefont {Wattecamps}}, \bibinfo {author}
  {\bibfnamefont {C.}~\bibnamefont {Yildirim}}, \bibinfo {author}
  {\bibfnamefont {H.~F.}\ \bibnamefont {Poulsen}}, \ and\ \bibinfo {author}
  {\bibfnamefont {C.}~\bibnamefont {Detlefs}},\ }\href {\doibase
  10.1088/1757-899x/580/1/012007} {\bibfield  {journal} {\bibinfo  {journal}
  {IOP Conference Series: Materials Science and Engineering}\ }\textbf
  {\bibinfo {volume} {580}},\ \bibinfo {pages} {012007} (\bibinfo {year}
  {2019})}\BibitemShut {NoStop}%
\bibitem [{\citenamefont {Holstad}\ \emph {et~al.}(2022)\citenamefont
  {Holstad}, \citenamefont {Ræder}, \citenamefont {Carlsen}, \citenamefont
  {Knudsen}, \citenamefont {Dresselhaus-Marais}, \citenamefont {Haldrup},
  \citenamefont {Simons}, \citenamefont {Nielsen},\ and\ \citenamefont
  {Poulsen}}]{HolstadJAC:22}%
  \BibitemOpen
  \bibfield  {author} {\bibinfo {author} {\bibfnamefont {T.~S.}\ \bibnamefont
  {Holstad}}, \bibinfo {author} {\bibfnamefont {T.~M.}\ \bibnamefont {Ræder}},
  \bibinfo {author} {\bibfnamefont {M.}~\bibnamefont {Carlsen}}, \bibinfo
  {author} {\bibfnamefont {E.~B.}\ \bibnamefont {Knudsen}}, \bibinfo {author}
  {\bibfnamefont {L.}~\bibnamefont {Dresselhaus-Marais}}, \bibinfo {author}
  {\bibfnamefont {K.}~\bibnamefont {Haldrup}}, \bibinfo {author} {\bibfnamefont
  {H.}~\bibnamefont {Simons}}, \bibinfo {author} {\bibfnamefont {M.~M.}\
  \bibnamefont {Nielsen}}, \ and\ \bibinfo {author} {\bibfnamefont {H.~F.}\
  \bibnamefont {Poulsen}},\ }\href {\doibase 10.1107/s1600576721012760}
  {\bibfield  {journal} {\bibinfo  {journal} {Journal of Applied
  Crystallography}\ }\textbf {\bibinfo {volume} {55}},\ \bibinfo {pages} {112}
  (\bibinfo {year} {2022})}\BibitemShut {NoStop}%
\bibitem [{\citenamefont {Dresselhaus-Marais}\ \emph
  {et~al.}(2022)\citenamefont {Dresselhaus-Marais}, \citenamefont
  {Kozioziemski}, \citenamefont {Holstad}, \citenamefont {Ræder},
  \citenamefont {Seaberg}, \citenamefont {Nam}, \citenamefont {Kim},
  \citenamefont {Breckling}, \citenamefont {Chollet}, \citenamefont {Cook},
  \citenamefont {Folsom}, \citenamefont {Galtier}, \citenamefont {Gavilan},
  \citenamefont {Gonzalez}, \citenamefont {Gorhover}, \citenamefont {Guillet},
  \citenamefont {Haldrup}, \citenamefont {Howard}, \citenamefont {Katagiri},
  \citenamefont {Kim}, \citenamefont {Kim}, \citenamefont {Kim}, \citenamefont
  {Kim}, \citenamefont {Knudsen}, \citenamefont {Kuschel}, \citenamefont {Lee},
  \citenamefont {Lin}, \citenamefont {McWilliams}, \citenamefont {Nagler},
  \citenamefont {Ozaki}, \citenamefont {Pal}, \citenamefont {Pedro},
  \citenamefont {Nielsen}, \citenamefont {Saunders}, \citenamefont {Schoofs},
  \citenamefont {Sekine}, \citenamefont {Simons}, \citenamefont {Driel},
  \citenamefont {Wang}, \citenamefont {Yang}, \citenamefont {Yildirim},
  \citenamefont {Poulsen},\ and\ \citenamefont
  {Eggert}}]{DresselhausMaraisARXIV:22}%
  \BibitemOpen
  \bibfield  {author} {\bibinfo {author} {\bibfnamefont {L.~E.}\ \bibnamefont
  {Dresselhaus-Marais}}, \bibinfo {author} {\bibfnamefont {B.}~\bibnamefont
  {Kozioziemski}}, \bibinfo {author} {\bibfnamefont {T.~S.}\ \bibnamefont
  {Holstad}}, \bibinfo {author} {\bibfnamefont {T.~M.}\ \bibnamefont {Ræder}},
  \bibinfo {author} {\bibfnamefont {M.}~\bibnamefont {Seaberg}}, \bibinfo
  {author} {\bibfnamefont {D.}~\bibnamefont {Nam}}, \bibinfo {author}
  {\bibfnamefont {S.}~\bibnamefont {Kim}}, \bibinfo {author} {\bibfnamefont
  {S.}~\bibnamefont {Breckling}}, \bibinfo {author} {\bibfnamefont
  {M.}~\bibnamefont {Chollet}}, \bibinfo {author} {\bibfnamefont {P.~K.}\
  \bibnamefont {Cook}}, \bibinfo {author} {\bibfnamefont {E.}~\bibnamefont
  {Folsom}}, \bibinfo {author} {\bibfnamefont {E.}~\bibnamefont {Galtier}},
  \bibinfo {author} {\bibfnamefont {L.}~\bibnamefont {Gavilan}}, \bibinfo
  {author} {\bibfnamefont {A.}~\bibnamefont {Gonzalez}}, \bibinfo {author}
  {\bibfnamefont {T.}~\bibnamefont {Gorhover}}, \bibinfo {author}
  {\bibfnamefont {S.}~\bibnamefont {Guillet}}, \bibinfo {author} {\bibfnamefont
  {K.}~\bibnamefont {Haldrup}}, \bibinfo {author} {\bibfnamefont
  {M.}~\bibnamefont {Howard}}, \bibinfo {author} {\bibfnamefont
  {K.}~\bibnamefont {Katagiri}}, \bibinfo {author} {\bibfnamefont
  {S.}~\bibnamefont {Kim}}, \bibinfo {author} {\bibfnamefont {S.}~\bibnamefont
  {Kim}}, \bibinfo {author} {\bibfnamefont {S.}~\bibnamefont {Kim}}, \bibinfo
  {author} {\bibfnamefont {H.}~\bibnamefont {Kim}}, \bibinfo {author}
  {\bibfnamefont {E.~B.}\ \bibnamefont {Knudsen}}, \bibinfo {author}
  {\bibfnamefont {S.}~\bibnamefont {Kuschel}}, \bibinfo {author} {\bibfnamefont
  {H.-J.}\ \bibnamefont {Lee}}, \bibinfo {author} {\bibfnamefont
  {C.}~\bibnamefont {Lin}}, \bibinfo {author} {\bibfnamefont {R.~S.}\
  \bibnamefont {McWilliams}}, \bibinfo {author} {\bibfnamefont
  {B.}~\bibnamefont {Nagler}}, \bibinfo {author} {\bibfnamefont
  {N.}~\bibnamefont {Ozaki}}, \bibinfo {author} {\bibfnamefont
  {D.}~\bibnamefont {Pal}}, \bibinfo {author} {\bibfnamefont {R.~P.}\
  \bibnamefont {Pedro}}, \bibinfo {author} {\bibfnamefont {M.~M.}\ \bibnamefont
  {Nielsen}}, \bibinfo {author} {\bibfnamefont {A.~M.}\ \bibnamefont
  {Saunders}}, \bibinfo {author} {\bibfnamefont {F.}~\bibnamefont {Schoofs}},
  \bibinfo {author} {\bibfnamefont {T.}~\bibnamefont {Sekine}}, \bibinfo
  {author} {\bibfnamefont {H.}~\bibnamefont {Simons}}, \bibinfo {author}
  {\bibfnamefont {T.~v.}\ \bibnamefont {Driel}}, \bibinfo {author}
  {\bibfnamefont {B.}~\bibnamefont {Wang}}, \bibinfo {author} {\bibfnamefont
  {W.}~\bibnamefont {Yang}}, \bibinfo {author} {\bibfnamefont {C.}~\bibnamefont
  {Yildirim}}, \bibinfo {author} {\bibfnamefont {H.~F.}\ \bibnamefont
  {Poulsen}}, \ and\ \bibinfo {author} {\bibfnamefont {J.~H.}\ \bibnamefont
  {Eggert}},\ }\href {\doibase 10.48550/arxiv.2210.08366} {\bibfield  {journal}
  {\bibinfo  {journal} {arXiv}\ } (\bibinfo {year} {2022}),\
  10.48550/arxiv.2210.08366},\ \Eprint {http://arxiv.org/abs/2210.08366}
  {2210.08366} \BibitemShut {NoStop}%
\bibitem [{\citenamefont {Chapman}(2009)}]{ChapmanNMat:09}%
  \BibitemOpen
  \bibfield  {author} {\bibinfo {author} {\bibfnamefont {H.~N.}\ \bibnamefont
  {Chapman}},\ }\href {\doibase 10.1038/nmat2402} {\bibfield  {journal}
  {\bibinfo  {journal} {Nature Materials}\ }\textbf {\bibinfo {volume} {8}},\
  \bibinfo {pages} {299 } (\bibinfo {year} {2009})}\BibitemShut {NoStop}%
\bibitem [{\citenamefont {Nugent}(2010)}]{NugentAdvPhys:10}%
  \BibitemOpen
  \bibfield  {author} {\bibinfo {author} {\bibfnamefont {K.~A.}\ \bibnamefont
  {Nugent}},\ }\href {\doibase 10.1080/00018730903270926} {\bibfield  {journal}
  {\bibinfo  {journal} {Advances in Physics}\ }\textbf {\bibinfo {volume}
  {59}},\ \bibinfo {pages} {1} (\bibinfo {year} {2010})},\ \Eprint
  {http://arxiv.org/abs/0908.3064} {0908.3064} \BibitemShut {NoStop}%
\bibitem [{\citenamefont {Miao}\ \emph {et~al.}(2015)\citenamefont {Miao},
  \citenamefont {Ishikawa}, \citenamefont {Robinson},\ and\ \citenamefont
  {Murnane}}]{MiaoScience:15}%
  \BibitemOpen
  \bibfield  {author} {\bibinfo {author} {\bibfnamefont {J.}~\bibnamefont
  {Miao}}, \bibinfo {author} {\bibfnamefont {T.}~\bibnamefont {Ishikawa}},
  \bibinfo {author} {\bibfnamefont {I.~K.}\ \bibnamefont {Robinson}}, \ and\
  \bibinfo {author} {\bibfnamefont {M.~M.}\ \bibnamefont {Murnane}},\ }\href
  {\doibase 10.1126/science.aaa1394} {\bibfield  {journal} {\bibinfo  {journal}
  {Science}\ }\textbf {\bibinfo {volume} {348}},\ \bibinfo {pages} {530}
  (\bibinfo {year} {2015})}\BibitemShut {NoStop}%
\bibitem [{\citenamefont {Chapman}\ \emph {et~al.}(2006)\citenamefont
  {Chapman}, \citenamefont {Barty}, \citenamefont {Bogan}, \citenamefont
  {Boutet}, \citenamefont {Frank}, \citenamefont {Hau-Riege}, \citenamefont
  {Marchesini}, \citenamefont {Woods}, \citenamefont {Bajt}, \citenamefont
  {Benner}, \citenamefont {London}, \citenamefont {Plönjes}, \citenamefont
  {Kuhlmann}, \citenamefont {Treusch}, \citenamefont {Düsterer}, \citenamefont
  {Tschentscher}, \citenamefont {Schneider}, \citenamefont {Spiller},
  \citenamefont {Möller}, \citenamefont {Bostedt}, \citenamefont {Hoener},
  \citenamefont {Shapiro}, \citenamefont {Hodgson}, \citenamefont {Spoel},
  \citenamefont {Burmeister}, \citenamefont {Bergh}, \citenamefont {Caleman},
  \citenamefont {Huldt}, \citenamefont {Seibert}, \citenamefont {Maia},
  \citenamefont {Lee}, \citenamefont {Szöke}, \citenamefont {Timneanu},\ and\
  \citenamefont {Hajdu}}]{ChapmanNP:06}%
  \BibitemOpen
  \bibfield  {author} {\bibinfo {author} {\bibfnamefont {H.~N.}\ \bibnamefont
  {Chapman}}, \bibinfo {author} {\bibfnamefont {A.}~\bibnamefont {Barty}},
  \bibinfo {author} {\bibfnamefont {M.~J.}\ \bibnamefont {Bogan}}, \bibinfo
  {author} {\bibfnamefont {S.}~\bibnamefont {Boutet}}, \bibinfo {author}
  {\bibfnamefont {M.}~\bibnamefont {Frank}}, \bibinfo {author} {\bibfnamefont
  {S.~P.}\ \bibnamefont {Hau-Riege}}, \bibinfo {author} {\bibfnamefont
  {S.}~\bibnamefont {Marchesini}}, \bibinfo {author} {\bibfnamefont {B.~W.}\
  \bibnamefont {Woods}}, \bibinfo {author} {\bibfnamefont {S.}~\bibnamefont
  {Bajt}}, \bibinfo {author} {\bibfnamefont {W.~H.}\ \bibnamefont {Benner}},
  \bibinfo {author} {\bibfnamefont {R.~A.}\ \bibnamefont {London}}, \bibinfo
  {author} {\bibfnamefont {E.}~\bibnamefont {Plönjes}}, \bibinfo {author}
  {\bibfnamefont {M.}~\bibnamefont {Kuhlmann}}, \bibinfo {author}
  {\bibfnamefont {R.}~\bibnamefont {Treusch}}, \bibinfo {author} {\bibfnamefont
  {S.}~\bibnamefont {Düsterer}}, \bibinfo {author} {\bibfnamefont
  {T.}~\bibnamefont {Tschentscher}}, \bibinfo {author} {\bibfnamefont {J.~R.}\
  \bibnamefont {Schneider}}, \bibinfo {author} {\bibfnamefont {E.}~\bibnamefont
  {Spiller}}, \bibinfo {author} {\bibfnamefont {T.}~\bibnamefont {Möller}},
  \bibinfo {author} {\bibfnamefont {C.}~\bibnamefont {Bostedt}}, \bibinfo
  {author} {\bibfnamefont {M.}~\bibnamefont {Hoener}}, \bibinfo {author}
  {\bibfnamefont {D.~A.}\ \bibnamefont {Shapiro}}, \bibinfo {author}
  {\bibfnamefont {K.~O.}\ \bibnamefont {Hodgson}}, \bibinfo {author}
  {\bibfnamefont {D.~v.~d.}\ \bibnamefont {Spoel}}, \bibinfo {author}
  {\bibfnamefont {F.}~\bibnamefont {Burmeister}}, \bibinfo {author}
  {\bibfnamefont {M.}~\bibnamefont {Bergh}}, \bibinfo {author} {\bibfnamefont
  {C.}~\bibnamefont {Caleman}}, \bibinfo {author} {\bibfnamefont
  {G.}~\bibnamefont {Huldt}}, \bibinfo {author} {\bibfnamefont {M.~M.}\
  \bibnamefont {Seibert}}, \bibinfo {author} {\bibfnamefont {F.~R. N.~C.}\
  \bibnamefont {Maia}}, \bibinfo {author} {\bibfnamefont {R.~W.}\ \bibnamefont
  {Lee}}, \bibinfo {author} {\bibfnamefont {A.}~\bibnamefont {Szöke}},
  \bibinfo {author} {\bibfnamefont {N.}~\bibnamefont {Timneanu}}, \ and\
  \bibinfo {author} {\bibfnamefont {J.}~\bibnamefont {Hajdu}},\ }\href
  {\doibase 10.1038/nphys461} {\bibfield  {journal} {\bibinfo  {journal}
  {Nature Physics}\ }\textbf {\bibinfo {volume} {2}},\ \bibinfo {pages} {839}
  (\bibinfo {year} {2006})}\BibitemShut {NoStop}%
\bibitem [{\citenamefont {Barty}\ \emph {et~al.}(2008)\citenamefont {Barty},
  \citenamefont {Boutet}, \citenamefont {Bogan}, \citenamefont {Hau-Riege},
  \citenamefont {Marchesini}, \citenamefont {Sokolowski-Tinten}, \citenamefont
  {Stojanovic}, \citenamefont {Tobey}, \citenamefont {Ehrke}, \citenamefont
  {Cavalleri}, \citenamefont {Düsterer}, \citenamefont {Frank}, \citenamefont
  {Bajt}, \citenamefont {Woods}, \citenamefont {Seibert}, \citenamefont
  {Hajdu}, \citenamefont {Treusch},\ and\ \citenamefont
  {Chapman}}]{BatyNPhot:08}%
  \BibitemOpen
  \bibfield  {author} {\bibinfo {author} {\bibfnamefont {A.}~\bibnamefont
  {Barty}}, \bibinfo {author} {\bibfnamefont {S.}~\bibnamefont {Boutet}},
  \bibinfo {author} {\bibfnamefont {M.~J.}\ \bibnamefont {Bogan}}, \bibinfo
  {author} {\bibfnamefont {S.}~\bibnamefont {Hau-Riege}}, \bibinfo {author}
  {\bibfnamefont {S.}~\bibnamefont {Marchesini}}, \bibinfo {author}
  {\bibfnamefont {K.}~\bibnamefont {Sokolowski-Tinten}}, \bibinfo {author}
  {\bibfnamefont {N.}~\bibnamefont {Stojanovic}}, \bibinfo {author}
  {\bibfnamefont {R.}~\bibnamefont {Tobey}}, \bibinfo {author} {\bibfnamefont
  {H.}~\bibnamefont {Ehrke}}, \bibinfo {author} {\bibfnamefont
  {A.}~\bibnamefont {Cavalleri}}, \bibinfo {author} {\bibfnamefont
  {S.}~\bibnamefont {Düsterer}}, \bibinfo {author} {\bibfnamefont
  {M.}~\bibnamefont {Frank}}, \bibinfo {author} {\bibfnamefont
  {S.}~\bibnamefont {Bajt}}, \bibinfo {author} {\bibfnamefont {B.~W.}\
  \bibnamefont {Woods}}, \bibinfo {author} {\bibfnamefont {M.~M.}\ \bibnamefont
  {Seibert}}, \bibinfo {author} {\bibfnamefont {J.}~\bibnamefont {Hajdu}},
  \bibinfo {author} {\bibfnamefont {R.}~\bibnamefont {Treusch}}, \ and\
  \bibinfo {author} {\bibfnamefont {H.~N.}\ \bibnamefont {Chapman}},\ }\href
  {\doibase 10.1038/nphoton.2008.128} {\bibfield  {journal} {\bibinfo
  {journal} {Nature Photonics}\ }\textbf {\bibinfo {volume} {2}},\ \bibinfo
  {pages} {415} (\bibinfo {year} {2008})}\BibitemShut {NoStop}%
\bibitem [{\citenamefont {Nicolas}\ \emph {et~al.}(2014)\citenamefont
  {Nicolas}, \citenamefont {Reusch}, \citenamefont {Osterhoff}, \citenamefont
  {Sprung}, \citenamefont {Schúlein}, \citenamefont {Krenner}, \citenamefont
  {Wixforth},\ and\ \citenamefont {Salditt}}]{NicolasJAC:14}%
  \BibitemOpen
  \bibfield  {author} {\bibinfo {author} {\bibfnamefont {J.-D.}\ \bibnamefont
  {Nicolas}}, \bibinfo {author} {\bibfnamefont {T.}~\bibnamefont {Reusch}},
  \bibinfo {author} {\bibfnamefont {M.}~\bibnamefont {Osterhoff}}, \bibinfo
  {author} {\bibfnamefont {M.}~\bibnamefont {Sprung}}, \bibinfo {author}
  {\bibfnamefont {F.}~\bibnamefont {Schúlein}}, \bibinfo {author}
  {\bibfnamefont {H.}~\bibnamefont {Krenner}}, \bibinfo {author} {\bibfnamefont
  {A.}~\bibnamefont {Wixforth}}, \ and\ \bibinfo {author} {\bibfnamefont
  {T.}~\bibnamefont {Salditt}},\ }\href {\doibase 10.1107/s1600576714016896}
  {\bibfield  {journal} {\bibinfo  {journal} {Journal of Applied
  Crystallography}\ }\textbf {\bibinfo {volume} {47}},\ \bibinfo {pages} {1596}
  (\bibinfo {year} {2014})}\BibitemShut {NoStop}%
\bibitem [{\citenamefont {Kovalchuk}\ and\ \citenamefont
  {Blagov}(2022)}]{KovalchukCR:22}%
  \BibitemOpen
  \bibfield  {author} {\bibinfo {author} {\bibfnamefont {M.~V.}\ \bibnamefont
  {Kovalchuk}}\ and\ \bibinfo {author} {\bibfnamefont {A.~E.}\ \bibnamefont
  {Blagov}},\ }\href {\doibase 10.1134/s1063774522050066} {\bibfield  {journal}
  {\bibinfo  {journal} {Crystallography Reports}\ }\textbf {\bibinfo {volume}
  {67}},\ \bibinfo {pages} {631} (\bibinfo {year} {2022})}\BibitemShut
  {NoStop}%
\bibitem [{\citenamefont {Singer}\ \emph {et~al.}(2018)\citenamefont {Singer},
  \citenamefont {Zhang}, \citenamefont {Hy}, \citenamefont {Cela},
  \citenamefont {Fang}, \citenamefont {Wynn}, \citenamefont {Qiu},
  \citenamefont {Xia}, \citenamefont {Liu}, \citenamefont {Ulvestad},
  \citenamefont {Hua}, \citenamefont {Wingert}, \citenamefont {Liu},
  \citenamefont {Sprung}, \citenamefont {Zozulya}, \citenamefont {Maxey},
  \citenamefont {Harder}, \citenamefont {Meng},\ and\ \citenamefont
  {Shpyrko}}]{SingerNatE:18}%
  \BibitemOpen
  \bibfield  {author} {\bibinfo {author} {\bibfnamefont {A.}~\bibnamefont
  {Singer}}, \bibinfo {author} {\bibfnamefont {M.}~\bibnamefont {Zhang}},
  \bibinfo {author} {\bibfnamefont {S.}~\bibnamefont {Hy}}, \bibinfo {author}
  {\bibfnamefont {D.}~\bibnamefont {Cela}}, \bibinfo {author} {\bibfnamefont
  {C.}~\bibnamefont {Fang}}, \bibinfo {author} {\bibfnamefont {T.~A.}\
  \bibnamefont {Wynn}}, \bibinfo {author} {\bibfnamefont {B.}~\bibnamefont
  {Qiu}}, \bibinfo {author} {\bibfnamefont {Y.}~\bibnamefont {Xia}}, \bibinfo
  {author} {\bibfnamefont {Z.}~\bibnamefont {Liu}}, \bibinfo {author}
  {\bibfnamefont {A.}~\bibnamefont {Ulvestad}}, \bibinfo {author}
  {\bibfnamefont {N.}~\bibnamefont {Hua}}, \bibinfo {author} {\bibfnamefont
  {J.}~\bibnamefont {Wingert}}, \bibinfo {author} {\bibfnamefont
  {H.}~\bibnamefont {Liu}}, \bibinfo {author} {\bibfnamefont {M.}~\bibnamefont
  {Sprung}}, \bibinfo {author} {\bibfnamefont {A.~V.}\ \bibnamefont {Zozulya}},
  \bibinfo {author} {\bibfnamefont {E.}~\bibnamefont {Maxey}}, \bibinfo
  {author} {\bibfnamefont {R.}~\bibnamefont {Harder}}, \bibinfo {author}
  {\bibfnamefont {Y.~S.}\ \bibnamefont {Meng}}, \ and\ \bibinfo {author}
  {\bibfnamefont {O.~G.}\ \bibnamefont {Shpyrko}},\ }\href {\doibase
  10.1038/s41560-018-0184-2} {\bibfield  {journal} {\bibinfo  {journal} {Nature
  Energy}\ }\textbf {\bibinfo {volume} {3}},\ \bibinfo {pages} {641} (\bibinfo
  {year} {2018})}\BibitemShut {NoStop}%
\bibitem [{\citenamefont {Liu}\ \emph {et~al.}(2022)\citenamefont {Liu},
  \citenamefont {Liu}, \citenamefont {Li}, \citenamefont {Yu}, \citenamefont
  {Diao}, \citenamefont {Zhou}, \citenamefont {Li}, \citenamefont {Dai},
  \citenamefont {Zhao}, \citenamefont {Xu}, \citenamefont {Ren}, \citenamefont
  {Wang}, \citenamefont {Wu}, \citenamefont {Qi}, \citenamefont {Xiao},
  \citenamefont {Zheng}, \citenamefont {Cha}, \citenamefont {Harder},
  \citenamefont {Robinson}, \citenamefont {Wen}, \citenamefont {Lu},
  \citenamefont {Pan},\ and\ \citenamefont {Amine}}]{LiuNature:22}%
  \BibitemOpen
  \bibfield  {author} {\bibinfo {author} {\bibfnamefont {T.}~\bibnamefont
  {Liu}}, \bibinfo {author} {\bibfnamefont {J.}~\bibnamefont {Liu}}, \bibinfo
  {author} {\bibfnamefont {L.}~\bibnamefont {Li}}, \bibinfo {author}
  {\bibfnamefont {L.}~\bibnamefont {Yu}}, \bibinfo {author} {\bibfnamefont
  {J.}~\bibnamefont {Diao}}, \bibinfo {author} {\bibfnamefont {T.}~\bibnamefont
  {Zhou}}, \bibinfo {author} {\bibfnamefont {S.}~\bibnamefont {Li}}, \bibinfo
  {author} {\bibfnamefont {A.}~\bibnamefont {Dai}}, \bibinfo {author}
  {\bibfnamefont {W.}~\bibnamefont {Zhao}}, \bibinfo {author} {\bibfnamefont
  {S.}~\bibnamefont {Xu}}, \bibinfo {author} {\bibfnamefont {Y.}~\bibnamefont
  {Ren}}, \bibinfo {author} {\bibfnamefont {L.}~\bibnamefont {Wang}}, \bibinfo
  {author} {\bibfnamefont {T.}~\bibnamefont {Wu}}, \bibinfo {author}
  {\bibfnamefont {R.}~\bibnamefont {Qi}}, \bibinfo {author} {\bibfnamefont
  {Y.}~\bibnamefont {Xiao}}, \bibinfo {author} {\bibfnamefont {J.}~\bibnamefont
  {Zheng}}, \bibinfo {author} {\bibfnamefont {W.}~\bibnamefont {Cha}}, \bibinfo
  {author} {\bibfnamefont {R.}~\bibnamefont {Harder}}, \bibinfo {author}
  {\bibfnamefont {I.}~\bibnamefont {Robinson}}, \bibinfo {author}
  {\bibfnamefont {J.}~\bibnamefont {Wen}}, \bibinfo {author} {\bibfnamefont
  {J.}~\bibnamefont {Lu}}, \bibinfo {author} {\bibfnamefont {F.}~\bibnamefont
  {Pan}}, \ and\ \bibinfo {author} {\bibfnamefont {K.}~\bibnamefont {Amine}},\
  }\href {\doibase 10.1038/s41586-022-04689-y} {\bibfield  {journal} {\bibinfo
  {journal} {Nature}\ }\textbf {\bibinfo {volume} {606}},\ \bibinfo {pages}
  {305} (\bibinfo {year} {2022})}\BibitemShut {NoStop}%
\bibitem [{\citenamefont {Simons}\ \emph {et~al.}(2018)\citenamefont {Simons},
  \citenamefont {Haugen}, \citenamefont {Jakobsen}, \citenamefont {Schmidt},
  \citenamefont {Stöhr}, \citenamefont {Majkut}, \citenamefont {Detlefs},
  \citenamefont {Daniels}, \citenamefont {Damjanovic},\ and\ \citenamefont
  {Poulsen}}]{SimonsNMat:18}%
  \BibitemOpen
  \bibfield  {author} {\bibinfo {author} {\bibfnamefont {H.}~\bibnamefont
  {Simons}}, \bibinfo {author} {\bibfnamefont {A.~B.}\ \bibnamefont {Haugen}},
  \bibinfo {author} {\bibfnamefont {A.~C.}\ \bibnamefont {Jakobsen}}, \bibinfo
  {author} {\bibfnamefont {S.}~\bibnamefont {Schmidt}}, \bibinfo {author}
  {\bibfnamefont {F.}~\bibnamefont {Stöhr}}, \bibinfo {author} {\bibfnamefont
  {M.}~\bibnamefont {Majkut}}, \bibinfo {author} {\bibfnamefont
  {C.}~\bibnamefont {Detlefs}}, \bibinfo {author} {\bibfnamefont {J.~E.}\
  \bibnamefont {Daniels}}, \bibinfo {author} {\bibfnamefont {D.}~\bibnamefont
  {Damjanovic}}, \ and\ \bibinfo {author} {\bibfnamefont {H.~F.}\ \bibnamefont
  {Poulsen}},\ }\href {\doibase 10.1038/s41563-018-0116-3} {\bibfield
  {journal} {\bibinfo  {journal} {Nature Materials}\ }\textbf {\bibinfo
  {volume} {17}},\ \bibinfo {pages} {814} (\bibinfo {year} {2018})}\BibitemShut
  {NoStop}%
\bibitem [{\citenamefont {Kim}\ \emph {et~al.}(2018)\citenamefont {Kim},
  \citenamefont {Chung}, \citenamefont {Carnis}, \citenamefont {Kim},
  \citenamefont {Yun}, \citenamefont {Kang}, \citenamefont {Cha}, \citenamefont
  {Cherukara}, \citenamefont {Maxey}, \citenamefont {Harder}, \citenamefont
  {Sasikumar}, \citenamefont {Sankaranarayanan}, \citenamefont {Zozulya},
  \citenamefont {Sprung}, \citenamefont {Riu},\ and\ \citenamefont
  {Kim}}]{KimNatComm:18}%
  \BibitemOpen
  \bibfield  {author} {\bibinfo {author} {\bibfnamefont {D.}~\bibnamefont
  {Kim}}, \bibinfo {author} {\bibfnamefont {M.}~\bibnamefont {Chung}}, \bibinfo
  {author} {\bibfnamefont {J.}~\bibnamefont {Carnis}}, \bibinfo {author}
  {\bibfnamefont {S.}~\bibnamefont {Kim}}, \bibinfo {author} {\bibfnamefont
  {K.}~\bibnamefont {Yun}}, \bibinfo {author} {\bibfnamefont {J.}~\bibnamefont
  {Kang}}, \bibinfo {author} {\bibfnamefont {W.}~\bibnamefont {Cha}}, \bibinfo
  {author} {\bibfnamefont {M.~J.}\ \bibnamefont {Cherukara}}, \bibinfo {author}
  {\bibfnamefont {E.}~\bibnamefont {Maxey}}, \bibinfo {author} {\bibfnamefont
  {R.}~\bibnamefont {Harder}}, \bibinfo {author} {\bibfnamefont
  {K.}~\bibnamefont {Sasikumar}}, \bibinfo {author} {\bibfnamefont {S.~K.
  R.~S.}\ \bibnamefont {Sankaranarayanan}}, \bibinfo {author} {\bibfnamefont
  {A.}~\bibnamefont {Zozulya}}, \bibinfo {author} {\bibfnamefont
  {M.}~\bibnamefont {Sprung}}, \bibinfo {author} {\bibfnamefont
  {D.}~\bibnamefont {Riu}}, \ and\ \bibinfo {author} {\bibfnamefont
  {H.}~\bibnamefont {Kim}},\ }\href {\doibase 10.1038/s41467-018-05464-2}
  {\bibfield  {journal} {\bibinfo  {journal} {Nature Communications}\ }\textbf
  {\bibinfo {volume} {9}},\ \bibinfo {pages} {3422} (\bibinfo {year}
  {2018})}\BibitemShut {NoStop}%
\bibitem [{\citenamefont {Clark}\ \emph {et~al.}(2013)\citenamefont {Clark},
  \citenamefont {Beitra}, \citenamefont {Xiong}, \citenamefont {Higginbotham},
  \citenamefont {Fritz}, \citenamefont {Lemke}, \citenamefont {Zhu},
  \citenamefont {Chollet}, \citenamefont {Williams}, \citenamefont
  {Messerschmidt}, \citenamefont {Abbey}, \citenamefont {Harder}, \citenamefont
  {Korsunsky}, \citenamefont {Wark},\ and\ \citenamefont
  {Robinson}}]{ClarkScience:13}%
  \BibitemOpen
  \bibfield  {author} {\bibinfo {author} {\bibfnamefont {J.~N.}\ \bibnamefont
  {Clark}}, \bibinfo {author} {\bibfnamefont {L.}~\bibnamefont {Beitra}},
  \bibinfo {author} {\bibfnamefont {G.}~\bibnamefont {Xiong}}, \bibinfo
  {author} {\bibfnamefont {A.}~\bibnamefont {Higginbotham}}, \bibinfo {author}
  {\bibfnamefont {D.~M.}\ \bibnamefont {Fritz}}, \bibinfo {author}
  {\bibfnamefont {H.~T.}\ \bibnamefont {Lemke}}, \bibinfo {author}
  {\bibfnamefont {D.}~\bibnamefont {Zhu}}, \bibinfo {author} {\bibfnamefont
  {M.}~\bibnamefont {Chollet}}, \bibinfo {author} {\bibfnamefont {G.~J.}\
  \bibnamefont {Williams}}, \bibinfo {author} {\bibfnamefont {M.}~\bibnamefont
  {Messerschmidt}}, \bibinfo {author} {\bibfnamefont {B.}~\bibnamefont
  {Abbey}}, \bibinfo {author} {\bibfnamefont {R.~J.}\ \bibnamefont {Harder}},
  \bibinfo {author} {\bibfnamefont {A.~M.}\ \bibnamefont {Korsunsky}}, \bibinfo
  {author} {\bibfnamefont {J.~S.}\ \bibnamefont {Wark}}, \ and\ \bibinfo
  {author} {\bibfnamefont {I.~K.}\ \bibnamefont {Robinson}},\ }\href {\doibase
  10.1126/science.1236034} {\bibfield  {journal} {\bibinfo  {journal}
  {Science}\ }\textbf {\bibinfo {volume} {341}},\ \bibinfo {pages} {56}
  (\bibinfo {year} {2013})}\BibitemShut {NoStop}%
\bibitem [{\citenamefont {Gao}\ \emph {et~al.}(2019)\citenamefont {Gao},
  \citenamefont {Harder}, \citenamefont {Southworth}, \citenamefont {Guest},
  \citenamefont {Huang}, \citenamefont {Yan}, \citenamefont {Ocola},
  \citenamefont {Yifat}, \citenamefont {Sule}, \citenamefont {Ho},
  \citenamefont {Pelton}, \citenamefont {Scherer},\ and\ \citenamefont
  {Young}}]{GaoPNAS:19}%
  \BibitemOpen
  \bibfield  {author} {\bibinfo {author} {\bibfnamefont {Y.}~\bibnamefont
  {Gao}}, \bibinfo {author} {\bibfnamefont {R.}~\bibnamefont {Harder}},
  \bibinfo {author} {\bibfnamefont {S.~H.}\ \bibnamefont {Southworth}},
  \bibinfo {author} {\bibfnamefont {J.~R.}\ \bibnamefont {Guest}}, \bibinfo
  {author} {\bibfnamefont {X.}~\bibnamefont {Huang}}, \bibinfo {author}
  {\bibfnamefont {Z.}~\bibnamefont {Yan}}, \bibinfo {author} {\bibfnamefont
  {L.~E.}\ \bibnamefont {Ocola}}, \bibinfo {author} {\bibfnamefont
  {Y.}~\bibnamefont {Yifat}}, \bibinfo {author} {\bibfnamefont
  {N.}~\bibnamefont {Sule}}, \bibinfo {author} {\bibfnamefont {P.~J.}\
  \bibnamefont {Ho}}, \bibinfo {author} {\bibfnamefont {M.}~\bibnamefont
  {Pelton}}, \bibinfo {author} {\bibfnamefont {N.~F.}\ \bibnamefont {Scherer}},
  \ and\ \bibinfo {author} {\bibfnamefont {L.}~\bibnamefont {Young}},\ }\href
  {\doibase 10.1073/pnas.1720785116} {\bibfield  {journal} {\bibinfo  {journal}
  {Proceedings of the National Academy of Sciences}\ }\textbf {\bibinfo
  {volume} {116}},\ \bibinfo {pages} {4018} (\bibinfo {year}
  {2019})}\BibitemShut {NoStop}%
\bibitem [{\citenamefont {Xiao}\ \emph {et~al.}(2014)\citenamefont {Xiao},
  \citenamefont {Xie}, \citenamefont {Deng},\ and\ \citenamefont {{et
  al.}}}]{CC1}%
  \BibitemOpen
  \bibfield  {author} {\bibinfo {author} {\bibfnamefont {T.~Q.}\ \bibnamefont
  {Xiao}}, \bibinfo {author} {\bibfnamefont {H.~L.}\ \bibnamefont {Xie}},
  \bibinfo {author} {\bibfnamefont {B.}~\bibnamefont {Deng}}, \ and\ \bibinfo
  {author} {\bibnamefont {{et al.}}},\ }\href@noop {} {\bibfield  {journal}
  {\bibinfo  {journal} {Acta Optica Sinica}\ }\textbf {\bibinfo {volume}
  {34}},\ \bibinfo {pages} {0100001} (\bibinfo {year} {2014})}\BibitemShut
  {NoStop}%
\bibitem [{\citenamefont {Artyukov}\ and\ \citenamefont
  {Irtuganov}(2019)}]{CC3}%
  \BibitemOpen
  \bibfield  {author} {\bibinfo {author} {\bibfnamefont {I.~A.}\ \bibnamefont
  {Artyukov}}\ and\ \bibinfo {author} {\bibfnamefont {N.~N.}\ \bibnamefont
  {Irtuganov}},\ }\href@noop {} {\bibfield  {journal} {\bibinfo  {journal} {J.
  Russ. Laser Res.}\ }\textbf {\bibinfo {volume} {40}},\ \bibinfo {pages} {150}
  (\bibinfo {year} {2019})}\BibitemShut {NoStop}%
\bibitem [{\citenamefont {Li}\ \emph {et~al.}(2021)\citenamefont {Li},
  \citenamefont {Gao}, \citenamefont {Zhang},\ and\ \citenamefont {{et
  al.}}}]{CC4}%
  \BibitemOpen
  \bibfield  {author} {\bibinfo {author} {\bibfnamefont {K.}~\bibnamefont
  {Li}}, \bibinfo {author} {\bibfnamefont {Y.}~\bibnamefont {Gao}}, \bibinfo
  {author} {\bibfnamefont {H.}~\bibnamefont {Zhang}}, \ and\ \bibinfo {author}
  {\bibnamefont {{et al.}}},\ }\href@noop {} {\bibfield  {journal} {\bibinfo
  {journal} {Chin. Opt. Lett.}\ }\textbf {\bibinfo {volume} {19}},\ \bibinfo
  {pages} {073401} (\bibinfo {year} {2021})}\BibitemShut {NoStop}%
\bibitem [{\citenamefont {Promdet}\ \emph {et~al.}(2018)\citenamefont
  {Promdet}, \citenamefont {Rodríguez-García}, \citenamefont {Henry},\ and\
  \citenamefont {{et al.}}}]{CC5}%
  \BibitemOpen
  \bibfield  {author} {\bibinfo {author} {\bibfnamefont {P.}~\bibnamefont
  {Promdet}}, \bibinfo {author} {\bibfnamefont {B.}~\bibnamefont
  {Rodríguez-García}}, \bibinfo {author} {\bibfnamefont {A.}~\bibnamefont
  {Henry}}, \ and\ \bibinfo {author} {\bibnamefont {{et al.}}},\ }\href@noop {}
  {\bibfield  {journal} {\bibinfo  {journal} {Dalton Transactions}\ }\textbf
  {\bibinfo {volume} {47}},\ \bibinfo {pages} {11960} (\bibinfo {year}
  {2018})}\BibitemShut {NoStop}%
\bibitem [{\citenamefont {Ju}\ \emph {et~al.}(2022)\citenamefont {Ju},
  \citenamefont {Deng}, \citenamefont {Li},\ and\ \citenamefont {{et
  al.}}}]{CC6}%
  \BibitemOpen
  \bibfield  {author} {\bibinfo {author} {\bibfnamefont {X.~L.}\ \bibnamefont
  {Ju}}, \bibinfo {author} {\bibfnamefont {B.}~\bibnamefont {Deng}}, \bibinfo
  {author} {\bibfnamefont {K.}~\bibnamefont {Li}}, \ and\ \bibinfo {author}
  {\bibnamefont {{et al.}}},\ }\href@noop {} {\bibfield  {journal} {\bibinfo
  {journal} {Nuclear Science and Techniques}\ }\textbf {\bibinfo {volume}
  {33}},\ \bibinfo {pages} {1} (\bibinfo {year} {2022})}\BibitemShut {NoStop}%
\bibitem [{\citenamefont {Yu}\ \emph {et~al.}(2022)\citenamefont {Yu},
  \citenamefont {Wang}, \citenamefont {Li},\ and\ \citenamefont {{et
  al.}}}]{CC7}%
  \BibitemOpen
  \bibfield  {author} {\bibinfo {author} {\bibfnamefont {F.~C.}\ \bibnamefont
  {Yu}}, \bibinfo {author} {\bibfnamefont {F.~X.}\ \bibnamefont {Wang}},
  \bibinfo {author} {\bibfnamefont {K.}~\bibnamefont {Li}}, \ and\ \bibinfo
  {author} {\bibnamefont {{et al.}}},\ }\href@noop {} {\bibfield  {journal}
  {\bibinfo  {journal} {Journal of Synchrotron Radiation}\ }\textbf {\bibinfo
  {volume} {29}},\ \bibinfo {pages} {239} (\bibinfo {year} {2022})}\BibitemShut
  {NoStop}%
\bibitem [{\citenamefont {Wang}\ \emph
  {et~al.}(2020{\natexlab{b}})\citenamefont {Wang}, \citenamefont {Zhou},
  \citenamefont {Li}, \citenamefont {Mamtilahun}, \citenamefont {Tang},
  \citenamefont {Du}, \citenamefont {Deng}, \citenamefont {Xie}, \citenamefont
  {Yang},\ and\ \citenamefont {Xiao}}]{WZL:2020}%
  \BibitemOpen
  \bibfield  {author} {\bibinfo {author} {\bibfnamefont {F.}~\bibnamefont
  {Wang}}, \bibinfo {author} {\bibfnamefont {P.}~\bibnamefont {Zhou}}, \bibinfo
  {author} {\bibfnamefont {K.}~\bibnamefont {Li}}, \bibinfo {author}
  {\bibfnamefont {M.}~\bibnamefont {Mamtilahun}}, \bibinfo {author}
  {\bibfnamefont {Y.}~\bibnamefont {Tang}}, \bibinfo {author} {\bibfnamefont
  {G.}~\bibnamefont {Du}}, \bibinfo {author} {\bibfnamefont {B.}~\bibnamefont
  {Deng}}, \bibinfo {author} {\bibfnamefont {H.}~\bibnamefont {Xie}}, \bibinfo
  {author} {\bibfnamefont {G.}~\bibnamefont {Yang}}, \ and\ \bibinfo {author}
  {\bibfnamefont {T.}~\bibnamefont {Xiao}},\ }\href@noop {} {\bibfield
  {journal} {\bibinfo  {journal} {IUCrJ.}\ }\textbf {\bibinfo {volume} {7}},\
  \bibinfo {pages} {793} (\bibinfo {year} {2020}{\natexlab{b}})}\BibitemShut
  {NoStop}%
\bibitem [{\citenamefont {Xu}\ \emph {et~al.}(2023)\citenamefont {Xu},
  \citenamefont {Li}, \citenamefont {Xue}, \citenamefont {Wang}, \citenamefont
  {Liu}, \citenamefont {Song},\ and\ \citenamefont {Xiao}}]{XLX:2023}%
  \BibitemOpen
  \bibfield  {author} {\bibinfo {author} {\bibfnamefont {M.}~\bibnamefont
  {Xu}}, \bibinfo {author} {\bibfnamefont {K.}~\bibnamefont {Li}}, \bibinfo
  {author} {\bibfnamefont {Y.}~\bibnamefont {Xue}}, \bibinfo {author}
  {\bibfnamefont {F.}~\bibnamefont {Wang}}, \bibinfo {author} {\bibfnamefont
  {Z.}~\bibnamefont {Liu}}, \bibinfo {author} {\bibfnamefont {Z.}~\bibnamefont
  {Song}}, \ and\ \bibinfo {author} {\bibfnamefont {T.}~\bibnamefont {Xiao}},\
  }\href {\doibase 10.3389/fphy.2023.1174387} {\bibfield  {journal} {\bibinfo
  {journal} {Front. Phys.}\ }\textbf {\bibinfo {volume} {11}} (\bibinfo {year}
  {2023}),\ 10.3389/fphy.2023.1174387}\BibitemShut {NoStop}%
\bibitem [{\citenamefont {Xiao-Lu}\ \emph {et~al.}(2022)\citenamefont
  {Xiao-Lu}, \citenamefont {Ke}, \citenamefont {Fu-Cheng}, \citenamefont
  {Ming-Wei}, \citenamefont {Biao}, \citenamefont {Bin},\ and\ \citenamefont
  {Xiao}}]{JLY:2022}%
  \BibitemOpen
  \bibfield  {author} {\bibinfo {author} {\bibfnamefont {J.}~\bibnamefont
  {Xiao-Lu}}, \bibinfo {author} {\bibfnamefont {L.}~\bibnamefont {Ke}},
  \bibinfo {author} {\bibfnamefont {Y.}~\bibnamefont {Fu-Cheng}}, \bibinfo
  {author} {\bibfnamefont {X.}~\bibnamefont {Ming-Wei}}, \bibinfo {author}
  {\bibfnamefont {D.}~\bibnamefont {Biao}}, \bibinfo {author} {\bibfnamefont
  {L.}~\bibnamefont {Bin}}, \ and\ \bibinfo {author} {\bibfnamefont
  {T.}~\bibnamefont {Xiao}},\ }\href {\doibase 10.7498/aps.71.20220339}
  {\bibfield  {journal} {\bibinfo  {journal} {Acta Phys. Sin.}\ }\textbf
  {\bibinfo {volume} {71}} (\bibinfo {year} {2022}),\
  10.7498/aps.71.20220339}\BibitemShut {NoStop}%
\bibitem [{\citenamefont {Bokman}\ \emph {et~al.}(2023)\citenamefont {Bokman},
  \citenamefont {Biasiori-Poulanges}, \citenamefont {Luki{\'c}}, \citenamefont
  {Bourquard}, \citenamefont {Meyer}, \citenamefont {Rack},\ and\ \citenamefont
  {Supponen}}]{BBP:2023}%
  \BibitemOpen
  \bibfield  {author} {\bibinfo {author} {\bibfnamefont {G.~T.}\ \bibnamefont
  {Bokman}}, \bibinfo {author} {\bibfnamefont {L.}~\bibnamefont
  {Biasiori-Poulanges}}, \bibinfo {author} {\bibfnamefont {B.}~\bibnamefont
  {Luki{\'c}}}, \bibinfo {author} {\bibfnamefont {C.}~\bibnamefont
  {Bourquard}}, \bibinfo {author} {\bibfnamefont {D.~W.}\ \bibnamefont
  {Meyer}}, \bibinfo {author} {\bibfnamefont {A.}~\bibnamefont {Rack}}, \ and\
  \bibinfo {author} {\bibfnamefont {O.}~\bibnamefont {Supponen}},\ }\href
  {\doibase 10.1063/5.0132104} {\bibfield  {journal} {\bibinfo  {journal}
  {Physics of Fluids}\ }\textbf {\bibinfo {volume} {35}},\ \bibinfo {pages}
  {013322} (\bibinfo {year} {2023})}\BibitemShut {NoStop}%
\bibitem [{\citenamefont {Sabzeghabae}\ \emph {et~al.}(2021)\citenamefont
  {Sabzeghabae}, \citenamefont {Devia-Cruz}, \citenamefont {Gutierrez-Herrera},
  \citenamefont {Camacho-Lopez},\ and\ \citenamefont {Aguilar}}]{CC14}%
  \BibitemOpen
  \bibfield  {author} {\bibinfo {author} {\bibfnamefont {A.~N.}\ \bibnamefont
  {Sabzeghabae}}, \bibinfo {author} {\bibfnamefont {L.~F.}\ \bibnamefont
  {Devia-Cruz}}, \bibinfo {author} {\bibfnamefont {E.}~\bibnamefont
  {Gutierrez-Herrera}}, \bibinfo {author} {\bibfnamefont {S.}~\bibnamefont
  {Camacho-Lopez}}, \ and\ \bibinfo {author} {\bibfnamefont {G.}~\bibnamefont
  {Aguilar}},\ }\href@noop {} {\bibfield  {journal} {\bibinfo  {journal}
  {Optics \& Laser Technology}\ }\textbf {\bibinfo {volume} {134}},\ \bibinfo
  {pages} {106621} (\bibinfo {year} {2021})}\BibitemShut {NoStop}%
\bibitem [{\citenamefont {Vogel}\ \emph {et~al.}(1989)\citenamefont {Vogel},
  \citenamefont {Lauterborn},\ and\ \citenamefont {Timm}}]{CC15}%
  \BibitemOpen
  \bibfield  {author} {\bibinfo {author} {\bibfnamefont {A.}~\bibnamefont
  {Vogel}}, \bibinfo {author} {\bibfnamefont {W.}~\bibnamefont {Lauterborn}}, \
  and\ \bibinfo {author} {\bibfnamefont {R.}~\bibnamefont {Timm}},\ }\href@noop
  {} {\bibfield  {journal} {\bibinfo  {journal} {Journal of Fluid Mechanics}\
  }\textbf {\bibinfo {volume} {206}},\ \bibinfo {pages} {299} (\bibinfo {year}
  {1989})}\BibitemShut {NoStop}%
\bibitem [{\citenamefont {Sadrozinski}\ \emph {et~al.}(2018)\citenamefont
  {Sadrozinski}, \citenamefont {Seiden},\ and\ \citenamefont
  {Cartiglia}}]{SSC:2017}%
  \BibitemOpen
  \bibfield  {author} {\bibinfo {author} {\bibfnamefont {H.~F.-W.}\
  \bibnamefont {Sadrozinski}}, \bibinfo {author} {\bibfnamefont
  {A.}~\bibnamefont {Seiden}}, \ and\ \bibinfo {author} {\bibfnamefont
  {N.}~\bibnamefont {Cartiglia}},\ }\href@noop {} {\bibfield  {journal}
  {\bibinfo  {journal} {Rep. Prog. Phys.}\ }\textbf {\bibinfo {volume} {81}},\
  \bibinfo {pages} {026101} (\bibinfo {year} {2018})}\BibitemShut {NoStop}%
\bibitem [{\citenamefont {Cartiglia}\ \emph {et~al.}(2017)\citenamefont
  {Cartiglia}, \citenamefont {Arcidiacono}, \citenamefont {Baldassarri},
  \citenamefont {Boscardin}, \citenamefont {Cenna}, \citenamefont {Dellacasa},
  \citenamefont {Betta}, \citenamefont {Ferrero}, \citenamefont {Fadeyev},
  \citenamefont {Galloway}, \citenamefont {Garbolino}, \citenamefont {Grabas},
  \citenamefont {Monaco}, \citenamefont {Obertino}, \citenamefont {Pancheri},
  \citenamefont {Paternoster}, \citenamefont {Rivetti}, \citenamefont {Rolo},
  \citenamefont {Sacchi}, \citenamefont {Sadrozinski}, \citenamefont {Seiden},
  \citenamefont {Sola}, \citenamefont {Solano}, \citenamefont {Staiano},
  \citenamefont {Ravera},\ and\ \citenamefont {Zatserklyaniy}}]{CAB:2017}%
  \BibitemOpen
  \bibfield  {author} {\bibinfo {author} {\bibfnamefont {N.}~\bibnamefont
  {Cartiglia}}, \bibinfo {author} {\bibfnamefont {R.}~\bibnamefont
  {Arcidiacono}}, \bibinfo {author} {\bibfnamefont {B.}~\bibnamefont
  {Baldassarri}}, \bibinfo {author} {\bibfnamefont {M.}~\bibnamefont
  {Boscardin}}, \bibinfo {author} {\bibfnamefont {F.}~\bibnamefont {Cenna}},
  \bibinfo {author} {\bibfnamefont {G.}~\bibnamefont {Dellacasa}}, \bibinfo
  {author} {\bibfnamefont {G.-F.~D.}\ \bibnamefont {Betta}}, \bibinfo {author}
  {\bibfnamefont {M.}~\bibnamefont {Ferrero}}, \bibinfo {author} {\bibfnamefont
  {V.}~\bibnamefont {Fadeyev}}, \bibinfo {author} {\bibfnamefont
  {Z.}~\bibnamefont {Galloway}}, \bibinfo {author} {\bibfnamefont
  {S.}~\bibnamefont {Garbolino}}, \bibinfo {author} {\bibfnamefont
  {H.}~\bibnamefont {Grabas}}, \bibinfo {author} {\bibfnamefont
  {V.}~\bibnamefont {Monaco}}, \bibinfo {author} {\bibfnamefont
  {M.}~\bibnamefont {Obertino}}, \bibinfo {author} {\bibfnamefont
  {L.}~\bibnamefont {Pancheri}}, \bibinfo {author} {\bibfnamefont
  {G.}~\bibnamefont {Paternoster}}, \bibinfo {author} {\bibfnamefont
  {A.}~\bibnamefont {Rivetti}}, \bibinfo {author} {\bibfnamefont
  {M.}~\bibnamefont {Rolo}}, \bibinfo {author} {\bibfnamefont {R.}~\bibnamefont
  {Sacchi}}, \bibinfo {author} {\bibfnamefont {H.}~\bibnamefont {Sadrozinski}},
  \bibinfo {author} {\bibfnamefont {A.}~\bibnamefont {Seiden}}, \bibinfo
  {author} {\bibfnamefont {V.}~\bibnamefont {Sola}}, \bibinfo {author}
  {\bibfnamefont {A.}~\bibnamefont {Solano}}, \bibinfo {author} {\bibfnamefont
  {A.}~\bibnamefont {Staiano}}, \bibinfo {author} {\bibfnamefont
  {F.}~\bibnamefont {Ravera}}, \ and\ \bibinfo {author} {\bibfnamefont
  {A.}~\bibnamefont {Zatserklyaniy}},\ }\href@noop {} {\bibfield  {journal}
  {\bibinfo  {journal} {Nucl. Instrum. Meth. A}\ }\textbf {\bibinfo {volume}
  {845}},\ \bibinfo {pages} {47} (\bibinfo {year} {2017})}\BibitemShut
  {NoStop}%
\bibitem [{\citenamefont {Wang}\ and\ \citenamefont {Morris}(2013)}]{WM:2013}%
  \BibitemOpen
  \bibfield  {author} {\bibinfo {author} {\bibfnamefont {Z.}~\bibnamefont
  {Wang}}\ and\ \bibinfo {author} {\bibfnamefont {C.~L.}\ \bibnamefont
  {Morris}},\ }\href@noop {} {\bibfield  {journal} {\bibinfo  {journal} {Nucl.
  Instrum. Meth. A}\ }\textbf {\bibinfo {volume} {726}},\ \bibinfo {pages}
  {145} (\bibinfo {year} {2013})}\BibitemShut {NoStop}%
\bibitem [{\citenamefont {Chu}\ \emph {et~al.}(2022)\citenamefont {Chu},
  \citenamefont {James},\ and\ \citenamefont {Wang}}]{CJW:2022}%
  \BibitemOpen
  \bibfield  {author} {\bibinfo {author} {\bibfnamefont {P.}~\bibnamefont
  {Chu}}, \bibinfo {author} {\bibfnamefont {M.~R.}\ \bibnamefont {James}}, \
  and\ \bibinfo {author} {\bibfnamefont {Z.}~\bibnamefont {Wang}},\ }\href@noop
  {} {\bibfield  {journal} {\bibinfo  {journal} {J. Nucl. Eng.}\ }\textbf
  {\bibinfo {volume} {3(2)}},\ \bibinfo {pages} {117} (\bibinfo {year}
  {2022})}\BibitemShut {NoStop}%
\bibitem [{xfe(2020)}]{xfel:2020}%
  \BibitemOpen
  \href@noop {} {\enquote {\bibinfo {title} {The next decade of {XFELs}},}\ }
  (\bibinfo {year} {2020}),\ \bibinfo {note} {{Nat. Rev. Phys.} {\bf 2}, 329.
  {\it https://doi.org/10.1038/s42254-020-0206-4}}\BibitemShut {NoStop}%
\bibitem [{\citenamefont {Kalinin}\ \emph {et~al.}(2015)\citenamefont
  {Kalinin}, \citenamefont {Sumpter},\ and\ \citenamefont
  {Archibald}}]{KSA:2015}%
  \BibitemOpen
  \bibfield  {author} {\bibinfo {author} {\bibfnamefont {S.~V.}\ \bibnamefont
  {Kalinin}}, \bibinfo {author} {\bibfnamefont {B.~G.}\ \bibnamefont
  {Sumpter}}, \ and\ \bibinfo {author} {\bibfnamefont {R.~K.}\ \bibnamefont
  {Archibald}},\ }\href@noop {} {\bibfield  {journal} {\bibinfo  {journal}
  {Nat. Mater.}\ }\textbf {\bibinfo {volume} {14}},\ \bibinfo {pages} {973 }
  (\bibinfo {year} {2015})}\BibitemShut {NoStop}%
\bibitem [{\citenamefont {Becher}\ \emph {et~al.}(2019)\citenamefont {Becher},
  \citenamefont {Sheppard}, \citenamefont {Fam}, \citenamefont {Baier},
  \citenamefont {Wang}, \citenamefont {Wang}, \citenamefont {Kulkarni},
  \citenamefont {Keller}, \citenamefont {Lyubomirskiy}, \citenamefont
  {Brueckner}, \citenamefont {Kahnt}, \citenamefont {Schropp}, \citenamefont
  {Schroer},\ and\ \citenamefont {Grunwaldt}}]{BSF:2019}%
  \BibitemOpen
  \bibfield  {author} {\bibinfo {author} {\bibfnamefont {J.}~\bibnamefont
  {Becher}}, \bibinfo {author} {\bibfnamefont {T.~L.}\ \bibnamefont
  {Sheppard}}, \bibinfo {author} {\bibfnamefont {Y.}~\bibnamefont {Fam}},
  \bibinfo {author} {\bibfnamefont {S.}~\bibnamefont {Baier}}, \bibinfo
  {author} {\bibfnamefont {W.}~\bibnamefont {Wang}}, \bibinfo {author}
  {\bibfnamefont {D.}~\bibnamefont {Wang}}, \bibinfo {author} {\bibfnamefont
  {S.}~\bibnamefont {Kulkarni}}, \bibinfo {author} {\bibfnamefont {T.~F.}\
  \bibnamefont {Keller}}, \bibinfo {author} {\bibfnamefont {M.}~\bibnamefont
  {Lyubomirskiy}}, \bibinfo {author} {\bibfnamefont {D.}~\bibnamefont
  {Brueckner}}, \bibinfo {author} {\bibfnamefont {M.}~\bibnamefont {Kahnt}},
  \bibinfo {author} {\bibfnamefont {A.}~\bibnamefont {Schropp}}, \bibinfo
  {author} {\bibfnamefont {C.~G.}\ \bibnamefont {Schroer}}, \ and\ \bibinfo
  {author} {\bibfnamefont {J.-D.}\ \bibnamefont {Grunwaldt}},\ }\href {\doibase
  10.1021/acs.jpcc.9b06541} {\bibfield  {journal} {\bibinfo  {journal} {J.
  Phys. Chem.}\ }\textbf {\bibinfo {volume} {123}},\ \bibinfo {pages} {25197}
  (\bibinfo {year} {2019})}\BibitemShut {NoStop}%
\bibitem [{\citenamefont {{X-Ray and Gamma-Ray Data}}(2023)}]{NIST:23}%
  \BibitemOpen
  \bibfield  {author} {\bibinfo {author} {\bibnamefont {{X-Ray and Gamma-Ray
  Data}}},\ }\href@noop {} {} (\bibinfo {year} {2023}),\ \bibinfo {note}
  {https://www.nist.gov/pml/x-ray-and-gamma-ray-data}\BibitemShut {NoStop}%
\bibitem [{\citenamefont {{NIF fun facts}}(2022)}]{NIF:22}%
  \BibitemOpen
  \bibfield  {author} {\bibinfo {author} {\bibnamefont {{NIF fun facts}}},\
  }\href@noop {} {} (\bibinfo {year} {2022}),\ \bibinfo {note}
  {https://lasers.llnl.gov/news/press-kit}\BibitemShut {NoStop}%
\bibitem [{\citenamefont {Lyubomirskiy}\ \emph {et~al.}(2022)\citenamefont
  {Lyubomirskiy}, \citenamefont {Wittwer}, \citenamefont {Kahnt}, \citenamefont
  {Koch}, \citenamefont {Kubec}, \citenamefont {Falch}, \citenamefont
  {Garrevoet}, \citenamefont {Seyrich}, \citenamefont {David},\ and\
  \citenamefont {Schroer}}]{LWK:2022}%
  \BibitemOpen
  \bibfield  {author} {\bibinfo {author} {\bibfnamefont {M.}~\bibnamefont
  {Lyubomirskiy}}, \bibinfo {author} {\bibfnamefont {F.}~\bibnamefont
  {Wittwer}}, \bibinfo {author} {\bibfnamefont {M.}~\bibnamefont {Kahnt}},
  \bibinfo {author} {\bibfnamefont {F.}~\bibnamefont {Koch}}, \bibinfo {author}
  {\bibfnamefont {A.}~\bibnamefont {Kubec}}, \bibinfo {author} {\bibfnamefont
  {K.~V.}\ \bibnamefont {Falch}}, \bibinfo {author} {\bibfnamefont
  {J.}~\bibnamefont {Garrevoet}}, \bibinfo {author} {\bibfnamefont
  {M.}~\bibnamefont {Seyrich}}, \bibinfo {author} {\bibfnamefont
  {C.}~\bibnamefont {David}}, \ and\ \bibinfo {author} {\bibfnamefont {C.~G.}\
  \bibnamefont {Schroer}},\ }\href {\doibase 10.1038/s41598-022-09466-5}
  {\bibfield  {journal} {\bibinfo  {journal} {Sci. Rep.}\ }\textbf {\bibinfo
  {volume} {12}},\ \bibinfo {pages} {6203} (\bibinfo {year}
  {2022})}\BibitemShut {NoStop}%
\bibitem [{\citenamefont {Herbst}\ \emph {et~al.}(2023)\citenamefont {Herbst},
  \citenamefont {Coffee}, \citenamefont {Fronk}, \citenamefont {Kim},
  \citenamefont {Kim}, \citenamefont {Ruckman},\ and\ \citenamefont
  {Russell}}]{HCF:2023}%
  \BibitemOpen
  \bibfield  {author} {\bibinfo {author} {\bibfnamefont {R.}~\bibnamefont
  {Herbst}}, \bibinfo {author} {\bibfnamefont {R.}~\bibnamefont {Coffee}},
  \bibinfo {author} {\bibfnamefont {N.}~\bibnamefont {Fronk}}, \bibinfo
  {author} {\bibfnamefont {K.}~\bibnamefont {Kim}}, \bibinfo {author}
  {\bibfnamefont {K.}~\bibnamefont {Kim}}, \bibinfo {author} {\bibfnamefont
  {L.}~\bibnamefont {Ruckman}}, \ and\ \bibinfo {author} {\bibfnamefont
  {J.~J.}\ \bibnamefont {Russell}},\ }in\ \href@noop {} {\emph {\bibinfo
  {booktitle} {Accelerating Science and Engineering Discoveries Through
  Integrated Research Infrastructure for Experiment, Big Data, Modeling and
  Simulation. SMC 2022. Communications in Computer and Information Science}}},\
  Vol.\ \bibinfo {volume} {1690},\ \bibinfo {editor} {edited by\ \bibinfo
  {editor} {\bibfnamefont {G.}~\bibnamefont {Doug}, \bibfnamefont {K.and~Al}},
  \bibinfo {editor} {\bibfnamefont {S.}~\bibnamefont {Pophale}}, \bibinfo
  {editor} {\bibfnamefont {H.}~\bibnamefont {Liu}}, \ and\ \bibinfo {editor}
  {\bibfnamefont {S.}~\bibnamefont {Parete-Koon}}}\ (\bibinfo  {publisher}
  {Springer},\ \bibinfo {address} {Switzerland},\ \bibinfo {year} {2023})\ p.\
  \bibinfo {pages} {120}\BibitemShut {NoStop}%
\bibitem [{\citenamefont {Hu}\ \emph {et~al.}(2018)\citenamefont {Hu},
  \citenamefont {Zhang}, \citenamefont {Zhu}, \citenamefont {Chen},
  \citenamefont {Wang}, \citenamefont {Ying},\ and\ \citenamefont
  {Yu}}]{HZZ:2018}%
  \BibitemOpen
  \bibfield  {author} {\bibinfo {author} {\bibfnamefont {C.}~\bibnamefont
  {Hu}}, \bibinfo {author} {\bibfnamefont {L.}~\bibnamefont {Zhang}}, \bibinfo
  {author} {\bibfnamefont {R.-Y.}\ \bibnamefont {Zhu}}, \bibinfo {author}
  {\bibfnamefont {A.}~\bibnamefont {Chen}}, \bibinfo {author} {\bibfnamefont
  {Z.}~\bibnamefont {Wang}}, \bibinfo {author} {\bibfnamefont {L.}~\bibnamefont
  {Ying}}, \ and\ \bibinfo {author} {\bibfnamefont {Z.}~\bibnamefont {Yu}},\
  }\href {\doibase 10.1109/TNS.2018.2808103} {\bibfield  {journal} {\bibinfo
  {journal} {IEEE Transactions on Nuclear Science}\ }\textbf {\bibinfo {volume}
  {65}},\ \bibinfo {pages} {2097} (\bibinfo {year} {2018})}\BibitemShut
  {NoStop}%
\bibitem [{\citenamefont {Hu}\ \emph {et~al.}(2019)\citenamefont {Hu},
  \citenamefont {Zhang}, \citenamefont {Zhu}, \citenamefont {Demarteau},
  \citenamefont {Wagner}, \citenamefont {Xia}, \citenamefont {Xie},
  \citenamefont {Li}, \citenamefont {Wang}, \citenamefont {Shih},\ and\
  \citenamefont {Smith}}]{HZZ:2019}%
  \BibitemOpen
  \bibfield  {author} {\bibinfo {author} {\bibfnamefont {C.}~\bibnamefont
  {Hu}}, \bibinfo {author} {\bibfnamefont {L.}~\bibnamefont {Zhang}}, \bibinfo
  {author} {\bibfnamefont {R.-Y.}\ \bibnamefont {Zhu}}, \bibinfo {author}
  {\bibfnamefont {M.}~\bibnamefont {Demarteau}}, \bibinfo {author}
  {\bibfnamefont {R.}~\bibnamefont {Wagner}}, \bibinfo {author} {\bibfnamefont
  {L.}~\bibnamefont {Xia}}, \bibinfo {author} {\bibfnamefont {J.}~\bibnamefont
  {Xie}}, \bibinfo {author} {\bibfnamefont {X.}~\bibnamefont {Li}}, \bibinfo
  {author} {\bibfnamefont {Z.}~\bibnamefont {Wang}}, \bibinfo {author}
  {\bibfnamefont {Y.}~\bibnamefont {Shih}}, \ and\ \bibinfo {author}
  {\bibfnamefont {T.}~\bibnamefont {Smith}},\ }\href {\doibase
  10.1016/j.nima.2019.06.011} {\bibfield  {journal} {\bibinfo  {journal} {Nucl.
  Instrum. Meth. Phys. Res. Sec. A}\ }\textbf {\bibinfo {volume} {940}},\
  \bibinfo {pages} {223} (\bibinfo {year} {2019})}\BibitemShut {NoStop}%
\bibitem [{\citenamefont {Turchetta}(2017)}]{Turch:2017}%
  \BibitemOpen
  \bibfield  {author} {\bibinfo {author} {\bibfnamefont {R.}~\bibnamefont
  {Turchetta}},\ }\href@noop {} {\enquote {\bibinfo {title} {{Towards Gfps CMOS
  image sensors}},}\ } (\bibinfo {year} {2017}),\ \bibinfo {note} {in Workshop
  on Computational Image Sensors and Smart Cameras (WASC) 2017}\BibitemShut
  {NoStop}%
\bibitem [{\citenamefont {Carini}\ \emph {et~al.}(2016)\citenamefont {Carini},
  \citenamefont {Alonso-Mori}, \citenamefont {Blaj}, \citenamefont {Caragiulo},
  \citenamefont {Chollet}, \citenamefont {Damiani}, \citenamefont {Dragone},
  \citenamefont {Feng}, \citenamefont {Haller}, \citenamefont {Hart},
  \citenamefont {Hasi}, \citenamefont {Herbst}, \citenamefont {Herrmann},
  \citenamefont {Kenney}, \citenamefont {Lemke}, \citenamefont {Manger},
  \citenamefont {Markovic}, \citenamefont {Mehta}, \citenamefont {Nelson},
  \citenamefont {Nishimura}, \citenamefont {Osier}, \citenamefont {Pines},
  \citenamefont {Reese}, \citenamefont {Robert}, \citenamefont {Segal},
  \citenamefont {Sikorski}, \citenamefont {Song}, \citenamefont {Thayer},
  \citenamefont {Tomada}, \citenamefont {Weaver},\ and\ \citenamefont
  {Zhu}}]{CAB:2016}%
  \BibitemOpen
  \bibfield  {author} {\bibinfo {author} {\bibfnamefont {G.~A.}\ \bibnamefont
  {Carini}}, \bibinfo {author} {\bibfnamefont {R.}~\bibnamefont {Alonso-Mori}},
  \bibinfo {author} {\bibfnamefont {G.}~\bibnamefont {Blaj}}, \bibinfo {author}
  {\bibfnamefont {P.}~\bibnamefont {Caragiulo}}, \bibinfo {author}
  {\bibfnamefont {M.}~\bibnamefont {Chollet}}, \bibinfo {author} {\bibfnamefont
  {D.}~\bibnamefont {Damiani}}, \bibinfo {author} {\bibfnamefont
  {A.}~\bibnamefont {Dragone}}, \bibinfo {author} {\bibfnamefont
  {Y.}~\bibnamefont {Feng}}, \bibinfo {author} {\bibfnamefont {G.}~\bibnamefont
  {Haller}}, \bibinfo {author} {\bibfnamefont {P.}~\bibnamefont {Hart}},
  \bibinfo {author} {\bibfnamefont {J.}~\bibnamefont {Hasi}}, \bibinfo {author}
  {\bibfnamefont {R.}~\bibnamefont {Herbst}}, \bibinfo {author} {\bibfnamefont
  {S.}~\bibnamefont {Herrmann}}, \bibinfo {author} {\bibfnamefont
  {C.}~\bibnamefont {Kenney}}, \bibinfo {author} {\bibfnamefont
  {H.}~\bibnamefont {Lemke}}, \bibinfo {author} {\bibfnamefont
  {L.}~\bibnamefont {Manger}}, \bibinfo {author} {\bibfnamefont
  {B.}~\bibnamefont {Markovic}}, \bibinfo {author} {\bibfnamefont
  {A.}~\bibnamefont {Mehta}}, \bibinfo {author} {\bibfnamefont
  {S.}~\bibnamefont {Nelson}}, \bibinfo {author} {\bibfnamefont
  {K.}~\bibnamefont {Nishimura}}, \bibinfo {author} {\bibfnamefont
  {S.}~\bibnamefont {Osier}}, \bibinfo {author} {\bibfnamefont
  {J.}~\bibnamefont {Pines}}, \bibinfo {author} {\bibfnamefont
  {B.}~\bibnamefont {Reese}}, \bibinfo {author} {\bibfnamefont
  {A.}~\bibnamefont {Robert}}, \bibinfo {author} {\bibfnamefont
  {J.}~\bibnamefont {Segal}}, \bibinfo {author} {\bibfnamefont
  {M.}~\bibnamefont {Sikorski}}, \bibinfo {author} {\bibfnamefont
  {S.}~\bibnamefont {Song}}, \bibinfo {author} {\bibfnamefont {J.}~\bibnamefont
  {Thayer}}, \bibinfo {author} {\bibfnamefont {A.}~\bibnamefont {Tomada}},
  \bibinfo {author} {\bibfnamefont {M.}~\bibnamefont {Weaver}}, \ and\ \bibinfo
  {author} {\bibfnamefont {D.}~\bibnamefont {Zhu}},\ }\href@noop {} {\bibfield
  {journal} {\bibinfo  {journal} {AIP Conf. Proc.}\ }\textbf {\bibinfo {volume}
  {1741}},\ \bibinfo {pages} {040008} (\bibinfo {year} {2016})}\BibitemShut
  {NoStop}%
\bibitem [{\citenamefont {Veale}\ \emph {et~al.}(2018)\citenamefont {Veale},
  \citenamefont {Seller}, \citenamefont {Wilson},\ and\ \citenamefont
  {Liotti}}]{VSW:2018}%
  \BibitemOpen
  \bibfield  {author} {\bibinfo {author} {\bibfnamefont {M.~C.}\ \bibnamefont
  {Veale}}, \bibinfo {author} {\bibfnamefont {P.}~\bibnamefont {Seller}},
  \bibinfo {author} {\bibfnamefont {M.}~\bibnamefont {Wilson}}, \ and\ \bibinfo
  {author} {\bibfnamefont {E.}~\bibnamefont {Liotti}},\ }\href@noop {}
  {\bibfield  {journal} {\bibinfo  {journal} {Synchr. Rad. News}\ }\textbf
  {\bibinfo {volume} {31(6)}},\ \bibinfo {pages} {28} (\bibinfo {year}
  {2018})}\BibitemShut {NoStop}%
\bibitem [{\citenamefont {Claus}\ \emph {et~al.}(2017)\citenamefont {Claus},
  \citenamefont {England}, \citenamefont {Fang}, \citenamefont {Robertson},
  \citenamefont {Sanchez}, \citenamefont {Trotter}, \citenamefont {Carpenter},
  \citenamefont {Dayton}, \citenamefont {Patel},\ and\ \citenamefont
  {Porter}}]{CEF:2017}%
  \BibitemOpen
  \bibfield  {author} {\bibinfo {author} {\bibfnamefont {L.}~\bibnamefont
  {Claus}}, \bibinfo {author} {\bibfnamefont {T.}~\bibnamefont {England}},
  \bibinfo {author} {\bibfnamefont {L.}~\bibnamefont {Fang}}, \bibinfo {author}
  {\bibfnamefont {G.}~\bibnamefont {Robertson}}, \bibinfo {author}
  {\bibfnamefont {M.}~\bibnamefont {Sanchez}}, \bibinfo {author} {\bibfnamefont
  {D.}~\bibnamefont {Trotter}}, \bibinfo {author} {\bibfnamefont
  {A.}~\bibnamefont {Carpenter}}, \bibinfo {author} {\bibfnamefont
  {M.}~\bibnamefont {Dayton}}, \bibinfo {author} {\bibfnamefont
  {P.}~\bibnamefont {Patel}}, \ and\ \bibinfo {author} {\bibfnamefont {J.~L.}\
  \bibnamefont {Porter}},\ }\href@noop {} {\bibfield  {journal} {\bibinfo
  {journal} {SPIE Proc.}\ }\textbf {\bibinfo {volume} {10390}},\ \bibinfo
  {pages} {103900A} (\bibinfo {year} {2017})}\BibitemShut {NoStop}%
\bibitem [{\citenamefont {Lewis}\ \emph {et~al.}(2021)\citenamefont {Lewis},
  \citenamefont {Baker}, \citenamefont {Corredor}, \citenamefont {Fitzpatrick},
  \citenamefont {Jones}, \citenamefont {O’Flarity}, \citenamefont {Walters},
  \citenamefont {Claus},\ and\ \citenamefont {Sanchez}}]{LBC:2021}%
  \BibitemOpen
  \bibfield  {author} {\bibinfo {author} {\bibfnamefont {A.}~\bibnamefont
  {Lewis}}, \bibinfo {author} {\bibfnamefont {S.}~\bibnamefont {Baker}},
  \bibinfo {author} {\bibfnamefont {A.}~\bibnamefont {Corredor}}, \bibinfo
  {author} {\bibfnamefont {L.~F.~Z.}\ \bibnamefont {Fitzpatrick}}, \bibinfo
  {author} {\bibfnamefont {M.}~\bibnamefont {Jones}}, \bibinfo {author}
  {\bibfnamefont {K.}~\bibnamefont {O’Flarity}}, \bibinfo {author}
  {\bibfnamefont {K.}~\bibnamefont {Walters}}, \bibinfo {author} {\bibfnamefont
  {L.}~\bibnamefont {Claus}}, \ and\ \bibinfo {author} {\bibfnamefont
  {M.}~\bibnamefont {Sanchez}},\ }\href@noop {} {\bibfield  {journal} {\bibinfo
   {journal} {Rev. Sci. Instrum.}\ }\textbf {\bibinfo {volume} {92}},\ \bibinfo
  {pages} {083103} (\bibinfo {year} {2021})},\ \bibinfo {note}
  {https://doi.org/10.1063/5.0049110}\BibitemShut {NoStop}%
\bibitem [{\citenamefont {Kwiatkowski}\ \emph {et~al.}(2014)\citenamefont
  {Kwiatkowski}, \citenamefont {V.~Douence}, \citenamefont {P.~Nedrow},
  \citenamefont {Merrill}, \citenamefont {Morris},\ and\ \citenamefont
  {Saunders}}]{KDB:2014}%
  \BibitemOpen
  \bibfield  {author} {\bibinfo {author} {\bibfnamefont {K.}~\bibnamefont
  {Kwiatkowski}}, \bibinfo {author} {\bibfnamefont {Y.~B.}\ \bibnamefont
  {V.~Douence}}, \bibinfo {author} {\bibfnamefont {F.~M.}\ \bibnamefont
  {P.~Nedrow}}, \bibinfo {author} {\bibfnamefont {F.}~\bibnamefont {Merrill}},
  \bibinfo {author} {\bibfnamefont {C.~L.}\ \bibnamefont {Morris}}, \ and\
  \bibinfo {author} {\bibfnamefont {A.}~\bibnamefont {Saunders}},\ }\href@noop
  {} {\bibfield  {journal} {\bibinfo  {journal} {SPIE Proc.}\ }\textbf
  {\bibinfo {volume} {9215}},\ \bibinfo {pages} {921506} (\bibinfo {year}
  {2014})}\BibitemShut {NoStop}%
\bibitem [{\citenamefont {Hirschman}\ \emph {et~al.}(2023)\citenamefont
  {Hirschman}, \citenamefont {Kamalov}, \citenamefont {Obaid}, \citenamefont
  {O’Shea},\ and\ \citenamefont {Coffee}}]{HKO:2023}%
  \BibitemOpen
  \bibfield  {author} {\bibinfo {author} {\bibfnamefont {J.}~\bibnamefont
  {Hirschman}}, \bibinfo {author} {\bibfnamefont {A.}~\bibnamefont {Kamalov}},
  \bibinfo {author} {\bibfnamefont {R.}~\bibnamefont {Obaid}}, \bibinfo
  {author} {\bibfnamefont {F.~H.}\ \bibnamefont {O’Shea}}, \ and\ \bibinfo
  {author} {\bibfnamefont {R.~N.}\ \bibnamefont {Coffee}},\ }in\ \href@noop {}
  {\emph {\bibinfo {booktitle} {Accelerating Science and Engineering
  Discoveries Through Integrated Research Infrastructure for Experiment, Big
  Data, Modeling and Simulation. SMC 2022. Communications in Computer and
  Information Science}}},\ Vol.\ \bibinfo {volume} {1690},\ \bibinfo {editor}
  {edited by\ \bibinfo {editor} {\bibfnamefont {G.}~\bibnamefont {Doug},
  \bibfnamefont {K.and~Al}}, \bibinfo {editor} {\bibfnamefont {S.}~\bibnamefont
  {Pophale}}, \bibinfo {editor} {\bibfnamefont {H.}~\bibnamefont {Liu}}, \ and\
  \bibinfo {editor} {\bibfnamefont {S.}~\bibnamefont {Parete-Koon}}}\ (\bibinfo
   {publisher} {Springer},\ \bibinfo {address} {Switzerland},\ \bibinfo {year}
  {2023})\ p.\ \bibinfo {pages} {101}\BibitemShut {NoStop}%
\bibitem [{\citenamefont {Kilic}\ \emph {et~al.}(2023)\citenamefont {Kilic},
  \citenamefont {Tran},\ and\ \citenamefont {Foster}}]{KTF:2023}%
  \BibitemOpen
  \bibfield  {author} {\bibinfo {author} {\bibfnamefont {V.}~\bibnamefont
  {Kilic}}, \bibinfo {author} {\bibfnamefont {T.~D.}\ \bibnamefont {Tran}}, \
  and\ \bibinfo {author} {\bibfnamefont {M.~A.}\ \bibnamefont {Foster}},\
  }\href@noop {} {\bibfield  {journal} {\bibinfo  {journal} {J. Opt. Soc. Am.
  B}\ }\textbf {\bibinfo {volume} {40}},\ \bibinfo {pages} {28} (\bibinfo
  {year} {2023})}\BibitemShut {NoStop}%
\bibitem [{\citenamefont {Candes}\ \emph {et~al.}(2006)\citenamefont {Candes},
  \citenamefont {Romberg},\ and\ \citenamefont {Tao}}]{CRT:2006}%
  \BibitemOpen
  \bibfield  {author} {\bibinfo {author} {\bibfnamefont {E.}~\bibnamefont
  {Candes}}, \bibinfo {author} {\bibfnamefont {J.}~\bibnamefont {Romberg}}, \
  and\ \bibinfo {author} {\bibfnamefont {T.}~\bibnamefont {Tao}},\ }\href@noop
  {} {\bibfield  {journal} {\bibinfo  {journal} {IEEE Trans. Inf. Theory}\
  }\textbf {\bibinfo {volume} {52}},\ \bibinfo {pages} {489} (\bibinfo {year}
  {2006})}\BibitemShut {NoStop}%
\bibitem [{\citenamefont {Donoho}(2006)}]{Don:2006}%
  \BibitemOpen
  \bibfield  {author} {\bibinfo {author} {\bibfnamefont {D.}~\bibnamefont
  {Donoho}},\ }\href@noop {} {\bibfield  {journal} {\bibinfo  {journal} {IEEE
  Trans. Info. Theory}\ }\textbf {\bibinfo {volume} {52}},\ \bibinfo {pages}
  {1289} (\bibinfo {year} {Apr. 2006})}\BibitemShut {NoStop}%
\bibitem [{\citenamefont {Candes}\ and\ \citenamefont {Tao}(2006)}]{CT:2006}%
  \BibitemOpen
  \bibfield  {author} {\bibinfo {author} {\bibfnamefont {E.}~\bibnamefont
  {Candes}}\ and\ \bibinfo {author} {\bibfnamefont {T.}~\bibnamefont {Tao}},\
  }\href@noop {} {\bibfield  {journal} {\bibinfo  {journal} {IEEE Trans. Inf.
  Theory}\ }\textbf {\bibinfo {volume} {52}},\ \bibinfo {pages} {5406}
  (\bibinfo {year} {2006})}\BibitemShut {NoStop}%
\bibitem [{\citenamefont {Baraniuk}(2007)}]{Bar:2007}%
  \BibitemOpen
  \bibfield  {author} {\bibinfo {author} {\bibfnamefont {R.}~\bibnamefont
  {Baraniuk}},\ }\href@noop {} {\bibfield  {journal} {\bibinfo  {journal} {IEEE
  Sig. Proc. Mag.}\ }\textbf {\bibinfo {volume} {24}},\ \bibinfo {pages} {118}
  (\bibinfo {year} {2007})}\BibitemShut {NoStop}%
\bibitem [{\citenamefont {Candes}\ and\ \citenamefont {Wakin}(2008)}]{CW:2008}%
  \BibitemOpen
  \bibfield  {author} {\bibinfo {author} {\bibfnamefont {E.~J.}\ \bibnamefont
  {Candes}}\ and\ \bibinfo {author} {\bibfnamefont {M.~B.}\ \bibnamefont
  {Wakin}},\ }\href@noop {} {\bibfield  {journal} {\bibinfo  {journal} {IEEE
  Sig. Proc. Mag.}\ }\textbf {\bibinfo {volume} {3}},\ \bibinfo {pages} {21}
  (\bibinfo {year} {2008})}\BibitemShut {NoStop}%
\bibitem [{\citenamefont {Wang}\ \emph {et~al.}(2018)\citenamefont {Wang},
  \citenamefont {Iaroshenko}, \citenamefont {Li}, \citenamefont {Liu},
  \citenamefont {Parab}, \citenamefont {Chen}, \citenamefont {Chu},
  \citenamefont {Kenyon}, \citenamefont {Lipton},\ and\ \citenamefont
  {Sun}}]{WIL:2018}%
  \BibitemOpen
  \bibfield  {author} {\bibinfo {author} {\bibfnamefont {Z.}~\bibnamefont
  {Wang}}, \bibinfo {author} {\bibfnamefont {O.}~\bibnamefont {Iaroshenko}},
  \bibinfo {author} {\bibfnamefont {S.}~\bibnamefont {Li}}, \bibinfo {author}
  {\bibfnamefont {T.}~\bibnamefont {Liu}}, \bibinfo {author} {\bibfnamefont
  {N.}~\bibnamefont {Parab}}, \bibinfo {author} {\bibfnamefont {W.~W.}\
  \bibnamefont {Chen}}, \bibinfo {author} {\bibfnamefont {P.}~\bibnamefont
  {Chu}}, \bibinfo {author} {\bibfnamefont {G.~T.}\ \bibnamefont {Kenyon}},
  \bibinfo {author} {\bibfnamefont {R.}~\bibnamefont {Lipton}}, \ and\ \bibinfo
  {author} {\bibfnamefont {K.-X.}\ \bibnamefont {Sun}},\ }\href@noop {}
  {\bibfield  {journal} {\bibinfo  {journal} {J. Instrum.}\ }\textbf {\bibinfo
  {volume} {13 (01)}},\ \bibinfo {pages} {C01035} (\bibinfo {year}
  {2018})}\BibitemShut {NoStop}%
\bibitem [{\citenamefont {Jeromin}\ \emph {et~al.}(2012)\citenamefont
  {Jeromin}, \citenamefont {Pattichis},\ and\ \citenamefont
  {Calhoun}}]{JPC:2012}%
  \BibitemOpen
  \bibfield  {author} {\bibinfo {author} {\bibfnamefont {O.}~\bibnamefont
  {Jeromin}}, \bibinfo {author} {\bibfnamefont {M.~S.}\ \bibnamefont
  {Pattichis}}, \ and\ \bibinfo {author} {\bibfnamefont {V.~D.}\ \bibnamefont
  {Calhoun}},\ }\href@noop {} {\bibfield  {journal} {\bibinfo  {journal}
  {BioMed. Eng. OnLine}\ }\textbf {\bibinfo {volume} {11}},\ \bibinfo {pages}
  {25} (\bibinfo {year} {2012})}\BibitemShut {NoStop}%
\bibitem [{\citenamefont {Tropp}\ and\ \citenamefont {Wright}(2010)}]{TW:2010}%
  \BibitemOpen
  \bibfield  {author} {\bibinfo {author} {\bibfnamefont {J.~A.}\ \bibnamefont
  {Tropp}}\ and\ \bibinfo {author} {\bibfnamefont {S.~J.}\ \bibnamefont
  {Wright}},\ }\href {\doibase 0.1109/JPROC.2010.2044010} {\bibfield  {journal}
  {\bibinfo  {journal} {Proc. IEEE}\ }\textbf {\bibinfo {volume} {98}},\
  \bibinfo {pages} {948} (\bibinfo {year} {2010})}\BibitemShut {NoStop}%
\bibitem [{\citenamefont {RANI}\ \emph {et~al.}(2018)\citenamefont {RANI},
  \citenamefont {DHOK},\ and\ \citenamefont {DESHMUKH}}]{RDD:2018}%
  \BibitemOpen
  \bibfield  {author} {\bibinfo {author} {\bibfnamefont {M.}~\bibnamefont
  {RANI}}, \bibinfo {author} {\bibfnamefont {S.~B.}\ \bibnamefont {DHOK}}, \
  and\ \bibinfo {author} {\bibfnamefont {R.~B.}\ \bibnamefont {DESHMUKH}},\
  }\href {\doibase 10.1109/ACCESS.2018.2793851} {\bibfield  {journal} {\bibinfo
   {journal} {IEEE Access}\ }\textbf {\bibinfo {volume} {6}},\ \bibinfo {pages}
  {4875} (\bibinfo {year} {2018})}\BibitemShut {NoStop}%
\bibitem [{\citenamefont {Thibault}\ \emph {et~al.}(2009)\citenamefont
  {Thibault}, \citenamefont {Dierolf}, \citenamefont {Bunk}, \citenamefont
  {Menzel},\ and\ \citenamefont {Pfeiffer}}]{TDB:2009}%
  \BibitemOpen
  \bibfield  {author} {\bibinfo {author} {\bibfnamefont {P.}~\bibnamefont
  {Thibault}}, \bibinfo {author} {\bibfnamefont {M.}~\bibnamefont {Dierolf}},
  \bibinfo {author} {\bibfnamefont {O.}~\bibnamefont {Bunk}}, \bibinfo {author}
  {\bibfnamefont {A.}~\bibnamefont {Menzel}}, \ and\ \bibinfo {author}
  {\bibfnamefont {F.}~\bibnamefont {Pfeiffer}},\ }\href@noop {} {\bibfield
  {journal} {\bibinfo  {journal} {Ultramicroscopy}\ }\textbf {\bibinfo {volume}
  {109}},\ \bibinfo {pages} {338} (\bibinfo {year} {2009})}\BibitemShut
  {NoStop}%
\bibitem [{\citenamefont {Candes}\ and\ \citenamefont {Tao}(2005)}]{CT:2005}%
  \BibitemOpen
  \bibfield  {author} {\bibinfo {author} {\bibfnamefont {E.}~\bibnamefont
  {Candes}}\ and\ \bibinfo {author} {\bibfnamefont {T.}~\bibnamefont {Tao}},\
  }\href@noop {} {\bibfield  {journal} {\bibinfo  {journal} {IEEE Trans. Inf.
  Theory}\ }\textbf {\bibinfo {volume} {51}},\ \bibinfo {pages} {4203}
  (\bibinfo {year} {2005})}\BibitemShut {NoStop}%
\bibitem [{\citenamefont {McCann}\ \emph {et~al.}(2017)\citenamefont {McCann},
  \citenamefont {Jin},\ and\ \citenamefont {Unser}}]{MJU:2017}%
  \BibitemOpen
  \bibfield  {author} {\bibinfo {author} {\bibfnamefont {M.~T.}\ \bibnamefont
  {McCann}}, \bibinfo {author} {\bibfnamefont {K.~H.}\ \bibnamefont {Jin}}, \
  and\ \bibinfo {author} {\bibfnamefont {M.}~\bibnamefont {Unser}},\
  }\href@noop {} {\bibfield  {journal} {\bibinfo  {journal} {IEEE Signal
  Process. Mag.}\ }\textbf {\bibinfo {volume} {34(6)}},\ \bibinfo {pages} {85}
  (\bibinfo {year} {2017})}\BibitemShut {NoStop}%
\bibitem [{\citenamefont {Machidon}\ and\ \citenamefont
  {Pejovi\'c}(2005)}]{MP:2022}%
  \BibitemOpen
  \bibfield  {author} {\bibinfo {author} {\bibfnamefont {A.~L.}\ \bibnamefont
  {Machidon}}\ and\ \bibinfo {author} {\bibfnamefont {V.}~\bibnamefont
  {Pejovi\'c}},\ }\href@noop {} {\bibfield  {journal} {\bibinfo  {journal}
  {IEEE Trans. Inf. Theory}\ }\textbf {\bibinfo {volume} {51}},\ \bibinfo
  {pages} {4203} (\bibinfo {year} {2005})}\BibitemShut {NoStop}%
\bibitem [{\citenamefont {Leary}\ \emph {et~al.}(2013)\citenamefont {Leary},
  \citenamefont {Saghi}, \citenamefont {Midgley},\ and\ \citenamefont
  {Holland}}]{LSM:2013}%
  \BibitemOpen
  \bibfield  {author} {\bibinfo {author} {\bibfnamefont {R.}~\bibnamefont
  {Leary}}, \bibinfo {author} {\bibfnamefont {Z.}~\bibnamefont {Saghi}},
  \bibinfo {author} {\bibfnamefont {P.~A.}\ \bibnamefont {Midgley}}, \ and\
  \bibinfo {author} {\bibfnamefont {D.~J.}\ \bibnamefont {Holland}},\
  }\href@noop {} {\bibfield  {journal} {\bibinfo  {journal} {Ultramicroscopy}\
  }\textbf {\bibinfo {volume} {131}},\ \bibinfo {pages} {70} (\bibinfo {year}
  {2013})}\BibitemShut {NoStop}%
\bibitem [{\citenamefont {Binev}\ \emph {et~al.}(2012)\citenamefont {Binev},
  \citenamefont {Dahmen}, \citenamefont {DeVore}, \citenamefont {Lamby},
  \citenamefont {Savu},\ and\ \citenamefont {Sharpley}}]{BDD:2012}%
  \BibitemOpen
  \bibfield  {author} {\bibinfo {author} {\bibfnamefont {P.}~\bibnamefont
  {Binev}}, \bibinfo {author} {\bibfnamefont {W.}~\bibnamefont {Dahmen}},
  \bibinfo {author} {\bibfnamefont {R.}~\bibnamefont {DeVore}}, \bibinfo
  {author} {\bibfnamefont {P.}~\bibnamefont {Lamby}}, \bibinfo {author}
  {\bibfnamefont {D.}~\bibnamefont {Savu}}, \ and\ \bibinfo {author}
  {\bibfnamefont {R.}~\bibnamefont {Sharpley}},\ }\enquote {\bibinfo {title}
  {Compressed sensing and electron microscopy},}\ in\ \href {\doibase
  10.1007/978-1-4614-2191-7_4} {\emph {\bibinfo {booktitle} {Modeling Nanoscale
  Imaging in Electron Microscopy}}},\ \bibinfo {editor} {edited by\ \bibinfo
  {editor} {\bibfnamefont {T.}~\bibnamefont {Vogt}}, \bibinfo {editor}
  {\bibfnamefont {W.}~\bibnamefont {Dahmen}}, \ and\ \bibinfo {editor}
  {\bibfnamefont {P.}~\bibnamefont {Binev}}}\ (\bibinfo  {publisher} {Springer
  US},\ \bibinfo {address} {Boston, MA},\ \bibinfo {year} {2012})\ pp.\
  \bibinfo {pages} {73--126}\BibitemShut {NoStop}%
\bibitem [{\citenamefont {Li}\ \emph {et~al.}(2018)\citenamefont {Li},
  \citenamefont {Dyck}, \citenamefont {Kalinin},\ and\ \citenamefont
  {Jesse}}]{LDK:2018}%
  \BibitemOpen
  \bibfield  {author} {\bibinfo {author} {\bibfnamefont {X.}~\bibnamefont
  {Li}}, \bibinfo {author} {\bibfnamefont {O.}~\bibnamefont {Dyck}}, \bibinfo
  {author} {\bibfnamefont {S.~V.}\ \bibnamefont {Kalinin}}, \ and\ \bibinfo
  {author} {\bibfnamefont {S.}~\bibnamefont {Jesse}},\ }\href@noop {}
  {\bibfield  {journal} {\bibinfo  {journal} {Microsc. Microanal.}\ }\textbf
  {\bibinfo {volume} {24(6)}},\ \bibinfo {pages} {623} (\bibinfo {year}
  {2018})}\BibitemShut {NoStop}%
\bibitem [{\citenamefont {Choi}\ \emph {et~al.}(2010)\citenamefont {Choi},
  \citenamefont {Wang}, \citenamefont {Zhu}, \citenamefont {Suh}, \citenamefont
  {Boyd},\ and\ \citenamefont {Xing}}]{CWZ:2010}%
  \BibitemOpen
  \bibfield  {author} {\bibinfo {author} {\bibfnamefont {K.}~\bibnamefont
  {Choi}}, \bibinfo {author} {\bibfnamefont {J.}~\bibnamefont {Wang}}, \bibinfo
  {author} {\bibfnamefont {L.}~\bibnamefont {Zhu}}, \bibinfo {author}
  {\bibfnamefont {T.-S.}\ \bibnamefont {Suh}}, \bibinfo {author} {\bibfnamefont
  {S.}~\bibnamefont {Boyd}}, \ and\ \bibinfo {author} {\bibfnamefont
  {L.}~\bibnamefont {Xing}},\ }\href@noop {} {\bibfield  {journal} {\bibinfo
  {journal} {Med. Phys.}\ }\textbf {\bibinfo {volume} {37(9)}},\ \bibinfo
  {pages} {5113 } (\bibinfo {year} {2010})}\BibitemShut {NoStop}%
\bibitem [{\citenamefont {Yu}\ and\ \citenamefont {Wang}(2009)}]{YW:2009}%
  \BibitemOpen
  \bibfield  {author} {\bibinfo {author} {\bibfnamefont {H.}~\bibnamefont
  {Yu}}\ and\ \bibinfo {author} {\bibfnamefont {G.}~\bibnamefont {Wang}},\
  }\href@noop {} {\bibfield  {journal} {\bibinfo  {journal} {Phys. Med. Biol.}\
  }\textbf {\bibinfo {volume} {54(9)}},\ \bibinfo {pages} {2791} (\bibinfo
  {year} {2009})}\BibitemShut {NoStop}%
\bibitem [{\citenamefont {Chen}\ \emph {et~al.}(2008)\citenamefont {Chen},
  \citenamefont {Tang},\ and\ \citenamefont {Leng}}]{CTL:2008}%
  \BibitemOpen
  \bibfield  {author} {\bibinfo {author} {\bibfnamefont {G.-H.}\ \bibnamefont
  {Chen}}, \bibinfo {author} {\bibfnamefont {J.}~\bibnamefont {Tang}}, \ and\
  \bibinfo {author} {\bibfnamefont {S.}~\bibnamefont {Leng}},\ }\href {\doibase
  10.1118/1.2836423} {\bibfield  {journal} {\bibinfo  {journal} {Med Phys.}\
  }\textbf {\bibinfo {volume} {35}},\ \bibinfo {pages} {660} (\bibinfo {year}
  {2008})}\BibitemShut {NoStop}%
\bibitem [{\citenamefont {Rangan}\ \emph {et~al.}(2012)\citenamefont {Rangan},
  \citenamefont {Fletcher},\ and\ \citenamefont {Goyal}}]{RFG:2012}%
  \BibitemOpen
  \bibfield  {author} {\bibinfo {author} {\bibfnamefont {S.}~\bibnamefont
  {Rangan}}, \bibinfo {author} {\bibfnamefont {A.~K.}\ \bibnamefont
  {Fletcher}}, \ and\ \bibinfo {author} {\bibfnamefont {V.~K.}\ \bibnamefont
  {Goyal}},\ }\href {\doibase 10.1109/TIT.2011.2177575} {\bibfield  {journal}
  {\bibinfo  {journal} {IEEE Trans. on Information Theory}\ }\textbf {\bibinfo
  {volume} {58}},\ \bibinfo {pages} {1903} (\bibinfo {year}
  {2012})}\BibitemShut {NoStop}%
\bibitem [{\citenamefont {Krzakala}\ and\ \citenamefont
  {Zdeborová}(2022)}]{KZ:2022}%
  \BibitemOpen
  \bibfield  {author} {\bibinfo {author} {\bibfnamefont {F.}~\bibnamefont
  {Krzakala}}\ and\ \bibinfo {author} {\bibfnamefont {L.}~\bibnamefont
  {Zdeborová}},\ }\href@noop {} {\enquote {\bibinfo {title} {Statistical
  physics methods in optimization and machine learning},}\ } (\bibinfo {year}
  {2022}),\ \Eprint {http://arxiv.org/abs/https://sphinxteam.github.io/ \\
  EPFLDoctoralLecture2021/Notes.pdf} {https://sphinxteam.github.io/ \\
  EPFLDoctoralLecture2021/Notes.pdf} \BibitemShut {NoStop}%
\bibitem [{\citenamefont {Bora}\ \emph {et~al.}(2017)\citenamefont {Bora},
  \citenamefont {Jalal}, \citenamefont {Price},\ and\ \citenamefont
  {Dimakis}}]{BJP:2017}%
  \BibitemOpen
  \bibfield  {author} {\bibinfo {author} {\bibfnamefont {A.}~\bibnamefont
  {Bora}}, \bibinfo {author} {\bibfnamefont {A.}~\bibnamefont {Jalal}},
  \bibinfo {author} {\bibfnamefont {E.}~\bibnamefont {Price}}, \ and\ \bibinfo
  {author} {\bibfnamefont {A.~G.}\ \bibnamefont {Dimakis}},\ }in\ \href@noop {}
  {\emph {\bibinfo {booktitle} {Proceedings of the 34th International
  Conference on Machine Learning}}},\ Vol.\ \bibinfo {volume} {PLMR 70}\
  (\bibinfo {address} {Sydney, Australia},\ \bibinfo {year} {2017})\ \bibinfo
  {note} {{`Compressed Sensing using Generative Models'}}\BibitemShut {NoStop}%
\bibitem [{\citenamefont {Boashash}(1988)}]{Boa:1988}%
  \BibitemOpen
  \bibfield  {author} {\bibinfo {author} {\bibfnamefont {B.}~\bibnamefont
  {Boashash}},\ }\href@noop {} {\bibfield  {journal} {\bibinfo  {journal} {IEEE
  Trans. Acoust. Speech Sig. Proc.}\ }\textbf {\bibinfo {volume} {36}},\
  \bibinfo {pages} {1518} (\bibinfo {year} {1988})}\BibitemShut {NoStop}%
\bibitem [{\citenamefont {Rodenburg}\ and\ \citenamefont
  {Bates}(1992)}]{RB:1992}%
  \BibitemOpen
  \bibfield  {author} {\bibinfo {author} {\bibfnamefont {J.~M.}\ \bibnamefont
  {Rodenburg}}\ and\ \bibinfo {author} {\bibfnamefont {R.~H.~T.}\ \bibnamefont
  {Bates}},\ }\href@noop {} {\bibfield  {journal} {\bibinfo  {journal} {Phil.
  Trans. R. Soc. Lond. A}\ }\textbf {\bibinfo {volume} {339}},\ \bibinfo
  {pages} {521} (\bibinfo {year} {1992})}\BibitemShut {NoStop}%
\bibitem [{\citenamefont {Pfeiffer}(2018)}]{Pfe:2018}%
  \BibitemOpen
  \bibfield  {author} {\bibinfo {author} {\bibfnamefont {F.}~\bibnamefont
  {Pfeiffer}},\ }\href@noop {} {\bibfield  {journal} {\bibinfo  {journal} {Nat.
  Photonics}\ }\textbf {\bibinfo {volume} {12}},\ \bibinfo {pages} {9}
  (\bibinfo {year} {2018})}\BibitemShut {NoStop}%
\bibitem [{\citenamefont {Rodenburg}\ and\ \citenamefont
  {Maiden}(2019)}]{RM:2019}%
  \BibitemOpen
  \bibfield  {author} {\bibinfo {author} {\bibfnamefont {J.~M.}\ \bibnamefont
  {Rodenburg}}\ and\ \bibinfo {author} {\bibfnamefont {A.~M.}\ \bibnamefont
  {Maiden}},\ }in\ \href@noop {} {\emph {\bibinfo {booktitle} {Springer
  Handbook of Microscopy}}},\ \bibinfo {editor} {edited by\ \bibinfo {editor}
  {\bibfnamefont {P.~W.}\ \bibnamefont {Hawkes}}\ and\ \bibinfo {editor}
  {\bibfnamefont {J.~C.~H.}\ \bibnamefont {Spence}}}\ (\bibinfo  {publisher}
  {Springer},\ \bibinfo {year} {2019})\ pp.\ \bibinfo {pages} {1--137},\
  \bibinfo {note} {{Ptychography}}\BibitemShut {NoStop}%
\bibitem [{\citenamefont {Chapman}(1996)}]{Cha:1996}%
  \BibitemOpen
  \bibfield  {author} {\bibinfo {author} {\bibfnamefont {H.~N.}\ \bibnamefont
  {Chapman}},\ }\href@noop {} {\bibfield  {journal} {\bibinfo  {journal}
  {Ultramicroscopy}\ }\textbf {\bibinfo {volume} {66}},\ \bibinfo {pages} {153}
  (\bibinfo {year} {1996})}\BibitemShut {NoStop}%
\bibitem [{\citenamefont {Xu}\ and\ \citenamefont {Mueller}(2007)}]{XM:2007}%
  \BibitemOpen
  \bibfield  {author} {\bibinfo {author} {\bibfnamefont {F.}~\bibnamefont
  {Xu}}\ and\ \bibinfo {author} {\bibfnamefont {K.}~\bibnamefont {Mueller}},\
  }\href@noop {} {\bibfield  {journal} {\bibinfo  {journal} {Phys. Med. Biol.}\
  }\textbf {\bibinfo {volume} {52(12)}},\ \bibinfo {pages} {3405} (\bibinfo
  {year} {2007})}\BibitemShut {NoStop}%
\bibitem [{\citenamefont {Dong}\ \emph {et~al.}(2014)\citenamefont {Dong},
  \citenamefont {Bian}, \citenamefont {Shiradkar},\ and\ \citenamefont
  {Zheng}}]{DBS:2014}%
  \BibitemOpen
  \bibfield  {author} {\bibinfo {author} {\bibfnamefont {S.}~\bibnamefont
  {Dong}}, \bibinfo {author} {\bibfnamefont {Z.}~\bibnamefont {Bian}}, \bibinfo
  {author} {\bibfnamefont {R.}~\bibnamefont {Shiradkar}}, \ and\ \bibinfo
  {author} {\bibfnamefont {G.}~\bibnamefont {Zheng}},\ }\href@noop {}
  {\bibfield  {journal} {\bibinfo  {journal} {Opt. Exp.}\ }\textbf {\bibinfo
  {volume} {22 (5)}},\ \bibinfo {pages} {5455} (\bibinfo {year}
  {2014})}\BibitemShut {NoStop}%
\bibitem [{\citenamefont {Robucci}\ \emph {et~al.}(2010)\citenamefont
  {Robucci}, \citenamefont {Gray}, \citenamefont {Chiu}, \citenamefont
  {Romberg},\ and\ \citenamefont {Haslwer}}]{RGC:2010}%
  \BibitemOpen
  \bibfield  {author} {\bibinfo {author} {\bibfnamefont {R.}~\bibnamefont
  {Robucci}}, \bibinfo {author} {\bibfnamefont {J.~D.}\ \bibnamefont {Gray}},
  \bibinfo {author} {\bibfnamefont {L.~K.}\ \bibnamefont {Chiu}}, \bibinfo
  {author} {\bibfnamefont {J.}~\bibnamefont {Romberg}}, \ and\ \bibinfo
  {author} {\bibfnamefont {P.}~\bibnamefont {Haslwer}},\ }\href@noop {}
  {\bibfield  {journal} {\bibinfo  {journal} {Proc. IEEE}\ }\textbf {\bibinfo
  {volume} {98(6)}},\ \bibinfo {pages} {1089} (\bibinfo {year}
  {2010})}\BibitemShut {NoStop}%
\bibitem [{\citenamefont {Shankar}\ \emph {et~al.}(2010)\citenamefont
  {Shankar}, \citenamefont {Pitsianis},\ and\ \citenamefont {Brady}}]{Dart:16}%
  \BibitemOpen
  \bibfield  {author} {\bibinfo {author} {\bibfnamefont {M.}~\bibnamefont
  {Shankar}}, \bibinfo {author} {\bibfnamefont {N.~P.}\ \bibnamefont
  {Pitsianis}}, \ and\ \bibinfo {author} {\bibfnamefont {D.~J.}\ \bibnamefont
  {Brady}},\ }\href@noop {} {\bibfield  {journal} {\bibinfo  {journal} {Appl.
  Opt.}\ }\textbf {\bibinfo {volume} {49}},\ \bibinfo {pages} {B9} (\bibinfo
  {year} {Feb. 2010})},\ \bibinfo {note} {doi:
  https://doi.org/10.1364/ao.49.0000b9}\BibitemShut {NoStop}%
\bibitem [{\citenamefont {Mochizuki}\ \emph {et~al.}(2015)\citenamefont
  {Mochizuki}, \citenamefont {Kagawa}, \citenamefont {ichiro Okihara},
  \citenamefont {Seo}, \citenamefont {Zhang}, \citenamefont {Takasawa},
  \citenamefont {Yasutomi},\ and\ \citenamefont {Kawahito}}]{Dart:17}%
  \BibitemOpen
  \bibfield  {author} {\bibinfo {author} {\bibfnamefont {F.}~\bibnamefont
  {Mochizuki}}, \bibinfo {author} {\bibfnamefont {K.}~\bibnamefont {Kagawa}},
  \bibinfo {author} {\bibfnamefont {S.}~\bibnamefont {ichiro Okihara}},
  \bibinfo {author} {\bibfnamefont {M.-W.}\ \bibnamefont {Seo}}, \bibinfo
  {author} {\bibfnamefont {B.}~\bibnamefont {Zhang}}, \bibinfo {author}
  {\bibfnamefont {T.}~\bibnamefont {Takasawa}}, \bibinfo {author}
  {\bibfnamefont {K.}~\bibnamefont {Yasutomi}}, \ and\ \bibinfo {author}
  {\bibfnamefont {S.}~\bibnamefont {Kawahito}},\ }\href@noop {} {\bibfield
  {journal} {\bibinfo  {journal} {IEEE Xplore}\ } (\bibinfo {year} {Feb.
  2015})},\ \bibinfo {note} {{https://ieeexplore.ieee.org/stamp/stamp.jsp? \\
  tp=\&arnumber=7062953} (accessed Mar. 09, 2023)}\BibitemShut {NoStop}%
\bibitem [{\citenamefont {Burvall}\ \emph {et~al.}(2011)\citenamefont
  {Burvall}, \citenamefont {Lundstr{\"o}m}, \citenamefont {Takman},
  \citenamefont {Larsson},\ and\ \citenamefont {Hertz}}]{BLT:2011}%
  \BibitemOpen
  \bibfield  {author} {\bibinfo {author} {\bibfnamefont {A.}~\bibnamefont
  {Burvall}}, \bibinfo {author} {\bibfnamefont {U.}~\bibnamefont
  {Lundstr{\"o}m}}, \bibinfo {author} {\bibfnamefont {P.~A.~C.}\ \bibnamefont
  {Takman}}, \bibinfo {author} {\bibfnamefont {D.~H.}\ \bibnamefont {Larsson}},
  \ and\ \bibinfo {author} {\bibfnamefont {H.~M.}\ \bibnamefont {Hertz}},\
  }\href@noop {} {\bibfield  {journal} {\bibinfo  {journal} {Opt. Exp.}\
  }\textbf {\bibinfo {volume} {19}},\ \bibinfo {pages} {10359} (\bibinfo {year}
  {2011})}\BibitemShut {NoStop}%
\bibitem [{\citenamefont {Jaganathan}\ \emph {et~al.}(2015)\citenamefont
  {Jaganathan}, \citenamefont {Eldar},\ and\ \citenamefont
  {Hassibi}}]{JEH:2015}%
  \BibitemOpen
  \bibfield  {author} {\bibinfo {author} {\bibfnamefont {K.}~\bibnamefont
  {Jaganathan}}, \bibinfo {author} {\bibfnamefont {Y.~C.}\ \bibnamefont
  {Eldar}}, \ and\ \bibinfo {author} {\bibfnamefont {B.}~\bibnamefont
  {Hassibi}},\ }\href@noop {} {\enquote {\bibinfo {title} {Phase retrieval: An
  overview of recent developments},}\ } (\bibinfo {year} {2015}),\ \Eprint
  {http://arxiv.org/abs/arXiv:1510.07713v1 [cs.IT]} {arXiv:1510.07713v1
  [cs.IT]} \BibitemShut {NoStop}%
\bibitem [{\citenamefont {Zuo}\ \emph {et~al.}(2022)\citenamefont {Zuo},
  \citenamefont {Qian}, \citenamefont {Feng}, \citenamefont {Yin},
  \citenamefont {Li}, \citenamefont {Fan}, \citenamefont {Han}, \citenamefont
  {Qian},\ and\ \citenamefont {Chen}}]{Zuo:2022}%
  \BibitemOpen
  \bibfield  {author} {\bibinfo {author} {\bibfnamefont {C.}~\bibnamefont
  {Zuo}}, \bibinfo {author} {\bibfnamefont {J.}~\bibnamefont {Qian}}, \bibinfo
  {author} {\bibfnamefont {S.}~\bibnamefont {Feng}}, \bibinfo {author}
  {\bibfnamefont {W.}~\bibnamefont {Yin}}, \bibinfo {author} {\bibfnamefont
  {Y.}~\bibnamefont {Li}}, \bibinfo {author} {\bibfnamefont {P.}~\bibnamefont
  {Fan}}, \bibinfo {author} {\bibfnamefont {J.}~\bibnamefont {Han}}, \bibinfo
  {author} {\bibfnamefont {K.}~\bibnamefont {Qian}}, \ and\ \bibinfo {author}
  {\bibfnamefont {Q.}~\bibnamefont {Chen}},\ }\href {\doibase
  10.1038/s41377-022-00714-x} {\bibfield  {journal} {\bibinfo  {journal}
  {Light: Science and Applications}\ }\textbf {\bibinfo {volume} {11}},\
  \bibinfo {pages} {1} (\bibinfo {year} {2022})}\BibitemShut {NoStop}%
\bibitem [{\citenamefont {Kitchen}\ \emph {et~al.}(2011)\citenamefont
  {Kitchen}, \citenamefont {Paganin}, \citenamefont {Uesugi}, \citenamefont
  {Allison}, \citenamefont {Lewis}, \citenamefont {Hooper},\ and\ \citenamefont
  {Pavlov}}]{Kitchen:2011}%
  \BibitemOpen
  \bibfield  {author} {\bibinfo {author} {\bibfnamefont {M.~J.}\ \bibnamefont
  {Kitchen}}, \bibinfo {author} {\bibfnamefont {D.~M.}\ \bibnamefont
  {Paganin}}, \bibinfo {author} {\bibfnamefont {K.}~\bibnamefont {Uesugi}},
  \bibinfo {author} {\bibfnamefont {B.~J.}\ \bibnamefont {Allison}}, \bibinfo
  {author} {\bibfnamefont {R.~A.}\ \bibnamefont {Lewis}}, \bibinfo {author}
  {\bibfnamefont {S.~B.}\ \bibnamefont {Hooper}}, \ and\ \bibinfo {author}
  {\bibfnamefont {K.~M.}\ \bibnamefont {Pavlov}},\ }\href {\doibase
  10.1088/0031-9155/56/3/001} {\bibfield  {journal} {\bibinfo  {journal}
  {Physics in medicine and biology}\ }\textbf {\bibinfo {volume} {56}},\
  \bibinfo {pages} {515} (\bibinfo {year} {2011})}\BibitemShut {NoStop}%
\bibitem [{\citenamefont {Wang}\ \emph {et~al.}(2019)\citenamefont {Wang},
  \citenamefont {Ren}, \citenamefont {Shi}, \citenamefont {Liu}, \citenamefont
  {Wu},\ and\ \citenamefont {Gao}}]{Wang:2019}%
  \BibitemOpen
  \bibfield  {author} {\bibinfo {author} {\bibfnamefont {Z.}~\bibnamefont
  {Wang}}, \bibinfo {author} {\bibfnamefont {K.}~\bibnamefont {Ren}}, \bibinfo
  {author} {\bibfnamefont {X.}~\bibnamefont {Shi}}, \bibinfo {author}
  {\bibfnamefont {D.}~\bibnamefont {Liu}}, \bibinfo {author} {\bibfnamefont
  {Z.}~\bibnamefont {Wu}}, \ and\ \bibinfo {author} {\bibfnamefont
  {K.}~\bibnamefont {Gao}},\ }\href {\doibase https://doi.org/10.1002/mp.13399}
  {\bibfield  {journal} {\bibinfo  {journal} {Medical Physics}\ }\textbf
  {\bibinfo {volume} {46}},\ \bibinfo {pages} {1317} (\bibinfo {year}
  {2019})}\BibitemShut {NoStop}%
\bibitem [{\citenamefont {Valdivia}\ \emph {et~al.}(2020)\citenamefont
  {Valdivia}, \citenamefont {Stutman}, \citenamefont {Stoeckl}, \citenamefont
  {Mileham}, \citenamefont {Zou}, \citenamefont {Muller}, \citenamefont
  {Kaiser}, \citenamefont {Sorce}, \citenamefont {Keiter}, \citenamefont
  {Fein}, \citenamefont {Trantham}, \citenamefont {Drake},\ and\ \citenamefont
  {Regan}}]{Valdivia:2020}%
  \BibitemOpen
  \bibfield  {author} {\bibinfo {author} {\bibfnamefont {M.~P.}\ \bibnamefont
  {Valdivia}}, \bibinfo {author} {\bibfnamefont {D.}~\bibnamefont {Stutman}},
  \bibinfo {author} {\bibfnamefont {C.}~\bibnamefont {Stoeckl}}, \bibinfo
  {author} {\bibfnamefont {C.}~\bibnamefont {Mileham}}, \bibinfo {author}
  {\bibfnamefont {J.}~\bibnamefont {Zou}}, \bibinfo {author} {\bibfnamefont
  {S.}~\bibnamefont {Muller}}, \bibinfo {author} {\bibfnamefont
  {K.}~\bibnamefont {Kaiser}}, \bibinfo {author} {\bibfnamefont
  {C.}~\bibnamefont {Sorce}}, \bibinfo {author} {\bibfnamefont {P.~A.}\
  \bibnamefont {Keiter}}, \bibinfo {author} {\bibfnamefont {J.~R.}\
  \bibnamefont {Fein}}, \bibinfo {author} {\bibfnamefont {M.}~\bibnamefont
  {Trantham}}, \bibinfo {author} {\bibfnamefont {R.~P.}\ \bibnamefont {Drake}},
  \ and\ \bibinfo {author} {\bibfnamefont {S.~P.}\ \bibnamefont {Regan}},\
  }\href {\doibase 10.1063/1.5123919} {\bibfield  {journal} {\bibinfo
  {journal} {Review of Scientific Instruments}\ }\textbf {\bibinfo {volume}
  {91}},\ \bibinfo {pages} {023511} (\bibinfo {year} {2020})}\BibitemShut
  {NoStop}%
\bibitem [{\citenamefont {Oh}\ \emph {et~al.}(2022)\citenamefont {Oh},
  \citenamefont {Kim}, \citenamefont {Kim}, \citenamefont {Hussey},\ and\
  \citenamefont {Lee}}]{Oh:2022}%
  \BibitemOpen
  \bibfield  {author} {\bibinfo {author} {\bibfnamefont {O.}~\bibnamefont
  {Oh}}, \bibinfo {author} {\bibfnamefont {Y.}~\bibnamefont {Kim}}, \bibinfo
  {author} {\bibfnamefont {D.}~\bibnamefont {Kim}}, \bibinfo {author}
  {\bibfnamefont {D.~S.}\ \bibnamefont {Hussey}}, \ and\ \bibinfo {author}
  {\bibfnamefont {S.~W.}\ \bibnamefont {Lee}},\ }\href {\doibase
  10.1038/s41598-022-10551-y} {\bibfield  {journal} {\bibinfo  {journal}
  {Scientific Reports}\ }\textbf {\bibinfo {volume} {12}},\ \bibinfo {pages}
  {1} (\bibinfo {year} {2022})}\BibitemShut {NoStop}%
\bibitem [{\citenamefont {Fouras}\ \emph {et~al.}(2009)\citenamefont {Fouras},
  \citenamefont {Kitchen}, \citenamefont {Dubsky}, \citenamefont {Lewis},
  \citenamefont {Hooper},\ and\ \citenamefont {Hourigan}}]{Fouras:2009}%
  \BibitemOpen
  \bibfield  {author} {\bibinfo {author} {\bibfnamefont {A.}~\bibnamefont
  {Fouras}}, \bibinfo {author} {\bibfnamefont {M.~J.}\ \bibnamefont {Kitchen}},
  \bibinfo {author} {\bibfnamefont {S.}~\bibnamefont {Dubsky}}, \bibinfo
  {author} {\bibfnamefont {R.~A.}\ \bibnamefont {Lewis}}, \bibinfo {author}
  {\bibfnamefont {S.~B.}\ \bibnamefont {Hooper}}, \ and\ \bibinfo {author}
  {\bibfnamefont {K.}~\bibnamefont {Hourigan}},\ }\href {\doibase
  10.1063/1.3115643} {\bibfield  {journal} {\bibinfo  {journal} {Journal of
  Applied Physics}\ }\textbf {\bibinfo {volume} {105}},\ \bibinfo {pages}
  {102009} (\bibinfo {year} {2009})}\BibitemShut {NoStop}%
\bibitem [{\citenamefont {Leong}\ \emph {et~al.}(2019)\citenamefont {Leong},
  \citenamefont {Asare}, \citenamefont {Rex}, \citenamefont {Xiao},
  \citenamefont {Ramesh},\ and\ \citenamefont {Hufnagel}}]{Leong:2019}%
  \BibitemOpen
  \bibfield  {author} {\bibinfo {author} {\bibfnamefont {A.~F.~T.}\
  \bibnamefont {Leong}}, \bibinfo {author} {\bibfnamefont {E.}~\bibnamefont
  {Asare}}, \bibinfo {author} {\bibfnamefont {R.}~\bibnamefont {Rex}}, \bibinfo
  {author} {\bibfnamefont {X.~H.}\ \bibnamefont {Xiao}}, \bibinfo {author}
  {\bibfnamefont {K.~T.}\ \bibnamefont {Ramesh}}, \ and\ \bibinfo {author}
  {\bibfnamefont {T.~C.}\ \bibnamefont {Hufnagel}},\ }\href {\doibase
  10.1364/OE.27.017322} {\bibfield  {journal} {\bibinfo  {journal} {Optics
  express}\ }\textbf {\bibinfo {volume} {27}},\ \bibinfo {pages} {17322}
  (\bibinfo {year} {2019})}\BibitemShut {NoStop}%
\bibitem [{\citenamefont {Paganin}\ and\ \citenamefont
  {Morgan}(2019)}]{Paganin:2019}%
  \BibitemOpen
  \bibfield  {author} {\bibinfo {author} {\bibfnamefont {D.~M.}\ \bibnamefont
  {Paganin}}\ and\ \bibinfo {author} {\bibfnamefont {K.~S.}\ \bibnamefont
  {Morgan}},\ }\href {\doibase 10.1038/s41598-019-52284-5} {\bibfield
  {journal} {\bibinfo  {journal} {Scientific Reports}\ }\textbf {\bibinfo
  {volume} {9}},\ \bibinfo {pages} {1} (\bibinfo {year} {2019})},\ \Eprint
  {http://arxiv.org/abs/1908.01473} {arXiv:1908.01473} \BibitemShut {NoStop}%
\bibitem [{\citenamefont {Wood}\ \emph
  {et~al.}(2018{\natexlab{a}})\citenamefont {Wood}, \citenamefont {Chapman},
  \citenamefont {Poder}, \citenamefont {Lopes}, \citenamefont {Rutherford},
  \citenamefont {White}, \citenamefont {Albert}, \citenamefont {Behm},
  \citenamefont {Booth}, \citenamefont {Bryant}, \citenamefont {Foster},
  \citenamefont {Glenzer}, \citenamefont {Hill}, \citenamefont {Krushelnick},
  \citenamefont {Najmudin}, \citenamefont {Pollock}, \citenamefont {Rose},
  \citenamefont {Schumaker}, \citenamefont {Scott}, \citenamefont {Sherlock},
  \citenamefont {Thomas}, \citenamefont {Zhao}, \citenamefont {Eakins},\ and\
  \citenamefont {Mangles}}]{Wood:2018}%
  \BibitemOpen
  \bibfield  {author} {\bibinfo {author} {\bibfnamefont {J.~C.}\ \bibnamefont
  {Wood}}, \bibinfo {author} {\bibfnamefont {D.~J.}\ \bibnamefont {Chapman}},
  \bibinfo {author} {\bibfnamefont {K.}~\bibnamefont {Poder}}, \bibinfo
  {author} {\bibfnamefont {N.~C.}\ \bibnamefont {Lopes}}, \bibinfo {author}
  {\bibfnamefont {M.~E.}\ \bibnamefont {Rutherford}}, \bibinfo {author}
  {\bibfnamefont {T.~G.}\ \bibnamefont {White}}, \bibinfo {author}
  {\bibfnamefont {F.}~\bibnamefont {Albert}}, \bibinfo {author} {\bibfnamefont
  {K.~T.}\ \bibnamefont {Behm}}, \bibinfo {author} {\bibfnamefont
  {N.}~\bibnamefont {Booth}}, \bibinfo {author} {\bibfnamefont {J.~S.}\
  \bibnamefont {Bryant}}, \bibinfo {author} {\bibfnamefont {P.~S.}\
  \bibnamefont {Foster}}, \bibinfo {author} {\bibfnamefont {S.}~\bibnamefont
  {Glenzer}}, \bibinfo {author} {\bibfnamefont {E.}~\bibnamefont {Hill}},
  \bibinfo {author} {\bibfnamefont {K.}~\bibnamefont {Krushelnick}}, \bibinfo
  {author} {\bibfnamefont {Z.}~\bibnamefont {Najmudin}}, \bibinfo {author}
  {\bibfnamefont {B.~B.}\ \bibnamefont {Pollock}}, \bibinfo {author}
  {\bibfnamefont {S.}~\bibnamefont {Rose}}, \bibinfo {author} {\bibfnamefont
  {W.}~\bibnamefont {Schumaker}}, \bibinfo {author} {\bibfnamefont {R.~H.}\
  \bibnamefont {Scott}}, \bibinfo {author} {\bibfnamefont {M.}~\bibnamefont
  {Sherlock}}, \bibinfo {author} {\bibfnamefont {A.~G.}\ \bibnamefont
  {Thomas}}, \bibinfo {author} {\bibfnamefont {Z.}~\bibnamefont {Zhao}},
  \bibinfo {author} {\bibfnamefont {D.~E.}\ \bibnamefont {Eakins}}, \ and\
  \bibinfo {author} {\bibfnamefont {S.~P.}\ \bibnamefont {Mangles}},\ }\href
  {\doibase 10.1038/s41598-018-29347-0} {\bibfield  {journal} {\bibinfo
  {journal} {Scientific Reports}\ }\textbf {\bibinfo {volume} {8}} (\bibinfo
  {year} {2018}{\natexlab{a}}),\ 10.1038/s41598-018-29347-0},\ \Eprint
  {http://arxiv.org/abs/1802.02119} {arXiv:1802.02119} \BibitemShut {NoStop}%
\bibitem [{\citenamefont {Latychevskaia}(2019)}]{A22b}%
  \BibitemOpen
  \bibfield  {author} {\bibinfo {author} {\bibfnamefont {T.}~\bibnamefont
  {Latychevskaia}},\ }\href {\doibase 10.1364/josaa.36.000d31} {\bibfield
  {journal} {\bibinfo  {journal} {Journal of the Optical Society of America A}\
  }\textbf {\bibinfo {volume} {36}},\ \bibinfo {pages} {D31} (\bibinfo {year}
  {2019})}\BibitemShut {NoStop}%
\bibitem [{\citenamefont {Hagemann}\ \emph {et~al.}(2021)\citenamefont
  {Hagemann}, \citenamefont {Vassholz}, \citenamefont {Hoeppe}, \citenamefont
  {Osterhoff}, \citenamefont {Rossell{\'{o}}}, \citenamefont {Mettin},
  \citenamefont {Seiboth}, \citenamefont {Schropp}, \citenamefont
  {M{\"{o}}ller}, \citenamefont {Hallmann}, \citenamefont {Kim}, \citenamefont
  {Scholz}, \citenamefont {Boesenberg}, \citenamefont {Schaffer}, \citenamefont
  {Zozulya}, \citenamefont {Lu}, \citenamefont {Shayduk}, \citenamefont
  {Madsen}, \citenamefont {Schroer},\ and\ \citenamefont {Salditt}}]{A23}%
  \BibitemOpen
  \bibfield  {author} {\bibinfo {author} {\bibfnamefont {J.}~\bibnamefont
  {Hagemann}}, \bibinfo {author} {\bibfnamefont {M.}~\bibnamefont {Vassholz}},
  \bibinfo {author} {\bibfnamefont {H.}~\bibnamefont {Hoeppe}}, \bibinfo
  {author} {\bibfnamefont {M.}~\bibnamefont {Osterhoff}}, \bibinfo {author}
  {\bibfnamefont {J.~M.}\ \bibnamefont {Rossell{\'{o}}}}, \bibinfo {author}
  {\bibfnamefont {R.}~\bibnamefont {Mettin}}, \bibinfo {author} {\bibfnamefont
  {F.}~\bibnamefont {Seiboth}}, \bibinfo {author} {\bibfnamefont
  {A.}~\bibnamefont {Schropp}}, \bibinfo {author} {\bibfnamefont
  {J.}~\bibnamefont {M{\"{o}}ller}}, \bibinfo {author} {\bibfnamefont
  {J.}~\bibnamefont {Hallmann}}, \bibinfo {author} {\bibfnamefont
  {C.}~\bibnamefont {Kim}}, \bibinfo {author} {\bibfnamefont {M.}~\bibnamefont
  {Scholz}}, \bibinfo {author} {\bibfnamefont {U.}~\bibnamefont {Boesenberg}},
  \bibinfo {author} {\bibfnamefont {R.}~\bibnamefont {Schaffer}}, \bibinfo
  {author} {\bibfnamefont {A.}~\bibnamefont {Zozulya}}, \bibinfo {author}
  {\bibfnamefont {W.}~\bibnamefont {Lu}}, \bibinfo {author} {\bibfnamefont
  {R.}~\bibnamefont {Shayduk}}, \bibinfo {author} {\bibfnamefont
  {A.}~\bibnamefont {Madsen}}, \bibinfo {author} {\bibfnamefont {C.~G.}\
  \bibnamefont {Schroer}}, \ and\ \bibinfo {author} {\bibfnamefont
  {T.}~\bibnamefont {Salditt}},\ }\href {\doibase 10.1107/S160057752001557X}
  {\bibfield  {journal} {\bibinfo  {journal} {Journal of Synchrotron
  Radiation}\ }\textbf {\bibinfo {volume} {28}},\ \bibinfo {pages} {52}
  (\bibinfo {year} {2021})}\BibitemShut {NoStop}%
\bibitem [{\citenamefont {Wittwer}\ \emph {et~al.}(2022)\citenamefont
  {Wittwer}, \citenamefont {Wittwer}, \citenamefont {Wittwer}, \citenamefont
  {Hagemann}, \citenamefont {Hagemann}, \citenamefont {Br{\"{u}}ckner},
  \citenamefont {Br{\"{u}}ckner}, \citenamefont {Br{\"{u}}ckner}, \citenamefont
  {Flenner}, \citenamefont {Schroer}, \citenamefont {Schroer},\ and\
  \citenamefont {Schroer}}]{A24}%
  \BibitemOpen
  \bibfield  {author} {\bibinfo {author} {\bibfnamefont {F.}~\bibnamefont
  {Wittwer}}, \bibinfo {author} {\bibfnamefont {F.}~\bibnamefont {Wittwer}},
  \bibinfo {author} {\bibfnamefont {F.}~\bibnamefont {Wittwer}}, \bibinfo
  {author} {\bibfnamefont {J.}~\bibnamefont {Hagemann}}, \bibinfo {author}
  {\bibfnamefont {J.}~\bibnamefont {Hagemann}}, \bibinfo {author}
  {\bibfnamefont {D.}~\bibnamefont {Br{\"{u}}ckner}}, \bibinfo {author}
  {\bibfnamefont {D.}~\bibnamefont {Br{\"{u}}ckner}}, \bibinfo {author}
  {\bibfnamefont {D.}~\bibnamefont {Br{\"{u}}ckner}}, \bibinfo {author}
  {\bibfnamefont {S.}~\bibnamefont {Flenner}}, \bibinfo {author} {\bibfnamefont
  {C.~G.}\ \bibnamefont {Schroer}}, \bibinfo {author} {\bibfnamefont {C.~G.}\
  \bibnamefont {Schroer}}, \ and\ \bibinfo {author} {\bibfnamefont {C.~G.}\
  \bibnamefont {Schroer}},\ }\href {\doibase 10.1364/OPTICA.447021} {\bibfield
  {journal} {\bibinfo  {journal} {Optica}\ }\textbf {\bibinfo {volume} {9}},\
  \bibinfo {pages} {295} (\bibinfo {year} {2022})}\BibitemShut {NoStop}%
\bibitem [{\citenamefont {Carroll}\ \emph {et~al.}(2017)\citenamefont
  {Carroll}, \citenamefont {{Van Riessen}}, \citenamefont {Balaur},
  \citenamefont {Dolbnya}, \citenamefont {Tran},\ and\ \citenamefont
  {Peele}}]{A26}%
  \BibitemOpen
  \bibfield  {author} {\bibinfo {author} {\bibfnamefont {A.~J.}\ \bibnamefont
  {Carroll}}, \bibinfo {author} {\bibfnamefont {G.~A.}\ \bibnamefont {{Van
  Riessen}}}, \bibinfo {author} {\bibfnamefont {E.}~\bibnamefont {Balaur}},
  \bibinfo {author} {\bibfnamefont {I.~P.}\ \bibnamefont {Dolbnya}}, \bibinfo
  {author} {\bibfnamefont {G.~N.}\ \bibnamefont {Tran}}, \ and\ \bibinfo
  {author} {\bibfnamefont {A.~G.}\ \bibnamefont {Peele}},\ }\href {\doibase
  10.1088/2040-8986/aa72c4} {\bibfield  {journal} {\bibinfo  {journal} {Journal
  of Optics}\ }\textbf {\bibinfo {volume} {19}} (\bibinfo {year} {2017}),\
  10.1088/2040-8986/aa72c4}\BibitemShut {NoStop}%
\bibitem [{\citenamefont {Zhang}\ \emph {et~al.}(2018)\citenamefont {Zhang},
  \citenamefont {Cao}, \citenamefont {Brady}, \citenamefont {Zhang},
  \citenamefont {Cang}, \citenamefont {Zhang},\ and\ \citenamefont
  {Jin}}]{A27}%
  \BibitemOpen
  \bibfield  {author} {\bibinfo {author} {\bibfnamefont {W.}~\bibnamefont
  {Zhang}}, \bibinfo {author} {\bibfnamefont {L.}~\bibnamefont {Cao}}, \bibinfo
  {author} {\bibfnamefont {D.~J.}\ \bibnamefont {Brady}}, \bibinfo {author}
  {\bibfnamefont {H.~H. H.~H.}\ \bibnamefont {Zhang}}, \bibinfo {author}
  {\bibfnamefont {J.}~\bibnamefont {Cang}}, \bibinfo {author} {\bibfnamefont
  {H.~H. H.~H.}\ \bibnamefont {Zhang}}, \ and\ \bibinfo {author} {\bibfnamefont
  {G.}~\bibnamefont {Jin}},\ }\href {\doibase 10.1103/PhysRevLett.121.093902}
  {\bibfield  {journal} {\bibinfo  {journal} {Physical Review Letters}\
  }\textbf {\bibinfo {volume} {121}},\ \bibinfo {pages} {093902} (\bibinfo
  {year} {2018})}\BibitemShut {NoStop}%
\bibitem [{\citenamefont {Rivenson}\ \emph {et~al.}(2018)\citenamefont
  {Rivenson}, \citenamefont {Zhang}, \citenamefont {G{\"{u}}naydın},
  \citenamefont {Teng},\ and\ \citenamefont {Ozcan}}]{A28}%
  \BibitemOpen
  \bibfield  {author} {\bibinfo {author} {\bibfnamefont {Y.}~\bibnamefont
  {Rivenson}}, \bibinfo {author} {\bibfnamefont {Y.}~\bibnamefont {Zhang}},
  \bibinfo {author} {\bibfnamefont {H.}~\bibnamefont {G{\"{u}}naydın}},
  \bibinfo {author} {\bibfnamefont {D.}~\bibnamefont {Teng}}, \ and\ \bibinfo
  {author} {\bibfnamefont {A.}~\bibnamefont {Ozcan}},\ }\href {\doibase
  10.1038/lsa.2017.141} {\bibfield  {journal} {\bibinfo  {journal} {Light:
  Science and Applications}\ }\textbf {\bibinfo {volume} {7}},\ \bibinfo
  {pages} {17141} (\bibinfo {year} {2018})},\ \Eprint
  {http://arxiv.org/abs/1705.04286} {arXiv:1705.04286} \BibitemShut {NoStop}%
\bibitem [{\citenamefont {Zhang}\ \emph {et~al.}(2021)\citenamefont {Zhang},
  \citenamefont {Liu}, \citenamefont {Guo}, \citenamefont {Lin}, \citenamefont
  {Jiang},\ and\ \citenamefont {Ji}}]{A29}%
  \BibitemOpen
  \bibfield  {author} {\bibinfo {author} {\bibfnamefont {F.}~\bibnamefont
  {Zhang}}, \bibinfo {author} {\bibfnamefont {X.}~\bibnamefont {Liu}}, \bibinfo
  {author} {\bibfnamefont {C.}~\bibnamefont {Guo}}, \bibinfo {author}
  {\bibfnamefont {S.}~\bibnamefont {Lin}}, \bibinfo {author} {\bibfnamefont
  {J.}~\bibnamefont {Jiang}}, \ and\ \bibinfo {author} {\bibfnamefont
  {X.}~\bibnamefont {Ji}},\ }in\ \href {\doibase 10.1109/CVPR46437.2021.01038}
  {\emph {\bibinfo {booktitle} {Proceedings of the IEEE Computer Society
  Conference on Computer Vision and Pattern Recognition}}}\ (\bibinfo {year}
  {2021})\ pp.\ \bibinfo {pages} {10518--10526}\BibitemShut {NoStop}%
\bibitem [{\citenamefont {Li}\ \emph {et~al.}(2020)\citenamefont {Li},
  \citenamefont {Chen}, \citenamefont {Chi}, \citenamefont {Mann},\ and\
  \citenamefont {Razi}}]{A30}%
  \BibitemOpen
  \bibfield  {author} {\bibinfo {author} {\bibfnamefont {H.}~\bibnamefont
  {Li}}, \bibinfo {author} {\bibfnamefont {X.}~\bibnamefont {Chen}}, \bibinfo
  {author} {\bibfnamefont {Z.}~\bibnamefont {Chi}}, \bibinfo {author}
  {\bibfnamefont {C.}~\bibnamefont {Mann}}, \ and\ \bibinfo {author}
  {\bibfnamefont {A.}~\bibnamefont {Razi}},\ }\href {\doibase
  10.1109/ACCESS.2020.3036380} {\bibfield  {journal} {\bibinfo  {journal} {IEEE
  Access}\ }\textbf {\bibinfo {volume} {8}},\ \bibinfo {pages} {202648}
  (\bibinfo {year} {2020})}\BibitemShut {NoStop}%
\bibitem [{\citenamefont {Galande}\ \emph {et~al.}(2023)\citenamefont
  {Galande}, \citenamefont {Thapa}, \citenamefont {Gurram},\ and\ \citenamefont
  {John}}]{NN31}%
  \BibitemOpen
  \bibfield  {author} {\bibinfo {author} {\bibfnamefont {A.~S.}\ \bibnamefont
  {Galande}}, \bibinfo {author} {\bibfnamefont {V.}~\bibnamefont {Thapa}},
  \bibinfo {author} {\bibfnamefont {H.~P.~R.}\ \bibnamefont {Gurram}}, \ and\
  \bibinfo {author} {\bibfnamefont {R.}~\bibnamefont {John}},\ }\href@noop {}
  {\bibfield  {journal} {\bibinfo  {journal} {Appl. Phys. Lett.}\ }\textbf
  {\bibinfo {volume} {122}},\ \bibinfo {pages} {133701} (\bibinfo {year}
  {2023})}\BibitemShut {NoStop}%
\bibitem [{\citenamefont {Smith}(2013)}]{smith2013uncertainty}%
  \BibitemOpen
  \bibfield  {author} {\bibinfo {author} {\bibfnamefont {R.~C.}\ \bibnamefont
  {Smith}},\ }\href@noop {} {\emph {\bibinfo {title} {Uncertainty
  quantification: theory, implementation, and applications}}},\ Vol.~\bibinfo
  {volume} {12}\ (\bibinfo  {publisher} {Siam},\ \bibinfo {year}
  {2013})\BibitemShut {NoStop}%
\bibitem [{\citenamefont {Jensen}(2017)}]{Jen:2017}%
  \BibitemOpen
  \bibfield  {author} {\bibinfo {author} {\bibfnamefont {F.}~\bibnamefont
  {Jensen}},\ }\href@noop {} {\emph {\bibinfo {title} {Introduction to
  Computational Chemistry}}}\ (\bibinfo  {publisher} {Wiley},\ \bibinfo {year}
  {2017})\BibitemShut {NoStop}%
\bibitem [{\citenamefont {Higdon}\ \emph {et~al.}(2008)\citenamefont {Higdon},
  \citenamefont {Gattiker}, \citenamefont {Williams},\ and\ \citenamefont
  {Rightley}}]{higdon2008cmc}%
  \BibitemOpen
  \bibfield  {author} {\bibinfo {author} {\bibfnamefont {D.}~\bibnamefont
  {Higdon}}, \bibinfo {author} {\bibfnamefont {J.}~\bibnamefont {Gattiker}},
  \bibinfo {author} {\bibfnamefont {B.}~\bibnamefont {Williams}}, \ and\
  \bibinfo {author} {\bibfnamefont {M.}~\bibnamefont {Rightley}},\ }\href@noop
  {} {\bibfield  {journal} {\bibinfo  {journal} {J. Am. Stat. Assoc.}\ }\textbf
  {\bibinfo {volume} {103}},\ \bibinfo {pages} {570} (\bibinfo {year}
  {2008})}\BibitemShut {NoStop}%
\bibitem [{\citenamefont {Calvetti}\ and\ \citenamefont
  {Somersalo}(2008)}]{CS:2008}%
  \BibitemOpen
  \bibfield  {author} {\bibinfo {author} {\bibfnamefont {D.}~\bibnamefont
  {Calvetti}}\ and\ \bibinfo {author} {\bibfnamefont {E.}~\bibnamefont
  {Somersalo}},\ }\href@noop {} {\bibfield  {journal} {\bibinfo  {journal}
  {Inverse Problems}\ }\textbf {\bibinfo {volume} {24}},\ \bibinfo {pages}
  {034013} (\bibinfo {year} {2008})}\BibitemShut {NoStop}%
\bibitem [{\citenamefont {Bardsley}(2012)}]{Bar:2012}%
  \BibitemOpen
  \bibfield  {author} {\bibinfo {author} {\bibfnamefont {J.~M.}\ \bibnamefont
  {Bardsley}},\ }\href@noop {} {\bibfield  {journal} {\bibinfo  {journal} {SIAM
  J. Sci. Comput.}\ }\textbf {\bibinfo {volume} {34}},\ \bibinfo {pages}
  {A1316} (\bibinfo {year} {2012})}\BibitemShut {NoStop}%
\bibitem [{\citenamefont {Baguer}\ \emph {et~al.}(2020)\citenamefont {Baguer},
  \citenamefont {Leuschner},\ and\ \citenamefont {Schmidt}}]{BLS:2020}%
  \BibitemOpen
  \bibfield  {author} {\bibinfo {author} {\bibfnamefont {D.~O.}\ \bibnamefont
  {Baguer}}, \bibinfo {author} {\bibfnamefont {J.}~\bibnamefont {Leuschner}}, \
  and\ \bibinfo {author} {\bibfnamefont {M.}~\bibnamefont {Schmidt}},\
  }\href@noop {} {\bibfield  {journal} {\bibinfo  {journal} {Inverse Problems}\
  }\textbf {\bibinfo {volume} {36}},\ \bibinfo {pages} {094004} (\bibinfo
  {year} {2020})}\BibitemShut {NoStop}%
\bibitem [{\citenamefont {Kennedy}\ and\ \citenamefont
  {O'{H}agan}(2001)}]{kenn:ohag:2001}%
  \BibitemOpen
  \bibfield  {author} {\bibinfo {author} {\bibfnamefont {M.}~\bibnamefont
  {Kennedy}}\ and\ \bibinfo {author} {\bibfnamefont {A.}~\bibnamefont
  {O'{H}agan}},\ }\href@noop {} {\bibfield  {journal} {\bibinfo  {journal}
  {Journal of the Royal Statistical Society (Series B)}\ }\textbf {\bibinfo
  {volume} {68}},\ \bibinfo {pages} {425} (\bibinfo {year} {2001})}\BibitemShut
  {NoStop}%
\bibitem [{\citenamefont {Abdar}\ \emph {et~al.}(2021)\citenamefont {Abdar},
  \citenamefont {Pourpanah}, \citenamefont {Hussain}, \citenamefont
  {Rezazadegan}, \citenamefont {Liu}, \citenamefont {Ghavamzadeh},
  \citenamefont {Fieguth}, \citenamefont {Cao}, \citenamefont {Khosravi},
  \citenamefont {Acharya}, \citenamefont {Makarenkov},\ and\ \citenamefont
  {Nahavandi}}]{APH:2021}%
  \BibitemOpen
  \bibfield  {author} {\bibinfo {author} {\bibfnamefont {M.}~\bibnamefont
  {Abdar}}, \bibinfo {author} {\bibfnamefont {F.}~\bibnamefont {Pourpanah}},
  \bibinfo {author} {\bibfnamefont {S.}~\bibnamefont {Hussain}}, \bibinfo
  {author} {\bibfnamefont {D.}~\bibnamefont {Rezazadegan}}, \bibinfo {author}
  {\bibfnamefont {L.}~\bibnamefont {Liu}}, \bibinfo {author} {\bibfnamefont
  {M.}~\bibnamefont {Ghavamzadeh}}, \bibinfo {author} {\bibfnamefont
  {P.}~\bibnamefont {Fieguth}}, \bibinfo {author} {\bibfnamefont
  {X.}~\bibnamefont {Cao}}, \bibinfo {author} {\bibfnamefont {A.}~\bibnamefont
  {Khosravi}}, \bibinfo {author} {\bibfnamefont {U.~R.}\ \bibnamefont
  {Acharya}}, \bibinfo {author} {\bibfnamefont {V.}~\bibnamefont {Makarenkov}},
  \ and\ \bibinfo {author} {\bibfnamefont {S.}~\bibnamefont {Nahavandi}},\
  }\href@noop {} {\bibfield  {journal} {\bibinfo  {journal} {Information
  fusion}\ }\textbf {\bibinfo {volume} {76}},\ \bibinfo {pages} {243} (\bibinfo
  {year} {2021})}\BibitemShut {NoStop}%
\bibitem [{\citenamefont {Barbano}\ \emph {et~al.}(2022)\citenamefont
  {Barbano}, \citenamefont {Arridge}, \citenamefont {Jin},\ and\ \citenamefont
  {Tanno}}]{BAJ:2022}%
  \BibitemOpen
  \bibfield  {author} {\bibinfo {author} {\bibfnamefont {R.}~\bibnamefont
  {Barbano}}, \bibinfo {author} {\bibfnamefont {S.}~\bibnamefont {Arridge}},
  \bibinfo {author} {\bibfnamefont {B.}~\bibnamefont {Jin}}, \ and\ \bibinfo
  {author} {\bibfnamefont {R.}~\bibnamefont {Tanno}},\ }\enquote {\bibinfo
  {title} {Uncertainty quantification in medical image synthesis},}\ in\ \href
  {https://doi.org/10.1016/B978-0-12-824349-7.00033-5} {\emph {\bibinfo
  {booktitle} {Biomedical image synthesis simulation: Methods and
  Applications}}},\ \bibinfo {editor} {edited by\ \bibinfo {editor}
  {\bibfnamefont {N.}~\bibnamefont {Burgos}}\ and\ \bibinfo {editor}
  {\bibfnamefont {D.}~\bibnamefont {Svoboda}}}\ (\bibinfo  {publisher}
  {Elsevier},\ \bibinfo {address} {Amsterdam, Netherlands},\ \bibinfo {year}
  {2022})\ Chap.~\bibinfo {chapter} {26}, pp.\ \bibinfo {pages}
  {601--641}\BibitemShut {NoStop}%
\bibitem [{\citenamefont {Kucukelbir}\ \emph {et~al.}(2017)\citenamefont
  {Kucukelbir}, \citenamefont {Tran}, \citenamefont {Ranganath}, \citenamefont
  {Gelman},\ and\ \citenamefont {Blei}}]{KTR:2017}%
  \BibitemOpen
  \bibfield  {author} {\bibinfo {author} {\bibfnamefont {A.}~\bibnamefont
  {Kucukelbir}}, \bibinfo {author} {\bibfnamefont {D.}~\bibnamefont {Tran}},
  \bibinfo {author} {\bibfnamefont {R.}~\bibnamefont {Ranganath}}, \bibinfo
  {author} {\bibfnamefont {A.}~\bibnamefont {Gelman}}, \ and\ \bibinfo {author}
  {\bibfnamefont {D.~M.}\ \bibnamefont {Blei}},\ }\href@noop {} {\bibfield
  {journal} {\bibinfo  {journal} {J. Mach. Learn. Res.}\ }\textbf {\bibinfo
  {volume} {18}},\ \bibinfo {pages} {1 } (\bibinfo {year} {2017})}\BibitemShut
  {NoStop}%
\bibitem [{\citenamefont {Xue}\ \emph {et~al.}(2019)\citenamefont {Xue},
  \citenamefont {Cheng}, \citenamefont {Li},\ and\ \citenamefont
  {Tian}}]{XCL:2019}%
  \BibitemOpen
  \bibfield  {author} {\bibinfo {author} {\bibfnamefont {Y.}~\bibnamefont
  {Xue}}, \bibinfo {author} {\bibfnamefont {S.}~\bibnamefont {Cheng}}, \bibinfo
  {author} {\bibfnamefont {Y.}~\bibnamefont {Li}}, \ and\ \bibinfo {author}
  {\bibfnamefont {L.}~\bibnamefont {Tian}},\ }\href@noop {} {\bibfield
  {journal} {\bibinfo  {journal} {Optica}\ }\textbf {\bibinfo {volume}
  {6(5)}},\ \bibinfo {pages} {618} (\bibinfo {year} {2019})}\BibitemShut
  {NoStop}%
\bibitem [{\citenamefont {Beisbart}\ and\ \citenamefont
  {Hartmann}(2011)}]{BH:2011}%
  \BibitemOpen
  \bibinfo {editor} {\bibfnamefont {C.}~\bibnamefont {Beisbart}}\ and\ \bibinfo
  {editor} {\bibfnamefont {S.}~\bibnamefont {Hartmann}},\ eds.,\ \href@noop {}
  {\emph {\bibinfo {title} {Probabilities in Physics}}}\ (\bibinfo  {publisher}
  {Oxford University Press},\ \bibinfo {address} {Oxford, UK},\ \bibinfo {year}
  {2011})\BibitemShut {NoStop}%
\bibitem [{\citenamefont {Olivier}\ \emph {et~al.}(2021)\citenamefont
  {Olivier}, \citenamefont {Shields},\ and\ \citenamefont
  {Graham-Brady}}]{OSG:2021}%
  \BibitemOpen
  \bibfield  {author} {\bibinfo {author} {\bibfnamefont {A.}~\bibnamefont
  {Olivier}}, \bibinfo {author} {\bibfnamefont {M.~D.}\ \bibnamefont
  {Shields}}, \ and\ \bibinfo {author} {\bibfnamefont {L.}~\bibnamefont
  {Graham-Brady}},\ }\href@noop {} {\bibfield  {journal} {\bibinfo  {journal}
  {Comp. Meth. Appl. Mech. Eng.}\ }\textbf {\bibinfo {volume} {386}},\ \bibinfo
  {pages} {114079} (\bibinfo {year} {2021})}\BibitemShut {NoStop}%
\bibitem [{\citenamefont {Spiegelhalter}(2015)}]{Spi:2014}%
  \BibitemOpen
  \bibfield  {author} {\bibinfo {author} {\bibfnamefont {D.~J.}\ \bibnamefont
  {Spiegelhalter}},\ }\href@noop {} {\bibfield  {journal} {\bibinfo  {journal}
  {Science}\ }\textbf {\bibinfo {volume} {345}},\ \bibinfo {pages} {264 }
  (\bibinfo {year} {2015})}\BibitemShut {NoStop}%
\bibitem [{\citenamefont {Rack}\ \emph
  {et~al.}(2010{\natexlab{b}})\citenamefont {Rack}, \citenamefont
  {Garc{\'i}a-Moreno}, \citenamefont {Schmitt}, \citenamefont {Betz},
  \citenamefont {Cecilia}, \citenamefont {Ershov}, \citenamefont {Rack},
  \citenamefont {Banhart},\ and\ \citenamefont {Zabler}}]{rack2010}%
  \BibitemOpen
  \bibfield  {author} {\bibinfo {author} {\bibfnamefont {A.}~\bibnamefont
  {Rack}}, \bibinfo {author} {\bibfnamefont {F.}~\bibnamefont
  {Garc{\'i}a-Moreno}}, \bibinfo {author} {\bibfnamefont {C.}~\bibnamefont
  {Schmitt}}, \bibinfo {author} {\bibfnamefont {O.}~\bibnamefont {Betz}},
  \bibinfo {author} {\bibfnamefont {A.}~\bibnamefont {Cecilia}}, \bibinfo
  {author} {\bibfnamefont {A.}~\bibnamefont {Ershov}}, \bibinfo {author}
  {\bibfnamefont {T.}~\bibnamefont {Rack}}, \bibinfo {author} {\bibfnamefont
  {J.}~\bibnamefont {Banhart}}, \ and\ \bibinfo {author} {\bibfnamefont
  {S.}~\bibnamefont {Zabler}},\ }\href {\doibase 10.3233/XST-2010-0273}
  {\bibfield  {journal} {\bibinfo  {journal} {J. X-Ray Sci. Tech.}\ }\textbf
  {\bibinfo {volume} {18}},\ \bibinfo {pages} {429} (\bibinfo {year}
  {2010}{\natexlab{b}})}\BibitemShut {NoStop}%
\bibitem [{\citenamefont {Rack}\ \emph {et~al.}(2014)\citenamefont {Rack},
  \citenamefont {Scheel}, \citenamefont {Hardy}, \citenamefont {Curfs},
  \citenamefont {Bonnin},\ and\ \citenamefont {Reichert}}]{rack2014}%
  \BibitemOpen
  \bibfield  {author} {\bibinfo {author} {\bibfnamefont {A.}~\bibnamefont
  {Rack}}, \bibinfo {author} {\bibfnamefont {M.}~\bibnamefont {Scheel}},
  \bibinfo {author} {\bibfnamefont {L.}~\bibnamefont {Hardy}}, \bibinfo
  {author} {\bibfnamefont {C.}~\bibnamefont {Curfs}}, \bibinfo {author}
  {\bibfnamefont {A.}~\bibnamefont {Bonnin}}, \ and\ \bibinfo {author}
  {\bibfnamefont {H.}~\bibnamefont {Reichert}},\ }\href {\doibase
  10.1107/S1600577514005852} {\bibfield  {journal} {\bibinfo  {journal} {J.
  Synchrotron Radiat.}\ }\textbf {\bibinfo {volume} {21}},\ \bibinfo {pages}
  {815} (\bibinfo {year} {2014})}\BibitemShut {NoStop}%
\bibitem [{\citenamefont {Olbinado}\ \emph {et~al.}(2017)\citenamefont
  {Olbinado}, \citenamefont {Just}, \citenamefont {Gelet}, \citenamefont
  {Lhuissier}, \citenamefont {Scheel}, \citenamefont {Vagovic}, \citenamefont
  {Sato}, \citenamefont {Graceffa}, \citenamefont {Schulz}, \citenamefont
  {Manusco}, \citenamefont {Morse},\ and\ \citenamefont {Rack}}]{olbinado2017}%
  \BibitemOpen
  \bibfield  {author} {\bibinfo {author} {\bibfnamefont {M.~P.}\ \bibnamefont
  {Olbinado}}, \bibinfo {author} {\bibfnamefont {X.}~\bibnamefont {Just}},
  \bibinfo {author} {\bibfnamefont {J.-L.}\ \bibnamefont {Gelet}}, \bibinfo
  {author} {\bibfnamefont {P.}~\bibnamefont {Lhuissier}}, \bibinfo {author}
  {\bibfnamefont {M.}~\bibnamefont {Scheel}}, \bibinfo {author} {\bibfnamefont
  {P.}~\bibnamefont {Vagovic}}, \bibinfo {author} {\bibfnamefont
  {T.}~\bibnamefont {Sato}}, \bibinfo {author} {\bibfnamefont {R.}~\bibnamefont
  {Graceffa}}, \bibinfo {author} {\bibfnamefont {J.}~\bibnamefont {Schulz}},
  \bibinfo {author} {\bibfnamefont {A.}~\bibnamefont {Manusco}}, \bibinfo
  {author} {\bibfnamefont {J.}~\bibnamefont {Morse}}, \ and\ \bibinfo {author}
  {\bibfnamefont {A.}~\bibnamefont {Rack}},\ }\href {\doibase
  10.1364/OE.25.013857} {\bibfield  {journal} {\bibinfo  {journal} {Optics
  Expr.}\ }\textbf {\bibinfo {volume} {25}},\ \bibinfo {pages} {13857}
  (\bibinfo {year} {2017})}\BibitemShut {NoStop}%
\bibitem [{\citenamefont {McCarthy}\ and\ \citenamefont
  {Reichert}(2022)}]{mccarthy2022}%
  \BibitemOpen
  \bibfield  {author} {\bibinfo {author} {\bibfnamefont {J.}~\bibnamefont
  {McCarthy}}\ and\ \bibinfo {author} {\bibfnamefont {H.}~\bibnamefont
  {Reichert}},\ }\href {\doibase 10.1080/08940886.2022.2064150} {\bibfield
  {journal} {\bibinfo  {journal} {Synchrotron Radiat. News}\ }\textbf {\bibinfo
  {volume} {35}},\ \bibinfo {pages} {52} (\bibinfo {year} {2022})}\BibitemShut
  {NoStop}%
\bibitem [{\citenamefont {Escauriza}\ \emph {et~al.}(2020)\citenamefont
  {Escauriza}, \citenamefont {Duarte}, \citenamefont {Chapman}, \citenamefont
  {Rutherford}, \citenamefont {Farbaniec}, \citenamefont {Jonsson},
  \citenamefont {Smith}, \citenamefont {Olbinado}, \citenamefont {Skidmore},
  \citenamefont {Foster}, \citenamefont {Ringrose}, \citenamefont {Rack},\ and\
  \citenamefont {Eakins}}]{escauriza2020}%
  \BibitemOpen
  \bibfield  {author} {\bibinfo {author} {\bibfnamefont {E.~M.}\ \bibnamefont
  {Escauriza}}, \bibinfo {author} {\bibfnamefont {J.~P.}\ \bibnamefont
  {Duarte}}, \bibinfo {author} {\bibfnamefont {D.~J.}\ \bibnamefont {Chapman}},
  \bibinfo {author} {\bibfnamefont {M.~E.}\ \bibnamefont {Rutherford}},
  \bibinfo {author} {\bibfnamefont {L.}~\bibnamefont {Farbaniec}}, \bibinfo
  {author} {\bibfnamefont {J.~C.}\ \bibnamefont {Jonsson}}, \bibinfo {author}
  {\bibfnamefont {L.~C.}\ \bibnamefont {Smith}}, \bibinfo {author}
  {\bibfnamefont {M.~P.}\ \bibnamefont {Olbinado}}, \bibinfo {author}
  {\bibfnamefont {J.}~\bibnamefont {Skidmore}}, \bibinfo {author}
  {\bibfnamefont {P.}~\bibnamefont {Foster}}, \bibinfo {author} {\bibfnamefont
  {T.}~\bibnamefont {Ringrose}}, \bibinfo {author} {\bibfnamefont
  {A.}~\bibnamefont {Rack}}, \ and\ \bibinfo {author} {\bibfnamefont {D.~E.}\
  \bibnamefont {Eakins}},\ }\href {\doibase 10.1038/s41598-020-64669-y}
  {\bibfield  {journal} {\bibinfo  {journal} {Sci. Rep.}\ }\textbf {\bibinfo
  {volume} {10}},\ \bibinfo {pages} {8455} (\bibinfo {year}
  {2020})}\BibitemShut {NoStop}%
\bibitem [{\citenamefont {Strucka}\ \emph {et~al.}(2023)\citenamefont
  {Strucka}, \citenamefont {Lukic}, \citenamefont {Koerner}, \citenamefont
  {Halliday}, \citenamefont {Yao}, \citenamefont {Mughal}, \citenamefont
  {Maler}, \citenamefont {Efimov}, \citenamefont {Skidmore}, \citenamefont
  {Rack}, \citenamefont {Krasik}, \citenamefont {Chittenden},\ and\
  \citenamefont {Bland}}]{strucka2023}%
  \BibitemOpen
  \bibfield  {author} {\bibinfo {author} {\bibfnamefont {J.}~\bibnamefont
  {Strucka}}, \bibinfo {author} {\bibfnamefont {B.}~\bibnamefont {Lukic}},
  \bibinfo {author} {\bibfnamefont {M.}~\bibnamefont {Koerner}}, \bibinfo
  {author} {\bibfnamefont {J.~W.~D.}\ \bibnamefont {Halliday}}, \bibinfo
  {author} {\bibfnamefont {Y.}~\bibnamefont {Yao}}, \bibinfo {author}
  {\bibfnamefont {K.}~\bibnamefont {Mughal}}, \bibinfo {author} {\bibfnamefont
  {D.}~\bibnamefont {Maler}}, \bibinfo {author} {\bibfnamefont
  {S.}~\bibnamefont {Efimov}}, \bibinfo {author} {\bibfnamefont
  {J.}~\bibnamefont {Skidmore}}, \bibinfo {author} {\bibfnamefont
  {A.}~\bibnamefont {Rack}}, \bibinfo {author} {\bibfnamefont {Y.}~\bibnamefont
  {Krasik}}, \bibinfo {author} {\bibfnamefont {J.}~\bibnamefont {Chittenden}},
  \ and\ \bibinfo {author} {\bibfnamefont {S.~N.}\ \bibnamefont {Bland}},\
  }\href {\doibase tba} {\bibfield  {journal} {\bibinfo  {journal} {Phys.
  Fluids}\ }\textbf {\bibinfo {volume} {35}},\ \bibinfo {pages} {in press}
  (\bibinfo {year} {2023})}\BibitemShut {NoStop}%
\bibitem [{\citenamefont {Ball}(2021)}]{ball2021}%
  \BibitemOpen
  \bibfield  {author} {\bibinfo {author} {\bibfnamefont {P.}~\bibnamefont
  {Ball}},\ }\href@noop {} {\bibfield  {journal} {\bibinfo  {journal} {Nature}\
  }\textbf {\bibinfo {volume} {599}},\ \bibinfo {pages} {362} (\bibinfo {year}
  {2021})}\BibitemShut {NoStop}%
\bibitem [{\citenamefont {Escauriza}\ \emph {et~al.}(2018)\citenamefont
  {Escauriza}, \citenamefont {Olbinado}, \citenamefont {Rutherford},
  \citenamefont {Chapman}, \citenamefont {Jonsson}, \citenamefont {Rack},\ and\
  \citenamefont {Eakins}}]{escauriza2018}%
  \BibitemOpen
  \bibfield  {author} {\bibinfo {author} {\bibfnamefont {E.~M.}\ \bibnamefont
  {Escauriza}}, \bibinfo {author} {\bibfnamefont {M.~P.}\ \bibnamefont
  {Olbinado}}, \bibinfo {author} {\bibfnamefont {M.~E.}\ \bibnamefont
  {Rutherford}}, \bibinfo {author} {\bibfnamefont {D.~J.}\ \bibnamefont
  {Chapman}}, \bibinfo {author} {\bibfnamefont {J.~C.~Z.}\ \bibnamefont
  {Jonsson}}, \bibinfo {author} {\bibfnamefont {A.}~\bibnamefont {Rack}}, \
  and\ \bibinfo {author} {\bibfnamefont {D.~E.}\ \bibnamefont {Eakins}},\
  }\href {\doibase 10.1364/AO.57.005004} {\bibfield  {journal} {\bibinfo
  {journal} {Appl. Opt.}\ }\textbf {\bibinfo {volume} {57}},\ \bibinfo {pages}
  {5004} (\bibinfo {year} {2018})}\BibitemShut {NoStop}%
\bibitem [{\citenamefont {Westneat}\ \emph {et~al.}(2003)\citenamefont
  {Westneat}, \citenamefont {Betz}, \citenamefont {Blob}, \citenamefont
  {Fezzaa}, \citenamefont {Cooper},\ and\ \citenamefont {Lee}}]{Fez1}%
  \BibitemOpen
  \bibfield  {author} {\bibinfo {author} {\bibfnamefont {M.~W.}\ \bibnamefont
  {Westneat}}, \bibinfo {author} {\bibfnamefont {O.}~\bibnamefont {Betz}},
  \bibinfo {author} {\bibfnamefont {R.~W.}\ \bibnamefont {Blob}}, \bibinfo
  {author} {\bibfnamefont {K.}~\bibnamefont {Fezzaa}}, \bibinfo {author}
  {\bibfnamefont {W.~J.}\ \bibnamefont {Cooper}}, \ and\ \bibinfo {author}
  {\bibfnamefont {W.-K.}\ \bibnamefont {Lee}},\ }\href@noop {} {\bibfield
  {journal} {\bibinfo  {journal} {Science}\ }\textbf {\bibinfo {volume}
  {229}},\ \bibinfo {pages} {558} (\bibinfo {year} {2003})}\BibitemShut
  {NoStop}%
\bibitem [{\citenamefont {Wang}\ \emph {et~al.}(2008)\citenamefont {Wang},
  \citenamefont {Liu}, \citenamefont {Im}, \citenamefont {Lee}, \citenamefont
  {Wang}, \citenamefont {Fezzaa}, \citenamefont {Hung},\ and\ \citenamefont
  {Winkelman}}]{Fez2}%
  \BibitemOpen
  \bibfield  {author} {\bibinfo {author} {\bibfnamefont {Y.}~\bibnamefont
  {Wang}}, \bibinfo {author} {\bibfnamefont {X.}~\bibnamefont {Liu}}, \bibinfo
  {author} {\bibfnamefont {K.-S.}\ \bibnamefont {Im}}, \bibinfo {author}
  {\bibfnamefont {W.-K.}\ \bibnamefont {Lee}}, \bibinfo {author} {\bibfnamefont
  {J.}~\bibnamefont {Wang}}, \bibinfo {author} {\bibfnamefont {K.}~\bibnamefont
  {Fezzaa}}, \bibinfo {author} {\bibfnamefont {D.}~\bibnamefont {Hung}}, \ and\
  \bibinfo {author} {\bibfnamefont {J.}~\bibnamefont {Winkelman}},\ }\href@noop
  {} {\bibfield  {journal} {\bibinfo  {journal} {Nature Physics}\ }\textbf
  {\bibinfo {volume} {4}},\ \bibinfo {pages} {305} (\bibinfo {year}
  {2008})}\BibitemShut {NoStop}%
\bibitem [{\citenamefont {Hudspeth}\ \emph {et~al.}(2015)\citenamefont
  {Hudspeth}, \citenamefont {Sun}, \citenamefont {Parab}, \citenamefont {Guo},
  \citenamefont {Fezzaa}, \citenamefont {Luo},\ and\ \citenamefont
  {Chen}}]{Fez4}%
  \BibitemOpen
  \bibfield  {author} {\bibinfo {author} {\bibfnamefont {M.}~\bibnamefont
  {Hudspeth}}, \bibinfo {author} {\bibfnamefont {T.}~\bibnamefont {Sun}},
  \bibinfo {author} {\bibfnamefont {N.}~\bibnamefont {Parab}}, \bibinfo
  {author} {\bibfnamefont {Z.}~\bibnamefont {Guo}}, \bibinfo {author}
  {\bibfnamefont {K.}~\bibnamefont {Fezzaa}}, \bibinfo {author} {\bibfnamefont
  {S.}~\bibnamefont {Luo}}, \ and\ \bibinfo {author} {\bibfnamefont
  {W.}~\bibnamefont {Chen}},\ }\href@noop {} {\bibfield  {journal} {\bibinfo
  {journal} {J. Synchrotron Rad.}\ }\textbf {\bibinfo {volume} {22}},\ \bibinfo
  {pages} {49} (\bibinfo {year} {2015})}\BibitemShut {NoStop}%
\bibitem [{\citenamefont {Jensen}\ \emph {et~al.}(2015)\citenamefont {Jensen},
  \citenamefont {Cherne}, \citenamefont {Prime}, \citenamefont {Fezzaa},
  \citenamefont {Iverson}, \citenamefont {Carlson}, \citenamefont {Yeager},
  \citenamefont {Ramos}, \citenamefont {Hooks}, \citenamefont {Cooley},\ and\
  \citenamefont {Dimonte}}]{Fez5}%
  \BibitemOpen
  \bibfield  {author} {\bibinfo {author} {\bibfnamefont {B.~J.}\ \bibnamefont
  {Jensen}}, \bibinfo {author} {\bibfnamefont {F.~J.}\ \bibnamefont {Cherne}},
  \bibinfo {author} {\bibfnamefont {M.~B.}\ \bibnamefont {Prime}}, \bibinfo
  {author} {\bibfnamefont {K.}~\bibnamefont {Fezzaa}}, \bibinfo {author}
  {\bibfnamefont {A.~J.}\ \bibnamefont {Iverson}}, \bibinfo {author}
  {\bibfnamefont {C.~A.}\ \bibnamefont {Carlson}}, \bibinfo {author}
  {\bibfnamefont {J.~D.}\ \bibnamefont {Yeager}}, \bibinfo {author}
  {\bibfnamefont {K.~J.}\ \bibnamefont {Ramos}}, \bibinfo {author}
  {\bibfnamefont {D.~E.}\ \bibnamefont {Hooks}}, \bibinfo {author}
  {\bibfnamefont {J.~C.}\ \bibnamefont {Cooley}}, \ and\ \bibinfo {author}
  {\bibfnamefont {G.}~\bibnamefont {Dimonte}},\ }\href@noop {} {\bibfield
  {journal} {\bibinfo  {journal} {J. Appl. Phys.}\ }\textbf {\bibinfo {volume}
  {118}},\ \bibinfo {pages} {195903} (\bibinfo {year} {2015})}\BibitemShut
  {NoStop}%
\bibitem [{\citenamefont {Lu}\ \emph {et~al.}(2016)\citenamefont {Lu},
  \citenamefont {Huang}, \citenamefont {Fan}, \citenamefont {Bie},
  \citenamefont {Sun}, \citenamefont {Fezzaa}, \citenamefont {Gong},\ and\
  \citenamefont {Luo}}]{Fez6}%
  \BibitemOpen
  \bibfield  {author} {\bibinfo {author} {\bibfnamefont {L.}~\bibnamefont
  {Lu}}, \bibinfo {author} {\bibfnamefont {J.~W.}\ \bibnamefont {Huang}},
  \bibinfo {author} {\bibfnamefont {D.}~\bibnamefont {Fan}}, \bibinfo {author}
  {\bibfnamefont {B.~X.}\ \bibnamefont {Bie}}, \bibinfo {author} {\bibfnamefont
  {T.}~\bibnamefont {Sun}}, \bibinfo {author} {\bibfnamefont {K.}~\bibnamefont
  {Fezzaa}}, \bibinfo {author} {\bibfnamefont {X.~L.}\ \bibnamefont {Gong}}, \
  and\ \bibinfo {author} {\bibfnamefont {S.~N.}\ \bibnamefont {Luo}},\
  }\href@noop {} {\bibfield  {journal} {\bibinfo  {journal} {Acta Materialia}\
  }\textbf {\bibinfo {volume} {120}},\ \bibinfo {pages} {86} (\bibinfo {year}
  {2016})}\BibitemShut {NoStop}%
\bibitem [{\citenamefont {Hammons}\ \emph {et~al.}(2019)\citenamefont
  {Hammons}, \citenamefont {Nielsen}, \citenamefont {Bagge-Hansen},
  \citenamefont {Bastea}, \citenamefont {Shaw}, \citenamefont {Lee},
  \citenamefont {Ilavsky}, \citenamefont {Sinclair}, \citenamefont {Fezzaa},
  \citenamefont {Lauderbach}, \citenamefont {Hodgin}, \citenamefont
  {Orlikowski}, \citenamefont {Fried},\ and\ \citenamefont {Willey}}]{Fez7}%
  \BibitemOpen
  \bibfield  {author} {\bibinfo {author} {\bibfnamefont {J.~A.}\ \bibnamefont
  {Hammons}}, \bibinfo {author} {\bibfnamefont {M.~H.}\ \bibnamefont
  {Nielsen}}, \bibinfo {author} {\bibfnamefont {M.}~\bibnamefont
  {Bagge-Hansen}}, \bibinfo {author} {\bibfnamefont {S.}~\bibnamefont
  {Bastea}}, \bibinfo {author} {\bibfnamefont {W.~L.}\ \bibnamefont {Shaw}},
  \bibinfo {author} {\bibfnamefont {J.~R.~I.}\ \bibnamefont {Lee}}, \bibinfo
  {author} {\bibfnamefont {J.}~\bibnamefont {Ilavsky}}, \bibinfo {author}
  {\bibfnamefont {N.}~\bibnamefont {Sinclair}}, \bibinfo {author}
  {\bibfnamefont {K.}~\bibnamefont {Fezzaa}}, \bibinfo {author} {\bibfnamefont
  {L.~M.}\ \bibnamefont {Lauderbach}}, \bibinfo {author} {\bibfnamefont
  {R.~L.}\ \bibnamefont {Hodgin}}, \bibinfo {author} {\bibfnamefont {D.~A.}\
  \bibnamefont {Orlikowski}}, \bibinfo {author} {\bibfnamefont {L.~E.}\
  \bibnamefont {Fried}}, \ and\ \bibinfo {author} {\bibfnamefont {T.~M.}\
  \bibnamefont {Willey}},\ }\href@noop {} {\bibfield  {journal} {\bibinfo
  {journal} {The Journal of Physical Chemistry C}\ }\textbf {\bibinfo {volume}
  {123}},\ \bibinfo {pages} {19153} (\bibinfo {year} {2019})}\BibitemShut
  {NoStop}%
\bibitem [{\citenamefont {Ren}\ \emph {et~al.}(2023)\citenamefont {Ren},
  \citenamefont {Gao}, \citenamefont {Clark}, \citenamefont {Fezzaa},
  \citenamefont {Shevchenko}, \citenamefont {Choi}, \citenamefont {Everhart},
  \citenamefont {Rollett}, \citenamefont {Chen},\ and\ \citenamefont
  {Sun}}]{Fez8}%
  \BibitemOpen
  \bibfield  {author} {\bibinfo {author} {\bibfnamefont {Z.}~\bibnamefont
  {Ren}}, \bibinfo {author} {\bibfnamefont {L.}~\bibnamefont {Gao}}, \bibinfo
  {author} {\bibfnamefont {S.~J.}\ \bibnamefont {Clark}}, \bibinfo {author}
  {\bibfnamefont {K.}~\bibnamefont {Fezzaa}}, \bibinfo {author} {\bibfnamefont
  {P.}~\bibnamefont {Shevchenko}}, \bibinfo {author} {\bibfnamefont
  {A.}~\bibnamefont {Choi}}, \bibinfo {author} {\bibfnamefont {W.}~\bibnamefont
  {Everhart}}, \bibinfo {author} {\bibfnamefont {A.}~\bibnamefont {Rollett}},
  \bibinfo {author} {\bibfnamefont {L.}~\bibnamefont {Chen}}, \ and\ \bibinfo
  {author} {\bibfnamefont {T.}~\bibnamefont {Sun}},\ }\href@noop {} {\bibfield
  {journal} {\bibinfo  {journal} {Science}\ }\textbf {\bibinfo {volume}
  {379}},\ \bibinfo {pages} {89} (\bibinfo {year} {2023})}\BibitemShut
  {NoStop}%
\bibitem [{\citenamefont {{L. D. Landau and E. M. Lifshitz}}(1987)}]{LL:1987}%
  \BibitemOpen
  \bibfield  {author} {\bibinfo {author} {\bibnamefont {{L. D. Landau and E. M.
  Lifshitz}}},\ }\href@noop {} {\emph {\bibinfo {title} {Fluid Mechanics}}},\
  \bibinfo {edition} {2nd}\ ed.\ (\bibinfo  {publisher} {Pergamon Press},\
  \bibinfo {year} {1987})\BibitemShut {NoStop}%
\bibitem [{\citenamefont {Davison}\ and\ \citenamefont
  {Graham}(1979)}]{DG:1979}%
  \BibitemOpen
  \bibfield  {author} {\bibinfo {author} {\bibfnamefont {L.}~\bibnamefont
  {Davison}}\ and\ \bibinfo {author} {\bibfnamefont {R.}~\bibnamefont
  {Graham}},\ }\href@noop {} {\bibfield  {journal} {\bibinfo  {journal} {Phys.
  Rep.}\ }\textbf {\bibinfo {volume} {55}},\ \bibinfo {pages} {255} (\bibinfo
  {year} {1979})}\BibitemShut {NoStop}%
\bibitem [{\citenamefont {{J. W. Forbes}}(2012)}]{For:2012}%
  \BibitemOpen
  \bibfield  {author} {\bibinfo {author} {\bibnamefont {{J. W. Forbes}}},\
  }\href@noop {} {\emph {\bibinfo {title} {Shockwave compression of condensed
  matter: a primer}}}\ (\bibinfo  {publisher} {Springer, Berlin, Heidelberg,
  Germany},\ \bibinfo {year} {2012})\BibitemShut {NoStop}%
\bibitem [{\citenamefont {{R. W. Fox and A. T. McDonald}}(1992)}]{FM:1992}%
  \BibitemOpen
  \bibfield  {author} {\bibinfo {author} {\bibnamefont {{R. W. Fox and A. T.
  McDonald}}},\ }\href@noop {} {\emph {\bibinfo {title} {Introduction to Fluid
  Mechanics}}},\ \bibinfo {edition} {4th}\ ed.\ (\bibinfo  {publisher}
  {Wiley},\ \bibinfo {year} {1992})\BibitemShut {NoStop}%
\bibitem [{\citenamefont {Dattelbaum}\ and\ \citenamefont
  {Coe}(2019)}]{DC:2019}%
  \BibitemOpen
  \bibfield  {author} {\bibinfo {author} {\bibfnamefont {D.~M.}\ \bibnamefont
  {Dattelbaum}}\ and\ \bibinfo {author} {\bibfnamefont {J.~D.}\ \bibnamefont
  {Coe}},\ }\href@noop {} {\bibfield  {journal} {\bibinfo  {journal}
  {Polymers}\ }\textbf {\bibinfo {volume} {11}},\ \bibinfo {pages} {493}
  (\bibinfo {year} {2019})}\BibitemShut {NoStop}%
\bibitem [{\citenamefont {Dattelbaum}\ \emph
  {et~al.}(2020{\natexlab{b}})\citenamefont {Dattelbaum}, \citenamefont
  {Ionita}, \citenamefont {Patterson}, \citenamefont {Branch},\ and\
  \citenamefont {Kuettner}}]{D1}%
  \BibitemOpen
  \bibfield  {author} {\bibinfo {author} {\bibfnamefont {D.~M.}\ \bibnamefont
  {Dattelbaum}}, \bibinfo {author} {\bibfnamefont {A.}~\bibnamefont {Ionita}},
  \bibinfo {author} {\bibfnamefont {B.~M.}\ \bibnamefont {Patterson}}, \bibinfo
  {author} {\bibfnamefont {B.~A.}\ \bibnamefont {Branch}}, \ and\ \bibinfo
  {author} {\bibfnamefont {L.}~\bibnamefont {Kuettner}},\ }\href@noop {}
  {\bibfield  {journal} {\bibinfo  {journal} {AIP Advances}\ }\textbf {\bibinfo
  {volume} {10}},\ \bibinfo {pages} {075016} (\bibinfo {year}
  {2020}{\natexlab{b}})}\BibitemShut {NoStop}%
\bibitem [{\citenamefont {Wood}\ \emph
  {et~al.}(2018{\natexlab{b}})\citenamefont {Wood}, \citenamefont {Kittell},
  \citenamefont {Yarrington},\ and\ \citenamefont {Thompson}}]{WKY:2018}%
  \BibitemOpen
  \bibfield  {author} {\bibinfo {author} {\bibfnamefont {M.~A.}\ \bibnamefont
  {Wood}}, \bibinfo {author} {\bibfnamefont {D.~E.}\ \bibnamefont {Kittell}},
  \bibinfo {author} {\bibfnamefont {C.~D.}\ \bibnamefont {Yarrington}}, \ and\
  \bibinfo {author} {\bibfnamefont {A.~P.}\ \bibnamefont {Thompson}},\
  }\href@noop {} {\bibfield  {journal} {\bibinfo  {journal} {Phys. Rev. B}\
  }\textbf {\bibinfo {volume} {97}},\ \bibinfo {pages} {014109} (\bibinfo
  {year} {2018}{\natexlab{b}})}\BibitemShut {NoStop}%
\bibitem [{\citenamefont {Dattelbaum}\ \emph {et~al.}(2022)\citenamefont
  {Dattelbaum}, \citenamefont {Kuettner}, \citenamefont {Patterson},
  \citenamefont {Huber}, \citenamefont {Ionita}, \citenamefont {Wang},
  \citenamefont {Campbell}, \citenamefont {Natan},\ and\ \citenamefont
  {MacNider}}]{D2}%
  \BibitemOpen
  \bibfield  {author} {\bibinfo {author} {\bibfnamefont {D.~M.}\ \bibnamefont
  {Dattelbaum}}, \bibinfo {author} {\bibfnamefont {L.}~\bibnamefont
  {Kuettner}}, \bibinfo {author} {\bibfnamefont {B.~M.}\ \bibnamefont
  {Patterson}}, \bibinfo {author} {\bibfnamefont {R.}~\bibnamefont {Huber}},
  \bibinfo {author} {\bibfnamefont {A.}~\bibnamefont {Ionita}}, \bibinfo
  {author} {\bibfnamefont {Z.}~\bibnamefont {Wang}}, \bibinfo {author}
  {\bibfnamefont {C.}~\bibnamefont {Campbell}}, \bibinfo {author}
  {\bibfnamefont {T.}~\bibnamefont {Natan}}, \ and\ \bibinfo {author}
  {\bibfnamefont {B.}~\bibnamefont {MacNider}},\ }\href@noop {} {\bibfield
  {journal} {\bibinfo  {journal} {DYMAT 22, 26th Technical Meeting Conference
  Proceedings}\ ,\ \bibinfo {pages} {221}} (\bibinfo {year}
  {2022})}\BibitemShut {NoStop}%
\bibitem [{\citenamefont {Xiong}\ \emph {et~al.}(2014)\citenamefont {Xiong},
  \citenamefont {Mountanabbir}, \citenamefont {Reiche}, \citenamefont
  {Harder},\ and\ \citenamefont {Robinson}}]{Xiong:2014}%
  \BibitemOpen
  \bibfield  {author} {\bibinfo {author} {\bibfnamefont {G.}~\bibnamefont
  {Xiong}}, \bibinfo {author} {\bibfnamefont {O.}~\bibnamefont {Mountanabbir}},
  \bibinfo {author} {\bibfnamefont {M.}~\bibnamefont {Reiche}}, \bibinfo
  {author} {\bibfnamefont {R.}~\bibnamefont {Harder}}, \ and\ \bibinfo {author}
  {\bibfnamefont {I.}~\bibnamefont {Robinson}},\ }\href@noop {} {\bibfield
  {journal} {\bibinfo  {journal} {Adv. Mater.}\ }\textbf {\bibinfo {volume}
  {26}},\ \bibinfo {pages} {7747} (\bibinfo {year} {2014})}\BibitemShut
  {NoStop}%
\bibitem [{\citenamefont {Milathianaki}\ \emph {et~al.}(2013)\citenamefont
  {Milathianaki}, \citenamefont {Boutet}, \citenamefont {Williams},
  \citenamefont {Higginbotham}, \citenamefont {Ratner}, \citenamefont
  {Gleason}, \citenamefont {Messerschmidt}, \citenamefont {Seibert},
  \citenamefont {Swift}, \citenamefont {Hering}, \citenamefont {Robinson},
  \citenamefont {White},\ and\ \citenamefont {Wark}}]{MBW:2013}%
  \BibitemOpen
  \bibfield  {author} {\bibinfo {author} {\bibfnamefont {D.}~\bibnamefont
  {Milathianaki}}, \bibinfo {author} {\bibfnamefont {S.}~\bibnamefont
  {Boutet}}, \bibinfo {author} {\bibfnamefont {G.~J.}\ \bibnamefont
  {Williams}}, \bibinfo {author} {\bibfnamefont {A.}~\bibnamefont
  {Higginbotham}}, \bibinfo {author} {\bibfnamefont {D.}~\bibnamefont
  {Ratner}}, \bibinfo {author} {\bibfnamefont {A.~E.}\ \bibnamefont {Gleason}},
  \bibinfo {author} {\bibfnamefont {M.}~\bibnamefont {Messerschmidt}}, \bibinfo
  {author} {\bibfnamefont {M.~M.}\ \bibnamefont {Seibert}}, \bibinfo {author}
  {\bibfnamefont {D.~C.}\ \bibnamefont {Swift}}, \bibinfo {author}
  {\bibfnamefont {P.}~\bibnamefont {Hering}}, \bibinfo {author} {\bibfnamefont
  {J.}~\bibnamefont {Robinson}}, \bibinfo {author} {\bibfnamefont {W.~E.}\
  \bibnamefont {White}}, \ and\ \bibinfo {author} {\bibfnamefont {J.~S.}\
  \bibnamefont {Wark}},\ }\href@noop {} {\bibfield  {journal} {\bibinfo
  {journal} {Science}\ }\textbf {\bibinfo {volume} {342}},\ \bibinfo {pages}
  {220} (\bibinfo {year} {2013})}\BibitemShut {NoStop}%
\bibitem [{\citenamefont {Nagler}\ and\ \citenamefont {{et.
  al.}}(2015)}]{Nagler:2015}%
  \BibitemOpen
  \bibfield  {author} {\bibinfo {author} {\bibfnamefont {B.}~\bibnamefont
  {Nagler}}\ and\ \bibinfo {author} {\bibnamefont {{et. al.}}},\ }\href@noop {}
  {\bibfield  {journal} {\bibinfo  {journal} {J. Synchr. Rad.}\ }\textbf
  {\bibinfo {volume} {22}},\ \bibinfo {pages} {520} (\bibinfo {year}
  {2015})}\BibitemShut {NoStop}%
\bibitem [{\citenamefont {Glenzer}\ \emph {et~al.}(2016)\citenamefont
  {Glenzer}, \citenamefont {Fletcher}, \citenamefont {Galtier}, \citenamefont
  {Nagler}, \citenamefont {Alonso-Mori}, \citenamefont {Barbrel}, \citenamefont
  {Brown}, \citenamefont {Chapman}, \citenamefont {Chen}, \citenamefont
  {Curry}, \citenamefont {Fiuza}, \citenamefont {Gamboa}, \citenamefont
  {Gauthier}, \citenamefont {Gericke}, \citenamefont {Gleason}, \citenamefont
  {Göde}, \citenamefont {Granados}, \citenamefont {Heimann}, \citenamefont
  {Kim}, \citenamefont {Kraus}, \citenamefont {MacDonald}, \citenamefont
  {Mackinnon}, \citenamefont {Mishra}, \citenamefont {Ravasio}, \citenamefont
  {Roedel}, \citenamefont {Sperling}, \citenamefont {Schumaker}, \citenamefont
  {Tsui}, \citenamefont {Vorberger}, \citenamefont {Zastrau}, \citenamefont
  {Fry}, \citenamefont {White}, \citenamefont {Hasting},\ and\ \citenamefont
  {Lee}}]{GFG:2016}%
  \BibitemOpen
  \bibfield  {author} {\bibinfo {author} {\bibfnamefont {S.~H.}\ \bibnamefont
  {Glenzer}}, \bibinfo {author} {\bibfnamefont {L.~B.}\ \bibnamefont
  {Fletcher}}, \bibinfo {author} {\bibfnamefont {E.}~\bibnamefont {Galtier}},
  \bibinfo {author} {\bibfnamefont {B.}~\bibnamefont {Nagler}}, \bibinfo
  {author} {\bibfnamefont {R.}~\bibnamefont {Alonso-Mori}}, \bibinfo {author}
  {\bibfnamefont {B.}~\bibnamefont {Barbrel}}, \bibinfo {author} {\bibfnamefont
  {S.~B.}\ \bibnamefont {Brown}}, \bibinfo {author} {\bibfnamefont {D.~A.}\
  \bibnamefont {Chapman}}, \bibinfo {author} {\bibfnamefont {Z.}~\bibnamefont
  {Chen}}, \bibinfo {author} {\bibfnamefont {C.~B.}\ \bibnamefont {Curry}},
  \bibinfo {author} {\bibfnamefont {F.}~\bibnamefont {Fiuza}}, \bibinfo
  {author} {\bibfnamefont {E.}~\bibnamefont {Gamboa}}, \bibinfo {author}
  {\bibfnamefont {M.}~\bibnamefont {Gauthier}}, \bibinfo {author}
  {\bibfnamefont {D.~O.}\ \bibnamefont {Gericke}}, \bibinfo {author}
  {\bibfnamefont {A.}~\bibnamefont {Gleason}}, \bibinfo {author} {\bibfnamefont
  {S.}~\bibnamefont {Göde}}, \bibinfo {author} {\bibfnamefont
  {E.}~\bibnamefont {Granados}}, \bibinfo {author} {\bibfnamefont
  {P.}~\bibnamefont {Heimann}}, \bibinfo {author} {\bibfnamefont
  {J.}~\bibnamefont {Kim}}, \bibinfo {author} {\bibfnamefont {D.}~\bibnamefont
  {Kraus}}, \bibinfo {author} {\bibfnamefont {M.~J.}\ \bibnamefont
  {MacDonald}}, \bibinfo {author} {\bibfnamefont {A.~J.}\ \bibnamefont
  {Mackinnon}}, \bibinfo {author} {\bibfnamefont {R.}~\bibnamefont {Mishra}},
  \bibinfo {author} {\bibfnamefont {A.}~\bibnamefont {Ravasio}}, \bibinfo
  {author} {\bibfnamefont {C.}~\bibnamefont {Roedel}}, \bibinfo {author}
  {\bibfnamefont {P.}~\bibnamefont {Sperling}}, \bibinfo {author}
  {\bibfnamefont {W.}~\bibnamefont {Schumaker}}, \bibinfo {author}
  {\bibfnamefont {Y.~Y.}\ \bibnamefont {Tsui}}, \bibinfo {author}
  {\bibfnamefont {J.}~\bibnamefont {Vorberger}}, \bibinfo {author}
  {\bibfnamefont {U.}~\bibnamefont {Zastrau}}, \bibinfo {author} {\bibfnamefont
  {A.}~\bibnamefont {Fry}}, \bibinfo {author} {\bibfnamefont {W.~E.}\
  \bibnamefont {White}}, \bibinfo {author} {\bibfnamefont {J.~B.}\ \bibnamefont
  {Hasting}}, \ and\ \bibinfo {author} {\bibfnamefont {H.~J.}\ \bibnamefont
  {Lee}},\ }\href@noop {} {\bibfield  {journal} {\bibinfo  {journal} {Journal
  of Physics B: Atomic, Molecular and Optical Physics}\ }\textbf {\bibinfo
  {volume} {49}},\ \bibinfo {pages} {092001} (\bibinfo {year}
  {2016})}\BibitemShut {NoStop}%
\bibitem [{\citenamefont {Gleason}\ \emph {et~al.}(2015)\citenamefont
  {Gleason}, \citenamefont {Bolme}, \citenamefont {Lee}, \citenamefont
  {Nagler}, \citenamefont {Galtier}, \citenamefont {Milathianaki},
  \citenamefont {Hawreliak}, \citenamefont {Kraus}, \citenamefont {Eggert},
  \citenamefont {Fratanduono}, \citenamefont {Collins}, \citenamefont
  {Sandberg}, \citenamefont {Yang},\ and\ \citenamefont {Mao}}]{Gleason:2015}%
  \BibitemOpen
  \bibfield  {author} {\bibinfo {author} {\bibfnamefont {A.}~\bibnamefont
  {Gleason}}, \bibinfo {author} {\bibfnamefont {C.}~\bibnamefont {Bolme}},
  \bibinfo {author} {\bibfnamefont {H.}~\bibnamefont {Lee}}, \bibinfo {author}
  {\bibfnamefont {B.}~\bibnamefont {Nagler}}, \bibinfo {author} {\bibfnamefont
  {E.}~\bibnamefont {Galtier}}, \bibinfo {author} {\bibfnamefont
  {D.}~\bibnamefont {Milathianaki}}, \bibinfo {author} {\bibfnamefont
  {J.}~\bibnamefont {Hawreliak}}, \bibinfo {author} {\bibfnamefont
  {R.}~\bibnamefont {Kraus}}, \bibinfo {author} {\bibfnamefont
  {J.}~\bibnamefont {Eggert}}, \bibinfo {author} {\bibfnamefont
  {D.}~\bibnamefont {Fratanduono}}, \bibinfo {author} {\bibfnamefont
  {G.}~\bibnamefont {Collins}}, \bibinfo {author} {\bibfnamefont
  {R.}~\bibnamefont {Sandberg}}, \bibinfo {author} {\bibfnamefont
  {W.}~\bibnamefont {Yang}}, \ and\ \bibinfo {author} {\bibfnamefont
  {W.}~\bibnamefont {Mao}},\ }\href@noop {} {\bibfield  {journal} {\bibinfo
  {journal} {Nat. Comm.}\ }\textbf {\bibinfo {volume} {6}},\ \bibinfo {pages}
  {8191} (\bibinfo {year} {2015})}\BibitemShut {NoStop}%
\bibitem [{\citenamefont {Gleason}\ \emph {et~al.}(2017)\citenamefont
  {Gleason}, \citenamefont {Bolme}, \citenamefont {Galtier}, \citenamefont
  {Lee}, \citenamefont {Granados}, \citenamefont {Dolan}, \citenamefont
  {Seagle}, \citenamefont {Ao}, \citenamefont {Ali}, \citenamefont {Lazicki},
  \citenamefont {Swift}, \citenamefont {Celliers},\ and\ \citenamefont
  {Mao}}]{Gleason:2017}%
  \BibitemOpen
  \bibfield  {author} {\bibinfo {author} {\bibfnamefont {A.~E.}\ \bibnamefont
  {Gleason}}, \bibinfo {author} {\bibfnamefont {C.~A.}\ \bibnamefont {Bolme}},
  \bibinfo {author} {\bibfnamefont {E.}~\bibnamefont {Galtier}}, \bibinfo
  {author} {\bibfnamefont {H.~J.}\ \bibnamefont {Lee}}, \bibinfo {author}
  {\bibfnamefont {E.}~\bibnamefont {Granados}}, \bibinfo {author}
  {\bibfnamefont {D.~H.}\ \bibnamefont {Dolan}}, \bibinfo {author}
  {\bibfnamefont {C.~T.}\ \bibnamefont {Seagle}}, \bibinfo {author}
  {\bibfnamefont {T.}~\bibnamefont {Ao}}, \bibinfo {author} {\bibfnamefont
  {S.}~\bibnamefont {Ali}}, \bibinfo {author} {\bibfnamefont {A.}~\bibnamefont
  {Lazicki}}, \bibinfo {author} {\bibfnamefont {D.}~\bibnamefont {Swift}},
  \bibinfo {author} {\bibfnamefont {P.}~\bibnamefont {Celliers}}, \ and\
  \bibinfo {author} {\bibfnamefont {W.~L.}\ \bibnamefont {Mao}},\ }\href@noop
  {} {\bibfield  {journal} {\bibinfo  {journal} {Phys. Rev. Lett.}\ }\textbf
  {\bibinfo {volume} {119}},\ \bibinfo {pages} {025701} (\bibinfo {year}
  {2017})}\BibitemShut {NoStop}%
\bibitem [{\citenamefont {Brown}\ \emph {et~al.}(2019)\citenamefont {Brown},
  \citenamefont {Gleason}, \citenamefont {Galtier}, \citenamefont
  {Higginbotham}, \citenamefont {Arnold}, \citenamefont {Fry}, \citenamefont
  {Granados}, \citenamefont {Hashim}, \citenamefont {Schroer}, \citenamefont
  {Schropp}, \citenamefont {Seiboth}, \citenamefont {Tavella}, \citenamefont
  {Xing}, \citenamefont {Mao}, \citenamefont {Lee},\ and\ \citenamefont
  {Nagler}}]{BGG:2019}%
  \BibitemOpen
  \bibfield  {author} {\bibinfo {author} {\bibfnamefont {S.~B.}\ \bibnamefont
  {Brown}}, \bibinfo {author} {\bibfnamefont {A.~E.}\ \bibnamefont {Gleason}},
  \bibinfo {author} {\bibfnamefont {E.}~\bibnamefont {Galtier}}, \bibinfo
  {author} {\bibfnamefont {A.}~\bibnamefont {Higginbotham}}, \bibinfo {author}
  {\bibfnamefont {B.}~\bibnamefont {Arnold}}, \bibinfo {author} {\bibfnamefont
  {A.}~\bibnamefont {Fry}}, \bibinfo {author} {\bibfnamefont {E.}~\bibnamefont
  {Granados}}, \bibinfo {author} {\bibfnamefont {A.}~\bibnamefont {Hashim}},
  \bibinfo {author} {\bibfnamefont {C.~G.}\ \bibnamefont {Schroer}}, \bibinfo
  {author} {\bibfnamefont {A.}~\bibnamefont {Schropp}}, \bibinfo {author}
  {\bibfnamefont {F.}~\bibnamefont {Seiboth}}, \bibinfo {author} {\bibfnamefont
  {F.}~\bibnamefont {Tavella}}, \bibinfo {author} {\bibfnamefont
  {Z.}~\bibnamefont {Xing}}, \bibinfo {author} {\bibfnamefont {W.}~\bibnamefont
  {Mao}}, \bibinfo {author} {\bibfnamefont {H.~J.}\ \bibnamefont {Lee}}, \ and\
  \bibinfo {author} {\bibfnamefont {B.}~\bibnamefont {Nagler}},\ }\href@noop {}
  {\bibfield  {journal} {\bibinfo  {journal} {Science Advances}\ }\textbf
  {\bibinfo {volume} {5}},\ \bibinfo {pages} {1} (\bibinfo {year}
  {2019})}\BibitemShut {NoStop}%
\bibitem [{\citenamefont {Merkel}\ \emph {et~al.}(2021)\citenamefont {Merkel},
  \citenamefont {Hok}, \citenamefont {Bolme}, \citenamefont {Rittman},
  \citenamefont {Ramos}, \citenamefont {Morrow}, \citenamefont {Lee},
  \citenamefont {Nagler}, \citenamefont {Galtier}, \citenamefont {Granados},
  \citenamefont {Hashim}, \citenamefont {Mao},\ and\ \citenamefont
  {Gleason}}]{MHB:2021}%
  \BibitemOpen
  \bibfield  {author} {\bibinfo {author} {\bibfnamefont {S.}~\bibnamefont
  {Merkel}}, \bibinfo {author} {\bibfnamefont {S.}~\bibnamefont {Hok}},
  \bibinfo {author} {\bibfnamefont {C.}~\bibnamefont {Bolme}}, \bibinfo
  {author} {\bibfnamefont {D.}~\bibnamefont {Rittman}}, \bibinfo {author}
  {\bibfnamefont {K.~J.}\ \bibnamefont {Ramos}}, \bibinfo {author}
  {\bibfnamefont {B.}~\bibnamefont {Morrow}}, \bibinfo {author} {\bibfnamefont
  {H.~J.}\ \bibnamefont {Lee}}, \bibinfo {author} {\bibfnamefont
  {B.}~\bibnamefont {Nagler}}, \bibinfo {author} {\bibfnamefont
  {E.}~\bibnamefont {Galtier}}, \bibinfo {author} {\bibfnamefont
  {E.}~\bibnamefont {Granados}}, \bibinfo {author} {\bibfnamefont
  {A.}~\bibnamefont {Hashim}}, \bibinfo {author} {\bibfnamefont {W.~L.}\
  \bibnamefont {Mao}}, \ and\ \bibinfo {author} {\bibfnamefont {A.~E.}\
  \bibnamefont {Gleason}},\ }\href@noop {} {\bibfield  {journal} {\bibinfo
  {journal} {Phys. Rev. Lett.}\ }\textbf {\bibinfo {volume} {127}},\ \bibinfo
  {pages} {205501} (\bibinfo {year} {2021})}\BibitemShut {NoStop}%
\bibitem [{\citenamefont {Wengrowicz}\ \emph {et~al.}(2019)\citenamefont
  {Wengrowicz}, \citenamefont {Peleg}, \citenamefont {Leovsky}, \citenamefont
  {Chen}, \citenamefont {Haham}, \citenamefont {Sainadh},\ and\ \citenamefont
  {Cohen}}]{WPL:2019}%
  \BibitemOpen
  \bibfield  {author} {\bibinfo {author} {\bibfnamefont {O.}~\bibnamefont
  {Wengrowicz}}, \bibinfo {author} {\bibfnamefont {O.}~\bibnamefont {Peleg}},
  \bibinfo {author} {\bibfnamefont {B.}~\bibnamefont {Leovsky}}, \bibinfo
  {author} {\bibfnamefont {B.~K.}\ \bibnamefont {Chen}}, \bibinfo {author}
  {\bibfnamefont {G.~I.}\ \bibnamefont {Haham}}, \bibinfo {author}
  {\bibfnamefont {U.~S.}\ \bibnamefont {Sainadh}}, \ and\ \bibinfo {author}
  {\bibfnamefont {O.}~\bibnamefont {Cohen}},\ }\href@noop {} {\bibfield
  {journal} {\bibinfo  {journal} {Opt. Exp.}\ }\textbf {\bibinfo {volume}
  {27}},\ \bibinfo {pages} {24568} (\bibinfo {year} {2019})}\BibitemShut
  {NoStop}%
\bibitem [{\citenamefont {Bernier}\ \emph {et~al.}(2020)\citenamefont
  {Bernier}, \citenamefont {Sutter}, \citenamefont {Rollett},\ and\
  \citenamefont {Almer}}]{BSR:2020}%
  \BibitemOpen
  \bibfield  {author} {\bibinfo {author} {\bibfnamefont {J.~V.}\ \bibnamefont
  {Bernier}}, \bibinfo {author} {\bibfnamefont {R.~M.}\ \bibnamefont {Sutter}},
  \bibinfo {author} {\bibfnamefont {A.~D.}\ \bibnamefont {Rollett}}, \ and\
  \bibinfo {author} {\bibfnamefont {J.~D.}\ \bibnamefont {Almer}},\ }\href@noop
  {} {\bibfield  {journal} {\bibinfo  {journal} {Ann. Rev. Mater. Res.}\
  }\textbf {\bibinfo {volume} {50}},\ \bibinfo {pages} {395} (\bibinfo {year}
  {2020})}\BibitemShut {NoStop}%
\bibitem [{\citenamefont {Morard}\ \emph {et~al.}(2020)\citenamefont {Morard},
  \citenamefont {Hernandez}, \citenamefont {Guarguaglini}, \citenamefont
  {Bolis}, \citenamefont {Benuzzi-Mounaix}, \citenamefont {Vinci},
  \citenamefont {Fiquet}, \citenamefont {Baron}, \citenamefont {Shim},
  \citenamefont {Ko}, \citenamefont {Gleason}, \citenamefont {Mao},
  \citenamefont {Alonso-Mori}, \citenamefont {Lee}, \citenamefont {Nagler},
  \citenamefont {Galtier}, \citenamefont {Sokaras}, \citenamefont {Glenzer},
  \citenamefont {Andrault}, \citenamefont {Garbarino}, \citenamefont {Mezouar},
  \citenamefont {Schuster},\ and\ \citenamefont {Ravasio}}]{MHG:2020}%
  \BibitemOpen
  \bibfield  {author} {\bibinfo {author} {\bibfnamefont {G.}~\bibnamefont
  {Morard}}, \bibinfo {author} {\bibfnamefont {J.-A.}\ \bibnamefont
  {Hernandez}}, \bibinfo {author} {\bibfnamefont {M.}~\bibnamefont
  {Guarguaglini}}, \bibinfo {author} {\bibfnamefont {R.}~\bibnamefont {Bolis}},
  \bibinfo {author} {\bibfnamefont {A.}~\bibnamefont {Benuzzi-Mounaix}},
  \bibinfo {author} {\bibfnamefont {T.}~\bibnamefont {Vinci}}, \bibinfo
  {author} {\bibfnamefont {G.}~\bibnamefont {Fiquet}}, \bibinfo {author}
  {\bibfnamefont {M.~A.}\ \bibnamefont {Baron}}, \bibinfo {author}
  {\bibfnamefont {S.~H.}\ \bibnamefont {Shim}}, \bibinfo {author}
  {\bibfnamefont {B.}~\bibnamefont {Ko}}, \bibinfo {author} {\bibfnamefont
  {A.~E.}\ \bibnamefont {Gleason}}, \bibinfo {author} {\bibfnamefont {W.~L.}\
  \bibnamefont {Mao}}, \bibinfo {author} {\bibfnamefont {R.}~\bibnamefont
  {Alonso-Mori}}, \bibinfo {author} {\bibfnamefont {H.~J.}\ \bibnamefont
  {Lee}}, \bibinfo {author} {\bibfnamefont {B.}~\bibnamefont {Nagler}},
  \bibinfo {author} {\bibfnamefont {E.}~\bibnamefont {Galtier}}, \bibinfo
  {author} {\bibfnamefont {D.}~\bibnamefont {Sokaras}}, \bibinfo {author}
  {\bibfnamefont {S.~H.}\ \bibnamefont {Glenzer}}, \bibinfo {author}
  {\bibfnamefont {D.}~\bibnamefont {Andrault}}, \bibinfo {author}
  {\bibfnamefont {G.}~\bibnamefont {Garbarino}}, \bibinfo {author}
  {\bibfnamefont {M.}~\bibnamefont {Mezouar}}, \bibinfo {author} {\bibfnamefont
  {A.~K.}\ \bibnamefont {Schuster}}, \ and\ \bibinfo {author} {\bibfnamefont
  {A.}~\bibnamefont {Ravasio}},\ }\href@noop {} {\bibfield  {journal} {\bibinfo
   {journal} {PNAS}\ }\textbf {\bibinfo {volume} {117}},\ \bibinfo {pages}
  {11981} (\bibinfo {year} {2020})}\BibitemShut {NoStop}%
\bibitem [{\citenamefont {Nuckolls}\ \emph {et~al.}(1972)\citenamefont
  {Nuckolls}, \citenamefont {Wood}, \citenamefont {Thiessen},\ and\
  \citenamefont {Zimmerman}}]{NWT:1972}%
  \BibitemOpen
  \bibfield  {author} {\bibinfo {author} {\bibfnamefont {J.}~\bibnamefont
  {Nuckolls}}, \bibinfo {author} {\bibfnamefont {L.}~\bibnamefont {Wood}},
  \bibinfo {author} {\bibfnamefont {A.}~\bibnamefont {Thiessen}}, \ and\
  \bibinfo {author} {\bibfnamefont {G.}~\bibnamefont {Zimmerman}},\ }\href@noop
  {} {\bibfield  {journal} {\bibinfo  {journal} {Nature}\ }\textbf {\bibinfo
  {volume} {239}},\ \bibinfo {pages} {139} (\bibinfo {year}
  {1972})}\BibitemShut {NoStop}%
\bibitem [{\citenamefont {Kritcher}\ \emph {et~al.}(2022)\citenamefont
  {Kritcher}, \citenamefont {Zylstra}, \citenamefont {Callahan}, \citenamefont
  {Hurricane}, \citenamefont {Weber}, \citenamefont {Clark}, \citenamefont
  {Young}, \citenamefont {Ralph}, \citenamefont {Casey}, \citenamefont {Pak},
  \citenamefont {Landen}, \citenamefont {Bachmann}, \citenamefont {Baker},
  \citenamefont {Hopkins}, \citenamefont {Bhandarkar}, \citenamefont {Biener},
  \citenamefont {Bionta}, \citenamefont {Birge}, \citenamefont {Braun},
  \citenamefont {Briggs}, \citenamefont {Celliers}, \citenamefont {Chen},
  \citenamefont {Choate}, \citenamefont {Divol}, \citenamefont {Döppner},
  \citenamefont {Fittinghoff}, \citenamefont {Edwards}, \citenamefont
  {Gharibyan}, \citenamefont {Haan}, \citenamefont {Hahn}, \citenamefont
  {Hartouni}, \citenamefont {Hinkel}, \citenamefont {Ho}, \citenamefont
  {Hohenberger}, \citenamefont {Holder}, \citenamefont {Huang}, \citenamefont
  {Izumi}, \citenamefont {Jeet}, \citenamefont {Jones}, \citenamefont {Kerr},
  \citenamefont {Khan}, \citenamefont {Kleinrath}, \citenamefont {Kleinrath},
  \citenamefont {Kong}, \citenamefont {Lamb}, \citenamefont {Pape},
  \citenamefont {Lemos}, \citenamefont {Lindl}, \citenamefont {MacGowan},
  \citenamefont {Mackinnon}, \citenamefont {MacPhee}, \citenamefont {Marley},
  \citenamefont {Meaney}, \citenamefont {Millot}, \citenamefont {Moore},
  \citenamefont {Newman}, \citenamefont {Nicola}, \citenamefont {Nikroo},
  \citenamefont {Nora}, \citenamefont {Patel}, \citenamefont {Rice},
  \citenamefont {Rubery}, \citenamefont {Sater}, \citenamefont {Schlossberg},
  \citenamefont {Sepke}, \citenamefont {Sequoia}, \citenamefont {Shin},
  \citenamefont {Stadermann}, \citenamefont {Stoupin}, \citenamefont {Strozzi},
  \citenamefont {Thomas}, \citenamefont {Tommasini}, \citenamefont
  {Trosseille}, \citenamefont {Tubman}, \citenamefont {Volegov}, \citenamefont
  {Wild}, \citenamefont {Woods},\ and\ \citenamefont {Yang}}]{KZC:2022}%
  \BibitemOpen
  \bibfield  {author} {\bibinfo {author} {\bibfnamefont {A.~L.}\ \bibnamefont
  {Kritcher}}, \bibinfo {author} {\bibfnamefont {A.~B.}\ \bibnamefont
  {Zylstra}}, \bibinfo {author} {\bibfnamefont {D.~A.}\ \bibnamefont
  {Callahan}}, \bibinfo {author} {\bibfnamefont {O.~A.}\ \bibnamefont
  {Hurricane}}, \bibinfo {author} {\bibfnamefont {C.~R.}\ \bibnamefont
  {Weber}}, \bibinfo {author} {\bibfnamefont {D.~S.}\ \bibnamefont {Clark}},
  \bibinfo {author} {\bibfnamefont {C.~V.}\ \bibnamefont {Young}}, \bibinfo
  {author} {\bibfnamefont {J.~E.}\ \bibnamefont {Ralph}}, \bibinfo {author}
  {\bibfnamefont {D.~T.}\ \bibnamefont {Casey}}, \bibinfo {author}
  {\bibfnamefont {A.}~\bibnamefont {Pak}}, \bibinfo {author} {\bibfnamefont
  {O.~L.}\ \bibnamefont {Landen}}, \bibinfo {author} {\bibfnamefont
  {B.}~\bibnamefont {Bachmann}}, \bibinfo {author} {\bibfnamefont {K.~L.}\
  \bibnamefont {Baker}}, \bibinfo {author} {\bibfnamefont {L.~B.}\ \bibnamefont
  {Hopkins}}, \bibinfo {author} {\bibfnamefont {S.~D.}\ \bibnamefont
  {Bhandarkar}}, \bibinfo {author} {\bibfnamefont {J.}~\bibnamefont {Biener}},
  \bibinfo {author} {\bibfnamefont {R.~M.}\ \bibnamefont {Bionta}}, \bibinfo
  {author} {\bibfnamefont {N.~W.}\ \bibnamefont {Birge}}, \bibinfo {author}
  {\bibfnamefont {T.}~\bibnamefont {Braun}}, \bibinfo {author} {\bibfnamefont
  {T.~M.}\ \bibnamefont {Briggs}}, \bibinfo {author} {\bibfnamefont {P.~M.}\
  \bibnamefont {Celliers}}, \bibinfo {author} {\bibfnamefont {H.}~\bibnamefont
  {Chen}}, \bibinfo {author} {\bibfnamefont {C.}~\bibnamefont {Choate}},
  \bibinfo {author} {\bibfnamefont {L.}~\bibnamefont {Divol}}, \bibinfo
  {author} {\bibfnamefont {T.}~\bibnamefont {Döppner}}, \bibinfo {author}
  {\bibfnamefont {D.}~\bibnamefont {Fittinghoff}}, \bibinfo {author}
  {\bibfnamefont {M.~J.}\ \bibnamefont {Edwards}}, \bibinfo {author}
  {\bibfnamefont {M.~G. J.~N.}\ \bibnamefont {Gharibyan}}, \bibinfo {author}
  {\bibfnamefont {S.}~\bibnamefont {Haan}}, \bibinfo {author} {\bibfnamefont
  {K.~D.}\ \bibnamefont {Hahn}}, \bibinfo {author} {\bibfnamefont
  {E.}~\bibnamefont {Hartouni}}, \bibinfo {author} {\bibfnamefont {D.~E.}\
  \bibnamefont {Hinkel}}, \bibinfo {author} {\bibfnamefont {D.~D.}\
  \bibnamefont {Ho}}, \bibinfo {author} {\bibfnamefont {M.}~\bibnamefont
  {Hohenberger}}, \bibinfo {author} {\bibfnamefont {J.~P.}\ \bibnamefont
  {Holder}}, \bibinfo {author} {\bibfnamefont {H.}~\bibnamefont {Huang}},
  \bibinfo {author} {\bibfnamefont {N.}~\bibnamefont {Izumi}}, \bibinfo
  {author} {\bibfnamefont {J.}~\bibnamefont {Jeet}}, \bibinfo {author}
  {\bibfnamefont {O.}~\bibnamefont {Jones}}, \bibinfo {author} {\bibfnamefont
  {S.~M.}\ \bibnamefont {Kerr}}, \bibinfo {author} {\bibfnamefont {S.~F.}\
  \bibnamefont {Khan}}, \bibinfo {author} {\bibfnamefont {H.~G.}\ \bibnamefont
  {Kleinrath}}, \bibinfo {author} {\bibfnamefont {V.~G.}\ \bibnamefont
  {Kleinrath}}, \bibinfo {author} {\bibfnamefont {C.}~\bibnamefont {Kong}},
  \bibinfo {author} {\bibfnamefont {K.~M.}\ \bibnamefont {Lamb}}, \bibinfo
  {author} {\bibfnamefont {S.~L.}\ \bibnamefont {Pape}}, \bibinfo {author}
  {\bibfnamefont {N.~C.}\ \bibnamefont {Lemos}}, \bibinfo {author}
  {\bibfnamefont {J.~D.}\ \bibnamefont {Lindl}}, \bibinfo {author}
  {\bibfnamefont {B.~J.}\ \bibnamefont {MacGowan}}, \bibinfo {author}
  {\bibfnamefont {A.~J.}\ \bibnamefont {Mackinnon}}, \bibinfo {author}
  {\bibfnamefont {A.~G.}\ \bibnamefont {MacPhee}}, \bibinfo {author}
  {\bibfnamefont {E.~V.}\ \bibnamefont {Marley}}, \bibinfo {author}
  {\bibfnamefont {K.}~\bibnamefont {Meaney}}, \bibinfo {author} {\bibfnamefont
  {M.}~\bibnamefont {Millot}}, \bibinfo {author} {\bibfnamefont {A.~S.}\
  \bibnamefont {Moore}}, \bibinfo {author} {\bibfnamefont {K.}~\bibnamefont
  {Newman}}, \bibinfo {author} {\bibfnamefont {J.-M. G.~D.}\ \bibnamefont
  {Nicola}}, \bibinfo {author} {\bibfnamefont {A.}~\bibnamefont {Nikroo}},
  \bibinfo {author} {\bibfnamefont {R.}~\bibnamefont {Nora}}, \bibinfo {author}
  {\bibfnamefont {P.~K.}\ \bibnamefont {Patel}}, \bibinfo {author}
  {\bibfnamefont {N.~G.}\ \bibnamefont {Rice}}, \bibinfo {author}
  {\bibfnamefont {M.~S.}\ \bibnamefont {Rubery}}, \bibinfo {author}
  {\bibfnamefont {J.}~\bibnamefont {Sater}}, \bibinfo {author} {\bibfnamefont
  {D.~J.}\ \bibnamefont {Schlossberg}}, \bibinfo {author} {\bibfnamefont
  {S.~M.}\ \bibnamefont {Sepke}}, \bibinfo {author} {\bibfnamefont
  {K.}~\bibnamefont {Sequoia}}, \bibinfo {author} {\bibfnamefont {S.~J.}\
  \bibnamefont {Shin}}, \bibinfo {author} {\bibfnamefont {M.}~\bibnamefont
  {Stadermann}}, \bibinfo {author} {\bibfnamefont {S.}~\bibnamefont {Stoupin}},
  \bibinfo {author} {\bibfnamefont {D.~J.}\ \bibnamefont {Strozzi}}, \bibinfo
  {author} {\bibfnamefont {C.~A.}\ \bibnamefont {Thomas}}, \bibinfo {author}
  {\bibfnamefont {R.}~\bibnamefont {Tommasini}}, \bibinfo {author}
  {\bibfnamefont {C.}~\bibnamefont {Trosseille}}, \bibinfo {author}
  {\bibfnamefont {E.~R.}\ \bibnamefont {Tubman}}, \bibinfo {author}
  {\bibfnamefont {P.~L.}\ \bibnamefont {Volegov}}, \bibinfo {author}
  {\bibfnamefont {C.}~\bibnamefont {Wild}}, \bibinfo {author} {\bibfnamefont
  {D.~T.}\ \bibnamefont {Woods}}, \ and\ \bibinfo {author} {\bibfnamefont
  {S.~T.}\ \bibnamefont {Yang}},\ }\href@noop {} {\bibfield  {journal}
  {\bibinfo  {journal} {Phys. Rev. E}\ }\textbf {\bibinfo {volume} {106}},\
  \bibinfo {pages} {025201} (\bibinfo {year} {2022})}\BibitemShut {NoStop}%
\bibitem [{\citenamefont {Ma}(2023)}]{Ma:2023}%
  \BibitemOpen
  \bibfield  {author} {\bibinfo {author} {\bibfnamefont {T.}~\bibnamefont
  {Ma}},\ }\href@noop {} {\emph {\bibinfo {title} {Ignition and the path toward
  an Inertial Fusion Energy Future}}},\ \bibinfo {type} {Tech. Rep.}\ \bibinfo
  {number} {LLNL-PRES-833900}\ (\bibinfo  {institution} {Lawrence Livermore
  National Laboratory},\ \bibinfo {year} {2023})\BibitemShut {NoStop}%
\bibitem [{\citenamefont {Pandolfi}\ \emph {et~al.}(2022)\citenamefont
  {Pandolfi}, \citenamefont {Carver}, \citenamefont {Hodge}, \citenamefont
  {Leong}, \citenamefont {Kurzer-Ogul}, \citenamefont {Galtier}, \citenamefont
  {Khaghani}, \citenamefont {Cunningham}, \citenamefont {Nagler}, \citenamefont
  {Lee}, \citenamefont {Bolme}, \citenamefont {Ramos}, \citenamefont {Li},
  \citenamefont {Liu}, \citenamefont {Sakdinawat}, \citenamefont {Marchesini},
  \citenamefont {Curry}, \citenamefont {Decker}, \citenamefont {Vetter},
  \citenamefont {Shang}, \citenamefont {Aluie}, \citenamefont {Dayton},
  \citenamefont {Montgomery}, \citenamefont {Sandberg},\ and\ \citenamefont
  {Gleason}}]{PCH:2022}%
  \BibitemOpen
  \bibfield  {author} {\bibinfo {author} {\bibfnamefont {S.}~\bibnamefont
  {Pandolfi}}, \bibinfo {author} {\bibfnamefont {T.}~\bibnamefont {Carver}},
  \bibinfo {author} {\bibfnamefont {D.}~\bibnamefont {Hodge}}, \bibinfo
  {author} {\bibfnamefont {A.~F.~T.}\ \bibnamefont {Leong}}, \bibinfo {author}
  {\bibfnamefont {K.}~\bibnamefont {Kurzer-Ogul}}, \bibinfo {author}
  {\bibfnamefont {P.~H.~E.}\ \bibnamefont {Galtier}}, \bibinfo {author}
  {\bibfnamefont {D.}~\bibnamefont {Khaghani}}, \bibinfo {author}
  {\bibfnamefont {E.}~\bibnamefont {Cunningham}}, \bibinfo {author}
  {\bibfnamefont {B.}~\bibnamefont {Nagler}}, \bibinfo {author} {\bibfnamefont
  {H.~J.}\ \bibnamefont {Lee}}, \bibinfo {author} {\bibfnamefont
  {C.}~\bibnamefont {Bolme}}, \bibinfo {author} {\bibfnamefont
  {K.}~\bibnamefont {Ramos}}, \bibinfo {author} {\bibfnamefont
  {K.}~\bibnamefont {Li}}, \bibinfo {author} {\bibfnamefont {Y.}~\bibnamefont
  {Liu}}, \bibinfo {author} {\bibfnamefont {A.}~\bibnamefont {Sakdinawat}},
  \bibinfo {author} {\bibfnamefont {S.}~\bibnamefont {Marchesini}}, \bibinfo
  {author} {\bibfnamefont {P.~M. K. C.~B.}\ \bibnamefont {Curry}}, \bibinfo
  {author} {\bibfnamefont {F.-J.}\ \bibnamefont {Decker}}, \bibinfo {author}
  {\bibfnamefont {S.}~\bibnamefont {Vetter}}, \bibinfo {author} {\bibfnamefont
  {J.}~\bibnamefont {Shang}}, \bibinfo {author} {\bibfnamefont
  {H.}~\bibnamefont {Aluie}}, \bibinfo {author} {\bibfnamefont
  {M.}~\bibnamefont {Dayton}}, \bibinfo {author} {\bibfnamefont {D.~S.}\
  \bibnamefont {Montgomery}}, \bibinfo {author} {\bibfnamefont {R.~L.}\
  \bibnamefont {Sandberg}}, \ and\ \bibinfo {author} {\bibfnamefont {A.~E.}\
  \bibnamefont {Gleason}},\ }\href@noop {} {\bibfield  {journal} {\bibinfo
  {journal} {Rev. Sci. Instrum.}\ }\textbf {\bibinfo {volume} {93}},\ \bibinfo
  {pages} {103502} (\bibinfo {year} {2022})}\BibitemShut {NoStop}%
\bibitem [{\citenamefont {Weber}\ \emph {et~al.}(2020)\citenamefont {Weber},
  \citenamefont {Clark}, \citenamefont {Pak}, \citenamefont {Alfonso},
  \citenamefont {Bachmann}, \citenamefont {Hopkins}, \citenamefont {Bunn},
  \citenamefont {Crippen}, \citenamefont {Divol}, \citenamefont {Dittrich},
  \citenamefont {Kritcher}, \citenamefont {Landen}, \citenamefont {Pape},
  \citenamefont {MacPhee}, \citenamefont {Marley}, \citenamefont {Masse},
  \citenamefont {Milovich}, \citenamefont {Nikroo}, \citenamefont {Patel},
  \citenamefont {Pickworth}, \citenamefont {Rice}, \citenamefont {Smalyuk},\
  and\ \citenamefont {Stadermann}}]{WCA:2020}%
  \BibitemOpen
  \bibfield  {author} {\bibinfo {author} {\bibfnamefont {C.~R.}\ \bibnamefont
  {Weber}}, \bibinfo {author} {\bibfnamefont {D.~S.}\ \bibnamefont {Clark}},
  \bibinfo {author} {\bibfnamefont {A.}~\bibnamefont {Pak}}, \bibinfo {author}
  {\bibfnamefont {N.}~\bibnamefont {Alfonso}}, \bibinfo {author} {\bibfnamefont
  {B.}~\bibnamefont {Bachmann}}, \bibinfo {author} {\bibfnamefont {L.~F.~B.}\
  \bibnamefont {Hopkins}}, \bibinfo {author} {\bibfnamefont {T.}~\bibnamefont
  {Bunn}}, \bibinfo {author} {\bibfnamefont {J.}~\bibnamefont {Crippen}},
  \bibinfo {author} {\bibfnamefont {L.}~\bibnamefont {Divol}}, \bibinfo
  {author} {\bibfnamefont {T.}~\bibnamefont {Dittrich}}, \bibinfo {author}
  {\bibfnamefont {A.~L.}\ \bibnamefont {Kritcher}}, \bibinfo {author}
  {\bibfnamefont {O.~L.}\ \bibnamefont {Landen}}, \bibinfo {author}
  {\bibfnamefont {S.~L.}\ \bibnamefont {Pape}}, \bibinfo {author}
  {\bibfnamefont {A.~G.}\ \bibnamefont {MacPhee}}, \bibinfo {author}
  {\bibfnamefont {E.}~\bibnamefont {Marley}}, \bibinfo {author} {\bibfnamefont
  {L.~P.}\ \bibnamefont {Masse}}, \bibinfo {author} {\bibfnamefont {J.~L.}\
  \bibnamefont {Milovich}}, \bibinfo {author} {\bibfnamefont {A.}~\bibnamefont
  {Nikroo}}, \bibinfo {author} {\bibfnamefont {P.~K.}\ \bibnamefont {Patel}},
  \bibinfo {author} {\bibfnamefont {L.~A.}\ \bibnamefont {Pickworth}}, \bibinfo
  {author} {\bibfnamefont {N.}~\bibnamefont {Rice}}, \bibinfo {author}
  {\bibfnamefont {V.~A.}\ \bibnamefont {Smalyuk}}, \ and\ \bibinfo {author}
  {\bibfnamefont {M.}~\bibnamefont {Stadermann}},\ }\href@noop {} {\bibfield
  {journal} {\bibinfo  {journal} {Phys. Plasmas}\ }\textbf {\bibinfo {volume}
  {27}},\ \bibinfo {pages} {032703} (\bibinfo {year} {2020})}\BibitemShut
  {NoStop}%
\bibitem [{\citenamefont {{Di Nicola}}\ \emph {et~al.}(2015)\citenamefont {{Di
  Nicola}}, \citenamefont {Yang}, \citenamefont {Boley}, \citenamefont {Crane},
  \citenamefont {Heebner}, \citenamefont {Spinka}, \citenamefont {Arnold},
  \citenamefont {Barty}, \citenamefont {andT. S.~Budge}, \citenamefont
  {Christensen}, \citenamefont {Dawson}, \citenamefont {Erbert}, \citenamefont
  {Feigenbaum}, \citenamefont {Guss}, \citenamefont {Haefner}, \citenamefont
  {Hermann}, \citenamefont {Homoelle}, \citenamefont {Jarboe}, \citenamefont
  {Lawson}, \citenamefont {Lowe-Webb}, \citenamefont {McCandless},
  \citenamefont {McHale}, \citenamefont {Pelz}, \citenamefont {Pham},
  \citenamefont {Prantil}, \citenamefont {Rehak}, \citenamefont {Rever},
  \citenamefont {Rushford}, \citenamefont {Sacks}, \citenamefont {Shaw},
  \citenamefont {Smauley}, \citenamefont {Smith}, \citenamefont {Speck},
  \citenamefont {Tietbohl}, \citenamefont {Wegner},\ and\ \citenamefont
  {Widmayer}}]{DNY:2015}%
  \BibitemOpen
  \bibfield  {author} {\bibinfo {author} {\bibfnamefont {J.~M.}\ \bibnamefont
  {{Di Nicola}}}, \bibinfo {author} {\bibfnamefont {S.~T.}\ \bibnamefont
  {Yang}}, \bibinfo {author} {\bibfnamefont {C.~D.}\ \bibnamefont {Boley}},
  \bibinfo {author} {\bibfnamefont {J.~K.}\ \bibnamefont {Crane}}, \bibinfo
  {author} {\bibfnamefont {J.~E.}\ \bibnamefont {Heebner}}, \bibinfo {author}
  {\bibfnamefont {T.~M.}\ \bibnamefont {Spinka}}, \bibinfo {author}
  {\bibfnamefont {P.}~\bibnamefont {Arnold}}, \bibinfo {author} {\bibfnamefont
  {C.~P.~J.}\ \bibnamefont {Barty}}, \bibinfo {author} {\bibfnamefont
  {M.~W.~B.}\ \bibnamefont {andT. S.~Budge}}, \bibinfo {author} {\bibfnamefont
  {K.}~\bibnamefont {Christensen}}, \bibinfo {author} {\bibfnamefont {J.~W.}\
  \bibnamefont {Dawson}}, \bibinfo {author} {\bibfnamefont {G.}~\bibnamefont
  {Erbert}}, \bibinfo {author} {\bibfnamefont {E.}~\bibnamefont {Feigenbaum}},
  \bibinfo {author} {\bibfnamefont {G.}~\bibnamefont {Guss}}, \bibinfo {author}
  {\bibfnamefont {C.}~\bibnamefont {Haefner}}, \bibinfo {author} {\bibfnamefont
  {M.~R.}\ \bibnamefont {Hermann}}, \bibinfo {author} {\bibfnamefont
  {D.}~\bibnamefont {Homoelle}}, \bibinfo {author} {\bibfnamefont {J.~A.}\
  \bibnamefont {Jarboe}}, \bibinfo {author} {\bibfnamefont {J.~K.}\
  \bibnamefont {Lawson}}, \bibinfo {author} {\bibfnamefont {R.}~\bibnamefont
  {Lowe-Webb}}, \bibinfo {author} {\bibfnamefont {K.}~\bibnamefont
  {McCandless}}, \bibinfo {author} {\bibfnamefont {B.}~\bibnamefont {McHale}},
  \bibinfo {author} {\bibfnamefont {L.~J.}\ \bibnamefont {Pelz}}, \bibinfo
  {author} {\bibfnamefont {P.}~\bibnamefont {Pham}}, \bibinfo {author}
  {\bibfnamefont {M.~A.}\ \bibnamefont {Prantil}}, \bibinfo {author}
  {\bibfnamefont {M.~L.}\ \bibnamefont {Rehak}}, \bibinfo {author}
  {\bibfnamefont {M.~A.}\ \bibnamefont {Rever}}, \bibinfo {author}
  {\bibfnamefont {M.~C.}\ \bibnamefont {Rushford}}, \bibinfo {author}
  {\bibfnamefont {R.~A.}\ \bibnamefont {Sacks}}, \bibinfo {author}
  {\bibfnamefont {M.}~\bibnamefont {Shaw}}, \bibinfo {author} {\bibfnamefont
  {D.}~\bibnamefont {Smauley}}, \bibinfo {author} {\bibfnamefont {L.~K.}\
  \bibnamefont {Smith}}, \bibinfo {author} {\bibfnamefont {R.}~\bibnamefont
  {Speck}}, \bibinfo {author} {\bibfnamefont {G.}~\bibnamefont {Tietbohl}},
  \bibinfo {author} {\bibfnamefont {P.~J.}\ \bibnamefont {Wegner}}, \ and\
  \bibinfo {author} {\bibfnamefont {C.}~\bibnamefont {Widmayer}},\ }\href@noop
  {} {\bibfield  {journal} {\bibinfo  {journal} {SPIE Proc.}\ }\textbf
  {\bibinfo {volume} {9345}},\ \bibinfo {pages} {93450I} (\bibinfo {year}
  {2015})}\BibitemShut {NoStop}%
\bibitem [{\citenamefont {Simpson}\ \emph {et~al.}(2021)\citenamefont
  {Simpson}, \citenamefont {Mariscal}, \citenamefont {Kim}, \citenamefont
  {Scott}, \citenamefont {Williams}, \citenamefont {Grace}, \citenamefont
  {McGuffey}, \citenamefont {Wilks}, \citenamefont {Kemp}, \citenamefont
  {Lemos}, \citenamefont {Djordjevic}, \citenamefont {Folsom}, \citenamefont
  {Kalantar}, \citenamefont {Zacharias}, \citenamefont {Pollock}, \citenamefont
  {Moody}, \citenamefont {Beg}, \citenamefont {Morace}, \citenamefont {Iwata},
  \citenamefont {Sentoku}, \citenamefont {Manuel}, \citenamefont {Mauldin},
  \citenamefont {Quinn}, \citenamefont {Youngblood}, \citenamefont
  {Gatu-Johnson}, \citenamefont {Lahmann}, \citenamefont {Haefner},
  \citenamefont {Neely},\ and\ \citenamefont {Ma}}]{SMK:2021}%
  \BibitemOpen
  \bibfield  {author} {\bibinfo {author} {\bibfnamefont {R.}~\bibnamefont
  {Simpson}}, \bibinfo {author} {\bibfnamefont {D.}~\bibnamefont {Mariscal}},
  \bibinfo {author} {\bibfnamefont {J.}~\bibnamefont {Kim}}, \bibinfo {author}
  {\bibfnamefont {G.}~\bibnamefont {Scott}}, \bibinfo {author} {\bibfnamefont
  {G.~J.}\ \bibnamefont {Williams}}, \bibinfo {author} {\bibfnamefont
  {E.}~\bibnamefont {Grace}}, \bibinfo {author} {\bibfnamefont
  {C.}~\bibnamefont {McGuffey}}, \bibinfo {author} {\bibfnamefont
  {S.}~\bibnamefont {Wilks}}, \bibinfo {author} {\bibfnamefont
  {A.}~\bibnamefont {Kemp}}, \bibinfo {author} {\bibfnamefont {N.}~\bibnamefont
  {Lemos}}, \bibinfo {author} {\bibfnamefont {B.}~\bibnamefont {Djordjevic}},
  \bibinfo {author} {\bibfnamefont {E.}~\bibnamefont {Folsom}}, \bibinfo
  {author} {\bibfnamefont {D.}~\bibnamefont {Kalantar}}, \bibinfo {author}
  {\bibfnamefont {R.}~\bibnamefont {Zacharias}}, \bibinfo {author}
  {\bibfnamefont {B.}~\bibnamefont {Pollock}}, \bibinfo {author} {\bibfnamefont
  {J.}~\bibnamefont {Moody}}, \bibinfo {author} {\bibfnamefont
  {F.}~\bibnamefont {Beg}}, \bibinfo {author} {\bibfnamefont {A.}~\bibnamefont
  {Morace}}, \bibinfo {author} {\bibfnamefont {N.}~\bibnamefont {Iwata}},
  \bibinfo {author} {\bibfnamefont {Y.}~\bibnamefont {Sentoku}}, \bibinfo
  {author} {\bibfnamefont {M.~J.-E.}\ \bibnamefont {Manuel}}, \bibinfo {author}
  {\bibfnamefont {M.}~\bibnamefont {Mauldin}}, \bibinfo {author} {\bibfnamefont
  {M.}~\bibnamefont {Quinn}}, \bibinfo {author} {\bibfnamefont
  {K.}~\bibnamefont {Youngblood}}, \bibinfo {author} {\bibfnamefont
  {M.}~\bibnamefont {Gatu-Johnson}}, \bibinfo {author} {\bibfnamefont
  {B.}~\bibnamefont {Lahmann}}, \bibinfo {author} {\bibfnamefont
  {C.}~\bibnamefont {Haefner}}, \bibinfo {author} {\bibfnamefont
  {D.}~\bibnamefont {Neely}}, \ and\ \bibinfo {author} {\bibfnamefont
  {T.}~\bibnamefont {Ma}},\ }\href@noop {} {\bibfield  {journal} {\bibinfo
  {journal} {Plasma Phys. Control. Fusion}\ }\textbf {\bibinfo {volume} {63}},\
  \bibinfo {pages} {124006} (\bibinfo {year} {2021})}\BibitemShut {NoStop}%
\bibitem [{\citenamefont {Williams}\ \emph {et~al.}(2021)\citenamefont
  {Williams}, \citenamefont {Crane}, \citenamefont {Alessi}, \citenamefont
  {Boley}, \citenamefont {Bowers}, \citenamefont {Conder}, \citenamefont
  {Nicola}, \citenamefont {Nicola}, \citenamefont {Haefner}, \citenamefont
  {Halpin}, \citenamefont {Hamamoto}, \citenamefont {Heebner}, \citenamefont
  {Hermann}, \citenamefont {Herriot}, \citenamefont {Homoelle}, \citenamefont
  {Kalantar}, \citenamefont {Lanier}, \citenamefont {LaFortune}, \citenamefont
  {Lawson}, \citenamefont {Lowe-Webb}, \citenamefont {Morrissey}, \citenamefont
  {Nguyen}, \citenamefont {Orth}, \citenamefont {Pelz}, \citenamefont
  {Prantil}, \citenamefont {Rushford}, \citenamefont {Sacks}, \citenamefont
  {Salmon}, \citenamefont {Seppala}, \citenamefont {Shaw}, \citenamefont
  {Sigurdsson}, \citenamefont {Wegner}, \citenamefont {Widmayer}, \citenamefont
  {Yang}, ,\ and\ \citenamefont {Zobrist}}]{WCA:2021}%
  \BibitemOpen
  \bibfield  {author} {\bibinfo {author} {\bibfnamefont {W.~H.}\ \bibnamefont
  {Williams}}, \bibinfo {author} {\bibfnamefont {J.~K.}\ \bibnamefont {Crane}},
  \bibinfo {author} {\bibfnamefont {D.~A.}\ \bibnamefont {Alessi}}, \bibinfo
  {author} {\bibfnamefont {C.~D.}\ \bibnamefont {Boley}}, \bibinfo {author}
  {\bibfnamefont {M.~W.}\ \bibnamefont {Bowers}}, \bibinfo {author}
  {\bibfnamefont {A.~D.}\ \bibnamefont {Conder}}, \bibinfo {author}
  {\bibfnamefont {J.-M. G.~D.}\ \bibnamefont {Nicola}}, \bibinfo {author}
  {\bibfnamefont {P.~D.}\ \bibnamefont {Nicola}}, \bibinfo {author}
  {\bibfnamefont {C.}~\bibnamefont {Haefner}}, \bibinfo {author} {\bibfnamefont
  {J.~M.}\ \bibnamefont {Halpin}}, \bibinfo {author} {\bibfnamefont {M.~Y.}\
  \bibnamefont {Hamamoto}}, \bibinfo {author} {\bibfnamefont {J.~E.}\
  \bibnamefont {Heebner}}, \bibinfo {author} {\bibfnamefont {M.~R.}\
  \bibnamefont {Hermann}}, \bibinfo {author} {\bibfnamefont {S.~I.}\
  \bibnamefont {Herriot}}, \bibinfo {author} {\bibfnamefont {D.~C.}\
  \bibnamefont {Homoelle}}, \bibinfo {author} {\bibfnamefont {D.~H.}\
  \bibnamefont {Kalantar}}, \bibinfo {author} {\bibfnamefont {T.~E.}\
  \bibnamefont {Lanier}}, \bibinfo {author} {\bibfnamefont {K.~N.}\
  \bibnamefont {LaFortune}}, \bibinfo {author} {\bibfnamefont {J.~K.}\
  \bibnamefont {Lawson}}, \bibinfo {author} {\bibfnamefont {R.~R.}\
  \bibnamefont {Lowe-Webb}}, \bibinfo {author} {\bibfnamefont {F.~X.}\
  \bibnamefont {Morrissey}}, \bibinfo {author} {\bibfnamefont {H.}~\bibnamefont
  {Nguyen}}, \bibinfo {author} {\bibfnamefont {C.~D.}\ \bibnamefont {Orth}},
  \bibinfo {author} {\bibfnamefont {L.~J.}\ \bibnamefont {Pelz}}, \bibinfo
  {author} {\bibfnamefont {M.~A.}\ \bibnamefont {Prantil}}, \bibinfo {author}
  {\bibfnamefont {M.~C.}\ \bibnamefont {Rushford}}, \bibinfo {author}
  {\bibfnamefont {R.~A.}\ \bibnamefont {Sacks}}, \bibinfo {author}
  {\bibfnamefont {J.~T.}\ \bibnamefont {Salmon}}, \bibinfo {author}
  {\bibfnamefont {L.~G.}\ \bibnamefont {Seppala}}, \bibinfo {author}
  {\bibfnamefont {M.~J.}\ \bibnamefont {Shaw}}, \bibinfo {author}
  {\bibfnamefont {R.~J.}\ \bibnamefont {Sigurdsson}}, \bibinfo {author}
  {\bibfnamefont {P.~J.}\ \bibnamefont {Wegner}}, \bibinfo {author}
  {\bibfnamefont {C.~C.}\ \bibnamefont {Widmayer}}, \bibinfo {author}
  {\bibfnamefont {S.~T.}\ \bibnamefont {Yang}}, , \ and\ \bibinfo {author}
  {\bibfnamefont {T.~L.}\ \bibnamefont {Zobrist}},\ }\href@noop {} {\bibfield
  {journal} {\bibinfo  {journal} {Appl. Opt.}\ }\textbf {\bibinfo {volume}
  {60}},\ \bibinfo {pages} {2288} (\bibinfo {year} {2021})},\ \bibinfo {note}
  {https://doi.org/10.1364/AO.416846}\BibitemShut {NoStop}%
\bibitem [{\citenamefont {Tommasini}\ \emph {et~al.}(2020)\citenamefont
  {Tommasini}, \citenamefont {Landen}, \citenamefont {Hopkins}, \citenamefont
  {Hatchett}, \citenamefont {Kalantar}, \citenamefont {Hsing}, \citenamefont
  {Alessi}, \citenamefont {Ayers}, \citenamefont {Bhandarkar}, \citenamefont
  {Bowers} \emph {et~al.}}]{tommasini2020time}%
  \BibitemOpen
  \bibfield  {author} {\bibinfo {author} {\bibfnamefont {R.}~\bibnamefont
  {Tommasini}}, \bibinfo {author} {\bibfnamefont {O.}~\bibnamefont {Landen}},
  \bibinfo {author} {\bibfnamefont {L.~B.}\ \bibnamefont {Hopkins}}, \bibinfo
  {author} {\bibfnamefont {S.}~\bibnamefont {Hatchett}}, \bibinfo {author}
  {\bibfnamefont {D.}~\bibnamefont {Kalantar}}, \bibinfo {author}
  {\bibfnamefont {W.}~\bibnamefont {Hsing}}, \bibinfo {author} {\bibfnamefont
  {D.}~\bibnamefont {Alessi}}, \bibinfo {author} {\bibfnamefont
  {S.}~\bibnamefont {Ayers}}, \bibinfo {author} {\bibfnamefont
  {S.}~\bibnamefont {Bhandarkar}}, \bibinfo {author} {\bibfnamefont
  {M.}~\bibnamefont {Bowers}},  \emph {et~al.},\ }\href@noop {} {\bibfield
  {journal} {\bibinfo  {journal} {Physical Review Letters}\ }\textbf {\bibinfo
  {volume} {125}},\ \bibinfo {pages} {155003} (\bibinfo {year}
  {2020})}\BibitemShut {NoStop}%
\bibitem [{\citenamefont {Schwarz}\ \emph {et~al.}(2008)\citenamefont
  {Schwarz}, \citenamefont {Rambo}, \citenamefont {Geissel}, \citenamefont
  {Edens}, \citenamefont {Smith}, \citenamefont {Brambrink}, \citenamefont
  {Kimmel},\ and\ \citenamefont {Atherton}}]{schwarz2008activation}%
  \BibitemOpen
  \bibfield  {author} {\bibinfo {author} {\bibfnamefont {J.}~\bibnamefont
  {Schwarz}}, \bibinfo {author} {\bibfnamefont {P.}~\bibnamefont {Rambo}},
  \bibinfo {author} {\bibfnamefont {M.}~\bibnamefont {Geissel}}, \bibinfo
  {author} {\bibfnamefont {A.}~\bibnamefont {Edens}}, \bibinfo {author}
  {\bibfnamefont {I.}~\bibnamefont {Smith}}, \bibinfo {author} {\bibfnamefont
  {E.}~\bibnamefont {Brambrink}}, \bibinfo {author} {\bibfnamefont
  {M.}~\bibnamefont {Kimmel}}, \ and\ \bibinfo {author} {\bibfnamefont
  {B.}~\bibnamefont {Atherton}},\ }in\ \href@noop {} {\emph {\bibinfo
  {booktitle} {Journal of Physics: Conference Series}}},\ Vol.\ \bibinfo
  {volume} {112}\ (\bibinfo {organization} {IOP Publishing},\ \bibinfo {year}
  {2008})\ p.\ \bibinfo {pages} {032020}\BibitemShut {NoStop}%
\bibitem [{\citenamefont {Maywar}\ \emph {et~al.}(2008)\citenamefont {Maywar},
  \citenamefont {Kelly}, \citenamefont {Waxer}, \citenamefont {Morse},
  \citenamefont {Begishev}, \citenamefont {Bromage}, \citenamefont {Dorrer},
  \citenamefont {Edwards}, \citenamefont {Folnsbee}, \citenamefont {Guardalben}
  \emph {et~al.}}]{maywar2008omega}%
  \BibitemOpen
  \bibfield  {author} {\bibinfo {author} {\bibfnamefont {D.}~\bibnamefont
  {Maywar}}, \bibinfo {author} {\bibfnamefont {J.}~\bibnamefont {Kelly}},
  \bibinfo {author} {\bibfnamefont {L.}~\bibnamefont {Waxer}}, \bibinfo
  {author} {\bibfnamefont {S.}~\bibnamefont {Morse}}, \bibinfo {author}
  {\bibfnamefont {I.}~\bibnamefont {Begishev}}, \bibinfo {author}
  {\bibfnamefont {J.}~\bibnamefont {Bromage}}, \bibinfo {author} {\bibfnamefont
  {C.}~\bibnamefont {Dorrer}}, \bibinfo {author} {\bibfnamefont
  {J.}~\bibnamefont {Edwards}}, \bibinfo {author} {\bibfnamefont
  {L.}~\bibnamefont {Folnsbee}}, \bibinfo {author} {\bibfnamefont
  {M.}~\bibnamefont {Guardalben}},  \emph {et~al.},\ }in\ \href@noop {} {\emph
  {\bibinfo {booktitle} {Journal of Physics: Conference Series}}},\ Vol.\
  \bibinfo {volume} {112}\ (\bibinfo {organization} {IOP Publishing},\ \bibinfo
  {year} {2008})\ p.\ \bibinfo {pages} {032007}\BibitemShut {NoStop}%
\bibitem [{\citenamefont {Tommasini}\ \emph {et~al.}(2008)\citenamefont
  {Tommasini}, \citenamefont {MacPhee}, \citenamefont {Hey}, \citenamefont
  {Ma}, \citenamefont {Chen}, \citenamefont {Izumi}, \citenamefont {Unites},
  \citenamefont {MacKinnon}, \citenamefont {Hatchett}, \citenamefont
  {Remington} \emph {et~al.}}]{tommasini2008development}%
  \BibitemOpen
  \bibfield  {author} {\bibinfo {author} {\bibfnamefont {R.}~\bibnamefont
  {Tommasini}}, \bibinfo {author} {\bibfnamefont {A.}~\bibnamefont {MacPhee}},
  \bibinfo {author} {\bibfnamefont {D.}~\bibnamefont {Hey}}, \bibinfo {author}
  {\bibfnamefont {T.}~\bibnamefont {Ma}}, \bibinfo {author} {\bibfnamefont
  {C.}~\bibnamefont {Chen}}, \bibinfo {author} {\bibfnamefont {N.}~\bibnamefont
  {Izumi}}, \bibinfo {author} {\bibfnamefont {W.}~\bibnamefont {Unites}},
  \bibinfo {author} {\bibfnamefont {A.}~\bibnamefont {MacKinnon}}, \bibinfo
  {author} {\bibfnamefont {S.}~\bibnamefont {Hatchett}}, \bibinfo {author}
  {\bibfnamefont {B.}~\bibnamefont {Remington}},  \emph {et~al.},\ }\href@noop
  {} {\bibfield  {journal} {\bibinfo  {journal} {Review of Scientific
  Instruments}\ }\textbf {\bibinfo {volume} {79}} (\bibinfo {year}
  {2008})}\BibitemShut {NoStop}%
\bibitem [{\citenamefont {Brambrink}\ \emph {et~al.}(2009)\citenamefont
  {Brambrink}, \citenamefont {Wei}, \citenamefont {Barbrel}, \citenamefont
  {Audebert}, \citenamefont {Benuzzi-Mounaix}, \citenamefont {Boehly},
  \citenamefont {Endo}, \citenamefont {Gregory}, \citenamefont {Kimura},
  \citenamefont {Kodama} \emph {et~al.}}]{brambrink2009direct}%
  \BibitemOpen
  \bibfield  {author} {\bibinfo {author} {\bibfnamefont {E.}~\bibnamefont
  {Brambrink}}, \bibinfo {author} {\bibfnamefont {H.}~\bibnamefont {Wei}},
  \bibinfo {author} {\bibfnamefont {B.}~\bibnamefont {Barbrel}}, \bibinfo
  {author} {\bibfnamefont {P.}~\bibnamefont {Audebert}}, \bibinfo {author}
  {\bibfnamefont {A.}~\bibnamefont {Benuzzi-Mounaix}}, \bibinfo {author}
  {\bibfnamefont {T.}~\bibnamefont {Boehly}}, \bibinfo {author} {\bibfnamefont
  {T.}~\bibnamefont {Endo}}, \bibinfo {author} {\bibfnamefont {C.}~\bibnamefont
  {Gregory}}, \bibinfo {author} {\bibfnamefont {T.}~\bibnamefont {Kimura}},
  \bibinfo {author} {\bibfnamefont {R.}~\bibnamefont {Kodama}},  \emph
  {et~al.},\ }\href@noop {} {\bibfield  {journal} {\bibinfo  {journal}
  {Physical Review E}\ }\textbf {\bibinfo {volume} {80}},\ \bibinfo {pages}
  {056407} (\bibinfo {year} {2009})}\BibitemShut {NoStop}%
\bibitem [{\citenamefont {Chu}\ \emph {et~al.}(2018)\citenamefont {Chu},
  \citenamefont {Xi}, \citenamefont {Yu}, \citenamefont {Fan}, \citenamefont
  {Zhao}, \citenamefont {Shui}, \citenamefont {He}, \citenamefont {Zhang},
  \citenamefont {Zhang}, \citenamefont {Wu} \emph {et~al.}}]{chu2018high}%
  \BibitemOpen
  \bibfield  {author} {\bibinfo {author} {\bibfnamefont {G.}~\bibnamefont
  {Chu}}, \bibinfo {author} {\bibfnamefont {T.}~\bibnamefont {Xi}}, \bibinfo
  {author} {\bibfnamefont {M.}~\bibnamefont {Yu}}, \bibinfo {author}
  {\bibfnamefont {W.}~\bibnamefont {Fan}}, \bibinfo {author} {\bibfnamefont
  {Y.}~\bibnamefont {Zhao}}, \bibinfo {author} {\bibfnamefont {M.}~\bibnamefont
  {Shui}}, \bibinfo {author} {\bibfnamefont {W.}~\bibnamefont {He}}, \bibinfo
  {author} {\bibfnamefont {T.}~\bibnamefont {Zhang}}, \bibinfo {author}
  {\bibfnamefont {B.}~\bibnamefont {Zhang}}, \bibinfo {author} {\bibfnamefont
  {Y.}~\bibnamefont {Wu}},  \emph {et~al.},\ }\href@noop {} {\bibfield
  {journal} {\bibinfo  {journal} {Review of Scientific Instruments}\ }\textbf
  {\bibinfo {volume} {89}} (\bibinfo {year} {2018})}\BibitemShut {NoStop}%
\bibitem [{\citenamefont {He}\ \emph {et~al.}(2019)\citenamefont {He},
  \citenamefont {Xi}, \citenamefont {Shui}, \citenamefont {Yu}, \citenamefont
  {Zhao}, \citenamefont {Wu}, \citenamefont {Gu}, \citenamefont {Chu},\ and\
  \citenamefont {Xin}}]{he2019high}%
  \BibitemOpen
  \bibfield  {author} {\bibinfo {author} {\bibfnamefont {W.}~\bibnamefont
  {He}}, \bibinfo {author} {\bibfnamefont {T.}~\bibnamefont {Xi}}, \bibinfo
  {author} {\bibfnamefont {M.}~\bibnamefont {Shui}}, \bibinfo {author}
  {\bibfnamefont {M.}~\bibnamefont {Yu}}, \bibinfo {author} {\bibfnamefont
  {Y.}~\bibnamefont {Zhao}}, \bibinfo {author} {\bibfnamefont {Y.}~\bibnamefont
  {Wu}}, \bibinfo {author} {\bibfnamefont {Y.}~\bibnamefont {Gu}}, \bibinfo
  {author} {\bibfnamefont {G.}~\bibnamefont {Chu}}, \ and\ \bibinfo {author}
  {\bibfnamefont {J.}~\bibnamefont {Xin}},\ }\href@noop {} {\bibfield
  {journal} {\bibinfo  {journal} {AIP Advances}\ }\textbf {\bibinfo {volume}
  {9}} (\bibinfo {year} {2019})}\BibitemShut {NoStop}%
\bibitem [{\citenamefont {Hatchett}\ \emph {et~al.}(2000)\citenamefont
  {Hatchett}, \citenamefont {Brown}, \citenamefont {Cowan}, \citenamefont
  {Henry}, \citenamefont {Johnson}, \citenamefont {Key}, \citenamefont {Koch},
  \citenamefont {Langdon}, \citenamefont {Lasinski}, \citenamefont {Lee} \emph
  {et~al.}}]{hatchett2000electron}%
  \BibitemOpen
  \bibfield  {author} {\bibinfo {author} {\bibfnamefont {S.~P.}\ \bibnamefont
  {Hatchett}}, \bibinfo {author} {\bibfnamefont {C.~G.}\ \bibnamefont {Brown}},
  \bibinfo {author} {\bibfnamefont {T.~E.}\ \bibnamefont {Cowan}}, \bibinfo
  {author} {\bibfnamefont {E.~A.}\ \bibnamefont {Henry}}, \bibinfo {author}
  {\bibfnamefont {J.~S.}\ \bibnamefont {Johnson}}, \bibinfo {author}
  {\bibfnamefont {M.~H.}\ \bibnamefont {Key}}, \bibinfo {author} {\bibfnamefont
  {J.~A.}\ \bibnamefont {Koch}}, \bibinfo {author} {\bibfnamefont {A.~B.}\
  \bibnamefont {Langdon}}, \bibinfo {author} {\bibfnamefont {B.~F.}\
  \bibnamefont {Lasinski}}, \bibinfo {author} {\bibfnamefont {R.~W.}\
  \bibnamefont {Lee}},  \emph {et~al.},\ }\href@noop {} {\bibfield  {journal}
  {\bibinfo  {journal} {Physics of Plasmas}\ }\textbf {\bibinfo {volume} {7}},\
  \bibinfo {pages} {2076} (\bibinfo {year} {2000})}\BibitemShut {NoStop}%
\bibitem [{\citenamefont {Xin}\ \emph {et~al.}(2019)\citenamefont {Xin},
  \citenamefont {He}, \citenamefont {Liu}, \citenamefont {Chu}, \citenamefont
  {Yu}, \citenamefont {Fan}, \citenamefont {Wu}, \citenamefont {Xi},
  \citenamefont {Shui}, \citenamefont {Zhao} \emph {et~al.}}]{xin2019x}%
  \BibitemOpen
  \bibfield  {author} {\bibinfo {author} {\bibfnamefont {J.}~\bibnamefont
  {Xin}}, \bibinfo {author} {\bibfnamefont {A.}~\bibnamefont {He}}, \bibinfo
  {author} {\bibfnamefont {W.}~\bibnamefont {Liu}}, \bibinfo {author}
  {\bibfnamefont {G.}~\bibnamefont {Chu}}, \bibinfo {author} {\bibfnamefont
  {M.}~\bibnamefont {Yu}}, \bibinfo {author} {\bibfnamefont {W.}~\bibnamefont
  {Fan}}, \bibinfo {author} {\bibfnamefont {Y.}~\bibnamefont {Wu}}, \bibinfo
  {author} {\bibfnamefont {T.}~\bibnamefont {Xi}}, \bibinfo {author}
  {\bibfnamefont {M.}~\bibnamefont {Shui}}, \bibinfo {author} {\bibfnamefont
  {Y.}~\bibnamefont {Zhao}},  \emph {et~al.},\ }\href@noop {} {\bibfield
  {journal} {\bibinfo  {journal} {Journal of Micromechanics and
  Microengineering}\ }\textbf {\bibinfo {volume} {29}},\ \bibinfo {pages}
  {095011} (\bibinfo {year} {2019})}\BibitemShut {NoStop}%
\bibitem [{\citenamefont {McCuistian}\ \emph {et~al.}(2008)\citenamefont
  {McCuistian}, \citenamefont {Moir}, \citenamefont {Rose}, \citenamefont
  {Bender}, \citenamefont {Carlson}, \citenamefont {Hollabaugh},\ and\
  \citenamefont {Trainham}}]{Darht:2008}%
  \BibitemOpen
  \bibfield  {author} {\bibinfo {author} {\bibfnamefont {B.~T.}\ \bibnamefont
  {McCuistian}}, \bibinfo {author} {\bibfnamefont {D.}~\bibnamefont {Moir}},
  \bibinfo {author} {\bibfnamefont {E.}~\bibnamefont {Rose}}, \bibinfo {author}
  {\bibfnamefont {H.}~\bibnamefont {Bender}}, \bibinfo {author} {\bibfnamefont
  {C.}~\bibnamefont {Carlson}}, \bibinfo {author} {\bibfnamefont
  {C.}~\bibnamefont {Hollabaugh}}, \ and\ \bibinfo {author} {\bibfnamefont
  {R.}~\bibnamefont {Trainham}},\ }in\ \href@noop {} {\emph {\bibinfo
  {booktitle} {Proceedings of EPAC08}}},\ Vol.\ \bibinfo {volume} {TUPC066}\
  (\bibinfo {address} {{Genoa, Italy}},\ \bibinfo {year} {2008})\ pp.\ \bibinfo
  {pages} {1206--1208},\ \bibinfo {note} {{\it `Temporal spot size evolution of
  the DARHT first axis radiographic source'}}\BibitemShut {NoStop}%
\bibitem [{\citenamefont {Nath}(2010)}]{Nath:2010}%
  \BibitemOpen
  \bibfield  {author} {\bibinfo {author} {\bibfnamefont {S.}~\bibnamefont
  {Nath}},\ }\href@noop {} {\emph {\bibinfo {title} {Linear induction
  accelerators at the Los Alamos National Laboratory DARHT facility}}},\
  \bibinfo {type} {Tech. Rep.}\ \bibinfo {number} {LA-UR-10-06001}\ (\bibinfo
  {institution} {Los Alamos National Laboratory},\ \bibinfo {year}
  {2010})\BibitemShut {NoStop}%
\bibitem [{\citenamefont {Palaniyappan}\ \emph {et~al.}(2018)\citenamefont
  {Palaniyappan}, \citenamefont {Gautier}, \citenamefont {Tobias},
  \citenamefont {Fernandez}, \citenamefont {Mendez}, \citenamefont
  {Burris-Mog}, \citenamefont {Huang}, \citenamefont {Favalli}, \citenamefont
  {Hunter}, \citenamefont {Espy} \emph {et~al.}}]{palaniyappan2018mev}%
  \BibitemOpen
  \bibfield  {author} {\bibinfo {author} {\bibfnamefont {S.}~\bibnamefont
  {Palaniyappan}}, \bibinfo {author} {\bibfnamefont {D.~C.}\ \bibnamefont
  {Gautier}}, \bibinfo {author} {\bibfnamefont {B.~J.}\ \bibnamefont {Tobias}},
  \bibinfo {author} {\bibfnamefont {J.}~\bibnamefont {Fernandez}}, \bibinfo
  {author} {\bibfnamefont {J.}~\bibnamefont {Mendez}}, \bibinfo {author}
  {\bibfnamefont {T.}~\bibnamefont {Burris-Mog}}, \bibinfo {author}
  {\bibfnamefont {C.}~\bibnamefont {Huang}}, \bibinfo {author} {\bibfnamefont
  {A.}~\bibnamefont {Favalli}}, \bibinfo {author} {\bibfnamefont
  {J.}~\bibnamefont {Hunter}}, \bibinfo {author} {\bibfnamefont
  {M.}~\bibnamefont {Espy}},  \emph {et~al.},\ }\href@noop {} {\bibfield
  {journal} {\bibinfo  {journal} {Laser and Particle Beams}\ }\textbf {\bibinfo
  {volume} {36}},\ \bibinfo {pages} {502} (\bibinfo {year} {2018})}\BibitemShut
  {NoStop}%
\bibitem [{\citenamefont {Tommasini}\ \emph {et~al.}(2011)\citenamefont
  {Tommasini}, \citenamefont {Hatchett}, \citenamefont {Hey}, \citenamefont
  {Iglesias}, \citenamefont {Izumi}, \citenamefont {Koch}, \citenamefont
  {Landen}, \citenamefont {MacKinnon}, \citenamefont {Sorce}, \citenamefont
  {Delettrez} \emph {et~al.}}]{tommasini2011development}%
  \BibitemOpen
  \bibfield  {author} {\bibinfo {author} {\bibfnamefont {R.}~\bibnamefont
  {Tommasini}}, \bibinfo {author} {\bibfnamefont {S.}~\bibnamefont {Hatchett}},
  \bibinfo {author} {\bibfnamefont {D.}~\bibnamefont {Hey}}, \bibinfo {author}
  {\bibfnamefont {C.}~\bibnamefont {Iglesias}}, \bibinfo {author}
  {\bibfnamefont {N.}~\bibnamefont {Izumi}}, \bibinfo {author} {\bibfnamefont
  {J.}~\bibnamefont {Koch}}, \bibinfo {author} {\bibfnamefont {O.}~\bibnamefont
  {Landen}}, \bibinfo {author} {\bibfnamefont {A.}~\bibnamefont {MacKinnon}},
  \bibinfo {author} {\bibfnamefont {C.}~\bibnamefont {Sorce}}, \bibinfo
  {author} {\bibfnamefont {J.}~\bibnamefont {Delettrez}},  \emph {et~al.},\
  }\href@noop {} {\bibfield  {journal} {\bibinfo  {journal} {Physics of
  Plasmas}\ }\textbf {\bibinfo {volume} {18}} (\bibinfo {year}
  {2011})}\BibitemShut {NoStop}%
\bibitem [{\citenamefont {Glinec}\ \emph {et~al.}(2005)\citenamefont {Glinec},
  \citenamefont {Faure}, \citenamefont {Le~Dain}, \citenamefont {Darbon},
  \citenamefont {Hosokai}, \citenamefont {Santos}, \citenamefont {Lefebvre},
  \citenamefont {Rousseau}, \citenamefont {Burgy}, \citenamefont {Mercier}
  \emph {et~al.}}]{glinec2005high}%
  \BibitemOpen
  \bibfield  {author} {\bibinfo {author} {\bibfnamefont {Y.}~\bibnamefont
  {Glinec}}, \bibinfo {author} {\bibfnamefont {J.}~\bibnamefont {Faure}},
  \bibinfo {author} {\bibfnamefont {L.}~\bibnamefont {Le~Dain}}, \bibinfo
  {author} {\bibfnamefont {S.}~\bibnamefont {Darbon}}, \bibinfo {author}
  {\bibfnamefont {T.}~\bibnamefont {Hosokai}}, \bibinfo {author} {\bibfnamefont
  {J.}~\bibnamefont {Santos}}, \bibinfo {author} {\bibfnamefont
  {E.}~\bibnamefont {Lefebvre}}, \bibinfo {author} {\bibfnamefont {J.-P.}\
  \bibnamefont {Rousseau}}, \bibinfo {author} {\bibfnamefont {F.}~\bibnamefont
  {Burgy}}, \bibinfo {author} {\bibfnamefont {B.}~\bibnamefont {Mercier}},
  \emph {et~al.},\ }\href@noop {} {\bibfield  {journal} {\bibinfo  {journal}
  {Physical review letters}\ }\textbf {\bibinfo {volume} {94}},\ \bibinfo
  {pages} {025003} (\bibinfo {year} {2005})}\BibitemShut {NoStop}%
\bibitem [{\citenamefont {Kerr}\ \emph {et~al.}(2023)\citenamefont {Kerr},
  \citenamefont {Rusby}, \citenamefont {Williams}, \citenamefont {Meaney},
  \citenamefont {Schlossberg}, \citenamefont {Aghedo}, \citenamefont {Alessi},
  \citenamefont {Ayers}, \citenamefont {Azhar}, \citenamefont {Aufderheide}
  \emph {et~al.}}]{kerr2023development}%
  \BibitemOpen
  \bibfield  {author} {\bibinfo {author} {\bibfnamefont {S.~M.}\ \bibnamefont
  {Kerr}}, \bibinfo {author} {\bibfnamefont {D.}~\bibnamefont {Rusby}},
  \bibinfo {author} {\bibfnamefont {G.~J.}\ \bibnamefont {Williams}}, \bibinfo
  {author} {\bibfnamefont {K.}~\bibnamefont {Meaney}}, \bibinfo {author}
  {\bibfnamefont {D.~J.}\ \bibnamefont {Schlossberg}}, \bibinfo {author}
  {\bibfnamefont {A.}~\bibnamefont {Aghedo}}, \bibinfo {author} {\bibfnamefont
  {D.}~\bibnamefont {Alessi}}, \bibinfo {author} {\bibfnamefont
  {J.}~\bibnamefont {Ayers}}, \bibinfo {author} {\bibfnamefont
  {S.}~\bibnamefont {Azhar}}, \bibinfo {author} {\bibfnamefont {M.~B.}\
  \bibnamefont {Aufderheide}},  \emph {et~al.},\ }\href@noop {} {\bibfield
  {journal} {\bibinfo  {journal} {Physics of Plasmas}\ }\textbf {\bibinfo
  {volume} {30}} (\bibinfo {year} {2023})}\BibitemShut {NoStop}%
\bibitem [{\citenamefont {Courtois}\ \emph {et~al.}(2011)\citenamefont
  {Courtois}, \citenamefont {Edwards}, \citenamefont {Compant La~Fontaine},
  \citenamefont {Aedy}, \citenamefont {Barbotin}, \citenamefont {Bazzoli},
  \citenamefont {Biddle}, \citenamefont {Brebion}, \citenamefont {Bourgade},
  \citenamefont {Drew} \emph {et~al.}}]{courtois2011high}%
  \BibitemOpen
  \bibfield  {author} {\bibinfo {author} {\bibfnamefont {C.}~\bibnamefont
  {Courtois}}, \bibinfo {author} {\bibfnamefont {R.}~\bibnamefont {Edwards}},
  \bibinfo {author} {\bibfnamefont {A.}~\bibnamefont {Compant La~Fontaine}},
  \bibinfo {author} {\bibfnamefont {C.}~\bibnamefont {Aedy}}, \bibinfo {author}
  {\bibfnamefont {M.}~\bibnamefont {Barbotin}}, \bibinfo {author}
  {\bibfnamefont {S.}~\bibnamefont {Bazzoli}}, \bibinfo {author} {\bibfnamefont
  {L.}~\bibnamefont {Biddle}}, \bibinfo {author} {\bibfnamefont
  {D.}~\bibnamefont {Brebion}}, \bibinfo {author} {\bibfnamefont
  {J.}~\bibnamefont {Bourgade}}, \bibinfo {author} {\bibfnamefont
  {D.}~\bibnamefont {Drew}},  \emph {et~al.},\ }\href@noop {} {\bibfield
  {journal} {\bibinfo  {journal} {Physics of Plasmas}\ }\textbf {\bibinfo
  {volume} {18}} (\bibinfo {year} {2011})}\BibitemShut {NoStop}%
\bibitem [{\citenamefont {Courtois}\ \emph {et~al.}(2009)\citenamefont
  {Courtois}, \citenamefont {Compant La~Fontaine}, \citenamefont {Landoas},
  \citenamefont {Lidove}, \citenamefont {M{\'e}ot}, \citenamefont {Morel},
  \citenamefont {Nuter}, \citenamefont {Lefebvre}, \citenamefont {Boscheron},
  \citenamefont {Grenier} \emph {et~al.}}]{courtois2009effect}%
  \BibitemOpen
  \bibfield  {author} {\bibinfo {author} {\bibfnamefont {C.}~\bibnamefont
  {Courtois}}, \bibinfo {author} {\bibfnamefont {A.}~\bibnamefont {Compant
  La~Fontaine}}, \bibinfo {author} {\bibfnamefont {O.}~\bibnamefont {Landoas}},
  \bibinfo {author} {\bibfnamefont {G.}~\bibnamefont {Lidove}}, \bibinfo
  {author} {\bibfnamefont {V.}~\bibnamefont {M{\'e}ot}}, \bibinfo {author}
  {\bibfnamefont {P.}~\bibnamefont {Morel}}, \bibinfo {author} {\bibfnamefont
  {R.}~\bibnamefont {Nuter}}, \bibinfo {author} {\bibfnamefont
  {E.}~\bibnamefont {Lefebvre}}, \bibinfo {author} {\bibfnamefont
  {A.}~\bibnamefont {Boscheron}}, \bibinfo {author} {\bibfnamefont
  {J.}~\bibnamefont {Grenier}},  \emph {et~al.},\ }\href@noop {} {\bibfield
  {journal} {\bibinfo  {journal} {Physics of Plasmas}\ }\textbf {\bibinfo
  {volume} {16}} (\bibinfo {year} {2009})}\BibitemShut {NoStop}%
\bibitem [{\citenamefont {Courtois}\ \emph {et~al.}(2013)\citenamefont
  {Courtois}, \citenamefont {Edwards}, \citenamefont {Compant La~Fontaine},
  \citenamefont {Aedy}, \citenamefont {Bazzoli}, \citenamefont {Bourgade},
  \citenamefont {Gazave}, \citenamefont {Lagrange}, \citenamefont {Landoas},
  \citenamefont {Dain} \emph {et~al.}}]{courtois2013characterisation}%
  \BibitemOpen
  \bibfield  {author} {\bibinfo {author} {\bibfnamefont {C.}~\bibnamefont
  {Courtois}}, \bibinfo {author} {\bibfnamefont {R.}~\bibnamefont {Edwards}},
  \bibinfo {author} {\bibfnamefont {A.}~\bibnamefont {Compant La~Fontaine}},
  \bibinfo {author} {\bibfnamefont {C.}~\bibnamefont {Aedy}}, \bibinfo {author}
  {\bibfnamefont {S.}~\bibnamefont {Bazzoli}}, \bibinfo {author} {\bibfnamefont
  {J.}~\bibnamefont {Bourgade}}, \bibinfo {author} {\bibfnamefont
  {J.}~\bibnamefont {Gazave}}, \bibinfo {author} {\bibfnamefont
  {J.}~\bibnamefont {Lagrange}}, \bibinfo {author} {\bibfnamefont
  {O.}~\bibnamefont {Landoas}}, \bibinfo {author} {\bibfnamefont {L.~L.}\
  \bibnamefont {Dain}},  \emph {et~al.},\ }\href@noop {} {\bibfield  {journal}
  {\bibinfo  {journal} {Physics of Plasmas}\ }\textbf {\bibinfo {volume} {20}}
  (\bibinfo {year} {2013})}\BibitemShut {NoStop}%
\bibitem [{\citenamefont {Tommasini}\ \emph {et~al.}(2017)\citenamefont
  {Tommasini}, \citenamefont {Bailey}, \citenamefont {Bradley}, \citenamefont
  {Bowers}, \citenamefont {Chen}, \citenamefont {Di~Nicola}, \citenamefont
  {Di~Nicola}, \citenamefont {Gururangan}, \citenamefont {Hall}, \citenamefont
  {Hardy} \emph {et~al.}}]{tommasini2017short}%
  \BibitemOpen
  \bibfield  {author} {\bibinfo {author} {\bibfnamefont {R.}~\bibnamefont
  {Tommasini}}, \bibinfo {author} {\bibfnamefont {C.}~\bibnamefont {Bailey}},
  \bibinfo {author} {\bibfnamefont {D.}~\bibnamefont {Bradley}}, \bibinfo
  {author} {\bibfnamefont {M.}~\bibnamefont {Bowers}}, \bibinfo {author}
  {\bibfnamefont {H.}~\bibnamefont {Chen}}, \bibinfo {author} {\bibfnamefont
  {J.}~\bibnamefont {Di~Nicola}}, \bibinfo {author} {\bibfnamefont
  {P.}~\bibnamefont {Di~Nicola}}, \bibinfo {author} {\bibfnamefont
  {G.}~\bibnamefont {Gururangan}}, \bibinfo {author} {\bibfnamefont
  {G.}~\bibnamefont {Hall}}, \bibinfo {author} {\bibfnamefont {C.}~\bibnamefont
  {Hardy}},  \emph {et~al.},\ }\href@noop {} {\bibfield  {journal} {\bibinfo
  {journal} {Physics of Plasmas}\ }\textbf {\bibinfo {volume} {24}} (\bibinfo
  {year} {2017})}\BibitemShut {NoStop}%
\bibitem [{\citenamefont {Park}\ \emph {et~al.}(2006)\citenamefont {Park},
  \citenamefont {Chambers}, \citenamefont {Chung}, \citenamefont {Clarke},
  \citenamefont {Eagleton}, \citenamefont {Giraldez}, \citenamefont {Goldsack},
  \citenamefont {Heathcote}, \citenamefont {Izumi}, \citenamefont {Key} \emph
  {et~al.}}]{park2006high}%
  \BibitemOpen
  \bibfield  {author} {\bibinfo {author} {\bibfnamefont {H.-S.}\ \bibnamefont
  {Park}}, \bibinfo {author} {\bibfnamefont {D.}~\bibnamefont {Chambers}},
  \bibinfo {author} {\bibfnamefont {H.-K.}\ \bibnamefont {Chung}}, \bibinfo
  {author} {\bibfnamefont {R.}~\bibnamefont {Clarke}}, \bibinfo {author}
  {\bibfnamefont {R.}~\bibnamefont {Eagleton}}, \bibinfo {author}
  {\bibfnamefont {E.}~\bibnamefont {Giraldez}}, \bibinfo {author}
  {\bibfnamefont {T.}~\bibnamefont {Goldsack}}, \bibinfo {author}
  {\bibfnamefont {R.}~\bibnamefont {Heathcote}}, \bibinfo {author}
  {\bibfnamefont {N.}~\bibnamefont {Izumi}}, \bibinfo {author} {\bibfnamefont
  {M.}~\bibnamefont {Key}},  \emph {et~al.},\ }\href@noop {} {\bibfield
  {journal} {\bibinfo  {journal} {Physics of plasmas}\ }\textbf {\bibinfo
  {volume} {13}} (\bibinfo {year} {2006})}\BibitemShut {NoStop}%
\bibitem [{\citenamefont {Workman}\ \emph {et~al.}(2010)\citenamefont
  {Workman}, \citenamefont {Cobble}, \citenamefont {Flippo}, \citenamefont
  {Gautier}, \citenamefont {Montgomery},\ and\ \citenamefont
  {Offermann}}]{workman2010phase}%
  \BibitemOpen
  \bibfield  {author} {\bibinfo {author} {\bibfnamefont {J.}~\bibnamefont
  {Workman}}, \bibinfo {author} {\bibfnamefont {J.}~\bibnamefont {Cobble}},
  \bibinfo {author} {\bibfnamefont {K.}~\bibnamefont {Flippo}}, \bibinfo
  {author} {\bibfnamefont {D.~C.}\ \bibnamefont {Gautier}}, \bibinfo {author}
  {\bibfnamefont {D.~S.}\ \bibnamefont {Montgomery}}, \ and\ \bibinfo {author}
  {\bibfnamefont {D.~T.}\ \bibnamefont {Offermann}},\ }\href@noop {} {\bibfield
   {journal} {\bibinfo  {journal} {Review of Scientific Instruments}\ }\textbf
  {\bibinfo {volume} {81}} (\bibinfo {year} {2010})}\BibitemShut {NoStop}%
\bibitem [{\citenamefont {Toth}\ \emph {et~al.}(2007)\citenamefont {Toth},
  \citenamefont {Fourmaux}, \citenamefont {Ozaki}, \citenamefont {Servol},
  \citenamefont {Kieffer}, \citenamefont {Kincaid},\ and\ \citenamefont
  {Krol}}]{toth2007evaluation}%
  \BibitemOpen
  \bibfield  {author} {\bibinfo {author} {\bibfnamefont {R.}~\bibnamefont
  {Toth}}, \bibinfo {author} {\bibfnamefont {S.}~\bibnamefont {Fourmaux}},
  \bibinfo {author} {\bibfnamefont {T.}~\bibnamefont {Ozaki}}, \bibinfo
  {author} {\bibfnamefont {M.}~\bibnamefont {Servol}}, \bibinfo {author}
  {\bibfnamefont {J.}~\bibnamefont {Kieffer}}, \bibinfo {author} {\bibfnamefont
  {R.}~\bibnamefont {Kincaid}}, \ and\ \bibinfo {author} {\bibfnamefont
  {A.}~\bibnamefont {Krol}},\ }\href@noop {} {\bibfield  {journal} {\bibinfo
  {journal} {Physics of plasmas}\ }\textbf {\bibinfo {volume} {14}} (\bibinfo
  {year} {2007})}\BibitemShut {NoStop}%
\bibitem [{\citenamefont {Antonelli}\ \emph {et~al.}(2019)\citenamefont
  {Antonelli}, \citenamefont {Barbato}, \citenamefont {Mancelli}, \citenamefont
  {Trela}, \citenamefont {Zeraouli}, \citenamefont {Boutoux}, \citenamefont
  {Neumayer}, \citenamefont {Atzeni}, \citenamefont {Schiavi}, \citenamefont
  {Volpe} \emph {et~al.}}]{antonelli2019x}%
  \BibitemOpen
  \bibfield  {author} {\bibinfo {author} {\bibfnamefont {L.}~\bibnamefont
  {Antonelli}}, \bibinfo {author} {\bibfnamefont {F.}~\bibnamefont {Barbato}},
  \bibinfo {author} {\bibfnamefont {D.}~\bibnamefont {Mancelli}}, \bibinfo
  {author} {\bibfnamefont {J.}~\bibnamefont {Trela}}, \bibinfo {author}
  {\bibfnamefont {G.}~\bibnamefont {Zeraouli}}, \bibinfo {author}
  {\bibfnamefont {G.}~\bibnamefont {Boutoux}}, \bibinfo {author} {\bibfnamefont
  {P.}~\bibnamefont {Neumayer}}, \bibinfo {author} {\bibfnamefont
  {S.}~\bibnamefont {Atzeni}}, \bibinfo {author} {\bibfnamefont
  {A.}~\bibnamefont {Schiavi}}, \bibinfo {author} {\bibfnamefont
  {L.}~\bibnamefont {Volpe}},  \emph {et~al.},\ }\href@noop {} {\bibfield
  {journal} {\bibinfo  {journal} {Europhysics Letters}\ }\textbf {\bibinfo
  {volume} {125}},\ \bibinfo {pages} {35002} (\bibinfo {year}
  {2019})}\BibitemShut {NoStop}%
\bibitem [{\citenamefont {Gambari}\ \emph {et~al.}(2020)\citenamefont
  {Gambari}, \citenamefont {Clady}, \citenamefont {Stolidi}, \citenamefont
  {Ut{\'e}za}, \citenamefont {Sentis},\ and\ \citenamefont
  {Ferr{\'e}}}]{gambari2020exploring}%
  \BibitemOpen
  \bibfield  {author} {\bibinfo {author} {\bibfnamefont {M.}~\bibnamefont
  {Gambari}}, \bibinfo {author} {\bibfnamefont {R.}~\bibnamefont {Clady}},
  \bibinfo {author} {\bibfnamefont {A.}~\bibnamefont {Stolidi}}, \bibinfo
  {author} {\bibfnamefont {O.}~\bibnamefont {Ut{\'e}za}}, \bibinfo {author}
  {\bibfnamefont {M.}~\bibnamefont {Sentis}}, \ and\ \bibinfo {author}
  {\bibfnamefont {A.}~\bibnamefont {Ferr{\'e}}},\ }\href@noop {} {\bibfield
  {journal} {\bibinfo  {journal} {Scientific Reports}\ }\textbf {\bibinfo
  {volume} {10}},\ \bibinfo {pages} {6766} (\bibinfo {year}
  {2020})}\BibitemShut {NoStop}%
\bibitem [{\citenamefont {Wood}\ \emph
  {et~al.}(2018{\natexlab{c}})\citenamefont {Wood}, \citenamefont {Chapman},
  \citenamefont {Poder}, \citenamefont {Lopes}, \citenamefont {Rutherford},
  \citenamefont {White}, \citenamefont {Albert}, \citenamefont {Behm},
  \citenamefont {Booth}, \citenamefont {Bryant} \emph
  {et~al.}}]{wood2018ultrafast}%
  \BibitemOpen
  \bibfield  {author} {\bibinfo {author} {\bibfnamefont {J.}~\bibnamefont
  {Wood}}, \bibinfo {author} {\bibfnamefont {D.}~\bibnamefont {Chapman}},
  \bibinfo {author} {\bibfnamefont {K.}~\bibnamefont {Poder}}, \bibinfo
  {author} {\bibfnamefont {N.}~\bibnamefont {Lopes}}, \bibinfo {author}
  {\bibfnamefont {M.}~\bibnamefont {Rutherford}}, \bibinfo {author}
  {\bibfnamefont {T.}~\bibnamefont {White}}, \bibinfo {author} {\bibfnamefont
  {F.}~\bibnamefont {Albert}}, \bibinfo {author} {\bibfnamefont
  {K.}~\bibnamefont {Behm}}, \bibinfo {author} {\bibfnamefont {N.}~\bibnamefont
  {Booth}}, \bibinfo {author} {\bibfnamefont {J.}~\bibnamefont {Bryant}},
  \emph {et~al.},\ }\href@noop {} {\bibfield  {journal} {\bibinfo  {journal}
  {Scientific reports}\ }\textbf {\bibinfo {volume} {8}},\ \bibinfo {pages} {1}
  (\bibinfo {year} {2018}{\natexlab{c}})}\BibitemShut {NoStop}%
\bibitem [{\citenamefont {Cipiccia}\ \emph {et~al.}(2011)\citenamefont
  {Cipiccia}, \citenamefont {Islam}, \citenamefont {Ersfeld}, \citenamefont
  {Shanks}, \citenamefont {Brunetti}, \citenamefont {Vieux}, \citenamefont
  {Yang}, \citenamefont {Issac}, \citenamefont {Wiggins}, \citenamefont {Welsh}
  \emph {et~al.}}]{cipiccia2011gamma}%
  \BibitemOpen
  \bibfield  {author} {\bibinfo {author} {\bibfnamefont {S.}~\bibnamefont
  {Cipiccia}}, \bibinfo {author} {\bibfnamefont {M.~R.}\ \bibnamefont {Islam}},
  \bibinfo {author} {\bibfnamefont {B.}~\bibnamefont {Ersfeld}}, \bibinfo
  {author} {\bibfnamefont {R.~P.}\ \bibnamefont {Shanks}}, \bibinfo {author}
  {\bibfnamefont {E.}~\bibnamefont {Brunetti}}, \bibinfo {author}
  {\bibfnamefont {G.}~\bibnamefont {Vieux}}, \bibinfo {author} {\bibfnamefont
  {X.}~\bibnamefont {Yang}}, \bibinfo {author} {\bibfnamefont {R.~C.}\
  \bibnamefont {Issac}}, \bibinfo {author} {\bibfnamefont {S.~M.}\ \bibnamefont
  {Wiggins}}, \bibinfo {author} {\bibfnamefont {G.~H.}\ \bibnamefont {Welsh}},
  \emph {et~al.},\ }\href@noop {} {\bibfield  {journal} {\bibinfo  {journal}
  {Nature Physics}\ }\textbf {\bibinfo {volume} {7}},\ \bibinfo {pages} {867}
  (\bibinfo {year} {2011})}\BibitemShut {NoStop}%
\bibitem [{\citenamefont {Albert}\ \emph {et~al.}(2017)\citenamefont {Albert},
  \citenamefont {Lemos}, \citenamefont {Shaw}, \citenamefont {Pollock},
  \citenamefont {Goyon}, \citenamefont {Schumaker}, \citenamefont {Saunders},
  \citenamefont {Marsh}, \citenamefont {Pak}, \citenamefont {Ralph} \emph
  {et~al.}}]{albert2017observation}%
  \BibitemOpen
  \bibfield  {author} {\bibinfo {author} {\bibfnamefont {F.}~\bibnamefont
  {Albert}}, \bibinfo {author} {\bibfnamefont {N.}~\bibnamefont {Lemos}},
  \bibinfo {author} {\bibfnamefont {J.}~\bibnamefont {Shaw}}, \bibinfo {author}
  {\bibfnamefont {B.}~\bibnamefont {Pollock}}, \bibinfo {author} {\bibfnamefont
  {C.}~\bibnamefont {Goyon}}, \bibinfo {author} {\bibfnamefont
  {W.}~\bibnamefont {Schumaker}}, \bibinfo {author} {\bibfnamefont
  {A.}~\bibnamefont {Saunders}}, \bibinfo {author} {\bibfnamefont
  {K.}~\bibnamefont {Marsh}}, \bibinfo {author} {\bibfnamefont
  {A.}~\bibnamefont {Pak}}, \bibinfo {author} {\bibfnamefont {J.}~\bibnamefont
  {Ralph}},  \emph {et~al.},\ }\href@noop {} {\bibfield  {journal} {\bibinfo
  {journal} {Physical review letters}\ }\textbf {\bibinfo {volume} {118}},\
  \bibinfo {pages} {134801} (\bibinfo {year} {2017})}\BibitemShut {NoStop}%
\bibitem [{\citenamefont {Albert}\ \emph {et~al.}(2018)\citenamefont {Albert},
  \citenamefont {Lemos}, \citenamefont {Shaw}, \citenamefont {King},
  \citenamefont {Pollock}, \citenamefont {Goyon}, \citenamefont {Schumaker},
  \citenamefont {Saunders}, \citenamefont {Marsh}, \citenamefont {Pak} \emph
  {et~al.}}]{albert2018betatron}%
  \BibitemOpen
  \bibfield  {author} {\bibinfo {author} {\bibfnamefont {F.}~\bibnamefont
  {Albert}}, \bibinfo {author} {\bibfnamefont {N.}~\bibnamefont {Lemos}},
  \bibinfo {author} {\bibfnamefont {J.}~\bibnamefont {Shaw}}, \bibinfo {author}
  {\bibfnamefont {P.}~\bibnamefont {King}}, \bibinfo {author} {\bibfnamefont
  {B.}~\bibnamefont {Pollock}}, \bibinfo {author} {\bibfnamefont
  {C.}~\bibnamefont {Goyon}}, \bibinfo {author} {\bibfnamefont
  {W.}~\bibnamefont {Schumaker}}, \bibinfo {author} {\bibfnamefont
  {A.}~\bibnamefont {Saunders}}, \bibinfo {author} {\bibfnamefont
  {K.}~\bibnamefont {Marsh}}, \bibinfo {author} {\bibfnamefont
  {A.}~\bibnamefont {Pak}},  \emph {et~al.},\ }\href@noop {} {\bibfield
  {journal} {\bibinfo  {journal} {Nuclear Fusion}\ }\textbf {\bibinfo {volume}
  {59}},\ \bibinfo {pages} {032003} (\bibinfo {year} {2018})}\BibitemShut
  {NoStop}%
\bibitem [{\citenamefont {Ferri}\ \emph {et~al.}(2016)\citenamefont {Ferri},
  \citenamefont {Davoine}, \citenamefont {Kalmykov},\ and\ \citenamefont
  {Lifschitz}}]{ferri2016electron}%
  \BibitemOpen
  \bibfield  {author} {\bibinfo {author} {\bibfnamefont {J.}~\bibnamefont
  {Ferri}}, \bibinfo {author} {\bibfnamefont {X.}~\bibnamefont {Davoine}},
  \bibinfo {author} {\bibfnamefont {S.}~\bibnamefont {Kalmykov}}, \ and\
  \bibinfo {author} {\bibfnamefont {A.}~\bibnamefont {Lifschitz}},\ }\href@noop
  {} {\bibfield  {journal} {\bibinfo  {journal} {Physical Review Accelerators
  and Beams}\ }\textbf {\bibinfo {volume} {19}},\ \bibinfo {pages} {101301}
  (\bibinfo {year} {2016})}\BibitemShut {NoStop}%
\bibitem [{\citenamefont {Rosmej}\ \emph {et~al.}(2021)\citenamefont {Rosmej},
  \citenamefont {Shen}, \citenamefont {Pukhov}, \citenamefont {Antonelli},
  \citenamefont {Barbato}, \citenamefont {Gyrdymov}, \citenamefont
  {G{\"u}nther}, \citenamefont {Z{\"a}hter}, \citenamefont {Popov},
  \citenamefont {Borisenko} \emph {et~al.}}]{rosmej2021bright}%
  \BibitemOpen
  \bibfield  {author} {\bibinfo {author} {\bibfnamefont {O.}~\bibnamefont
  {Rosmej}}, \bibinfo {author} {\bibfnamefont {X.}~\bibnamefont {Shen}},
  \bibinfo {author} {\bibfnamefont {A.}~\bibnamefont {Pukhov}}, \bibinfo
  {author} {\bibfnamefont {L.}~\bibnamefont {Antonelli}}, \bibinfo {author}
  {\bibfnamefont {F.}~\bibnamefont {Barbato}}, \bibinfo {author} {\bibfnamefont
  {M.}~\bibnamefont {Gyrdymov}}, \bibinfo {author} {\bibfnamefont
  {M.}~\bibnamefont {G{\"u}nther}}, \bibinfo {author} {\bibfnamefont
  {S.}~\bibnamefont {Z{\"a}hter}}, \bibinfo {author} {\bibfnamefont
  {V.}~\bibnamefont {Popov}}, \bibinfo {author} {\bibfnamefont
  {N.}~\bibnamefont {Borisenko}},  \emph {et~al.},\ }\href@noop {} {\bibfield
  {journal} {\bibinfo  {journal} {Matter and Radiation at Extremes}\ }\textbf
  {\bibinfo {volume} {6}} (\bibinfo {year} {2021})}\BibitemShut {NoStop}%
\bibitem [{\citenamefont {Espy}\ \emph {et~al.}(2021)\citenamefont {Espy},
  \citenamefont {Klasky}, \citenamefont {James}, \citenamefont {Moir},
  \citenamefont {Mendez}, \citenamefont {Morneau}, \citenamefont {Shurter},
  \citenamefont {Sedillo}, \citenamefont {Volegov},\ and\ \citenamefont
  {Gehring}}]{espy2021spectral}%
  \BibitemOpen
  \bibfield  {author} {\bibinfo {author} {\bibfnamefont {M.}~\bibnamefont
  {Espy}}, \bibinfo {author} {\bibfnamefont {M.}~\bibnamefont {Klasky}},
  \bibinfo {author} {\bibfnamefont {M.}~\bibnamefont {James}}, \bibinfo
  {author} {\bibfnamefont {D.}~\bibnamefont {Moir}}, \bibinfo {author}
  {\bibfnamefont {J.}~\bibnamefont {Mendez}}, \bibinfo {author} {\bibfnamefont
  {R.}~\bibnamefont {Morneau}}, \bibinfo {author} {\bibfnamefont
  {R.}~\bibnamefont {Shurter}}, \bibinfo {author} {\bibfnamefont
  {R.}~\bibnamefont {Sedillo}}, \bibinfo {author} {\bibfnamefont
  {P.}~\bibnamefont {Volegov}}, \ and\ \bibinfo {author} {\bibfnamefont
  {A.}~\bibnamefont {Gehring}},\ }\href@noop {} {\bibfield  {journal} {\bibinfo
   {journal} {Review of Scientific Instruments}\ }\textbf {\bibinfo {volume}
  {92}} (\bibinfo {year} {2021})}\BibitemShut {NoStop}%
\bibitem [{\citenamefont {Chen}\ \emph {et~al.}(2017)\citenamefont {Chen},
  \citenamefont {Hermann}, \citenamefont {Kalantar}, \citenamefont {Martinez},
  \citenamefont {Di~Nicola}, \citenamefont {Tommasini}, \citenamefont {Landen},
  \citenamefont {Alessi}, \citenamefont {Bowers}, \citenamefont {Browning}
  \emph {et~al.}}]{chen2017high}%
  \BibitemOpen
  \bibfield  {author} {\bibinfo {author} {\bibfnamefont {H.}~\bibnamefont
  {Chen}}, \bibinfo {author} {\bibfnamefont {M.}~\bibnamefont {Hermann}},
  \bibinfo {author} {\bibfnamefont {D.}~\bibnamefont {Kalantar}}, \bibinfo
  {author} {\bibfnamefont {D.}~\bibnamefont {Martinez}}, \bibinfo {author}
  {\bibfnamefont {P.}~\bibnamefont {Di~Nicola}}, \bibinfo {author}
  {\bibfnamefont {R.}~\bibnamefont {Tommasini}}, \bibinfo {author}
  {\bibfnamefont {O.}~\bibnamefont {Landen}}, \bibinfo {author} {\bibfnamefont
  {D.}~\bibnamefont {Alessi}}, \bibinfo {author} {\bibfnamefont
  {M.}~\bibnamefont {Bowers}}, \bibinfo {author} {\bibfnamefont
  {D.}~\bibnamefont {Browning}},  \emph {et~al.},\ }\href@noop {} {\bibfield
  {journal} {\bibinfo  {journal} {Physics of Plasmas}\ }\textbf {\bibinfo
  {volume} {24}} (\bibinfo {year} {2017})}\BibitemShut {NoStop}%
\bibitem [{\citenamefont {Meaney}\ \emph {et~al.}(2021)\citenamefont {Meaney},
  \citenamefont {Kerr}, \citenamefont {Williams}, \citenamefont
  {Geppert-Kleinrath}, \citenamefont {Kim}, \citenamefont {Herrmann},
  \citenamefont {Kalantar}, \citenamefont {Mackinnon}, \citenamefont {Bowers},
  \citenamefont {Pelz} \emph {et~al.}}]{meaney2021multi}%
  \BibitemOpen
  \bibfield  {author} {\bibinfo {author} {\bibfnamefont {K.~D.}\ \bibnamefont
  {Meaney}}, \bibinfo {author} {\bibfnamefont {S.}~\bibnamefont {Kerr}},
  \bibinfo {author} {\bibfnamefont {G.}~\bibnamefont {Williams}}, \bibinfo
  {author} {\bibfnamefont {H.}~\bibnamefont {Geppert-Kleinrath}}, \bibinfo
  {author} {\bibfnamefont {Y.}~\bibnamefont {Kim}}, \bibinfo {author}
  {\bibfnamefont {H.~W.}\ \bibnamefont {Herrmann}}, \bibinfo {author}
  {\bibfnamefont {D.}~\bibnamefont {Kalantar}}, \bibinfo {author}
  {\bibfnamefont {A.}~\bibnamefont {Mackinnon}}, \bibinfo {author}
  {\bibfnamefont {M.}~\bibnamefont {Bowers}}, \bibinfo {author} {\bibfnamefont
  {L.}~\bibnamefont {Pelz}},  \emph {et~al.},\ }\href@noop {} {\bibfield
  {journal} {\bibinfo  {journal} {Physics of Plasmas}\ }\textbf {\bibinfo
  {volume} {28}} (\bibinfo {year} {2021})}\BibitemShut {NoStop}%
\bibitem [{\citenamefont {Ping}\ \emph {et~al.}(2008)\citenamefont {Ping},
  \citenamefont {Shepherd}, \citenamefont {Lasinski}, \citenamefont {Tabak},
  \citenamefont {Chen}, \citenamefont {Chung}, \citenamefont {Fournier},
  \citenamefont {Hansen}, \citenamefont {Kemp}, \citenamefont {Liedahl} \emph
  {et~al.}}]{ping2008absorption}%
  \BibitemOpen
  \bibfield  {author} {\bibinfo {author} {\bibfnamefont {Y.}~\bibnamefont
  {Ping}}, \bibinfo {author} {\bibfnamefont {R.}~\bibnamefont {Shepherd}},
  \bibinfo {author} {\bibfnamefont {B.}~\bibnamefont {Lasinski}}, \bibinfo
  {author} {\bibfnamefont {M.}~\bibnamefont {Tabak}}, \bibinfo {author}
  {\bibfnamefont {H.}~\bibnamefont {Chen}}, \bibinfo {author} {\bibfnamefont
  {H.}~\bibnamefont {Chung}}, \bibinfo {author} {\bibfnamefont
  {K.}~\bibnamefont {Fournier}}, \bibinfo {author} {\bibfnamefont
  {S.}~\bibnamefont {Hansen}}, \bibinfo {author} {\bibfnamefont
  {A.}~\bibnamefont {Kemp}}, \bibinfo {author} {\bibfnamefont {D.}~\bibnamefont
  {Liedahl}},  \emph {et~al.},\ }\href@noop {} {\bibfield  {journal} {\bibinfo
  {journal} {Physical review letters}\ }\textbf {\bibinfo {volume} {100}},\
  \bibinfo {pages} {085004} (\bibinfo {year} {2008})}\BibitemShut {NoStop}%
\bibitem [{\citenamefont {Park}\ \emph {et~al.}(2021)\citenamefont {Park},
  \citenamefont {Tommasini}, \citenamefont {Shepherd}, \citenamefont {London},
  \citenamefont {Bargsten}, \citenamefont {Hollinger}, \citenamefont
  {Capeluto}, \citenamefont {Shlyaptsev}, \citenamefont {Hill}, \citenamefont
  {Kaymak} \emph {et~al.}}]{park2021absolute}%
  \BibitemOpen
  \bibfield  {author} {\bibinfo {author} {\bibfnamefont {J.}~\bibnamefont
  {Park}}, \bibinfo {author} {\bibfnamefont {R.}~\bibnamefont {Tommasini}},
  \bibinfo {author} {\bibfnamefont {R.}~\bibnamefont {Shepherd}}, \bibinfo
  {author} {\bibfnamefont {R.}~\bibnamefont {London}}, \bibinfo {author}
  {\bibfnamefont {C.}~\bibnamefont {Bargsten}}, \bibinfo {author}
  {\bibfnamefont {R.}~\bibnamefont {Hollinger}}, \bibinfo {author}
  {\bibfnamefont {M.~G.}\ \bibnamefont {Capeluto}}, \bibinfo {author}
  {\bibfnamefont {V.}~\bibnamefont {Shlyaptsev}}, \bibinfo {author}
  {\bibfnamefont {M.}~\bibnamefont {Hill}}, \bibinfo {author} {\bibfnamefont
  {V.}~\bibnamefont {Kaymak}},  \emph {et~al.},\ }\href@noop {} {\bibfield
  {journal} {\bibinfo  {journal} {Physics of Plasmas}\ }\textbf {\bibinfo
  {volume} {28}} (\bibinfo {year} {2021})}\BibitemShut {NoStop}%
\bibitem [{\citenamefont {Ceurvorst}\ \emph {et~al.}(2016)\citenamefont
  {Ceurvorst}, \citenamefont {Ratan}, \citenamefont {Levy}, \citenamefont
  {Kasim}, \citenamefont {Sadler}, \citenamefont {Scott}, \citenamefont
  {Trines}, \citenamefont {Huang}, \citenamefont {Skramic}, \citenamefont
  {Vranic} \emph {et~al.}}]{ceurvorst2016mitigating}%
  \BibitemOpen
  \bibfield  {author} {\bibinfo {author} {\bibfnamefont {L.}~\bibnamefont
  {Ceurvorst}}, \bibinfo {author} {\bibfnamefont {N.}~\bibnamefont {Ratan}},
  \bibinfo {author} {\bibfnamefont {M.}~\bibnamefont {Levy}}, \bibinfo {author}
  {\bibfnamefont {M.}~\bibnamefont {Kasim}}, \bibinfo {author} {\bibfnamefont
  {J.}~\bibnamefont {Sadler}}, \bibinfo {author} {\bibfnamefont
  {R.}~\bibnamefont {Scott}}, \bibinfo {author} {\bibfnamefont
  {R.}~\bibnamefont {Trines}}, \bibinfo {author} {\bibfnamefont
  {T.}~\bibnamefont {Huang}}, \bibinfo {author} {\bibfnamefont
  {M.}~\bibnamefont {Skramic}}, \bibinfo {author} {\bibfnamefont
  {M.}~\bibnamefont {Vranic}},  \emph {et~al.},\ }\href@noop {} {\bibfield
  {journal} {\bibinfo  {journal} {New Journal of Physics}\ }\textbf {\bibinfo
  {volume} {18}},\ \bibinfo {pages} {053023} (\bibinfo {year}
  {2016})}\BibitemShut {NoStop}%
\end{thebibliography}%

\end{document}